%% file: paper.tex
\documentclass[sigplan,screen]{acmart}

\renewenvironment{acks}{%
  \makeatletter\if@ACM@anonymous\makeatother
  \else\makeatother 
  \begingroup
  \section*{Acknowledgments}
  \phantomsection\addcontentsline{toc}{section}{Acknowledgments}
}{%
  \endgroup
  \fi
}

\input{bib/macros}

\renewcommand\paragraph[1]{\subsubsection*{\em #1}}

\usepackage{stackengine} 

\usepackage{amsmath}
\usepackage{amsthm}
\usepackage{amssymb}
\usepackage{array}
\usepackage{changepage} 
\usepackage{mathtools} 
\usepackage{MnSymbol} 
\usepackage{arydshln} 

\newcolumntype{H}{>{\setbox0=\hbox\bgroup}c<{\egroup}@{}} 
\newcolumntype{Z}{>{\setbox0=\hbox\bgroup}c<{\egroup}@{\hspace*{-\tabcolsep}}} 

\usepackage{enumerate}
\usepackage{enumitem} 

\usepackage[utf8]{inputenc}
\usepackage[T1]{fontenc}
\usepackage{microtype} 
\theoremstyle{definition}
\newtheorem{definition}{Definition}
\newtheorem*{definition*}{Definition}
\newtheorem{inv}{Invariant}

\usepackage[font=small,labelfont=bf]{caption}

\usepackage{algorithm}
\usepackage[noend]{algpseudocode} 
\usepackage{multicol}
\algnewcommand{\LineComment}[1]{\hfill \(\triangleright\) #1} 
\makeatletter
\algnewcommand{\LineCommentx}[1]{\Statex \hskip\ALG@thistlm \(\triangleright\) #1}
\algnewcommand{\LineCommentxx}[1]{\Statex \hskip\ALG@tlm \(\triangleright\) #1}
\makeatother
\algnewcommand{\CaseComment}[1]{\hfill #1}
\algnewcommand{\lIf}[2] {\State \algorithmicif\ #1 \algorithmicthen\ #2} 
\algnewcommand{\lIfElse}[3] {\State \algorithmicif\ #1 \algorithmicthen\ #2 \algorithmicelse\ #3} 
\algnewcommand{\lElse}[1] {\State \algorithmicelse\ #1}
\algnewcommand{\lElsIf}[2] {\State \algorithmicelse\ \algorithmicif\ #1 \algorithmicthen\ #2}
\algnewcommand{\lForAll}[2]{\State \algorithmicforall\ #1 \algorithmicdo\ #2} 
\algnewcommand\algorithmicforeach{\textbf{foreach}} 
\algdef{S}[FOR]{ForEach}[1]{\algorithmicforeach\ #1\ \algorithmicdo}
\algnewcommand{\lForEach}[2] {\State \algorithmicforeach\ #1 \algorithmicdo\ #2} 

\usepackage{tikz}
\usetikzlibrary{decorations.pathmorphing}
\newcommand{\tikzmark}[1]{\tikz[remember picture, baseline] \node[inner sep=0pt, outer sep=0pt] (#1){};}
\newcommand{\link}[2]{%
\begin{tikzpicture}[remember picture, overlay, >=stealth, shift={(0,0)}] 
	\draw[arrows=->] (#1) to (#2);
\end{tikzpicture}}

\newcommand{\linkdash}[2]{%
\begin{tikzpicture}[dashed, remember picture, overlay, >=stealth, shift={(0,0)}] 
	\draw[arrows=->] (#1) to (#2);
\end{tikzpicture}}
\newcommand{\undertextlinkdash}[3]{%
\begin{tikzpicture}[dashed, remember picture, overlay, >=stealth, shift={(0,0)}]
	\draw[arrows=->] (#1) to node[sloped,anchor=center,below] {{\smaller #3}} (#2);
\end{tikzpicture}}
\newcommand{\textlink}[3]{%
\begin{tikzpicture}[remember picture, overlay, >=stealth, shift={(0,0)}] 
	\draw[arrows=->] (#1) to node[sloped,anchor=center,above] {{\smaller #3}} (#2);
\end{tikzpicture}}
\newcommand{\undertextlink}[3]{%
\begin{tikzpicture}[remember picture, overlay, >=stealth, shift={(0,0)}] 
	\draw[arrows=->] (#1) to node[sloped,anchor=center,below] {{\smaller #3}} (#2);
\end{tikzpicture}}
\newcommand{\undertextcurve}[5]{%
\begin{tikzpicture}[remember picture, overlay, >=stealth, shift={(0,0)}] 
	\draw[arrows=->] (#1) to [out=#3,in=#4] node[sloped,anchor=center,below] {{\smaller #5}} (#2); 
\end{tikzpicture}}

\newcommand{\num}[1]{}

\usepackage{xcolor} 
\usetikzlibrary{fit,calc}
\newcommand*{\tikzhighlightmk}[1]{\tikz[remember picture,overlay,] \node (#1) {};\ignorespaces}
\newcommand{\boxit}[1]{\tikz[remember picture,overlay]{\node[yshift=-2pt,xshift=-3pt,fill=#1,opacity=.25,fit={($(A)+(-0.02\textwidth,.8\baselineskip)$)($(B)+(0,.4\baselineskip)$)}] {};}\ignorespaces}
\newcommand{\boxittaller}[1]{\tikz[remember picture,overlay]{\node[yshift=-2pt,xshift=-3pt,fill=#1,opacity=.25,fit={($(A)+(-0.02\textwidth,1.3\baselineskip)$)($(B)+(0,.4\baselineskip)$)}] {};}\ignorespaces}
\newcommand{\boxforeachit}[1]{\tikz[remember picture,overlay]{\node[yshift=-2pt,xshift=-3pt,fill=#1,opacity=.25,fit={($(A)+(-0.07\textwidth,.8\baselineskip)$)($(B)+(0,.4\baselineskip)$)}] {};}\ignorespaces}
\colorlet{pink}{red!40}
\colorlet{grey}{black!60}

\newcommand{\smalltablerowheight}{0.7}

\newcommand{\Rule}[1]{rule~(#1)\xspace}
\newcommand{\CS}[1]{\textcolor{red}{\ensuremath{\mathit{CS(}#1)}}}
\newcommand{\conflicts}[2]{\ensuremath{#1 \asymp #2}} 

\newcommand{\event}[2]{\ensuremath{#1#2}\xspace}
\newcommand{\race}[1]{#1-race} 
\newcommand{\races}[1]{#1-races}

\newcommand{\conflictingEvents}{conflicting events\xspace}

\newcommand{\A}[1]{\ensuremath{A_x^{#1}}} 

\newcommand{\PredTrace}{Predicted Trace\xspace}

\newcommand{\predtrace}{predicted trace\xspace}

\newcommand{\epoch}[2]{\ensuremath{#1@#2}\xspace}

\newcommand{\ltpartial}{\ensuremath{\sqsubseteq}\xspace}

\newcommand{\tr}{\ensuremath{\mathit{tr}}\xspace}
\newcommand{\trPrime}{\ensuremath{\mathit{tr'}}\xspace}
\newcommand{\PO}{PO\xspace}

\newcommand{\PoFull}{Program-order\xspace}
\newcommand{\poFull}{program-order\xspace}
\newcommand{\HB}{HB\xspace}
\newcommand{\HBFull}{Happens-Before\xspace}
\newcommand{\HbFull}{Happens-before\xspace}

\newcommand{\FT}{FT\xspace}
\newcommand{\FTFull}{FastTrack\xspace}
\newcommand{\CP}{CP\xspace}

\newcommand{\cpFull}{causally-precedes\xspace}
\newcommand{\WCP}{WCP\xspace}

\newcommand{\wcpFull}{weak-causally-precedes\xspace}
\newcommand{\DC}{DC\xspace}

\newcommand{\DcFull}{Doesn't-commute\xspace}
\newcommand{\dcFull}{doesn't-commute\xspace}
\newcommand{\CAPO}{WDC\xspace}
\newcommand{\CAPOFull}{Weak-Doesn't-Commute\xspace}

\newcommand{\capoFull}{weak-doesn't-commute\xspace}

\newcommand{\REAbbrv}{ST\xspace}
\newcommand{\RE}{SmartTrack\xspace}
\newcommand{\REFull}{\RE}
\newcommand{\ST}{ST\xspace}
\newcommand{\STFull}{SmartTrack\xspace}
\newcommand{\FTOFull}{FastTrack-Ownership\xspace}

\newcommand{\FTTwoAbbrv}{FT2\xspace}
\newcommand{\FTTwo}{FastTrack2\xspace}

\newcommand{\REWCP}{\RE-\WCP}
\newcommand{\REDC}{\RE-\DC}
\newcommand{\RECAPO}{\RE-\CAPO}

\newcommand{\FTO}{FTO\xspace}
\newcommand{\HBO}{\FTO-\HB}
\newcommand{\WCPO}{\FTO-\WCP}
\newcommand{\DCO}{\FTO-\DC}
\newcommand{\CAPOO}{\FTO-\CAPO}

\newcommand{\VCHB}{\HB}
\newcommand{\VCWCP}{\WCP}
\newcommand{\VCDC}{\DC}
\newcommand{\VCCAPO}{\CAPO}

\newcommand{\initA}{\ensuremath{\bot}\xspace}

\newcommand{\factor}[1]{#1\ensuremath{\;\!\times}}

\newcommand{\case}[1]{\textsc{\small [#1]}}
\newcommand{\ReadSameEpoch}{\case{Read Same Epoch}\xspace}
\newcommand{\WriteSameEpoch}{\case{Write Same Epoch}\xspace}
\newcommand{\SharedSameEpoch}{\case{Shared Same Epoch}\xspace}
\newcommand{\ReadOwned}{\case{Read Owned}\xspace}
\newcommand{\ReadExclusive}{\case{Read Exclusive}\xspace}
\newcommand{\ReadShare}{\case{Read Share}\xspace}
\newcommand{\ReadSharedOwned}{\case{Read Shared Owned}\xspace}
\newcommand{\ReadShared}{\case{Read Shared}\xspace}
\newcommand{\WriteOwned}{\case{Write Owned}\xspace}
\newcommand{\WriteExclusive}{\case{Write Exclusive}\xspace}
\newcommand{\WriteShared}{\case{Write Shared}\xspace}

\newcommand{\epLeq}{\ensuremath{\preceq}\xspace}
\newcommand{\epochLeqVC}[2]{\ensuremath{#1 \epLeq #2}}
\newcommand{\epochLeqVCThr}[3]{\ensuremath{#1 \epLeq #2}}

\newcommand{\ltTR}{\ensuremath{<_\textsc{\tr}}\xspace}

\newcommand{\ltPO}{\ensuremath{\prec_\textsc{\tiny{\PO}}}\xspace}

\newcommand{\ltHB}{\ensuremath{\prec_\textsc{\tiny{\HB}}}\xspace}

\newcommand{\ltWCP}{\ensuremath{\prec_\textsc{\tiny{\WCP}}}\xspace}
\newcommand{\nltWCP}{\ensuremath{\not\prec_\textsc{\tiny{\WCP}}}\xspace}
\newcommand{\ltDC}{\ensuremath{\prec_\textsc{\tiny{\DC}}}\xspace}

\newcommand{\nltDC}{\ensuremath{\not\prec_\textsc{\tiny{\DC}}}\xspace}

\newcommand{\nltCAPO}{\ensuremath{\not\prec_\textsc{\tiny{\CAPO}}}\xspace}

\newcommand{\Ordered}[3]{\ensuremath{#1 #2 #3}}
\newcommand{\TROrdered}[2]{\Ordered{#1}{\ltTR}{#2}}

\newcommand{\POOrdered}[2]{\Ordered{#1}{\ltPO}{#2}}

\newcommand{\HBOrdered}[2]{\Ordered{#1}{\ltHB}{#2}}

\newcommand{\WCPOrdered}[2]{\Ordered{#1}{\ltWCP}{#2}}
\newcommand{\nWCPOrdered}[2]{\Ordered{#1}{\nltWCP}{#2}}
\newcommand{\DCOrdered}[2]{\Ordered{#1}{\ltDC}{#2}}

\newcommand{\nDCOrdered}[2]{\Ordered{#1}{\nltDC}{#2}}

\newcommand{\nCAPOOrdered}[2]{\Ordered{#1}{\nltCAPO}{#2}}

\newcommand{\Write}[1]{\ensuremath{\code{wr(#1)}}}
\newcommand{\Read}[1]{\ensuremath{\code{rd(#1)}}}
\newcommand{\Acquire}[1]{\ensuremath{\code{acq(#1)}}}
\newcommand{\Release}[1]{\ensuremath{\code{rel(#1)}}}
\newcommand{\Sync}[1]{\ensuremath{\code{sync(#1)}}}

\newcommand\AcqT[2]{\Acquire{#1}\ensuremath{^\thr{#2}}}
\newcommand\RelT[2]{\Release{#1}\ensuremath{^\thr{#2}}}
\newcommand\ReadT[2]{\Read{#1}\ensuremath{^\thr{#2}}}
\newcommand\WriteT[2]{\Write{#1}\ensuremath{^\thr{#2}}}

\newcommand{\AcqMQ}[3]{\ensuremath{\mathit{Acq}_{#1,#2}(#3)}\xspace}
\newcommand{\RelMQ}[3]{\ensuremath{\mathit{Rel}_{#1,#2}(#3)}\xspace}
\newcommand{\LockVarQ}[3]{\ensuremath{L_{#1,#2}^{#3}}\xspace}
\newcommand{\LockVarQC}[3]{\ensuremath{L_{\code{#1},\code{#2}}^{#3}}\xspace}

\newcommand{\ThreadVC}[1]{\ensuremath{C_\thr{#1}}\xspace}
\newcommand{\WrVar}[1]{\ensuremath{W_\code{#1}}\xspace}
\newcommand{\RdVar}[1]{\ensuremath{R_\code{#1}}\xspace}

\newcommand{\VCExample}[3]{\ensuremath{\textnormal{<}#1,#2,#3\textnormal{>}}\xspace}
\newcommand{\VarCSList}[2]{\ensuremath{L_\code{#1}^{#2}}\xspace}
\newcommand{\ThrCSList}[1]{\ensuremath{H_\thr{#1}}\xspace}
\newcommand{\VCShort}[3]{\ensuremath{#1_\code{#2}^\thr{#3}}\xspace}
\newcommand{\CSListSet}[1]{\ensuremath{\langle #1 \rangle}\xspace}

\newcommand\notes[1]{\begin{quote}\textcolor{darkgreen}{\textbackslash \textbf{notes\{}} #1 \textcolor{darkgreen}{\}}\end{quote}}

\newcommand\later[1]{\begin{quote}\textcolor{darkgreen}{\textbackslash \textbf{later\{}} #1 \textcolor{darkgreen}{\}}\end{quote}} 

\newtoggle{includeRaceResults}
\newtoggle{twoColumnText}
\newtoggle{techReport}
\input{toggle-comments}
\begin{document}

\setcopyright{acmlicensed}
\acmPrice{15.00}
\acmDOI{10.1145/3385412.3385993}
\acmYear{2020}
\copyrightyear{2020}
\acmSubmissionID{pldi20main-p228-p}
\acmISBN{978-1-4503-7613-6/20/06}
\acmConference[PLDI '20]{Proceedings of the 41st ACM SIGPLAN International Conference on Programming Language Design and Implementation}{June 15--20, 2020}{London, UK}
\acmBooktitle{Proceedings of the 41st ACM SIGPLAN International Conference on Programming Language Design and Implementation (PLDI '20), June 15--20, 2020, London, UK}



\title{\RE: Efficient Predictive Race Detection}

\author{Jake Roemer}
\affiliation{
	\institution{Ohio State University}
	\country{USA}
}
\email{roemer.37@osu.edu}

\author{Kaan Gen\c{c}}
\affiliation{
	\institution{Ohio State University}
	\country{USA}
}
\email{genc.5@osu.edu}

\author{Michael D. Bond}
\affiliation{
	\institution{Ohio State University}
	\country{USA}
}
\email{mikebond@cse.ohio-state.edu}

\iftoggle{techReport}{
  \begin{teaserfigure}
  \begin{center}
  \vspace*{-1em}
  \framebox{\Large Extended arXiv version of PLDI 2020 paper (adds Appendices~\ref{appendix:detailed-performance-results}-\ref{appendix:smarttrack-stats})}
  \vspace*{1em}
  \end{center}
  \end{teaserfigure}
}{}

\input{abstract}

\maketitle

\input{intro}

\input{background}

\input{capo}

\input{algorithm}
\input{evaluation}

\input{relatedwork}

\input{conclusion}

\input{acks}

\iftoggle{techReport}{}{\balance}

\newcommand{\showDOI}[1]{\unskip}
\bibliography{bib/conf-abbrv,bib/plass}

\iftoggle{techReport}{
\appendix

\notes{
\input{correctness}
}

\balance

\input{ci-results}

\input{re-stats-table}
}{}

\end{document}

%% file: bib/macros.tex
\makeatletter
\@ifpackageloaded{subfig}
{}
{\usepackage[labelformat=simple,subrefformat=parens,caption=false]{subfig}
 
 }
\makeatother
\usepackage{graphicx}

\usepackage{color}
\usepackage{listings}
\usepackage{relsize}
\usepackage{multirow}
\usepackage[normalem]{ulem} 

\usepackage{xspace}

\newcommand{\originalgrumbler}[2]{\begin{quote}\textcolor{blue}{\sl{\bf #1 says:} #2}\end{quote}}
\newcommand{\grumbler}[2]{\originalgrumbler{#1}{#2}}

\newcommand{\mike}[1]{\grumbler{Mike}{#1}}

\newcommand{\jake}[1]{\grumbler{Jake}{#1}}

\definecolor{darkgreen}{rgb}{0,0.4,0}

\newcommand{\sfsmaller}{}

\newcommand{\bench}[1]{\textsf{\sfsmaller#1}}
\newcommand{\code}[1]{\textsf{\sfsmaller#1}}

\newcommand{\mc}[3]{\multicolumn{#1}{#2}{#3}}

\newcommand{\eg}{e.g.\xspace}
\newcommand{\ie}{i.e.\xspace}
\newcommand{\cf}{cf.\xspace}
\newcommand{\etal}{et al.\xspace}

\newcommand{\thr}[1]{\textsf{\sfsmaller#1}}

\makeatletter
\@ifundefined{state}{%
}{}
\makeatother

\usepackage{etoolbox}



%% file: toggle-comments.tex
\renewcommand{\grumbler}[2]{}
\renewcommand{\notes}[1]{}
\renewcommand{\later}[1]{}

\togglefalse{includeRaceResults}

\toggletrue{techReport}

\toggletrue{twoColumnText}


%% file: abstract.tex
\begin{abstract}


Widely used data race detectors,
including the state-of-the-art \emph{FastTrack} algorithm,
incur performance costs that are acceptable for regular in-house testing,
but miss races detectable from the analyzed execution.
\emph{Predictive analyses} detect more data races in an analyzed execution than FastTrack detects,
but at significantly higher performance cost.

This paper presents \emph{\REFull}, an algorithm that
optimizes predictive race detection analyses, including two analyses from prior work
and a new analysis introduced in this paper.
SmartTrack incorporates two main optimizations:
(1) \emph{epoch and ownership optimizations} from prior work, applied to predictive analysis for the first time, and
(2) novel \emph{conflicting critical section optimizations} introduced by this paper.
%
Our evaluation
shows that \REFull
achieves performance competitive with FastTrack---a
qualitative improvement in the state of the art for data race detection.

\end{abstract}

%% file: intro.tex
\section{Introduction}


Data races are common concurrent programming errors that lead to crashes, hangs, and data
corruption~\cite{boehm-miscompile-hotpar-11, portend-toplas15,
conc-bug-study-2008, portend-asplos12, benign-races-2007, prescient-memory,
adversarial-memory, racefuzzer, relaxer},
incurring significant monetary and human costs~\cite{cost-of-software-errors,blackout-2003-tr,therac-25,nasdaq-facebook}.
Data races also cause shared-memory programs to have weak or undefined semantics~\cite{java-memory-model,c++-memory-model-2008,memory-models-cacm-2010}.

Data races are hard to detect. They occur nondeterministically under specific thread
interleavings, program inputs, and execution environments, and can stay hidden
even for extensively tested programs~\cite{blackout-2003-tr,zhou-hard,microsoft-exploratory-survey,conc-bug-study-2008}.
The prevailing approach for detecting data races is to use dynamic analysis---usually
\emph{happens-before (\HB) analysis}~\cite{happens-before}---during in-house testing.
\emph{FastTrack}~\cite{fasttrack} is a state-of-the-art algorithm for \HB analysis
that is implemented by commercial detectors~\cite{google-tsan-v1,google-tsan-v2,intel-inspector}.
However, \HB analysis misses races that are detectable in the observed execution (Section~\ref{sec:background}).


To detect more races than \HB analysis detects,
researchers have developed dynamic
\emph{predictive analyses}~\cite{causally-precedes,rvpredict-pldi-2014,jpredictor,maximal-causal-models,rdit-oopsla-2016,said-nfm-2011,ipa,wcp,vindicator,dighr,pavlogiannis-2019,depaware}.
SMT-based predictive analyses are powerful but fail to scale beyond bounded windows of execution (Section~\ref{Sec:related}).
In contrast, recently introduced \emph{partial-order-based} predictive analyses scale to full program executions.
Notably, \emph{\wcpFull (WCP)} and \emph{\dcFull (\DC)} analyses detect more races than \HB analysis~\cite{wcp,vindicator},
but they are substantially slower than FastTrack-optimized \HB analysis:
\factor{27--50} vs.\ \factor{6--8} according to prior work~\cite{vindicator,wcp,fasttrack,fasttrack2} and our evaluation (Section~\ref{sec:eval}).

\begin{table*}[t]
\small
\begin{tabular}{@{}ll@{\;}ll@{}}
Source of poor performance & Contribution & & Speedup \\\hline
Release--release ordering & \CAPO relation and analysis & & 1.04--\factor{1.25} \\
Frequent vector clock operations & Epoch and ownership optimizations & \multirow{2}{*}{\Big\} SmartTrack} & 2.15--\factor{2.62} \\
Detecting conflicting critical sections (CCSs) & CCS optimizations & & 1.51--\factor{1.74} \\
\end{tabular}
\caption{Sources of poor performance for existing partial-order-based predictive analyses
(\WCP and \DC~\cite{wcp,vindicator}),
and corresponding solutions introduced in this paper.
Speedups associated with each solution are the geomean across evaluated programs.
The second and third optimizations constitute this paper's \emph{\RE} algorithm,
with speedups ranging across predictive analyses.
\emph{\CAPO} is this paper's \emph{\capoFull},
with speedups ranging across multiple optimization levels.}
\label{tab:contributions}
\end{table*}


Why are the \WCP and \DC predictive analyses significantly slower than FastTrack-optimized \HB analysis?
Can FastTrack's optimizations be applied to \emph{predictive} analyses to achieve significant speedups?
In a nutshell, as we show, FastTrack's optimizations \emph{can} be applied to predictive analyses,
but there still remains a significant performance gap between predictive and \HB analyses.
This gap exists because predictive partial orders such as \WCP and \DC are inherently more complex than
\HB. Chiefly, predictive partial orders, in contrast with \HB, order
critical sections on the same lock only if they contain conflicting accesses,\footnote{Conflicting accesses
are accesses to the same variable by different threads such that at least one is a write.}
which we call \emph{conflicting critical sections (CCSs)}.
In addition, \WCP and \DC order releases of the same lock
if any part of their critical sections are ordered with each other.
These sources of predictive analysis complexity---especially detecting CCSs---present nontrivial performance challenges
with non-obvious solutions.

\paragraph{Contributions.}

This paper introduces novel contributions that enable predictive analysis
to perform competitively with optimized \HB analysis.
Table~\ref{tab:contributions} summarizes our contributions,
in the same order that Sections~\ref{sec:CAPO}--\ref{sec:analysis} present them.
Our principal technical contribution is \emph{conflicting critical section (CCS) optimizations}
(last row of Table~\ref{tab:contributions}).
These CCS optimizations introduce novel analysis state
and techniques to avoid computing redundant CCS ordering.
A novel but smaller contribution is a new predictive analysis, \emph{\capoFull (\CAPO) analysis} (first row),
that elides release--release ordering from \DC analysis,
a strength--complexity tradeoff that proves worthwhile in practice.
In addition, this work applies FastTrack's \emph{epoch and ownership optimizations} (middle row) to predictive analysis for the first time.

The CCS optimizations and epoch and ownership optimizations together constitute the new \emph{\RE} algorithm,
which applies to the \WCP, \DC, and \CAPO predictive analyses.

\begin{table}[t]
\newcommand{\roh}[1]{\ifthenelse{\equal{#1}{\rna}}{\rna}{#1$\;\!\times$}} 
\newcommand{\rna}{N/A}
\newcommand{\mem}[1]{\ifthenelse{\equal{#1}{\memna}}{\memna}{#1$\;\!\times$}} 
\newcommand{\memna}{N/A} 
\input{result-macros/PIP_slowTool_noCoresSet}

\input{result-macros/PIP_fastTool_extraOpt2Quiet}

\small
\centering
%

\newcommand\sep{\;\;\hfill}
\begin{tabular}{@{}l@{\;\;}l|l|l@{}}
      & & Prior work & This work \\\hline
\mc{2}{@{}l|}{Non-predictive analysis: \HB} & \roh{\FASTFTOHBTimeGeoMean} \sep (\roh{\FASTFTOHBMemGeoMean}) & \mc{1}{c}{---} \\\hline
\multirow{3}{*}{Predictive analysis \; \Bigg\{} & \WCP  & \roh{\SLOWWCPTimeGeoMean} \sep (\roh{\SLOWWCPMemGeoMean}) & \roh{\FASTREWCPTimeGeoMean} \sep (\roh{\FASTREWCPMemGeoMean}) \\
& \DC   & \roh{\SLOWDCnoGExcTimeGeoMean} \sep (\roh{\SLOWDCnoGExcMemGeoMean}) & \roh{\FASTREDCTimeGeoMean} \sep (\roh{\FASTREDCMemGeoMean}) \\
& \CAPO & \mc{1}{c|}{---} & \roh{\FASTRECAPOTimeGeoMean} \sep (\roh{\FASTRECAPOMemGeoMean}) \\
\end{tabular}


\caption{Slowdowns and (in parentheses) relative memory usage over native execution,
for state-of-the-art analyses without and with this paper's contributions.
Lower is better.
Each value is the geomean across the evaluated programs.}
\label{tab:performance:headline}
\end{table}


This paper's contributions, evaluated on large, real Java programs,
improve the performance of predictive analyses substantially,
as Table~\ref{tab:performance:headline} summarizes (based on Section~\ref{sec:eval}'s results).
The \emph{Predictive analysis} rows show
that SmartTrack optimizations substantially improve the
performance of predictive analyses compared with prior work.
\later{reducing execution time by \factor{3.3--4.1} and
reducing memory usage by \factor{4.2--6.4} compared with prior work's unoptimized predictive analyses.}%
The last row shows that the new \CAPO analysis is cheaper than prior predictive analyses.
Furthermore,
the table shows that the optimized predictive analyses perform nearly as well as high-performance \HB analysis.

Predictive analysis thus not only finds more races than \HB analysis for an observed execution,
but this paper shows how predictive analysis can close the performance gap with \HB analysis.
This result suggests the potential for using predictive analysis instead of \HB analysis
as the prevailing approach for detecting data races.

%% file: result-macros/PIP_slowTool_noCoresSet.tex
\newcommand{\SLOWavroraEvents}{0}
\newcommand{\SLOWavroraNoFPEvents}{0}
\newcommand{\SLOWavroraMaxLiveThreads}{7}
\newcommand{\SLOWavroraTotalThreads}{7}
\newcommand{\SLOWavroraBaseTime}{2.4}
\newcommand{\SLOWavroraBaseTimeCI}{16}
\newcommand{\SLOWavroraEmptyTime}{\rna}
\newcommand{\SLOWavroraEmptyTimeCI}{\rna}
\newcommand{\SLOWavroraEmptyTimeCIMIN}{\rna}
\newcommand{\SLOWavroraEmptyTimeCIMAX}{\rna}
\newcommand{\SLOWavroraFTTime}{7.0}
\newcommand{\SLOWavroraFTTimeCI}{0.028}
\newcommand{\SLOWavroraHBTime}{17}
\newcommand{\SLOWavroraHBTimeCI}{1.2}
\newcommand{\SLOWavroraWCPTime}{23}
\newcommand{\SLOWavroraWCPTimeCI}{0.54}
\newcommand{\SLOWavroraDCnoGExcTime}{21}
\newcommand{\SLOWavroraDCnoGExcTimeCI}{0.23}
\newcommand{\SLOWavroraDCnoGTime}{\rna}
\newcommand{\SLOWavroraDCnoGTimeCI}{\rna}
\newcommand{\SLOWavroraDCnoGTimeCIMIN}{\rna}
\newcommand{\SLOWavroraDCnoGTimeCIMAX}{\rna}
\newcommand{\SLOWavroraDCExcTime}{24}
\newcommand{\SLOWavroraDCExcTimeCI}{0.66}
\newcommand{\SLOWavroraDCTime}{\rna}
\newcommand{\SLOWavroraDCTimeCI}{\rna}
\newcommand{\SLOWavroraDCTimeCIMIN}{\rna}
\newcommand{\SLOWavroraDCTimeCIMAX}{\rna}
\newcommand{\SLOWavroraCAPOnoGExcTime}{18}
\newcommand{\SLOWavroraCAPOnoGExcTimeCI}{0.13}
\newcommand{\SLOWavroraCAPOnoGTime}{\rna}
\newcommand{\SLOWavroraCAPOnoGTimeCI}{\rna}
\newcommand{\SLOWavroraCAPOnoGTimeCIMIN}{\rna}
\newcommand{\SLOWavroraCAPOnoGTimeCIMAX}{\rna}
\newcommand{\SLOWavroraCAPOExcTime}{22}
\newcommand{\SLOWavroraCAPOExcTimeCI}{2.0}
\newcommand{\SLOWavroraCAPOTime}{\rna}
\newcommand{\SLOWavroraCAPOTimeCI}{\rna}
\newcommand{\SLOWavroraCAPOTimeCIMIN}{\rna}
\newcommand{\SLOWavroraCAPOTimeCIMAX}{\rna}
\newcommand{\SLOWavroraStaticTime}{\rzero}
\newcommand{\SLOWavroraDynamicTime}{\rzero}
\newcommand{\SLOWavroraBaseMem}{150}
\newcommand{\SLOWavroraBaseMemCI}{0.31}
\newcommand{\SLOWavroraHBMem}{32}
\newcommand{\SLOWavroraHBMemCI}{0.021}
\newcommand{\SLOWavroraFTMem}{22}
\newcommand{\SLOWavroraFTMemCI}{13.0}
\newcommand{\SLOWavroraWCPMem}{99}
\newcommand{\SLOWavroraWCPMemCI}{0.60}
\newcommand{\SLOWavroraDCnoGExcMem}{42}
\newcommand{\SLOWavroraDCnoGExcMemCI}{0.13}
\newcommand{\SLOWavroraDCnoGMem}{\memna}
\newcommand{\SLOWavroraDCnoGMemCI}{\memna}
\newcommand{\SLOWavroraDCnoGMemCIMIN}{\memna}
\newcommand{\SLOWavroraDCnoGMemCIMAX}{\memna}
\newcommand{\SLOWavroraDCExcMem}{72}
\newcommand{\SLOWavroraDCExcMemCI}{0.060}
\newcommand{\SLOWavroraDCMem}{\memna}
\newcommand{\SLOWavroraDCMemCI}{\memna}
\newcommand{\SLOWavroraDCMemCIMIN}{\memna}
\newcommand{\SLOWavroraDCMemCIMAX}{\memna}
\newcommand{\SLOWavroraCAPOnoGExcMem}{37}
\newcommand{\SLOWavroraCAPOnoGExcMemCI}{3.9}
\newcommand{\SLOWavroraCAPOnoGMem}{\memna}
\newcommand{\SLOWavroraCAPOnoGMemCI}{\memna}
\newcommand{\SLOWavroraCAPOnoGMemCIMIN}{\memna}
\newcommand{\SLOWavroraCAPOnoGMemCIMAX}{\memna}
\newcommand{\SLOWavroraCAPOExcMem}{72}
\newcommand{\SLOWavroraCAPOExcMemCI}{0.35}
\newcommand{\SLOWavroraCAPOMem}{\memna}
\newcommand{\SLOWavroraCAPOMemCI}{\memna}
\newcommand{\SLOWavroraCAPOMemCIMIN}{\memna}
\newcommand{\SLOWavroraCAPOMemCIMAX}{\memna}
\newcommand{\SLOWavroraEventsCI}{0}
\newcommand{\SLOWavroraEventsCIMIN}{0}
\newcommand{\SLOWavroraEventsCIMAX}{0}
\newcommand{\SLOWavroraNoFPEventsCI}{0}
\newcommand{\SLOWavroraNoFPEventsCIMIN}{0}
\newcommand{\SLOWavroraNoFPEventsCIMAX}{0}
\newcommand{\SLOWavroraHB}{6}
\newcommand{\SLOWavroraHBCI}{0.0}
\newcommand{\SLOWavroraHBCIMIN}{6}
\newcommand{\SLOWavroraHBCIMAX}{6}
\newcommand{\SLOWavroraHBDynamic}{426,723}
\newcommand{\SLOWavroraHBDynamicCI}{7,794}
\newcommand{\SLOWavroraHBDynamicCIMIN}{418,929}
\newcommand{\SLOWavroraHBDynamicCIMAX}{434,517}
\newcommand{\SLOWavroraFT}{3}
\newcommand{\SLOWavroraFTCI}{0.0}
\newcommand{\SLOWavroraFTCIMIN}{3}
\newcommand{\SLOWavroraFTCIMAX}{3}
\newcommand{\SLOWavroraFTDynamic}{754,302}
\newcommand{\SLOWavroraFTDynamicCI}{9,248}
\newcommand{\SLOWavroraFTDynamicCIMIN}{745,054}
\newcommand{\SLOWavroraFTDynamicCIMAX}{763,550}
\newcommand{\SLOWavroraWCP}{6}
\newcommand{\SLOWavroraWCPCI}{0.0}
\newcommand{\SLOWavroraWCPCIMIN}{6}
\newcommand{\SLOWavroraWCPCIMAX}{6}
\newcommand{\SLOWavroraWCPDynamic}{423,373}
\newcommand{\SLOWavroraWCPDynamicCI}{432}
\newcommand{\SLOWavroraWCPDynamicCIMIN}{422,941}
\newcommand{\SLOWavroraWCPDynamicCIMAX}{423,805}
\newcommand{\SLOWavroraDCnoGExc}{6}
\newcommand{\SLOWavroraDCnoGExcCI}{0.0}
\newcommand{\SLOWavroraDCnoGExcCIMIN}{6}
\newcommand{\SLOWavroraDCnoGExcCIMAX}{6}
\newcommand{\SLOWavroraDCnoGExcDynamic}{443,731}
\newcommand{\SLOWavroraDCnoGExcDynamicCI}{1,117}
\newcommand{\SLOWavroraDCnoGExcDynamicCIMIN}{442,614}
\newcommand{\SLOWavroraDCnoGExcDynamicCIMAX}{444,848}
\newcommand{\SLOWavroraDCnoG}{\rna}
\newcommand{\SLOWavroraDCnoGCI}{\rna}
\newcommand{\SLOWavroraDCnoGCIMIN}{\rna}
\newcommand{\SLOWavroraDCnoGCIMAX}{\rna}
\newcommand{\SLOWavroraDCnoGDynamic}{\rna}
\newcommand{\SLOWavroraDCnoGDynamicCI}{\rna}
\newcommand{\SLOWavroraDCnoGDynamicCIMIN}{\rna}
\newcommand{\SLOWavroraDCnoGDynamicCIMAX}{\rna}
\newcommand{\SLOWavroraDCExc}{6}
\newcommand{\SLOWavroraDCExcCI}{0.0}
\newcommand{\SLOWavroraDCExcCIMIN}{6}
\newcommand{\SLOWavroraDCExcCIMAX}{6}
\newcommand{\SLOWavroraDCExcDynamic}{153,338}
\newcommand{\SLOWavroraDCExcDynamicCI}{6,918}
\newcommand{\SLOWavroraDCExcDynamicCIMIN}{146,420}
\newcommand{\SLOWavroraDCExcDynamicCIMAX}{160,256}
\newcommand{\SLOWavroraDC}{\rna}
\newcommand{\SLOWavroraDCCI}{\rna}
\newcommand{\SLOWavroraDCCIMIN}{\rna}
\newcommand{\SLOWavroraDCCIMAX}{\rna}
\newcommand{\SLOWavroraDCDynamic}{\rna}
\newcommand{\SLOWavroraDCDynamicCI}{\rna}
\newcommand{\SLOWavroraDCDynamicCIMIN}{\rna}
\newcommand{\SLOWavroraDCDynamicCIMAX}{\rna}
\newcommand{\SLOWavroraCAPOnoGExc}{6}
\newcommand{\SLOWavroraCAPOnoGExcCI}{0.0}
\newcommand{\SLOWavroraCAPOnoGExcCIMIN}{6}
\newcommand{\SLOWavroraCAPOnoGExcCIMAX}{6}
\newcommand{\SLOWavroraCAPOnoGExcDynamic}{440,519}
\newcommand{\SLOWavroraCAPOnoGExcDynamicCI}{646}
\newcommand{\SLOWavroraCAPOnoGExcDynamicCIMIN}{439,873}
\newcommand{\SLOWavroraCAPOnoGExcDynamicCIMAX}{441,165}
\newcommand{\SLOWavroraCAPOnoG}{\rna}
\newcommand{\SLOWavroraCAPOnoGCI}{\rna}
\newcommand{\SLOWavroraCAPOnoGCIMIN}{\rna}
\newcommand{\SLOWavroraCAPOnoGCIMAX}{\rna}
\newcommand{\SLOWavroraCAPOnoGDynamic}{\rna}
\newcommand{\SLOWavroraCAPOnoGDynamicCI}{\rna}
\newcommand{\SLOWavroraCAPOnoGDynamicCIMIN}{\rna}
\newcommand{\SLOWavroraCAPOnoGDynamicCIMAX}{\rna}
\newcommand{\SLOWavroraCAPOExc}{6}
\newcommand{\SLOWavroraCAPOExcCI}{0.0}
\newcommand{\SLOWavroraCAPOExcCIMIN}{6}
\newcommand{\SLOWavroraCAPOExcCIMAX}{6}
\newcommand{\SLOWavroraCAPOExcDynamic}{221,622}
\newcommand{\SLOWavroraCAPOExcDynamicCI}{374}
\newcommand{\SLOWavroraCAPOExcDynamicCIMIN}{221,248}
\newcommand{\SLOWavroraCAPOExcDynamicCIMAX}{221,996}
\newcommand{\SLOWavroraCAPO}{\rna}
\newcommand{\SLOWavroraCAPOCI}{\rna}
\newcommand{\SLOWavroraCAPOCIMIN}{\rna}
\newcommand{\SLOWavroraCAPOCIMAX}{\rna}
\newcommand{\SLOWavroraCAPODynamic}{\rna}
\newcommand{\SLOWavroraCAPODynamicCI}{\rna}
\newcommand{\SLOWavroraCAPODynamicCIMIN}{\rna}
\newcommand{\SLOWavroraCAPODynamicCIMAX}{\rna}
\newcommand{\SLOWavroraPIP}{\rna}
\newcommand{\SLOWavroraPIPCI}{\rna}
\newcommand{\SLOWavroraPIPCIMIN}{\rna}
\newcommand{\SLOWavroraPIPCIMAX}{\rna}
\newcommand{\SLOWavroraPIPDynamic}{\rna}
\newcommand{\SLOWavroraPIPDynamicCI}{\rna}
\newcommand{\SLOWavroraPIPDynamicCIMIN}{\rna}
\newcommand{\SLOWavroraPIPDynamicCIMAX}{\rna}
\newcommand{\SLOWavroraHBUP}{3}
\newcommand{\SLOWavroraHBUPCI}{0.0}
\newcommand{\SLOWavroraHBUPCIMIN}{3}
\newcommand{\SLOWavroraHBUPCIMAX}{3}
\newcommand{\SLOWavroraHBDynamicUP}{426,723}
\newcommand{\SLOWavroraHBDynamicUPCI}{7,794}
\newcommand{\SLOWavroraHBDynamicUPCIMIN}{418,929}
\newcommand{\SLOWavroraHBDynamicUPCIMAX}{434,517}
\newcommand{\SLOWavroraWCPUP}{3}
\newcommand{\SLOWavroraWCPUPCI}{0.0}
\newcommand{\SLOWavroraWCPUPCIMIN}{3}
\newcommand{\SLOWavroraWCPUPCIMAX}{3}
\newcommand{\SLOWavroraWCPDynamicUP}{423,373}
\newcommand{\SLOWavroraWCPDynamicUPCI}{432}
\newcommand{\SLOWavroraWCPDynamicUPCIMIN}{422,941}
\newcommand{\SLOWavroraWCPDynamicUPCIMAX}{423,805}
\newcommand{\SLOWavroraWDCnoGUP}{\rna}
\newcommand{\SLOWavroraWDCnoGUPCI}{\rna}
\newcommand{\SLOWavroraWDCnoGUPCIMIN}{\rna}
\newcommand{\SLOWavroraWDCnoGUPCIMAX}{\rna}
\newcommand{\SLOWavroraWDCnoGDynamicUP}{\rna}
\newcommand{\SLOWavroraWDCnoGDynamicUPCI}{\rna}
\newcommand{\SLOWavroraWDCnoGDynamicUPCIMIN}{\rna}
\newcommand{\SLOWavroraWDCnoGDynamicUPCIMAX}{\rna}
\newcommand{\SLOWavroraWDCUP}{\rna}
\newcommand{\SLOWavroraWDCUPCI}{\rna}
\newcommand{\SLOWavroraWDCUPCIMIN}{\rna}
\newcommand{\SLOWavroraWDCUPCIMAX}{\rna}
\newcommand{\SLOWavroraWDCDynamicUP}{\rna}
\newcommand{\SLOWavroraWDCDynamicUPCI}{\rna}
\newcommand{\SLOWavroraWDCDynamicUPCIMIN}{\rna}
\newcommand{\SLOWavroraWDCDynamicUPCIMAX}{\rna}
\newcommand{\SLOWavroraCAPOnoGUP}{\rna}
\newcommand{\SLOWavroraCAPOnoGUPCI}{\rna}
\newcommand{\SLOWavroraCAPOnoGUPCIMIN}{\rna}
\newcommand{\SLOWavroraCAPOnoGUPCIMAX}{\rna}
\newcommand{\SLOWavroraCAPOnoGDynamicUP}{\rna}
\newcommand{\SLOWavroraCAPOnoGDynamicUPCI}{\rna}
\newcommand{\SLOWavroraCAPOnoGDynamicUPCIMIN}{\rna}
\newcommand{\SLOWavroraCAPOnoGDynamicUPCIMAX}{\rna}
\newcommand{\SLOWavroraCAPOUP}{\rna}
\newcommand{\SLOWavroraCAPOUPCI}{\rna}
\newcommand{\SLOWavroraCAPOUPCIMIN}{\rna}
\newcommand{\SLOWavroraCAPOUPCIMAX}{\rna}
\newcommand{\SLOWavroraCAPODynamicUP}{\rna}
\newcommand{\SLOWavroraCAPODynamicUPCI}{\rna}
\newcommand{\SLOWavroraCAPODynamicUPCIMIN}{\rna}
\newcommand{\SLOWavroraCAPODynamicUPCIMAX}{\rna}
\newcommand{\SLOWavroraPIPUP}{\rna}
\newcommand{\SLOWavroraPIPUPCI}{\rna}
\newcommand{\SLOWavroraPIPUPCIMIN}{\rna}
\newcommand{\SLOWavroraPIPUPCIMAX}{\rna}
\newcommand{\SLOWavroraPIPDynamicUP}{\rna}
\newcommand{\SLOWavroraPIPDynamicUPCI}{\rna}
\newcommand{\SLOWavroraPIPDynamicUPCIMIN}{\rna}
\newcommand{\SLOWavroraPIPDynamicUPCIMAX}{\rna}
\newcommand{\SLOWavroraPIPHB}{\rna}
\newcommand{\SLOWavroraPIPHBCI}{\rna}
\newcommand{\SLOWavroraPIPHBCIMIN}{\rna}
\newcommand{\SLOWavroraPIPHBCIMAX}{\rna}
\newcommand{\SLOWavroraPIPHBDynamic}{\rna}
\newcommand{\SLOWavroraPIPHBDynamicCI}{\rna}
\newcommand{\SLOWavroraPIPHBDynamicCIMIN}{\rna}
\newcommand{\SLOWavroraPIPHBDynamicCIMAX}{\rna}
\newcommand{\SLOWavroraPIPWCP}{\rna}
\newcommand{\SLOWavroraPIPWCPCI}{\rna}
\newcommand{\SLOWavroraPIPWCPCIMIN}{\rna}
\newcommand{\SLOWavroraPIPWCPCIMAX}{\rna}
\newcommand{\SLOWavroraPIPWCPDynamic}{\rna}
\newcommand{\SLOWavroraPIPWCPDynamicCI}{\rna}
\newcommand{\SLOWavroraPIPWCPDynamicCIMIN}{\rna}
\newcommand{\SLOWavroraPIPWCPDynamicCIMAX}{\rna}
\newcommand{\SLOWavroraPIPWDC}{\rna}
\newcommand{\SLOWavroraPIPWDCCI}{\rna}
\newcommand{\SLOWavroraPIPWDCCIMIN}{\rna}
\newcommand{\SLOWavroraPIPWDCCIMAX}{\rna}
\newcommand{\SLOWavroraPIPWDCDynamic}{\rna}
\newcommand{\SLOWavroraPIPWDCDynamicCI}{\rna}
\newcommand{\SLOWavroraPIPWDCDynamicCIMIN}{\rna}
\newcommand{\SLOWavroraPIPWDCDynamicCIMAX}{\rna}
\newcommand{\SLOWavroraPIPCAPO}{\rna}
\newcommand{\SLOWavroraPIPCAPOCI}{\rna}
\newcommand{\SLOWavroraPIPCAPOCIMIN}{\rna}
\newcommand{\SLOWavroraPIPCAPOCIMAX}{\rna}
\newcommand{\SLOWavroraPIPCAPODynamic}{\rna}
\newcommand{\SLOWavroraPIPCAPODynamicCI}{\rna}
\newcommand{\SLOWavroraPIPCAPODynamicCIMIN}{\rna}
\newcommand{\SLOWavroraPIPCAPODynamicCIMAX}{\rna}
\newcommand{\SLOWavroraPIPPIP}{\rna}
\newcommand{\SLOWavroraPIPPIPCI}{\rna}
\newcommand{\SLOWavroraPIPPIPCIMIN}{\rna}
\newcommand{\SLOWavroraPIPPIPCIMAX}{\rna}
\newcommand{\SLOWavroraPIPPIPDynamic}{\rna}
\newcommand{\SLOWavroraPIPPIPDynamicCI}{\rna}
\newcommand{\SLOWavroraPIPPIPDynamicCIMIN}{\rna}
\newcommand{\SLOWavroraPIPPIPDynamicCIMAX}{\rna}
\newcommand{\SLOWbatikEvents}{0}
\newcommand{\SLOWbatikNoFPEvents}{0}
\newcommand{\SLOWbatikMaxLiveThreads}{7}
\newcommand{\SLOWbatikTotalThreads}{7}
\newcommand{\SLOWbatikBaseTime}{2.6}
\newcommand{\SLOWbatikBaseTimeCI}{34}
\newcommand{\SLOWbatikEmptyTime}{\rna}
\newcommand{\SLOWbatikEmptyTimeCI}{\rna}
\newcommand{\SLOWbatikEmptyTimeCIMIN}{\rna}
\newcommand{\SLOWbatikEmptyTimeCIMAX}{\rna}
\newcommand{\SLOWbatikFTTime}{4.2}
\newcommand{\SLOWbatikFTTimeCI}{0.17}
\newcommand{\SLOWbatikHBTime}{7.8}
\newcommand{\SLOWbatikHBTimeCI}{0.068}
\newcommand{\SLOWbatikWCPTime}{12}
\newcommand{\SLOWbatikWCPTimeCI}{0.40}
\newcommand{\SLOWbatikDCnoGExcTime}{10}
\newcommand{\SLOWbatikDCnoGExcTimeCI}{0.41}
\newcommand{\SLOWbatikDCnoGTime}{\rna}
\newcommand{\SLOWbatikDCnoGTimeCI}{\rna}
\newcommand{\SLOWbatikDCnoGTimeCIMIN}{\rna}
\newcommand{\SLOWbatikDCnoGTimeCIMAX}{\rna}
\newcommand{\SLOWbatikDCExcTime}{11}
\newcommand{\SLOWbatikDCExcTimeCI}{0.55}
\newcommand{\SLOWbatikDCTime}{\rna}
\newcommand{\SLOWbatikDCTimeCI}{\rna}
\newcommand{\SLOWbatikDCTimeCIMIN}{\rna}
\newcommand{\SLOWbatikDCTimeCIMAX}{\rna}
\newcommand{\SLOWbatikCAPOnoGExcTime}{10}
\newcommand{\SLOWbatikCAPOnoGExcTimeCI}{0.37}
\newcommand{\SLOWbatikCAPOnoGTime}{\rna}
\newcommand{\SLOWbatikCAPOnoGTimeCI}{\rna}
\newcommand{\SLOWbatikCAPOnoGTimeCIMIN}{\rna}
\newcommand{\SLOWbatikCAPOnoGTimeCIMAX}{\rna}
\newcommand{\SLOWbatikCAPOExcTime}{11}
\newcommand{\SLOWbatikCAPOExcTimeCI}{0.46}
\newcommand{\SLOWbatikCAPOTime}{\rna}
\newcommand{\SLOWbatikCAPOTimeCI}{\rna}
\newcommand{\SLOWbatikCAPOTimeCIMIN}{\rna}
\newcommand{\SLOWbatikCAPOTimeCIMAX}{\rna}
\newcommand{\SLOWbatikStaticTime}{\rzero}
\newcommand{\SLOWbatikDynamicTime}{\rzero}
\newcommand{\SLOWbatikBaseMem}{220}
\newcommand{\SLOWbatikBaseMemCI}{2.5}
\newcommand{\SLOWbatikHBMem}{30}
\newcommand{\SLOWbatikHBMemCI}{0.53}
\newcommand{\SLOWbatikFTMem}{5.5}
\newcommand{\SLOWbatikFTMemCI}{0.15}
\newcommand{\SLOWbatikWCPMem}{46}
\newcommand{\SLOWbatikWCPMemCI}{0.48}
\newcommand{\SLOWbatikDCnoGExcMem}{43}
\newcommand{\SLOWbatikDCnoGExcMemCI}{10}
\newcommand{\SLOWbatikDCnoGMem}{\memna}
\newcommand{\SLOWbatikDCnoGMemCI}{\memna}
\newcommand{\SLOWbatikDCnoGMemCIMIN}{\memna}
\newcommand{\SLOWbatikDCnoGMemCIMAX}{\memna}
\newcommand{\SLOWbatikDCExcMem}{46}
\newcommand{\SLOWbatikDCExcMemCI}{3.2}
\newcommand{\SLOWbatikDCMem}{\memna}
\newcommand{\SLOWbatikDCMemCI}{\memna}
\newcommand{\SLOWbatikDCMemCIMIN}{\memna}
\newcommand{\SLOWbatikDCMemCIMAX}{\memna}
\newcommand{\SLOWbatikCAPOnoGExcMem}{42}
\newcommand{\SLOWbatikCAPOnoGExcMemCI}{9.7}
\newcommand{\SLOWbatikCAPOnoGMem}{\memna}
\newcommand{\SLOWbatikCAPOnoGMemCI}{\memna}
\newcommand{\SLOWbatikCAPOnoGMemCIMIN}{\memna}
\newcommand{\SLOWbatikCAPOnoGMemCIMAX}{\memna}
\newcommand{\SLOWbatikCAPOExcMem}{44}
\newcommand{\SLOWbatikCAPOExcMemCI}{2.1}
\newcommand{\SLOWbatikCAPOMem}{\memna}
\newcommand{\SLOWbatikCAPOMemCI}{\memna}
\newcommand{\SLOWbatikCAPOMemCIMIN}{\memna}
\newcommand{\SLOWbatikCAPOMemCIMAX}{\memna}
\newcommand{\SLOWbatikEventsCI}{0}
\newcommand{\SLOWbatikEventsCIMIN}{0}
\newcommand{\SLOWbatikEventsCIMAX}{0}
\newcommand{\SLOWbatikNoFPEventsCI}{0}
\newcommand{\SLOWbatikNoFPEventsCIMIN}{0}
\newcommand{\SLOWbatikNoFPEventsCIMAX}{0}
\newcommand{\SLOWbatikHB}{0}
\newcommand{\SLOWbatikHBCI}{0.0}
\newcommand{\SLOWbatikHBCIMIN}{0}
\newcommand{\SLOWbatikHBCIMAX}{0}
\newcommand{\SLOWbatikHBDynamic}{0}
\newcommand{\SLOWbatikHBDynamicCI}{0.0}
\newcommand{\SLOWbatikHBDynamicCIMIN}{0}
\newcommand{\SLOWbatikHBDynamicCIMAX}{0}
\newcommand{\SLOWbatikFT}{0}
\newcommand{\SLOWbatikFTCI}{0.0}
\newcommand{\SLOWbatikFTCIMIN}{0}
\newcommand{\SLOWbatikFTCIMAX}{0}
\newcommand{\SLOWbatikFTDynamic}{0}
\newcommand{\SLOWbatikFTDynamicCI}{0.0}
\newcommand{\SLOWbatikFTDynamicCIMIN}{0}
\newcommand{\SLOWbatikFTDynamicCIMAX}{0}
\newcommand{\SLOWbatikWCP}{0}
\newcommand{\SLOWbatikWCPCI}{0.0}
\newcommand{\SLOWbatikWCPCIMIN}{0}
\newcommand{\SLOWbatikWCPCIMAX}{0}
\newcommand{\SLOWbatikWCPDynamic}{0}
\newcommand{\SLOWbatikWCPDynamicCI}{0.0}
\newcommand{\SLOWbatikWCPDynamicCIMIN}{0}
\newcommand{\SLOWbatikWCPDynamicCIMAX}{0}
\newcommand{\SLOWbatikDCnoGExc}{0}
\newcommand{\SLOWbatikDCnoGExcCI}{0.0}
\newcommand{\SLOWbatikDCnoGExcCIMIN}{0}
\newcommand{\SLOWbatikDCnoGExcCIMAX}{0}
\newcommand{\SLOWbatikDCnoGExcDynamic}{0}
\newcommand{\SLOWbatikDCnoGExcDynamicCI}{0.0}
\newcommand{\SLOWbatikDCnoGExcDynamicCIMIN}{0}
\newcommand{\SLOWbatikDCnoGExcDynamicCIMAX}{0}
\newcommand{\SLOWbatikDCnoG}{\rna}
\newcommand{\SLOWbatikDCnoGCI}{\rna}
\newcommand{\SLOWbatikDCnoGCIMIN}{\rna}
\newcommand{\SLOWbatikDCnoGCIMAX}{\rna}
\newcommand{\SLOWbatikDCnoGDynamic}{\rna}
\newcommand{\SLOWbatikDCnoGDynamicCI}{\rna}
\newcommand{\SLOWbatikDCnoGDynamicCIMIN}{\rna}
\newcommand{\SLOWbatikDCnoGDynamicCIMAX}{\rna}
\newcommand{\SLOWbatikDCExc}{0}
\newcommand{\SLOWbatikDCExcCI}{0.0}
\newcommand{\SLOWbatikDCExcCIMIN}{0}
\newcommand{\SLOWbatikDCExcCIMAX}{0}
\newcommand{\SLOWbatikDCExcDynamic}{0}
\newcommand{\SLOWbatikDCExcDynamicCI}{0.0}
\newcommand{\SLOWbatikDCExcDynamicCIMIN}{0}
\newcommand{\SLOWbatikDCExcDynamicCIMAX}{0}
\newcommand{\SLOWbatikDC}{\rna}
\newcommand{\SLOWbatikDCCI}{\rna}
\newcommand{\SLOWbatikDCCIMIN}{\rna}
\newcommand{\SLOWbatikDCCIMAX}{\rna}
\newcommand{\SLOWbatikDCDynamic}{\rna}
\newcommand{\SLOWbatikDCDynamicCI}{\rna}
\newcommand{\SLOWbatikDCDynamicCIMIN}{\rna}
\newcommand{\SLOWbatikDCDynamicCIMAX}{\rna}
\newcommand{\SLOWbatikCAPOnoGExc}{0}
\newcommand{\SLOWbatikCAPOnoGExcCI}{0.0}
\newcommand{\SLOWbatikCAPOnoGExcCIMIN}{0}
\newcommand{\SLOWbatikCAPOnoGExcCIMAX}{0}
\newcommand{\SLOWbatikCAPOnoGExcDynamic}{0}
\newcommand{\SLOWbatikCAPOnoGExcDynamicCI}{0.0}
\newcommand{\SLOWbatikCAPOnoGExcDynamicCIMIN}{0}
\newcommand{\SLOWbatikCAPOnoGExcDynamicCIMAX}{0}
\newcommand{\SLOWbatikCAPOnoG}{\rna}
\newcommand{\SLOWbatikCAPOnoGCI}{\rna}
\newcommand{\SLOWbatikCAPOnoGCIMIN}{\rna}
\newcommand{\SLOWbatikCAPOnoGCIMAX}{\rna}
\newcommand{\SLOWbatikCAPOnoGDynamic}{\rna}
\newcommand{\SLOWbatikCAPOnoGDynamicCI}{\rna}
\newcommand{\SLOWbatikCAPOnoGDynamicCIMIN}{\rna}
\newcommand{\SLOWbatikCAPOnoGDynamicCIMAX}{\rna}
\newcommand{\SLOWbatikCAPOExc}{0}
\newcommand{\SLOWbatikCAPOExcCI}{0.0}
\newcommand{\SLOWbatikCAPOExcCIMIN}{0}
\newcommand{\SLOWbatikCAPOExcCIMAX}{0}
\newcommand{\SLOWbatikCAPOExcDynamic}{0}
\newcommand{\SLOWbatikCAPOExcDynamicCI}{0.0}
\newcommand{\SLOWbatikCAPOExcDynamicCIMIN}{0}
\newcommand{\SLOWbatikCAPOExcDynamicCIMAX}{0}
\newcommand{\SLOWbatikCAPO}{\rna}
\newcommand{\SLOWbatikCAPOCI}{\rna}
\newcommand{\SLOWbatikCAPOCIMIN}{\rna}
\newcommand{\SLOWbatikCAPOCIMAX}{\rna}
\newcommand{\SLOWbatikCAPODynamic}{\rna}
\newcommand{\SLOWbatikCAPODynamicCI}{\rna}
\newcommand{\SLOWbatikCAPODynamicCIMIN}{\rna}
\newcommand{\SLOWbatikCAPODynamicCIMAX}{\rna}
\newcommand{\SLOWbatikPIP}{\rna}
\newcommand{\SLOWbatikPIPCI}{\rna}
\newcommand{\SLOWbatikPIPCIMIN}{\rna}
\newcommand{\SLOWbatikPIPCIMAX}{\rna}
\newcommand{\SLOWbatikPIPDynamic}{\rna}
\newcommand{\SLOWbatikPIPDynamicCI}{\rna}
\newcommand{\SLOWbatikPIPDynamicCIMIN}{\rna}
\newcommand{\SLOWbatikPIPDynamicCIMAX}{\rna}
\newcommand{\SLOWbatikHBUP}{0}
\newcommand{\SLOWbatikHBUPCI}{0.0}
\newcommand{\SLOWbatikHBUPCIMIN}{0}
\newcommand{\SLOWbatikHBUPCIMAX}{0}
\newcommand{\SLOWbatikHBDynamicUP}{0}
\newcommand{\SLOWbatikHBDynamicUPCI}{0.0}
\newcommand{\SLOWbatikHBDynamicUPCIMIN}{0}
\newcommand{\SLOWbatikHBDynamicUPCIMAX}{0}
\newcommand{\SLOWbatikWCPUP}{0}
\newcommand{\SLOWbatikWCPUPCI}{0.0}
\newcommand{\SLOWbatikWCPUPCIMIN}{0}
\newcommand{\SLOWbatikWCPUPCIMAX}{0}
\newcommand{\SLOWbatikWCPDynamicUP}{0}
\newcommand{\SLOWbatikWCPDynamicUPCI}{0.0}
\newcommand{\SLOWbatikWCPDynamicUPCIMIN}{0}
\newcommand{\SLOWbatikWCPDynamicUPCIMAX}{0}
\newcommand{\SLOWbatikWDCnoGUP}{\rna}
\newcommand{\SLOWbatikWDCnoGUPCI}{\rna}
\newcommand{\SLOWbatikWDCnoGUPCIMIN}{\rna}
\newcommand{\SLOWbatikWDCnoGUPCIMAX}{\rna}
\newcommand{\SLOWbatikWDCnoGDynamicUP}{\rna}
\newcommand{\SLOWbatikWDCnoGDynamicUPCI}{\rna}
\newcommand{\SLOWbatikWDCnoGDynamicUPCIMIN}{\rna}
\newcommand{\SLOWbatikWDCnoGDynamicUPCIMAX}{\rna}
\newcommand{\SLOWbatikWDCUP}{\rna}
\newcommand{\SLOWbatikWDCUPCI}{\rna}
\newcommand{\SLOWbatikWDCUPCIMIN}{\rna}
\newcommand{\SLOWbatikWDCUPCIMAX}{\rna}
\newcommand{\SLOWbatikWDCDynamicUP}{\rna}
\newcommand{\SLOWbatikWDCDynamicUPCI}{\rna}
\newcommand{\SLOWbatikWDCDynamicUPCIMIN}{\rna}
\newcommand{\SLOWbatikWDCDynamicUPCIMAX}{\rna}
\newcommand{\SLOWbatikCAPOnoGUP}{\rna}
\newcommand{\SLOWbatikCAPOnoGUPCI}{\rna}
\newcommand{\SLOWbatikCAPOnoGUPCIMIN}{\rna}
\newcommand{\SLOWbatikCAPOnoGUPCIMAX}{\rna}
\newcommand{\SLOWbatikCAPOnoGDynamicUP}{\rna}
\newcommand{\SLOWbatikCAPOnoGDynamicUPCI}{\rna}
\newcommand{\SLOWbatikCAPOnoGDynamicUPCIMIN}{\rna}
\newcommand{\SLOWbatikCAPOnoGDynamicUPCIMAX}{\rna}
\newcommand{\SLOWbatikCAPOUP}{\rna}
\newcommand{\SLOWbatikCAPOUPCI}{\rna}
\newcommand{\SLOWbatikCAPOUPCIMIN}{\rna}
\newcommand{\SLOWbatikCAPOUPCIMAX}{\rna}
\newcommand{\SLOWbatikCAPODynamicUP}{\rna}
\newcommand{\SLOWbatikCAPODynamicUPCI}{\rna}
\newcommand{\SLOWbatikCAPODynamicUPCIMIN}{\rna}
\newcommand{\SLOWbatikCAPODynamicUPCIMAX}{\rna}
\newcommand{\SLOWbatikPIPUP}{\rna}
\newcommand{\SLOWbatikPIPUPCI}{\rna}
\newcommand{\SLOWbatikPIPUPCIMIN}{\rna}
\newcommand{\SLOWbatikPIPUPCIMAX}{\rna}
\newcommand{\SLOWbatikPIPDynamicUP}{\rna}
\newcommand{\SLOWbatikPIPDynamicUPCI}{\rna}
\newcommand{\SLOWbatikPIPDynamicUPCIMIN}{\rna}
\newcommand{\SLOWbatikPIPDynamicUPCIMAX}{\rna}
\newcommand{\SLOWbatikPIPHB}{\rna}
\newcommand{\SLOWbatikPIPHBCI}{\rna}
\newcommand{\SLOWbatikPIPHBCIMIN}{\rna}
\newcommand{\SLOWbatikPIPHBCIMAX}{\rna}
\newcommand{\SLOWbatikPIPHBDynamic}{\rna}
\newcommand{\SLOWbatikPIPHBDynamicCI}{\rna}
\newcommand{\SLOWbatikPIPHBDynamicCIMIN}{\rna}
\newcommand{\SLOWbatikPIPHBDynamicCIMAX}{\rna}
\newcommand{\SLOWbatikPIPWCP}{\rna}
\newcommand{\SLOWbatikPIPWCPCI}{\rna}
\newcommand{\SLOWbatikPIPWCPCIMIN}{\rna}
\newcommand{\SLOWbatikPIPWCPCIMAX}{\rna}
\newcommand{\SLOWbatikPIPWCPDynamic}{\rna}
\newcommand{\SLOWbatikPIPWCPDynamicCI}{\rna}
\newcommand{\SLOWbatikPIPWCPDynamicCIMIN}{\rna}
\newcommand{\SLOWbatikPIPWCPDynamicCIMAX}{\rna}
\newcommand{\SLOWbatikPIPWDC}{\rna}
\newcommand{\SLOWbatikPIPWDCCI}{\rna}
\newcommand{\SLOWbatikPIPWDCCIMIN}{\rna}
\newcommand{\SLOWbatikPIPWDCCIMAX}{\rna}
\newcommand{\SLOWbatikPIPWDCDynamic}{\rna}
\newcommand{\SLOWbatikPIPWDCDynamicCI}{\rna}
\newcommand{\SLOWbatikPIPWDCDynamicCIMIN}{\rna}
\newcommand{\SLOWbatikPIPWDCDynamicCIMAX}{\rna}
\newcommand{\SLOWbatikPIPCAPO}{\rna}
\newcommand{\SLOWbatikPIPCAPOCI}{\rna}
\newcommand{\SLOWbatikPIPCAPOCIMIN}{\rna}
\newcommand{\SLOWbatikPIPCAPOCIMAX}{\rna}
\newcommand{\SLOWbatikPIPCAPODynamic}{\rna}
\newcommand{\SLOWbatikPIPCAPODynamicCI}{\rna}
\newcommand{\SLOWbatikPIPCAPODynamicCIMIN}{\rna}
\newcommand{\SLOWbatikPIPCAPODynamicCIMAX}{\rna}
\newcommand{\SLOWbatikPIPPIP}{\rna}
\newcommand{\SLOWbatikPIPPIPCI}{\rna}
\newcommand{\SLOWbatikPIPPIPCIMIN}{\rna}
\newcommand{\SLOWbatikPIPPIPCIMAX}{\rna}
\newcommand{\SLOWbatikPIPPIPDynamic}{\rna}
\newcommand{\SLOWbatikPIPPIPDynamicCI}{\rna}
\newcommand{\SLOWbatikPIPPIPDynamicCIMIN}{\rna}
\newcommand{\SLOWbatikPIPPIPDynamicCIMAX}{\rna}
\newcommand{\SLOWhtwoEvents}{0}
\newcommand{\SLOWhtwoNoFPEvents}{0}
\newcommand{\SLOWhtwoMaxLiveThreads}{15}
\newcommand{\SLOWhtwoTotalThreads}{16}
\newcommand{\SLOWhtwoBaseTime}{4.8}
\newcommand{\SLOWhtwoBaseTimeCI}{59}
\newcommand{\SLOWhtwoEmptyTime}{\rna}
\newcommand{\SLOWhtwoEmptyTimeCI}{\rna}
\newcommand{\SLOWhtwoEmptyTimeCIMIN}{\rna}
\newcommand{\SLOWhtwoEmptyTimeCIMAX}{\rna}
\newcommand{\SLOWhtwoFTTime}{9.6}
\newcommand{\SLOWhtwoFTTimeCI}{0.77}
\newcommand{\SLOWhtwoHBTime}{31}
\newcommand{\SLOWhtwoHBTimeCI}{1.1}
\newcommand{\SLOWhtwoWCPTime}{83}
\newcommand{\SLOWhtwoWCPTimeCI}{6.4}
\newcommand{\SLOWhtwoDCnoGExcTime}{85}
\newcommand{\SLOWhtwoDCnoGExcTimeCI}{6.0}
\newcommand{\SLOWhtwoDCnoGTime}{\rna}
\newcommand{\SLOWhtwoDCnoGTimeCI}{\rna}
\newcommand{\SLOWhtwoDCnoGTimeCIMIN}{\rna}
\newcommand{\SLOWhtwoDCnoGTimeCIMAX}{\rna}
\newcommand{\SLOWhtwoDCExcTime}{79}
\newcommand{\SLOWhtwoDCExcTimeCI}{5.2}
\newcommand{\SLOWhtwoDCTime}{\rna}
\newcommand{\SLOWhtwoDCTimeCI}{\rna}
\newcommand{\SLOWhtwoDCTimeCIMIN}{\rna}
\newcommand{\SLOWhtwoDCTimeCIMAX}{\rna}
\newcommand{\SLOWhtwoCAPOnoGExcTime}{89}
\newcommand{\SLOWhtwoCAPOnoGExcTimeCI}{1.7}
\newcommand{\SLOWhtwoCAPOnoGTime}{\rna}
\newcommand{\SLOWhtwoCAPOnoGTimeCI}{\rna}
\newcommand{\SLOWhtwoCAPOnoGTimeCIMIN}{\rna}
\newcommand{\SLOWhtwoCAPOnoGTimeCIMAX}{\rna}
\newcommand{\SLOWhtwoCAPOExcTime}{80}
\newcommand{\SLOWhtwoCAPOExcTimeCI}{1.5}
\newcommand{\SLOWhtwoCAPOTime}{\rna}
\newcommand{\SLOWhtwoCAPOTimeCI}{\rna}
\newcommand{\SLOWhtwoCAPOTimeCIMIN}{\rna}
\newcommand{\SLOWhtwoCAPOTimeCIMAX}{\rna}
\newcommand{\SLOWhtwoStaticTime}{\rzero}
\newcommand{\SLOWhtwoDynamicTime}{\rzero}
\newcommand{\SLOWhtwoBaseMem}{1,700}
\newcommand{\SLOWhtwoBaseMemCI}{71.0}
\newcommand{\SLOWhtwoHBMem}{15}
\newcommand{\SLOWhtwoHBMemCI}{0.95}
\newcommand{\SLOWhtwoFTMem}{3.2}
\newcommand{\SLOWhtwoFTMemCI}{0.14}
\newcommand{\SLOWhtwoWCPMem}{65}
\newcommand{\SLOWhtwoWCPMemCI}{0.67}
\newcommand{\SLOWhtwoDCnoGExcMem}{57}
\newcommand{\SLOWhtwoDCnoGExcMemCI}{0.014}
\newcommand{\SLOWhtwoDCnoGMem}{\memna}
\newcommand{\SLOWhtwoDCnoGMemCI}{\memna}
\newcommand{\SLOWhtwoDCnoGMemCIMIN}{\memna}
\newcommand{\SLOWhtwoDCnoGMemCIMAX}{\memna}
\newcommand{\SLOWhtwoDCExcMem}{59}
\newcommand{\SLOWhtwoDCExcMemCI}{0.29}
\newcommand{\SLOWhtwoDCMem}{\memna}
\newcommand{\SLOWhtwoDCMemCI}{\memna}
\newcommand{\SLOWhtwoDCMemCIMIN}{\memna}
\newcommand{\SLOWhtwoDCMemCIMAX}{\memna}
\newcommand{\SLOWhtwoCAPOnoGExcMem}{65}
\newcommand{\SLOWhtwoCAPOnoGExcMemCI}{2.8}
\newcommand{\SLOWhtwoCAPOnoGMem}{\memna}
\newcommand{\SLOWhtwoCAPOnoGMemCI}{\memna}
\newcommand{\SLOWhtwoCAPOnoGMemCIMIN}{\memna}
\newcommand{\SLOWhtwoCAPOnoGMemCIMAX}{\memna}
\newcommand{\SLOWhtwoCAPOExcMem}{57}
\newcommand{\SLOWhtwoCAPOExcMemCI}{6.7}
\newcommand{\SLOWhtwoCAPOMem}{\memna}
\newcommand{\SLOWhtwoCAPOMemCI}{\memna}
\newcommand{\SLOWhtwoCAPOMemCIMIN}{\memna}
\newcommand{\SLOWhtwoCAPOMemCIMAX}{\memna}
\newcommand{\SLOWhtwoEventsCI}{0}
\newcommand{\SLOWhtwoEventsCIMIN}{0}
\newcommand{\SLOWhtwoEventsCIMAX}{0}
\newcommand{\SLOWhtwoNoFPEventsCI}{0}
\newcommand{\SLOWhtwoNoFPEventsCIMIN}{0}
\newcommand{\SLOWhtwoNoFPEventsCIMAX}{0}
\newcommand{\SLOWhtwoHB}{13}
\newcommand{\SLOWhtwoHBCI}{0.0}
\newcommand{\SLOWhtwoHBCIMIN}{13}
\newcommand{\SLOWhtwoHBCIMAX}{13}
\newcommand{\SLOWhtwoHBDynamic}{89,105}
\newcommand{\SLOWhtwoHBDynamicCI}{464}
\newcommand{\SLOWhtwoHBDynamicCIMIN}{88,641}
\newcommand{\SLOWhtwoHBDynamicCIMAX}{89,569}
\newcommand{\SLOWhtwoFT}{11}
\newcommand{\SLOWhtwoFTCI}{0.0}
\newcommand{\SLOWhtwoFTCIMIN}{11}
\newcommand{\SLOWhtwoFTCIMAX}{11}
\newcommand{\SLOWhtwoFTDynamic}{106,302}
\newcommand{\SLOWhtwoFTDynamicCI}{206}
\newcommand{\SLOWhtwoFTDynamicCIMIN}{106,096}
\newcommand{\SLOWhtwoFTDynamicCIMAX}{106,508}
\newcommand{\SLOWhtwoWCP}{13}
\newcommand{\SLOWhtwoWCPCI}{0.0}
\newcommand{\SLOWhtwoWCPCIMIN}{13}
\newcommand{\SLOWhtwoWCPCIMAX}{13}
\newcommand{\SLOWhtwoWCPDynamic}{89,916}
\newcommand{\SLOWhtwoWCPDynamicCI}{560}
\newcommand{\SLOWhtwoWCPDynamicCIMIN}{89,356}
\newcommand{\SLOWhtwoWCPDynamicCIMAX}{90,476}
\newcommand{\SLOWhtwoDCnoGExc}{13}
\newcommand{\SLOWhtwoDCnoGExcCI}{0.0}
\newcommand{\SLOWhtwoDCnoGExcCIMIN}{13}
\newcommand{\SLOWhtwoDCnoGExcCIMAX}{13}
\newcommand{\SLOWhtwoDCnoGExcDynamic}{91,074}
\newcommand{\SLOWhtwoDCnoGExcDynamicCI}{114}
\newcommand{\SLOWhtwoDCnoGExcDynamicCIMIN}{90,960}
\newcommand{\SLOWhtwoDCnoGExcDynamicCIMAX}{91,188}
\newcommand{\SLOWhtwoDCnoG}{\rna}
\newcommand{\SLOWhtwoDCnoGCI}{\rna}
\newcommand{\SLOWhtwoDCnoGCIMIN}{\rna}
\newcommand{\SLOWhtwoDCnoGCIMAX}{\rna}
\newcommand{\SLOWhtwoDCnoGDynamic}{\rna}
\newcommand{\SLOWhtwoDCnoGDynamicCI}{\rna}
\newcommand{\SLOWhtwoDCnoGDynamicCIMIN}{\rna}
\newcommand{\SLOWhtwoDCnoGDynamicCIMAX}{\rna}
\newcommand{\SLOWhtwoDCExc}{13}
\newcommand{\SLOWhtwoDCExcCI}{0.0}
\newcommand{\SLOWhtwoDCExcCIMIN}{13}
\newcommand{\SLOWhtwoDCExcCIMAX}{13}
\newcommand{\SLOWhtwoDCExcDynamic}{44,972}
\newcommand{\SLOWhtwoDCExcDynamicCI}{331}
\newcommand{\SLOWhtwoDCExcDynamicCIMIN}{44,641}
\newcommand{\SLOWhtwoDCExcDynamicCIMAX}{45,303}
\newcommand{\SLOWhtwoDC}{\rna}
\newcommand{\SLOWhtwoDCCI}{\rna}
\newcommand{\SLOWhtwoDCCIMIN}{\rna}
\newcommand{\SLOWhtwoDCCIMAX}{\rna}
\newcommand{\SLOWhtwoDCDynamic}{\rna}
\newcommand{\SLOWhtwoDCDynamicCI}{\rna}
\newcommand{\SLOWhtwoDCDynamicCIMIN}{\rna}
\newcommand{\SLOWhtwoDCDynamicCIMAX}{\rna}
\newcommand{\SLOWhtwoCAPOnoGExc}{13}
\newcommand{\SLOWhtwoCAPOnoGExcCI}{0.0}
\newcommand{\SLOWhtwoCAPOnoGExcCIMIN}{13}
\newcommand{\SLOWhtwoCAPOnoGExcCIMAX}{13}
\newcommand{\SLOWhtwoCAPOnoGExcDynamic}{91,092}
\newcommand{\SLOWhtwoCAPOnoGExcDynamicCI}{202}
\newcommand{\SLOWhtwoCAPOnoGExcDynamicCIMIN}{90,890}
\newcommand{\SLOWhtwoCAPOnoGExcDynamicCIMAX}{91,294}
\newcommand{\SLOWhtwoCAPOnoG}{\rna}
\newcommand{\SLOWhtwoCAPOnoGCI}{\rna}
\newcommand{\SLOWhtwoCAPOnoGCIMIN}{\rna}
\newcommand{\SLOWhtwoCAPOnoGCIMAX}{\rna}
\newcommand{\SLOWhtwoCAPOnoGDynamic}{\rna}
\newcommand{\SLOWhtwoCAPOnoGDynamicCI}{\rna}
\newcommand{\SLOWhtwoCAPOnoGDynamicCIMIN}{\rna}
\newcommand{\SLOWhtwoCAPOnoGDynamicCIMAX}{\rna}
\newcommand{\SLOWhtwoCAPOExc}{13}
\newcommand{\SLOWhtwoCAPOExcCI}{0.0}
\newcommand{\SLOWhtwoCAPOExcCIMIN}{13}
\newcommand{\SLOWhtwoCAPOExcCIMAX}{13}
\newcommand{\SLOWhtwoCAPOExcDynamic}{45,101}
\newcommand{\SLOWhtwoCAPOExcDynamicCI}{189}
\newcommand{\SLOWhtwoCAPOExcDynamicCIMIN}{44,912}
\newcommand{\SLOWhtwoCAPOExcDynamicCIMAX}{45,290}
\newcommand{\SLOWhtwoCAPO}{\rna}
\newcommand{\SLOWhtwoCAPOCI}{\rna}
\newcommand{\SLOWhtwoCAPOCIMIN}{\rna}
\newcommand{\SLOWhtwoCAPOCIMAX}{\rna}
\newcommand{\SLOWhtwoCAPODynamic}{\rna}
\newcommand{\SLOWhtwoCAPODynamicCI}{\rna}
\newcommand{\SLOWhtwoCAPODynamicCIMIN}{\rna}
\newcommand{\SLOWhtwoCAPODynamicCIMAX}{\rna}
\newcommand{\SLOWhtwoPIP}{\rna}
\newcommand{\SLOWhtwoPIPCI}{\rna}
\newcommand{\SLOWhtwoPIPCIMIN}{\rna}
\newcommand{\SLOWhtwoPIPCIMAX}{\rna}
\newcommand{\SLOWhtwoPIPDynamic}{\rna}
\newcommand{\SLOWhtwoPIPDynamicCI}{\rna}
\newcommand{\SLOWhtwoPIPDynamicCIMIN}{\rna}
\newcommand{\SLOWhtwoPIPDynamicCIMAX}{\rna}
\newcommand{\SLOWhtwoHBUP}{9}
\newcommand{\SLOWhtwoHBUPCI}{0.0}
\newcommand{\SLOWhtwoHBUPCIMIN}{9}
\newcommand{\SLOWhtwoHBUPCIMAX}{9}
\newcommand{\SLOWhtwoHBDynamicUP}{89,106}
\newcommand{\SLOWhtwoHBDynamicUPCI}{461}
\newcommand{\SLOWhtwoHBDynamicUPCIMIN}{88,645}
\newcommand{\SLOWhtwoHBDynamicUPCIMAX}{89,567}
\newcommand{\SLOWhtwoWCPUP}{10}
\newcommand{\SLOWhtwoWCPUPCI}{0.98}
\newcommand{\SLOWhtwoWCPUPCIMIN}{9}
\newcommand{\SLOWhtwoWCPUPCIMAX}{11}
\newcommand{\SLOWhtwoWCPDynamicUP}{89,915}
\newcommand{\SLOWhtwoWCPDynamicUPCI}{559}
\newcommand{\SLOWhtwoWCPDynamicUPCIMIN}{89,356}
\newcommand{\SLOWhtwoWCPDynamicUPCIMAX}{90,474}
\newcommand{\SLOWhtwoWDCnoGUP}{\rna}
\newcommand{\SLOWhtwoWDCnoGUPCI}{\rna}
\newcommand{\SLOWhtwoWDCnoGUPCIMIN}{\rna}
\newcommand{\SLOWhtwoWDCnoGUPCIMAX}{\rna}
\newcommand{\SLOWhtwoWDCnoGDynamicUP}{\rna}
\newcommand{\SLOWhtwoWDCnoGDynamicUPCI}{\rna}
\newcommand{\SLOWhtwoWDCnoGDynamicUPCIMIN}{\rna}
\newcommand{\SLOWhtwoWDCnoGDynamicUPCIMAX}{\rna}
\newcommand{\SLOWhtwoWDCUP}{\rna}
\newcommand{\SLOWhtwoWDCUPCI}{\rna}
\newcommand{\SLOWhtwoWDCUPCIMIN}{\rna}
\newcommand{\SLOWhtwoWDCUPCIMAX}{\rna}
\newcommand{\SLOWhtwoWDCDynamicUP}{\rna}
\newcommand{\SLOWhtwoWDCDynamicUPCI}{\rna}
\newcommand{\SLOWhtwoWDCDynamicUPCIMIN}{\rna}
\newcommand{\SLOWhtwoWDCDynamicUPCIMAX}{\rna}
\newcommand{\SLOWhtwoCAPOnoGUP}{\rna}
\newcommand{\SLOWhtwoCAPOnoGUPCI}{\rna}
\newcommand{\SLOWhtwoCAPOnoGUPCIMIN}{\rna}
\newcommand{\SLOWhtwoCAPOnoGUPCIMAX}{\rna}
\newcommand{\SLOWhtwoCAPOnoGDynamicUP}{\rna}
\newcommand{\SLOWhtwoCAPOnoGDynamicUPCI}{\rna}
\newcommand{\SLOWhtwoCAPOnoGDynamicUPCIMIN}{\rna}
\newcommand{\SLOWhtwoCAPOnoGDynamicUPCIMAX}{\rna}
\newcommand{\SLOWhtwoCAPOUP}{\rna}
\newcommand{\SLOWhtwoCAPOUPCI}{\rna}
\newcommand{\SLOWhtwoCAPOUPCIMIN}{\rna}
\newcommand{\SLOWhtwoCAPOUPCIMAX}{\rna}
\newcommand{\SLOWhtwoCAPODynamicUP}{\rna}
\newcommand{\SLOWhtwoCAPODynamicUPCI}{\rna}
\newcommand{\SLOWhtwoCAPODynamicUPCIMIN}{\rna}
\newcommand{\SLOWhtwoCAPODynamicUPCIMAX}{\rna}
\newcommand{\SLOWhtwoPIPUP}{\rna}
\newcommand{\SLOWhtwoPIPUPCI}{\rna}
\newcommand{\SLOWhtwoPIPUPCIMIN}{\rna}
\newcommand{\SLOWhtwoPIPUPCIMAX}{\rna}
\newcommand{\SLOWhtwoPIPDynamicUP}{\rna}
\newcommand{\SLOWhtwoPIPDynamicUPCI}{\rna}
\newcommand{\SLOWhtwoPIPDynamicUPCIMIN}{\rna}
\newcommand{\SLOWhtwoPIPDynamicUPCIMAX}{\rna}
\newcommand{\SLOWhtwoPIPHB}{\rna}
\newcommand{\SLOWhtwoPIPHBCI}{\rna}
\newcommand{\SLOWhtwoPIPHBCIMIN}{\rna}
\newcommand{\SLOWhtwoPIPHBCIMAX}{\rna}
\newcommand{\SLOWhtwoPIPHBDynamic}{\rna}
\newcommand{\SLOWhtwoPIPHBDynamicCI}{\rna}
\newcommand{\SLOWhtwoPIPHBDynamicCIMIN}{\rna}
\newcommand{\SLOWhtwoPIPHBDynamicCIMAX}{\rna}
\newcommand{\SLOWhtwoPIPWCP}{\rna}
\newcommand{\SLOWhtwoPIPWCPCI}{\rna}
\newcommand{\SLOWhtwoPIPWCPCIMIN}{\rna}
\newcommand{\SLOWhtwoPIPWCPCIMAX}{\rna}
\newcommand{\SLOWhtwoPIPWCPDynamic}{\rna}
\newcommand{\SLOWhtwoPIPWCPDynamicCI}{\rna}
\newcommand{\SLOWhtwoPIPWCPDynamicCIMIN}{\rna}
\newcommand{\SLOWhtwoPIPWCPDynamicCIMAX}{\rna}
\newcommand{\SLOWhtwoPIPWDC}{\rna}
\newcommand{\SLOWhtwoPIPWDCCI}{\rna}
\newcommand{\SLOWhtwoPIPWDCCIMIN}{\rna}
\newcommand{\SLOWhtwoPIPWDCCIMAX}{\rna}
\newcommand{\SLOWhtwoPIPWDCDynamic}{\rna}
\newcommand{\SLOWhtwoPIPWDCDynamicCI}{\rna}
\newcommand{\SLOWhtwoPIPWDCDynamicCIMIN}{\rna}
\newcommand{\SLOWhtwoPIPWDCDynamicCIMAX}{\rna}
\newcommand{\SLOWhtwoPIPCAPO}{\rna}
\newcommand{\SLOWhtwoPIPCAPOCI}{\rna}
\newcommand{\SLOWhtwoPIPCAPOCIMIN}{\rna}
\newcommand{\SLOWhtwoPIPCAPOCIMAX}{\rna}
\newcommand{\SLOWhtwoPIPCAPODynamic}{\rna}
\newcommand{\SLOWhtwoPIPCAPODynamicCI}{\rna}
\newcommand{\SLOWhtwoPIPCAPODynamicCIMIN}{\rna}
\newcommand{\SLOWhtwoPIPCAPODynamicCIMAX}{\rna}
\newcommand{\SLOWhtwoPIPPIP}{\rna}
\newcommand{\SLOWhtwoPIPPIPCI}{\rna}
\newcommand{\SLOWhtwoPIPPIPCIMIN}{\rna}
\newcommand{\SLOWhtwoPIPPIPCIMAX}{\rna}
\newcommand{\SLOWhtwoPIPPIPDynamic}{\rna}
\newcommand{\SLOWhtwoPIPPIPDynamicCI}{\rna}
\newcommand{\SLOWhtwoPIPPIPDynamicCIMIN}{\rna}
\newcommand{\SLOWhtwoPIPPIPDynamicCIMAX}{\rna}
\newcommand{\SLOWjythonEvents}{0}
\newcommand{\SLOWjythonNoFPEvents}{0}
\newcommand{\SLOWjythonMaxLiveThreads}{2}
\newcommand{\SLOWjythonTotalThreads}{2}
\newcommand{\SLOWjythonBaseTime}{3.8}
\newcommand{\SLOWjythonBaseTimeCI}{290}
\newcommand{\SLOWjythonEmptyTime}{\rna}
\newcommand{\SLOWjythonEmptyTimeCI}{\rna}
\newcommand{\SLOWjythonEmptyTimeCIMIN}{\rna}
\newcommand{\SLOWjythonEmptyTimeCIMAX}{\rna}
\newcommand{\SLOWjythonFTTime}{8.6}
\newcommand{\SLOWjythonFTTimeCI}{0.25}
\newcommand{\SLOWjythonHBTime}{22}
\newcommand{\SLOWjythonHBTimeCI}{2.5}
\newcommand{\SLOWjythonWCPTime}{33}
\newcommand{\SLOWjythonWCPTimeCI}{1.3}
\newcommand{\SLOWjythonDCnoGExcTime}{25}
\newcommand{\SLOWjythonDCnoGExcTimeCI}{1.2}
\newcommand{\SLOWjythonDCnoGTime}{\rna}
\newcommand{\SLOWjythonDCnoGTimeCI}{\rna}
\newcommand{\SLOWjythonDCnoGTimeCIMIN}{\rna}
\newcommand{\SLOWjythonDCnoGTimeCIMAX}{\rna}
\newcommand{\SLOWjythonDCExcTime}{31}
\newcommand{\SLOWjythonDCExcTimeCI}{1.8}
\newcommand{\SLOWjythonDCTime}{\rna}
\newcommand{\SLOWjythonDCTimeCI}{\rna}
\newcommand{\SLOWjythonDCTimeCIMIN}{\rna}
\newcommand{\SLOWjythonDCTimeCIMAX}{\rna}
\newcommand{\SLOWjythonCAPOnoGExcTime}{22}
\newcommand{\SLOWjythonCAPOnoGExcTimeCI}{1.4}
\newcommand{\SLOWjythonCAPOnoGTime}{\rna}
\newcommand{\SLOWjythonCAPOnoGTimeCI}{\rna}
\newcommand{\SLOWjythonCAPOnoGTimeCIMIN}{\rna}
\newcommand{\SLOWjythonCAPOnoGTimeCIMAX}{\rna}
\newcommand{\SLOWjythonCAPOExcTime}{27}
\newcommand{\SLOWjythonCAPOExcTimeCI}{1.8}
\newcommand{\SLOWjythonCAPOTime}{\rna}
\newcommand{\SLOWjythonCAPOTimeCI}{\rna}
\newcommand{\SLOWjythonCAPOTimeCIMIN}{\rna}
\newcommand{\SLOWjythonCAPOTimeCIMAX}{\rna}
\newcommand{\SLOWjythonStaticTime}{\rzero}
\newcommand{\SLOWjythonDynamicTime}{\rzero}
\newcommand{\SLOWjythonBaseMem}{730}
\newcommand{\SLOWjythonBaseMemCI}{0.37}
\newcommand{\SLOWjythonHBMem}{21}
\newcommand{\SLOWjythonHBMemCI}{1.7}
\newcommand{\SLOWjythonFTMem}{5.8}
\newcommand{\SLOWjythonFTMemCI}{0.62}
\newcommand{\SLOWjythonWCPMem}{21}
\newcommand{\SLOWjythonWCPMemCI}{0.74}
\newcommand{\SLOWjythonDCnoGExcMem}{18}
\newcommand{\SLOWjythonDCnoGExcMemCI}{1.1}
\newcommand{\SLOWjythonDCnoGMem}{\memna}
\newcommand{\SLOWjythonDCnoGMemCI}{\memna}
\newcommand{\SLOWjythonDCnoGMemCIMIN}{\memna}
\newcommand{\SLOWjythonDCnoGMemCIMAX}{\memna}
\newcommand{\SLOWjythonDCExcMem}{32}
\newcommand{\SLOWjythonDCExcMemCI}{2.2}
\newcommand{\SLOWjythonDCMem}{\memna}
\newcommand{\SLOWjythonDCMemCI}{\memna}
\newcommand{\SLOWjythonDCMemCIMIN}{\memna}
\newcommand{\SLOWjythonDCMemCIMAX}{\memna}
\newcommand{\SLOWjythonCAPOnoGExcMem}{16}
\newcommand{\SLOWjythonCAPOnoGExcMemCI}{0.89}
\newcommand{\SLOWjythonCAPOnoGMem}{\memna}
\newcommand{\SLOWjythonCAPOnoGMemCI}{\memna}
\newcommand{\SLOWjythonCAPOnoGMemCIMIN}{\memna}
\newcommand{\SLOWjythonCAPOnoGMemCIMAX}{\memna}
\newcommand{\SLOWjythonCAPOExcMem}{25}
\newcommand{\SLOWjythonCAPOExcMemCI}{1.1}
\newcommand{\SLOWjythonCAPOMem}{\memna}
\newcommand{\SLOWjythonCAPOMemCI}{\memna}
\newcommand{\SLOWjythonCAPOMemCIMIN}{\memna}
\newcommand{\SLOWjythonCAPOMemCIMAX}{\memna}
\newcommand{\SLOWjythonEventsCI}{0}
\newcommand{\SLOWjythonEventsCIMIN}{0}
\newcommand{\SLOWjythonEventsCIMAX}{0}
\newcommand{\SLOWjythonNoFPEventsCI}{0}
\newcommand{\SLOWjythonNoFPEventsCIMIN}{0}
\newcommand{\SLOWjythonNoFPEventsCIMAX}{0}
\newcommand{\SLOWjythonHB}{21}
\newcommand{\SLOWjythonHBCI}{2.9}
\newcommand{\SLOWjythonHBCIMIN}{18}
\newcommand{\SLOWjythonHBCIMAX}{24}
\newcommand{\SLOWjythonHBDynamic}{22}
\newcommand{\SLOWjythonHBDynamicCI}{4.9}
\newcommand{\SLOWjythonHBDynamicCIMIN}{17}
\newcommand{\SLOWjythonHBDynamicCIMAX}{27}
\newcommand{\SLOWjythonFT}{25}
\newcommand{\SLOWjythonFTCI}{0.0}
\newcommand{\SLOWjythonFTCIMIN}{25}
\newcommand{\SLOWjythonFTCIMAX}{25}
\newcommand{\SLOWjythonFTDynamic}{49}
\newcommand{\SLOWjythonFTDynamicCI}{0.0}
\newcommand{\SLOWjythonFTDynamicCIMIN}{49}
\newcommand{\SLOWjythonFTDynamicCIMAX}{49}
\newcommand{\SLOWjythonWCP}{22}
\newcommand{\SLOWjythonWCPCI}{0.0}
\newcommand{\SLOWjythonWCPCIMIN}{22}
\newcommand{\SLOWjythonWCPCIMAX}{22}
\newcommand{\SLOWjythonWCPDynamic}{25}
\newcommand{\SLOWjythonWCPDynamicCI}{0.98}
\newcommand{\SLOWjythonWCPDynamicCIMIN}{24}
\newcommand{\SLOWjythonWCPDynamicCIMAX}{26}
\newcommand{\SLOWjythonDCnoGExc}{31}
\newcommand{\SLOWjythonDCnoGExcCI}{0.0}
\newcommand{\SLOWjythonDCnoGExcCIMIN}{31}
\newcommand{\SLOWjythonDCnoGExcCIMAX}{31}
\newcommand{\SLOWjythonDCnoGExcDynamic}{35}
\newcommand{\SLOWjythonDCnoGExcDynamicCI}{0.0}
\newcommand{\SLOWjythonDCnoGExcDynamicCIMIN}{35}
\newcommand{\SLOWjythonDCnoGExcDynamicCIMAX}{35}
\newcommand{\SLOWjythonDCnoG}{\rna}
\newcommand{\SLOWjythonDCnoGCI}{\rna}
\newcommand{\SLOWjythonDCnoGCIMIN}{\rna}
\newcommand{\SLOWjythonDCnoGCIMAX}{\rna}
\newcommand{\SLOWjythonDCnoGDynamic}{\rna}
\newcommand{\SLOWjythonDCnoGDynamicCI}{\rna}
\newcommand{\SLOWjythonDCnoGDynamicCIMIN}{\rna}
\newcommand{\SLOWjythonDCnoGDynamicCIMAX}{\rna}
\newcommand{\SLOWjythonDCExc}{3}
\newcommand{\SLOWjythonDCExcCI}{0.98}
\newcommand{\SLOWjythonDCExcCIMIN}{2}
\newcommand{\SLOWjythonDCExcCIMAX}{4}
\newcommand{\SLOWjythonDCExcDynamic}{4}
\newcommand{\SLOWjythonDCExcDynamicCI}{2.0}
\newcommand{\SLOWjythonDCExcDynamicCIMIN}{2}
\newcommand{\SLOWjythonDCExcDynamicCIMAX}{6}
\newcommand{\SLOWjythonDC}{\rna}
\newcommand{\SLOWjythonDCCI}{\rna}
\newcommand{\SLOWjythonDCCIMIN}{\rna}
\newcommand{\SLOWjythonDCCIMAX}{\rna}
\newcommand{\SLOWjythonDCDynamic}{\rna}
\newcommand{\SLOWjythonDCDynamicCI}{\rna}
\newcommand{\SLOWjythonDCDynamicCIMIN}{\rna}
\newcommand{\SLOWjythonDCDynamicCIMAX}{\rna}
\newcommand{\SLOWjythonCAPOnoGExc}{31}
\newcommand{\SLOWjythonCAPOnoGExcCI}{0.0}
\newcommand{\SLOWjythonCAPOnoGExcCIMIN}{31}
\newcommand{\SLOWjythonCAPOnoGExcCIMAX}{31}
\newcommand{\SLOWjythonCAPOnoGExcDynamic}{35}
\newcommand{\SLOWjythonCAPOnoGExcDynamicCI}{0.0}
\newcommand{\SLOWjythonCAPOnoGExcDynamicCIMIN}{35}
\newcommand{\SLOWjythonCAPOnoGExcDynamicCIMAX}{35}
\newcommand{\SLOWjythonCAPOnoG}{\rna}
\newcommand{\SLOWjythonCAPOnoGCI}{\rna}
\newcommand{\SLOWjythonCAPOnoGCIMIN}{\rna}
\newcommand{\SLOWjythonCAPOnoGCIMAX}{\rna}
\newcommand{\SLOWjythonCAPOnoGDynamic}{\rna}
\newcommand{\SLOWjythonCAPOnoGDynamicCI}{\rna}
\newcommand{\SLOWjythonCAPOnoGDynamicCIMIN}{\rna}
\newcommand{\SLOWjythonCAPOnoGDynamicCIMAX}{\rna}
\newcommand{\SLOWjythonCAPOExc}{3}
\newcommand{\SLOWjythonCAPOExcCI}{0.98}
\newcommand{\SLOWjythonCAPOExcCIMIN}{2}
\newcommand{\SLOWjythonCAPOExcCIMAX}{4}
\newcommand{\SLOWjythonCAPOExcDynamic}{4}
\newcommand{\SLOWjythonCAPOExcDynamicCI}{2.0}
\newcommand{\SLOWjythonCAPOExcDynamicCIMIN}{2}
\newcommand{\SLOWjythonCAPOExcDynamicCIMAX}{6}
\newcommand{\SLOWjythonCAPO}{\rna}
\newcommand{\SLOWjythonCAPOCI}{\rna}
\newcommand{\SLOWjythonCAPOCIMIN}{\rna}
\newcommand{\SLOWjythonCAPOCIMAX}{\rna}
\newcommand{\SLOWjythonCAPODynamic}{\rna}
\newcommand{\SLOWjythonCAPODynamicCI}{\rna}
\newcommand{\SLOWjythonCAPODynamicCIMIN}{\rna}
\newcommand{\SLOWjythonCAPODynamicCIMAX}{\rna}
\newcommand{\SLOWjythonPIP}{\rna}
\newcommand{\SLOWjythonPIPCI}{\rna}
\newcommand{\SLOWjythonPIPCIMIN}{\rna}
\newcommand{\SLOWjythonPIPCIMAX}{\rna}
\newcommand{\SLOWjythonPIPDynamic}{\rna}
\newcommand{\SLOWjythonPIPDynamicCI}{\rna}
\newcommand{\SLOWjythonPIPDynamicCIMIN}{\rna}
\newcommand{\SLOWjythonPIPDynamicCIMAX}{\rna}
\newcommand{\SLOWjythonHBUP}{21}
\newcommand{\SLOWjythonHBUPCI}{2.9}
\newcommand{\SLOWjythonHBUPCIMIN}{18}
\newcommand{\SLOWjythonHBUPCIMAX}{24}
\newcommand{\SLOWjythonHBDynamicUP}{22}
\newcommand{\SLOWjythonHBDynamicUPCI}{4.9}
\newcommand{\SLOWjythonHBDynamicUPCIMIN}{17}
\newcommand{\SLOWjythonHBDynamicUPCIMAX}{27}
\newcommand{\SLOWjythonWCPUP}{22}
\newcommand{\SLOWjythonWCPUPCI}{0.0}
\newcommand{\SLOWjythonWCPUPCIMIN}{22}
\newcommand{\SLOWjythonWCPUPCIMAX}{22}
\newcommand{\SLOWjythonWCPDynamicUP}{25}
\newcommand{\SLOWjythonWCPDynamicUPCI}{0.98}
\newcommand{\SLOWjythonWCPDynamicUPCIMIN}{24}
\newcommand{\SLOWjythonWCPDynamicUPCIMAX}{26}
\newcommand{\SLOWjythonWDCnoGUP}{\rna}
\newcommand{\SLOWjythonWDCnoGUPCI}{\rna}
\newcommand{\SLOWjythonWDCnoGUPCIMIN}{\rna}
\newcommand{\SLOWjythonWDCnoGUPCIMAX}{\rna}
\newcommand{\SLOWjythonWDCnoGDynamicUP}{\rna}
\newcommand{\SLOWjythonWDCnoGDynamicUPCI}{\rna}
\newcommand{\SLOWjythonWDCnoGDynamicUPCIMIN}{\rna}
\newcommand{\SLOWjythonWDCnoGDynamicUPCIMAX}{\rna}
\newcommand{\SLOWjythonWDCUP}{\rna}
\newcommand{\SLOWjythonWDCUPCI}{\rna}
\newcommand{\SLOWjythonWDCUPCIMIN}{\rna}
\newcommand{\SLOWjythonWDCUPCIMAX}{\rna}
\newcommand{\SLOWjythonWDCDynamicUP}{\rna}
\newcommand{\SLOWjythonWDCDynamicUPCI}{\rna}
\newcommand{\SLOWjythonWDCDynamicUPCIMIN}{\rna}
\newcommand{\SLOWjythonWDCDynamicUPCIMAX}{\rna}
\newcommand{\SLOWjythonCAPOnoGUP}{\rna}
\newcommand{\SLOWjythonCAPOnoGUPCI}{\rna}
\newcommand{\SLOWjythonCAPOnoGUPCIMIN}{\rna}
\newcommand{\SLOWjythonCAPOnoGUPCIMAX}{\rna}
\newcommand{\SLOWjythonCAPOnoGDynamicUP}{\rna}
\newcommand{\SLOWjythonCAPOnoGDynamicUPCI}{\rna}
\newcommand{\SLOWjythonCAPOnoGDynamicUPCIMIN}{\rna}
\newcommand{\SLOWjythonCAPOnoGDynamicUPCIMAX}{\rna}
\newcommand{\SLOWjythonCAPOUP}{\rna}
\newcommand{\SLOWjythonCAPOUPCI}{\rna}
\newcommand{\SLOWjythonCAPOUPCIMIN}{\rna}
\newcommand{\SLOWjythonCAPOUPCIMAX}{\rna}
\newcommand{\SLOWjythonCAPODynamicUP}{\rna}
\newcommand{\SLOWjythonCAPODynamicUPCI}{\rna}
\newcommand{\SLOWjythonCAPODynamicUPCIMIN}{\rna}
\newcommand{\SLOWjythonCAPODynamicUPCIMAX}{\rna}
\newcommand{\SLOWjythonPIPUP}{\rna}
\newcommand{\SLOWjythonPIPUPCI}{\rna}
\newcommand{\SLOWjythonPIPUPCIMIN}{\rna}
\newcommand{\SLOWjythonPIPUPCIMAX}{\rna}
\newcommand{\SLOWjythonPIPDynamicUP}{\rna}
\newcommand{\SLOWjythonPIPDynamicUPCI}{\rna}
\newcommand{\SLOWjythonPIPDynamicUPCIMIN}{\rna}
\newcommand{\SLOWjythonPIPDynamicUPCIMAX}{\rna}
\newcommand{\SLOWjythonPIPHB}{\rna}
\newcommand{\SLOWjythonPIPHBCI}{\rna}
\newcommand{\SLOWjythonPIPHBCIMIN}{\rna}
\newcommand{\SLOWjythonPIPHBCIMAX}{\rna}
\newcommand{\SLOWjythonPIPHBDynamic}{\rna}
\newcommand{\SLOWjythonPIPHBDynamicCI}{\rna}
\newcommand{\SLOWjythonPIPHBDynamicCIMIN}{\rna}
\newcommand{\SLOWjythonPIPHBDynamicCIMAX}{\rna}
\newcommand{\SLOWjythonPIPWCP}{\rna}
\newcommand{\SLOWjythonPIPWCPCI}{\rna}
\newcommand{\SLOWjythonPIPWCPCIMIN}{\rna}
\newcommand{\SLOWjythonPIPWCPCIMAX}{\rna}
\newcommand{\SLOWjythonPIPWCPDynamic}{\rna}
\newcommand{\SLOWjythonPIPWCPDynamicCI}{\rna}
\newcommand{\SLOWjythonPIPWCPDynamicCIMIN}{\rna}
\newcommand{\SLOWjythonPIPWCPDynamicCIMAX}{\rna}
\newcommand{\SLOWjythonPIPWDC}{\rna}
\newcommand{\SLOWjythonPIPWDCCI}{\rna}
\newcommand{\SLOWjythonPIPWDCCIMIN}{\rna}
\newcommand{\SLOWjythonPIPWDCCIMAX}{\rna}
\newcommand{\SLOWjythonPIPWDCDynamic}{\rna}
\newcommand{\SLOWjythonPIPWDCDynamicCI}{\rna}
\newcommand{\SLOWjythonPIPWDCDynamicCIMIN}{\rna}
\newcommand{\SLOWjythonPIPWDCDynamicCIMAX}{\rna}
\newcommand{\SLOWjythonPIPCAPO}{\rna}
\newcommand{\SLOWjythonPIPCAPOCI}{\rna}
\newcommand{\SLOWjythonPIPCAPOCIMIN}{\rna}
\newcommand{\SLOWjythonPIPCAPOCIMAX}{\rna}
\newcommand{\SLOWjythonPIPCAPODynamic}{\rna}
\newcommand{\SLOWjythonPIPCAPODynamicCI}{\rna}
\newcommand{\SLOWjythonPIPCAPODynamicCIMIN}{\rna}
\newcommand{\SLOWjythonPIPCAPODynamicCIMAX}{\rna}
\newcommand{\SLOWjythonPIPPIP}{\rna}
\newcommand{\SLOWjythonPIPPIPCI}{\rna}
\newcommand{\SLOWjythonPIPPIPCIMIN}{\rna}
\newcommand{\SLOWjythonPIPPIPCIMAX}{\rna}
\newcommand{\SLOWjythonPIPPIPDynamic}{\rna}
\newcommand{\SLOWjythonPIPPIPDynamicCI}{\rna}
\newcommand{\SLOWjythonPIPPIPDynamicCIMIN}{\rna}
\newcommand{\SLOWjythonPIPPIPDynamicCIMAX}{\rna}
\newcommand{\SLOWluindexEvents}{0}
\newcommand{\SLOWluindexNoFPEvents}{0}
\newcommand{\SLOWluindexMaxLiveThreads}{3}
\newcommand{\SLOWluindexTotalThreads}{3}
\newcommand{\SLOWluindexBaseTime}{1.2}
\newcommand{\SLOWluindexBaseTimeCI}{130}
\newcommand{\SLOWluindexEmptyTime}{\rna}
\newcommand{\SLOWluindexEmptyTimeCI}{\rna}
\newcommand{\SLOWluindexEmptyTimeCIMIN}{\rna}
\newcommand{\SLOWluindexEmptyTimeCIMAX}{\rna}
\newcommand{\SLOWluindexFTTime}{7.5}
\newcommand{\SLOWluindexFTTimeCI}{0.82}
\newcommand{\SLOWluindexHBTime}{25}
\newcommand{\SLOWluindexHBTimeCI}{2.5}
\newcommand{\SLOWluindexWCPTime}{42}
\newcommand{\SLOWluindexWCPTimeCI}{3.6}
\newcommand{\SLOWluindexDCnoGExcTime}{36}
\newcommand{\SLOWluindexDCnoGExcTimeCI}{2.4}
\newcommand{\SLOWluindexDCnoGTime}{\rna}
\newcommand{\SLOWluindexDCnoGTimeCI}{\rna}
\newcommand{\SLOWluindexDCnoGTimeCIMIN}{\rna}
\newcommand{\SLOWluindexDCnoGTimeCIMAX}{\rna}
\newcommand{\SLOWluindexDCExcTime}{41}
\newcommand{\SLOWluindexDCExcTimeCI}{0.13}
\newcommand{\SLOWluindexDCTime}{\rna}
\newcommand{\SLOWluindexDCTimeCI}{\rna}
\newcommand{\SLOWluindexDCTimeCIMIN}{\rna}
\newcommand{\SLOWluindexDCTimeCIMAX}{\rna}
\newcommand{\SLOWluindexCAPOnoGExcTime}{36}
\newcommand{\SLOWluindexCAPOnoGExcTimeCI}{4.6}
\newcommand{\SLOWluindexCAPOnoGTime}{\rna}
\newcommand{\SLOWluindexCAPOnoGTimeCI}{\rna}
\newcommand{\SLOWluindexCAPOnoGTimeCIMIN}{\rna}
\newcommand{\SLOWluindexCAPOnoGTimeCIMAX}{\rna}
\newcommand{\SLOWluindexCAPOExcTime}{39}
\newcommand{\SLOWluindexCAPOExcTimeCI}{2.4}
\newcommand{\SLOWluindexCAPOTime}{\rna}
\newcommand{\SLOWluindexCAPOTimeCI}{\rna}
\newcommand{\SLOWluindexCAPOTimeCIMIN}{\rna}
\newcommand{\SLOWluindexCAPOTimeCIMAX}{\rna}
\newcommand{\SLOWluindexStaticTime}{\rzero}
\newcommand{\SLOWluindexDynamicTime}{\rzero}
\newcommand{\SLOWluindexBaseMem}{110}
\newcommand{\SLOWluindexBaseMemCI}{6.8}
\newcommand{\SLOWluindexHBMem}{37}
\newcommand{\SLOWluindexHBMemCI}{2.3}
\newcommand{\SLOWluindexFTMem}{5.3}
\newcommand{\SLOWluindexFTMemCI}{0.25}
\newcommand{\SLOWluindexWCPMem}{75}
\newcommand{\SLOWluindexWCPMemCI}{4.1}
\newcommand{\SLOWluindexDCnoGExcMem}{53}
\newcommand{\SLOWluindexDCnoGExcMemCI}{3.4}
\newcommand{\SLOWluindexDCnoGMem}{\memna}
\newcommand{\SLOWluindexDCnoGMemCI}{\memna}
\newcommand{\SLOWluindexDCnoGMemCIMIN}{\memna}
\newcommand{\SLOWluindexDCnoGMemCIMAX}{\memna}
\newcommand{\SLOWluindexDCExcMem}{68}
\newcommand{\SLOWluindexDCExcMemCI}{3.4}
\newcommand{\SLOWluindexDCMem}{\memna}
\newcommand{\SLOWluindexDCMemCI}{\memna}
\newcommand{\SLOWluindexDCMemCIMIN}{\memna}
\newcommand{\SLOWluindexDCMemCIMAX}{\memna}
\newcommand{\SLOWluindexCAPOnoGExcMem}{53}
\newcommand{\SLOWluindexCAPOnoGExcMemCI}{3.5}
\newcommand{\SLOWluindexCAPOnoGMem}{\memna}
\newcommand{\SLOWluindexCAPOnoGMemCI}{\memna}
\newcommand{\SLOWluindexCAPOnoGMemCIMIN}{\memna}
\newcommand{\SLOWluindexCAPOnoGMemCIMAX}{\memna}
\newcommand{\SLOWluindexCAPOExcMem}{69}
\newcommand{\SLOWluindexCAPOExcMemCI}{3.8}
\newcommand{\SLOWluindexCAPOMem}{\memna}
\newcommand{\SLOWluindexCAPOMemCI}{\memna}
\newcommand{\SLOWluindexCAPOMemCIMIN}{\memna}
\newcommand{\SLOWluindexCAPOMemCIMAX}{\memna}
\newcommand{\SLOWluindexEventsCI}{0}
\newcommand{\SLOWluindexEventsCIMIN}{0}
\newcommand{\SLOWluindexEventsCIMAX}{0}
\newcommand{\SLOWluindexNoFPEventsCI}{0}
\newcommand{\SLOWluindexNoFPEventsCIMIN}{0}
\newcommand{\SLOWluindexNoFPEventsCIMAX}{0}
\newcommand{\SLOWluindexHB}{1}
\newcommand{\SLOWluindexHBCI}{0.0}
\newcommand{\SLOWluindexHBCIMIN}{1}
\newcommand{\SLOWluindexHBCIMAX}{1}
\newcommand{\SLOWluindexHBDynamic}{1}
\newcommand{\SLOWluindexHBDynamicCI}{0.0}
\newcommand{\SLOWluindexHBDynamicCIMIN}{1}
\newcommand{\SLOWluindexHBDynamicCIMAX}{1}
\newcommand{\SLOWluindexFT}{1}
\newcommand{\SLOWluindexFTCI}{0.0}
\newcommand{\SLOWluindexFTCIMIN}{1}
\newcommand{\SLOWluindexFTCIMAX}{1}
\newcommand{\SLOWluindexFTDynamic}{1}
\newcommand{\SLOWluindexFTDynamicCI}{0.0}
\newcommand{\SLOWluindexFTDynamicCIMIN}{1}
\newcommand{\SLOWluindexFTDynamicCIMAX}{1}
\newcommand{\SLOWluindexWCP}{1}
\newcommand{\SLOWluindexWCPCI}{0.0}
\newcommand{\SLOWluindexWCPCIMIN}{1}
\newcommand{\SLOWluindexWCPCIMAX}{1}
\newcommand{\SLOWluindexWCPDynamic}{1}
\newcommand{\SLOWluindexWCPDynamicCI}{0.0}
\newcommand{\SLOWluindexWCPDynamicCIMIN}{1}
\newcommand{\SLOWluindexWCPDynamicCIMAX}{1}
\newcommand{\SLOWluindexDCnoGExc}{1}
\newcommand{\SLOWluindexDCnoGExcCI}{0.0}
\newcommand{\SLOWluindexDCnoGExcCIMIN}{1}
\newcommand{\SLOWluindexDCnoGExcCIMAX}{1}
\newcommand{\SLOWluindexDCnoGExcDynamic}{1}
\newcommand{\SLOWluindexDCnoGExcDynamicCI}{0.0}
\newcommand{\SLOWluindexDCnoGExcDynamicCIMIN}{1}
\newcommand{\SLOWluindexDCnoGExcDynamicCIMAX}{1}
\newcommand{\SLOWluindexDCnoG}{\rna}
\newcommand{\SLOWluindexDCnoGCI}{\rna}
\newcommand{\SLOWluindexDCnoGCIMIN}{\rna}
\newcommand{\SLOWluindexDCnoGCIMAX}{\rna}
\newcommand{\SLOWluindexDCnoGDynamic}{\rna}
\newcommand{\SLOWluindexDCnoGDynamicCI}{\rna}
\newcommand{\SLOWluindexDCnoGDynamicCIMIN}{\rna}
\newcommand{\SLOWluindexDCnoGDynamicCIMAX}{\rna}
\newcommand{\SLOWluindexDCExc}{1}
\newcommand{\SLOWluindexDCExcCI}{0.0}
\newcommand{\SLOWluindexDCExcCIMIN}{1}
\newcommand{\SLOWluindexDCExcCIMAX}{1}
\newcommand{\SLOWluindexDCExcDynamic}{1}
\newcommand{\SLOWluindexDCExcDynamicCI}{0.0}
\newcommand{\SLOWluindexDCExcDynamicCIMIN}{1}
\newcommand{\SLOWluindexDCExcDynamicCIMAX}{1}
\newcommand{\SLOWluindexDC}{\rna}
\newcommand{\SLOWluindexDCCI}{\rna}
\newcommand{\SLOWluindexDCCIMIN}{\rna}
\newcommand{\SLOWluindexDCCIMAX}{\rna}
\newcommand{\SLOWluindexDCDynamic}{\rna}
\newcommand{\SLOWluindexDCDynamicCI}{\rna}
\newcommand{\SLOWluindexDCDynamicCIMIN}{\rna}
\newcommand{\SLOWluindexDCDynamicCIMAX}{\rna}
\newcommand{\SLOWluindexCAPOnoGExc}{1}
\newcommand{\SLOWluindexCAPOnoGExcCI}{0.0}
\newcommand{\SLOWluindexCAPOnoGExcCIMIN}{1}
\newcommand{\SLOWluindexCAPOnoGExcCIMAX}{1}
\newcommand{\SLOWluindexCAPOnoGExcDynamic}{1}
\newcommand{\SLOWluindexCAPOnoGExcDynamicCI}{0.0}
\newcommand{\SLOWluindexCAPOnoGExcDynamicCIMIN}{1}
\newcommand{\SLOWluindexCAPOnoGExcDynamicCIMAX}{1}
\newcommand{\SLOWluindexCAPOnoG}{\rna}
\newcommand{\SLOWluindexCAPOnoGCI}{\rna}
\newcommand{\SLOWluindexCAPOnoGCIMIN}{\rna}
\newcommand{\SLOWluindexCAPOnoGCIMAX}{\rna}
\newcommand{\SLOWluindexCAPOnoGDynamic}{\rna}
\newcommand{\SLOWluindexCAPOnoGDynamicCI}{\rna}
\newcommand{\SLOWluindexCAPOnoGDynamicCIMIN}{\rna}
\newcommand{\SLOWluindexCAPOnoGDynamicCIMAX}{\rna}
\newcommand{\SLOWluindexCAPOExc}{1}
\newcommand{\SLOWluindexCAPOExcCI}{0.0}
\newcommand{\SLOWluindexCAPOExcCIMIN}{1}
\newcommand{\SLOWluindexCAPOExcCIMAX}{1}
\newcommand{\SLOWluindexCAPOExcDynamic}{1}
\newcommand{\SLOWluindexCAPOExcDynamicCI}{0.0}
\newcommand{\SLOWluindexCAPOExcDynamicCIMIN}{1}
\newcommand{\SLOWluindexCAPOExcDynamicCIMAX}{1}
\newcommand{\SLOWluindexCAPO}{\rna}
\newcommand{\SLOWluindexCAPOCI}{\rna}
\newcommand{\SLOWluindexCAPOCIMIN}{\rna}
\newcommand{\SLOWluindexCAPOCIMAX}{\rna}
\newcommand{\SLOWluindexCAPODynamic}{\rna}
\newcommand{\SLOWluindexCAPODynamicCI}{\rna}
\newcommand{\SLOWluindexCAPODynamicCIMIN}{\rna}
\newcommand{\SLOWluindexCAPODynamicCIMAX}{\rna}
\newcommand{\SLOWluindexPIP}{\rna}
\newcommand{\SLOWluindexPIPCI}{\rna}
\newcommand{\SLOWluindexPIPCIMIN}{\rna}
\newcommand{\SLOWluindexPIPCIMAX}{\rna}
\newcommand{\SLOWluindexPIPDynamic}{\rna}
\newcommand{\SLOWluindexPIPDynamicCI}{\rna}
\newcommand{\SLOWluindexPIPDynamicCIMIN}{\rna}
\newcommand{\SLOWluindexPIPDynamicCIMAX}{\rna}
\newcommand{\SLOWluindexHBUP}{1}
\newcommand{\SLOWluindexHBUPCI}{0.0}
\newcommand{\SLOWluindexHBUPCIMIN}{1}
\newcommand{\SLOWluindexHBUPCIMAX}{1}
\newcommand{\SLOWluindexHBDynamicUP}{1}
\newcommand{\SLOWluindexHBDynamicUPCI}{0.0}
\newcommand{\SLOWluindexHBDynamicUPCIMIN}{1}
\newcommand{\SLOWluindexHBDynamicUPCIMAX}{1}
\newcommand{\SLOWluindexWCPUP}{1}
\newcommand{\SLOWluindexWCPUPCI}{0.0}
\newcommand{\SLOWluindexWCPUPCIMIN}{1}
\newcommand{\SLOWluindexWCPUPCIMAX}{1}
\newcommand{\SLOWluindexWCPDynamicUP}{1}
\newcommand{\SLOWluindexWCPDynamicUPCI}{0.0}
\newcommand{\SLOWluindexWCPDynamicUPCIMIN}{1}
\newcommand{\SLOWluindexWCPDynamicUPCIMAX}{1}
\newcommand{\SLOWluindexWDCnoGUP}{\rna}
\newcommand{\SLOWluindexWDCnoGUPCI}{\rna}
\newcommand{\SLOWluindexWDCnoGUPCIMIN}{\rna}
\newcommand{\SLOWluindexWDCnoGUPCIMAX}{\rna}
\newcommand{\SLOWluindexWDCnoGDynamicUP}{\rna}
\newcommand{\SLOWluindexWDCnoGDynamicUPCI}{\rna}
\newcommand{\SLOWluindexWDCnoGDynamicUPCIMIN}{\rna}
\newcommand{\SLOWluindexWDCnoGDynamicUPCIMAX}{\rna}
\newcommand{\SLOWluindexWDCUP}{\rna}
\newcommand{\SLOWluindexWDCUPCI}{\rna}
\newcommand{\SLOWluindexWDCUPCIMIN}{\rna}
\newcommand{\SLOWluindexWDCUPCIMAX}{\rna}
\newcommand{\SLOWluindexWDCDynamicUP}{\rna}
\newcommand{\SLOWluindexWDCDynamicUPCI}{\rna}
\newcommand{\SLOWluindexWDCDynamicUPCIMIN}{\rna}
\newcommand{\SLOWluindexWDCDynamicUPCIMAX}{\rna}
\newcommand{\SLOWluindexCAPOnoGUP}{\rna}
\newcommand{\SLOWluindexCAPOnoGUPCI}{\rna}
\newcommand{\SLOWluindexCAPOnoGUPCIMIN}{\rna}
\newcommand{\SLOWluindexCAPOnoGUPCIMAX}{\rna}
\newcommand{\SLOWluindexCAPOnoGDynamicUP}{\rna}
\newcommand{\SLOWluindexCAPOnoGDynamicUPCI}{\rna}
\newcommand{\SLOWluindexCAPOnoGDynamicUPCIMIN}{\rna}
\newcommand{\SLOWluindexCAPOnoGDynamicUPCIMAX}{\rna}
\newcommand{\SLOWluindexCAPOUP}{\rna}
\newcommand{\SLOWluindexCAPOUPCI}{\rna}
\newcommand{\SLOWluindexCAPOUPCIMIN}{\rna}
\newcommand{\SLOWluindexCAPOUPCIMAX}{\rna}
\newcommand{\SLOWluindexCAPODynamicUP}{\rna}
\newcommand{\SLOWluindexCAPODynamicUPCI}{\rna}
\newcommand{\SLOWluindexCAPODynamicUPCIMIN}{\rna}
\newcommand{\SLOWluindexCAPODynamicUPCIMAX}{\rna}
\newcommand{\SLOWluindexPIPUP}{\rna}
\newcommand{\SLOWluindexPIPUPCI}{\rna}
\newcommand{\SLOWluindexPIPUPCIMIN}{\rna}
\newcommand{\SLOWluindexPIPUPCIMAX}{\rna}
\newcommand{\SLOWluindexPIPDynamicUP}{\rna}
\newcommand{\SLOWluindexPIPDynamicUPCI}{\rna}
\newcommand{\SLOWluindexPIPDynamicUPCIMIN}{\rna}
\newcommand{\SLOWluindexPIPDynamicUPCIMAX}{\rna}
\newcommand{\SLOWluindexPIPHB}{\rna}
\newcommand{\SLOWluindexPIPHBCI}{\rna}
\newcommand{\SLOWluindexPIPHBCIMIN}{\rna}
\newcommand{\SLOWluindexPIPHBCIMAX}{\rna}
\newcommand{\SLOWluindexPIPHBDynamic}{\rna}
\newcommand{\SLOWluindexPIPHBDynamicCI}{\rna}
\newcommand{\SLOWluindexPIPHBDynamicCIMIN}{\rna}
\newcommand{\SLOWluindexPIPHBDynamicCIMAX}{\rna}
\newcommand{\SLOWluindexPIPWCP}{\rna}
\newcommand{\SLOWluindexPIPWCPCI}{\rna}
\newcommand{\SLOWluindexPIPWCPCIMIN}{\rna}
\newcommand{\SLOWluindexPIPWCPCIMAX}{\rna}
\newcommand{\SLOWluindexPIPWCPDynamic}{\rna}
\newcommand{\SLOWluindexPIPWCPDynamicCI}{\rna}
\newcommand{\SLOWluindexPIPWCPDynamicCIMIN}{\rna}
\newcommand{\SLOWluindexPIPWCPDynamicCIMAX}{\rna}
\newcommand{\SLOWluindexPIPWDC}{\rna}
\newcommand{\SLOWluindexPIPWDCCI}{\rna}
\newcommand{\SLOWluindexPIPWDCCIMIN}{\rna}
\newcommand{\SLOWluindexPIPWDCCIMAX}{\rna}
\newcommand{\SLOWluindexPIPWDCDynamic}{\rna}
\newcommand{\SLOWluindexPIPWDCDynamicCI}{\rna}
\newcommand{\SLOWluindexPIPWDCDynamicCIMIN}{\rna}
\newcommand{\SLOWluindexPIPWDCDynamicCIMAX}{\rna}
\newcommand{\SLOWluindexPIPCAPO}{\rna}
\newcommand{\SLOWluindexPIPCAPOCI}{\rna}
\newcommand{\SLOWluindexPIPCAPOCIMIN}{\rna}
\newcommand{\SLOWluindexPIPCAPOCIMAX}{\rna}
\newcommand{\SLOWluindexPIPCAPODynamic}{\rna}
\newcommand{\SLOWluindexPIPCAPODynamicCI}{\rna}
\newcommand{\SLOWluindexPIPCAPODynamicCIMIN}{\rna}
\newcommand{\SLOWluindexPIPCAPODynamicCIMAX}{\rna}
\newcommand{\SLOWluindexPIPPIP}{\rna}
\newcommand{\SLOWluindexPIPPIPCI}{\rna}
\newcommand{\SLOWluindexPIPPIPCIMIN}{\rna}
\newcommand{\SLOWluindexPIPPIPCIMAX}{\rna}
\newcommand{\SLOWluindexPIPPIPDynamic}{\rna}
\newcommand{\SLOWluindexPIPPIPDynamicCI}{\rna}
\newcommand{\SLOWluindexPIPPIPDynamicCIMIN}{\rna}
\newcommand{\SLOWluindexPIPPIPDynamicCIMAX}{\rna}
\newcommand{\SLOWlusearchEvents}{0}
\newcommand{\SLOWlusearchNoFPEvents}{0}
\newcommand{\SLOWlusearchMaxLiveThreads}{16}
\newcommand{\SLOWlusearchTotalThreads}{16}
\newcommand{\SLOWlusearchBaseTime}{0.98}
\newcommand{\SLOWlusearchBaseTimeCI}{15}
\newcommand{\SLOWlusearchEmptyTime}{\rna}
\newcommand{\SLOWlusearchEmptyTimeCI}{\rna}
\newcommand{\SLOWlusearchEmptyTimeCIMIN}{\rna}
\newcommand{\SLOWlusearchEmptyTimeCIMAX}{\rna}
\newcommand{\SLOWlusearchFTTime}{10}
\newcommand{\SLOWlusearchFTTimeCI}{0.84}
\newcommand{\SLOWlusearchHBTime}{22}
\newcommand{\SLOWlusearchHBTimeCI}{1.3}
\newcommand{\SLOWlusearchWCPTime}{40}
\newcommand{\SLOWlusearchWCPTimeCI}{12}
\newcommand{\SLOWlusearchDCnoGExcTime}{26}
\newcommand{\SLOWlusearchDCnoGExcTimeCI}{0.25}
\newcommand{\SLOWlusearchDCnoGTime}{\rna}
\newcommand{\SLOWlusearchDCnoGTimeCI}{\rna}
\newcommand{\SLOWlusearchDCnoGTimeCIMIN}{\rna}
\newcommand{\SLOWlusearchDCnoGTimeCIMAX}{\rna}
\newcommand{\SLOWlusearchDCExcTime}{28}
\newcommand{\SLOWlusearchDCExcTimeCI}{2.6}
\newcommand{\SLOWlusearchDCTime}{\rna}
\newcommand{\SLOWlusearchDCTimeCI}{\rna}
\newcommand{\SLOWlusearchDCTimeCIMIN}{\rna}
\newcommand{\SLOWlusearchDCTimeCIMAX}{\rna}
\newcommand{\SLOWlusearchCAPOnoGExcTime}{29}
\newcommand{\SLOWlusearchCAPOnoGExcTimeCI}{11}
\newcommand{\SLOWlusearchCAPOnoGTime}{\rna}
\newcommand{\SLOWlusearchCAPOnoGTimeCI}{\rna}
\newcommand{\SLOWlusearchCAPOnoGTimeCIMIN}{\rna}
\newcommand{\SLOWlusearchCAPOnoGTimeCIMAX}{\rna}
\newcommand{\SLOWlusearchCAPOExcTime}{26}
\newcommand{\SLOWlusearchCAPOExcTimeCI}{0.59}
\newcommand{\SLOWlusearchCAPOTime}{\rna}
\newcommand{\SLOWlusearchCAPOTimeCI}{\rna}
\newcommand{\SLOWlusearchCAPOTimeCIMIN}{\rna}
\newcommand{\SLOWlusearchCAPOTimeCIMAX}{\rna}
\newcommand{\SLOWlusearchStaticTime}{\rzero}
\newcommand{\SLOWlusearchDynamicTime}{\rzero}
\newcommand{\SLOWlusearchBaseMem}{1,600}
\newcommand{\SLOWlusearchBaseMemCI}{0.82}
\newcommand{\SLOWlusearchHBMem}{13}
\newcommand{\SLOWlusearchHBMemCI}{1.2}
\newcommand{\SLOWlusearchFTMem}{8.2}
\newcommand{\SLOWlusearchFTMemCI}{0.27}
\newcommand{\SLOWlusearchWCPMem}{20}
\newcommand{\SLOWlusearchWCPMemCI}{2.9}
\newcommand{\SLOWlusearchDCnoGExcMem}{13}
\newcommand{\SLOWlusearchDCnoGExcMemCI}{0.96}
\newcommand{\SLOWlusearchDCnoGMem}{\memna}
\newcommand{\SLOWlusearchDCnoGMemCI}{\memna}
\newcommand{\SLOWlusearchDCnoGMemCIMIN}{\memna}
\newcommand{\SLOWlusearchDCnoGMemCIMAX}{\memna}
\newcommand{\SLOWlusearchDCExcMem}{15}
\newcommand{\SLOWlusearchDCExcMemCI}{0.34}
\newcommand{\SLOWlusearchDCMem}{\memna}
\newcommand{\SLOWlusearchDCMemCI}{\memna}
\newcommand{\SLOWlusearchDCMemCIMIN}{\memna}
\newcommand{\SLOWlusearchDCMemCIMAX}{\memna}
\newcommand{\SLOWlusearchCAPOnoGExcMem}{13}
\newcommand{\SLOWlusearchCAPOnoGExcMemCI}{0.084}
\newcommand{\SLOWlusearchCAPOnoGMem}{\memna}
\newcommand{\SLOWlusearchCAPOnoGMemCI}{\memna}
\newcommand{\SLOWlusearchCAPOnoGMemCIMIN}{\memna}
\newcommand{\SLOWlusearchCAPOnoGMemCIMAX}{\memna}
\newcommand{\SLOWlusearchCAPOExcMem}{15}
\newcommand{\SLOWlusearchCAPOExcMemCI}{0.69}
\newcommand{\SLOWlusearchCAPOMem}{\memna}
\newcommand{\SLOWlusearchCAPOMemCI}{\memna}
\newcommand{\SLOWlusearchCAPOMemCIMIN}{\memna}
\newcommand{\SLOWlusearchCAPOMemCIMAX}{\memna}
\newcommand{\SLOWlusearchEventsCI}{0}
\newcommand{\SLOWlusearchEventsCIMIN}{0}
\newcommand{\SLOWlusearchEventsCIMAX}{0}
\newcommand{\SLOWlusearchNoFPEventsCI}{0}
\newcommand{\SLOWlusearchNoFPEventsCIMIN}{0}
\newcommand{\SLOWlusearchNoFPEventsCIMAX}{0}
\newcommand{\SLOWlusearchHB}{0}
\newcommand{\SLOWlusearchHBCI}{0.0}
\newcommand{\SLOWlusearchHBCIMIN}{0}
\newcommand{\SLOWlusearchHBCIMAX}{0}
\newcommand{\SLOWlusearchHBDynamic}{0}
\newcommand{\SLOWlusearchHBDynamicCI}{0.0}
\newcommand{\SLOWlusearchHBDynamicCIMIN}{0}
\newcommand{\SLOWlusearchHBDynamicCIMAX}{0}
\newcommand{\SLOWlusearchFT}{0}
\newcommand{\SLOWlusearchFTCI}{0.0}
\newcommand{\SLOWlusearchFTCIMIN}{0}
\newcommand{\SLOWlusearchFTCIMAX}{0}
\newcommand{\SLOWlusearchFTDynamic}{0}
\newcommand{\SLOWlusearchFTDynamicCI}{0.0}
\newcommand{\SLOWlusearchFTDynamicCIMIN}{0}
\newcommand{\SLOWlusearchFTDynamicCIMAX}{0}
\newcommand{\SLOWlusearchWCP}{0}
\newcommand{\SLOWlusearchWCPCI}{0.0}
\newcommand{\SLOWlusearchWCPCIMIN}{0}
\newcommand{\SLOWlusearchWCPCIMAX}{0}
\newcommand{\SLOWlusearchWCPDynamic}{0}
\newcommand{\SLOWlusearchWCPDynamicCI}{0.0}
\newcommand{\SLOWlusearchWCPDynamicCIMIN}{0}
\newcommand{\SLOWlusearchWCPDynamicCIMAX}{0}
\newcommand{\SLOWlusearchDCnoGExc}{0}
\newcommand{\SLOWlusearchDCnoGExcCI}{0.0}
\newcommand{\SLOWlusearchDCnoGExcCIMIN}{0}
\newcommand{\SLOWlusearchDCnoGExcCIMAX}{0}
\newcommand{\SLOWlusearchDCnoGExcDynamic}{0}
\newcommand{\SLOWlusearchDCnoGExcDynamicCI}{0.0}
\newcommand{\SLOWlusearchDCnoGExcDynamicCIMIN}{0}
\newcommand{\SLOWlusearchDCnoGExcDynamicCIMAX}{0}
\newcommand{\SLOWlusearchDCnoG}{\rna}
\newcommand{\SLOWlusearchDCnoGCI}{\rna}
\newcommand{\SLOWlusearchDCnoGCIMIN}{\rna}
\newcommand{\SLOWlusearchDCnoGCIMAX}{\rna}
\newcommand{\SLOWlusearchDCnoGDynamic}{\rna}
\newcommand{\SLOWlusearchDCnoGDynamicCI}{\rna}
\newcommand{\SLOWlusearchDCnoGDynamicCIMIN}{\rna}
\newcommand{\SLOWlusearchDCnoGDynamicCIMAX}{\rna}
\newcommand{\SLOWlusearchDCExc}{0}
\newcommand{\SLOWlusearchDCExcCI}{0.0}
\newcommand{\SLOWlusearchDCExcCIMIN}{0}
\newcommand{\SLOWlusearchDCExcCIMAX}{0}
\newcommand{\SLOWlusearchDCExcDynamic}{0}
\newcommand{\SLOWlusearchDCExcDynamicCI}{0.0}
\newcommand{\SLOWlusearchDCExcDynamicCIMIN}{0}
\newcommand{\SLOWlusearchDCExcDynamicCIMAX}{0}
\newcommand{\SLOWlusearchDC}{\rna}
\newcommand{\SLOWlusearchDCCI}{\rna}
\newcommand{\SLOWlusearchDCCIMIN}{\rna}
\newcommand{\SLOWlusearchDCCIMAX}{\rna}
\newcommand{\SLOWlusearchDCDynamic}{\rna}
\newcommand{\SLOWlusearchDCDynamicCI}{\rna}
\newcommand{\SLOWlusearchDCDynamicCIMIN}{\rna}
\newcommand{\SLOWlusearchDCDynamicCIMAX}{\rna}
\newcommand{\SLOWlusearchCAPOnoGExc}{0}
\newcommand{\SLOWlusearchCAPOnoGExcCI}{0.0}
\newcommand{\SLOWlusearchCAPOnoGExcCIMIN}{0}
\newcommand{\SLOWlusearchCAPOnoGExcCIMAX}{0}
\newcommand{\SLOWlusearchCAPOnoGExcDynamic}{0}
\newcommand{\SLOWlusearchCAPOnoGExcDynamicCI}{0.0}
\newcommand{\SLOWlusearchCAPOnoGExcDynamicCIMIN}{0}
\newcommand{\SLOWlusearchCAPOnoGExcDynamicCIMAX}{0}
\newcommand{\SLOWlusearchCAPOnoG}{\rna}
\newcommand{\SLOWlusearchCAPOnoGCI}{\rna}
\newcommand{\SLOWlusearchCAPOnoGCIMIN}{\rna}
\newcommand{\SLOWlusearchCAPOnoGCIMAX}{\rna}
\newcommand{\SLOWlusearchCAPOnoGDynamic}{\rna}
\newcommand{\SLOWlusearchCAPOnoGDynamicCI}{\rna}
\newcommand{\SLOWlusearchCAPOnoGDynamicCIMIN}{\rna}
\newcommand{\SLOWlusearchCAPOnoGDynamicCIMAX}{\rna}
\newcommand{\SLOWlusearchCAPOExc}{0}
\newcommand{\SLOWlusearchCAPOExcCI}{0.0}
\newcommand{\SLOWlusearchCAPOExcCIMIN}{0}
\newcommand{\SLOWlusearchCAPOExcCIMAX}{0}
\newcommand{\SLOWlusearchCAPOExcDynamic}{0}
\newcommand{\SLOWlusearchCAPOExcDynamicCI}{0.0}
\newcommand{\SLOWlusearchCAPOExcDynamicCIMIN}{0}
\newcommand{\SLOWlusearchCAPOExcDynamicCIMAX}{0}
\newcommand{\SLOWlusearchCAPO}{\rna}
\newcommand{\SLOWlusearchCAPOCI}{\rna}
\newcommand{\SLOWlusearchCAPOCIMIN}{\rna}
\newcommand{\SLOWlusearchCAPOCIMAX}{\rna}
\newcommand{\SLOWlusearchCAPODynamic}{\rna}
\newcommand{\SLOWlusearchCAPODynamicCI}{\rna}
\newcommand{\SLOWlusearchCAPODynamicCIMIN}{\rna}
\newcommand{\SLOWlusearchCAPODynamicCIMAX}{\rna}
\newcommand{\SLOWlusearchPIP}{\rna}
\newcommand{\SLOWlusearchPIPCI}{\rna}
\newcommand{\SLOWlusearchPIPCIMIN}{\rna}
\newcommand{\SLOWlusearchPIPCIMAX}{\rna}
\newcommand{\SLOWlusearchPIPDynamic}{\rna}
\newcommand{\SLOWlusearchPIPDynamicCI}{\rna}
\newcommand{\SLOWlusearchPIPDynamicCIMIN}{\rna}
\newcommand{\SLOWlusearchPIPDynamicCIMAX}{\rna}
\newcommand{\SLOWlusearchHBUP}{0}
\newcommand{\SLOWlusearchHBUPCI}{0.0}
\newcommand{\SLOWlusearchHBUPCIMIN}{0}
\newcommand{\SLOWlusearchHBUPCIMAX}{0}
\newcommand{\SLOWlusearchHBDynamicUP}{0}
\newcommand{\SLOWlusearchHBDynamicUPCI}{0.0}
\newcommand{\SLOWlusearchHBDynamicUPCIMIN}{0}
\newcommand{\SLOWlusearchHBDynamicUPCIMAX}{0}
\newcommand{\SLOWlusearchWCPUP}{0}
\newcommand{\SLOWlusearchWCPUPCI}{0.0}
\newcommand{\SLOWlusearchWCPUPCIMIN}{0}
\newcommand{\SLOWlusearchWCPUPCIMAX}{0}
\newcommand{\SLOWlusearchWCPDynamicUP}{0}
\newcommand{\SLOWlusearchWCPDynamicUPCI}{0.0}
\newcommand{\SLOWlusearchWCPDynamicUPCIMIN}{0}
\newcommand{\SLOWlusearchWCPDynamicUPCIMAX}{0}
\newcommand{\SLOWlusearchWDCnoGUP}{\rna}
\newcommand{\SLOWlusearchWDCnoGUPCI}{\rna}
\newcommand{\SLOWlusearchWDCnoGUPCIMIN}{\rna}
\newcommand{\SLOWlusearchWDCnoGUPCIMAX}{\rna}
\newcommand{\SLOWlusearchWDCnoGDynamicUP}{\rna}
\newcommand{\SLOWlusearchWDCnoGDynamicUPCI}{\rna}
\newcommand{\SLOWlusearchWDCnoGDynamicUPCIMIN}{\rna}
\newcommand{\SLOWlusearchWDCnoGDynamicUPCIMAX}{\rna}
\newcommand{\SLOWlusearchWDCUP}{\rna}
\newcommand{\SLOWlusearchWDCUPCI}{\rna}
\newcommand{\SLOWlusearchWDCUPCIMIN}{\rna}
\newcommand{\SLOWlusearchWDCUPCIMAX}{\rna}
\newcommand{\SLOWlusearchWDCDynamicUP}{\rna}
\newcommand{\SLOWlusearchWDCDynamicUPCI}{\rna}
\newcommand{\SLOWlusearchWDCDynamicUPCIMIN}{\rna}
\newcommand{\SLOWlusearchWDCDynamicUPCIMAX}{\rna}
\newcommand{\SLOWlusearchCAPOnoGUP}{\rna}
\newcommand{\SLOWlusearchCAPOnoGUPCI}{\rna}
\newcommand{\SLOWlusearchCAPOnoGUPCIMIN}{\rna}
\newcommand{\SLOWlusearchCAPOnoGUPCIMAX}{\rna}
\newcommand{\SLOWlusearchCAPOnoGDynamicUP}{\rna}
\newcommand{\SLOWlusearchCAPOnoGDynamicUPCI}{\rna}
\newcommand{\SLOWlusearchCAPOnoGDynamicUPCIMIN}{\rna}
\newcommand{\SLOWlusearchCAPOnoGDynamicUPCIMAX}{\rna}
\newcommand{\SLOWlusearchCAPOUP}{\rna}
\newcommand{\SLOWlusearchCAPOUPCI}{\rna}
\newcommand{\SLOWlusearchCAPOUPCIMIN}{\rna}
\newcommand{\SLOWlusearchCAPOUPCIMAX}{\rna}
\newcommand{\SLOWlusearchCAPODynamicUP}{\rna}
\newcommand{\SLOWlusearchCAPODynamicUPCI}{\rna}
\newcommand{\SLOWlusearchCAPODynamicUPCIMIN}{\rna}
\newcommand{\SLOWlusearchCAPODynamicUPCIMAX}{\rna}
\newcommand{\SLOWlusearchPIPUP}{\rna}
\newcommand{\SLOWlusearchPIPUPCI}{\rna}
\newcommand{\SLOWlusearchPIPUPCIMIN}{\rna}
\newcommand{\SLOWlusearchPIPUPCIMAX}{\rna}
\newcommand{\SLOWlusearchPIPDynamicUP}{\rna}
\newcommand{\SLOWlusearchPIPDynamicUPCI}{\rna}
\newcommand{\SLOWlusearchPIPDynamicUPCIMIN}{\rna}
\newcommand{\SLOWlusearchPIPDynamicUPCIMAX}{\rna}
\newcommand{\SLOWlusearchPIPHB}{\rna}
\newcommand{\SLOWlusearchPIPHBCI}{\rna}
\newcommand{\SLOWlusearchPIPHBCIMIN}{\rna}
\newcommand{\SLOWlusearchPIPHBCIMAX}{\rna}
\newcommand{\SLOWlusearchPIPHBDynamic}{\rna}
\newcommand{\SLOWlusearchPIPHBDynamicCI}{\rna}
\newcommand{\SLOWlusearchPIPHBDynamicCIMIN}{\rna}
\newcommand{\SLOWlusearchPIPHBDynamicCIMAX}{\rna}
\newcommand{\SLOWlusearchPIPWCP}{\rna}
\newcommand{\SLOWlusearchPIPWCPCI}{\rna}
\newcommand{\SLOWlusearchPIPWCPCIMIN}{\rna}
\newcommand{\SLOWlusearchPIPWCPCIMAX}{\rna}
\newcommand{\SLOWlusearchPIPWCPDynamic}{\rna}
\newcommand{\SLOWlusearchPIPWCPDynamicCI}{\rna}
\newcommand{\SLOWlusearchPIPWCPDynamicCIMIN}{\rna}
\newcommand{\SLOWlusearchPIPWCPDynamicCIMAX}{\rna}
\newcommand{\SLOWlusearchPIPWDC}{\rna}
\newcommand{\SLOWlusearchPIPWDCCI}{\rna}
\newcommand{\SLOWlusearchPIPWDCCIMIN}{\rna}
\newcommand{\SLOWlusearchPIPWDCCIMAX}{\rna}
\newcommand{\SLOWlusearchPIPWDCDynamic}{\rna}
\newcommand{\SLOWlusearchPIPWDCDynamicCI}{\rna}
\newcommand{\SLOWlusearchPIPWDCDynamicCIMIN}{\rna}
\newcommand{\SLOWlusearchPIPWDCDynamicCIMAX}{\rna}
\newcommand{\SLOWlusearchPIPCAPO}{\rna}
\newcommand{\SLOWlusearchPIPCAPOCI}{\rna}
\newcommand{\SLOWlusearchPIPCAPOCIMIN}{\rna}
\newcommand{\SLOWlusearchPIPCAPOCIMAX}{\rna}
\newcommand{\SLOWlusearchPIPCAPODynamic}{\rna}
\newcommand{\SLOWlusearchPIPCAPODynamicCI}{\rna}
\newcommand{\SLOWlusearchPIPCAPODynamicCIMIN}{\rna}
\newcommand{\SLOWlusearchPIPCAPODynamicCIMAX}{\rna}
\newcommand{\SLOWlusearchPIPPIP}{\rna}
\newcommand{\SLOWlusearchPIPPIPCI}{\rna}
\newcommand{\SLOWlusearchPIPPIPCIMIN}{\rna}
\newcommand{\SLOWlusearchPIPPIPCIMAX}{\rna}
\newcommand{\SLOWlusearchPIPPIPDynamic}{\rna}
\newcommand{\SLOWlusearchPIPPIPDynamicCI}{\rna}
\newcommand{\SLOWlusearchPIPPIPDynamicCIMIN}{\rna}
\newcommand{\SLOWlusearchPIPPIPDynamicCIMAX}{\rna}
\newcommand{\SLOWpmdEvents}{0}
\newcommand{\SLOWpmdNoFPEvents}{0}
\newcommand{\SLOWpmdMaxLiveThreads}{15}
\newcommand{\SLOWpmdTotalThreads}{15}
\newcommand{\SLOWpmdBaseTime}{1.4}
\newcommand{\SLOWpmdBaseTimeCI}{35}
\newcommand{\SLOWpmdEmptyTime}{\rna}
\newcommand{\SLOWpmdEmptyTimeCI}{\rna}
\newcommand{\SLOWpmdEmptyTimeCIMIN}{\rna}
\newcommand{\SLOWpmdEmptyTimeCIMAX}{\rna}
\newcommand{\SLOWpmdFTTime}{6.5}
\newcommand{\SLOWpmdFTTimeCI}{0.56}
\newcommand{\SLOWpmdHBTime}{14}
\newcommand{\SLOWpmdHBTimeCI}{0.74}
\newcommand{\SLOWpmdWCPTime}{18}
\newcommand{\SLOWpmdWCPTimeCI}{0.026}
\newcommand{\SLOWpmdDCnoGExcTime}{16}
\newcommand{\SLOWpmdDCnoGExcTimeCI}{1.6}
\newcommand{\SLOWpmdDCnoGTime}{\rna}
\newcommand{\SLOWpmdDCnoGTimeCI}{\rna}
\newcommand{\SLOWpmdDCnoGTimeCIMIN}{\rna}
\newcommand{\SLOWpmdDCnoGTimeCIMAX}{\rna}
\newcommand{\SLOWpmdDCExcTime}{17}
\newcommand{\SLOWpmdDCExcTimeCI}{2.1}
\newcommand{\SLOWpmdDCTime}{\rna}
\newcommand{\SLOWpmdDCTimeCI}{\rna}
\newcommand{\SLOWpmdDCTimeCIMIN}{\rna}
\newcommand{\SLOWpmdDCTimeCIMAX}{\rna}
\newcommand{\SLOWpmdCAPOnoGExcTime}{15}
\newcommand{\SLOWpmdCAPOnoGExcTimeCI}{1.4}
\newcommand{\SLOWpmdCAPOnoGTime}{\rna}
\newcommand{\SLOWpmdCAPOnoGTimeCI}{\rna}
\newcommand{\SLOWpmdCAPOnoGTimeCIMIN}{\rna}
\newcommand{\SLOWpmdCAPOnoGTimeCIMAX}{\rna}
\newcommand{\SLOWpmdCAPOExcTime}{15}
\newcommand{\SLOWpmdCAPOExcTimeCI}{14}
\newcommand{\SLOWpmdCAPOTime}{\rna}
\newcommand{\SLOWpmdCAPOTimeCI}{\rna}
\newcommand{\SLOWpmdCAPOTimeCIMIN}{\rna}
\newcommand{\SLOWpmdCAPOTimeCIMAX}{\rna}
\newcommand{\SLOWpmdStaticTime}{\rzero}
\newcommand{\SLOWpmdDynamicTime}{\rzero}
\newcommand{\SLOWpmdBaseMem}{590}
\newcommand{\SLOWpmdBaseMemCI}{36.0}
\newcommand{\SLOWpmdHBMem}{14}
\newcommand{\SLOWpmdHBMemCI}{0.46}
\newcommand{\SLOWpmdFTMem}{2.6}
\newcommand{\SLOWpmdFTMemCI}{0.24}
\newcommand{\SLOWpmdWCPMem}{22}
\newcommand{\SLOWpmdWCPMemCI}{0.84}
\newcommand{\SLOWpmdDCnoGExcMem}{14}
\newcommand{\SLOWpmdDCnoGExcMemCI}{1.4}
\newcommand{\SLOWpmdDCnoGMem}{\memna}
\newcommand{\SLOWpmdDCnoGMemCI}{\memna}
\newcommand{\SLOWpmdDCnoGMemCIMIN}{\memna}
\newcommand{\SLOWpmdDCnoGMemCIMAX}{\memna}
\newcommand{\SLOWpmdDCExcMem}{15}
\newcommand{\SLOWpmdDCExcMemCI}{0.44}
\newcommand{\SLOWpmdDCMem}{\memna}
\newcommand{\SLOWpmdDCMemCI}{\memna}
\newcommand{\SLOWpmdDCMemCIMIN}{\memna}
\newcommand{\SLOWpmdDCMemCIMAX}{\memna}
\newcommand{\SLOWpmdCAPOnoGExcMem}{14}
\newcommand{\SLOWpmdCAPOnoGExcMemCI}{0.89}
\newcommand{\SLOWpmdCAPOnoGMem}{\memna}
\newcommand{\SLOWpmdCAPOnoGMemCI}{\memna}
\newcommand{\SLOWpmdCAPOnoGMemCIMIN}{\memna}
\newcommand{\SLOWpmdCAPOnoGMemCIMAX}{\memna}
\newcommand{\SLOWpmdCAPOExcMem}{14}
\newcommand{\SLOWpmdCAPOExcMemCI}{13}
\newcommand{\SLOWpmdCAPOMem}{\memna}
\newcommand{\SLOWpmdCAPOMemCI}{\memna}
\newcommand{\SLOWpmdCAPOMemCIMIN}{\memna}
\newcommand{\SLOWpmdCAPOMemCIMAX}{\memna}
\newcommand{\SLOWpmdEventsCI}{0}
\newcommand{\SLOWpmdEventsCIMIN}{0}
\newcommand{\SLOWpmdEventsCIMAX}{0}
\newcommand{\SLOWpmdNoFPEventsCI}{0}
\newcommand{\SLOWpmdNoFPEventsCIMIN}{0}
\newcommand{\SLOWpmdNoFPEventsCIMAX}{0}
\newcommand{\SLOWpmdHB}{7}
\newcommand{\SLOWpmdHBCI}{0.98}
\newcommand{\SLOWpmdHBCIMIN}{6}
\newcommand{\SLOWpmdHBCIMAX}{8}
\newcommand{\SLOWpmdHBDynamic}{320}
\newcommand{\SLOWpmdHBDynamicCI}{121}
\newcommand{\SLOWpmdHBDynamicCIMIN}{199}
\newcommand{\SLOWpmdHBDynamicCIMAX}{441}
\newcommand{\SLOWpmdFT}{16}
\newcommand{\SLOWpmdFTCI}{0.0}
\newcommand{\SLOWpmdFTCIMIN}{16}
\newcommand{\SLOWpmdFTCIMAX}{16}
\newcommand{\SLOWpmdFTDynamic}{3,820}
\newcommand{\SLOWpmdFTDynamicCI}{2.9}
\newcommand{\SLOWpmdFTDynamicCIMIN}{3,817}
\newcommand{\SLOWpmdFTDynamicCIMAX}{3,823}
\newcommand{\SLOWpmdWCP}{6}
\newcommand{\SLOWpmdWCPCI}{0.0}
\newcommand{\SLOWpmdWCPCIMIN}{6}
\newcommand{\SLOWpmdWCPCIMAX}{6}
\newcommand{\SLOWpmdWCPDynamic}{443}
\newcommand{\SLOWpmdWCPDynamicCI}{431}
\newcommand{\SLOWpmdWCPDynamicCIMIN}{12}
\newcommand{\SLOWpmdWCPDynamicCIMAX}{874}
\newcommand{\SLOWpmdDCnoGExc}{11}
\newcommand{\SLOWpmdDCnoGExcCI}{0.0}
\newcommand{\SLOWpmdDCnoGExcCIMIN}{11}
\newcommand{\SLOWpmdDCnoGExcCIMAX}{11}
\newcommand{\SLOWpmdDCnoGExcDynamic}{1,623}
\newcommand{\SLOWpmdDCnoGExcDynamicCI}{42}
\newcommand{\SLOWpmdDCnoGExcDynamicCIMIN}{1,581}
\newcommand{\SLOWpmdDCnoGExcDynamicCIMAX}{1,665}
\newcommand{\SLOWpmdDCnoG}{\rna}
\newcommand{\SLOWpmdDCnoGCI}{\rna}
\newcommand{\SLOWpmdDCnoGCIMIN}{\rna}
\newcommand{\SLOWpmdDCnoGCIMAX}{\rna}
\newcommand{\SLOWpmdDCnoGDynamic}{\rna}
\newcommand{\SLOWpmdDCnoGDynamicCI}{\rna}
\newcommand{\SLOWpmdDCnoGDynamicCIMIN}{\rna}
\newcommand{\SLOWpmdDCnoGDynamicCIMAX}{\rna}
\newcommand{\SLOWpmdDCExc}{6}
\newcommand{\SLOWpmdDCExcCI}{0.98}
\newcommand{\SLOWpmdDCExcCIMIN}{5}
\newcommand{\SLOWpmdDCExcCIMAX}{7}
\newcommand{\SLOWpmdDCExcDynamic}{73}
\newcommand{\SLOWpmdDCExcDynamicCI}{19}
\newcommand{\SLOWpmdDCExcDynamicCIMIN}{54}
\newcommand{\SLOWpmdDCExcDynamicCIMAX}{92}
\newcommand{\SLOWpmdDC}{\rna}
\newcommand{\SLOWpmdDCCI}{\rna}
\newcommand{\SLOWpmdDCCIMIN}{\rna}
\newcommand{\SLOWpmdDCCIMAX}{\rna}
\newcommand{\SLOWpmdDCDynamic}{\rna}
\newcommand{\SLOWpmdDCDynamicCI}{\rna}
\newcommand{\SLOWpmdDCDynamicCIMIN}{\rna}
\newcommand{\SLOWpmdDCDynamicCIMAX}{\rna}
\newcommand{\SLOWpmdCAPOnoGExc}{10}
\newcommand{\SLOWpmdCAPOnoGExcCI}{0.0}
\newcommand{\SLOWpmdCAPOnoGExcCIMIN}{10}
\newcommand{\SLOWpmdCAPOnoGExcCIMAX}{10}
\newcommand{\SLOWpmdCAPOnoGExcDynamic}{1,555}
\newcommand{\SLOWpmdCAPOnoGExcDynamicCI}{517}
\newcommand{\SLOWpmdCAPOnoGExcDynamicCIMIN}{1,038}
\newcommand{\SLOWpmdCAPOnoGExcDynamicCIMAX}{2,072}
\newcommand{\SLOWpmdCAPOnoG}{\rna}
\newcommand{\SLOWpmdCAPOnoGCI}{\rna}
\newcommand{\SLOWpmdCAPOnoGCIMIN}{\rna}
\newcommand{\SLOWpmdCAPOnoGCIMAX}{\rna}
\newcommand{\SLOWpmdCAPOnoGDynamic}{\rna}
\newcommand{\SLOWpmdCAPOnoGDynamicCI}{\rna}
\newcommand{\SLOWpmdCAPOnoGDynamicCIMIN}{\rna}
\newcommand{\SLOWpmdCAPOnoGDynamicCIMAX}{\rna}
\newcommand{\SLOWpmdCAPOExc}{5}
\newcommand{\SLOWpmdCAPOExcCI}{5.9}
\newcommand{\SLOWpmdCAPOExcCIMIN}{-1}
\newcommand{\SLOWpmdCAPOExcCIMAX}{11}
\newcommand{\SLOWpmdCAPOExcDynamic}{65}
\newcommand{\SLOWpmdCAPOExcDynamicCI}{65}
\newcommand{\SLOWpmdCAPOExcDynamicCIMIN}{0}
\newcommand{\SLOWpmdCAPOExcDynamicCIMAX}{130}
\newcommand{\SLOWpmdCAPO}{\rna}
\newcommand{\SLOWpmdCAPOCI}{\rna}
\newcommand{\SLOWpmdCAPOCIMIN}{\rna}
\newcommand{\SLOWpmdCAPOCIMAX}{\rna}
\newcommand{\SLOWpmdCAPODynamic}{\rna}
\newcommand{\SLOWpmdCAPODynamicCI}{\rna}
\newcommand{\SLOWpmdCAPODynamicCIMIN}{\rna}
\newcommand{\SLOWpmdCAPODynamicCIMAX}{\rna}
\newcommand{\SLOWpmdPIP}{\rna}
\newcommand{\SLOWpmdPIPCI}{\rna}
\newcommand{\SLOWpmdPIPCIMIN}{\rna}
\newcommand{\SLOWpmdPIPCIMAX}{\rna}
\newcommand{\SLOWpmdPIPDynamic}{\rna}
\newcommand{\SLOWpmdPIPDynamicCI}{\rna}
\newcommand{\SLOWpmdPIPDynamicCIMIN}{\rna}
\newcommand{\SLOWpmdPIPDynamicCIMAX}{\rna}
\newcommand{\SLOWpmdHBUP}{6}
\newcommand{\SLOWpmdHBUPCI}{0.0}
\newcommand{\SLOWpmdHBUPCIMIN}{6}
\newcommand{\SLOWpmdHBUPCIMAX}{6}
\newcommand{\SLOWpmdHBDynamicUP}{324}
\newcommand{\SLOWpmdHBDynamicUPCI}{128}
\newcommand{\SLOWpmdHBDynamicUPCIMIN}{196}
\newcommand{\SLOWpmdHBDynamicUPCIMAX}{452}
\newcommand{\SLOWpmdWCPUP}{6}
\newcommand{\SLOWpmdWCPUPCI}{0.0}
\newcommand{\SLOWpmdWCPUPCIMIN}{6}
\newcommand{\SLOWpmdWCPUPCIMAX}{6}
\newcommand{\SLOWpmdWCPDynamicUP}{443}
\newcommand{\SLOWpmdWCPDynamicUPCI}{431}
\newcommand{\SLOWpmdWCPDynamicUPCIMIN}{12}
\newcommand{\SLOWpmdWCPDynamicUPCIMAX}{874}
\newcommand{\SLOWpmdWDCnoGUP}{\rna}
\newcommand{\SLOWpmdWDCnoGUPCI}{\rna}
\newcommand{\SLOWpmdWDCnoGUPCIMIN}{\rna}
\newcommand{\SLOWpmdWDCnoGUPCIMAX}{\rna}
\newcommand{\SLOWpmdWDCnoGDynamicUP}{\rna}
\newcommand{\SLOWpmdWDCnoGDynamicUPCI}{\rna}
\newcommand{\SLOWpmdWDCnoGDynamicUPCIMIN}{\rna}
\newcommand{\SLOWpmdWDCnoGDynamicUPCIMAX}{\rna}
\newcommand{\SLOWpmdWDCUP}{\rna}
\newcommand{\SLOWpmdWDCUPCI}{\rna}
\newcommand{\SLOWpmdWDCUPCIMIN}{\rna}
\newcommand{\SLOWpmdWDCUPCIMAX}{\rna}
\newcommand{\SLOWpmdWDCDynamicUP}{\rna}
\newcommand{\SLOWpmdWDCDynamicUPCI}{\rna}
\newcommand{\SLOWpmdWDCDynamicUPCIMIN}{\rna}
\newcommand{\SLOWpmdWDCDynamicUPCIMAX}{\rna}
\newcommand{\SLOWpmdCAPOnoGUP}{\rna}
\newcommand{\SLOWpmdCAPOnoGUPCI}{\rna}
\newcommand{\SLOWpmdCAPOnoGUPCIMIN}{\rna}
\newcommand{\SLOWpmdCAPOnoGUPCIMAX}{\rna}
\newcommand{\SLOWpmdCAPOnoGDynamicUP}{\rna}
\newcommand{\SLOWpmdCAPOnoGDynamicUPCI}{\rna}
\newcommand{\SLOWpmdCAPOnoGDynamicUPCIMIN}{\rna}
\newcommand{\SLOWpmdCAPOnoGDynamicUPCIMAX}{\rna}
\newcommand{\SLOWpmdCAPOUP}{\rna}
\newcommand{\SLOWpmdCAPOUPCI}{\rna}
\newcommand{\SLOWpmdCAPOUPCIMIN}{\rna}
\newcommand{\SLOWpmdCAPOUPCIMAX}{\rna}
\newcommand{\SLOWpmdCAPODynamicUP}{\rna}
\newcommand{\SLOWpmdCAPODynamicUPCI}{\rna}
\newcommand{\SLOWpmdCAPODynamicUPCIMIN}{\rna}
\newcommand{\SLOWpmdCAPODynamicUPCIMAX}{\rna}
\newcommand{\SLOWpmdPIPUP}{\rna}
\newcommand{\SLOWpmdPIPUPCI}{\rna}
\newcommand{\SLOWpmdPIPUPCIMIN}{\rna}
\newcommand{\SLOWpmdPIPUPCIMAX}{\rna}
\newcommand{\SLOWpmdPIPDynamicUP}{\rna}
\newcommand{\SLOWpmdPIPDynamicUPCI}{\rna}
\newcommand{\SLOWpmdPIPDynamicUPCIMIN}{\rna}
\newcommand{\SLOWpmdPIPDynamicUPCIMAX}{\rna}
\newcommand{\SLOWpmdPIPHB}{\rna}
\newcommand{\SLOWpmdPIPHBCI}{\rna}
\newcommand{\SLOWpmdPIPHBCIMIN}{\rna}
\newcommand{\SLOWpmdPIPHBCIMAX}{\rna}
\newcommand{\SLOWpmdPIPHBDynamic}{\rna}
\newcommand{\SLOWpmdPIPHBDynamicCI}{\rna}
\newcommand{\SLOWpmdPIPHBDynamicCIMIN}{\rna}
\newcommand{\SLOWpmdPIPHBDynamicCIMAX}{\rna}
\newcommand{\SLOWpmdPIPWCP}{\rna}
\newcommand{\SLOWpmdPIPWCPCI}{\rna}
\newcommand{\SLOWpmdPIPWCPCIMIN}{\rna}
\newcommand{\SLOWpmdPIPWCPCIMAX}{\rna}
\newcommand{\SLOWpmdPIPWCPDynamic}{\rna}
\newcommand{\SLOWpmdPIPWCPDynamicCI}{\rna}
\newcommand{\SLOWpmdPIPWCPDynamicCIMIN}{\rna}
\newcommand{\SLOWpmdPIPWCPDynamicCIMAX}{\rna}
\newcommand{\SLOWpmdPIPWDC}{\rna}
\newcommand{\SLOWpmdPIPWDCCI}{\rna}
\newcommand{\SLOWpmdPIPWDCCIMIN}{\rna}
\newcommand{\SLOWpmdPIPWDCCIMAX}{\rna}
\newcommand{\SLOWpmdPIPWDCDynamic}{\rna}
\newcommand{\SLOWpmdPIPWDCDynamicCI}{\rna}
\newcommand{\SLOWpmdPIPWDCDynamicCIMIN}{\rna}
\newcommand{\SLOWpmdPIPWDCDynamicCIMAX}{\rna}
\newcommand{\SLOWpmdPIPCAPO}{\rna}
\newcommand{\SLOWpmdPIPCAPOCI}{\rna}
\newcommand{\SLOWpmdPIPCAPOCIMIN}{\rna}
\newcommand{\SLOWpmdPIPCAPOCIMAX}{\rna}
\newcommand{\SLOWpmdPIPCAPODynamic}{\rna}
\newcommand{\SLOWpmdPIPCAPODynamicCI}{\rna}
\newcommand{\SLOWpmdPIPCAPODynamicCIMIN}{\rna}
\newcommand{\SLOWpmdPIPCAPODynamicCIMAX}{\rna}
\newcommand{\SLOWpmdPIPPIP}{\rna}
\newcommand{\SLOWpmdPIPPIPCI}{\rna}
\newcommand{\SLOWpmdPIPPIPCIMIN}{\rna}
\newcommand{\SLOWpmdPIPPIPCIMAX}{\rna}
\newcommand{\SLOWpmdPIPPIPDynamic}{\rna}
\newcommand{\SLOWpmdPIPPIPDynamicCI}{\rna}
\newcommand{\SLOWpmdPIPPIPDynamicCIMIN}{\rna}
\newcommand{\SLOWpmdPIPPIPDynamicCIMAX}{\rna}
\newcommand{\SLOWsunflowEvents}{0}
\newcommand{\SLOWsunflowNoFPEvents}{0}
\newcommand{\SLOWsunflowMaxLiveThreads}{7}
\newcommand{\SLOWsunflowTotalThreads}{14}
\newcommand{\SLOWsunflowBaseTime}{1.4}
\newcommand{\SLOWsunflowBaseTimeCI}{140}
\newcommand{\SLOWsunflowEmptyTime}{\rna}
\newcommand{\SLOWsunflowEmptyTimeCI}{\rna}
\newcommand{\SLOWsunflowEmptyTimeCIMIN}{\rna}
\newcommand{\SLOWsunflowEmptyTimeCIMAX}{\rna}
\newcommand{\SLOWsunflowFTTime}{16}
\newcommand{\SLOWsunflowFTTimeCI}{0.64}
\newcommand{\SLOWsunflowHBTime}{67}
\newcommand{\SLOWsunflowHBTimeCI}{5.0}
\newcommand{\SLOWsunflowWCPTime}{83}
\newcommand{\SLOWsunflowWCPTimeCI}{12}
\newcommand{\SLOWsunflowDCnoGExcTime}{70}
\newcommand{\SLOWsunflowDCnoGExcTimeCI}{5.1}
\newcommand{\SLOWsunflowDCnoGTime}{\rna}
\newcommand{\SLOWsunflowDCnoGTimeCI}{\rna}
\newcommand{\SLOWsunflowDCnoGTimeCIMIN}{\rna}
\newcommand{\SLOWsunflowDCnoGTimeCIMAX}{\rna}
\newcommand{\SLOWsunflowDCExcTime}{74}
\newcommand{\SLOWsunflowDCExcTimeCI}{69}
\newcommand{\SLOWsunflowDCTime}{\rna}
\newcommand{\SLOWsunflowDCTimeCI}{\rna}
\newcommand{\SLOWsunflowDCTimeCIMIN}{\rna}
\newcommand{\SLOWsunflowDCTimeCIMAX}{\rna}
\newcommand{\SLOWsunflowCAPOnoGExcTime}{69}
\newcommand{\SLOWsunflowCAPOnoGExcTimeCI}{7.4}
\newcommand{\SLOWsunflowCAPOnoGTime}{\rna}
\newcommand{\SLOWsunflowCAPOnoGTimeCI}{\rna}
\newcommand{\SLOWsunflowCAPOnoGTimeCIMIN}{\rna}
\newcommand{\SLOWsunflowCAPOnoGTimeCIMAX}{\rna}
\newcommand{\SLOWsunflowCAPOExcTime}{74}
\newcommand{\SLOWsunflowCAPOExcTimeCI}{69}
\newcommand{\SLOWsunflowCAPOTime}{\rna}
\newcommand{\SLOWsunflowCAPOTimeCI}{\rna}
\newcommand{\SLOWsunflowCAPOTimeCIMIN}{\rna}
\newcommand{\SLOWsunflowCAPOTimeCIMAX}{\rna}
\newcommand{\SLOWsunflowStaticTime}{\rzero}
\newcommand{\SLOWsunflowDynamicTime}{\rzero}
\newcommand{\SLOWsunflowBaseMem}{610}
\newcommand{\SLOWsunflowBaseMemCI}{2.6}
\newcommand{\SLOWsunflowHBMem}{57}
\newcommand{\SLOWsunflowHBMemCI}{10}
\newcommand{\SLOWsunflowFTMem}{8.5}
\newcommand{\SLOWsunflowFTMemCI}{0.075}
\newcommand{\SLOWsunflowWCPMem}{87}
\newcommand{\SLOWsunflowWCPMemCI}{3.9}
\newcommand{\SLOWsunflowDCnoGExcMem}{39}
\newcommand{\SLOWsunflowDCnoGExcMemCI}{0.43}
\newcommand{\SLOWsunflowDCnoGMem}{\memna}
\newcommand{\SLOWsunflowDCnoGMemCI}{\memna}
\newcommand{\SLOWsunflowDCnoGMemCIMIN}{\memna}
\newcommand{\SLOWsunflowDCnoGMemCIMAX}{\memna}
\newcommand{\SLOWsunflowDCExcMem}{43}
\newcommand{\SLOWsunflowDCExcMemCI}{41}
\newcommand{\SLOWsunflowDCMem}{\memna}
\newcommand{\SLOWsunflowDCMemCI}{\memna}
\newcommand{\SLOWsunflowDCMemCIMIN}{\memna}
\newcommand{\SLOWsunflowDCMemCIMAX}{\memna}
\newcommand{\SLOWsunflowCAPOnoGExcMem}{38}
\newcommand{\SLOWsunflowCAPOnoGExcMemCI}{1.8}
\newcommand{\SLOWsunflowCAPOnoGMem}{\memna}
\newcommand{\SLOWsunflowCAPOnoGMemCI}{\memna}
\newcommand{\SLOWsunflowCAPOnoGMemCIMIN}{\memna}
\newcommand{\SLOWsunflowCAPOnoGMemCIMAX}{\memna}
\newcommand{\SLOWsunflowCAPOExcMem}{41}
\newcommand{\SLOWsunflowCAPOExcMemCI}{41}
\newcommand{\SLOWsunflowCAPOMem}{\memna}
\newcommand{\SLOWsunflowCAPOMemCI}{\memna}
\newcommand{\SLOWsunflowCAPOMemCIMIN}{\memna}
\newcommand{\SLOWsunflowCAPOMemCIMAX}{\memna}
\newcommand{\SLOWsunflowEventsCI}{0}
\newcommand{\SLOWsunflowEventsCIMIN}{0}
\newcommand{\SLOWsunflowEventsCIMAX}{0}
\newcommand{\SLOWsunflowNoFPEventsCI}{0}
\newcommand{\SLOWsunflowNoFPEventsCIMIN}{0}
\newcommand{\SLOWsunflowNoFPEventsCIMAX}{0}
\newcommand{\SLOWsunflowHB}{6}
\newcommand{\SLOWsunflowHBCI}{0.0}
\newcommand{\SLOWsunflowHBCIMIN}{6}
\newcommand{\SLOWsunflowHBCIMAX}{6}
\newcommand{\SLOWsunflowHBDynamic}{49}
\newcommand{\SLOWsunflowHBDynamicCI}{6.9}
\newcommand{\SLOWsunflowHBDynamicCIMIN}{42}
\newcommand{\SLOWsunflowHBDynamicCIMAX}{56}
\newcommand{\SLOWsunflowFT}{5}
\newcommand{\SLOWsunflowFTCI}{0.0}
\newcommand{\SLOWsunflowFTCIMIN}{5}
\newcommand{\SLOWsunflowFTCIMAX}{5}
\newcommand{\SLOWsunflowFTDynamic}{117}
\newcommand{\SLOWsunflowFTDynamicCI}{13}
\newcommand{\SLOWsunflowFTDynamicCIMIN}{104}
\newcommand{\SLOWsunflowFTDynamicCIMAX}{130}
\newcommand{\SLOWsunflowWCP}{18}
\newcommand{\SLOWsunflowWCPCI}{0.0}
\newcommand{\SLOWsunflowWCPCIMIN}{18}
\newcommand{\SLOWsunflowWCPCIMAX}{18}
\newcommand{\SLOWsunflowWCPDynamic}{288}
\newcommand{\SLOWsunflowWCPDynamicCI}{25}
\newcommand{\SLOWsunflowWCPDynamicCIMIN}{263}
\newcommand{\SLOWsunflowWCPDynamicCIMAX}{313}
\newcommand{\SLOWsunflowDCnoGExc}{19}
\newcommand{\SLOWsunflowDCnoGExcCI}{0.0}
\newcommand{\SLOWsunflowDCnoGExcCIMIN}{19}
\newcommand{\SLOWsunflowDCnoGExcCIMAX}{19}
\newcommand{\SLOWsunflowDCnoGExcDynamic}{734}
\newcommand{\SLOWsunflowDCnoGExcDynamicCI}{97}
\newcommand{\SLOWsunflowDCnoGExcDynamicCIMIN}{637}
\newcommand{\SLOWsunflowDCnoGExcDynamicCIMAX}{831}
\newcommand{\SLOWsunflowDCnoG}{\rna}
\newcommand{\SLOWsunflowDCnoGCI}{\rna}
\newcommand{\SLOWsunflowDCnoGCIMIN}{\rna}
\newcommand{\SLOWsunflowDCnoGCIMAX}{\rna}
\newcommand{\SLOWsunflowDCnoGDynamic}{\rna}
\newcommand{\SLOWsunflowDCnoGDynamicCI}{\rna}
\newcommand{\SLOWsunflowDCnoGDynamicCIMIN}{\rna}
\newcommand{\SLOWsunflowDCnoGDynamicCIMAX}{\rna}
\newcommand{\SLOWsunflowDCExc}{3}
\newcommand{\SLOWsunflowDCExcCI}{3.9}
\newcommand{\SLOWsunflowDCExcCIMIN}{-1}
\newcommand{\SLOWsunflowDCExcCIMAX}{7}
\newcommand{\SLOWsunflowDCExcDynamic}{22}
\newcommand{\SLOWsunflowDCExcDynamicCI}{23}
\newcommand{\SLOWsunflowDCExcDynamicCIMIN}{-1}
\newcommand{\SLOWsunflowDCExcDynamicCIMAX}{45}
\newcommand{\SLOWsunflowDC}{\rna}
\newcommand{\SLOWsunflowDCCI}{\rna}
\newcommand{\SLOWsunflowDCCIMIN}{\rna}
\newcommand{\SLOWsunflowDCCIMAX}{\rna}
\newcommand{\SLOWsunflowDCDynamic}{\rna}
\newcommand{\SLOWsunflowDCDynamicCI}{\rna}
\newcommand{\SLOWsunflowDCDynamicCIMIN}{\rna}
\newcommand{\SLOWsunflowDCDynamicCIMAX}{\rna}
\newcommand{\SLOWsunflowCAPOnoGExc}{19}
\newcommand{\SLOWsunflowCAPOnoGExcCI}{0.0}
\newcommand{\SLOWsunflowCAPOnoGExcCIMIN}{19}
\newcommand{\SLOWsunflowCAPOnoGExcCIMAX}{19}
\newcommand{\SLOWsunflowCAPOnoGExcDynamic}{677}
\newcommand{\SLOWsunflowCAPOnoGExcDynamicCI}{293}
\newcommand{\SLOWsunflowCAPOnoGExcDynamicCIMIN}{384}
\newcommand{\SLOWsunflowCAPOnoGExcDynamicCIMAX}{970}
\newcommand{\SLOWsunflowCAPOnoG}{\rna}
\newcommand{\SLOWsunflowCAPOnoGCI}{\rna}
\newcommand{\SLOWsunflowCAPOnoGCIMIN}{\rna}
\newcommand{\SLOWsunflowCAPOnoGCIMAX}{\rna}
\newcommand{\SLOWsunflowCAPOnoGDynamic}{\rna}
\newcommand{\SLOWsunflowCAPOnoGDynamicCI}{\rna}
\newcommand{\SLOWsunflowCAPOnoGDynamicCIMIN}{\rna}
\newcommand{\SLOWsunflowCAPOnoGDynamicCIMAX}{\rna}
\newcommand{\SLOWsunflowCAPOExc}{3}
\newcommand{\SLOWsunflowCAPOExcCI}{3.9}
\newcommand{\SLOWsunflowCAPOExcCIMIN}{-1}
\newcommand{\SLOWsunflowCAPOExcCIMAX}{7}
\newcommand{\SLOWsunflowCAPOExcDynamic}{24}
\newcommand{\SLOWsunflowCAPOExcDynamicCI}{24}
\newcommand{\SLOWsunflowCAPOExcDynamicCIMIN}{0}
\newcommand{\SLOWsunflowCAPOExcDynamicCIMAX}{48}
\newcommand{\SLOWsunflowCAPO}{\rna}
\newcommand{\SLOWsunflowCAPOCI}{\rna}
\newcommand{\SLOWsunflowCAPOCIMIN}{\rna}
\newcommand{\SLOWsunflowCAPOCIMAX}{\rna}
\newcommand{\SLOWsunflowCAPODynamic}{\rna}
\newcommand{\SLOWsunflowCAPODynamicCI}{\rna}
\newcommand{\SLOWsunflowCAPODynamicCIMIN}{\rna}
\newcommand{\SLOWsunflowCAPODynamicCIMAX}{\rna}
\newcommand{\SLOWsunflowPIP}{\rna}
\newcommand{\SLOWsunflowPIPCI}{\rna}
\newcommand{\SLOWsunflowPIPCIMIN}{\rna}
\newcommand{\SLOWsunflowPIPCIMAX}{\rna}
\newcommand{\SLOWsunflowPIPDynamic}{\rna}
\newcommand{\SLOWsunflowPIPDynamicCI}{\rna}
\newcommand{\SLOWsunflowPIPDynamicCIMIN}{\rna}
\newcommand{\SLOWsunflowPIPDynamicCIMAX}{\rna}
\newcommand{\SLOWsunflowHBUP}{5}
\newcommand{\SLOWsunflowHBUPCI}{0.0}
\newcommand{\SLOWsunflowHBUPCIMIN}{5}
\newcommand{\SLOWsunflowHBUPCIMAX}{5}
\newcommand{\SLOWsunflowHBDynamicUP}{48}
\newcommand{\SLOWsunflowHBDynamicUPCI}{5.9}
\newcommand{\SLOWsunflowHBDynamicUPCIMIN}{42}
\newcommand{\SLOWsunflowHBDynamicUPCIMAX}{54}
\newcommand{\SLOWsunflowWCPUP}{17}
\newcommand{\SLOWsunflowWCPUPCI}{0.0}
\newcommand{\SLOWsunflowWCPUPCIMIN}{17}
\newcommand{\SLOWsunflowWCPUPCIMAX}{17}
\newcommand{\SLOWsunflowWCPDynamicUP}{289}
\newcommand{\SLOWsunflowWCPDynamicUPCI}{26}
\newcommand{\SLOWsunflowWCPDynamicUPCIMIN}{263}
\newcommand{\SLOWsunflowWCPDynamicUPCIMAX}{315}
\newcommand{\SLOWsunflowWDCnoGUP}{\rna}
\newcommand{\SLOWsunflowWDCnoGUPCI}{\rna}
\newcommand{\SLOWsunflowWDCnoGUPCIMIN}{\rna}
\newcommand{\SLOWsunflowWDCnoGUPCIMAX}{\rna}
\newcommand{\SLOWsunflowWDCnoGDynamicUP}{\rna}
\newcommand{\SLOWsunflowWDCnoGDynamicUPCI}{\rna}
\newcommand{\SLOWsunflowWDCnoGDynamicUPCIMIN}{\rna}
\newcommand{\SLOWsunflowWDCnoGDynamicUPCIMAX}{\rna}
\newcommand{\SLOWsunflowWDCUP}{\rna}
\newcommand{\SLOWsunflowWDCUPCI}{\rna}
\newcommand{\SLOWsunflowWDCUPCIMIN}{\rna}
\newcommand{\SLOWsunflowWDCUPCIMAX}{\rna}
\newcommand{\SLOWsunflowWDCDynamicUP}{\rna}
\newcommand{\SLOWsunflowWDCDynamicUPCI}{\rna}
\newcommand{\SLOWsunflowWDCDynamicUPCIMIN}{\rna}
\newcommand{\SLOWsunflowWDCDynamicUPCIMAX}{\rna}
\newcommand{\SLOWsunflowCAPOnoGUP}{\rna}
\newcommand{\SLOWsunflowCAPOnoGUPCI}{\rna}
\newcommand{\SLOWsunflowCAPOnoGUPCIMIN}{\rna}
\newcommand{\SLOWsunflowCAPOnoGUPCIMAX}{\rna}
\newcommand{\SLOWsunflowCAPOnoGDynamicUP}{\rna}
\newcommand{\SLOWsunflowCAPOnoGDynamicUPCI}{\rna}
\newcommand{\SLOWsunflowCAPOnoGDynamicUPCIMIN}{\rna}
\newcommand{\SLOWsunflowCAPOnoGDynamicUPCIMAX}{\rna}
\newcommand{\SLOWsunflowCAPOUP}{\rna}
\newcommand{\SLOWsunflowCAPOUPCI}{\rna}
\newcommand{\SLOWsunflowCAPOUPCIMIN}{\rna}
\newcommand{\SLOWsunflowCAPOUPCIMAX}{\rna}
\newcommand{\SLOWsunflowCAPODynamicUP}{\rna}
\newcommand{\SLOWsunflowCAPODynamicUPCI}{\rna}
\newcommand{\SLOWsunflowCAPODynamicUPCIMIN}{\rna}
\newcommand{\SLOWsunflowCAPODynamicUPCIMAX}{\rna}
\newcommand{\SLOWsunflowPIPUP}{\rna}
\newcommand{\SLOWsunflowPIPUPCI}{\rna}
\newcommand{\SLOWsunflowPIPUPCIMIN}{\rna}
\newcommand{\SLOWsunflowPIPUPCIMAX}{\rna}
\newcommand{\SLOWsunflowPIPDynamicUP}{\rna}
\newcommand{\SLOWsunflowPIPDynamicUPCI}{\rna}
\newcommand{\SLOWsunflowPIPDynamicUPCIMIN}{\rna}
\newcommand{\SLOWsunflowPIPDynamicUPCIMAX}{\rna}
\newcommand{\SLOWsunflowPIPHB}{\rna}
\newcommand{\SLOWsunflowPIPHBCI}{\rna}
\newcommand{\SLOWsunflowPIPHBCIMIN}{\rna}
\newcommand{\SLOWsunflowPIPHBCIMAX}{\rna}
\newcommand{\SLOWsunflowPIPHBDynamic}{\rna}
\newcommand{\SLOWsunflowPIPHBDynamicCI}{\rna}
\newcommand{\SLOWsunflowPIPHBDynamicCIMIN}{\rna}
\newcommand{\SLOWsunflowPIPHBDynamicCIMAX}{\rna}
\newcommand{\SLOWsunflowPIPWCP}{\rna}
\newcommand{\SLOWsunflowPIPWCPCI}{\rna}
\newcommand{\SLOWsunflowPIPWCPCIMIN}{\rna}
\newcommand{\SLOWsunflowPIPWCPCIMAX}{\rna}
\newcommand{\SLOWsunflowPIPWCPDynamic}{\rna}
\newcommand{\SLOWsunflowPIPWCPDynamicCI}{\rna}
\newcommand{\SLOWsunflowPIPWCPDynamicCIMIN}{\rna}
\newcommand{\SLOWsunflowPIPWCPDynamicCIMAX}{\rna}
\newcommand{\SLOWsunflowPIPWDC}{\rna}
\newcommand{\SLOWsunflowPIPWDCCI}{\rna}
\newcommand{\SLOWsunflowPIPWDCCIMIN}{\rna}
\newcommand{\SLOWsunflowPIPWDCCIMAX}{\rna}
\newcommand{\SLOWsunflowPIPWDCDynamic}{\rna}
\newcommand{\SLOWsunflowPIPWDCDynamicCI}{\rna}
\newcommand{\SLOWsunflowPIPWDCDynamicCIMIN}{\rna}
\newcommand{\SLOWsunflowPIPWDCDynamicCIMAX}{\rna}
\newcommand{\SLOWsunflowPIPCAPO}{\rna}
\newcommand{\SLOWsunflowPIPCAPOCI}{\rna}
\newcommand{\SLOWsunflowPIPCAPOCIMIN}{\rna}
\newcommand{\SLOWsunflowPIPCAPOCIMAX}{\rna}
\newcommand{\SLOWsunflowPIPCAPODynamic}{\rna}
\newcommand{\SLOWsunflowPIPCAPODynamicCI}{\rna}
\newcommand{\SLOWsunflowPIPCAPODynamicCIMIN}{\rna}
\newcommand{\SLOWsunflowPIPCAPODynamicCIMAX}{\rna}
\newcommand{\SLOWsunflowPIPPIP}{\rna}
\newcommand{\SLOWsunflowPIPPIPCI}{\rna}
\newcommand{\SLOWsunflowPIPPIPCIMIN}{\rna}
\newcommand{\SLOWsunflowPIPPIPCIMAX}{\rna}
\newcommand{\SLOWsunflowPIPPIPDynamic}{\rna}
\newcommand{\SLOWsunflowPIPPIPDynamicCI}{\rna}
\newcommand{\SLOWsunflowPIPPIPDynamicCIMIN}{\rna}
\newcommand{\SLOWsunflowPIPPIPDynamicCIMAX}{\rna}
\newcommand{\SLOWtomcatEvents}{0}
\newcommand{\SLOWtomcatNoFPEvents}{0}
\newcommand{\SLOWtomcatMaxLiveThreads}{53}
\newcommand{\SLOWtomcatTotalThreads}{53}
\newcommand{\SLOWtomcatBaseTime}{0.83}
\newcommand{\SLOWtomcatBaseTimeCI}{2.0}
\newcommand{\SLOWtomcatEmptyTime}{\rna}
\newcommand{\SLOWtomcatEmptyTimeCI}{\rna}
\newcommand{\SLOWtomcatEmptyTimeCIMIN}{\rna}
\newcommand{\SLOWtomcatEmptyTimeCIMAX}{\rna}
\newcommand{\SLOWtomcatFTTime}{30}
\newcommand{\SLOWtomcatFTTimeCI}{1.5}
\newcommand{\SLOWtomcatHBTime}{12}
\newcommand{\SLOWtomcatHBTimeCI}{0.93}
\newcommand{\SLOWtomcatWCPTime}{20}
\newcommand{\SLOWtomcatWCPTimeCI}{0.92}
\newcommand{\SLOWtomcatDCnoGExcTime}{17}
\newcommand{\SLOWtomcatDCnoGExcTimeCI}{0.94}
\newcommand{\SLOWtomcatDCnoGTime}{\rna}
\newcommand{\SLOWtomcatDCnoGTimeCI}{\rna}
\newcommand{\SLOWtomcatDCnoGTimeCIMIN}{\rna}
\newcommand{\SLOWtomcatDCnoGTimeCIMAX}{\rna}
\newcommand{\SLOWtomcatDCExcTime}{15}
\newcommand{\SLOWtomcatDCExcTimeCI}{0.11}
\newcommand{\SLOWtomcatDCTime}{\rna}
\newcommand{\SLOWtomcatDCTimeCI}{\rna}
\newcommand{\SLOWtomcatDCTimeCIMIN}{\rna}
\newcommand{\SLOWtomcatDCTimeCIMAX}{\rna}
\newcommand{\SLOWtomcatCAPOnoGExcTime}{12}
\newcommand{\SLOWtomcatCAPOnoGExcTimeCI}{1.4}
\newcommand{\SLOWtomcatCAPOnoGTime}{\rna}
\newcommand{\SLOWtomcatCAPOnoGTimeCI}{\rna}
\newcommand{\SLOWtomcatCAPOnoGTimeCIMIN}{\rna}
\newcommand{\SLOWtomcatCAPOnoGTimeCIMAX}{\rna}
\newcommand{\SLOWtomcatCAPOExcTime}{9.8}
\newcommand{\SLOWtomcatCAPOExcTimeCI}{0.26}
\newcommand{\SLOWtomcatCAPOTime}{\rna}
\newcommand{\SLOWtomcatCAPOTimeCI}{\rna}
\newcommand{\SLOWtomcatCAPOTimeCIMIN}{\rna}
\newcommand{\SLOWtomcatCAPOTimeCIMAX}{\rna}
\newcommand{\SLOWtomcatStaticTime}{\rzero}
\newcommand{\SLOWtomcatDynamicTime}{\rzero}
\newcommand{\SLOWtomcatBaseMem}{610}
\newcommand{\SLOWtomcatBaseMemCI}{13.0}
\newcommand{\SLOWtomcatHBMem}{33}
\newcommand{\SLOWtomcatHBMemCI}{1.9}
\newcommand{\SLOWtomcatFTMem}{110}
\newcommand{\SLOWtomcatFTMemCI}{2.6}
\newcommand{\SLOWtomcatWCPMem}{41}
\newcommand{\SLOWtomcatWCPMemCI}{5.1}
\newcommand{\SLOWtomcatDCnoGExcMem}{29}
\newcommand{\SLOWtomcatDCnoGExcMemCI}{3.0}
\newcommand{\SLOWtomcatDCnoGMem}{\memna}
\newcommand{\SLOWtomcatDCnoGMemCI}{\memna}
\newcommand{\SLOWtomcatDCnoGMemCIMIN}{\memna}
\newcommand{\SLOWtomcatDCnoGMemCIMAX}{\memna}
\newcommand{\SLOWtomcatDCExcMem}{28}
\newcommand{\SLOWtomcatDCExcMemCI}{0.62}
\newcommand{\SLOWtomcatDCMem}{\memna}
\newcommand{\SLOWtomcatDCMemCI}{\memna}
\newcommand{\SLOWtomcatDCMemCIMIN}{\memna}
\newcommand{\SLOWtomcatDCMemCIMAX}{\memna}
\newcommand{\SLOWtomcatCAPOnoGExcMem}{25}
\newcommand{\SLOWtomcatCAPOnoGExcMemCI}{0.21}
\newcommand{\SLOWtomcatCAPOnoGMem}{\memna}
\newcommand{\SLOWtomcatCAPOnoGMemCI}{\memna}
\newcommand{\SLOWtomcatCAPOnoGMemCIMIN}{\memna}
\newcommand{\SLOWtomcatCAPOnoGMemCIMAX}{\memna}
\newcommand{\SLOWtomcatCAPOExcMem}{21}
\newcommand{\SLOWtomcatCAPOExcMemCI}{0.35}
\newcommand{\SLOWtomcatCAPOMem}{\memna}
\newcommand{\SLOWtomcatCAPOMemCI}{\memna}
\newcommand{\SLOWtomcatCAPOMemCIMIN}{\memna}
\newcommand{\SLOWtomcatCAPOMemCIMAX}{\memna}
\newcommand{\SLOWtomcatEventsCI}{0}
\newcommand{\SLOWtomcatEventsCIMIN}{0}
\newcommand{\SLOWtomcatEventsCIMAX}{0}
\newcommand{\SLOWtomcatNoFPEventsCI}{0}
\newcommand{\SLOWtomcatNoFPEventsCIMIN}{0}
\newcommand{\SLOWtomcatNoFPEventsCIMAX}{0}
\newcommand{\SLOWtomcatHB}{637}
\newcommand{\SLOWtomcatHBCI}{41}
\newcommand{\SLOWtomcatHBCIMIN}{596}
\newcommand{\SLOWtomcatHBCIMAX}{678}
\newcommand{\SLOWtomcatHBDynamic}{1,782,554}
\newcommand{\SLOWtomcatHBDynamicCI}{46,253}
\newcommand{\SLOWtomcatHBDynamicCIMIN}{1,736,301}
\newcommand{\SLOWtomcatHBDynamicCIMAX}{1,828,807}
\newcommand{\SLOWtomcatFT}{422}
\newcommand{\SLOWtomcatFTCI}{19}
\newcommand{\SLOWtomcatFTCIMIN}{403}
\newcommand{\SLOWtomcatFTCIMAX}{441}
\newcommand{\SLOWtomcatFTDynamic}{3,563,576}
\newcommand{\SLOWtomcatFTDynamicCI}{254,208}
\newcommand{\SLOWtomcatFTDynamicCIMIN}{3,309,368}
\newcommand{\SLOWtomcatFTDynamicCIMAX}{3,817,784}
\newcommand{\SLOWtomcatWCP}{597}
\newcommand{\SLOWtomcatWCPCI}{38}
\newcommand{\SLOWtomcatWCPCIMIN}{559}
\newcommand{\SLOWtomcatWCPCIMAX}{635}
\newcommand{\SLOWtomcatWCPDynamic}{1,635,042}
\newcommand{\SLOWtomcatWCPDynamicCI}{7,885}
\newcommand{\SLOWtomcatWCPDynamicCIMIN}{1,627,157}
\newcommand{\SLOWtomcatWCPDynamicCIMAX}{1,642,927}
\newcommand{\SLOWtomcatDCnoGExc}{629}
\newcommand{\SLOWtomcatDCnoGExcCI}{83}
\newcommand{\SLOWtomcatDCnoGExcCIMIN}{546}
\newcommand{\SLOWtomcatDCnoGExcCIMAX}{712}
\newcommand{\SLOWtomcatDCnoGExcDynamic}{1,559,537}
\newcommand{\SLOWtomcatDCnoGExcDynamicCI}{73,962}
\newcommand{\SLOWtomcatDCnoGExcDynamicCIMIN}{1,485,575}
\newcommand{\SLOWtomcatDCnoGExcDynamicCIMAX}{1,633,499}
\newcommand{\SLOWtomcatDCnoG}{\rna}
\newcommand{\SLOWtomcatDCnoGCI}{\rna}
\newcommand{\SLOWtomcatDCnoGCIMIN}{\rna}
\newcommand{\SLOWtomcatDCnoGCIMAX}{\rna}
\newcommand{\SLOWtomcatDCnoGDynamic}{\rna}
\newcommand{\SLOWtomcatDCnoGDynamicCI}{\rna}
\newcommand{\SLOWtomcatDCnoGDynamicCIMIN}{\rna}
\newcommand{\SLOWtomcatDCnoGDynamicCIMAX}{\rna}
\newcommand{\SLOWtomcatDCExc}{67}
\newcommand{\SLOWtomcatDCExcCI}{5.9}
\newcommand{\SLOWtomcatDCExcCIMIN}{61}
\newcommand{\SLOWtomcatDCExcCIMAX}{73}
\newcommand{\SLOWtomcatDCExcDynamic}{13,325}
\newcommand{\SLOWtomcatDCExcDynamicCI}{134}
\newcommand{\SLOWtomcatDCExcDynamicCIMIN}{13,191}
\newcommand{\SLOWtomcatDCExcDynamicCIMAX}{13,459}
\newcommand{\SLOWtomcatDC}{\rna}
\newcommand{\SLOWtomcatDCCI}{\rna}
\newcommand{\SLOWtomcatDCCIMIN}{\rna}
\newcommand{\SLOWtomcatDCCIMAX}{\rna}
\newcommand{\SLOWtomcatDCDynamic}{\rna}
\newcommand{\SLOWtomcatDCDynamicCI}{\rna}
\newcommand{\SLOWtomcatDCDynamicCIMIN}{\rna}
\newcommand{\SLOWtomcatDCDynamicCIMAX}{\rna}
\newcommand{\SLOWtomcatCAPOnoGExc}{656}
\newcommand{\SLOWtomcatCAPOnoGExcCI}{50}
\newcommand{\SLOWtomcatCAPOnoGExcCIMIN}{606}
\newcommand{\SLOWtomcatCAPOnoGExcCIMAX}{706}
\newcommand{\SLOWtomcatCAPOnoGExcDynamic}{1,833,028}
\newcommand{\SLOWtomcatCAPOnoGExcDynamicCI}{142,330}
\newcommand{\SLOWtomcatCAPOnoGExcDynamicCIMIN}{1,690,698}
\newcommand{\SLOWtomcatCAPOnoGExcDynamicCIMAX}{1,975,358}
\newcommand{\SLOWtomcatCAPOnoG}{\rna}
\newcommand{\SLOWtomcatCAPOnoGCI}{\rna}
\newcommand{\SLOWtomcatCAPOnoGCIMIN}{\rna}
\newcommand{\SLOWtomcatCAPOnoGCIMAX}{\rna}
\newcommand{\SLOWtomcatCAPOnoGDynamic}{\rna}
\newcommand{\SLOWtomcatCAPOnoGDynamicCI}{\rna}
\newcommand{\SLOWtomcatCAPOnoGDynamicCIMIN}{\rna}
\newcommand{\SLOWtomcatCAPOnoGDynamicCIMAX}{\rna}
\newcommand{\SLOWtomcatCAPOExc}{64}
\newcommand{\SLOWtomcatCAPOExcCI}{0.98}
\newcommand{\SLOWtomcatCAPOExcCIMIN}{63}
\newcommand{\SLOWtomcatCAPOExcCIMAX}{65}
\newcommand{\SLOWtomcatCAPOExcDynamic}{12,530}
\newcommand{\SLOWtomcatCAPOExcDynamicCI}{608}
\newcommand{\SLOWtomcatCAPOExcDynamicCIMIN}{11,922}
\newcommand{\SLOWtomcatCAPOExcDynamicCIMAX}{13,138}
\newcommand{\SLOWtomcatCAPO}{\rna}
\newcommand{\SLOWtomcatCAPOCI}{\rna}
\newcommand{\SLOWtomcatCAPOCIMIN}{\rna}
\newcommand{\SLOWtomcatCAPOCIMAX}{\rna}
\newcommand{\SLOWtomcatCAPODynamic}{\rna}
\newcommand{\SLOWtomcatCAPODynamicCI}{\rna}
\newcommand{\SLOWtomcatCAPODynamicCIMIN}{\rna}
\newcommand{\SLOWtomcatCAPODynamicCIMAX}{\rna}
\newcommand{\SLOWtomcatPIP}{\rna}
\newcommand{\SLOWtomcatPIPCI}{\rna}
\newcommand{\SLOWtomcatPIPCIMIN}{\rna}
\newcommand{\SLOWtomcatPIPCIMAX}{\rna}
\newcommand{\SLOWtomcatPIPDynamic}{\rna}
\newcommand{\SLOWtomcatPIPDynamicCI}{\rna}
\newcommand{\SLOWtomcatPIPDynamicCIMIN}{\rna}
\newcommand{\SLOWtomcatPIPDynamicCIMAX}{\rna}
\newcommand{\SLOWtomcatHBUP}{831}
\newcommand{\SLOWtomcatHBUPCI}{15}
\newcommand{\SLOWtomcatHBUPCIMIN}{816}
\newcommand{\SLOWtomcatHBUPCIMAX}{846}
\newcommand{\SLOWtomcatHBDynamicUP}{1,772,130}
\newcommand{\SLOWtomcatHBDynamicUPCI}{65,730}
\newcommand{\SLOWtomcatHBDynamicUPCIMIN}{1,706,400}
\newcommand{\SLOWtomcatHBDynamicUPCIMAX}{1,837,860}
\newcommand{\SLOWtomcatWCPUP}{799}
\newcommand{\SLOWtomcatWCPUPCI}{46}
\newcommand{\SLOWtomcatWCPUPCIMIN}{753}
\newcommand{\SLOWtomcatWCPUPCIMAX}{845}
\newcommand{\SLOWtomcatWCPDynamicUP}{1,633,951}
\newcommand{\SLOWtomcatWCPDynamicUPCI}{6,435}
\newcommand{\SLOWtomcatWCPDynamicUPCIMIN}{1,627,516}
\newcommand{\SLOWtomcatWCPDynamicUPCIMAX}{1,640,386}
\newcommand{\SLOWtomcatWDCnoGUP}{\rna}
\newcommand{\SLOWtomcatWDCnoGUPCI}{\rna}
\newcommand{\SLOWtomcatWDCnoGUPCIMIN}{\rna}
\newcommand{\SLOWtomcatWDCnoGUPCIMAX}{\rna}
\newcommand{\SLOWtomcatWDCnoGDynamicUP}{\rna}
\newcommand{\SLOWtomcatWDCnoGDynamicUPCI}{\rna}
\newcommand{\SLOWtomcatWDCnoGDynamicUPCIMIN}{\rna}
\newcommand{\SLOWtomcatWDCnoGDynamicUPCIMAX}{\rna}
\newcommand{\SLOWtomcatWDCUP}{\rna}
\newcommand{\SLOWtomcatWDCUPCI}{\rna}
\newcommand{\SLOWtomcatWDCUPCIMIN}{\rna}
\newcommand{\SLOWtomcatWDCUPCIMAX}{\rna}
\newcommand{\SLOWtomcatWDCDynamicUP}{\rna}
\newcommand{\SLOWtomcatWDCDynamicUPCI}{\rna}
\newcommand{\SLOWtomcatWDCDynamicUPCIMIN}{\rna}
\newcommand{\SLOWtomcatWDCDynamicUPCIMAX}{\rna}
\newcommand{\SLOWtomcatCAPOnoGUP}{\rna}
\newcommand{\SLOWtomcatCAPOnoGUPCI}{\rna}
\newcommand{\SLOWtomcatCAPOnoGUPCIMIN}{\rna}
\newcommand{\SLOWtomcatCAPOnoGUPCIMAX}{\rna}
\newcommand{\SLOWtomcatCAPOnoGDynamicUP}{\rna}
\newcommand{\SLOWtomcatCAPOnoGDynamicUPCI}{\rna}
\newcommand{\SLOWtomcatCAPOnoGDynamicUPCIMIN}{\rna}
\newcommand{\SLOWtomcatCAPOnoGDynamicUPCIMAX}{\rna}
\newcommand{\SLOWtomcatCAPOUP}{\rna}
\newcommand{\SLOWtomcatCAPOUPCI}{\rna}
\newcommand{\SLOWtomcatCAPOUPCIMIN}{\rna}
\newcommand{\SLOWtomcatCAPOUPCIMAX}{\rna}
\newcommand{\SLOWtomcatCAPODynamicUP}{\rna}
\newcommand{\SLOWtomcatCAPODynamicUPCI}{\rna}
\newcommand{\SLOWtomcatCAPODynamicUPCIMIN}{\rna}
\newcommand{\SLOWtomcatCAPODynamicUPCIMAX}{\rna}
\newcommand{\SLOWtomcatPIPUP}{\rna}
\newcommand{\SLOWtomcatPIPUPCI}{\rna}
\newcommand{\SLOWtomcatPIPUPCIMIN}{\rna}
\newcommand{\SLOWtomcatPIPUPCIMAX}{\rna}
\newcommand{\SLOWtomcatPIPDynamicUP}{\rna}
\newcommand{\SLOWtomcatPIPDynamicUPCI}{\rna}
\newcommand{\SLOWtomcatPIPDynamicUPCIMIN}{\rna}
\newcommand{\SLOWtomcatPIPDynamicUPCIMAX}{\rna}
\newcommand{\SLOWtomcatPIPHB}{\rna}
\newcommand{\SLOWtomcatPIPHBCI}{\rna}
\newcommand{\SLOWtomcatPIPHBCIMIN}{\rna}
\newcommand{\SLOWtomcatPIPHBCIMAX}{\rna}
\newcommand{\SLOWtomcatPIPHBDynamic}{\rna}
\newcommand{\SLOWtomcatPIPHBDynamicCI}{\rna}
\newcommand{\SLOWtomcatPIPHBDynamicCIMIN}{\rna}
\newcommand{\SLOWtomcatPIPHBDynamicCIMAX}{\rna}
\newcommand{\SLOWtomcatPIPWCP}{\rna}
\newcommand{\SLOWtomcatPIPWCPCI}{\rna}
\newcommand{\SLOWtomcatPIPWCPCIMIN}{\rna}
\newcommand{\SLOWtomcatPIPWCPCIMAX}{\rna}
\newcommand{\SLOWtomcatPIPWCPDynamic}{\rna}
\newcommand{\SLOWtomcatPIPWCPDynamicCI}{\rna}
\newcommand{\SLOWtomcatPIPWCPDynamicCIMIN}{\rna}
\newcommand{\SLOWtomcatPIPWCPDynamicCIMAX}{\rna}
\newcommand{\SLOWtomcatPIPWDC}{\rna}
\newcommand{\SLOWtomcatPIPWDCCI}{\rna}
\newcommand{\SLOWtomcatPIPWDCCIMIN}{\rna}
\newcommand{\SLOWtomcatPIPWDCCIMAX}{\rna}
\newcommand{\SLOWtomcatPIPWDCDynamic}{\rna}
\newcommand{\SLOWtomcatPIPWDCDynamicCI}{\rna}
\newcommand{\SLOWtomcatPIPWDCDynamicCIMIN}{\rna}
\newcommand{\SLOWtomcatPIPWDCDynamicCIMAX}{\rna}
\newcommand{\SLOWtomcatPIPCAPO}{\rna}
\newcommand{\SLOWtomcatPIPCAPOCI}{\rna}
\newcommand{\SLOWtomcatPIPCAPOCIMIN}{\rna}
\newcommand{\SLOWtomcatPIPCAPOCIMAX}{\rna}
\newcommand{\SLOWtomcatPIPCAPODynamic}{\rna}
\newcommand{\SLOWtomcatPIPCAPODynamicCI}{\rna}
\newcommand{\SLOWtomcatPIPCAPODynamicCIMIN}{\rna}
\newcommand{\SLOWtomcatPIPCAPODynamicCIMAX}{\rna}
\newcommand{\SLOWtomcatPIPPIP}{\rna}
\newcommand{\SLOWtomcatPIPPIPCI}{\rna}
\newcommand{\SLOWtomcatPIPPIPCIMIN}{\rna}
\newcommand{\SLOWtomcatPIPPIPCIMAX}{\rna}
\newcommand{\SLOWtomcatPIPPIPDynamic}{\rna}
\newcommand{\SLOWtomcatPIPPIPDynamicCI}{\rna}
\newcommand{\SLOWtomcatPIPPIPDynamicCIMIN}{\rna}
\newcommand{\SLOWtomcatPIPPIPDynamicCIMAX}{\rna}
\newcommand{\SLOWxalanEvents}{0}
\newcommand{\SLOWxalanNoFPEvents}{0}
\newcommand{\SLOWxalanMaxLiveThreads}{7}
\newcommand{\SLOWxalanTotalThreads}{7}
\newcommand{\SLOWxalanBaseTime}{2.0}
\newcommand{\SLOWxalanBaseTimeCI}{130}
\newcommand{\SLOWxalanEmptyTime}{\rna}
\newcommand{\SLOWxalanEmptyTimeCI}{\rna}
\newcommand{\SLOWxalanEmptyTimeCIMIN}{\rna}
\newcommand{\SLOWxalanEmptyTimeCIMAX}{\rna}
\newcommand{\SLOWxalanFTTime}{4.7}
\newcommand{\SLOWxalanFTTimeCI}{0.069}
\newcommand{\SLOWxalanHBTime}{11}
\newcommand{\SLOWxalanHBTimeCI}{0.041}
\newcommand{\SLOWxalanWCPTime}{47}
\newcommand{\SLOWxalanWCPTimeCI}{0.25}
\newcommand{\SLOWxalanDCnoGExcTime}{45}
\newcommand{\SLOWxalanDCnoGExcTimeCI}{2.4}
\newcommand{\SLOWxalanDCnoGTime}{\rna}
\newcommand{\SLOWxalanDCnoGTimeCI}{\rna}
\newcommand{\SLOWxalanDCnoGTimeCIMIN}{\rna}
\newcommand{\SLOWxalanDCnoGTimeCIMAX}{\rna}
\newcommand{\SLOWxalanDCExcTime}{54}
\newcommand{\SLOWxalanDCExcTimeCI}{55}
\newcommand{\SLOWxalanDCTime}{\rna}
\newcommand{\SLOWxalanDCTimeCI}{\rna}
\newcommand{\SLOWxalanDCTimeCIMIN}{\rna}
\newcommand{\SLOWxalanDCTimeCIMAX}{\rna}
\newcommand{\SLOWxalanCAPOnoGExcTime}{39}
\newcommand{\SLOWxalanCAPOnoGExcTimeCI}{5.1}
\newcommand{\SLOWxalanCAPOnoGTime}{\rna}
\newcommand{\SLOWxalanCAPOnoGTimeCI}{\rna}
\newcommand{\SLOWxalanCAPOnoGTimeCIMIN}{\rna}
\newcommand{\SLOWxalanCAPOnoGTimeCIMAX}{\rna}
\newcommand{\SLOWxalanCAPOExcTime}{48}
\newcommand{\SLOWxalanCAPOExcTimeCI}{49}
\newcommand{\SLOWxalanCAPOTime}{\rna}
\newcommand{\SLOWxalanCAPOTimeCI}{\rna}
\newcommand{\SLOWxalanCAPOTimeCIMIN}{\rna}
\newcommand{\SLOWxalanCAPOTimeCIMAX}{\rna}
\newcommand{\SLOWxalanStaticTime}{\rzero}
\newcommand{\SLOWxalanDynamicTime}{\rzero}
\newcommand{\SLOWxalanBaseMem}{680}
\newcommand{\SLOWxalanBaseMemCI}{14.0}
\newcommand{\SLOWxalanHBMem}{25}
\newcommand{\SLOWxalanHBMemCI}{0.27}
\newcommand{\SLOWxalanFTMem}{6.5}
\newcommand{\SLOWxalanFTMemCI}{0.13}
\newcommand{\SLOWxalanWCPMem}{67}
\newcommand{\SLOWxalanWCPMemCI}{1.5}
\newcommand{\SLOWxalanDCnoGExcMem}{59}
\newcommand{\SLOWxalanDCnoGExcMemCI}{0.21}
\newcommand{\SLOWxalanDCnoGMem}{\memna}
\newcommand{\SLOWxalanDCnoGMemCI}{\memna}
\newcommand{\SLOWxalanDCnoGMemCIMIN}{\memna}
\newcommand{\SLOWxalanDCnoGMemCIMAX}{\memna}
\newcommand{\SLOWxalanDCExcMem}{65}
\newcommand{\SLOWxalanDCExcMemCI}{64}
\newcommand{\SLOWxalanDCMem}{\memna}
\newcommand{\SLOWxalanDCMemCI}{\memna}
\newcommand{\SLOWxalanDCMemCIMIN}{\memna}
\newcommand{\SLOWxalanDCMemCIMAX}{\memna}
\newcommand{\SLOWxalanCAPOnoGExcMem}{59}
\newcommand{\SLOWxalanCAPOnoGExcMemCI}{7.4}
\newcommand{\SLOWxalanCAPOnoGMem}{\memna}
\newcommand{\SLOWxalanCAPOnoGMemCI}{\memna}
\newcommand{\SLOWxalanCAPOnoGMemCIMIN}{\memna}
\newcommand{\SLOWxalanCAPOnoGMemCIMAX}{\memna}
\newcommand{\SLOWxalanCAPOExcMem}{62}
\newcommand{\SLOWxalanCAPOExcMemCI}{62}
\newcommand{\SLOWxalanCAPOMem}{\memna}
\newcommand{\SLOWxalanCAPOMemCI}{\memna}
\newcommand{\SLOWxalanCAPOMemCIMIN}{\memna}
\newcommand{\SLOWxalanCAPOMemCIMAX}{\memna}
\newcommand{\SLOWxalanEventsCI}{0}
\newcommand{\SLOWxalanEventsCIMIN}{0}
\newcommand{\SLOWxalanEventsCIMAX}{0}
\newcommand{\SLOWxalanNoFPEventsCI}{0}
\newcommand{\SLOWxalanNoFPEventsCIMIN}{0}
\newcommand{\SLOWxalanNoFPEventsCIMAX}{0}
\newcommand{\SLOWxalanHB}{8}
\newcommand{\SLOWxalanHBCI}{0.0}
\newcommand{\SLOWxalanHBCIMIN}{8}
\newcommand{\SLOWxalanHBCIMAX}{8}
\newcommand{\SLOWxalanHBDynamic}{2,324}
\newcommand{\SLOWxalanHBDynamicCI}{75}
\newcommand{\SLOWxalanHBDynamicCIMIN}{2,249}
\newcommand{\SLOWxalanHBDynamicCIMAX}{2,399}
\newcommand{\SLOWxalanFT}{8}
\newcommand{\SLOWxalanFTCI}{0.0}
\newcommand{\SLOWxalanFTCIMIN}{8}
\newcommand{\SLOWxalanFTCIMAX}{8}
\newcommand{\SLOWxalanFTDynamic}{2,810}
\newcommand{\SLOWxalanFTDynamicCI}{68}
\newcommand{\SLOWxalanFTDynamicCIMIN}{2,742}
\newcommand{\SLOWxalanFTDynamicCIMAX}{2,878}
\newcommand{\SLOWxalanWCP}{64}
\newcommand{\SLOWxalanWCPCI}{0.0}
\newcommand{\SLOWxalanWCPCIMIN}{64}
\newcommand{\SLOWxalanWCPCIMAX}{64}
\newcommand{\SLOWxalanWCPDynamic}{5,307,857}
\newcommand{\SLOWxalanWCPDynamicCI}{252}
\newcommand{\SLOWxalanWCPDynamicCIMIN}{5,307,605}
\newcommand{\SLOWxalanWCPDynamicCIMAX}{5,308,109}
\newcommand{\SLOWxalanDCnoGExc}{78}
\newcommand{\SLOWxalanDCnoGExcCI}{0.98}
\newcommand{\SLOWxalanDCnoGExcCIMIN}{77}
\newcommand{\SLOWxalanDCnoGExcCIMAX}{79}
\newcommand{\SLOWxalanDCnoGExcDynamic}{7,444,803}
\newcommand{\SLOWxalanDCnoGExcDynamicCI}{13,902}
\newcommand{\SLOWxalanDCnoGExcDynamicCIMIN}{7,430,901}
\newcommand{\SLOWxalanDCnoGExcDynamicCIMAX}{7,458,705}
\newcommand{\SLOWxalanDCnoG}{\rna}
\newcommand{\SLOWxalanDCnoGCI}{\rna}
\newcommand{\SLOWxalanDCnoGCIMIN}{\rna}
\newcommand{\SLOWxalanDCnoGCIMAX}{\rna}
\newcommand{\SLOWxalanDCnoGDynamic}{\rna}
\newcommand{\SLOWxalanDCnoGDynamicCI}{\rna}
\newcommand{\SLOWxalanDCnoGDynamicCIMIN}{\rna}
\newcommand{\SLOWxalanDCnoGDynamicCIMAX}{\rna}
\newcommand{\SLOWxalanDCExc}{9}
\newcommand{\SLOWxalanDCExcCI}{10}
\newcommand{\SLOWxalanDCExcCIMIN}{-1}
\newcommand{\SLOWxalanDCExcCIMAX}{19}
\newcommand{\SLOWxalanDCExcDynamic}{313,191}
\newcommand{\SLOWxalanDCExcDynamicCI}{306,922}
\newcommand{\SLOWxalanDCExcDynamicCIMIN}{6,269}
\newcommand{\SLOWxalanDCExcDynamicCIMAX}{620,113}
\newcommand{\SLOWxalanDC}{\rna}
\newcommand{\SLOWxalanDCCI}{\rna}
\newcommand{\SLOWxalanDCCIMIN}{\rna}
\newcommand{\SLOWxalanDCCIMAX}{\rna}
\newcommand{\SLOWxalanDCDynamic}{\rna}
\newcommand{\SLOWxalanDCDynamicCI}{\rna}
\newcommand{\SLOWxalanDCDynamicCIMIN}{\rna}
\newcommand{\SLOWxalanDCDynamicCIMAX}{\rna}
\newcommand{\SLOWxalanCAPOnoGExc}{78}
\newcommand{\SLOWxalanCAPOnoGExcCI}{0.0}
\newcommand{\SLOWxalanCAPOnoGExcCIMIN}{78}
\newcommand{\SLOWxalanCAPOnoGExcCIMAX}{78}
\newcommand{\SLOWxalanCAPOnoGExcDynamic}{7,437,373}
\newcommand{\SLOWxalanCAPOnoGExcDynamicCI}{33,554}
\newcommand{\SLOWxalanCAPOnoGExcDynamicCIMIN}{7,403,819}
\newcommand{\SLOWxalanCAPOnoGExcDynamicCIMAX}{7,470,927}
\newcommand{\SLOWxalanCAPOnoG}{\rna}
\newcommand{\SLOWxalanCAPOnoGCI}{\rna}
\newcommand{\SLOWxalanCAPOnoGCIMIN}{\rna}
\newcommand{\SLOWxalanCAPOnoGCIMAX}{\rna}
\newcommand{\SLOWxalanCAPOnoGDynamic}{\rna}
\newcommand{\SLOWxalanCAPOnoGDynamicCI}{\rna}
\newcommand{\SLOWxalanCAPOnoGDynamicCIMIN}{\rna}
\newcommand{\SLOWxalanCAPOnoGDynamicCIMAX}{\rna}
\newcommand{\SLOWxalanCAPOExc}{10}
\newcommand{\SLOWxalanCAPOExcCI}{11}
\newcommand{\SLOWxalanCAPOExcCIMIN}{-1}
\newcommand{\SLOWxalanCAPOExcCIMAX}{21}
\newcommand{\SLOWxalanCAPOExcDynamic}{307,427}
\newcommand{\SLOWxalanCAPOExcDynamicCI}{301,273}
\newcommand{\SLOWxalanCAPOExcDynamicCIMIN}{6,154}
\newcommand{\SLOWxalanCAPOExcDynamicCIMAX}{608,700}
\newcommand{\SLOWxalanCAPO}{\rna}
\newcommand{\SLOWxalanCAPOCI}{\rna}
\newcommand{\SLOWxalanCAPOCIMIN}{\rna}
\newcommand{\SLOWxalanCAPOCIMAX}{\rna}
\newcommand{\SLOWxalanCAPODynamic}{\rna}
\newcommand{\SLOWxalanCAPODynamicCI}{\rna}
\newcommand{\SLOWxalanCAPODynamicCIMIN}{\rna}
\newcommand{\SLOWxalanCAPODynamicCIMAX}{\rna}
\newcommand{\SLOWxalanPIP}{\rna}
\newcommand{\SLOWxalanPIPCI}{\rna}
\newcommand{\SLOWxalanPIPCIMIN}{\rna}
\newcommand{\SLOWxalanPIPCIMAX}{\rna}
\newcommand{\SLOWxalanPIPDynamic}{\rna}
\newcommand{\SLOWxalanPIPDynamicCI}{\rna}
\newcommand{\SLOWxalanPIPDynamicCIMIN}{\rna}
\newcommand{\SLOWxalanPIPDynamicCIMAX}{\rna}
\newcommand{\SLOWxalanHBUP}{33}
\newcommand{\SLOWxalanHBUPCI}{0.0}
\newcommand{\SLOWxalanHBUPCIMIN}{33}
\newcommand{\SLOWxalanHBUPCIMAX}{33}
\newcommand{\SLOWxalanHBDynamicUP}{2,305}
\newcommand{\SLOWxalanHBDynamicUPCI}{53}
\newcommand{\SLOWxalanHBDynamicUPCIMIN}{2,252}
\newcommand{\SLOWxalanHBDynamicUPCIMAX}{2,358}
\newcommand{\SLOWxalanWCPUP}{164}
\newcommand{\SLOWxalanWCPUPCI}{0.0}
\newcommand{\SLOWxalanWCPUPCIMIN}{164}
\newcommand{\SLOWxalanWCPUPCIMAX}{164}
\newcommand{\SLOWxalanWCPDynamicUP}{5,314,305}
\newcommand{\SLOWxalanWCPDynamicUPCI}{2,200}
\newcommand{\SLOWxalanWCPDynamicUPCIMIN}{5,312,105}
\newcommand{\SLOWxalanWCPDynamicUPCIMAX}{5,316,505}
\newcommand{\SLOWxalanWDCnoGUP}{\rna}
\newcommand{\SLOWxalanWDCnoGUPCI}{\rna}
\newcommand{\SLOWxalanWDCnoGUPCIMIN}{\rna}
\newcommand{\SLOWxalanWDCnoGUPCIMAX}{\rna}
\newcommand{\SLOWxalanWDCnoGDynamicUP}{\rna}
\newcommand{\SLOWxalanWDCnoGDynamicUPCI}{\rna}
\newcommand{\SLOWxalanWDCnoGDynamicUPCIMIN}{\rna}
\newcommand{\SLOWxalanWDCnoGDynamicUPCIMAX}{\rna}
\newcommand{\SLOWxalanWDCUP}{\rna}
\newcommand{\SLOWxalanWDCUPCI}{\rna}
\newcommand{\SLOWxalanWDCUPCIMIN}{\rna}
\newcommand{\SLOWxalanWDCUPCIMAX}{\rna}
\newcommand{\SLOWxalanWDCDynamicUP}{\rna}
\newcommand{\SLOWxalanWDCDynamicUPCI}{\rna}
\newcommand{\SLOWxalanWDCDynamicUPCIMIN}{\rna}
\newcommand{\SLOWxalanWDCDynamicUPCIMAX}{\rna}
\newcommand{\SLOWxalanCAPOnoGUP}{\rna}
\newcommand{\SLOWxalanCAPOnoGUPCI}{\rna}
\newcommand{\SLOWxalanCAPOnoGUPCIMIN}{\rna}
\newcommand{\SLOWxalanCAPOnoGUPCIMAX}{\rna}
\newcommand{\SLOWxalanCAPOnoGDynamicUP}{\rna}
\newcommand{\SLOWxalanCAPOnoGDynamicUPCI}{\rna}
\newcommand{\SLOWxalanCAPOnoGDynamicUPCIMIN}{\rna}
\newcommand{\SLOWxalanCAPOnoGDynamicUPCIMAX}{\rna}
\newcommand{\SLOWxalanCAPOUP}{\rna}
\newcommand{\SLOWxalanCAPOUPCI}{\rna}
\newcommand{\SLOWxalanCAPOUPCIMIN}{\rna}
\newcommand{\SLOWxalanCAPOUPCIMAX}{\rna}
\newcommand{\SLOWxalanCAPODynamicUP}{\rna}
\newcommand{\SLOWxalanCAPODynamicUPCI}{\rna}
\newcommand{\SLOWxalanCAPODynamicUPCIMIN}{\rna}
\newcommand{\SLOWxalanCAPODynamicUPCIMAX}{\rna}
\newcommand{\SLOWxalanPIPUP}{\rna}
\newcommand{\SLOWxalanPIPUPCI}{\rna}
\newcommand{\SLOWxalanPIPUPCIMIN}{\rna}
\newcommand{\SLOWxalanPIPUPCIMAX}{\rna}
\newcommand{\SLOWxalanPIPDynamicUP}{\rna}
\newcommand{\SLOWxalanPIPDynamicUPCI}{\rna}
\newcommand{\SLOWxalanPIPDynamicUPCIMIN}{\rna}
\newcommand{\SLOWxalanPIPDynamicUPCIMAX}{\rna}
\newcommand{\SLOWxalanPIPHB}{\rna}
\newcommand{\SLOWxalanPIPHBCI}{\rna}
\newcommand{\SLOWxalanPIPHBCIMIN}{\rna}
\newcommand{\SLOWxalanPIPHBCIMAX}{\rna}
\newcommand{\SLOWxalanPIPHBDynamic}{\rna}
\newcommand{\SLOWxalanPIPHBDynamicCI}{\rna}
\newcommand{\SLOWxalanPIPHBDynamicCIMIN}{\rna}
\newcommand{\SLOWxalanPIPHBDynamicCIMAX}{\rna}
\newcommand{\SLOWxalanPIPWCP}{\rna}
\newcommand{\SLOWxalanPIPWCPCI}{\rna}
\newcommand{\SLOWxalanPIPWCPCIMIN}{\rna}
\newcommand{\SLOWxalanPIPWCPCIMAX}{\rna}
\newcommand{\SLOWxalanPIPWCPDynamic}{\rna}
\newcommand{\SLOWxalanPIPWCPDynamicCI}{\rna}
\newcommand{\SLOWxalanPIPWCPDynamicCIMIN}{\rna}
\newcommand{\SLOWxalanPIPWCPDynamicCIMAX}{\rna}
\newcommand{\SLOWxalanPIPWDC}{\rna}
\newcommand{\SLOWxalanPIPWDCCI}{\rna}
\newcommand{\SLOWxalanPIPWDCCIMIN}{\rna}
\newcommand{\SLOWxalanPIPWDCCIMAX}{\rna}
\newcommand{\SLOWxalanPIPWDCDynamic}{\rna}
\newcommand{\SLOWxalanPIPWDCDynamicCI}{\rna}
\newcommand{\SLOWxalanPIPWDCDynamicCIMIN}{\rna}
\newcommand{\SLOWxalanPIPWDCDynamicCIMAX}{\rna}
\newcommand{\SLOWxalanPIPCAPO}{\rna}
\newcommand{\SLOWxalanPIPCAPOCI}{\rna}
\newcommand{\SLOWxalanPIPCAPOCIMIN}{\rna}
\newcommand{\SLOWxalanPIPCAPOCIMAX}{\rna}
\newcommand{\SLOWxalanPIPCAPODynamic}{\rna}
\newcommand{\SLOWxalanPIPCAPODynamicCI}{\rna}
\newcommand{\SLOWxalanPIPCAPODynamicCIMIN}{\rna}
\newcommand{\SLOWxalanPIPCAPODynamicCIMAX}{\rna}
\newcommand{\SLOWxalanPIPPIP}{\rna}
\newcommand{\SLOWxalanPIPPIPCI}{\rna}
\newcommand{\SLOWxalanPIPPIPCIMIN}{\rna}
\newcommand{\SLOWxalanPIPPIPCIMAX}{\rna}
\newcommand{\SLOWxalanPIPPIPDynamic}{\rna}
\newcommand{\SLOWxalanPIPPIPDynamicCI}{\rna}
\newcommand{\SLOWxalanPIPPIPDynamicCIMIN}{\rna}
\newcommand{\SLOWxalanPIPPIPDynamicCIMAX}{\rna}
\newcommand{\SLOWBaseTimeGeoMean}{1800}
\newcommand{\SLOWEmptyTimeGeoMean}{\rna}
\newcommand{\SLOWFTTimeGeoMean}{8.8}
\newcommand{\SLOWHBTimeGeoMean}{19}
\newcommand{\SLOWWCPTimeGeoMean}{34}
\newcommand{\SLOWDCnoGExcTimeGeoMean}{28}
\newcommand{\SLOWDCnoGTimeGeoMean}{\rna}
\newcommand{\SLOWDCExcTimeGeoMean}{31}
\newcommand{\SLOWDCTimeGeoMean}{\rna}
\newcommand{\SLOWCAPOnoGExcTimeGeoMean}{27}
\newcommand{\SLOWCAPOnoGTimeGeoMean}{\rna}
\newcommand{\SLOWCAPOExcTimeGeoMean}{28}
\newcommand{\SLOWCAPOTimeGeoMean}{\rna}
\newcommand{\SLOWBaseMemGeoMean}{500}
\newcommand{\SLOWEmptyMemGeoMean}{0.0}
\newcommand{\SLOWFTMemGeoMean}{8.3}
\newcommand{\SLOWHBMemGeoMean}{25}
\newcommand{\SLOWWCPMemGeoMean}{47}
\newcommand{\SLOWDCnoGExcMemGeoMean}{32}
\newcommand{\SLOWDCnoGMemGeoMean}{\memna}
\newcommand{\SLOWDCExcMemGeoMean}{38}
\newcommand{\SLOWDCMemGeoMean}{\memna}
\newcommand{\SLOWCAPOnoGExcMemGeoMean}{31}
\newcommand{\SLOWCAPOnoGMemGeoMean}{\memna}
\newcommand{\SLOWCAPOExcMemGeoMean}{36}
\newcommand{\SLOWCAPOMemGeoMean}{\memna}
\newcommand{\SLOWWCPDynamicTotal}{7,456,945}
\newcommand{\SLOWHBDynamicUPTotal}{2,290,659}
\newcommand{\SLOWPIPWDCDynamicTotal}{0}
\newcommand{\SLOWCAPOTotal}{0}
\newcommand{\SLOWPIPWCPTotal}{0}
\newcommand{\SLOWWCPTotal}{727}
\newcommand{\SLOWWDCnoGDynamicUPTotal}{0}
\newcommand{\SLOWCAPOnoGExcDynamicTotal}{9,804,280}
\newcommand{\SLOWCAPOExcTotal}{105}
\newcommand{\SLOWDCExcTotal}{108}
\newcommand{\SLOWDCnoGExcTotal}{788}
\newcommand{\SLOWPIPDynamicUPTotal}{0}
\newcommand{\SLOWPIPPIPTotal}{0}
\newcommand{\SLOWFTTotal}{491}
\newcommand{\SLOWHBDynamicTotal}{2,301,098}
\newcommand{\SLOWCAPOnoGDynamicUPTotal}{0}
\newcommand{\SLOWDCnoGExcDynamicTotal}{9,541,538}
\newcommand{\SLOWCAPOnoGTotal}{0}
\newcommand{\SLOWPIPWDCTotal}{0}
\newcommand{\SLOWCAPOnoGExcTotal}{814}
\newcommand{\SLOWPIPHBDynamicTotal}{0}
\newcommand{\SLOWWDCUPTotal}{0}
\newcommand{\SLOWDCnoGTotal}{0}
\newcommand{\SLOWWCPDynamicUPTotal}{7,462,302}
\newcommand{\SLOWPIPUPTotal}{0}
\newcommand{\SLOWCAPODynamicTotal}{0}
\newcommand{\SLOWCAPOnoGUPTotal}{0}
\newcommand{\SLOWPIPCAPOTotal}{0}
\newcommand{\SLOWCAPOnoGDynamicTotal}{0}
\newcommand{\SLOWFTDynamicTotal}{4,430,977}
\newcommand{\SLOWPIPCAPODynamicTotal}{0}
\newcommand{\SLOWDCnoGDynamicTotal}{0}
\newcommand{\SLOWWDCnoGUPTotal}{0}
\newcommand{\SLOWDCExcDynamicTotal}{524,926}
\newcommand{\SLOWWDCDynamicUPTotal}{0}
\newcommand{\SLOWHBUPTotal}{909}
\newcommand{\SLOWCAPOExcDynamicTotal}{586,774}
\newcommand{\SLOWPIPTotal}{0}
\newcommand{\SLOWWCPUPTotal}{1022}
\newcommand{\SLOWCAPOUPTotal}{0}
\newcommand{\SLOWPIPHBTotal}{0}
\newcommand{\SLOWHBTotal}{699}
\newcommand{\SLOWDCDynamicTotal}{0}
\newcommand{\SLOWPIPDynamicTotal}{0}
\newcommand{\SLOWCAPODynamicUPTotal}{0}
\newcommand{\SLOWPIPPIPDynamicTotal}{0}
\newcommand{\SLOWPIPWCPDynamicTotal}{0}
\newcommand{\SLOWDCTotal}{0}

%% file: result-macros/PIP_fastTool_extraOpt2Quiet.tex
\newcommand{\FASTavroraEvents}{0}
\newcommand{\FASTavroraNoFPEvents}{0}
\newcommand{\FASTavroraMaxLiveThreads}{6}
\newcommand{\FASTavroraTotalThreads}{6}
\newcommand{\FASTavroraBaseTime}{2.5}
\newcommand{\FASTavroraBaseTimeCI}{17}
\newcommand{\FASTavroraEmptyTime}{\rna}
\newcommand{\FASTavroraEmptyTimeCI}{\rna}
\newcommand{\FASTavroraEmptyTimeCIMIN}{\rna}
\newcommand{\FASTavroraEmptyTimeCIMAX}{\rna}
\newcommand{\FASTavroraFTTime}{6.3}
\newcommand{\FASTavroraFTTimeCI}{0.11}
\newcommand{\FASTavroraHBTime}{4.0}
\newcommand{\FASTavroraHBTimeCI}{0.079}
\newcommand{\FASTavroraFTOHBTime}{3.9}
\newcommand{\FASTavroraFTOHBTimeCI}{0.085}
\newcommand{\FASTavroraWCPTime}{\rna}
\newcommand{\FASTavroraWCPTimeCI}{\rna}
\newcommand{\FASTavroraWCPTimeCIMIN}{\rna}
\newcommand{\FASTavroraWCPTimeCIMAX}{\rna}
\newcommand{\FASTavroraFTOWCPTime}{7.8}
\newcommand{\FASTavroraFTOWCPTimeCI}{0.14}
\newcommand{\FASTavroraREWCPTime}{5.9}
\newcommand{\FASTavroraREWCPTimeCI}{0.064}
\newcommand{\FASTavroraDCTime}{\rna}
\newcommand{\FASTavroraDCTimeCI}{\rna}
\newcommand{\FASTavroraDCTimeCIMIN}{\rna}
\newcommand{\FASTavroraDCTimeCIMAX}{\rna}
\newcommand{\FASTavroraFTODCTime}{8.0}
\newcommand{\FASTavroraFTODCTimeCI}{0.12}
\newcommand{\FASTavroraREDCTime}{6.4}
\newcommand{\FASTavroraREDCTimeCI}{0.11}
\newcommand{\FASTavroraCAPOTime}{\rna}
\newcommand{\FASTavroraCAPOTimeCI}{\rna}
\newcommand{\FASTavroraCAPOTimeCIMIN}{\rna}
\newcommand{\FASTavroraCAPOTimeCIMAX}{\rna}
\newcommand{\FASTavroraFTOCAPOTime}{6.1}
\newcommand{\FASTavroraFTOCAPOTimeCI}{0.13}
\newcommand{\FASTavroraRECAPOTime}{4.5}
\newcommand{\FASTavroraRECAPOTimeCI}{0.072}
\newcommand{\FASTavroraAGGCAPOTime}{\rna}
\newcommand{\FASTavroraAGGCAPOTimeCI}{\rna}
\newcommand{\FASTavroraAGGCAPOTimeCIMIN}{\rna}
\newcommand{\FASTavroraAGGCAPOTimeCIMAX}{\rna}
\newcommand{\FASTavroraStaticTime}{\rzero}
\newcommand{\FASTavroraDynamicTime}{\rzero}
\newcommand{\FASTavroraBaseMem}{180}
\newcommand{\FASTavroraBaseMemCI}{5.6}
\newcommand{\FASTavroraFTMem}{21}
\newcommand{\FASTavroraFTMemCI}{3.5}
\newcommand{\FASTavroraHBMem}{4.1}
\newcommand{\FASTavroraHBMemCI}{0.15}
\newcommand{\FASTavroraFTOHBMem}{4.1}
\newcommand{\FASTavroraFTOHBMemCI}{0.15}
\newcommand{\FASTavroraWCPMem}{\memna}
\newcommand{\FASTavroraWCPMemCI}{\memna}
\newcommand{\FASTavroraWCPMemCIMIN}{\memna}
\newcommand{\FASTavroraWCPMemCIMAX}{\memna}
\newcommand{\FASTavroraFTOWCPMem}{11}
\newcommand{\FASTavroraFTOWCPMemCI}{0.36}
\newcommand{\FASTavroraREWCPMem}{7.2}
\newcommand{\FASTavroraREWCPMemCI}{0.26}
\newcommand{\FASTavroraDCMem}{\memna}
\newcommand{\FASTavroraDCMemCI}{\memna}
\newcommand{\FASTavroraDCMemCIMIN}{\memna}
\newcommand{\FASTavroraDCMemCIMAX}{\memna}
\newcommand{\FASTavroraFTODCMem}{11}
\newcommand{\FASTavroraFTODCMemCI}{0.34}
\newcommand{\FASTavroraREDCMem}{7.2}
\newcommand{\FASTavroraREDCMemCI}{0.27}
\newcommand{\FASTavroraCAPOMem}{\memna}
\newcommand{\FASTavroraCAPOMemCI}{\memna}
\newcommand{\FASTavroraCAPOMemCIMIN}{\memna}
\newcommand{\FASTavroraCAPOMemCIMAX}{\memna}
\newcommand{\FASTavroraFTOCAPOMem}{8.3}
\newcommand{\FASTavroraFTOCAPOMemCI}{0.27}
\newcommand{\FASTavroraRECAPOMem}{4.7}
\newcommand{\FASTavroraRECAPOMemCI}{0.16}
\newcommand{\FASTavroraAGGCAPOMem}{\memna}
\newcommand{\FASTavroraAGGCAPOMemCI}{\memna}
\newcommand{\FASTavroraAGGCAPOMemCIMIN}{\memna}
\newcommand{\FASTavroraAGGCAPOMemCIMAX}{\memna}
\newcommand{\FASTavroraEventsCI}{0}
\newcommand{\FASTavroraEventsCIMIN}{0}
\newcommand{\FASTavroraEventsCIMAX}{0}
\newcommand{\FASTavroraNoFPEventsCI}{0}
\newcommand{\FASTavroraNoFPEventsCIMIN}{0}
\newcommand{\FASTavroraNoFPEventsCIMAX}{0}
\newcommand{\FASTavroraFT}{3}
\newcommand{\FASTavroraFTCI}{0.0}
\newcommand{\FASTavroraFTCIMIN}{3}
\newcommand{\FASTavroraFTCIMAX}{3}
\newcommand{\FASTavroraFTDynamic}{755,061}
\newcommand{\FASTavroraFTDynamicCI}{1,927}
\newcommand{\FASTavroraFTDynamicCIMIN}{753,134}
\newcommand{\FASTavroraFTDynamicCIMAX}{756,988}
\newcommand{\FASTavroraHB}{6}
\newcommand{\FASTavroraHBCI}{0.0}
\newcommand{\FASTavroraHBCIMIN}{6}
\newcommand{\FASTavroraHBCIMAX}{6}
\newcommand{\FASTavroraHBDynamic}{426,824}
\newcommand{\FASTavroraHBDynamicCI}{445}
\newcommand{\FASTavroraHBDynamicCIMIN}{426,379}
\newcommand{\FASTavroraHBDynamicCIMAX}{427,269}
\newcommand{\FASTavroraFTOHB}{6}
\newcommand{\FASTavroraFTOHBCI}{0.20}
\newcommand{\FASTavroraFTOHBCIMIN}{6}
\newcommand{\FASTavroraFTOHBCIMAX}{6}
\newcommand{\FASTavroraFTOHBDynamic}{405,573}
\newcommand{\FASTavroraFTOHBDynamicCI}{285}
\newcommand{\FASTavroraFTOHBDynamicCIMIN}{405,288}
\newcommand{\FASTavroraFTOHBDynamicCIMAX}{405,858}
\newcommand{\FASTavroraWCP}{\rna}
\newcommand{\FASTavroraWCPCI}{\rna}
\newcommand{\FASTavroraWCPCIMIN}{\rna}
\newcommand{\FASTavroraWCPCIMAX}{\rna}
\newcommand{\FASTavroraWCPDynamic}{\rna}
\newcommand{\FASTavroraWCPDynamicCI}{\rna}
\newcommand{\FASTavroraWCPDynamicCIMIN}{\rna}
\newcommand{\FASTavroraWCPDynamicCIMAX}{\rna}
\newcommand{\FASTavroraFTOWCP}{6}
\newcommand{\FASTavroraFTOWCPCI}{0.0}
\newcommand{\FASTavroraFTOWCPCIMIN}{6}
\newcommand{\FASTavroraFTOWCPCIMAX}{6}
\newcommand{\FASTavroraFTOWCPDynamic}{404,954}
\newcommand{\FASTavroraFTOWCPDynamicCI}{208}
\newcommand{\FASTavroraFTOWCPDynamicCIMIN}{404,746}
\newcommand{\FASTavroraFTOWCPDynamicCIMAX}{405,162}
\newcommand{\FASTavroraREWCP}{6}
\newcommand{\FASTavroraREWCPCI}{0.20}
\newcommand{\FASTavroraREWCPCIMIN}{6}
\newcommand{\FASTavroraREWCPCIMAX}{6}
\newcommand{\FASTavroraREWCPDynamic}{406,346}
\newcommand{\FASTavroraREWCPDynamicCI}{399}
\newcommand{\FASTavroraREWCPDynamicCIMIN}{405,947}
\newcommand{\FASTavroraREWCPDynamicCIMAX}{406,745}
\newcommand{\FASTavroraDC}{\rna}
\newcommand{\FASTavroraDCCI}{\rna}
\newcommand{\FASTavroraDCCIMIN}{\rna}
\newcommand{\FASTavroraDCCIMAX}{\rna}
\newcommand{\FASTavroraDCDynamic}{\rna}
\newcommand{\FASTavroraDCDynamicCI}{\rna}
\newcommand{\FASTavroraDCDynamicCIMIN}{\rna}
\newcommand{\FASTavroraDCDynamicCIMAX}{\rna}
\newcommand{\FASTavroraFTODC}{6}
\newcommand{\FASTavroraFTODCCI}{0.0}
\newcommand{\FASTavroraFTODCCIMIN}{6}
\newcommand{\FASTavroraFTODCCIMAX}{6}
\newcommand{\FASTavroraFTODCDynamic}{404,498}
\newcommand{\FASTavroraFTODCDynamicCI}{220}
\newcommand{\FASTavroraFTODCDynamicCIMIN}{404,278}
\newcommand{\FASTavroraFTODCDynamicCIMAX}{404,718}
\newcommand{\FASTavroraREDC}{6}
\newcommand{\FASTavroraREDCCI}{0.0}
\newcommand{\FASTavroraREDCCIMIN}{6}
\newcommand{\FASTavroraREDCCIMAX}{6}
\newcommand{\FASTavroraREDCDynamic}{406,902}
\newcommand{\FASTavroraREDCDynamicCI}{228}
\newcommand{\FASTavroraREDCDynamicCIMIN}{406,674}
\newcommand{\FASTavroraREDCDynamicCIMAX}{407,130}
\newcommand{\FASTavroraCAPO}{\rna}
\newcommand{\FASTavroraCAPOCI}{\rna}
\newcommand{\FASTavroraCAPOCIMIN}{\rna}
\newcommand{\FASTavroraCAPOCIMAX}{\rna}
\newcommand{\FASTavroraCAPODynamic}{\rna}
\newcommand{\FASTavroraCAPODynamicCI}{\rna}
\newcommand{\FASTavroraCAPODynamicCIMIN}{\rna}
\newcommand{\FASTavroraCAPODynamicCIMAX}{\rna}
\newcommand{\FASTavroraFTOCAPO}{6}
\newcommand{\FASTavroraFTOCAPOCI}{0.20}
\newcommand{\FASTavroraFTOCAPOCIMIN}{6}
\newcommand{\FASTavroraFTOCAPOCIMAX}{6}
\newcommand{\FASTavroraFTOCAPODynamic}{406,375}
\newcommand{\FASTavroraFTOCAPODynamicCI}{162}
\newcommand{\FASTavroraFTOCAPODynamicCIMIN}{406,213}
\newcommand{\FASTavroraFTOCAPODynamicCIMAX}{406,537}
\newcommand{\FASTavroraRECAPO}{6}
\newcommand{\FASTavroraRECAPOCI}{0.0}
\newcommand{\FASTavroraRECAPOCIMIN}{6}
\newcommand{\FASTavroraRECAPOCIMAX}{6}
\newcommand{\FASTavroraRECAPODynamic}{408,624}
\newcommand{\FASTavroraRECAPODynamicCI}{204}
\newcommand{\FASTavroraRECAPODynamicCIMIN}{408,420}
\newcommand{\FASTavroraRECAPODynamicCIMAX}{408,828}
\newcommand{\FASTavroraAGGCAPO}{\rna}
\newcommand{\FASTavroraAGGCAPOCI}{\rna}
\newcommand{\FASTavroraAGGCAPOCIMIN}{\rna}
\newcommand{\FASTavroraAGGCAPOCIMAX}{\rna}
\newcommand{\FASTavroraAGGCAPODynamic}{\rna}
\newcommand{\FASTavroraAGGCAPODynamicCI}{\rna}
\newcommand{\FASTavroraAGGCAPODynamicCIMIN}{\rna}
\newcommand{\FASTavroraAGGCAPODynamicCIMAX}{\rna}
\newcommand{\FASTbatikEvents}{0}
\newcommand{\FASTbatikNoFPEvents}{0}
\newcommand{\FASTbatikMaxLiveThreads}{2}
\newcommand{\FASTbatikTotalThreads}{6}
\newcommand{\FASTbatikBaseTime}{2.6}
\newcommand{\FASTbatikBaseTimeCI}{34}
\newcommand{\FASTbatikEmptyTime}{\rna}
\newcommand{\FASTbatikEmptyTimeCI}{\rna}
\newcommand{\FASTbatikEmptyTimeCIMIN}{\rna}
\newcommand{\FASTbatikEmptyTimeCIMAX}{\rna}
\newcommand{\FASTbatikFTTime}{4.1}
\newcommand{\FASTbatikFTTimeCI}{0.078}
\newcommand{\FASTbatikHBTime}{4.1}
\newcommand{\FASTbatikHBTimeCI}{0.065}
\newcommand{\FASTbatikFTOHBTime}{4.0}
\newcommand{\FASTbatikFTOHBTimeCI}{0.077}
\newcommand{\FASTbatikWCPTime}{\rna}
\newcommand{\FASTbatikWCPTimeCI}{\rna}
\newcommand{\FASTbatikWCPTimeCIMIN}{\rna}
\newcommand{\FASTbatikWCPTimeCIMAX}{\rna}
\newcommand{\FASTbatikFTOWCPTime}{6.9}
\newcommand{\FASTbatikFTOWCPTimeCI}{0.099}
\newcommand{\FASTbatikREWCPTime}{4.3}
\newcommand{\FASTbatikREWCPTimeCI}{0.099}
\newcommand{\FASTbatikDCTime}{\rna}
\newcommand{\FASTbatikDCTimeCI}{\rna}
\newcommand{\FASTbatikDCTimeCIMIN}{\rna}
\newcommand{\FASTbatikDCTimeCIMAX}{\rna}
\newcommand{\FASTbatikFTODCTime}{6.9}
\newcommand{\FASTbatikFTODCTimeCI}{0.11}
\newcommand{\FASTbatikREDCTime}{4.2}
\newcommand{\FASTbatikREDCTimeCI}{0.076}
\newcommand{\FASTbatikCAPOTime}{\rna}
\newcommand{\FASTbatikCAPOTimeCI}{\rna}
\newcommand{\FASTbatikCAPOTimeCIMIN}{\rna}
\newcommand{\FASTbatikCAPOTimeCIMAX}{\rna}
\newcommand{\FASTbatikFTOCAPOTime}{6.8}
\newcommand{\FASTbatikFTOCAPOTimeCI}{0.13}
\newcommand{\FASTbatikRECAPOTime}{4.1}
\newcommand{\FASTbatikRECAPOTimeCI}{0.070}
\newcommand{\FASTbatikAGGCAPOTime}{\rna}
\newcommand{\FASTbatikAGGCAPOTimeCI}{\rna}
\newcommand{\FASTbatikAGGCAPOTimeCIMIN}{\rna}
\newcommand{\FASTbatikAGGCAPOTimeCIMAX}{\rna}
\newcommand{\FASTbatikStaticTime}{\rzero}
\newcommand{\FASTbatikDynamicTime}{\rzero}
\newcommand{\FASTbatikBaseMem}{250}
\newcommand{\FASTbatikBaseMemCI}{1.6}
\newcommand{\FASTbatikFTMem}{4.9}
\newcommand{\FASTbatikFTMemCI}{0.033}
\newcommand{\FASTbatikHBMem}{4.9}
\newcommand{\FASTbatikHBMemCI}{0.048}
\newcommand{\FASTbatikFTOHBMem}{4.9}
\newcommand{\FASTbatikFTOHBMemCI}{0.044}
\newcommand{\FASTbatikWCPMem}{\memna}
\newcommand{\FASTbatikWCPMemCI}{\memna}
\newcommand{\FASTbatikWCPMemCIMIN}{\memna}
\newcommand{\FASTbatikWCPMemCIMAX}{\memna}
\newcommand{\FASTbatikFTOWCPMem}{13}
\newcommand{\FASTbatikFTOWCPMemCI}{0.083}
\newcommand{\FASTbatikREWCPMem}{6.1}
\newcommand{\FASTbatikREWCPMemCI}{0.083}
\newcommand{\FASTbatikDCMem}{\memna}
\newcommand{\FASTbatikDCMemCI}{\memna}
\newcommand{\FASTbatikDCMemCIMIN}{\memna}
\newcommand{\FASTbatikDCMemCIMAX}{\memna}
\newcommand{\FASTbatikFTODCMem}{13}
\newcommand{\FASTbatikFTODCMemCI}{0.10}
\newcommand{\FASTbatikREDCMem}{5.7}
\newcommand{\FASTbatikREDCMemCI}{0.10}
\newcommand{\FASTbatikCAPOMem}{\memna}
\newcommand{\FASTbatikCAPOMemCI}{\memna}
\newcommand{\FASTbatikCAPOMemCIMIN}{\memna}
\newcommand{\FASTbatikCAPOMemCIMAX}{\memna}
\newcommand{\FASTbatikFTOCAPOMem}{13}
\newcommand{\FASTbatikFTOCAPOMemCI}{0.31}
\newcommand{\FASTbatikRECAPOMem}{5.7}
\newcommand{\FASTbatikRECAPOMemCI}{0.045}
\newcommand{\FASTbatikAGGCAPOMem}{\memna}
\newcommand{\FASTbatikAGGCAPOMemCI}{\memna}
\newcommand{\FASTbatikAGGCAPOMemCIMIN}{\memna}
\newcommand{\FASTbatikAGGCAPOMemCIMAX}{\memna}
\newcommand{\FASTbatikEventsCI}{0}
\newcommand{\FASTbatikEventsCIMIN}{0}
\newcommand{\FASTbatikEventsCIMAX}{0}
\newcommand{\FASTbatikNoFPEventsCI}{0}
\newcommand{\FASTbatikNoFPEventsCIMIN}{0}
\newcommand{\FASTbatikNoFPEventsCIMAX}{0}
\newcommand{\FASTbatikFT}{0}
\newcommand{\FASTbatikFTCI}{0.0}
\newcommand{\FASTbatikFTCIMIN}{0}
\newcommand{\FASTbatikFTCIMAX}{0}
\newcommand{\FASTbatikFTDynamic}{0}
\newcommand{\FASTbatikFTDynamicCI}{0.0}
\newcommand{\FASTbatikFTDynamicCIMIN}{0}
\newcommand{\FASTbatikFTDynamicCIMAX}{0}
\newcommand{\FASTbatikHB}{0}
\newcommand{\FASTbatikHBCI}{0.0}
\newcommand{\FASTbatikHBCIMIN}{0}
\newcommand{\FASTbatikHBCIMAX}{0}
\newcommand{\FASTbatikHBDynamic}{0}
\newcommand{\FASTbatikHBDynamicCI}{0.0}
\newcommand{\FASTbatikHBDynamicCIMIN}{0}
\newcommand{\FASTbatikHBDynamicCIMAX}{0}
\newcommand{\FASTbatikFTOHB}{0}
\newcommand{\FASTbatikFTOHBCI}{0.0}
\newcommand{\FASTbatikFTOHBCIMIN}{0}
\newcommand{\FASTbatikFTOHBCIMAX}{0}
\newcommand{\FASTbatikFTOHBDynamic}{0}
\newcommand{\FASTbatikFTOHBDynamicCI}{0.0}
\newcommand{\FASTbatikFTOHBDynamicCIMIN}{0}
\newcommand{\FASTbatikFTOHBDynamicCIMAX}{0}
\newcommand{\FASTbatikWCP}{\rna}
\newcommand{\FASTbatikWCPCI}{\rna}
\newcommand{\FASTbatikWCPCIMIN}{\rna}
\newcommand{\FASTbatikWCPCIMAX}{\rna}
\newcommand{\FASTbatikWCPDynamic}{\rna}
\newcommand{\FASTbatikWCPDynamicCI}{\rna}
\newcommand{\FASTbatikWCPDynamicCIMIN}{\rna}
\newcommand{\FASTbatikWCPDynamicCIMAX}{\rna}
\newcommand{\FASTbatikFTOWCP}{0}
\newcommand{\FASTbatikFTOWCPCI}{0.0}
\newcommand{\FASTbatikFTOWCPCIMIN}{0}
\newcommand{\FASTbatikFTOWCPCIMAX}{0}
\newcommand{\FASTbatikFTOWCPDynamic}{0}
\newcommand{\FASTbatikFTOWCPDynamicCI}{0.0}
\newcommand{\FASTbatikFTOWCPDynamicCIMIN}{0}
\newcommand{\FASTbatikFTOWCPDynamicCIMAX}{0}
\newcommand{\FASTbatikREWCP}{0}
\newcommand{\FASTbatikREWCPCI}{0.0}
\newcommand{\FASTbatikREWCPCIMIN}{0}
\newcommand{\FASTbatikREWCPCIMAX}{0}
\newcommand{\FASTbatikREWCPDynamic}{0}
\newcommand{\FASTbatikREWCPDynamicCI}{0.0}
\newcommand{\FASTbatikREWCPDynamicCIMIN}{0}
\newcommand{\FASTbatikREWCPDynamicCIMAX}{0}
\newcommand{\FASTbatikDC}{\rna}
\newcommand{\FASTbatikDCCI}{\rna}
\newcommand{\FASTbatikDCCIMIN}{\rna}
\newcommand{\FASTbatikDCCIMAX}{\rna}
\newcommand{\FASTbatikDCDynamic}{\rna}
\newcommand{\FASTbatikDCDynamicCI}{\rna}
\newcommand{\FASTbatikDCDynamicCIMIN}{\rna}
\newcommand{\FASTbatikDCDynamicCIMAX}{\rna}
\newcommand{\FASTbatikFTODC}{0}
\newcommand{\FASTbatikFTODCCI}{0.0}
\newcommand{\FASTbatikFTODCCIMIN}{0}
\newcommand{\FASTbatikFTODCCIMAX}{0}
\newcommand{\FASTbatikFTODCDynamic}{0}
\newcommand{\FASTbatikFTODCDynamicCI}{0.0}
\newcommand{\FASTbatikFTODCDynamicCIMIN}{0}
\newcommand{\FASTbatikFTODCDynamicCIMAX}{0}
\newcommand{\FASTbatikREDC}{0}
\newcommand{\FASTbatikREDCCI}{0.0}
\newcommand{\FASTbatikREDCCIMIN}{0}
\newcommand{\FASTbatikREDCCIMAX}{0}
\newcommand{\FASTbatikREDCDynamic}{0}
\newcommand{\FASTbatikREDCDynamicCI}{0.0}
\newcommand{\FASTbatikREDCDynamicCIMIN}{0}
\newcommand{\FASTbatikREDCDynamicCIMAX}{0}
\newcommand{\FASTbatikCAPO}{\rna}
\newcommand{\FASTbatikCAPOCI}{\rna}
\newcommand{\FASTbatikCAPOCIMIN}{\rna}
\newcommand{\FASTbatikCAPOCIMAX}{\rna}
\newcommand{\FASTbatikCAPODynamic}{\rna}
\newcommand{\FASTbatikCAPODynamicCI}{\rna}
\newcommand{\FASTbatikCAPODynamicCIMIN}{\rna}
\newcommand{\FASTbatikCAPODynamicCIMAX}{\rna}
\newcommand{\FASTbatikFTOCAPO}{0}
\newcommand{\FASTbatikFTOCAPOCI}{0.0}
\newcommand{\FASTbatikFTOCAPOCIMIN}{0}
\newcommand{\FASTbatikFTOCAPOCIMAX}{0}
\newcommand{\FASTbatikFTOCAPODynamic}{0}
\newcommand{\FASTbatikFTOCAPODynamicCI}{0.0}
\newcommand{\FASTbatikFTOCAPODynamicCIMIN}{0}
\newcommand{\FASTbatikFTOCAPODynamicCIMAX}{0}
\newcommand{\FASTbatikRECAPO}{0}
\newcommand{\FASTbatikRECAPOCI}{0.0}
\newcommand{\FASTbatikRECAPOCIMIN}{0}
\newcommand{\FASTbatikRECAPOCIMAX}{0}
\newcommand{\FASTbatikRECAPODynamic}{0}
\newcommand{\FASTbatikRECAPODynamicCI}{0.0}
\newcommand{\FASTbatikRECAPODynamicCIMIN}{0}
\newcommand{\FASTbatikRECAPODynamicCIMAX}{0}
\newcommand{\FASTbatikAGGCAPO}{\rna}
\newcommand{\FASTbatikAGGCAPOCI}{\rna}
\newcommand{\FASTbatikAGGCAPOCIMIN}{\rna}
\newcommand{\FASTbatikAGGCAPOCIMAX}{\rna}
\newcommand{\FASTbatikAGGCAPODynamic}{\rna}
\newcommand{\FASTbatikAGGCAPODynamicCI}{\rna}
\newcommand{\FASTbatikAGGCAPODynamicCIMIN}{\rna}
\newcommand{\FASTbatikAGGCAPODynamicCIMAX}{\rna}
\newcommand{\FASThtwoEvents}{0}
\newcommand{\FASThtwoNoFPEvents}{0}
\newcommand{\FASThtwoMaxLiveThreads}{14}
\newcommand{\FASThtwoTotalThreads}{15}
\newcommand{\FASThtwoBaseTime}{4.8}
\newcommand{\FASThtwoBaseTimeCI}{44}
\newcommand{\FASThtwoEmptyTime}{\rna}
\newcommand{\FASThtwoEmptyTimeCI}{\rna}
\newcommand{\FASThtwoEmptyTimeCIMIN}{\rna}
\newcommand{\FASThtwoEmptyTimeCIMAX}{\rna}
\newcommand{\FASThtwoFTTime}{9.3}
\newcommand{\FASThtwoFTTimeCI}{0.53}
\newcommand{\FASThtwoHBTime}{8.3}
\newcommand{\FASThtwoHBTimeCI}{0.11}
\newcommand{\FASThtwoFTOHBTime}{8.3}
\newcommand{\FASThtwoFTOHBTimeCI}{0.13}
\newcommand{\FASThtwoWCPTime}{\rna}
\newcommand{\FASThtwoWCPTimeCI}{\rna}
\newcommand{\FASThtwoWCPTimeCIMIN}{\rna}
\newcommand{\FASThtwoWCPTimeCIMAX}{\rna}
\newcommand{\FASThtwoFTOWCPTime}{55}
\newcommand{\FASThtwoFTOWCPTimeCI}{4.0}
\newcommand{\FASThtwoREWCPTime}{11}
\newcommand{\FASThtwoREWCPTimeCI}{0.36}
\newcommand{\FASThtwoDCTime}{\rna}
\newcommand{\FASThtwoDCTimeCI}{\rna}
\newcommand{\FASThtwoDCTimeCIMIN}{\rna}
\newcommand{\FASThtwoDCTimeCIMAX}{\rna}
\newcommand{\FASThtwoFTODCTime}{55}
\newcommand{\FASThtwoFTODCTimeCI}{6.2}
\newcommand{\FASThtwoREDCTime}{10}
\newcommand{\FASThtwoREDCTimeCI}{0.34}
\newcommand{\FASThtwoCAPOTime}{\rna}
\newcommand{\FASThtwoCAPOTimeCI}{\rna}
\newcommand{\FASThtwoCAPOTimeCIMIN}{\rna}
\newcommand{\FASThtwoCAPOTimeCIMAX}{\rna}
\newcommand{\FASThtwoFTOCAPOTime}{52}
\newcommand{\FASThtwoFTOCAPOTimeCI}{5.8}
\newcommand{\FASThtwoRECAPOTime}{9.5}
\newcommand{\FASThtwoRECAPOTimeCI}{0.39}
\newcommand{\FASThtwoAGGCAPOTime}{\rna}
\newcommand{\FASThtwoAGGCAPOTimeCI}{\rna}
\newcommand{\FASThtwoAGGCAPOTimeCIMIN}{\rna}
\newcommand{\FASThtwoAGGCAPOTimeCIMAX}{\rna}
\newcommand{\FASThtwoStaticTime}{\rzero}
\newcommand{\FASThtwoDynamicTime}{\rzero}
\newcommand{\FASThtwoBaseMem}{1,800}
\newcommand{\FASThtwoBaseMemCI}{31.0}
\newcommand{\FASThtwoFTMem}{3.0}
\newcommand{\FASThtwoFTMemCI}{0.060}
\newcommand{\FASThtwoHBMem}{3.1}
\newcommand{\FASThtwoHBMemCI}{0.078}
\newcommand{\FASThtwoFTOHBMem}{3.1}
\newcommand{\FASThtwoFTOHBMemCI}{0.062}
\newcommand{\FASThtwoWCPMem}{\memna}
\newcommand{\FASThtwoWCPMemCI}{\memna}
\newcommand{\FASThtwoWCPMemCIMIN}{\memna}
\newcommand{\FASThtwoWCPMemCIMAX}{\memna}
\newcommand{\FASThtwoFTOWCPMem}{40}
\newcommand{\FASThtwoFTOWCPMemCI}{2.0}
\newcommand{\FASThtwoREWCPMem}{4.9}
\newcommand{\FASThtwoREWCPMemCI}{0.10}
\newcommand{\FASThtwoDCMem}{\memna}
\newcommand{\FASThtwoDCMemCI}{\memna}
\newcommand{\FASThtwoDCMemCIMIN}{\memna}
\newcommand{\FASThtwoDCMemCIMAX}{\memna}
\newcommand{\FASThtwoFTODCMem}{39}
\newcommand{\FASThtwoFTODCMemCI}{3.1}
\newcommand{\FASThtwoREDCMem}{4.9}
\newcommand{\FASThtwoREDCMemCI}{0.19}
\newcommand{\FASThtwoCAPOMem}{\memna}
\newcommand{\FASThtwoCAPOMemCI}{\memna}
\newcommand{\FASThtwoCAPOMemCIMIN}{\memna}
\newcommand{\FASThtwoCAPOMemCIMAX}{\memna}
\newcommand{\FASThtwoFTOCAPOMem}{40}
\newcommand{\FASThtwoFTOCAPOMemCI}{1.9}
\newcommand{\FASThtwoRECAPOMem}{4.4}
\newcommand{\FASThtwoRECAPOMemCI}{0.16}
\newcommand{\FASThtwoAGGCAPOMem}{\memna}
\newcommand{\FASThtwoAGGCAPOMemCI}{\memna}
\newcommand{\FASThtwoAGGCAPOMemCIMIN}{\memna}
\newcommand{\FASThtwoAGGCAPOMemCIMAX}{\memna}
\newcommand{\FASThtwoEventsCI}{0}
\newcommand{\FASThtwoEventsCIMIN}{0}
\newcommand{\FASThtwoEventsCIMAX}{0}
\newcommand{\FASThtwoNoFPEventsCI}{0}
\newcommand{\FASThtwoNoFPEventsCIMIN}{0}
\newcommand{\FASThtwoNoFPEventsCIMAX}{0}
\newcommand{\FASThtwoFT}{11}
\newcommand{\FASThtwoFTCI}{0.0}
\newcommand{\FASThtwoFTCIMIN}{11}
\newcommand{\FASThtwoFTCIMAX}{11}
\newcommand{\FASThtwoFTDynamic}{106,139}
\newcommand{\FASThtwoFTDynamicCI}{1,366}
\newcommand{\FASThtwoFTDynamicCIMIN}{104,773}
\newcommand{\FASThtwoFTDynamicCIMAX}{107,505}
\newcommand{\FASThtwoHB}{13}
\newcommand{\FASThtwoHBCI}{0.0}
\newcommand{\FASThtwoHBCIMIN}{13}
\newcommand{\FASThtwoHBCIMAX}{13}
\newcommand{\FASThtwoHBDynamic}{69,171}
\newcommand{\FASThtwoHBDynamicCI}{1,703}
\newcommand{\FASThtwoHBDynamicCIMIN}{67,468}
\newcommand{\FASThtwoHBDynamicCIMAX}{70,874}
\newcommand{\FASThtwoFTOHB}{13}
\newcommand{\FASThtwoFTOHBCI}{0.0}
\newcommand{\FASThtwoFTOHBCIMIN}{13}
\newcommand{\FASThtwoFTOHBCIMAX}{13}
\newcommand{\FASThtwoFTOHBDynamic}{44,928}
\newcommand{\FASThtwoFTOHBDynamicCI}{2,971}
\newcommand{\FASThtwoFTOHBDynamicCIMIN}{41,957}
\newcommand{\FASThtwoFTOHBDynamicCIMAX}{47,899}
\newcommand{\FASThtwoWCP}{\rna}
\newcommand{\FASThtwoWCPCI}{\rna}
\newcommand{\FASThtwoWCPCIMIN}{\rna}
\newcommand{\FASThtwoWCPCIMAX}{\rna}
\newcommand{\FASThtwoWCPDynamic}{\rna}
\newcommand{\FASThtwoWCPDynamicCI}{\rna}
\newcommand{\FASThtwoWCPDynamicCIMIN}{\rna}
\newcommand{\FASThtwoWCPDynamicCIMAX}{\rna}
\newcommand{\FASThtwoFTOWCP}{13}
\newcommand{\FASThtwoFTOWCPCI}{0.0}
\newcommand{\FASThtwoFTOWCPCIMIN}{13}
\newcommand{\FASThtwoFTOWCPCIMAX}{13}
\newcommand{\FASThtwoFTOWCPDynamic}{85,936}
\newcommand{\FASThtwoFTOWCPDynamicCI}{155}
\newcommand{\FASThtwoFTOWCPDynamicCIMIN}{85,781}
\newcommand{\FASThtwoFTOWCPDynamicCIMAX}{86,091}
\newcommand{\FASThtwoREWCP}{13}
\newcommand{\FASThtwoREWCPCI}{0.26}
\newcommand{\FASThtwoREWCPCIMIN}{13}
\newcommand{\FASThtwoREWCPCIMAX}{13}
\newcommand{\FASThtwoREWCPDynamic}{66,303}
\newcommand{\FASThtwoREWCPDynamicCI}{4,880}
\newcommand{\FASThtwoREWCPDynamicCIMIN}{61,423}
\newcommand{\FASThtwoREWCPDynamicCIMAX}{71,183}
\newcommand{\FASThtwoDC}{\rna}
\newcommand{\FASThtwoDCCI}{\rna}
\newcommand{\FASThtwoDCCIMIN}{\rna}
\newcommand{\FASThtwoDCCIMAX}{\rna}
\newcommand{\FASThtwoDCDynamic}{\rna}
\newcommand{\FASThtwoDCDynamicCI}{\rna}
\newcommand{\FASThtwoDCDynamicCIMIN}{\rna}
\newcommand{\FASThtwoDCDynamicCIMAX}{\rna}
\newcommand{\FASThtwoFTODC}{13}
\newcommand{\FASThtwoFTODCCI}{0.0}
\newcommand{\FASThtwoFTODCCIMIN}{13}
\newcommand{\FASThtwoFTODCCIMAX}{13}
\newcommand{\FASThtwoFTODCDynamic}{86,383}
\newcommand{\FASThtwoFTODCDynamicCI}{168}
\newcommand{\FASThtwoFTODCDynamicCIMIN}{86,215}
\newcommand{\FASThtwoFTODCDynamicCIMAX}{86,551}
\newcommand{\FASThtwoREDC}{13}
\newcommand{\FASThtwoREDCCI}{0.0}
\newcommand{\FASThtwoREDCCIMIN}{13}
\newcommand{\FASThtwoREDCCIMAX}{13}
\newcommand{\FASThtwoREDCDynamic}{70,557}
\newcommand{\FASThtwoREDCDynamicCI}{5,686}
\newcommand{\FASThtwoREDCDynamicCIMIN}{64,871}
\newcommand{\FASThtwoREDCDynamicCIMAX}{76,243}
\newcommand{\FASThtwoCAPO}{\rna}
\newcommand{\FASThtwoCAPOCI}{\rna}
\newcommand{\FASThtwoCAPOCIMIN}{\rna}
\newcommand{\FASThtwoCAPOCIMAX}{\rna}
\newcommand{\FASThtwoCAPODynamic}{\rna}
\newcommand{\FASThtwoCAPODynamicCI}{\rna}
\newcommand{\FASThtwoCAPODynamicCIMIN}{\rna}
\newcommand{\FASThtwoCAPODynamicCIMAX}{\rna}
\newcommand{\FASThtwoFTOCAPO}{13}
\newcommand{\FASThtwoFTOCAPOCI}{0.0}
\newcommand{\FASThtwoFTOCAPOCIMIN}{13}
\newcommand{\FASThtwoFTOCAPOCIMAX}{13}
\newcommand{\FASThtwoFTOCAPODynamic}{86,657}
\newcommand{\FASThtwoFTOCAPODynamicCI}{126}
\newcommand{\FASThtwoFTOCAPODynamicCIMIN}{86,531}
\newcommand{\FASThtwoFTOCAPODynamicCIMAX}{86,783}
\newcommand{\FASThtwoRECAPO}{13}
\newcommand{\FASThtwoRECAPOCI}{0.0}
\newcommand{\FASThtwoRECAPOCIMIN}{13}
\newcommand{\FASThtwoRECAPOCIMAX}{13}
\newcommand{\FASThtwoRECAPODynamic}{50,618}
\newcommand{\FASThtwoRECAPODynamicCI}{4,178}
\newcommand{\FASThtwoRECAPODynamicCIMIN}{46,440}
\newcommand{\FASThtwoRECAPODynamicCIMAX}{54,796}
\newcommand{\FASThtwoAGGCAPO}{\rna}
\newcommand{\FASThtwoAGGCAPOCI}{\rna}
\newcommand{\FASThtwoAGGCAPOCIMIN}{\rna}
\newcommand{\FASThtwoAGGCAPOCIMAX}{\rna}
\newcommand{\FASThtwoAGGCAPODynamic}{\rna}
\newcommand{\FASThtwoAGGCAPODynamicCI}{\rna}
\newcommand{\FASThtwoAGGCAPODynamicCIMIN}{\rna}
\newcommand{\FASThtwoAGGCAPODynamicCIMAX}{\rna}
\newcommand{\FASTjythonEvents}{0}
\newcommand{\FASTjythonNoFPEvents}{0}
\newcommand{\FASTjythonMaxLiveThreads}{0}
\newcommand{\FASTjythonTotalThreads}{0}
\newcommand{\FASTjythonBaseTime}{4.0}
\newcommand{\FASTjythonBaseTimeCI}{130}
\newcommand{\FASTjythonEmptyTime}{\rna}
\newcommand{\FASTjythonEmptyTimeCI}{\rna}
\newcommand{\FASTjythonEmptyTimeCIMIN}{\rna}
\newcommand{\FASTjythonEmptyTimeCIMAX}{\rna}
\newcommand{\FASTjythonFTTime}{7.6}
\newcommand{\FASTjythonFTTimeCI}{0.23}
\newcommand{\FASTjythonHBTime}{7.8}
\newcommand{\FASTjythonHBTimeCI}{0.30}
\newcommand{\FASTjythonFTOHBTime}{7.7}
\newcommand{\FASTjythonFTOHBTimeCI}{0.24}
\newcommand{\FASTjythonWCPTime}{\rna}
\newcommand{\FASTjythonWCPTimeCI}{\rna}
\newcommand{\FASTjythonWCPTimeCIMIN}{\rna}
\newcommand{\FASTjythonWCPTimeCIMAX}{\rna}
\newcommand{\FASTjythonFTOWCPTime}{9.9}
\newcommand{\FASTjythonFTOWCPTimeCI}{0.29}
\newcommand{\FASTjythonREWCPTime}{11}
\newcommand{\FASTjythonREWCPTimeCI}{0.30}
\newcommand{\FASTjythonDCTime}{\rna}
\newcommand{\FASTjythonDCTimeCI}{\rna}
\newcommand{\FASTjythonDCTimeCIMIN}{\rna}
\newcommand{\FASTjythonDCTimeCIMAX}{\rna}
\newcommand{\FASTjythonFTODCTime}{10}
\newcommand{\FASTjythonFTODCTimeCI}{0.33}
\newcommand{\FASTjythonREDCTime}{10}
\newcommand{\FASTjythonREDCTimeCI}{0.37}
\newcommand{\FASTjythonCAPOTime}{\rna}
\newcommand{\FASTjythonCAPOTimeCI}{\rna}
\newcommand{\FASTjythonCAPOTimeCIMIN}{\rna}
\newcommand{\FASTjythonCAPOTimeCIMAX}{\rna}
\newcommand{\FASTjythonFTOCAPOTime}{8.2}
\newcommand{\FASTjythonFTOCAPOTimeCI}{0.27}
\newcommand{\FASTjythonRECAPOTime}{8.2}
\newcommand{\FASTjythonRECAPOTimeCI}{0.22}
\newcommand{\FASTjythonAGGCAPOTime}{\rna}
\newcommand{\FASTjythonAGGCAPOTimeCI}{\rna}
\newcommand{\FASTjythonAGGCAPOTimeCIMIN}{\rna}
\newcommand{\FASTjythonAGGCAPOTimeCIMAX}{\rna}
\newcommand{\FASTjythonStaticTime}{\rzero}
\newcommand{\FASTjythonDynamicTime}{\rzero}
\newcommand{\FASTjythonBaseMem}{760}
\newcommand{\FASTjythonBaseMemCI}{5.1}
\newcommand{\FASTjythonFTMem}{6.6}
\newcommand{\FASTjythonFTMemCI}{0.57}
\newcommand{\FASTjythonHBMem}{7.0}
\newcommand{\FASTjythonHBMemCI}{0.36}
\newcommand{\FASTjythonFTOHBMem}{7.0}
\newcommand{\FASTjythonFTOHBMemCI}{0.28}
\newcommand{\FASTjythonWCPMem}{\memna}
\newcommand{\FASTjythonWCPMemCI}{\memna}
\newcommand{\FASTjythonWCPMemCIMIN}{\memna}
\newcommand{\FASTjythonWCPMemCIMAX}{\memna}
\newcommand{\FASTjythonFTOWCPMem}{13}
\newcommand{\FASTjythonFTOWCPMemCI}{0.41}
\newcommand{\FASTjythonREWCPMem}{10}
\newcommand{\FASTjythonREWCPMemCI}{0.30}
\newcommand{\FASTjythonDCMem}{\memna}
\newcommand{\FASTjythonDCMemCI}{\memna}
\newcommand{\FASTjythonDCMemCIMIN}{\memna}
\newcommand{\FASTjythonDCMemCIMAX}{\memna}
\newcommand{\FASTjythonFTODCMem}{12}
\newcommand{\FASTjythonFTODCMemCI}{0.37}
\newcommand{\FASTjythonREDCMem}{11}
\newcommand{\FASTjythonREDCMemCI}{0.64}
\newcommand{\FASTjythonCAPOMem}{\memna}
\newcommand{\FASTjythonCAPOMemCI}{\memna}
\newcommand{\FASTjythonCAPOMemCIMIN}{\memna}
\newcommand{\FASTjythonCAPOMemCIMAX}{\memna}
\newcommand{\FASTjythonFTOCAPOMem}{8.7}
\newcommand{\FASTjythonFTOCAPOMemCI}{0.063}
\newcommand{\FASTjythonRECAPOMem}{7.5}
\newcommand{\FASTjythonRECAPOMemCI}{0.40}
\newcommand{\FASTjythonAGGCAPOMem}{\memna}
\newcommand{\FASTjythonAGGCAPOMemCI}{\memna}
\newcommand{\FASTjythonAGGCAPOMemCIMIN}{\memna}
\newcommand{\FASTjythonAGGCAPOMemCIMAX}{\memna}
\newcommand{\FASTjythonEventsCI}{0}
\newcommand{\FASTjythonEventsCIMIN}{0}
\newcommand{\FASTjythonEventsCIMAX}{0}
\newcommand{\FASTjythonNoFPEventsCI}{0}
\newcommand{\FASTjythonNoFPEventsCIMIN}{0}
\newcommand{\FASTjythonNoFPEventsCIMAX}{0}
\newcommand{\FASTjythonFT}{23}
\newcommand{\FASTjythonFTCI}{1.3}
\newcommand{\FASTjythonFTCIMIN}{22}
\newcommand{\FASTjythonFTCIMAX}{24}
\newcommand{\FASTjythonFTDynamic}{47}
\newcommand{\FASTjythonFTDynamicCI}{1.2}
\newcommand{\FASTjythonFTDynamicCIMIN}{46}
\newcommand{\FASTjythonFTDynamicCIMAX}{48}
\newcommand{\FASTjythonHB}{24}
\newcommand{\FASTjythonHBCI}{1.3}
\newcommand{\FASTjythonHBCIMIN}{23}
\newcommand{\FASTjythonHBCIMAX}{25}
\newcommand{\FASTjythonHBDynamic}{27}
\newcommand{\FASTjythonHBDynamicCI}{1.1}
\newcommand{\FASTjythonHBDynamicCIMIN}{26}
\newcommand{\FASTjythonHBDynamicCIMAX}{28}
\newcommand{\FASTjythonFTOHB}{24}
\newcommand{\FASTjythonFTOHBCI}{1.3}
\newcommand{\FASTjythonFTOHBCIMIN}{23}
\newcommand{\FASTjythonFTOHBCIMAX}{25}
\newcommand{\FASTjythonFTOHBDynamic}{26}
\newcommand{\FASTjythonFTOHBDynamicCI}{1.2}
\newcommand{\FASTjythonFTOHBDynamicCIMIN}{25}
\newcommand{\FASTjythonFTOHBDynamicCIMAX}{27}
\newcommand{\FASTjythonWCP}{\rna}
\newcommand{\FASTjythonWCPCI}{\rna}
\newcommand{\FASTjythonWCPCIMIN}{\rna}
\newcommand{\FASTjythonWCPCIMAX}{\rna}
\newcommand{\FASTjythonWCPDynamic}{\rna}
\newcommand{\FASTjythonWCPDynamicCI}{\rna}
\newcommand{\FASTjythonWCPDynamicCIMIN}{\rna}
\newcommand{\FASTjythonWCPDynamicCIMAX}{\rna}
\newcommand{\FASTjythonFTOWCP}{19}
\newcommand{\FASTjythonFTOWCPCI}{0.89}
\newcommand{\FASTjythonFTOWCPCIMIN}{18}
\newcommand{\FASTjythonFTOWCPCIMAX}{20}
\newcommand{\FASTjythonFTOWCPDynamic}{20}
\newcommand{\FASTjythonFTOWCPDynamicCI}{1.4}
\newcommand{\FASTjythonFTOWCPDynamicCIMIN}{19}
\newcommand{\FASTjythonFTOWCPDynamicCIMAX}{21}
\newcommand{\FASTjythonREWCP}{25}
\newcommand{\FASTjythonREWCPCI}{0.96}
\newcommand{\FASTjythonREWCPCIMIN}{24}
\newcommand{\FASTjythonREWCPCIMAX}{26}
\newcommand{\FASTjythonREWCPDynamic}{26}
\newcommand{\FASTjythonREWCPDynamicCI}{1.6}
\newcommand{\FASTjythonREWCPDynamicCIMIN}{24}
\newcommand{\FASTjythonREWCPDynamicCIMAX}{28}
\newcommand{\FASTjythonDC}{\rna}
\newcommand{\FASTjythonDCCI}{\rna}
\newcommand{\FASTjythonDCCIMIN}{\rna}
\newcommand{\FASTjythonDCCIMAX}{\rna}
\newcommand{\FASTjythonDCDynamic}{\rna}
\newcommand{\FASTjythonDCDynamicCI}{\rna}
\newcommand{\FASTjythonDCDynamicCIMIN}{\rna}
\newcommand{\FASTjythonDCDynamicCIMAX}{\rna}
\newcommand{\FASTjythonFTODC}{27}
\newcommand{\FASTjythonFTODCCI}{0.0}
\newcommand{\FASTjythonFTODCCIMIN}{27}
\newcommand{\FASTjythonFTODCCIMAX}{27}
\newcommand{\FASTjythonFTODCDynamic}{30}
\newcommand{\FASTjythonFTODCDynamicCI}{0.32}
\newcommand{\FASTjythonFTODCDynamicCIMIN}{30}
\newcommand{\FASTjythonFTODCDynamicCIMAX}{30}
\newcommand{\FASTjythonREDC}{31}
\newcommand{\FASTjythonREDCCI}{0.78}
\newcommand{\FASTjythonREDCCIMIN}{30}
\newcommand{\FASTjythonREDCCIMAX}{32}
\newcommand{\FASTjythonREDCDynamic}{34}
\newcommand{\FASTjythonREDCDynamicCI}{0.78}
\newcommand{\FASTjythonREDCDynamicCIMIN}{33}
\newcommand{\FASTjythonREDCDynamicCIMAX}{35}
\newcommand{\FASTjythonCAPO}{\rna}
\newcommand{\FASTjythonCAPOCI}{\rna}
\newcommand{\FASTjythonCAPOCIMIN}{\rna}
\newcommand{\FASTjythonCAPOCIMAX}{\rna}
\newcommand{\FASTjythonCAPODynamic}{\rna}
\newcommand{\FASTjythonCAPODynamicCI}{\rna}
\newcommand{\FASTjythonCAPODynamicCIMIN}{\rna}
\newcommand{\FASTjythonCAPODynamicCIMAX}{\rna}
\newcommand{\FASTjythonFTOCAPO}{29}
\newcommand{\FASTjythonFTOCAPOCI}{1.3}
\newcommand{\FASTjythonFTOCAPOCIMIN}{28}
\newcommand{\FASTjythonFTOCAPOCIMAX}{30}
\newcommand{\FASTjythonFTOCAPODynamic}{31}
\newcommand{\FASTjythonFTOCAPODynamicCI}{1.2}
\newcommand{\FASTjythonFTOCAPODynamicCIMIN}{30}
\newcommand{\FASTjythonFTOCAPODynamicCIMAX}{32}
\newcommand{\FASTjythonRECAPO}{27}
\newcommand{\FASTjythonRECAPOCI}{0.0}
\newcommand{\FASTjythonRECAPOCIMIN}{27}
\newcommand{\FASTjythonRECAPOCIMAX}{27}
\newcommand{\FASTjythonRECAPODynamic}{30}
\newcommand{\FASTjythonRECAPODynamicCI}{0.0}
\newcommand{\FASTjythonRECAPODynamicCIMIN}{30}
\newcommand{\FASTjythonRECAPODynamicCIMAX}{30}
\newcommand{\FASTjythonAGGCAPO}{\rna}
\newcommand{\FASTjythonAGGCAPOCI}{\rna}
\newcommand{\FASTjythonAGGCAPOCIMIN}{\rna}
\newcommand{\FASTjythonAGGCAPOCIMAX}{\rna}
\newcommand{\FASTjythonAGGCAPODynamic}{\rna}
\newcommand{\FASTjythonAGGCAPODynamicCI}{\rna}
\newcommand{\FASTjythonAGGCAPODynamicCIMIN}{\rna}
\newcommand{\FASTjythonAGGCAPODynamicCIMAX}{\rna}
\newcommand{\FASTluindexEvents}{0}
\newcommand{\FASTluindexNoFPEvents}{0}
\newcommand{\FASTluindexMaxLiveThreads}{1}
\newcommand{\FASTluindexTotalThreads}{1}
\newcommand{\FASTluindexBaseTime}{1.2}
\newcommand{\FASTluindexBaseTimeCI}{110}
\newcommand{\FASTluindexEmptyTime}{\rna}
\newcommand{\FASTluindexEmptyTimeCI}{\rna}
\newcommand{\FASTluindexEmptyTimeCIMIN}{\rna}
\newcommand{\FASTluindexEmptyTimeCIMAX}{\rna}
\newcommand{\FASTluindexFTTime}{7.3}
\newcommand{\FASTluindexFTTimeCI}{0.43}
\newcommand{\FASTluindexHBTime}{6.9}
\newcommand{\FASTluindexHBTimeCI}{0.53}
\newcommand{\FASTluindexFTOHBTime}{6.9}
\newcommand{\FASTluindexFTOHBTimeCI}{0.40}
\newcommand{\FASTluindexWCPTime}{\rna}
\newcommand{\FASTluindexWCPTimeCI}{\rna}
\newcommand{\FASTluindexWCPTimeCIMIN}{\rna}
\newcommand{\FASTluindexWCPTimeCIMAX}{\rna}
\newcommand{\FASTluindexFTOWCPTime}{20}
\newcommand{\FASTluindexFTOWCPTimeCI}{1.3}
\newcommand{\FASTluindexREWCPTime}{7.7}
\newcommand{\FASTluindexREWCPTimeCI}{0.45}
\newcommand{\FASTluindexDCTime}{\rna}
\newcommand{\FASTluindexDCTimeCI}{\rna}
\newcommand{\FASTluindexDCTimeCIMIN}{\rna}
\newcommand{\FASTluindexDCTimeCIMAX}{\rna}
\newcommand{\FASTluindexFTODCTime}{20}
\newcommand{\FASTluindexFTODCTimeCI}{1.4}
\newcommand{\FASTluindexREDCTime}{7.6}
\newcommand{\FASTluindexREDCTimeCI}{0.51}
\newcommand{\FASTluindexCAPOTime}{\rna}
\newcommand{\FASTluindexCAPOTimeCI}{\rna}
\newcommand{\FASTluindexCAPOTimeCIMIN}{\rna}
\newcommand{\FASTluindexCAPOTimeCIMAX}{\rna}
\newcommand{\FASTluindexFTOCAPOTime}{19}
\newcommand{\FASTluindexFTOCAPOTimeCI}{1.2}
\newcommand{\FASTluindexRECAPOTime}{7.5}
\newcommand{\FASTluindexRECAPOTimeCI}{0.52}
\newcommand{\FASTluindexAGGCAPOTime}{\rna}
\newcommand{\FASTluindexAGGCAPOTimeCI}{\rna}
\newcommand{\FASTluindexAGGCAPOTimeCIMIN}{\rna}
\newcommand{\FASTluindexAGGCAPOTimeCIMAX}{\rna}
\newcommand{\FASTluindexStaticTime}{\rzero}
\newcommand{\FASTluindexDynamicTime}{\rzero}
\newcommand{\FASTluindexBaseMem}{150}
\newcommand{\FASTluindexBaseMemCI}{1.1}
\newcommand{\FASTluindexFTMem}{4.3}
\newcommand{\FASTluindexFTMemCI}{0.075}
\newcommand{\FASTluindexHBMem}{4.4}
\newcommand{\FASTluindexHBMemCI}{0.039}
\newcommand{\FASTluindexFTOHBMem}{4.4}
\newcommand{\FASTluindexFTOHBMemCI}{0.067}
\newcommand{\FASTluindexWCPMem}{\memna}
\newcommand{\FASTluindexWCPMemCI}{\memna}
\newcommand{\FASTluindexWCPMemCIMIN}{\memna}
\newcommand{\FASTluindexWCPMemCIMAX}{\memna}
\newcommand{\FASTluindexFTOWCPMem}{20}
\newcommand{\FASTluindexFTOWCPMemCI}{0.21}
\newcommand{\FASTluindexREWCPMem}{6.2}
\newcommand{\FASTluindexREWCPMemCI}{0.064}
\newcommand{\FASTluindexDCMem}{\memna}
\newcommand{\FASTluindexDCMemCI}{\memna}
\newcommand{\FASTluindexDCMemCIMIN}{\memna}
\newcommand{\FASTluindexDCMemCIMAX}{\memna}
\newcommand{\FASTluindexFTODCMem}{20}
\newcommand{\FASTluindexFTODCMemCI}{0.23}
\newcommand{\FASTluindexREDCMem}{6.1}
\newcommand{\FASTluindexREDCMemCI}{0.052}
\newcommand{\FASTluindexCAPOMem}{\memna}
\newcommand{\FASTluindexCAPOMemCI}{\memna}
\newcommand{\FASTluindexCAPOMemCIMIN}{\memna}
\newcommand{\FASTluindexCAPOMemCIMAX}{\memna}
\newcommand{\FASTluindexFTOCAPOMem}{20}
\newcommand{\FASTluindexFTOCAPOMemCI}{0.21}
\newcommand{\FASTluindexRECAPOMem}{6.1}
\newcommand{\FASTluindexRECAPOMemCI}{0.080}
\newcommand{\FASTluindexAGGCAPOMem}{\memna}
\newcommand{\FASTluindexAGGCAPOMemCI}{\memna}
\newcommand{\FASTluindexAGGCAPOMemCIMIN}{\memna}
\newcommand{\FASTluindexAGGCAPOMemCIMAX}{\memna}
\newcommand{\FASTluindexEventsCI}{0}
\newcommand{\FASTluindexEventsCIMIN}{0}
\newcommand{\FASTluindexEventsCIMAX}{0}
\newcommand{\FASTluindexNoFPEventsCI}{0}
\newcommand{\FASTluindexNoFPEventsCIMIN}{0}
\newcommand{\FASTluindexNoFPEventsCIMAX}{0}
\newcommand{\FASTluindexFT}{1}
\newcommand{\FASTluindexFTCI}{0.0}
\newcommand{\FASTluindexFTCIMIN}{1}
\newcommand{\FASTluindexFTCIMAX}{1}
\newcommand{\FASTluindexFTDynamic}{1}
\newcommand{\FASTluindexFTDynamicCI}{0.0}
\newcommand{\FASTluindexFTDynamicCIMIN}{1}
\newcommand{\FASTluindexFTDynamicCIMAX}{1}
\newcommand{\FASTluindexHB}{1}
\newcommand{\FASTluindexHBCI}{0.0}
\newcommand{\FASTluindexHBCIMIN}{1}
\newcommand{\FASTluindexHBCIMAX}{1}
\newcommand{\FASTluindexHBDynamic}{1}
\newcommand{\FASTluindexHBDynamicCI}{0.0}
\newcommand{\FASTluindexHBDynamicCIMIN}{1}
\newcommand{\FASTluindexHBDynamicCIMAX}{1}
\newcommand{\FASTluindexFTOHB}{1}
\newcommand{\FASTluindexFTOHBCI}{0.0}
\newcommand{\FASTluindexFTOHBCIMIN}{1}
\newcommand{\FASTluindexFTOHBCIMAX}{1}
\newcommand{\FASTluindexFTOHBDynamic}{1}
\newcommand{\FASTluindexFTOHBDynamicCI}{0.0}
\newcommand{\FASTluindexFTOHBDynamicCIMIN}{1}
\newcommand{\FASTluindexFTOHBDynamicCIMAX}{1}
\newcommand{\FASTluindexWCP}{\rna}
\newcommand{\FASTluindexWCPCI}{\rna}
\newcommand{\FASTluindexWCPCIMIN}{\rna}
\newcommand{\FASTluindexWCPCIMAX}{\rna}
\newcommand{\FASTluindexWCPDynamic}{\rna}
\newcommand{\FASTluindexWCPDynamicCI}{\rna}
\newcommand{\FASTluindexWCPDynamicCIMIN}{\rna}
\newcommand{\FASTluindexWCPDynamicCIMAX}{\rna}
\newcommand{\FASTluindexFTOWCP}{1}
\newcommand{\FASTluindexFTOWCPCI}{0.0}
\newcommand{\FASTluindexFTOWCPCIMIN}{1}
\newcommand{\FASTluindexFTOWCPCIMAX}{1}
\newcommand{\FASTluindexFTOWCPDynamic}{1}
\newcommand{\FASTluindexFTOWCPDynamicCI}{0.0}
\newcommand{\FASTluindexFTOWCPDynamicCIMIN}{1}
\newcommand{\FASTluindexFTOWCPDynamicCIMAX}{1}
\newcommand{\FASTluindexREWCP}{1}
\newcommand{\FASTluindexREWCPCI}{0.0}
\newcommand{\FASTluindexREWCPCIMIN}{1}
\newcommand{\FASTluindexREWCPCIMAX}{1}
\newcommand{\FASTluindexREWCPDynamic}{1}
\newcommand{\FASTluindexREWCPDynamicCI}{0.0}
\newcommand{\FASTluindexREWCPDynamicCIMIN}{1}
\newcommand{\FASTluindexREWCPDynamicCIMAX}{1}
\newcommand{\FASTluindexDC}{\rna}
\newcommand{\FASTluindexDCCI}{\rna}
\newcommand{\FASTluindexDCCIMIN}{\rna}
\newcommand{\FASTluindexDCCIMAX}{\rna}
\newcommand{\FASTluindexDCDynamic}{\rna}
\newcommand{\FASTluindexDCDynamicCI}{\rna}
\newcommand{\FASTluindexDCDynamicCIMIN}{\rna}
\newcommand{\FASTluindexDCDynamicCIMAX}{\rna}
\newcommand{\FASTluindexFTODC}{1}
\newcommand{\FASTluindexFTODCCI}{0.0}
\newcommand{\FASTluindexFTODCCIMIN}{1}
\newcommand{\FASTluindexFTODCCIMAX}{1}
\newcommand{\FASTluindexFTODCDynamic}{1}
\newcommand{\FASTluindexFTODCDynamicCI}{0.0}
\newcommand{\FASTluindexFTODCDynamicCIMIN}{1}
\newcommand{\FASTluindexFTODCDynamicCIMAX}{1}
\newcommand{\FASTluindexREDC}{1}
\newcommand{\FASTluindexREDCCI}{0.0}
\newcommand{\FASTluindexREDCCIMIN}{1}
\newcommand{\FASTluindexREDCCIMAX}{1}
\newcommand{\FASTluindexREDCDynamic}{1}
\newcommand{\FASTluindexREDCDynamicCI}{0.0}
\newcommand{\FASTluindexREDCDynamicCIMIN}{1}
\newcommand{\FASTluindexREDCDynamicCIMAX}{1}
\newcommand{\FASTluindexCAPO}{\rna}
\newcommand{\FASTluindexCAPOCI}{\rna}
\newcommand{\FASTluindexCAPOCIMIN}{\rna}
\newcommand{\FASTluindexCAPOCIMAX}{\rna}
\newcommand{\FASTluindexCAPODynamic}{\rna}
\newcommand{\FASTluindexCAPODynamicCI}{\rna}
\newcommand{\FASTluindexCAPODynamicCIMIN}{\rna}
\newcommand{\FASTluindexCAPODynamicCIMAX}{\rna}
\newcommand{\FASTluindexFTOCAPO}{1}
\newcommand{\FASTluindexFTOCAPOCI}{0.0}
\newcommand{\FASTluindexFTOCAPOCIMIN}{1}
\newcommand{\FASTluindexFTOCAPOCIMAX}{1}
\newcommand{\FASTluindexFTOCAPODynamic}{1}
\newcommand{\FASTluindexFTOCAPODynamicCI}{0.0}
\newcommand{\FASTluindexFTOCAPODynamicCIMIN}{1}
\newcommand{\FASTluindexFTOCAPODynamicCIMAX}{1}
\newcommand{\FASTluindexRECAPO}{1}
\newcommand{\FASTluindexRECAPOCI}{0.0}
\newcommand{\FASTluindexRECAPOCIMIN}{1}
\newcommand{\FASTluindexRECAPOCIMAX}{1}
\newcommand{\FASTluindexRECAPODynamic}{1}
\newcommand{\FASTluindexRECAPODynamicCI}{0.0}
\newcommand{\FASTluindexRECAPODynamicCIMIN}{1}
\newcommand{\FASTluindexRECAPODynamicCIMAX}{1}
\newcommand{\FASTluindexAGGCAPO}{\rna}
\newcommand{\FASTluindexAGGCAPOCI}{\rna}
\newcommand{\FASTluindexAGGCAPOCIMIN}{\rna}
\newcommand{\FASTluindexAGGCAPOCIMAX}{\rna}
\newcommand{\FASTluindexAGGCAPODynamic}{\rna}
\newcommand{\FASTluindexAGGCAPODynamicCI}{\rna}
\newcommand{\FASTluindexAGGCAPODynamicCIMIN}{\rna}
\newcommand{\FASTluindexAGGCAPODynamicCIMAX}{\rna}
\newcommand{\FASTlusearchEvents}{0}
\newcommand{\FASTlusearchNoFPEvents}{0}
\newcommand{\FASTlusearchMaxLiveThreads}{14}
\newcommand{\FASTlusearchTotalThreads}{14}
\newcommand{\FASTlusearchBaseTime}{1.0}
\newcommand{\FASTlusearchBaseTimeCI}{94}
\newcommand{\FASTlusearchEmptyTime}{\rna}
\newcommand{\FASTlusearchEmptyTimeCI}{\rna}
\newcommand{\FASTlusearchEmptyTimeCIMIN}{\rna}
\newcommand{\FASTlusearchEmptyTimeCIMAX}{\rna}
\newcommand{\FASTlusearchFTTime}{10}
\newcommand{\FASTlusearchFTTimeCI}{0.96}
\newcommand{\FASTlusearchHBTime}{9.8}
\newcommand{\FASTlusearchHBTimeCI}{0.92}
\newcommand{\FASTlusearchFTOHBTime}{9.8}
\newcommand{\FASTlusearchFTOHBTimeCI}{0.73}
\newcommand{\FASTlusearchWCPTime}{\rna}
\newcommand{\FASTlusearchWCPTimeCI}{\rna}
\newcommand{\FASTlusearchWCPTimeCIMIN}{\rna}
\newcommand{\FASTlusearchWCPTimeCIMAX}{\rna}
\newcommand{\FASTlusearchFTOWCPTime}{13}
\newcommand{\FASTlusearchFTOWCPTimeCI}{1.4}
\newcommand{\FASTlusearchREWCPTime}{12}
\newcommand{\FASTlusearchREWCPTimeCI}{1.2}
\newcommand{\FASTlusearchDCTime}{\rna}
\newcommand{\FASTlusearchDCTimeCI}{\rna}
\newcommand{\FASTlusearchDCTimeCIMIN}{\rna}
\newcommand{\FASTlusearchDCTimeCIMAX}{\rna}
\newcommand{\FASTlusearchFTODCTime}{14}
\newcommand{\FASTlusearchFTODCTimeCI}{1.4}
\newcommand{\FASTlusearchREDCTime}{13}
\newcommand{\FASTlusearchREDCTimeCI}{1.5}
\newcommand{\FASTlusearchCAPOTime}{\rna}
\newcommand{\FASTlusearchCAPOTimeCI}{\rna}
\newcommand{\FASTlusearchCAPOTimeCIMIN}{\rna}
\newcommand{\FASTlusearchCAPOTimeCIMAX}{\rna}
\newcommand{\FASTlusearchFTOCAPOTime}{12}
\newcommand{\FASTlusearchFTOCAPOTimeCI}{1.2}
\newcommand{\FASTlusearchRECAPOTime}{11}
\newcommand{\FASTlusearchRECAPOTimeCI}{1.0}
\newcommand{\FASTlusearchAGGCAPOTime}{\rna}
\newcommand{\FASTlusearchAGGCAPOTimeCI}{\rna}
\newcommand{\FASTlusearchAGGCAPOTimeCIMIN}{\rna}
\newcommand{\FASTlusearchAGGCAPOTimeCIMAX}{\rna}
\newcommand{\FASTlusearchStaticTime}{\rzero}
\newcommand{\FASTlusearchDynamicTime}{\rzero}
\newcommand{\FASTlusearchBaseMem}{1,600}
\newcommand{\FASTlusearchBaseMemCI}{60.0}
\newcommand{\FASTlusearchFTMem}{9.9}
\newcommand{\FASTlusearchFTMemCI}{0.60}
\newcommand{\FASTlusearchHBMem}{9.3}
\newcommand{\FASTlusearchHBMemCI}{0.45}
\newcommand{\FASTlusearchFTOHBMem}{9.3}
\newcommand{\FASTlusearchFTOHBMemCI}{0.67}
\newcommand{\FASTlusearchWCPMem}{\memna}
\newcommand{\FASTlusearchWCPMemCI}{\memna}
\newcommand{\FASTlusearchWCPMemCIMIN}{\memna}
\newcommand{\FASTlusearchWCPMemCIMAX}{\memna}
\newcommand{\FASTlusearchFTOWCPMem}{9.9}
\newcommand{\FASTlusearchFTOWCPMemCI}{0.48}
\newcommand{\FASTlusearchREWCPMem}{9.9}
\newcommand{\FASTlusearchREWCPMemCI}{0.64}
\newcommand{\FASTlusearchDCMem}{\memna}
\newcommand{\FASTlusearchDCMemCI}{\memna}
\newcommand{\FASTlusearchDCMemCIMIN}{\memna}
\newcommand{\FASTlusearchDCMemCIMAX}{\memna}
\newcommand{\FASTlusearchFTODCMem}{9.9}
\newcommand{\FASTlusearchFTODCMemCI}{0.37}
\newcommand{\FASTlusearchREDCMem}{9.9}
\newcommand{\FASTlusearchREDCMemCI}{0.60}
\newcommand{\FASTlusearchCAPOMem}{\memna}
\newcommand{\FASTlusearchCAPOMemCI}{\memna}
\newcommand{\FASTlusearchCAPOMemCIMIN}{\memna}
\newcommand{\FASTlusearchCAPOMemCIMAX}{\memna}
\newcommand{\FASTlusearchFTOCAPOMem}{11}
\newcommand{\FASTlusearchFTOCAPOMemCI}{0.73}
\newcommand{\FASTlusearchRECAPOMem}{11}
\newcommand{\FASTlusearchRECAPOMemCI}{0.63}
\newcommand{\FASTlusearchAGGCAPOMem}{\memna}
\newcommand{\FASTlusearchAGGCAPOMemCI}{\memna}
\newcommand{\FASTlusearchAGGCAPOMemCIMIN}{\memna}
\newcommand{\FASTlusearchAGGCAPOMemCIMAX}{\memna}
\newcommand{\FASTlusearchEventsCI}{0}
\newcommand{\FASTlusearchEventsCIMIN}{0}
\newcommand{\FASTlusearchEventsCIMAX}{0}
\newcommand{\FASTlusearchNoFPEventsCI}{0}
\newcommand{\FASTlusearchNoFPEventsCIMIN}{0}
\newcommand{\FASTlusearchNoFPEventsCIMAX}{0}
\newcommand{\FASTlusearchFT}{0}
\newcommand{\FASTlusearchFTCI}{0.0}
\newcommand{\FASTlusearchFTCIMIN}{0}
\newcommand{\FASTlusearchFTCIMAX}{0}
\newcommand{\FASTlusearchFTDynamic}{0}
\newcommand{\FASTlusearchFTDynamicCI}{0.0}
\newcommand{\FASTlusearchFTDynamicCIMIN}{0}
\newcommand{\FASTlusearchFTDynamicCIMAX}{0}
\newcommand{\FASTlusearchHB}{0}
\newcommand{\FASTlusearchHBCI}{0.0}
\newcommand{\FASTlusearchHBCIMIN}{0}
\newcommand{\FASTlusearchHBCIMAX}{0}
\newcommand{\FASTlusearchHBDynamic}{0}
\newcommand{\FASTlusearchHBDynamicCI}{0.0}
\newcommand{\FASTlusearchHBDynamicCIMIN}{0}
\newcommand{\FASTlusearchHBDynamicCIMAX}{0}
\newcommand{\FASTlusearchFTOHB}{0}
\newcommand{\FASTlusearchFTOHBCI}{0.0}
\newcommand{\FASTlusearchFTOHBCIMIN}{0}
\newcommand{\FASTlusearchFTOHBCIMAX}{0}
\newcommand{\FASTlusearchFTOHBDynamic}{0}
\newcommand{\FASTlusearchFTOHBDynamicCI}{0.0}
\newcommand{\FASTlusearchFTOHBDynamicCIMIN}{0}
\newcommand{\FASTlusearchFTOHBDynamicCIMAX}{0}
\newcommand{\FASTlusearchWCP}{\rna}
\newcommand{\FASTlusearchWCPCI}{\rna}
\newcommand{\FASTlusearchWCPCIMIN}{\rna}
\newcommand{\FASTlusearchWCPCIMAX}{\rna}
\newcommand{\FASTlusearchWCPDynamic}{\rna}
\newcommand{\FASTlusearchWCPDynamicCI}{\rna}
\newcommand{\FASTlusearchWCPDynamicCIMIN}{\rna}
\newcommand{\FASTlusearchWCPDynamicCIMAX}{\rna}
\newcommand{\FASTlusearchFTOWCP}{0}
\newcommand{\FASTlusearchFTOWCPCI}{0.0}
\newcommand{\FASTlusearchFTOWCPCIMIN}{0}
\newcommand{\FASTlusearchFTOWCPCIMAX}{0}
\newcommand{\FASTlusearchFTOWCPDynamic}{0}
\newcommand{\FASTlusearchFTOWCPDynamicCI}{0.0}
\newcommand{\FASTlusearchFTOWCPDynamicCIMIN}{0}
\newcommand{\FASTlusearchFTOWCPDynamicCIMAX}{0}
\newcommand{\FASTlusearchREWCP}{0}
\newcommand{\FASTlusearchREWCPCI}{0.0}
\newcommand{\FASTlusearchREWCPCIMIN}{0}
\newcommand{\FASTlusearchREWCPCIMAX}{0}
\newcommand{\FASTlusearchREWCPDynamic}{0}
\newcommand{\FASTlusearchREWCPDynamicCI}{0.0}
\newcommand{\FASTlusearchREWCPDynamicCIMIN}{0}
\newcommand{\FASTlusearchREWCPDynamicCIMAX}{0}
\newcommand{\FASTlusearchDC}{\rna}
\newcommand{\FASTlusearchDCCI}{\rna}
\newcommand{\FASTlusearchDCCIMIN}{\rna}
\newcommand{\FASTlusearchDCCIMAX}{\rna}
\newcommand{\FASTlusearchDCDynamic}{\rna}
\newcommand{\FASTlusearchDCDynamicCI}{\rna}
\newcommand{\FASTlusearchDCDynamicCIMIN}{\rna}
\newcommand{\FASTlusearchDCDynamicCIMAX}{\rna}
\newcommand{\FASTlusearchFTODC}{0}
\newcommand{\FASTlusearchFTODCCI}{0.0}
\newcommand{\FASTlusearchFTODCCIMIN}{0}
\newcommand{\FASTlusearchFTODCCIMAX}{0}
\newcommand{\FASTlusearchFTODCDynamic}{0}
\newcommand{\FASTlusearchFTODCDynamicCI}{0.0}
\newcommand{\FASTlusearchFTODCDynamicCIMIN}{0}
\newcommand{\FASTlusearchFTODCDynamicCIMAX}{0}
\newcommand{\FASTlusearchREDC}{0}
\newcommand{\FASTlusearchREDCCI}{0.0}
\newcommand{\FASTlusearchREDCCIMIN}{0}
\newcommand{\FASTlusearchREDCCIMAX}{0}
\newcommand{\FASTlusearchREDCDynamic}{0}
\newcommand{\FASTlusearchREDCDynamicCI}{0.0}
\newcommand{\FASTlusearchREDCDynamicCIMIN}{0}
\newcommand{\FASTlusearchREDCDynamicCIMAX}{0}
\newcommand{\FASTlusearchCAPO}{\rna}
\newcommand{\FASTlusearchCAPOCI}{\rna}
\newcommand{\FASTlusearchCAPOCIMIN}{\rna}
\newcommand{\FASTlusearchCAPOCIMAX}{\rna}
\newcommand{\FASTlusearchCAPODynamic}{\rna}
\newcommand{\FASTlusearchCAPODynamicCI}{\rna}
\newcommand{\FASTlusearchCAPODynamicCIMIN}{\rna}
\newcommand{\FASTlusearchCAPODynamicCIMAX}{\rna}
\newcommand{\FASTlusearchFTOCAPO}{0}
\newcommand{\FASTlusearchFTOCAPOCI}{0.0}
\newcommand{\FASTlusearchFTOCAPOCIMIN}{0}
\newcommand{\FASTlusearchFTOCAPOCIMAX}{0}
\newcommand{\FASTlusearchFTOCAPODynamic}{0}
\newcommand{\FASTlusearchFTOCAPODynamicCI}{0.0}
\newcommand{\FASTlusearchFTOCAPODynamicCIMIN}{0}
\newcommand{\FASTlusearchFTOCAPODynamicCIMAX}{0}
\newcommand{\FASTlusearchRECAPO}{0}
\newcommand{\FASTlusearchRECAPOCI}{0.0}
\newcommand{\FASTlusearchRECAPOCIMIN}{0}
\newcommand{\FASTlusearchRECAPOCIMAX}{0}
\newcommand{\FASTlusearchRECAPODynamic}{0}
\newcommand{\FASTlusearchRECAPODynamicCI}{0.0}
\newcommand{\FASTlusearchRECAPODynamicCIMIN}{0}
\newcommand{\FASTlusearchRECAPODynamicCIMAX}{0}
\newcommand{\FASTlusearchAGGCAPO}{\rna}
\newcommand{\FASTlusearchAGGCAPOCI}{\rna}
\newcommand{\FASTlusearchAGGCAPOCIMIN}{\rna}
\newcommand{\FASTlusearchAGGCAPOCIMAX}{\rna}
\newcommand{\FASTlusearchAGGCAPODynamic}{\rna}
\newcommand{\FASTlusearchAGGCAPODynamicCI}{\rna}
\newcommand{\FASTlusearchAGGCAPODynamicCIMIN}{\rna}
\newcommand{\FASTlusearchAGGCAPODynamicCIMAX}{\rna}
\newcommand{\FASTpmdEvents}{0}
\newcommand{\FASTpmdNoFPEvents}{0}
\newcommand{\FASTpmdMaxLiveThreads}{0}
\newcommand{\FASTpmdTotalThreads}{0}
\newcommand{\FASTpmdBaseTime}{1.2}
\newcommand{\FASTpmdBaseTimeCI}{16}
\newcommand{\FASTpmdEmptyTime}{\rna}
\newcommand{\FASTpmdEmptyTimeCI}{\rna}
\newcommand{\FASTpmdEmptyTimeCIMIN}{\rna}
\newcommand{\FASTpmdEmptyTimeCIMAX}{\rna}
\newcommand{\FASTpmdFTTime}{6.9}
\newcommand{\FASTpmdFTTimeCI}{0.25}
\newcommand{\FASTpmdHBTime}{6.3}
\newcommand{\FASTpmdHBTimeCI}{0.21}
\newcommand{\FASTpmdFTOHBTime}{6.4}
\newcommand{\FASTpmdFTOHBTimeCI}{0.21}
\newcommand{\FASTpmdWCPTime}{\rna}
\newcommand{\FASTpmdWCPTimeCI}{\rna}
\newcommand{\FASTpmdWCPTimeCIMIN}{\rna}
\newcommand{\FASTpmdWCPTimeCIMAX}{\rna}
\newcommand{\FASTpmdFTOWCPTime}{6.7}
\newcommand{\FASTpmdFTOWCPTimeCI}{0.16}
\newcommand{\FASTpmdREWCPTime}{6.6}
\newcommand{\FASTpmdREWCPTimeCI}{0.12}
\newcommand{\FASTpmdDCTime}{\rna}
\newcommand{\FASTpmdDCTimeCI}{\rna}
\newcommand{\FASTpmdDCTimeCIMIN}{\rna}
\newcommand{\FASTpmdDCTimeCIMAX}{\rna}
\newcommand{\FASTpmdFTODCTime}{6.6}
\newcommand{\FASTpmdFTODCTimeCI}{0.24}
\newcommand{\FASTpmdREDCTime}{6.5}
\newcommand{\FASTpmdREDCTimeCI}{0.15}
\newcommand{\FASTpmdCAPOTime}{\rna}
\newcommand{\FASTpmdCAPOTimeCI}{\rna}
\newcommand{\FASTpmdCAPOTimeCIMIN}{\rna}
\newcommand{\FASTpmdCAPOTimeCIMAX}{\rna}
\newcommand{\FASTpmdFTOCAPOTime}{6.5}
\newcommand{\FASTpmdFTOCAPOTimeCI}{0.16}
\newcommand{\FASTpmdRECAPOTime}{6.5}
\newcommand{\FASTpmdRECAPOTimeCI}{0.20}
\newcommand{\FASTpmdAGGCAPOTime}{\rna}
\newcommand{\FASTpmdAGGCAPOTimeCI}{\rna}
\newcommand{\FASTpmdAGGCAPOTimeCIMIN}{\rna}
\newcommand{\FASTpmdAGGCAPOTimeCIMAX}{\rna}
\newcommand{\FASTpmdStaticTime}{\rzero}
\newcommand{\FASTpmdDynamicTime}{\rzero}
\newcommand{\FASTpmdBaseMem}{630}
\newcommand{\FASTpmdBaseMemCI}{8.4}
\newcommand{\FASTpmdFTMem}{2.9}
\newcommand{\FASTpmdFTMemCI}{0.094}
\newcommand{\FASTpmdHBMem}{2.9}
\newcommand{\FASTpmdHBMemCI}{0.12}
\newcommand{\FASTpmdFTOHBMem}{2.9}
\newcommand{\FASTpmdFTOHBMemCI}{0.040}
\newcommand{\FASTpmdWCPMem}{\memna}
\newcommand{\FASTpmdWCPMemCI}{\memna}
\newcommand{\FASTpmdWCPMemCIMIN}{\memna}
\newcommand{\FASTpmdWCPMemCIMAX}{\memna}
\newcommand{\FASTpmdFTOWCPMem}{3.0}
\newcommand{\FASTpmdFTOWCPMemCI}{0.13}
\newcommand{\FASTpmdREWCPMem}{3.0}
\newcommand{\FASTpmdREWCPMemCI}{0.11}
\newcommand{\FASTpmdDCMem}{\memna}
\newcommand{\FASTpmdDCMemCI}{\memna}
\newcommand{\FASTpmdDCMemCIMIN}{\memna}
\newcommand{\FASTpmdDCMemCIMAX}{\memna}
\newcommand{\FASTpmdFTODCMem}{3.0}
\newcommand{\FASTpmdFTODCMemCI}{0.15}
\newcommand{\FASTpmdREDCMem}{2.9}
\newcommand{\FASTpmdREDCMemCI}{0.18}
\newcommand{\FASTpmdCAPOMem}{\memna}
\newcommand{\FASTpmdCAPOMemCI}{\memna}
\newcommand{\FASTpmdCAPOMemCIMIN}{\memna}
\newcommand{\FASTpmdCAPOMemCIMAX}{\memna}
\newcommand{\FASTpmdFTOCAPOMem}{3.0}
\newcommand{\FASTpmdFTOCAPOMemCI}{0.14}
\newcommand{\FASTpmdRECAPOMem}{2.9}
\newcommand{\FASTpmdRECAPOMemCI}{0.18}
\newcommand{\FASTpmdAGGCAPOMem}{\memna}
\newcommand{\FASTpmdAGGCAPOMemCI}{\memna}
\newcommand{\FASTpmdAGGCAPOMemCIMIN}{\memna}
\newcommand{\FASTpmdAGGCAPOMemCIMAX}{\memna}
\newcommand{\FASTpmdEventsCI}{0}
\newcommand{\FASTpmdEventsCIMIN}{0}
\newcommand{\FASTpmdEventsCIMAX}{0}
\newcommand{\FASTpmdNoFPEventsCI}{0}
\newcommand{\FASTpmdNoFPEventsCIMIN}{0}
\newcommand{\FASTpmdNoFPEventsCIMAX}{0}
\newcommand{\FASTpmdFT}{18}
\newcommand{\FASTpmdFTCI}{0.52}
\newcommand{\FASTpmdFTCIMIN}{17}
\newcommand{\FASTpmdFTCIMAX}{19}
\newcommand{\FASTpmdFTDynamic}{4,332}
\newcommand{\FASTpmdFTDynamicCI}{679}
\newcommand{\FASTpmdFTDynamicCIMIN}{3,653}
\newcommand{\FASTpmdFTDynamicCIMAX}{5,011}
\newcommand{\FASTpmdHB}{18}
\newcommand{\FASTpmdHBCI}{0.39}
\newcommand{\FASTpmdHBCIMIN}{18}
\newcommand{\FASTpmdHBCIMAX}{18}
\newcommand{\FASTpmdHBDynamic}{1,799}
\newcommand{\FASTpmdHBDynamicCI}{104}
\newcommand{\FASTpmdHBDynamicCIMIN}{1,695}
\newcommand{\FASTpmdHBDynamicCIMAX}{1,903}
\newcommand{\FASTpmdFTOHB}{18}
\newcommand{\FASTpmdFTOHBCI}{0.70}
\newcommand{\FASTpmdFTOHBCIMIN}{17}
\newcommand{\FASTpmdFTOHBCIMAX}{19}
\newcommand{\FASTpmdFTOHBDynamic}{1,668}
\newcommand{\FASTpmdFTOHBDynamicCI}{104}
\newcommand{\FASTpmdFTOHBDynamicCIMIN}{1,564}
\newcommand{\FASTpmdFTOHBDynamicCIMAX}{1,772}
\newcommand{\FASTpmdWCP}{\rna}
\newcommand{\FASTpmdWCPCI}{\rna}
\newcommand{\FASTpmdWCPCIMIN}{\rna}
\newcommand{\FASTpmdWCPCIMAX}{\rna}
\newcommand{\FASTpmdWCPDynamic}{\rna}
\newcommand{\FASTpmdWCPDynamicCI}{\rna}
\newcommand{\FASTpmdWCPDynamicCIMIN}{\rna}
\newcommand{\FASTpmdWCPDynamicCIMAX}{\rna}
\newcommand{\FASTpmdFTOWCP}{18}
\newcommand{\FASTpmdFTOWCPCI}{0.30}
\newcommand{\FASTpmdFTOWCPCIMIN}{18}
\newcommand{\FASTpmdFTOWCPCIMAX}{18}
\newcommand{\FASTpmdFTOWCPDynamic}{1,770}
\newcommand{\FASTpmdFTOWCPDynamicCI}{128}
\newcommand{\FASTpmdFTOWCPDynamicCIMIN}{1,642}
\newcommand{\FASTpmdFTOWCPDynamicCIMAX}{1,898}
\newcommand{\FASTpmdREWCP}{18}
\newcommand{\FASTpmdREWCPCI}{0.32}
\newcommand{\FASTpmdREWCPCIMIN}{18}
\newcommand{\FASTpmdREWCPCIMAX}{18}
\newcommand{\FASTpmdREWCPDynamic}{1,659}
\newcommand{\FASTpmdREWCPDynamicCI}{87}
\newcommand{\FASTpmdREWCPDynamicCIMIN}{1,572}
\newcommand{\FASTpmdREWCPDynamicCIMAX}{1,746}
\newcommand{\FASTpmdDC}{\rna}
\newcommand{\FASTpmdDCCI}{\rna}
\newcommand{\FASTpmdDCCIMIN}{\rna}
\newcommand{\FASTpmdDCCIMAX}{\rna}
\newcommand{\FASTpmdDCDynamic}{\rna}
\newcommand{\FASTpmdDCDynamicCI}{\rna}
\newcommand{\FASTpmdDCDynamicCIMIN}{\rna}
\newcommand{\FASTpmdDCDynamicCIMAX}{\rna}
\newcommand{\FASTpmdFTODC}{18}
\newcommand{\FASTpmdFTODCCI}{0.30}
\newcommand{\FASTpmdFTODCCIMIN}{18}
\newcommand{\FASTpmdFTODCCIMAX}{18}
\newcommand{\FASTpmdFTODCDynamic}{3,511}
\newcommand{\FASTpmdFTODCDynamicCI}{104}
\newcommand{\FASTpmdFTODCDynamicCIMIN}{3,407}
\newcommand{\FASTpmdFTODCDynamicCIMAX}{3,615}
\newcommand{\FASTpmdREDC}{19}
\newcommand{\FASTpmdREDCCI}{0.43}
\newcommand{\FASTpmdREDCCIMIN}{19}
\newcommand{\FASTpmdREDCCIMAX}{19}
\newcommand{\FASTpmdREDCDynamic}{3,835}
\newcommand{\FASTpmdREDCDynamicCI}{295}
\newcommand{\FASTpmdREDCDynamicCIMIN}{3,540}
\newcommand{\FASTpmdREDCDynamicCIMAX}{4,130}
\newcommand{\FASTpmdCAPO}{\rna}
\newcommand{\FASTpmdCAPOCI}{\rna}
\newcommand{\FASTpmdCAPOCIMIN}{\rna}
\newcommand{\FASTpmdCAPOCIMAX}{\rna}
\newcommand{\FASTpmdCAPODynamic}{\rna}
\newcommand{\FASTpmdCAPODynamicCI}{\rna}
\newcommand{\FASTpmdCAPODynamicCIMIN}{\rna}
\newcommand{\FASTpmdCAPODynamicCIMAX}{\rna}
\newcommand{\FASTpmdFTOCAPO}{18}
\newcommand{\FASTpmdFTOCAPOCI}{0.30}
\newcommand{\FASTpmdFTOCAPOCIMIN}{18}
\newcommand{\FASTpmdFTOCAPOCIMAX}{18}
\newcommand{\FASTpmdFTOCAPODynamic}{3,671}
\newcommand{\FASTpmdFTOCAPODynamicCI}{283}
\newcommand{\FASTpmdFTOCAPODynamicCIMIN}{3,388}
\newcommand{\FASTpmdFTOCAPODynamicCIMAX}{3,954}
\newcommand{\FASTpmdRECAPO}{18}
\newcommand{\FASTpmdRECAPOCI}{0.20}
\newcommand{\FASTpmdRECAPOCIMIN}{18}
\newcommand{\FASTpmdRECAPOCIMAX}{18}
\newcommand{\FASTpmdRECAPODynamic}{3,537}
\newcommand{\FASTpmdRECAPODynamicCI}{179}
\newcommand{\FASTpmdRECAPODynamicCIMIN}{3,358}
\newcommand{\FASTpmdRECAPODynamicCIMAX}{3,716}
\newcommand{\FASTpmdAGGCAPO}{\rna}
\newcommand{\FASTpmdAGGCAPOCI}{\rna}
\newcommand{\FASTpmdAGGCAPOCIMIN}{\rna}
\newcommand{\FASTpmdAGGCAPOCIMAX}{\rna}
\newcommand{\FASTpmdAGGCAPODynamic}{\rna}
\newcommand{\FASTpmdAGGCAPODynamicCI}{\rna}
\newcommand{\FASTpmdAGGCAPODynamicCIMIN}{\rna}
\newcommand{\FASTpmdAGGCAPODynamicCIMAX}{\rna}
\newcommand{\FASTsunflowEvents}{0}
\newcommand{\FASTsunflowNoFPEvents}{0}
\newcommand{\FASTsunflowMaxLiveThreads}{14}
\newcommand{\FASTsunflowTotalThreads}{28}
\newcommand{\FASTsunflowBaseTime}{1.6}
\newcommand{\FASTsunflowBaseTimeCI}{140}
\newcommand{\FASTsunflowEmptyTime}{\rna}
\newcommand{\FASTsunflowEmptyTimeCI}{\rna}
\newcommand{\FASTsunflowEmptyTimeCIMIN}{\rna}
\newcommand{\FASTsunflowEmptyTimeCIMAX}{\rna}
\newcommand{\FASTsunflowFTTime}{14}
\newcommand{\FASTsunflowFTTimeCI}{1.3}
\newcommand{\FASTsunflowHBTime}{15}
\newcommand{\FASTsunflowHBTimeCI}{1.5}
\newcommand{\FASTsunflowFTOHBTime}{14}
\newcommand{\FASTsunflowFTOHBTimeCI}{1.3}
\newcommand{\FASTsunflowWCPTime}{\rna}
\newcommand{\FASTsunflowWCPTimeCI}{\rna}
\newcommand{\FASTsunflowWCPTimeCIMIN}{\rna}
\newcommand{\FASTsunflowWCPTimeCIMAX}{\rna}
\newcommand{\FASTsunflowFTOWCPTime}{16}
\newcommand{\FASTsunflowFTOWCPTimeCI}{1.6}
\newcommand{\FASTsunflowREWCPTime}{14}
\newcommand{\FASTsunflowREWCPTimeCI}{1.3}
\newcommand{\FASTsunflowDCTime}{\rna}
\newcommand{\FASTsunflowDCTimeCI}{\rna}
\newcommand{\FASTsunflowDCTimeCIMIN}{\rna}
\newcommand{\FASTsunflowDCTimeCIMAX}{\rna}
\newcommand{\FASTsunflowFTODCTime}{15}
\newcommand{\FASTsunflowFTODCTimeCI}{1.4}
\newcommand{\FASTsunflowREDCTime}{15}
\newcommand{\FASTsunflowREDCTimeCI}{1.4}
\newcommand{\FASTsunflowCAPOTime}{\rna}
\newcommand{\FASTsunflowCAPOTimeCI}{\rna}
\newcommand{\FASTsunflowCAPOTimeCIMIN}{\rna}
\newcommand{\FASTsunflowCAPOTimeCIMAX}{\rna}
\newcommand{\FASTsunflowFTOCAPOTime}{15}
\newcommand{\FASTsunflowFTOCAPOTimeCI}{1.8}
\newcommand{\FASTsunflowRECAPOTime}{15}
\newcommand{\FASTsunflowRECAPOTimeCI}{1.4}
\newcommand{\FASTsunflowAGGCAPOTime}{\rna}
\newcommand{\FASTsunflowAGGCAPOTimeCI}{\rna}
\newcommand{\FASTsunflowAGGCAPOTimeCIMIN}{\rna}
\newcommand{\FASTsunflowAGGCAPOTimeCIMAX}{\rna}
\newcommand{\FASTsunflowStaticTime}{\rzero}
\newcommand{\FASTsunflowDynamicTime}{\rzero}
\newcommand{\FASTsunflowBaseMem}{630}
\newcommand{\FASTsunflowBaseMemCI}{2.9}
\newcommand{\FASTsunflowFTMem}{8.4}
\newcommand{\FASTsunflowFTMemCI}{0.056}
\newcommand{\FASTsunflowHBMem}{8.4}
\newcommand{\FASTsunflowHBMemCI}{0.031}
\newcommand{\FASTsunflowFTOHBMem}{8.4}
\newcommand{\FASTsunflowFTOHBMemCI}{0.047}
\newcommand{\FASTsunflowWCPMem}{\memna}
\newcommand{\FASTsunflowWCPMemCI}{\memna}
\newcommand{\FASTsunflowWCPMemCIMIN}{\memna}
\newcommand{\FASTsunflowWCPMemCIMAX}{\memna}
\newcommand{\FASTsunflowFTOWCPMem}{9.0}
\newcommand{\FASTsunflowFTOWCPMemCI}{0.038}
\newcommand{\FASTsunflowREWCPMem}{15}
\newcommand{\FASTsunflowREWCPMemCI}{0.085}
\newcommand{\FASTsunflowDCMem}{\memna}
\newcommand{\FASTsunflowDCMemCI}{\memna}
\newcommand{\FASTsunflowDCMemCIMIN}{\memna}
\newcommand{\FASTsunflowDCMemCIMAX}{\memna}
\newcommand{\FASTsunflowFTODCMem}{9.0}
\newcommand{\FASTsunflowFTODCMemCI}{0.084}
\newcommand{\FASTsunflowREDCMem}{15}
\newcommand{\FASTsunflowREDCMemCI}{0.091}
\newcommand{\FASTsunflowCAPOMem}{\memna}
\newcommand{\FASTsunflowCAPOMemCI}{\memna}
\newcommand{\FASTsunflowCAPOMemCIMIN}{\memna}
\newcommand{\FASTsunflowCAPOMemCIMAX}{\memna}
\newcommand{\FASTsunflowFTOCAPOMem}{9.0}
\newcommand{\FASTsunflowFTOCAPOMemCI}{0.034}
\newcommand{\FASTsunflowRECAPOMem}{15}
\newcommand{\FASTsunflowRECAPOMemCI}{0.044}
\newcommand{\FASTsunflowAGGCAPOMem}{\memna}
\newcommand{\FASTsunflowAGGCAPOMemCI}{\memna}
\newcommand{\FASTsunflowAGGCAPOMemCIMIN}{\memna}
\newcommand{\FASTsunflowAGGCAPOMemCIMAX}{\memna}
\newcommand{\FASTsunflowEventsCI}{0}
\newcommand{\FASTsunflowEventsCIMIN}{0}
\newcommand{\FASTsunflowEventsCIMAX}{0}
\newcommand{\FASTsunflowNoFPEventsCI}{0}
\newcommand{\FASTsunflowNoFPEventsCIMIN}{0}
\newcommand{\FASTsunflowNoFPEventsCIMAX}{0}
\newcommand{\FASTsunflowFT}{5}
\newcommand{\FASTsunflowFTCI}{0.0}
\newcommand{\FASTsunflowFTCIMIN}{5}
\newcommand{\FASTsunflowFTCIMAX}{5}
\newcommand{\FASTsunflowFTDynamic}{125}
\newcommand{\FASTsunflowFTDynamicCI}{10}
\newcommand{\FASTsunflowFTDynamicCIMIN}{115}
\newcommand{\FASTsunflowFTDynamicCIMAX}{135}
\newcommand{\FASTsunflowHB}{6}
\newcommand{\FASTsunflowHBCI}{0.0}
\newcommand{\FASTsunflowHBCIMIN}{6}
\newcommand{\FASTsunflowHBCIMAX}{6}
\newcommand{\FASTsunflowHBDynamic}{56}
\newcommand{\FASTsunflowHBDynamicCI}{5.7}
\newcommand{\FASTsunflowHBDynamicCIMIN}{50}
\newcommand{\FASTsunflowHBDynamicCIMAX}{62}
\newcommand{\FASTsunflowFTOHB}{6}
\newcommand{\FASTsunflowFTOHBCI}{0.0}
\newcommand{\FASTsunflowFTOHBCIMIN}{6}
\newcommand{\FASTsunflowFTOHBCIMAX}{6}
\newcommand{\FASTsunflowFTOHBDynamic}{51}
\newcommand{\FASTsunflowFTOHBDynamicCI}{3.1}
\newcommand{\FASTsunflowFTOHBDynamicCIMIN}{48}
\newcommand{\FASTsunflowFTOHBDynamicCIMAX}{54}
\newcommand{\FASTsunflowWCP}{\rna}
\newcommand{\FASTsunflowWCPCI}{\rna}
\newcommand{\FASTsunflowWCPCIMIN}{\rna}
\newcommand{\FASTsunflowWCPCIMAX}{\rna}
\newcommand{\FASTsunflowWCPDynamic}{\rna}
\newcommand{\FASTsunflowWCPDynamicCI}{\rna}
\newcommand{\FASTsunflowWCPDynamicCIMIN}{\rna}
\newcommand{\FASTsunflowWCPDynamicCIMAX}{\rna}
\newcommand{\FASTsunflowFTOWCP}{18}
\newcommand{\FASTsunflowFTOWCPCI}{0.0}
\newcommand{\FASTsunflowFTOWCPCIMIN}{18}
\newcommand{\FASTsunflowFTOWCPCIMAX}{18}
\newcommand{\FASTsunflowFTOWCPDynamic}{219}
\newcommand{\FASTsunflowFTOWCPDynamicCI}{4.5}
\newcommand{\FASTsunflowFTOWCPDynamicCIMIN}{215}
\newcommand{\FASTsunflowFTOWCPDynamicCIMAX}{223}
\newcommand{\FASTsunflowREWCP}{19}
\newcommand{\FASTsunflowREWCPCI}{0.0}
\newcommand{\FASTsunflowREWCPCIMIN}{19}
\newcommand{\FASTsunflowREWCPCIMAX}{19}
\newcommand{\FASTsunflowREWCPDynamic}{250}
\newcommand{\FASTsunflowREWCPDynamicCI}{6.1}
\newcommand{\FASTsunflowREWCPDynamicCIMIN}{244}
\newcommand{\FASTsunflowREWCPDynamicCIMAX}{256}
\newcommand{\FASTsunflowDC}{\rna}
\newcommand{\FASTsunflowDCCI}{\rna}
\newcommand{\FASTsunflowDCCIMIN}{\rna}
\newcommand{\FASTsunflowDCCIMAX}{\rna}
\newcommand{\FASTsunflowDCDynamic}{\rna}
\newcommand{\FASTsunflowDCDynamicCI}{\rna}
\newcommand{\FASTsunflowDCDynamicCIMIN}{\rna}
\newcommand{\FASTsunflowDCDynamicCIMAX}{\rna}
\newcommand{\FASTsunflowFTODC}{19}
\newcommand{\FASTsunflowFTODCCI}{0.0}
\newcommand{\FASTsunflowFTODCCIMIN}{19}
\newcommand{\FASTsunflowFTODCCIMAX}{19}
\newcommand{\FASTsunflowFTODCDynamic}{406}
\newcommand{\FASTsunflowFTODCDynamicCI}{5.2}
\newcommand{\FASTsunflowFTODCDynamicCIMIN}{401}
\newcommand{\FASTsunflowFTODCDynamicCIMAX}{411}
\newcommand{\FASTsunflowREDC}{19}
\newcommand{\FASTsunflowREDCCI}{0.0}
\newcommand{\FASTsunflowREDCCIMIN}{19}
\newcommand{\FASTsunflowREDCCIMAX}{19}
\newcommand{\FASTsunflowREDCDynamic}{406}
\newcommand{\FASTsunflowREDCDynamicCI}{6.7}
\newcommand{\FASTsunflowREDCDynamicCIMIN}{399}
\newcommand{\FASTsunflowREDCDynamicCIMAX}{413}
\newcommand{\FASTsunflowCAPO}{\rna}
\newcommand{\FASTsunflowCAPOCI}{\rna}
\newcommand{\FASTsunflowCAPOCIMIN}{\rna}
\newcommand{\FASTsunflowCAPOCIMAX}{\rna}
\newcommand{\FASTsunflowCAPODynamic}{\rna}
\newcommand{\FASTsunflowCAPODynamicCI}{\rna}
\newcommand{\FASTsunflowCAPODynamicCIMIN}{\rna}
\newcommand{\FASTsunflowCAPODynamicCIMAX}{\rna}
\newcommand{\FASTsunflowFTOCAPO}{19}
\newcommand{\FASTsunflowFTOCAPOCI}{0.0}
\newcommand{\FASTsunflowFTOCAPOCIMIN}{19}
\newcommand{\FASTsunflowFTOCAPOCIMAX}{19}
\newcommand{\FASTsunflowFTOCAPODynamic}{402}
\newcommand{\FASTsunflowFTOCAPODynamicCI}{13}
\newcommand{\FASTsunflowFTOCAPODynamicCIMIN}{389}
\newcommand{\FASTsunflowFTOCAPODynamicCIMAX}{415}
\newcommand{\FASTsunflowRECAPO}{19}
\newcommand{\FASTsunflowRECAPOCI}{0.0}
\newcommand{\FASTsunflowRECAPOCIMIN}{19}
\newcommand{\FASTsunflowRECAPOCIMAX}{19}
\newcommand{\FASTsunflowRECAPODynamic}{413}
\newcommand{\FASTsunflowRECAPODynamicCI}{5.9}
\newcommand{\FASTsunflowRECAPODynamicCIMIN}{407}
\newcommand{\FASTsunflowRECAPODynamicCIMAX}{419}
\newcommand{\FASTsunflowAGGCAPO}{\rna}
\newcommand{\FASTsunflowAGGCAPOCI}{\rna}
\newcommand{\FASTsunflowAGGCAPOCIMIN}{\rna}
\newcommand{\FASTsunflowAGGCAPOCIMAX}{\rna}
\newcommand{\FASTsunflowAGGCAPODynamic}{\rna}
\newcommand{\FASTsunflowAGGCAPODynamicCI}{\rna}
\newcommand{\FASTsunflowAGGCAPODynamicCIMIN}{\rna}
\newcommand{\FASTsunflowAGGCAPODynamicCIMAX}{\rna}
\newcommand{\FASTtomcatEvents}{0}
\newcommand{\FASTtomcatNoFPEvents}{0}
\newcommand{\FASTtomcatMaxLiveThreads}{37}
\newcommand{\FASTtomcatTotalThreads}{37}
\newcommand{\FASTtomcatBaseTime}{0.89}
\newcommand{\FASTtomcatBaseTimeCI}{3.5}
\newcommand{\FASTtomcatEmptyTime}{\rna}
\newcommand{\FASTtomcatEmptyTimeCI}{\rna}
\newcommand{\FASTtomcatEmptyTimeCIMIN}{\rna}
\newcommand{\FASTtomcatEmptyTimeCIMAX}{\rna}
\newcommand{\FASTtomcatFTTime}{30}
\newcommand{\FASTtomcatFTTimeCI}{2.0}
\newcommand{\FASTtomcatHBTime}{4.2}
\newcommand{\FASTtomcatHBTimeCI}{0.14}
\newcommand{\FASTtomcatFTOHBTime}{4.0}
\newcommand{\FASTtomcatFTOHBTimeCI}{0.16}
\newcommand{\FASTtomcatWCPTime}{\rna}
\newcommand{\FASTtomcatWCPTimeCI}{\rna}
\newcommand{\FASTtomcatWCPTimeCIMIN}{\rna}
\newcommand{\FASTtomcatWCPTimeCIMAX}{\rna}
\newcommand{\FASTtomcatFTOWCPTime}{7.1}
\newcommand{\FASTtomcatFTOWCPTimeCI}{0.35}
\newcommand{\FASTtomcatREWCPTime}{7.3}
\newcommand{\FASTtomcatREWCPTimeCI}{0.35}
\newcommand{\FASTtomcatDCTime}{\rna}
\newcommand{\FASTtomcatDCTimeCI}{\rna}
\newcommand{\FASTtomcatDCTimeCIMIN}{\rna}
\newcommand{\FASTtomcatDCTimeCIMAX}{\rna}
\newcommand{\FASTtomcatFTODCTime}{7.9}
\newcommand{\FASTtomcatFTODCTimeCI}{0.16}
\newcommand{\FASTtomcatREDCTime}{8.5}
\newcommand{\FASTtomcatREDCTimeCI}{0.17}
\newcommand{\FASTtomcatCAPOTime}{\rna}
\newcommand{\FASTtomcatCAPOTimeCI}{\rna}
\newcommand{\FASTtomcatCAPOTimeCIMIN}{\rna}
\newcommand{\FASTtomcatCAPOTimeCIMAX}{\rna}
\newcommand{\FASTtomcatFTOCAPOTime}{4.3}
\newcommand{\FASTtomcatFTOCAPOTimeCI}{0.26}
\newcommand{\FASTtomcatRECAPOTime}{4.4}
\newcommand{\FASTtomcatRECAPOTimeCI}{0.28}
\newcommand{\FASTtomcatAGGCAPOTime}{\rna}
\newcommand{\FASTtomcatAGGCAPOTimeCI}{\rna}
\newcommand{\FASTtomcatAGGCAPOTimeCIMIN}{\rna}
\newcommand{\FASTtomcatAGGCAPOTimeCIMAX}{\rna}
\newcommand{\FASTtomcatStaticTime}{\rzero}
\newcommand{\FASTtomcatDynamicTime}{\rzero}
\newcommand{\FASTtomcatBaseMem}{600}
\newcommand{\FASTtomcatBaseMemCI}{4.2}
\newcommand{\FASTtomcatFTMem}{88}
\newcommand{\FASTtomcatFTMemCI}{10.0}
\newcommand{\FASTtomcatHBMem}{2.7}
\newcommand{\FASTtomcatHBMemCI}{0.20}
\newcommand{\FASTtomcatFTOHBMem}{2.7}
\newcommand{\FASTtomcatFTOHBMemCI}{0.15}
\newcommand{\FASTtomcatWCPMem}{\memna}
\newcommand{\FASTtomcatWCPMemCI}{\memna}
\newcommand{\FASTtomcatWCPMemCIMIN}{\memna}
\newcommand{\FASTtomcatWCPMemCIMAX}{\memna}
\newcommand{\FASTtomcatFTOWCPMem}{6.8}
\newcommand{\FASTtomcatFTOWCPMemCI}{0.13}
\newcommand{\FASTtomcatREWCPMem}{6.0}
\newcommand{\FASTtomcatREWCPMemCI}{0.45}
\newcommand{\FASTtomcatDCMem}{\memna}
\newcommand{\FASTtomcatDCMemCI}{\memna}
\newcommand{\FASTtomcatDCMemCIMIN}{\memna}
\newcommand{\FASTtomcatDCMemCIMAX}{\memna}
\newcommand{\FASTtomcatFTODCMem}{7.6}
\newcommand{\FASTtomcatFTODCMemCI}{0.16}
\newcommand{\FASTtomcatREDCMem}{7.6}
\newcommand{\FASTtomcatREDCMemCI}{0.35}
\newcommand{\FASTtomcatCAPOMem}{\memna}
\newcommand{\FASTtomcatCAPOMemCI}{\memna}
\newcommand{\FASTtomcatCAPOMemCIMIN}{\memna}
\newcommand{\FASTtomcatCAPOMemCIMAX}{\memna}
\newcommand{\FASTtomcatFTOCAPOMem}{3.7}
\newcommand{\FASTtomcatFTOCAPOMemCI}{0.079}
\newcommand{\FASTtomcatRECAPOMem}{3.5}
\newcommand{\FASTtomcatRECAPOMemCI}{0.097}
\newcommand{\FASTtomcatAGGCAPOMem}{\memna}
\newcommand{\FASTtomcatAGGCAPOMemCI}{\memna}
\newcommand{\FASTtomcatAGGCAPOMemCIMIN}{\memna}
\newcommand{\FASTtomcatAGGCAPOMemCIMAX}{\memna}
\newcommand{\FASTtomcatEventsCI}{0}
\newcommand{\FASTtomcatEventsCIMIN}{0}
\newcommand{\FASTtomcatEventsCIMAX}{0}
\newcommand{\FASTtomcatNoFPEventsCI}{0}
\newcommand{\FASTtomcatNoFPEventsCIMIN}{0}
\newcommand{\FASTtomcatNoFPEventsCIMAX}{0}
\newcommand{\FASTtomcatFT}{422}
\newcommand{\FASTtomcatFTCI}{5.1}
\newcommand{\FASTtomcatFTCIMIN}{417}
\newcommand{\FASTtomcatFTCIMAX}{427}
\newcommand{\FASTtomcatFTDynamic}{3,563,840}
\newcommand{\FASTtomcatFTDynamicCI}{166,126}
\newcommand{\FASTtomcatFTDynamicCIMIN}{3,397,714}
\newcommand{\FASTtomcatFTDynamicCIMAX}{3,729,966}
\newcommand{\FASTtomcatHB}{641}
\newcommand{\FASTtomcatHBCI}{4.1}
\newcommand{\FASTtomcatHBCIMIN}{637}
\newcommand{\FASTtomcatHBCIMAX}{645}
\newcommand{\FASTtomcatHBDynamic}{1,563,262}
\newcommand{\FASTtomcatHBDynamicCI}{56,007}
\newcommand{\FASTtomcatHBDynamicCIMIN}{1,507,255}
\newcommand{\FASTtomcatHBDynamicCIMAX}{1,619,269}
\newcommand{\FASTtomcatFTOHB}{601}
\newcommand{\FASTtomcatFTOHBCI}{2.8}
\newcommand{\FASTtomcatFTOHBCIMIN}{598}
\newcommand{\FASTtomcatFTOHBCIMAX}{604}
\newcommand{\FASTtomcatFTOHBDynamic}{1,489,958}
\newcommand{\FASTtomcatFTOHBDynamicCI}{68,667}
\newcommand{\FASTtomcatFTOHBDynamicCIMIN}{1,421,291}
\newcommand{\FASTtomcatFTOHBDynamicCIMAX}{1,558,625}
\newcommand{\FASTtomcatWCP}{\rna}
\newcommand{\FASTtomcatWCPCI}{\rna}
\newcommand{\FASTtomcatWCPCIMIN}{\rna}
\newcommand{\FASTtomcatWCPCIMAX}{\rna}
\newcommand{\FASTtomcatWCPDynamic}{\rna}
\newcommand{\FASTtomcatWCPDynamicCI}{\rna}
\newcommand{\FASTtomcatWCPDynamicCIMIN}{\rna}
\newcommand{\FASTtomcatWCPDynamicCIMAX}{\rna}
\newcommand{\FASTtomcatFTOWCP}{620}
\newcommand{\FASTtomcatFTOWCPCI}{3.5}
\newcommand{\FASTtomcatFTOWCPCIMIN}{616}
\newcommand{\FASTtomcatFTOWCPCIMAX}{624}
\newcommand{\FASTtomcatFTOWCPDynamic}{1,465,152}
\newcommand{\FASTtomcatFTOWCPDynamicCI}{12,392}
\newcommand{\FASTtomcatFTOWCPDynamicCIMIN}{1,452,760}
\newcommand{\FASTtomcatFTOWCPDynamicCIMAX}{1,477,544}
\newcommand{\FASTtomcatREWCP}{612}
\newcommand{\FASTtomcatREWCPCI}{7.2}
\newcommand{\FASTtomcatREWCPCIMIN}{605}
\newcommand{\FASTtomcatREWCPCIMAX}{619}
\newcommand{\FASTtomcatREWCPDynamic}{1,499,621}
\newcommand{\FASTtomcatREWCPDynamicCI}{52,391}
\newcommand{\FASTtomcatREWCPDynamicCIMIN}{1,447,230}
\newcommand{\FASTtomcatREWCPDynamicCIMAX}{1,552,012}
\newcommand{\FASTtomcatDC}{\rna}
\newcommand{\FASTtomcatDCCI}{\rna}
\newcommand{\FASTtomcatDCCIMIN}{\rna}
\newcommand{\FASTtomcatDCCIMAX}{\rna}
\newcommand{\FASTtomcatDCDynamic}{\rna}
\newcommand{\FASTtomcatDCDynamicCI}{\rna}
\newcommand{\FASTtomcatDCDynamicCIMIN}{\rna}
\newcommand{\FASTtomcatDCDynamicCIMAX}{\rna}
\newcommand{\FASTtomcatFTODC}{623}
\newcommand{\FASTtomcatFTODCCI}{2.9}
\newcommand{\FASTtomcatFTODCCIMIN}{620}
\newcommand{\FASTtomcatFTODCCIMAX}{626}
\newcommand{\FASTtomcatFTODCDynamic}{1,438,653}
\newcommand{\FASTtomcatFTODCDynamicCI}{18,141}
\newcommand{\FASTtomcatFTODCDynamicCIMIN}{1,420,512}
\newcommand{\FASTtomcatFTODCDynamicCIMAX}{1,456,794}
\newcommand{\FASTtomcatREDC}{613}
\newcommand{\FASTtomcatREDCCI}{10}
\newcommand{\FASTtomcatREDCCIMIN}{603}
\newcommand{\FASTtomcatREDCCIMAX}{623}
\newcommand{\FASTtomcatREDCDynamic}{1,493,252}
\newcommand{\FASTtomcatREDCDynamicCI}{86,333}
\newcommand{\FASTtomcatREDCDynamicCIMIN}{1,406,919}
\newcommand{\FASTtomcatREDCDynamicCIMAX}{1,579,585}
\newcommand{\FASTtomcatCAPO}{\rna}
\newcommand{\FASTtomcatCAPOCI}{\rna}
\newcommand{\FASTtomcatCAPOCIMIN}{\rna}
\newcommand{\FASTtomcatCAPOCIMAX}{\rna}
\newcommand{\FASTtomcatCAPODynamic}{\rna}
\newcommand{\FASTtomcatCAPODynamicCI}{\rna}
\newcommand{\FASTtomcatCAPODynamicCIMIN}{\rna}
\newcommand{\FASTtomcatCAPODynamicCIMAX}{\rna}
\newcommand{\FASTtomcatFTOCAPO}{622}
\newcommand{\FASTtomcatFTOCAPOCI}{4.4}
\newcommand{\FASTtomcatFTOCAPOCIMIN}{618}
\newcommand{\FASTtomcatFTOCAPOCIMAX}{626}
\newcommand{\FASTtomcatFTOCAPODynamic}{1,450,638}
\newcommand{\FASTtomcatFTOCAPODynamicCI}{61,419}
\newcommand{\FASTtomcatFTOCAPODynamicCIMIN}{1,389,219}
\newcommand{\FASTtomcatFTOCAPODynamicCIMAX}{1,512,057}
\newcommand{\FASTtomcatRECAPO}{608}
\newcommand{\FASTtomcatRECAPOCI}{2.7}
\newcommand{\FASTtomcatRECAPOCIMIN}{605}
\newcommand{\FASTtomcatRECAPOCIMAX}{611}
\newcommand{\FASTtomcatRECAPODynamic}{1,485,273}
\newcommand{\FASTtomcatRECAPODynamicCI}{43,529}
\newcommand{\FASTtomcatRECAPODynamicCIMIN}{1,441,744}
\newcommand{\FASTtomcatRECAPODynamicCIMAX}{1,528,802}
\newcommand{\FASTtomcatAGGCAPO}{\rna}
\newcommand{\FASTtomcatAGGCAPOCI}{\rna}
\newcommand{\FASTtomcatAGGCAPOCIMIN}{\rna}
\newcommand{\FASTtomcatAGGCAPOCIMAX}{\rna}
\newcommand{\FASTtomcatAGGCAPODynamic}{\rna}
\newcommand{\FASTtomcatAGGCAPODynamicCI}{\rna}
\newcommand{\FASTtomcatAGGCAPODynamicCIMIN}{\rna}
\newcommand{\FASTtomcatAGGCAPODynamicCIMAX}{\rna}
\newcommand{\FASTxalanEvents}{0}
\newcommand{\FASTxalanNoFPEvents}{0}
\newcommand{\FASTxalanMaxLiveThreads}{14}
\newcommand{\FASTxalanTotalThreads}{14}
\newcommand{\FASTxalanBaseTime}{2.1}
\newcommand{\FASTxalanBaseTimeCI}{130}
\newcommand{\FASTxalanEmptyTime}{\rna}
\newcommand{\FASTxalanEmptyTimeCI}{\rna}
\newcommand{\FASTxalanEmptyTimeCIMIN}{\rna}
\newcommand{\FASTxalanEmptyTimeCIMAX}{\rna}
\newcommand{\FASTxalanFTTime}{4.0}
\newcommand{\FASTxalanFTTimeCI}{0.40}
\newcommand{\FASTxalanHBTime}{4.0}
\newcommand{\FASTxalanHBTimeCI}{0.33}
\newcommand{\FASTxalanFTOHBTime}{4.0}
\newcommand{\FASTxalanFTOHBTimeCI}{0.34}
\newcommand{\FASTxalanWCPTime}{\rna}
\newcommand{\FASTxalanWCPTimeCI}{\rna}
\newcommand{\FASTxalanWCPTimeCIMIN}{\rna}
\newcommand{\FASTxalanWCPTimeCIMAX}{\rna}
\newcommand{\FASTxalanFTOWCPTime}{30}
\newcommand{\FASTxalanFTOWCPTimeCI}{2.5}
\newcommand{\FASTxalanREWCPTime}{8.2}
\newcommand{\FASTxalanREWCPTimeCI}{0.52}
\newcommand{\FASTxalanDCTime}{\rna}
\newcommand{\FASTxalanDCTimeCI}{\rna}
\newcommand{\FASTxalanDCTimeCIMIN}{\rna}
\newcommand{\FASTxalanDCTimeCIMAX}{\rna}
\newcommand{\FASTxalanFTODCTime}{31}
\newcommand{\FASTxalanFTODCTimeCI}{1.9}
\newcommand{\FASTxalanREDCTime}{9.5}
\newcommand{\FASTxalanREDCTimeCI}{0.60}
\newcommand{\FASTxalanCAPOTime}{\rna}
\newcommand{\FASTxalanCAPOTimeCI}{\rna}
\newcommand{\FASTxalanCAPOTimeCIMIN}{\rna}
\newcommand{\FASTxalanCAPOTimeCIMAX}{\rna}
\newcommand{\FASTxalanFTOCAPOTime}{28}
\newcommand{\FASTxalanFTOCAPOTimeCI}{2.4}
\newcommand{\FASTxalanRECAPOTime}{4.8}
\newcommand{\FASTxalanRECAPOTimeCI}{0.35}
\newcommand{\FASTxalanAGGCAPOTime}{\rna}
\newcommand{\FASTxalanAGGCAPOTimeCI}{\rna}
\newcommand{\FASTxalanAGGCAPOTimeCIMIN}{\rna}
\newcommand{\FASTxalanAGGCAPOTimeCIMAX}{\rna}
\newcommand{\FASTxalanStaticTime}{\rzero}
\newcommand{\FASTxalanDynamicTime}{\rzero}
\newcommand{\FASTxalanBaseMem}{700}
\newcommand{\FASTxalanBaseMemCI}{8.6}
\newcommand{\FASTxalanFTMem}{6.3}
\newcommand{\FASTxalanFTMemCI}{0.098}
\newcommand{\FASTxalanHBMem}{6.5}
\newcommand{\FASTxalanHBMemCI}{0.093}
\newcommand{\FASTxalanFTOHBMem}{6.5}
\newcommand{\FASTxalanFTOHBMemCI}{0.087}
\newcommand{\FASTxalanWCPMem}{\memna}
\newcommand{\FASTxalanWCPMemCI}{\memna}
\newcommand{\FASTxalanWCPMemCIMIN}{\memna}
\newcommand{\FASTxalanWCPMemCIMAX}{\memna}
\newcommand{\FASTxalanFTOWCPMem}{36}
\newcommand{\FASTxalanFTOWCPMemCI}{1.6}
\newcommand{\FASTxalanREWCPMem}{14}
\newcommand{\FASTxalanREWCPMemCI}{0.36}
\newcommand{\FASTxalanDCMem}{\memna}
\newcommand{\FASTxalanDCMemCI}{\memna}
\newcommand{\FASTxalanDCMemCIMIN}{\memna}
\newcommand{\FASTxalanDCMemCIMAX}{\memna}
\newcommand{\FASTxalanFTODCMem}{33}
\newcommand{\FASTxalanFTODCMemCI}{1.2}
\newcommand{\FASTxalanREDCMem}{14}
\newcommand{\FASTxalanREDCMemCI}{0.17}
\newcommand{\FASTxalanCAPOMem}{\memna}
\newcommand{\FASTxalanCAPOMemCI}{\memna}
\newcommand{\FASTxalanCAPOMemCIMIN}{\memna}
\newcommand{\FASTxalanCAPOMemCIMAX}{\memna}
\newcommand{\FASTxalanFTOCAPOMem}{33}
\newcommand{\FASTxalanFTOCAPOMemCI}{1.5}
\newcommand{\FASTxalanRECAPOMem}{9.2}
\newcommand{\FASTxalanRECAPOMemCI}{0.21}
\newcommand{\FASTxalanAGGCAPOMem}{\memna}
\newcommand{\FASTxalanAGGCAPOMemCI}{\memna}
\newcommand{\FASTxalanAGGCAPOMemCIMIN}{\memna}
\newcommand{\FASTxalanAGGCAPOMemCIMAX}{\memna}
\newcommand{\FASTxalanEventsCI}{0}
\newcommand{\FASTxalanEventsCIMIN}{0}
\newcommand{\FASTxalanEventsCIMAX}{0}
\newcommand{\FASTxalanNoFPEventsCI}{0}
\newcommand{\FASTxalanNoFPEventsCIMIN}{0}
\newcommand{\FASTxalanNoFPEventsCIMAX}{0}
\newcommand{\FASTxalanFT}{8}
\newcommand{\FASTxalanFTCI}{0.0}
\newcommand{\FASTxalanFTCIMIN}{8}
\newcommand{\FASTxalanFTCIMAX}{8}
\newcommand{\FASTxalanFTDynamic}{2,816}
\newcommand{\FASTxalanFTDynamicCI}{44}
\newcommand{\FASTxalanFTDynamicCIMIN}{2,772}
\newcommand{\FASTxalanFTDynamicCIMAX}{2,860}
\newcommand{\FASTxalanHB}{8}
\newcommand{\FASTxalanHBCI}{0.20}
\newcommand{\FASTxalanHBCIMIN}{8}
\newcommand{\FASTxalanHBCIMAX}{8}
\newcommand{\FASTxalanHBDynamic}{2,606}
\newcommand{\FASTxalanHBDynamicCI}{0.0}
\newcommand{\FASTxalanHBDynamicCIMIN}{2,606}
\newcommand{\FASTxalanHBDynamicCIMAX}{2,606}
\newcommand{\FASTxalanFTOHB}{8}
\newcommand{\FASTxalanFTOHBCI}{0.26}
\newcommand{\FASTxalanFTOHBCIMIN}{8}
\newcommand{\FASTxalanFTOHBCIMAX}{8}
\newcommand{\FASTxalanFTOHBDynamic}{2,584}
\newcommand{\FASTxalanFTOHBDynamicCI}{38}
\newcommand{\FASTxalanFTOHBDynamicCIMIN}{2,546}
\newcommand{\FASTxalanFTOHBDynamicCIMAX}{2,622}
\newcommand{\FASTxalanWCP}{\rna}
\newcommand{\FASTxalanWCPCI}{\rna}
\newcommand{\FASTxalanWCPCIMIN}{\rna}
\newcommand{\FASTxalanWCPCIMAX}{\rna}
\newcommand{\FASTxalanWCPDynamic}{\rna}
\newcommand{\FASTxalanWCPDynamicCI}{\rna}
\newcommand{\FASTxalanWCPDynamicCIMIN}{\rna}
\newcommand{\FASTxalanWCPDynamicCIMAX}{\rna}
\newcommand{\FASTxalanFTOWCP}{43}
\newcommand{\FASTxalanFTOWCPCI}{0.90}
\newcommand{\FASTxalanFTOWCPCIMIN}{42}
\newcommand{\FASTxalanFTOWCPCIMAX}{44}
\newcommand{\FASTxalanFTOWCPDynamic}{3,589,223}
\newcommand{\FASTxalanFTOWCPDynamicCI}{16,250}
\newcommand{\FASTxalanFTOWCPDynamicCIMIN}{3,572,973}
\newcommand{\FASTxalanFTOWCPDynamicCIMAX}{3,605,473}
\newcommand{\FASTxalanREWCP}{50}
\newcommand{\FASTxalanREWCPCI}{0.26}
\newcommand{\FASTxalanREWCPCIMIN}{50}
\newcommand{\FASTxalanREWCPCIMAX}{50}
\newcommand{\FASTxalanREWCPDynamic}{3,578,681}
\newcommand{\FASTxalanREWCPDynamicCI}{26,803}
\newcommand{\FASTxalanREWCPDynamicCIMIN}{3,551,878}
\newcommand{\FASTxalanREWCPDynamicCIMAX}{3,605,484}
\newcommand{\FASTxalanDC}{\rna}
\newcommand{\FASTxalanDCCI}{\rna}
\newcommand{\FASTxalanDCCIMIN}{\rna}
\newcommand{\FASTxalanDCCIMAX}{\rna}
\newcommand{\FASTxalanDCDynamic}{\rna}
\newcommand{\FASTxalanDCDynamicCI}{\rna}
\newcommand{\FASTxalanDCDynamicCIMIN}{\rna}
\newcommand{\FASTxalanDCDynamicCIMAX}{\rna}
\newcommand{\FASTxalanFTODC}{53}
\newcommand{\FASTxalanFTODCCI}{0.26}
\newcommand{\FASTxalanFTODCCIMIN}{53}
\newcommand{\FASTxalanFTODCCIMAX}{53}
\newcommand{\FASTxalanFTODCDynamic}{3,965,177}
\newcommand{\FASTxalanFTODCDynamicCI}{12,932}
\newcommand{\FASTxalanFTODCDynamicCIMIN}{3,952,245}
\newcommand{\FASTxalanFTODCDynamicCIMAX}{3,978,109}
\newcommand{\FASTxalanREDC}{52}
\newcommand{\FASTxalanREDCCI}{1.1}
\newcommand{\FASTxalanREDCCIMIN}{51}
\newcommand{\FASTxalanREDCCIMAX}{53}
\newcommand{\FASTxalanREDCDynamic}{3,984,338}
\newcommand{\FASTxalanREDCDynamicCI}{18,372}
\newcommand{\FASTxalanREDCDynamicCIMIN}{3,965,966}
\newcommand{\FASTxalanREDCDynamicCIMAX}{4,002,710}
\newcommand{\FASTxalanCAPO}{\rna}
\newcommand{\FASTxalanCAPOCI}{\rna}
\newcommand{\FASTxalanCAPOCIMIN}{\rna}
\newcommand{\FASTxalanCAPOCIMAX}{\rna}
\newcommand{\FASTxalanCAPODynamic}{\rna}
\newcommand{\FASTxalanCAPODynamicCI}{\rna}
\newcommand{\FASTxalanCAPODynamicCIMIN}{\rna}
\newcommand{\FASTxalanCAPODynamicCIMAX}{\rna}
\newcommand{\FASTxalanFTOCAPO}{53}
\newcommand{\FASTxalanFTOCAPOCI}{0.46}
\newcommand{\FASTxalanFTOCAPOCIMIN}{53}
\newcommand{\FASTxalanFTOCAPOCIMAX}{53}
\newcommand{\FASTxalanFTOCAPODynamic}{3,892,399}
\newcommand{\FASTxalanFTOCAPODynamicCI}{17,671}
\newcommand{\FASTxalanFTOCAPODynamicCIMIN}{3,874,728}
\newcommand{\FASTxalanFTOCAPODynamicCIMAX}{3,910,070}
\newcommand{\FASTxalanRECAPO}{53}
\newcommand{\FASTxalanRECAPOCI}{1.2}
\newcommand{\FASTxalanRECAPOCIMIN}{52}
\newcommand{\FASTxalanRECAPOCIMAX}{54}
\newcommand{\FASTxalanRECAPODynamic}{3,987,647}
\newcommand{\FASTxalanRECAPODynamicCI}{17,233}
\newcommand{\FASTxalanRECAPODynamicCIMIN}{3,970,414}
\newcommand{\FASTxalanRECAPODynamicCIMAX}{4,004,880}
\newcommand{\FASTxalanAGGCAPO}{\rna}
\newcommand{\FASTxalanAGGCAPOCI}{\rna}
\newcommand{\FASTxalanAGGCAPOCIMIN}{\rna}
\newcommand{\FASTxalanAGGCAPOCIMAX}{\rna}
\newcommand{\FASTxalanAGGCAPODynamic}{\rna}
\newcommand{\FASTxalanAGGCAPODynamicCI}{\rna}
\newcommand{\FASTxalanAGGCAPODynamicCIMIN}{\rna}
\newcommand{\FASTxalanAGGCAPODynamicCIMAX}{\rna}
\newcommand{\FASTBaseTimeGeoMean}{1900}
\newcommand{\FASTEmptyTimeGeoMean}{\rna}
\newcommand{\FASTFTTimeGeoMean}{8.4}
\newcommand{\FASTHBTimeGeoMean}{6.4}
\newcommand{\FASTFTOHBTimeGeoMean}{6.3}
\newcommand{\FASTWCPTimeGeoMean}{\rna}
\newcommand{\FASTFTOWCPTimeGeoMean}{13}
\newcommand{\FASTREWCPTimeGeoMean}{8.3}
\newcommand{\FASTDCTimeGeoMean}{\rna}
\newcommand{\FASTFTODCTimeGeoMean}{13}
\newcommand{\FASTREDCTimeGeoMean}{8.6}
\newcommand{\FASTCAPOTimeGeoMean}{\rna}
\newcommand{\FASTFTOCAPOTimeGeoMean}{12}
\newcommand{\FASTRECAPOTimeGeoMean}{6.9}
\newcommand{\FASTAGGCAPOTimeGeoMean}{\rna}
\newcommand{\FASTBaseMemGeoMean}{550}
\newcommand{\FASTEmptyMemGeoMean}{0.0}
\newcommand{\FASTFTMemGeoMean}{8.1}
\newcommand{\FASTHBMemGeoMean}{4.9}
\newcommand{\FASTFTOHBMemGeoMean}{4.9}
\newcommand{\FASTWCPMemGeoMean}{\memna}
\newcommand{\FASTFTOWCPMemGeoMean}{13}
\newcommand{\FASTREWCPMemGeoMean}{7.5}
\newcommand{\FASTDCMemGeoMean}{\memna}
\newcommand{\FASTFTODCMemGeoMean}{12}
\newcommand{\FASTREDCMemGeoMean}{7.6}
\newcommand{\FASTCAPOMemGeoMean}{\memna}
\newcommand{\FASTFTOCAPOMemGeoMean}{11}
\newcommand{\FASTRECAPOMemGeoMean}{6.2}
\newcommand{\FASTAGGCAPOMemGeoMean}{\memna}
\newcommand{\FASTWCPDynamicTotal}{0}
\newcommand{\FASTFTODCTotal}{760}
\newcommand{\FASTCAPOTotal}{0}
\newcommand{\FASTWCPTotal}{0}
\newcommand{\FASTAGGCAPODynamicTotal}{0}
\newcommand{\FASTRECAPOTotal}{745}
\newcommand{\FASTFTTotal}{491}
\newcommand{\FASTHBDynamicTotal}{2,063,746}
\newcommand{\FASTREDCTotal}{754}
\newcommand{\FASTREWCPTotal}{744}
\newcommand{\FASTFTOWCPDynamicTotal}{5,547,275}
\newcommand{\FASTRECAPODynamicTotal}{5,936,143}
\newcommand{\FASTFTOWCPTotal}{738}
\newcommand{\FASTAGGCAPOTotal}{0}
\newcommand{\FASTCAPODynamicTotal}{0}
\newcommand{\FASTFTOCAPOTotal}{761}
\newcommand{\FASTFTODCDynamicTotal}{5,898,659}
\newcommand{\FASTFTDynamicTotal}{4,432,361}
\newcommand{\FASTREWCPDynamicTotal}{5,552,887}
\newcommand{\FASTREDCDynamicTotal}{5,959,325}
\newcommand{\FASTFTOCAPODynamicTotal}{5,840,174}
\newcommand{\FASTFTOHBDynamicTotal}{1,944,789}
\newcommand{\FASTHBTotal}{717}
\newcommand{\FASTFTOHBTotal}{677}
\newcommand{\FASTDCDynamicTotal}{0}
\newcommand{\FASTDCTotal}{0}

%% file: background.tex
\section{Background and Motivation}
\label{sec:background}

This section describes non-predictive and predictive analyses that detect data races
and explains their limitations.
Some notation and terminology follow prior work's~\cite{wcp, vindicator}.

\subsection{Execution Traces and Other Preliminaries}

An execution trace \tr is a totally ordered list of events, denoted by \TROrdered{}{},
that represents a linearization of events in a multithreaded
execution.\footnote{Data-race-free programs have sequential consistency (SC) semantics
under the Java and C++ memory models~\cite{java-memory-model,c++-memory-model-2008}.
An execution of a program with a data race may have non-SC behavior~\cite{memory-models-cacm-2010,dolan-bounding-races},
but instrumentation added by dynamic race detection analysis typically ensures SC for every execution.}
Each event consists of a thread identifier (\eg, \thr{T1} or \thr{T2}) and an
operation with the form \Write{x}{}, \Read{x}{}, \Acquire{m}{}, or \Release{m}{},
where \code{x} is a variable and \code{m} is a lock.
(Other synchronization events, such as Java \code{volatile} and C++ \code{atomic} accesses and thread fork/join,
are straightforward for our analysis implementations to handle; Section~\ref{subsec:impl}.)
Throughout the paper, we often denote events simply by their operation (\eg, \Write{x}{} or \Acquire{m}{}).

An execution trace must be \emph{well formed}:
a thread only acquires an un-held lock and only releases a lock it holds.

Figure~\ref{fig:example:simple-predictive-race:original} shows an example execution trace,
in which (as for all example traces in the paper)
top-to-bottom ordering denotes observed execution order \ltTR, and
column placement denotes which thread executes each event.

For convenience, we define \emph{\poFull (\PO)}, a strict partial order over events in the same thread:




\begin{definition*}[\PoFull]

Given a trace \tr, \POOrdered{}{} is the smallest relation such that, for
two events \event{e}{} and \event{e}{'}, 
\POOrdered{\event{e}{}}{\event{e}{'}} if both
$\TROrdered{\event{e}{}}{\event{e}{'}}$
and \event{e}{} and \event{e}{'} are executed by the same thread.

\end{definition*}


\noindent
Throughout the paper, ordering notation such as \POOrdered{\event{e}{}}{\event{e}{'}} that omits \emph{which trace} the ordering applies to,
generally refers to ordering
\emph{in the observed execution trace \tr} (not some trace \trPrime predicted from \tr---a concept explained next).

\subsection{\PredTrace{}s and Predictable Races}


A trace \trPrime is a \emph{\predtrace} of \tr if
\trPrime is a feasible execution derived from the existence of \tr.
In a \predtrace \trPrime,
every event is also present in \tr (but not every event in \tr is present in \trPrime in general);
event order preserves \tr's \PO ordering;
every read in \trPrime has the same last writer (or lack of a preceding writer) as in \tr;
and \trPrime is well formed (\ie, obeys locking rules).\footnote{
Prior work provides formal definitions of \predtrace{}s~\cite{wcp,vindicator,rvpredict-pldi-2014}.}

The execution in Figure~\ref{fig:example:simple-predictive-race:reordered} is a \predtrace of
the execution in Figure~\ref{fig:example:simple-predictive-race:original}:
its events are a subset of the observed trace's events,
it preserves the original trace's \PO and last-writer ordering,
and it is well formed.

\begin{figure}
  \renewcommand{\arraystretch}{\smalltablerowheight}
  \small
  \centering
  \iftoggle{twoColumnText}{\hfill}{}
  \subfloat[An execution trace with a predictable race]{
  \begin{minipage}{\iftoggle{twoColumnText}{0.3}{0.5}\linewidth}
  \centering 
  \sf
  \begin{tabular}{@{}ll@{}} 
  \thr{Thread 1} & \textnormal{Thread 2} \\\hline
  \Read{x} \\
  \Acquire{m} \\
  \Write{y} \\
  \Release{m}\tikzmark{1} \\
  & \tikzmark{2}\Acquire{m} \\
  & \Read{z} \\
  & \Release{m} \\
  & \Write{x}
  \end{tabular}%
  \textlink{1}{2}{\HB}
  \end{minipage}
  \label{fig:example:simple-predictive-race:original}
  }
  \iftoggle{twoColumnText}{\hfill\hfill}{}
  \subfloat[A \predtrace of (a) exposing the race]{
  \begin{minipage}{\iftoggle{twoColumnText}{0.3}{0.5}\linewidth}
  \centering 
  \sf
  \begin{tabular}{@{}ll@{}} 
  \textnormal{Thread 1} & \textnormal{Thread 2} \\\hline
  & \Acquire{m} \\
  & \Read{z} \\
  & \Release{m} \\
  \Read{x} \\
  & \Write{x} \\ \\ \\ \\
  \end{tabular}%
  \end{minipage}
  \label{fig:example:simple-predictive-race:reordered}
  }
  \iftoggle{twoColumnText}{\hfill\hfill}{} 
  \caption{The execution in (a) has no \HB-race
  ($\protect\HBOrdered{\Read{x}}{\Write{x}}$),
  but it has a predictable race,
  as the \predtrace in (b) demonstrates.}
  \label{fig:example:simple-predictive-race}
\end{figure}

An execution trace \tr has a \emph{predictable race} if
some \predtrace of \tr, \trPrime, contains
conflicting events
that are consecutive (no intervening event).
Events \event{e}{} and \event{e}{'} are \emph{conflicting},
denoted \conflicts{\event{e}{}}{\event{e}{'}},
if they are accesses to the same variable by different threads,
and at least one is a write.

By definition,
Figure~\ref{fig:example:simple-predictive-race:original} has a predictable race (involving accesses to \code{x})
as demonstrated by Figure~\ref{fig:example:simple-predictive-race:reordered}.
Intuitively,
it is \emph{knowable from the observed execution alone} that the conflicting accesses
\Read{x} and \Write{x} can execute simultaneously in \emph{another} execution.

Note that if we replaced \Read{z} with \Read{y} in Figure~\ref{fig:example:simple-predictive-race:original},
the execution would \emph{not} have a predictable race. The insight is that executing \Read{y} \emph{before} \Write{y}
might see a different value, which could alter control flow to \emph{not} execute \Write{x}.

%
%

A race detection analysis is \emph{sound} if every reported race is a (true) predictable race.\footnote{This definition of soundness
follows the predictive race detection literature (\eg,~\cite{causally-precedes,wcp,rvpredict-pldi-2014,vindicator}).}
Soundness is an important property because each reported data race, whether true or false,
takes hours or days to investigate~\cite{microsoft-exploratory-survey,conc-bug-study-2008,
pacer-2010,literace,benign-races-2007,fasttrack,billion-lines-later}.
%
%
%

\subsection{\HBFull Analysis}

\emph{\HbFull (\HB)}~\cite{happens-before} is a strict partial order that orders events by \PO and synchronization order:

\input{HB-definition}

\noindent
\emph{\HB analysis} is a dynamic analysis that computes \HB over an executing program and detects \HB-races.

An execution trace has an \emph{\HB-race} if it has two \conflictingEvents
unordered by \HB.
\HB analysis is sound: An \HB-race indicates a predictable race~\cite{happens-before}.

Classical \HB analysis uses vector clocks~\cite{vector-clocks} to record variables' last-access times.
\emph{FastTrack} and follow-up work perform optimized, state-of-the-art \HB analysis,
using a lightweight representation of read and write metadata~\cite{fasttrack,fasttrack2,fib}.
FastTrack's optimizations result in an average \factor{3} speedup over vector-clock-based \HB analysis
(Section~\ref{sec:eval:fasttrack}).
FastTrack's optimized \HB analysis is widely used in data race detectors including
Google's ThreadSanitizer~\cite{google-tsan-v1,google-tsan-v2} and Intel Inspector~\cite{intel-inspector}.

Optimzed \HB analysis achieves performance acceptable for regular in-house testing---roughly \factor{6--8} slowdown
according to prior work~\cite{fasttrack,fasttrack2} and our evaluation---but it misses predictable races.
Consider Figure~\ref{fig:example:simple-predictive-race:original}: the observed execution has no \HB-race,
despite having a predictable race.

\subsection{Predictive Analyses}
\label{sec:background:predictive-analyses}

A \emph{predictive analysis} is a dynamic analysis that detects predictable races
in an observed trace, including races that are not \HB-races.
(This definition distinguishes \HB analysis from predictive analyses.)


\notes{
Some existing approaches detect predictable races by encoding the conditions for a predictable race as SMT constraints~\cite{rvpredict-pldi-2014,said-nfm-2011,ipa,rdit-oopsla-2016,jpredictor,maximal-causal-models}.
These approaches cannot scale to full executions
and instead analyze bounded windows of execution, missing predictable races
(Section~\ref{Sec:related}).

In contrast, predictive analyses that compute partial orders can scale to full executions.
\mike{Above is already mentioned in Intro.}}%
Recent work introduces two strict partial orders weaker than \HB, \emph{\wcpFull (\WCP)} and \emph{\dcFull (\DC)},
and corresponding
analyses~\cite{wcp,vindicator}.
For simplicity of exposition, the paper generally shows details only for \DC analysis,
which is reasonable because
\WCP analysis is inefficient for the same reasons as \DC analysis, and
our optimizations to \DC analysis apply directly to \WCP analysis.

\DC is a strict partial order with the following definition:




\input{DC-definition}

\noindent
\WCP
differs from \DC in one way: it composes with \HB instead of \PO,
by replacing \DC rules~(c) and (d) with a rule that \WCP left- and right-composes with \HB~\cite{wcp}.
That is, \WCPOrdered{\event{e}{}}{\event{e}{'}} if 
$\exists \event{e}{''} \mid \HBOrdered{\event{e}{}}{\WCPOrdered{\event{e}{''}}{\event{e}{'}}} \lor
\WCPOrdered{\event{e}{}}{\HBOrdered{\event{e}{''}}{\event{e}{'}}}$.

An execution has a \emph{\WCP-race} or \emph{\DC-race} if it has two
conflicting accesses unordered by \ltWCP or \ltDC, respectively.
The execution from Figure~\ref{fig:example:simple-predictive-race:original}
has a \WCP-race and a \DC-race:
\WCP and \DC do not order the critical sections on lock \code{m}
because the critical sections do not contain conflicting accesses,
resulting in
\nWCPOrdered{\Read{x}}{\Write{x}} and \nDCOrdered{\Read{x}}{\Write{x}}.
Figure~\ref{fig:example:dc:original}, on the other hand,
has a \DC-race but no \WCP-race
(since \WCP composes with \HB).

\iftoggle{twoColumnText}{
\begin{figure}
\renewcommand{\arraystretch}{\smalltablerowheight}
\small
\centering
\hfill
\subfloat[An execution trace with a predictable race]{
\begin{minipage}{0.45\linewidth}
\centering
\sf
\begin{tabular}{@{}lll@{}} 
\textnormal{Thread 1} & \textnormal{Thread 2} & \textnormal{Thread 3} \\\hline
\Read{x} \\
\Acquire{m} \\
\Write{y} \\
\Release{m}\tikzmark{7}\tikzmark{1} \\
& \Acquire{m} \\
& \tikzmark{2}\Read{y} \\
& \Release{m} \\
& \Acquire{n} \\
& \Release{n}\tikzmark{5} \\
& & \tikzmark{8}\tikzmark{6}\Acquire{n} \\
& & \Release{n} \\
& & \Write{x}
\end{tabular}%
\undertextlink{1}{2}{\WCP}
\textlink{1}{2}{\DC}
\textlink{5}{6}{\HB}
\undertextcurve{7}{8}{250}{200}{\WCP}
\end{minipage}
\label{fig:example:dc:original}
}
\hfill\hfill
\subfloat[A \predtrace of (a) exposing the race]{
\begin{minipage}{0.35\linewidth}
\centering
\sf
\begin{tabular}{@{}lll@{}} 
\textnormal{Thread 1} & & \textnormal{Thread 3} \\\hline
& & \tikzmark{6}\Acquire{n} \\
& & \Release{n} \\
\Read{x} \\
& & \Write{x} \\ \\ \\ \\ \\ \\ \\ \\ \\ \\
\end{tabular}%
\end{minipage}
\label{fig:example:dc:reordered}
}
\hfill\ 
\caption{The execution in (a) has a predictable race and a \DC-race
($\protect\nDCOrdered{\ReadT{x}{T1}}{\WriteT{x}{T3}}$), 
but no \WCP-race
($\protect\WCPOrdered{\ReadT{x}{T1}}{\WriteT{x}{T3}}$).
Arrows show cross-thread ordering as labeled.}
\label{fig:example:dc}
\end{figure}
}{
\begin{figure}
\renewcommand{\arraystretch}{\smalltablerowheight}
\small
\centering 
\ \hfill
\subfloat[An execution trace with a predictable race]{
\hspace{1cm}
\centering
\sf
\begin{tabular}{@{}lll@{}} 
\textnormal{Thread 1} & \textnormal{Thread 2} & \textnormal{Thread 3} \\\hline
\Read{x} \\
\Acquire{m} \\
\Write{y} \\
\Release{m}\tikzmark{7}\tikzmark{1} \\
& \Acquire{m} \\
& \tikzmark{2}\Read{y} \\
& \Release{m} \\
& \Acquire{n} \\
& \Release{n}\tikzmark{5} \\
& & \tikzmark{8}\tikzmark{6}\Acquire{n} \\
& & \Release{n} \\
& & \Write{x}
\end{tabular}%
\undertextlink{1}{2}{\WCP}
\textlink{1}{2}{\DC}
\textlink{5}{6}{\HB}
\undertextcurve{7}{8}{250}{200}{\WCP}
\hspace{1cm}
\label{fig:example:dc:original}
}
\hfill\hfill
\subfloat[A \predtrace of (a) exposing the race]{
\hspace{1.5cm}
\centering
\sf
\begin{tabular}{@{}lll@{}} 
\textnormal{Thread 1} & & \textnormal{Thread 3} \\\hline
& & \tikzmark{6}\Acquire{n} \\
& & \Release{n} \\
\Read{x} \\
& & \Write{x} \\ \\ \\ \\ \\ \\ \\ \\ \\ \\
\end{tabular}%
\hspace{1.5cm}
\label{fig:example:dc:reordered}
}
\hfill\ 
\caption{The execution in (a) has a predictable race and a \DC-race
($\protect\nDCOrdered{\ReadT{x}{T1}}{\WriteT{x}{T3}}$), 
but no \WCP-race
($\protect\WCPOrdered{\ReadT{x}{T1}}{\WriteT{x}{T3}}$).
Arrows show cross-thread ordering as labeled.}
\label{fig:example:dc}
\end{figure}
}

\emph{\WCP analysis} and \emph{\DC analysis}
compute \WCP and \DC for an execution and detect \WCP- and \DC-races, respectively.
\WCP analysis is sound: every \WCP-race indicates a predictable race~\cite{wcp}.\footnote{Technically,
an execution with a \WCP-race has a predictable race or a predictable deadlock~\cite{wcp}.}
\DC, which is strictly weaker than \WCP,\footnote{\WCP in turn
is strictly weaker than prior work's \emph{\cpFull} (\CP) relation~\cite{causally-precedes,raptor,dighr} and thus predicts more races than \CP.}
is unsound:
it may report a race when no predictable race (or deadlock) exists.
However, \DC analysis reports few if any false races in practice;
furthermore, a \emph{vindication} algorithm can rule out false races, providing soundness overall~\cite{vindicator}.

\paragraph{\DC analysis details.}

Algorithm~\ref{alg:DC-VC} shows the details of an algorithm for \DC analysis based closely on prior work's
algorithms for \WCP and \DC analyses~\cite{vindicator,wcp}.
We refer to this algorithm as \emph{unoptimized \DC analysis} to distinguish it from optimized algorithms introduced in this paper.




The algorithm computes \DC using vector clocks that represent logical time.
A vector clock $C : \mathit{Tid} \mapsto \mathit{Val}$ maps
each thread to a nonnegative integer~\cite{vector-clocks}. Operations on vector
clocks are pointwise comparison ($\sqsubseteq$) and pointwise join ($\sqcup$):
\begin{eqnarray*}
  C_1 \sqsubseteq C_2 & \iff & \forall t . C_1(t) \leq C_2(t) \\
  C_1 \sqcup C_2 & \equiv & \lambda t . \mathit{max(C_1(t), C_2(t))}
\end{eqnarray*}
The algorithm maintains the following analysis state:

\begin{itemize}[leftmargin=*]

\item a vector clock $C_t$ for each thread $t$ that represents $t$'s current time;

\item vector clocks $R_x$ and $W_x$ for each program variable $x$
that represent times of reads and writes, respectively, to $x$;

\item vector clocks \LockVarQ{m}{x}{r} and \LockVarQ{m}{x}{w} that represent
the times of critical sections on lock $m$ containing reads and writes, respectively, to $x$;

\item sets $R_m$ and $W_m$ of variables read and written, respectively,
by each lock $m$'s ongoing critical section (if any); and

\item queues
\AcqMQ{m}{t}{t'} and \RelMQ{m}{t}{t'}, explained below.

\end{itemize}

\input{DC-analysis}

\noindent
Initially,
every set and queue is empty, and
every vector clock maps all threads to 0,
except $C_t(t)$ is 1 for every $t$.

\label{subsec:dc-analysis-costs}

A significant and challenging source of performance costs
is the logic for detecting \emph{conflicting critical sections}
to provide \DC \Rule{a}---a cost not present in \HB analysis.
At each release of a lock $m$,
the algorithm updates \LockVarQ{m}{x}{r} and \LockVarQ{m}{x}{w}
based on the variables accessed in the ending critical section on $m$
(lines~\ref{alg:DC-VC:TrackRuleARead}--\ref{alg:DC-VC:TrackRuleAWrite} in Algorithm~\ref{alg:DC-VC}).
At a read or write to $x$ by $t$,
the algorithm uses \LockVarQ{m}{x}{r} and \LockVarQ{m}{x}{w} to join $C_t$ with all
prior critical sections on $m$ that performed conflicting accesses to $x$
(lines~\ref{alg:DC-VC:RuleAWrite} and \ref{alg:DC-VC:RuleARead}).

The algorithm checks for \races{\DC} by checking for \DC ordering with prior conflicting accesses to $x$;
a failed check indicates a \race{\DC}
(lines~\ref{alg:DC-VC:write-write-race}, \ref{alg:DC-VC:read-write-race}, and \ref{alg:DC-VC:write-read-race}).
The algorithm
updates the logical time of the current thread's last write or read to $x$ (lines~\ref{alg:DC-VC:updateWx} and \ref{alg:DC-VC:updateRx}).

Finally, we explain how unoptimized \DC analysis orders events by \DC \Rule{b} (release events are ordered if critical sections are ordered);
the details are not important for understanding this paper.
The algorithm uses \AcqMQ{m}{t}{t'} and \RelMQ{m}{t}{t'}
to detect acquire--release ordering between critical sections and add release--release ordering.
Each vector clock in the queue \AcqMQ{m}{t}{t'} represents the time of an \Acquire{m} by $t'$ that
has \emph{not} been determined to be \DC ordered to the most recent release of $m$ by $t$.
Vector clocks in \RelMQ{m}{t}{t'} represent the corresponding \Release{m} times for clocks in \AcqMQ{m}{t}{t'}.
At \Release{m} by $t$,
the algorithm checks whether the release is ordered to a prior acquire of $m$ by any thread $t'$
(line~\ref{alg:DC-VC:RuleBCheck}).
If so, the algorithm orders the release corresponding to
the prior acquire to the current \Release{m} (line~\ref{alg:DC-VC:RelMQDeque}).

\input{analysis-state-example}

\paragraph{Running example.}
\label{subsec:unopt-dc-running-example}

To illustrate how unoptimized \DC analysis works and how it compares with optimized algorithms
introduced in this paper, Figure~\ref{fig:smarttrack-examples:analysis-state:unoptdc}
shows an example execution and the corresponding analysis state updates after each event in the execution---focusing on
the subset of analysis state relevant for detecting and ordering conflicting critical sections (\DC \Rule{a}).

At Thread~1's \Release{m},
the algorithm updates \LockVarQC{m}{x}{w} to reflect the fact that \code{x} was written in the critical section on \code{m}
(line~\ref{alg:DC-VC:TrackRuleAWrite} in Algorithm~\ref{alg:DC-VC}).
Similarly, the algorithm updates \LockVarQC{m}{x}{r} or \LockVarQC{p}{x}{w} at subsequent lock releases.

At Thread~2's \Read{x}, unoptimized \DC analysis updates \ThreadVC{T2} to establish
ordering with the prior conflicting critical section (line~\ref{alg:DC-VC:RuleARead}).
Likewise, at Thread~3's \Write{x}, the algorithm updates \ThreadVC{T3} to establish
ordering with both prior conflicting critical sections (line~\ref{alg:DC-VC:RuleAWrite}).
(Thread~3 is already transitively ordered with Thread~2's prior conflicting critical section
because of the \Sync{o} events.)
As a result, the checks at both threads' accesses to \code{x} correctly detect no race
(lines~\ref{alg:DC-VC:write-write-race}, \ref{alg:DC-VC:read-write-race}, and \ref{alg:DC-VC:write-read-race}).

\subsection{Performance Costs of Predictive Analyses}
\label{subsec:analysis-costs}

Unoptimized \DC (and \WCP) analyses~\cite{wcp,vindicator} incur three costs over
FastTrack-optimized \HB analysis~\cite{fasttrack,fasttrack2,fib}.

\paragraph{Conflicting critical section (CCS) ordering.}

Tracking \DC \Rule{a} requires $O(T \times L)$ time 
(lines~\ref{line:DC-VC:writeHeld}--\ref{line:DC-VC:writeHeld:end}
and \ref{line:DC-VC:readHeld}--\ref{line:DC-VC:readHeld:end} in Algorithm~\ref{alg:DC-VC})
for each access inside of critical sections on $L$ locks, where $T$ is the thread count;
we find that many of our evaluated real programs have 
a high proportion of accesses executing inside one or more critical sections
(Section~\ref{sec:eval}).
Furthermore, \LockVarQ{m}{x}{r} and \LockVarQ{m}{x}{w}
store information for lock--variable pairs, requiring indirect metadata lookups.
Note that \LockVarQ{m}{x}{r} and \LockVarQ{m}{x}{w} cannot be represented using epochs, and
applying FastTrack's epoch optimizations to last-access metadata does not optimize detecting CCS ordering.

\paragraph{Vector clocks.}

Unoptimized \DC analysis uses full vector clock operations to update write and read metadata and check for races
(lines~\ref{alg:DC-VC:write-write-race}--\ref{alg:DC-VC:updateWx} 
and \ref{alg:DC-VC:write-read-race}--\ref{alg:DC-VC:updateRx}).




\paragraph{Release--release ordering.}

Computing \DC \Rule{b} requires complex queue operations at every synchronization operation
(lines~\ref{alg:DC-VC:AcqMQEnque} and \ref{alg:DC-VC:AcqMQFrontLoop}--\ref{alg:DC-VC:RelMQEnque}).\footnote{\emph{\WCP} analysis
provides the same property
at lower cost because it can maintain per-lock queues for each thread, instead of each pair of threads,
as a consequence of \WCP composing with \HB~\cite{wcp}.}

\medskip
\noindent
The next two sections describe our optimizations for these challenges, starting with release--release ordering.

%% file: HB-definition.tex
\begin{definition*}[\HbFull] 

Given a trace \tr,
\ltHB is the smallest relation that satisfies the following properties:

\begin{itemize}[leftmargin=*]

  \item Two events are ordered by \HB if they are ordered by \PO.
  That is,
  \HBOrdered{e}{e'} if \POOrdered{e}{e'}.
   
  \item Release and acquire events on the same lock are ordered by \HB.
  That is, \HBOrdered{\event{r}{}}{\event{a}{}} if
  \event{r}{} and \event{a}{} are release and acquire events on the same lock and \TROrdered{r}{a}.
  
  \item \HB is transitively closed.
  That is,
  \HBOrdered{e}{e'} if $\exists e'' \mid \HBOrdered{e}{e''} \land \HBOrdered{e''}{e'}$.

\end{itemize} 

\end{definition*}

%% file: DC-definition.tex
\begin{definition}[\DcFull]
\label{def:dc}

Given a trace \tr,
\ltDC is the smallest relation that satisfies the following properties:


\begin{enumerate}[leftmargin=*,label=(\alph*)]

\renewcommand{\CS}[1]{\ensuremath{\mathit{CS}(#1)}}

  \item If two critical sections on the same lock contain \conflictingEvents,
  then the first critical section is ordered by \DC to the second event.
  That is, \DCOrdered{\event{r}{_1}}{\event{e}{_2}} if
  \event{r}{_1} and \event{r}{_2} are release events on the same lock,
  \TROrdered{\event{r}{_1}}{\event{r}{_2}},
  $\event{e}{_1} \in \CS{\event{r}{_1}}$, $\event{e}{_2} \in \CS{\event{r}{_2}}$, and
  \conflicts{\event{e}{_1}}{\event{e}{_2}}.
  (\CS{\event{r}{}} returns the set of events in the critical section ended by
  release event \event{r}{}, including \event{r}{} and the corresponding acquire event.)
  
  \item Release events on the same lock are ordered by \DC
  if their critical sections contain \DC-ordered events.
  Because of the next two rules, this rule can be expressed simply as:
  \DCOrdered{\event{r}{_1}}{\event{r}{_2}} if
  \event{r}{_1} and \event{r}{_2} are release events on the same lock and
  \DCOrdered{\event{a}{_1}}{\event{r}{_2}}
  where \event{a}{_1} is the acquire event that starts the critical section ended by \event{r}{_1}.
  
  \item Two events are ordered by \DC if they are ordered by \PO.
  That is, \DCOrdered{\event{e}{}}{\event{e}{'}}
  if \POOrdered{\event{e}{}}{\event{e}{'}}.
  
  \item \DC is transitively closed.
  That is, \DCOrdered{\event{e}{}}{\event{e}{'}} 
  if $\exists e'' \mid \DCOrdered{\event{e}{}}{\event{e}{''}} \land \DCOrdered{\event{e}{''}}{\event{e}{'}}$.
\end{enumerate}

\end{definition}

%% file: DC-analysis.tex
\begin{algorithm}[t]
\caption{\hfill Unoptimized \DC analysis}
\linespread{0.8}


\newcommand\RaceCheck[2]{\State \textbf{check} $#1 \sqsubseteq C_t$}

\small
\begin{algorithmic}[1]

	\Procedure{Acquire}{$t,m$}
		\lForEach{$t' \neq t$}{$\AcqMQ{m}{t'}{t}$.Enque($C_t$)} \label{alg:DC-VC:AcqMQEnque} \Comment{\DC \Rule{b}}
	\EndProcedure
	
	\Procedure{Release}{$t,m$} \CaseComment{{\smaller (\code{rel}--\code{rel} ordering)}}
		\ForEach{$t' \neq t$} \label{alg:DC-VC:AcqMQFrontLoop}
			\While{$\AcqMQ{m}{t}{t'}\textnormal{.Front()} \sqsubseteq C_t$} \label{alg:DC-VC:RuleBCheck}
				\State $\AcqMQ{m}{t}{t'}$.Deque() \label{alg:DC-VC:AcqMQDeque}
\hspace*{5em}\rlap{\smash{$\left.\begin{array}{@{}c@{}}\\{}\\{}\\{}\\{}\\{}\end{array}\right\}\begin{tabular}{l}\end{tabular}$}}
\Comment{\DC \Rule{b}}
								\State $C_t \gets C_t \sqcup \RelMQ{m}{t}{t'}$.Deque() \label{alg:DC-VC:RelMQDeque} \CaseComment{{\smaller (\code{rel}--\code{rel} ordering)}}
				\EndWhile
		\EndFor
		\lForEach{$t' \neq t$}{$\RelMQ{m}{t'}{t}$.Enque($C_t$)} \label{alg:DC-VC:RelMQEnque}
		\lForEach{$x \in R_m$}{$\LockVarQ{m}{x}{r} \gets \LockVarQ{m}{x}{r} \sqcup C_t$} \label{alg:DC-VC:TrackRuleARead}
		\lForEach{$x \in W_m$}{$\LockVarQ{m}{x}{w} \gets \LockVarQ{m}{x}{w} \sqcup C_t$} \label{alg:DC-VC:TrackRuleAWrite}
\rlap{\smash{$\left.\begin{array}{@{}c@{}}\\{}\\{}\end{array}\right\}\begin{tabular}{l}\end{tabular}$}}
\Comment{\DC \Rule{a}}
		\State $R_m \gets W_m \gets \emptyset$ \CaseComment{\iftoggle{twoColumnText}{{\smaller (CCS ordering)}}{(conflicting critical sections)}}
		\State $C_t(t) \gets C_t(t) + 1$ \label{line:DC-VC:Ctinc} \Comment{\DC \Rule{c}}
	\EndProcedure
	
	\Procedure{Write}{$t,x$} \CaseComment{{\smaller (\PO ordering)}}
		\ForEach {$m \in \textnormal{HeldLocks}(t)$} \label{line:DC-VC:writeHeld}
			\State $C_t \gets C_t \sqcup \big( \LockVarQ{m}{x}{r} \sqcup \LockVarQ{m}{x}{w}\big)$ \label{alg:DC-VC:RuleAWrite}
\hspace*{0.5em}\rlap{\smash{$\left.\begin{array}{@{}c@{}}\\{}\\{}\end{array}\right\}\begin{tabular}{l}\end{tabular}$}}
\Comment{\DC \Rule{a}}
			\State $W_m \gets W_m \cup \{x\}$ \label{line:DC-VC:writeHeld:end} \CaseComment{\iftoggle{twoColumnText}{{\smaller(CCS ordering)}}{(conflicting critical sections)}}
		\EndFor
		\RaceCheck{W_x}{\DC} \label{alg:DC-VC:write-write-race}
		\RaceCheck{R_x}{\DC} \label{alg:DC-VC:read-write-race}
		\State $W_x(t) \gets C_t(t)$ \label{alg:DC-VC:updateWx}
	\EndProcedure

	\Procedure{Read}{$t,x$}
		\ForEach {$m \in \textnormal{HeldLocks}(t)$} \label{line:DC-VC:readHeld}
			\State $C_t \gets C_t \sqcup \LockVarQ{m}{x}{w}$ \label{alg:DC-VC:RuleARead}
\hspace*{4.5em}\rlap{\smash{$\left.\begin{array}{@{}c@{}}\\{}\\{}\end{array}\right\}\begin{tabular}{l}\end{tabular}$}}
\label{line:DC-VC:readHeld:ruleA}\Comment{\DC \Rule{a}}
			\State $R_m \gets R_m \cup \{x\}$ \label{line:DC-VC:readHeld:end} \CaseComment{\iftoggle{twoColumnText}{{\smaller(CCS ordering)}}{(conflicting critical sections)}}
		\EndFor
		\RaceCheck{W_x}{\DC} \label{alg:DC-VC:write-read-race}
		\State $R_x(t) \gets C_t(t)$ \label{alg:DC-VC:updateRx}
	\EndProcedure
	
\end{algorithmic}
\label{alg:DC-VC}
\end{algorithm}

%% file: analysis-state-example.tex
\begin{figure*}
\centering
\captionsetup[subfigure]{justification=centering}
\newcommand\mythrs{\mc{1}{c}{\thr{T1}} & \mc{1}{c}{\thr{T2}} & \mc{1}{c|}{\thr{T3}}}
\newcommand\myheading[1]{\mc{3}{c|}{Execution events} & \mc{#1}{c}{Analysis state after event}}
\subfloat[Analysis state for \textbf{unoptimized \DC analysis} (Algorithm~\ref{alg:DC-VC}).]{
  \renewcommand{\arraystretch}{.85}
  \small
  \begin{tabular}{@{}l@{\;}l@{\;}l@{\;}|@{\;}c@{\;}c@{\;}c:c@{\;}c:cc:cc@{}}
  \myheading{9} \\
	\mythrs & \ThreadVC{T1} & \ThreadVC{T2} & \ThreadVC{T3} & \WrVar{x} & \RdVar{x} & \LockVarQC{m}{x}{w} & \LockVarQC{m}{x}{r} & \LockVarQC{p}{x}{w} & \LockVarQC{p}{x}{r} \\\hline
	& & 						& \VCExample{1}{0}{0} & \VCExample{0}{1}{0} & \VCExample{0}{0}{1} & \VCExample{0}{0}{0} & \VCExample{0}{0}{0} & \VCExample{0}{0}{0} & \VCExample{0}{0}{0} & \VCExample{0}{0}{0} & \VCExample{0}{0}{0} \\ 
	\Acquire{p} & & 			& & & & & & & & \\
	\Acquire{m} & & 			& & & & & & & & \\
	\Write{x} & & 				& & & & \VCExample{1}{0}{0} & & & & \\
	\Release{m}\tikzmark{1} &&	& \VCExample{2}{0}{0} & & & & & \VCExample{1}{0}{0} & \\
	& \Acquire{m} & 			& & & & & & & & \\
	& \tikzmark{2}\Read{x} & 	& & \VCExample{1}{1}{0} & & & \VCExample{0}{1}{0} & & \\
	\Release{p}\tikzmark{5} &&	& \VCExample{3}{0}{0} & & & & & & & \VCExample{2}{0}{0} \\ 
	& \Release{m} & 			& & \VCExample{1}{2}{0} & & & & & \VCExample{1}{1}{0} \\
	& \Sync{o}\tikzmark{3} & 	& & \VCExample{1}{3}{0} & & & & & & \\
	& & \tikzmark{4}\Sync{o} 	& & & \VCExample{1}{2}{2} & & & & & \\
	& & \Acquire{p} 			& & & & & & & & \\
	& & \tikzmark{6}\Write{x} 	& & & \VCExample{2}{2}{2} & \VCExample{1}{0}{2} & & & & \\
	& & \Release{p} 			& & & \VCExample{2}{2}{3} & & & & & \VCExample{2}{2}{2} \\
  \end{tabular}%
  \link{1}{2}%
  \link{3}{4}%
  \link{5}{6}%
  \label{fig:smarttrack-examples:analysis-state:unoptdc}
}\\\bigskip\bigskip
\subfloat[Analysis state for \textbf{\FTO-based \DC analysis} (Algorithm~\ref{alg:DC-Ownership-VC}).]{
  \renewcommand{\arraystretch}{.85}
  \small
  \begin{tabular}{@{}l@{\;}l@{\;}l@{\;}|@{\;}c@{\;}c@{\;}c:cc:cc:cc@{}} 
  \myheading{9} \\
  \mythrs & \ThreadVC{T1} & \ThreadVC{T2} & \ThreadVC{T3} & \WrVar{x} & \RdVar{x} & \LockVarQC{m}{x}{w} & \LockVarQC{m}{x}{r} & \LockVarQC{p}{x}{w} & \LockVarQC{p}{x}{r} \\\hline
	& &							& \VCExample{1}{0}{0} & \VCExample{0}{1}{0} & \VCExample{0}{0}{1} & \initA & \initA & \VCExample{0}{0}{0} & \VCExample{0}{0}{0} & \VCExample{0}{0}{0} & \VCExample{0}{0}{0} \\ 
	\Acquire{p} & & 			& \VCExample{2}{0}{0} & & & & & & \\
	\Acquire{m} & & 			& \VCExample{3}{0}{0} & & & & & & \\
	\Write{x} & & 				& & & & 3@\thr{T1} & 3@\thr{T1} & & & & \\
	\Release{m}\tikzmark{1} &&	& \VCExample{4}{0}{0} & & & & & \VCExample{3}{0}{0} & \VCExample{3}{0}{0} & & \\
	& \Acquire{m} & 			& & \VCExample{0}{2}{0} & & & & & & \\
	& \tikzmark{2}\Read{x} & 	& & \VCExample{3}{2}{0} & & & 2@\thr{T2} & & & & \\
	\Release{p}\tikzmark{5} &&	& \VCExample{5}{0}{0} & & & & & & & \VCExample{4}{0}{0} & \VCExample{4}{0}{0} \\ 
	& \Release{m} & 			& & \VCExample{3}{3}{0} & & & & & \VCExample{3}{2}{0} & & \\
	& \Sync{o}\tikzmark{3} & 	& & \VCExample{3}{5}{0} & & & & & \\
	& & \tikzmark{4}\Sync{o} 	& & & \VCExample{3}{4}{3} & & & & \\
	& & \Acquire{p} 			& & & \VCExample{3}{4}{4} & & & & \\
	& & \tikzmark{6}\Write{x} 	& & & \VCExample{4}{4}{4} & 4@\thr{T3} & 4@\thr{T3} & & & & \\
	& & \Release{p} 			& & & \VCExample{4}{4}{5} & & & & & \VCExample{4}{4}{4} & \VCExample{4}{4}{4} \\
  \end{tabular}%
  \link{1}{2}%
  \link{3}{4}%
  \link{5}{6}%
  \label{fig:smarttrack-examples:analysis-state:ftodc}
}\\\bigskip\bigskip
\subfloat[Analysis state for \textbf{\RE-based \DC analysis} (Algorithm~\ref{alg:RE-VC}).]{
  \renewcommand{\arraystretch}{.95}
  \small
  \newcommand\myepoch[2]{%
    \ifthenelse{\equal{#2}{T1}}
               {\VCExample{#1}{0}{0}}
               {\ifthenelse{\equal{#2}{T2}}
                           {\VCExample{0}{#1}{0}}
                           {\VCExample{0}{0}{#1}}}}
  \begin{tabular}{@{}l@{\;}l@{\;}l@{\;}|@{\;}c@{\;}c@{\;}c:c@{\;\;}c:p{4.25cm}@{\;}p{3.75cm}ZZZ}
  \myheading{7} \\
  \mythrs & \ThreadVC{T1} & \ThreadVC{T2} & \ThreadVC{T3} & \WrVar{x} & \RdVar{x} & \mc{1}{c}{\VarCSList{x}{w}} & \mc{1}{c}{\VarCSList{x}{r}} & \ThrCSList{T1} & \ThrCSList{T2} & \ThrCSList{T3} \\\hline
	& &							& \VCExample{1}{0}{0} & \VCExample{0}{1}{0} & \VCExample{0}{0}{1} & \initA & \initA & \CSListSet{} & \CSListSet{} & \CSListSet{} & \CSListSet{} & \CSListSet{} \\ 
	\Acquire{p} & & 			& \VCExample{2}{0}{0} & & & & & & & \CSListSet{\VCShort{\infty}{p}{T1}} \\
	\Acquire{m} & & 			& \VCExample{3}{0}{0} & & & & & & & \CSListSet{\VCShort{\infty}{m,p}{T1}} \\
	\Write{x} & & 				& & & & 3@\thr{T1} & 3@\thr{T1} & \CSListSet{\langle \myepoch{\infty}{T1}, \code{m} \rangle, \langle \myepoch{\infty}{T1}, \code{p} \rangle} & \CSListSet{\langle \myepoch{\infty}{T1}, \code{m} \rangle, \langle \myepoch{\infty}{T1}, \code{p} \rangle} \\
	\Release{m}\tikzmark{1} &&	& \VCExample{4}{0}{0} & & & & & \CSListSet{\langle \myepoch{3}{T1}, \code{m} \rangle, \langle \myepoch{\infty}{T1}, \code{p} \rangle} & \CSListSet{\langle \myepoch{3}{T1}, \code{m} \rangle, \langle \myepoch{\infty}{T1}, \code{p} \rangle} & \CSListSet{\VCShort{\infty}{p}{T1}} \\
	& \Acquire{m} & 			& & \VCExample{0}{2}{0} & & & & & & & \CSListSet{\VCShort{\infty}{m}{T2}} \\
  	& \tikzmark{2}\Read{x} & 	& & \VCExample{3}{2}{0} & & & \VCExample{3}{2}{0} & & \VCExample{\CSListSet{\langle \myepoch{3}{T1}, \code{m} \rangle, \langle \myepoch{\infty}{T1}, \code{p} \rangle}}{\allowbreak\hspace*{0.4em}\CSListSet{\langle \myepoch{\infty}{T2}, \code{m} \rangle}}{\CSListSet{}} \\
	\Release{p}\tikzmark{5} &&	& \VCExample{5}{0}{0} & & & & & \CSListSet{\langle \myepoch{3}{T1}, \code{m} \rangle, \langle \myepoch{4}{T1}, \code{p} \rangle} & \VCExample{\CSListSet{\langle \myepoch{3}{T1}, \code{m} \rangle, \langle \myepoch{4}{T1}, \code{p} \rangle}}{\allowbreak\hspace*{0.4em}\CSListSet{\langle \myepoch{\infty}{T2}, \code{m} \rangle}}{\CSListSet{}} & \CSListSet{} \\
	& \Release{m} & 			& & \VCExample{3}{3}{0} & & & & & \VCExample{\CSListSet{\langle \myepoch{3}{T1}, \code{m} \rangle, \langle \myepoch{4}{T1}, \code{p} \rangle}}{\allowbreak\hspace*{0.4em}\CSListSet{\langle \VCExample{3}{2}{0}, \code{m} \rangle}}{\CSListSet{}} & & \CSListSet{} \\
	& \Sync{o}\tikzmark{3} & 	& & \VCExample{3}{5}{0} & & & & & \\
	& & \tikzmark{4}\Sync{o} 	& & & \VCExample{3}{4}{3} & & & & \\
	& & \Acquire{p} 			& & & \VCExample{3}{4}{4} & & & & & & & \CSListSet{\VCShort{\infty}{p}{T3}} \\
	& & \tikzmark{6}\Write{x} 	& & & \VCExample{4}{4}{4} & 4@\thr{T3} & 4@\thr{T3} & \CSListSet{\langle \myepoch{\infty}{T3}, \code{p} \rangle} & \CSListSet{\langle \myepoch{\infty}{T3}, \code{p} \rangle} \\
	& & \Release{p} 			& & & \VCExample{4}{4}{5} & & & \CSListSet{\langle {\VCExample{4}{4}{4}}, \code{p} \rangle} & \CSListSet{\langle {\VCExample{4}{4}{4}}, \code{p} \rangle} & & & \CSListSet{} \\
  \end{tabular}%
  \link{1}{2}%
  \link{3}{4}%
  \link{5}{6}%
  \label{fig:smarttrack-examples:analysis-state:stdc}
}

\caption{The operation of three \DC analysis algorithms
  (Algorithms~\ref{alg:DC-VC}--\ref{alg:RE-VC}) on the same example execution.
  The entries show the analysis state (right side) updated after each executed event.
  The event \Sync{o} represents the event sequence \Acquire{o}; \Read{oVar}; \Write{oVar}; \Release{o}
  and serves to establish \DC ordering between two threads.
  Arrows show cross-thread \DC ordering.}
  \label{fig:smarttrack-examples:analysis-state}
\end{figure*}

%% file: capo.tex
\section{\CAPOFull}
\label{sec:CAPO}


This section introduces a new \emph{\capoFull (\CAPO)} relation, and a \emph{\CAPO analysis} that detects \emph{\CAPO-races}.

\CAPO is a strict partial order that has the same definition as \DC except that it \emph{omits \DC \Rule{b}} (Definition~\ref{def:dc}).\footnote{Weakening
\WCP in the same way would result in an unsound relation, giving up a key property of \WCP. In contrast, \DC is already unsound.}
Removing lines~\ref{alg:DC-VC:AcqMQEnque} and
\ref{alg:DC-VC:AcqMQFrontLoop}--\ref{alg:DC-VC:RelMQEnque}
from unoptimized \DC analysis (Algorithm~\ref{alg:DC-VC}) yields unoptimized \CAPO analysis.
This change addresses the ``release--release ordering'' cost
explained in Section~\ref{subsec:analysis-costs}.
The \DC-races in Figures~\ref{fig:example:simple-predictive-race} and \ref{fig:example:dc}
are by definition \CAPO-races.

%
%
The motivation for \CAPO is that it is simpler than \DC and thus cheaper to compute.
\CAPO is strictly weaker than \DC and thus finds some races that \DC does not---but they
are generally false races (\ie, not predictable races).
\input{capo-false-race}

To ensure soundness,
the prior work's \emph{vindication} algorithm for \DC analysis~\cite{vindicator}
can be used without modification to verify \CAPO-races as predictable races.
Section~\ref{subsec:record-replay} discusses vindication and its costs.
However, like \DC analysis, \CAPO analysis detects few if any false races in practice.
In our evaluation, despite \CAPO being weaker than \DC,
\CAPO analysis does \emph{not} report more races than \DC analysis.

The next section's optimizations apply to \WCP, \DC, and \CAPO analyses alike.


%% file: capo-false-race.tex
\begin{figure}
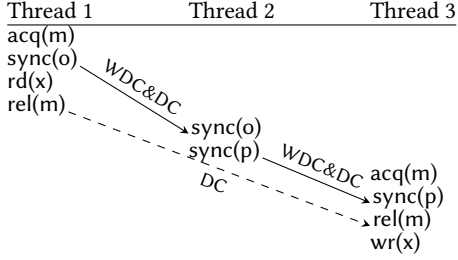

\renewcommand{\arraystretch}{\smalltablerowheight}
\small
\centering 
\sf
\begin{tabular}{@{}l@{\qquad\qquad}l@{\qquad\qquad}l@{}} 
\textnormal{Thread 1} & \textnormal{Thread 2} & \textnormal{Thread 3} \\\hline
\Acquire{m} \\
\Sync{o}\tikzmark{1} \\
\Read{x} \\
\Release{m}\tikzmark{5} \\
& \tikzmark{2}\Sync{o} \\
& \Sync{p}\tikzmark{3} \\
& & \Acquire{m} \\
& & \tikzmark{4}\Sync{p} \\
& & \tikzmark{6}\Release{m} \\
& & \Write{x} \\ \\
\end{tabular}\\
\textnormal{Note: $\Sync{o}$ represents $\Acquire{o}; \Read{oVar}; \Write{oVar}; \Release{o}$}
\textlink{1}{2}{\CAPO{}\&\DC}
\textlink{3}{4}{\CAPO{}\&\DC}
\undertextlinkdash{5}{6}{\DC}
\caption{An execution that has no predictable race and no \DC-race (\DCOrdered{\ReadT{x}{T1}}{\WriteT{x}{T3}}), 
but does have a \CAPO-race (\protect\nCAPOOrdered{\ReadT{x}{T1}}{\WriteT{x}{T3}}).
Arrows show cross-thread ordering.}
\label{fig:example:capo}
\end{figure}

Figure~\ref{fig:example:capo} shows an execution
with a \CAPO-race but no \DC-race or predictable race.
The execution has no \DC-race because 
\DCOrdered{\AcqT{m}{T1}}{\RelT{m}{T3}}
implies \DCOrdered{\RelT{m}{T1}}{\RelT{m}{T3}} by \DC \Rule{b}.
In contrast, \nCAPOOrdered{\RelT{m}{T1}}{\RelT{m}{T3}}.
Thus \nCAPOOrdered{\ReadT{x}{T1}}{\WriteT{x}{T3}}.

%% file: algorithm.tex
\section{\REFull}
\label{sec:analysis}

This section introduces \emph{\REFull},
a set of analysis optimizations applicable to predictive analyses:
\begin{itemize}[leftmargin=*]
  \item \emph{Epoch and ownership optimizations} are from prior work that optimizes \HB analysis~\cite{fasttrack,fasttrack2,fib}.
  We apply them to predictive analysis for the first time
  (Section~\ref{subsec:fto-dc-analysis}).

  \item \emph{Conflicting critical section (CCS) optimizations} are novel analysis
  optimizations that represent the paper's most significant technical contribution
  (Section~\ref{subsec:re-analysis}).
\end{itemize}

\subsection{Epoch and Ownership Optimizations}
\label{subsec:fto-dc-analysis}

In 2009,
Flanagan and Freund introduced \emph{epoch optimizations} to \HB analysis, 
realized in the \emph{FastTrack} algorithm~\cite{fasttrack}.
The core idea is that \HB analysis only needs to track the latest write to a variable $x$, 
and in some cases only needs to track the latest read to $x$, to detect the first race.
So FastTrack replaces the use of a vector clock with an \emph{epoch}, 
\epoch{c}{t}, to represent the latest write or read, 
where $c$ is an integer clock value and $t$ is a thread ID.
The lightweight epoch representation is sufficient for detecting the first race soundly
because whenever an access races with a prior write \emph{not} represented by the last-write epoch,
then it must also race with the last write (similarly for reads in some cases).
That is, if the current access does not race with the last write, then either 
(1) the current access does not race with any earlier write or 
(2) the last write races with an earlier write (which would have been detected earlier). 
A similar argument applies to reads.
\notes{In FastTrack, $W_x$ is an epoch that represents the latest write;
representing only the latest write is correct because
any future access that races with an earlier write must also race with the latest write
(assuming data-race-free execution up to that point)~\cite{fasttrack}.
Similarly, FastTrack can track only the latest read, by using a read epoch $R_x = \epoch{c}{t}$,
if the latest read is \HB ordered after all prior reads (otherwise $R_x$ is a vector clock
representing all reads to $x$)---which is correct
because any future write that races with an earlier read must race with the latest read.}%

\notes{
FastTrack's epoch optimizations enable race checks that often take $O(1)$ time instead of $O(T)$ time
($T$ is the number of threads), and use $O(1)$ instead of $O(T)$ space for every $W_x$ and some $R_x$.
In contrast, \emph{non-FastTrack} \HB analysis, as well as the unoptimized \DC analysis, shown in Algorithm~\ref{alg:DC-VC},
track every thread's last read and write to $x$ by using vector clocks for $R_x$ and $W_x$.
\mike{We don't talk about asymptotic time later for \RE.}
}



It is straightforward to adapt FastTrack's epoch optimizations 
to \emph{predictive} analysis's last-access metadata updates:
changes to $R_x$ and $W_x$'s representations will not affect
the logic for detecting CCSs.
We apply epoch optimizations together with \emph{ownership optimizations}
from Wood \etal's \emph{FastTrack-Ownership (\FTO)} algorithm~\cite{fib}.
\FTO's invariants enable a more elegant formulation for \REFull.
We explain \FTO shortly, in the context of applying it to \DC analysis.


Algorithm~\ref{alg:DC-Ownership-VC} shows \emph{\DCO},
which applies \FTO's optimizations to  unoptimized \DC analysis (Algorithm~\ref{alg:DC-VC}).
Differences between Algorithms~\ref{alg:DC-VC} and \ref{alg:DC-Ownership-VC} are highlighted in gray.
Optimizing \WCP and \CAPO analyses similarly is straightforward.

As mentioned above briefly, an epoch is a scalar \epoch{c}{t},
where $c$ is a nonnegative integer, and
the leading bits represent $t$, a thread ID.
For simplicity of exposition,
for the rest of the paper, we redefine vector clocks to map to epochs instead of integers,
$C : \mathit{Tid} \mapsto \mathit{Epoch}$,
and redefine $C_1 \sqsubseteq C_2$
and $C_1 \sqcup C_2$ in terms of epochs.
The notation \epochLeqVC{e}{C} checks whether an epoch $e = \epoch{c}{t}$ is ordered before a vector clock $C$,
and evaluates to $c \le c'$
where $\epoch{c'}{t} = C(t)$.
An ``uninitialized'' epoch representing no prior access is denoted as \initA,
and $\epochLeqVC{\initA}{C}$ for every vector clock $C$.

\DCO modifies the metadata used by unoptimized \DC analysis (Algorithm~\ref{alg:DC-VC})
in the following ways:


\begin{itemize}[leftmargin=*]

\item $W_x$ is an epoch representing the latest write to $x$.

\item $R_x$ is either an epoch or a vector clock representing the latest reads \emph{and write} to $x$.
\end{itemize}
Initially, every $R_x$ and $W_x$ is \initA.

Additionally, although \DCO does not change the representations of
$L_{m,x}^r$ and $R_m$ from unoptimized \DC analysis, in \DCO they represent reads \emph{and writes}, not just reads, within a critical section on $m$.

\input{DC-Ownership-analysis}


\notes{
\DCO maintains the following high-level invariants,
adapted from prior work~\cite{fib}:

\begin{enumerate}

\item Every write to $x$, and every to read to $x$ before the latest write to $x$, is \DC ordered before $W_x$.
\item Every read and write to $x$ is \DC ordered before at least one epoch in $R_x$.

\end{enumerate}
\DCO provides the first invariant by maintaining the latest write epoch $W_x$.
To provide the second invariant,
\DCO treats $R_x$ as representing prior reads \emph{and writes}.
}

%



Compared with unoptimized \DC analysis,
\DCO
significantly changes the maintenance and checking of $R_x$ and $W_x$,
by using a set of increasingly complex cases:

\paragraph{Same-epoch cases.}

At a write (or read) to $x$ by $t$,
if $t$ has already written (or read or written) $x$ since the last synchronization event,
then the access is effectively redundant (it cannot introduce a race or change last-access metadata).
\DCO checks these cases by comparing the current thread's epoch with $R_x$ or $W_x$,
shown in the \ReadSameEpoch, \SharedSameEpoch, and \WriteSameEpoch
cases in Algorithm~\ref{alg:DC-Ownership-VC}.

\DCO's same-epoch check works because a thread increments its logical clock $C_t(t)$ at not only release events
but also acquire events
(line~\ref{line:DC-Ownership-VC:CtincAcq} in Algorithm~\ref{alg:DC-Ownership-VC}).
The same-epoch check thus succeeds only for accesses redundant since the last synchronization operation.

If a same-epoch case does not apply,
then \DCO adds ordering from prior conflicting critical sections
(lines~\ref{line:DC-Ownership-VC:wrRuleAStart}--\ref{line:DC-Ownership-VC:wrRuleAEnd} and
\ref{line:DC-Ownership-VC:rdRuleAStart}--\ref{line:DC-Ownership-VC:rdRuleAEnd}),
just as in unoptimized \DC analysis,
before checking other \DCO cases.
Because $R_x$, $R_m$, and $L_{m,x}^r$ represent last reads \emph{and writes},
at writes \DCO updates $R_x$ as well as $W_x$ (line~\ref{line:DC-Ownership-VC:writeAccessUpdate})
and $R_m$ as well as $W_m$ (line~\ref{line:DC-Ownership-VC:rdAtWrRuleASource}).

\paragraph{Owned cases.}

At a read or write to $x$ by $t$,
if $R_x$ represents a prior access by $t$ (\ie, $R_x = \epoch{c}{t}$ or $R_x(t) \ne \initA$),
then the current access cannot race with any prior accesses.
\notes{because of the second \DCO invariant above.}%
The \ReadOwned, \ReadSharedOwned, and \WriteOwned cases
thus skip race check(s) and proceed to update $R_x$ and/or $W_x$.

\paragraph{Exclusive cases.}

If an owned case does not apply and $R_x$ is an epoch,
\DCO compares the current time with $R_x$.
If the current access is a write, this comparison acts as a race check \WriteExclusive.
If the current access is a read, then the comparison determines whether $R_x$ can remain an epoch or must become a vector clock.
If $R_x$ is \DC ordered before the current access, then $R_x$ remains an epoch \ReadExclusive.
Otherwise, the algorithm checks for a write--read race by comparing the current access with $W_x$,
and upgrades $R_x$ to a vector clock representing both the current read and prior read or write \ReadShare.

\paragraph{Shared cases.}

Finally, if an owned case does not apply and $R_x$ is a vector clock,
a shared case handles the access.
Since $R_x$ may \emph{not} be \DC ordered before the current access,
\ReadShared checks for a race by comparing with $W_x$,
while \WriteShared checks for a race by comparing with $R_x$ (comparing with $W_x$ is unnecessary since \epochLeqVC{W_x}{R_x}).

\paragraph{Running example.}

Figure~\ref{fig:smarttrack-examples:analysis-state:ftodc} shows how \DCO works,
using the same execution that we used to show how unoptimized \DC works
(Figure~\ref{fig:smarttrack-examples:analysis-state:unoptdc}, described in Section~\ref{subsec:unopt-dc-running-example}).
Here we focus on the differences between the two algorithms.

First, unlike unoptimized \DC analysis, \DCO increments thread vector clocks at acquire events,
leading to larger vector clock times.
Second, \DCO uses epochs instead of vector clocks to represent last-access times when possible,
as illustrated by the \WrVar{x} and \RdVar{x} columns in Figure~\ref{fig:smarttrack-examples:analysis-state:ftodc}.
Third, \DCO essentially treats each write to \code{x} as both a write and read to \code{x}.
As a result, at the execution's \Write{x} events, the algorithm updates \RdVar{x} as well as \WrVar{x};
and at all release events for a critical section containing a \Write{x},
the algorithm updates \LockVarQC{m}{x}{r} or \LockVarQC{p}{x}{r} in addition to updating \LockVarQC{m}{x}{w} or \LockVarQC{p}{x}{w}.

\subsection{Conflicting Critical Section Optimizations}
\label{subsec:re-analysis}

\input{RE-analysis}

While epoch and ownership optimizations improve the performance of predictive analyses,
they cannot optimize
detecting \emph{conflicting critical sections (CCSs)}
to compute \DC (or \WCP or \CAPO) \Rule{a}.
\notes{Applying epoch optimizations directly to CCSs would
not accurately track \DC (or \CAPO or \WCP).}%

Instead, our insight for efficiently detecting CCSs is that,
in common cases, an algorithm can unify how it maintains CCS metadata and last-access metadata for each variable $x$.
Our \emph{CCS optimizations}
use new analysis state $L_x^w$ and $L_x^r$, which have a correspondence with $W_x$ and $R_x$ at all times.
$L_x^w$ represents critical sections containing the write represented by $W_x$.
$L_x^r$ represents critical sections containing the read or write represented by $R_x$ if $R_x$ is an epoch,
or a vector of critical sections containing the reads and/or writes represented by $R_x$ if $R_x$ is a vecor clock.
Representing CCSs in this manner leads to cheaper logic than prior algorithms for predictive analysis in the common case.

The idea is that if an access within a critical section
conflicts with a prior access in a critical section on the same lock
\emph{not} represented by $L_x^w$ and $L_x^r$,
then it must conflict with the last access within a critical section,
represented by $L_x^w$ and $L_x^r$,
or else it races with the last access.
Furthermore, CCS optimizations exploit the synergy between CCS and last-access metadata,
often avoiding a race check after detecting CCSs.

\emph{\RE} is our new algorithm that combines CCS optimizations with epoch and ownership optimizations.
Algorithm~\ref{alg:RE-VC} shows \emph{\REDC},
which applies the \REFull algorithm to \DC analysis.
\REDC modifies \FTO-\DC (Algorithm~\ref{alg:DC-Ownership-VC})
by integrating CCS optimizations;
differences between the algorithms are highlighted in gray.
(Applying \RE to \WCP or \CAPO analysis is analogous.)
In particular, removing lines~\ref{line:RE-VC:AcqMQEnque} and
\ref{line:RE-VC:AcqMQFrontLoop}--\ref{line:RE-VC:RelMQEnque}
from Algorithm~\ref{alg:RE-VC}
yields \RECAPO.


\paragraph{Analysis state.}

\REFull introduces a new data type:
the \emph{critical section (CS) list},
which represents the logical times for releases of active critical sections by thread $t$ at some point in the execution.
A CS list has the following form:
\[
\langle \langle C_1, m_1 \rangle, \; \ldots, \; \langle C_n, m_n \rangle \rangle
\]
where $m_1 \dots m_n$ are
locks held by $t$, in innermost to outermost order;
and $C_1 \dots C_n$ are \emph{references} to (equivalently, shallow copies of) vector clocks representing the release time
of each critical section, in innermost to outermost order.
CS lists store \emph{references} to vector clocks
in order to allow the update of $C_i$ to be deferred until the release of $m_i$ executes.

\input{smarttrack-example}

\REDC maintains analysis state similar to
Algorithm~\ref{alg:DC-Ownership-VC} with the following additions and changes:

\begin{itemize}[leftmargin=*]

\item $H_t$ for each thread $t$, which is a current CS list for $t$;

\item $L_{x}^{w}$ for each variable $x$ (replaces \DCO's $L_{m,x}^w$), which is a CS list
for the last write access to $x$;

\item $L_{x}^{r}$ (replaces \DCO's $L_{m,x}^r$) has a form dependent on $R_x$:
\begin{itemize}
  \item if $R_x$ is an epoch, $L_{x}^{r}$ is a CS list for the last access to $x$;
  \item  if $R_x$ is a vector clock, $L_{x}^{r}$ is a thread-indexed vector of CS lists
  ($\mathit{Tid} \mapsto \mathit{CS\ list}$),
  with $L_{x}^{r}(t)$ representing the CS list for the last access to $x$ by $t$;
\end{itemize}


\item \A{w} and \A{r} (``ancillary'' metadata) for each variable $x$,
which are vectors of maps from locks to \emph{references to} vector clocks
($\mathit{Tid} \mapsto \mathit{Lock} \mapsto \mathit{VC}$).
\A{w} and \A{r} represent critical sections containing accesses to $x$
that are \emph{not necessarily} captured by $L_x^w$ and $L_x^r$, respectively.

\end{itemize}
In addition to the above changes to integrate CCS optimizations,
\REDC makes the following change to \DCO as a small optimization:
\begin{itemize}[leftmargin=*]

\item \AcqMQ{m}{t}{t'} is now a queue of \emph{epochs}.

\end{itemize}
Initially all CS lists are empty; \A{w} and \A{r} are empty maps.

\later{
\REDC analysis maintains the following invariants, which are an extension of \DCO's invariants:

\begin{enumerate}

\item Every critical section time represented in $L_{x}^w$ and $L_{x}^r$\ldots


\end{enumerate}
}

\paragraph{Maintaining CS lists.}

\REDC uses the same analysis cases as \DCO.
At each read or write to $x$,
\REDC maintains CCS metadata in $L_x^w$ and $L_x^r$ that corresponds to last-access metadata in $W_x$ and $R_x$.
At an access, the algorithm updates $L_x^r$ and/or $L_x^w$
to represent the current thread's active critical sections.

\REDC obtains the CS list representing the current thread's active critical sections from $H_t$,
which the algorithm maintains at each acquire and release event.
At an acquire, the algorithm prepends a new entry $\langle C, m \rangle$ to $H_t$
representing the new innermost critical section (lines~\ref{line:RE-VC:newCm}--\ref{line:RE-VC:addHt}).
$C$ is a reference to (\ie, shallow copy of) a newly allocated vector clock that represents the critical section's release time,
which is not yet known and will be updated at the release.
In the meantime, another thread $u$ may query whether $t$'s release of $m$ is \DC ordered before $u$'s current time
(line~\ref{line:RE-VC:multiCheck:rel-check}; explained later).
To ensure that this query returns false before $t$'s release of $m$,
the algorithm initializes $C(t)$ to $\infty$ (line~\ref{line:RE-VC:inftyCm}).
When the release of $m$ happens,
the algorithm removes the first element $\langle C, m \rangle$ of $H_t$, 
representing the critical section on $m$,
and updates the vector clock referenced by $C$ with the release time (lines~\ref{line:RE-VC:headHt}--\ref{line:RE-VC:removeHt}).

\paragraph{Checking for CCSs and races.}

At a read or write that may conflict with prior access(es),
\REDC combines the CCS check with the race check.
To perform this combined check,
the algorithm calls the helper function \textsc{MultiCheck}.
\textsc{MultiCheck} traverses a CS list in reverse (outermost-to-innermost) order,
looking for a prior critical section that is ordered to the current access or that conflicts with one of the current access's held locks
(lines~\ref{line:RE-VC:multiCheck:foreach-CS}--\ref{line:RE-VC:multiCheck:end-foreach-CS}).
If a critical section matches, it subsumes checking for inner critical sections or a \DC-race, so \textsc{MultiCheck} returns.
If no critical section matches, \textsc{MultiCheck} performs the race check (line~\ref{line:RE-VC:multiCheck:race-check}).

\paragraph{Running example.}

Figure~\ref{fig:smarttrack-examples:analysis-state:stdc} shows
how \REDC works, focusing on differences with \DCO.

Unique to \REDC are \VarCSList{x}{w} and \VarCSList{x}{r}.
At each access to \code{x} by a thread $t$, the algorithm updates \VarCSList{x}{r} and/or \VarCSList{x}{w}
using the current value of $H_t$, the CS list representing $t$'s ongoing critical sections
(line~\ref{line:RE-VC:multiCheck:update-thread-vc} in Algorithm~\ref{alg:RE-VC}).
Note that $H_t$ and thus \VarCSList{x}{r} and/or \VarCSList{x}{w} contain \emph{references to} (\ie, shallow copies of) vector clocks.
At each release of a lock, the algorithm updates vector clocks \emph{referenced by}
\VarCSList{x}{r} and/or \VarCSList{x}{w}.

\REDC uses \VarCSList{x}{w} and \VarCSList{x}{r} to detect and order conflicting critical sections
and to detect races.
At Thread~2's \Read{x},
the algorithm takes the \ReadShare case after detecting
that Thread~1's critical section on \code{p}
is \emph{not} fully \DC ordered before the current time
(lines~\ref{line:RE-VC:check-read1}--\ref{line:RE-VC:check-read2}).
(Below we explain why \REDC must take the \ReadShare in this situation.)
The \ReadShare case inflates both \RdVar{x} and \VarCSList{x}{r} to vectors;
\VarCSList{x}{r} represents Thread~1 and 2's prior accesses to \code{x} within critical sections.

At Thread~3's \Write{x}, \REDC takes the \WriteShared case,
which first checks ordering with Thread~1's \Write{x};
it detects the conflicting critical sections on $p$, so it adds ordering from \Release{p} to the current access (line~\ref{line:RE-VC:multiCheck:update-thread-vc}).
The algorithm then checks ordering with Thread~2's \Read{x};
the check succeeds immediately (line~\ref{line:RE-VC:multiCheck:rel-check})
because the events are already \DC ordered due to the \Sync{o} events.

%

\paragraph{\RE's \ReadShare behavior.}

\RE's CCS optimizations unify
the representations of critical section and last-access metadata.
To handle this unification correctly, \REDC takes the \ReadShare case in some situations---such as
Thread~2's \Read{x} in Figure~\ref{fig:smarttrack-examples:analysis-state}---when \DCO would take \ReadExclusive.
\notes{For example, for Figure~\ref{fig:smarttrack-examples:basic},
\REDC takes the \ReadShare case at Thread~2's \Read{x}, although \FTO would take \ReadExclusive.
\REDC takes \ReadShare because Thread~2's \Read{x} is not ordered after all of the last access's critical sections;
taking the \ReadExclusive case would lose information about Thread~1's critical section on \Release{p} containing $x$.}%

Figure~\ref{fig:smarttrack-examples:read-share} shows an execution that motivates the need for this behavior.
If \REDC were to take the \ReadExclusive case at Thread~2's \Read{x},
then the algorithm would lose information about Thread~1's \Read{x} being inside of the critical section on $m$.
As a result, \REDC would miss adding ordering from Thread~1's \Release{m} to Thread~3's \Write{x} (dashed arrow),
leading to unsound tracking of \DC and potentially reporting a false race later.
\REDC thus takes \ReadShare in situations like Thread~2's \Read{x} when the prior access's critical sections
(represented by the CS list $L_x^r$) are not all ordered before the current access.

\paragraph{Using ``ancillary'' metadata.}

Partly as a result of its \ReadShare behavior,
\REDC loses \emph{no} needed CCS information at \emph{reads}.
\notes{Essentially, the algorithm only overwrites a CS list in $L_x^r$
for a thread $t$ because of a new access to $x$ by $t$
that naturally subsumes critical sections of the prior access to $x$ by $t$.}%
However, as described so far, \REDC can lose needed CCS information at \emph{writes} to $x$,
by overwriting information about critical sections in $L_x^r$ and $L_x^w$ that are not ordered before the current write.
Figures~\ref{fig:smarttrack-examples:write-write-read-extra} and \ref{fig:smarttrack-examples:read-write-write-extra}
show two executions in which this situation occurs.
In each execution, at Thread~2's \Write{x},
\REDC updates $L_x^r$ and $L_x^w$ to $\langle \rangle$ (representing the access's lack of active critical sections)---which
loses information about Thread~1's critical section on $m$ containing an access to $x$.
As a result, in each execution, when Thread~3 accesses $x$,
\REDC cannot use $L_x^r$ or $L_x^w$ to detect the ordering from Thread~1's \Release{m} to the current access.

To ensure sound tracking of \DC,
\REDC uses the ancillary metadata \A{r} and \A{w} to track CCS information
lost from $L_x^r$ and $L_x^w$ at writes to $x$.
$\A{r}(t)(m)$ and $\A{w}(t)(m)$ each represent the release time of a critical section on $m$ by $t$
containing a read or write (\A{r}) or write (\A{w}) to $x$.
\textsc{MultiCheck} computes a ``residual'' map $A$ of critical sections
that are not ordered to the current access (line~\ref{line:RE-VC:multiCheck:compute-residual}),
which \REDC assigns to \A{r} or \A{w}.
At a write or read not handled by a same-epoch case,
if \A{r} or \A{w}, respectively, is non-empty,
the analysis adds ordering for CCSs represented in \A{r} (lines~\ref{line:RE-VC:rdExtraAtWrStart}--\ref{line:RE-VC:rdExtraAtWrEnd})
or \A{w} (lines~\ref{line:RE-VC:wrExtraAtRdStart}--\ref{line:RE-VC:wrExtraAtRdEnd}), respectively.

In essence, \REDC uses per-variable CCS metadata ($L_x^r$ and $L_x^w$) that mimics last-access metadata ($R_x$ and $W_x$)
when feasible, and otherwise falls back to CCS metadata (\A{r} and \A{w})
analogous to non-\RE metadata (\ie, \LockVarQ{m}{x}{r} and \LockVarQ{m}{x}{w} in Algorithms~\ref{alg:DC-VC} and \ref{alg:DC-Ownership-VC}).
\RE's performance improvement over \FTO relies on \A{r} and \A{w} being empty in most cases.

\paragraph{Optimizing $\boldsymbol{\AcqMQ{m}{t}{t'}}$.}

A final optimization
that we include as part of \REDC is to change \AcqMQ{m}{t}{t'}
from a vector clock (used in \DCO) to an epoch.
This optimization
is correct because an epoch is sufficient for
checking if ordering has been established from an \Acquire{m} on $t'$ to a \Release{m} on $t$,
since \REDC increments $C_t(t)$ after every acquire operation.

\begin{table*}
	\small
	\begin{tabular}{@{}l|Hllll@{}}
			 & Unopt w/$G$                         & Unopt					                     & Epochs                       & + Ownership                            		& + CS optimizations \\\hline
	HB       & N/A                                 & Unopt-\VCHB                                 & \FTTwo~\cite{fasttrack2}	 	& \HBO~\cite{fib}			             		& N/A \\
	WCP      & N/A                                 & Unopt-\VCWCP~\cite{wcp}			         & ---                          & \WCPO                                  		& \REWCP \\
	DC       & Unopt-\VCDC w/$G$~\cite{vindicator} & Unopt-\VCDC (Algorithm~\ref{alg:DC-VC})	 & ---                          & \DCO (Algorithm~\ref{alg:DC-Ownership-VC})	& \REDC (Algorithm~\ref{alg:RE-VC})\\
	\CAPO    & Unopt-\VCCAPO w/$G$                 & Unopt-\VCCAPO                               & ---                          & \CAPOO                                 		& \RECAPO \\
	\end{tabular}
	\caption{Implemented and evaluated analyses. Optimizations increase from left to right,
	and relations weaken from top to bottom.}
	\label{tab:configs}
\end{table*}

\subsection{Vindication: Performance Cost of Soundness}
\label{subsec:record-replay}


A final significant cost of \DC analysis is supporting a \emph{vindication} algorithm that checks
whether a \race{\DC} is a predictable race (similarly for \CAPO analysis and \races{\CAPO}).
Vindication operates on a constraint graph $G$, constructed during \DC analysis,
which adds significant time and space overhead.

To avoid the cost of constructing a constraint graph,
an implementation of \DC analysis
can either
(1) report all \DC-races, which are almost never false races in practice, or
(2) \emph{replay} any execution that detects a new (\ie, previously unknown) \DC-race---and construct a constraint graph
and perform vindication during the replayed execution only.
Recent multithreaded record \& replay approaches add very low (3\%) run-time overhead to record an
execution~\cite{ireplayer,castor}.\footnote{We have not implemented or tested an approach using record \& replay,
which is beyond the scope of this paper.
The recent practical multithreaded record \& replay tools iReplayer~\cite{ireplayer}
and Castor~\cite{castor} both target C/C++ programs, while our implementation targets Java programs.}
Replay failures caused by undetected \HB-races~\cite{respec-2010} are a non-issue since \DC analysis detects all \HB-races.

Our optimized \DC and \CAPO analyses do \emph{not} construct a constraint graph
and thus do \emph{not} perform vindication.

%% file: DC-Ownership-analysis.tex
\begin{algorithm}[t]
\caption{\hfill \DCO (\FTO-based \DC analysis)}
\linespread{0.8}


\small
\raggedright Differences with unoptimized \DC analysis (Algorithm~\ref{alg:DC-VC}) are highlighted gray.\\\smallskip\hrule\smallskip
\begin{algorithmic}[1]

	\Procedure{Acquire}{$t$, $m$}
		\lForEach{$t' \neq t$}{$\AcqMQ{m}{t'}{t}$.Enque($C_t$)} \label{alg:DC-Ownership-VC:AcqMQEnque}
		\State \tikzhighlightmk{A} $C_t(t) \gets C_t(t) + 1$ \label{line:DC-Ownership-VC:CtincAcq} \Comment{Supports same-epoch checks} \tikzhighlightmk{B}\boxit{grey}
	\EndProcedure
	
	\Procedure{Release}{$t$, $m$}
		\ForEach{$t' \neq t$} \label{alg:DC-Ownership-VC:AcqMQFrontLoop}
			\While{$\AcqMQ{m}{t}{t'}\textnormal{.Front()} \sqsubseteq C_t$}
				\State $\AcqMQ{m}{t}{t'}$.Deque()
				\State $C_t \gets C_t \sqcup \RelMQ{m}{t}{t'}$.Deque() \label{alg:DC-Ownership-VC:RelMQDeque}
			\EndWhile
		\EndFor
		\lForEach{$t' \neq t$}{$\RelMQ{m}{t'}{t}$.Enque($C_t$)} \label{alg:DC-Ownership-VC:RelMQEnque}
		\lForEach{$x \in R_m$}{$\LockVarQ{m}{x}{r} \gets \LockVarQ{m}{x}{r} \sqcup C_t$} \label{alg:DC-Ownership-VC:TrackRuleARead}
		\lForEach{$x \in W_m$}{$\LockVarQ{m}{x}{w} \gets \LockVarQ{m}{x}{w} \sqcup C_t$} \label{alg:DC-Ownership-VC:TrackRuleAWrite}
		\State $R_m \gets W_m \gets \emptyset$
		\State $C_t(t) \gets C_t(t) + 1$ \label{line:DC-Ownership-VC:CtincRel}
	\EndProcedure
	
	\Procedure{Write}{$t$, $x$}
		\lIf {\tikzhighlightmk{A} $W_x = C_t(t)$} \textbf{return} \CaseComment{\WriteSameEpoch} \label{line:DC-Ownership-VC:wrSameEpoch} \tikzhighlightmk{B}\boxit{grey}
		\ForEach {$m \in \textnormal{HeldLocks}(t)$} \label{line:DC-Ownership-VC:wrRuleAStart}
			\State $C_t \gets C_t \sqcup \big( \LockVarQ{m}{x}{r} \sqcup \LockVarQ{m}{x}{w}\big)$ \label{line:DC-Ownership-VC:RuleAWrite}
			\State $W_m \gets W_m \cup \{x\}$ \label{line:DC-Ownership-VC:wrAtWrRuleASource}
			\State \tikzhighlightmk{A} $R_m \gets R_m \cup \{x\}$ \label{line:DC-Ownership-VC:rdAtWrRuleASource} \CaseComment{}\tikzhighlightmk{B}\boxit{grey}
		\EndFor \label{line:DC-Ownership-VC:wrRuleAEnd}
		\If {\tikzhighlightmk{A} $R_x = \epoch{c}{t}$} \textbf{skip} \CaseComment{\WriteOwned}
		\ElsIf {$R_x = \epoch{c}{u}$} \CaseComment{\WriteExclusive} \label{line:DC-Ownership-VC:writeExclusiveCase}
			\State \textbf{check} \epochLeqVCThr{R_x}{C_t}{u} \label{line:DC-Ownership-VC:readWriteRace}
		\Else \CaseComment{\WriteShared}
		    \State \textbf{check} $R_x \ltpartial C_t$ \label{line:DC-Ownership-VC:readShrWriteRace}
		\EndIf
		\State $W_x \gets R_x \gets C_t(t)$ \label{line:DC-Ownership-VC:writeAccessUpdate}
		\CaseComment{}\tikzhighlightmk{B}\boxit{grey}
	\EndProcedure

	\Procedure{Read}{$t$, $x$}
		\lIf {\tikzhighlightmk{A} $R_x = C_t(t)$}{\textbf{return}} \CaseComment{\ReadSameEpoch} \label{line:DC-Ownership-VC:rdSameEpoch}
		\lIf {$R_x(t) = C_t(t)$}{\textbf{return}} \CaseComment{\SharedSameEpoch} \label{line:DC-Ownership-VC:rdShrSameEpoch} \tikzhighlightmk{B}\boxit{grey}
		\ForEach {$m \in \textnormal{HeldLocks}(t)$} \label{line:DC-Ownership-VC:rdRuleAStart}
			\State $C_t \gets C_t \sqcup \LockVarQ{m}{x}{w}$ \label{line:DC-Ownership-VC:RuleARead}
			\State $R_m \gets R_m \cup \{x\}$
		\EndFor \label{line:DC-Ownership-VC:rdRuleAEnd}
		\If {\tikzhighlightmk{A} $R_x = \epoch{c}{t}$} \CaseComment{\ReadOwned}
			\State $R_x \gets C_t(t)$ \label{line:DC-Ownership-VC:readOwned}
		\ElsIf {$R_x = \epoch{c}{u}$} \label{line:DC-Ownership-VC:readExclusive:ifcase}
			\If {\epochLeqVCThr{R_x}{C_t}{u}} \CaseComment{\ReadExclusive}
				\State $R_x \gets C_t(t)$ \label{line:DC-Ownership-VC:readExclusive}
			\Else \CaseComment{\ReadShare}
				\State \textbf{check} \epochLeqVC{W_x}{C_t} \label{line:DC-Ownership-VC:writeReadRace}
				\State $R_x \gets \{\epoch{c}{u}, C_t(t)\}$ \label{line:DC-Ownership-VC:readShare}
			\EndIf
		\ElsIf {$R_x(t) = \epoch{c}{t}$} \CaseComment{\ReadSharedOwned}
			\State $R_x(t) \gets C_t(t)$ \label{line:DC-Ownership-VC:readSharedOwned}
		\Else \CaseComment{\ReadShared}
			\State \textbf{check} \epochLeqVC{W_x}{C_t} \label{line:DC-Ownership-VC:writeReadShrRace}
			\State $R_x(t) \gets C_t(t)$ \label{line:DC-Ownership-VC:readShared}
		\EndIf
		\CaseComment{}\tikzhighlightmk{B}\boxit{grey}
	\EndProcedure
	
\end{algorithmic}
\label{alg:DC-Ownership-VC}
\end{algorithm}

%% file: RE-analysis.tex
\begin{algorithm*}
\caption{\hfill \REDC (\RE-based \DC analysis)}
\linespread{0.8}

\newcommand\Ht{\ensuremath{H_t}\xspace}
\newcommand\Rc{\ensuremath{C}\xspace}
\newcommand\dereference[1]{\ensuremath{#1}\xspace}

\def\NoNumber#1{{\def\alglinenumber##1{}\State #1}\addtocounter{ALG@line}{-1}}

\small
\raggedright Differences with \FTO-based \DC analysis (Algorithm~\ref{alg:DC-Ownership-VC}) are highlighted gray.\\\smallskip\hrule\smallskip
\vspace*{-1.5em}
\begin{multicols}{2}
\begin{algorithmic}[1]
	\Procedure{Acquire}{$t$, $m$}
		\lForEach{\tikzhighlightmk{A} $t' \neq t$}{$\AcqMQ{m}{t'}{t}$.Enque($C_t(t)$)} \label{line:RE-VC:AcqMQEnque}
		\State \textbf{let} $\Rc =$ reference to new vector clock \label{line:RE-VC:newCm}
		\State $\dereference{\Rc}(t) \gets \infty$ \label{line:RE-VC:inftyCm}
		\State $\Ht \gets \textnormal{prepend}(\langle \Rc, m \rangle, \Ht)$ \Comment{Add $\langle \Rc, m \rangle$ as head of list}\label{line:RE-VC:addHt} \tikzhighlightmk{B}\boxforeachit{grey}
		\State $C_t(t) \gets C_t(t) + 1$ \label{line:RE-VC:CtincAcq}
	\EndProcedure
	
	\Procedure{Release}{$t$, $m$}
		\ForEach{$t' \neq t$} \label{line:RE-VC:AcqMQFrontLoop}
			\While{\epochLeqVCThr{\AcqMQ{m}{t}{t'}\textnormal{.Front()}}{C_t}{t'}}
				\State $\AcqMQ{m}{t}{t'}$.Deque()
				\State $C_t \gets C_t \sqcup \RelMQ{m}{t}{t'}$.Deque() \label{line:RE-VC:RelMQDeque}
			\EndWhile
		\EndFor
		\lForEach{$t' \neq t$}{$\RelMQ{m}{t'}{t}$.Enque($C_t$)} \label{line:RE-VC:RelMQEnque}
		\State \tikzhighlightmk{A} \textbf{let} $\langle \Rc, \_ \rangle = \textnormal{head}(\Ht)$ \Comment{head() returns first element} \label{line:RE-VC:headHt}
		\State $\dereference{\Rc} \gets C_t$ \Comment{Update vector clock referenced by $C$} \label{line:RE-VC:setCm}
		\State $\Ht \gets$ rest(\Ht) \Comment{rest() returns list without first element} \label{line:RE-VC:removeHt} \tikzhighlightmk{B}\boxit{grey}
		\State $C_t(t) \gets C_t(t) + 1$ \label{line:RE-VC:CtincRel}
	\EndProcedure

	\Procedure{Write}{$t$, $x$}
		\lIf {$W_x = C_t(t)$} \textbf{return} \CaseComment{\WriteSameEpoch} \label{line:RE-VC:wrSameEpoch}
		\If {\tikzhighlightmk{A} $\A{r} \ne \emptyset$} \label{line:RE-VC:rdExtraAtWrStart}
			\ForEach {$m \in \textnormal{HeldLocks}(t)$}
				\State $C_t \gets C_t \sqcup \big(\bigsqcup_{u \ne t} \A{r}(u)(m)\big)$ \label{line:RE-VC:rdExtraAtWrUpdate}
				\lForEach {$u \ne t$}{$\A{r}(u)(m) \gets \A{w}(u)(m) \gets \emptyset$}
			\EndFor
			\State $\A{r}(t) \gets \A{w}(t) \gets \emptyset$ 
		\EndIf  \label{line:RE-VC:rdExtraAtWrEnd} \CaseComment{}\tikzhighlightmk{B}\boxit{grey}
		\If {$R_x = \epoch{c}{t}$} \textbf{skip} \CaseComment{\WriteOwned} 
		\ElsIf {$R_x = \epoch{c}{u}$} \CaseComment{\WriteExclusive}
			\State \tikzhighlightmk{A} \textbf{let} $A = \textsc{MultiCheck}(L_x^r, u, R_x)$
			\If {$A \ne \emptyset$}
				\State $\A{r}(u) \gets A$
				\State $\A{w}(u) \gets \textsc{MultiCheck}(L_x^w, u, \initA)$
			\EndIf \CaseComment{}\tikzhighlightmk{B}\boxit{grey}
		\Else \CaseComment{\WriteShared}
			\ForEach {\tikzhighlightmk{A} $u \neq t$}
				\State \textbf{let} $A = \textsc{MultiCheck}(L_x^r(u), u, R_x(u))$
				\If {$A \ne \emptyset$}
					\State $\A{r}(u) \gets A$
					\State $\A{w}(u) \gets \textsc{MultiCheck}(L_x^w(u), u, \initA)$ \CaseComment{}\tikzhighlightmk{B}\boxforeachit{grey}
				\EndIf 
			\EndFor
		\EndIf 
		\State \tikzhighlightmk{A} $L_{x}^{w} \gets L_{x}^{r} \gets \Ht$ \label{line:RE-VC:wrOwned} \label{line:RE-VC:wrOwned:RuleA} \CaseComment{}\tikzhighlightmk{B}\boxit{grey}
		\State $W_x \gets R_x \gets C_t(t)$
	\EndProcedure
	\newpage
	\Procedure{Read}{$t$, $x$}
		\lIf {$R_x = C_t(t)$}{\textbf{return}} \CaseComment{\ReadSameEpoch} \label{line:RE-VC:rdSameEpoch}
		\lIf {$R_x(t) = C_t(t)$}{\textbf{return}} \CaseComment{\SharedSameEpoch} \label{line:RE-VC:rdShrSameEpoch}
		\If {\tikzhighlightmk{A} $\A{w} \ne \emptyset$}  \label{line:RE-VC:wrExtraAtRdStart}
			\ForEach {$m \in \textnormal{HeldLocks}(t)$}
				\State $C_t \gets C_t \sqcup \big(\bigsqcup_{u \ne t} \A{w}(u)(m)\big)$ \label{line:RE-VC:wrExtraAtRdUpdate}
			\EndFor
		\EndIf  \label{line:RE-VC:wrExtraAtRdEnd} \CaseComment{}\tikzhighlightmk{B}\boxit{grey}
		\If {$R_x = \epoch{c}{t}$} \CaseComment{\ReadOwned}
			\State \tikzhighlightmk{A} $L_{x}^{r} \gets \Ht$ \label{line:RE-VC:rdOwned:RuleA} \CaseComment{}\tikzhighlightmk{B}\boxit{grey}
			\State $R_x \gets C_t(t)$ \label{line:RE-VC:rdOwned}
		\ElsIf {$R_x = \epoch{c}{u}$}
			\State \tikzhighlightmk{A} \textbf{let} $\epoch{c'}{u} = \left\{\begin{tabular}{@{}l@{\;\;}l@{}} $C'(u)$ s.t.\ $\langle C', \_ \rangle = \textnormal{tail}(L_x^r)$ & if $L_x^r \ne \langle\rangle$ \\
 			                                                  $R_x$ & otherwise \\\end{tabular}\right.$\label{line:RE-VC:check-read1}
 			\If {\epochLeqVCThr{\epoch{c'}{u}}{C_t}{u}} \CaseComment{\ReadExclusive}\label{line:RE-VC:check-read2}
				\State $L_{x}^{r} \gets \Ht$ \label{line:RE-VC:rdExclusive:RuleA} \CaseComment{}\tikzhighlightmk{B}\boxittaller{grey}
				\State $R_x \gets C_t(t)$ \label{line:RE-VC:rdExclusive} 
			\Else \CaseComment{\ReadShare}
				\State \tikzhighlightmk{A} \textsc{MultiCheck}($L_x^w$, tid($W_x$), $W_x$)
				\State $L_{x}^{r} \gets \{L_{x}^{r}, \Ht\}$ \label{line:RE-VC:rdShare:RuleA} \CaseComment{}\tikzhighlightmk{B}\boxit{grey}
				\State $R_x \gets \{\epoch{c}{u}, C_t(t)\}$ \label{line:RE-VC:rdShare}
			\EndIf
		\ElsIf {$R_x(t) = \epoch{c}{t}$} \CaseComment{\ReadSharedOwned}
			\State \tikzhighlightmk{A} $L_{x}^{r}(t) \gets \Ht$ \label{line:RE-VC:rdSharedOwned:RuleA} \CaseComment{}\tikzhighlightmk{B}\boxit{grey}
			\State $R_x(t) \gets C_t(t)$ \label{line:RE-VC:rdSharedOwned}
		\Else \CaseComment{\ReadShared}
			\State \tikzhighlightmk{A} \textsc{MultiCheck}($L_x^w$, tid($W_x$), $W_x$)
			\State $L_{x}^{r}(t) \gets \Ht$ \label{line:RE-VC:rdShared:RuleA} \CaseComment{}\tikzhighlightmk{B}\boxit{grey}
			\State $R_x(t) \gets C_t(t)$ \label{line:RE-VC:rdShared}
		\EndIf
	\EndProcedure

	\Procedure{MultiCheck}{$L$, $u$, $a$}
		\State \tikzhighlightmk{A} \textbf{let} $A = \emptyset$ \Comment{Empty map}
				\ForEach {$\langle \Rc, m \rangle \textnormal{ in } L$ in tail-to-head order} \label{line:RE-VC:multiCheck:foreach-CS}
			\lIf {\epochLeqVCThr{\dereference{\Rc}(u)}{C_t}{u}} \textbf{return} $A$ \label{line:RE-VC:multiCheck:rel-check}
			\If {$m \in \textnormal{heldby}(t)$}
 				\State $C_t \gets C_t \sqcup \dereference{\Rc}$ \label{line:RE-VC:multiCheck:update-thread-vc}
				\State \textbf{return} $A$
			\EndIf
			\State $A(m) \gets \dereference{\Rc}$ \label{line:RE-VC:multiCheck:compute-residual}
		\EndFor \label{line:RE-VC:multiCheck:end-foreach-CS}
		\State \textbf{check} \epochLeqVC{a}{C_t} \label{line:RE-VC:multiCheck:race-check}
		\State \textbf{return} $A$
	\EndProcedure
	\CaseComment{}\tikzhighlightmk{B}\boxit{grey}

\end{algorithmic}
\end{multicols}
\vspace*{-1em}
\label{alg:RE-VC}
\end{algorithm*}

%% file: smarttrack-example.tex
\begin{figure*}[t]
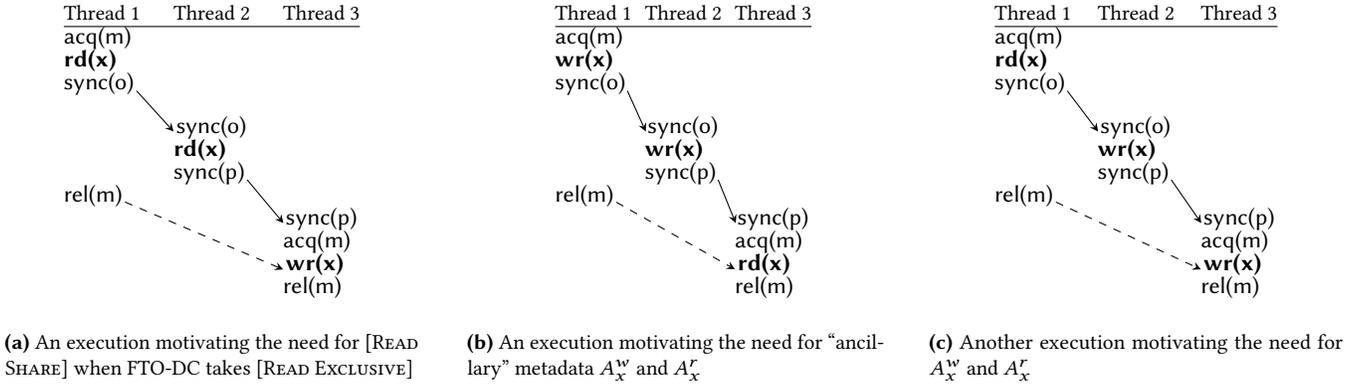

\renewcommand{\arraystretch}{\smalltablerowheight}
\newcommand{\MyThr}[1]{\textnormal{\footnotesize Thread #1}}
\small
\centering
\sf
\renewcommand{\case}[1]{\textsc{\footnotesize [#1]}}%
\subfloat[An execution motivating the need for \case{Read Share} when \DCO takes \case{Read Exclusive}]{
\begin{minipage}{0.3\linewidth}
\centering
\begin{tabular}{@{}l@{\;\;\;\;\;}l@{\;\;\;\;\;}l@{}} 
\MyThr{1} & \MyThr{2} & \MyThr{3} \\\hline
\Acquire{m} \\
\bf \Read{x} \\
\Sync{o}\tikzmark{1} \\
\\
& \tikzmark{2}\Sync{o} \\
& \bf \Read{x} \\
& \Sync{p}\tikzmark{3} \\
\Release{m}\tikzmark{5} \\
& & \tikzmark{4}\Sync{p} \\
& & \Acquire{m} \\
& & \bf \tikzmark{6}\Write{x} \\
& & \Release{m} \\
\\
\end{tabular}%
\link{1}{2}%
\link{3}{4}%
\linkdash{5}{6}%
\end{minipage}
\label{fig:smarttrack-examples:read-share}
}
\hfill
\subfloat[An execution motivating the need for ``ancillary'' metadata \A{w} and \A{r}]{
\begin{minipage}{0.3\linewidth}
\centering
\begin{tabular}{@{}l@{\;\;}l@{\;\;}l@{}} 
\MyThr{1} & \MyThr{2} & \MyThr{3} \\\hline
\Acquire{m} \\
\bf \Write{x} \\
\Sync{o}\tikzmark{1} \\
\\
& \tikzmark{2}\Sync{o} \\
& \bf \Write{x} \\
& \Sync{p}\tikzmark{3} \\
\Release{m}\tikzmark{5} \\
& & \tikzmark{4}\Sync{p} \\
& & \Acquire{m} \\
& & \bf \tikzmark{6}\Read{x} \\
& & \Release{m} \\
\\
\end{tabular}%
\link{1}{2}%
\link{3}{4}%
\linkdash{5}{6}%
\hspace{-0.25cm}
\end{minipage}
\label{fig:smarttrack-examples:write-write-read-extra}
}
\hfill
\subfloat[Another execution motivating the need for \A{w} and \A{r}]{
\begin{minipage}{0.3\linewidth}
\centering
\begin{tabular}{@{}l@{\;\;\;\;}l@{\;\;\;\;}l@{}} 
\MyThr{1} & \MyThr{2} & \MyThr{3} \\\hline
\Acquire{m} \\
\bf \Read{x} \\
\Sync{o}\tikzmark{1} \\
\\
& \tikzmark{2}\Sync{o} \\
& \bf \Write{x} \\
& \Sync{p}\tikzmark{3} \\
\Release{m}\tikzmark{5} \\
& & \tikzmark{4}\Sync{p} \\
& & \Acquire{m} \\
& & \bf \tikzmark{6}\Write{x} \\
& & \Release{m} \\
\\
\end{tabular}%
\link{1}{2}%
\link{3}{4}%
\linkdash{5}{6}%
\end{minipage}
\label{fig:smarttrack-examples:read-write-write-extra}
}


\caption{Example executions used by the text to illustrate how \REDC computes \DC accurately.
All arrows show \DC ordering.
Dashed arrows represent ordering that would be missed without specific \RE features.
\Sync{o} represents the sequence \Acquire{o}; \Read{oVar}; \Write{oVar}; \Release{o}.}
\label{fig:smarttrack-examples}
\end{figure*}

%% file: evaluation.tex
\section{Evaluation}
\label{sec:eval}

\newcommand{\col}[1]{\emph{#1}}

This section evaluates the effectiveness of this paper's predictive analysis optimizations.

\subsection{Implementation}
\label{subsec:impl}

Table~\ref{tab:configs} presents the analyses that we have implemented and evaluated,
categorized by analysis type (row headings) and optimization level (column headings).
Each cell in the table (\eg, \emph{\CAPOO}) is an analysis that
represents the application of an algorithm (\emph{\FTO})
to an analysis type (\emph{\CAPO analysis}).

We have made all of these analysis implementations open source and publicly available.\footnote{%
\url{https://github.com/PLaSSticity/SmartTrack-pldi20}}
\mike{As of 24 March, the repository's README describes an artifact (the OOPSLA 2019 artifact, in fact)
rather than the implementation.}

We implemented the optimized analyses (\col{+~Ownership} and \col{+~CS optimizations} columns in Table~\ref{tab:configs})
based on the default \emph{FastTrack2} analysis~\cite{fasttrack2} in \emph{RoadRunner},
a dynamic analysis framework for concurrent Java programs~\cite{roadrunner}.\footnote{%
\url{https://github.com/stephenfreund/RoadRunner/releases/tag/v0.5}}
Our optimized analysis implementations minimally extend the existing FastTrack
analysis that is part of the publicly available RoadRunner implementation.

For the unoptimized analyses (\col{Unopt} column),
we used our RoadRunner-based \emph{Vindicator} implementation\footnote{%
\url{https://github.com/PLaSSticity/Vindicator}}
which implements vector-clock-based \HB, \WCP, and \DC analyses
and the vindication algorithm for checking \DC-races~\cite{vindicator}.
We extended Unopt-\DC to implement Unopt-\CAPO.

All analyses are online and detect races synchronously;
none of them build a constraint graph or perform vindication.
\notes{(We configured the \col{Unopt} analyses to \emph{not} build a constraint graph nor perform vindication.)}%
\iftoggle{techReport}{%
Appendix~\ref{appendix:baseline-confidence-interval-results} shows the cost of supporting vindication.
}{%
\notes{Our extended arXiv paper shows the cost of supporting vindication~\cite{smarttrack-extended-arxiv}.}%
}%

\paragraph{Handling events.}

In addition to handling read, write, acquire, and release events as described so far,
every analysis supports additional synchronization primitives.
Each analysis establishes order on thread fork and join;
between conflicting \code{volatile} variable accesses; and
from ``class initialized'' to ``class accessed'' events.
Each analysis treats \code{wait()} as a release
followed by an acquire.

Every analysis maintains last-access metadata at the granularity of Java memory accesses,
\ie, each object field, static field, and array element has its own last-access metadata.

\paragraph{Same-epoch cases.}

The Unopt-$\ast$ analysis implementations perform a \SharedSameEpoch-like check at reads and writes
(not shown in Algorithm~\ref{alg:DC-VC}).
Thus, the unoptimized predictive analysis implementations (Unopt-\{\WCP, \DC, \CAPO{}\})
increment $C_t(t)$ at acquires as well as releases,
just as for the optimized predictive analyses.

\paragraph{Handling races.}

In theory, the analyses handle executions up to the first race.
In practice, similar to industry-standard race detectors~\cite{google-tsan-v1,google-tsan-v2,intel-inspector},
our analysis implementations continue analyzing executions after the first race
in order to report more races to users and collect performance results for full executions.
At a race, an analysis reports the race with the static program location
that detected the race. If an analysis detects multiple races at an access
(\eg, a write races with multiple last readers),
we still count it as a single race.
After the analysis detects a race, it continues normally.

 

\paragraph{Analysis metadata.}

Each analysis processes events correctly in parallel by using
fine-grained synchronization on analysis metadata.
An analysis can forgo synchronization for an access if
a same-epoch check succeeds.
To synchronize this lock-free check correctly (\ie, fence semantics),
the read and write epochs in all analyses are \code{volatile} variables.


\begin{table*}[t]
\newcommand{\rzero}{0\xspace}
\newcommand{\ena}{\cna}
\newcommand{\roh}[1]{\ifthenelse{\equal{#1}{\rzero}}{0}{#1$\;\!\times$}} 
\newcommand{\rna}{N/A}
\newcommand{\memna}{N/A} 
\newcommand{\st}[1]{(#1~s)} 
\newcommand{\dt}[1]{#1~s} 
\newcommand{\mem}[1]{\ifthenelse{\equal{#1}{\memna}}{\rna}{#1$\;\!\times$}}
\newcommand{\et}[1]{#1M} 
\newcommand{\enfp}[1]{(#1M)} 
\newcommand{\nfpet}[2]{#2M} 
\newcommand{\base}[1]{#1~s} 
\newcommand{\tot}[1]{#1~M}
\newcommand{\per}[1]{\ifthenelse{\equal{#1}{\cna}}{\cna}{#1\%}}
\newcommand{\cna}{0} 
\newcommand{\loc}[2]{#1~K} 
\input{result-macros/PIP_fastTool_extraOpt2Stat}

\small
\centering
\begin{tabular}{@{}l|rZ|r|rr|rrr|rr||rr|rr@{}}
& & & \mc{1}{l|}{Size} & \mc{2}{c|}{Events} & \mc{3}{c|}{Locks held at NSEAs} & \mc{2}{c||}{``Ancillary'' metadata} & \mc{2}{c|}{Run time} & \mc{2}{c@{}}{Memory usage} \\
Program & \#Thr & & \mc{1}{l|}{(LoC)} & All & NSEAs & $\geq 1$ & $\geq 2$ & $\geq 3$ & Check & Use & \FTTwoAbbrv & \FTO-\HB & \FTTwoAbbrv & \FTO-\HB \\\hline 
\bench{avrora} 	& \FASTavroraTotalThreads	& (\FASTavroraMaxLiveThreads)	& \loc{69}{68,874}	 & \et{\FASTavroraEvents}   & \nfpet{\avroraRECAPONoFPOtherTotal}{\avroraRECAPONoFPEventTotal} 		& \per{\avroraRECAPOOneLockHeld} 	& \per{\avroraRECAPOTwoNestedLocks} 	& \per{\avroraRECAPOThreeNestedLocks}	& \per{\avroraRECAPOExTotalCheck} 	& \per{\avroraRECAPOExTotalUpdate} & \roh{\FASTavroraHBTime}	& \roh{\FASTavroraFTOHBTime} & \mem{\FASTavroraHBMem}	& \mem{\FASTavroraFTOHBMem} \\
\bench{batik} 	& \FASTbatikTotalThreads 	& (\FASTbatikMaxLiveThreads)	& \loc{188}{187,715} & \et{\FASTbatikEvents}    & \nfpet{\batikRECAPONoFPOtherTotal}{\batikRECAPONoFPEventTotal}		& \per{\batikRECAPOOneLockHeld} 	& \per{\batikRECAPOTwoNestedLocks} 		& \per{\batikRECAPOThreeNestedLocks} 	& \per{\batikRECAPOExTotalCheck} 	& \per{\batikRECAPOExTotalUpdate} & \roh{\FASTbatikHBTime}   	& \roh{\FASTbatikFTOHBTime} & \mem{\FASTbatikHBMem}   	& \mem{\FASTbatikFTOHBMem} \\
\bench{h2} 		& \FASThtwoTotalThreads 	& (\FASThtwoMaxLiveThreads)	    & \loc{116}{115,865} & \et{\FASThtwoEvents}     & \nfpet{\htwoRECAPONoFPOtherTotal}{\htwoRECAPONoFPEventTotal}			& \per{\htwoRECAPOOneLockHeld} 		& \per{\htwoRECAPOTwoNestedLocks} 		& \per{\htwoRECAPOThreeNestedLocks} 	& \per{\htwoRECAPOExTotalCheck}		& \per{\htwoRECAPOExTotalUpdate} & \roh{\FASThtwoHBTime}    	& \roh{\FASThtwoFTOHBTime} & \mem{\FASThtwoHBMem}    	& \mem{\FASThtwoFTOHBMem} \\
\bench{jython} 	& \FASTjythonTotalThreads	& (\FASTjythonMaxLiveThreads)	& \loc{212}{211,905} & \et{\FASTjythonEvents}   & \nfpet{\jythonRECAPONoFPOtherTotal}{\jythonRECAPONoFPEventTotal}		& \per{\jythonRECAPOOneLockHeld} 	& \per{\jythonRECAPOTwoNestedLocks} 	& \per{\jythonRECAPOThreeNestedLocks} 	& \per{\jythonRECAPOExTotalCheck} 	& \per{\jythonRECAPOExTotalUpdate} & \roh{\FASTjythonHBTime} 	& \roh{\FASTjythonFTOHBTime} & \mem{\FASTjythonHBMem} 	& \mem{\FASTjythonFTOHBMem} \\
\bench{luindex} & \FASTluindexTotalThreads	& (\FASTluindexMaxLiveThreads)	& \loc{126}{126,209} & \et{\FASTluindexEvents}  & \nfpet{\luindexRECAPONoFPOtherTotal}{\luindexRECAPONoFPEventTotal}	& \per{\luindexRECAPOOneLockHeld} 	& \per{\luindexRECAPOTwoNestedLocks} 	& \per{\luindexRECAPOThreeNestedLocks} 	& \per{\luindexRECAPOExTotalCheck} 	& \per{\luindexRECAPOExTotalUpdate} & \roh{\FASTluindexHBTime} 	& \roh{\FASTluindexFTOHBTime} & \mem{\FASTluindexHBMem} 	& \mem{\FASTluindexFTOHBMem} \\
\bench{lusearch}& \FASTlusearchTotalThreads	& (\FASTlusearchMaxLiveThreads)	& \loc{126}{126,386} & \et{\FASTlusearchEvents} & \nfpet{\lusearchRECAPONoFPOtherTotal}{\lusearchRECAPONoFPEventTotal}	& \per{\lusearchRECAPOOneLockHeld}	& \per{\lusearchRECAPOTwoNestedLocks} 	& \per{\lusearchRECAPOThreeNestedLocks} & \per{\lusearchRECAPOExTotalCheck} & \per{\lusearchRECAPOExTotalUpdate} & \roh{\FASTlusearchHBTime}	& \roh{\FASTlusearchFTOHBTime} & \mem{\FASTlusearchHBMem}	& \mem{\FASTlusearchFTOHBMem} \\
\bench{pmd}		& \FASTpmdTotalThreads	    & (\FASTpmdMaxLiveThreads)	    & \loc{61}{60,747} 	 & \et{\FASTpmdEvents}      & \nfpet{\pmdRECAPONoFPOtherTotal}{\pmdRECAPONoFPEventTotal}			& \per{\pmdRECAPOOneLockHeld} 		& \per{\pmdRECAPOTwoNestedLocks} 		& \per{\pmdRECAPOThreeNestedLocks} 		& \per{\pmdRECAPOExTotalCheck} 		& \per{\pmdRECAPOExTotalUpdate} & \roh{\FASTpmdHBTime}		& \roh{\FASTpmdFTOHBTime} & \mem{\FASTpmdHBMem}		& \mem{\FASTpmdFTOHBMem} \\
\bench{sunflow} & \FASTsunflowTotalThreads	& (\FASTsunflowMaxLiveThreads)	& \loc{22}{21,970} 	 & \et{\FASTsunflowEvents}  & \nfpet{\sunflowRECAPONoFPOtherTotal}{\sunflowRECAPONoFPEventTotal}	& \per{\sunflowRECAPOOneLockHeld} 	& \per{\sunflowRECAPOTwoNestedLocks} 	& \per{\sunflowRECAPOThreeNestedLocks} 	& \per{\sunflowRECAPOExTotalCheck} 	& \per{\sunflowRECAPOExTotalUpdate} & \roh{\FASTsunflowHBTime}	& \roh{\FASTsunflowFTOHBTime} & \mem{\FASTsunflowHBMem}	& \mem{\FASTsunflowFTOHBMem} \\
\bench{tomcat} 	& \FASTtomcatTotalThreads	& (\FASTtomcatMaxLiveThreads)	& \loc{159}{159,028} & \et{\FASTtomcatEvents}   & \nfpet{\tomcatRECAPONoFPOtherTotal}{\tomcatRECAPONoFPEventTotal}		& \per{\tomcatRECAPOOneLockHeld} 	& \per{\tomcatRECAPOTwoNestedLocks} 	& \per{\tomcatRECAPOThreeNestedLocks} 	& \per{\tomcatRECAPOExTotalCheck} 	& \per{\tomcatRECAPOExTotalUpdate} & \roh{\FASTtomcatHBTime} 	& \roh{\FASTtomcatFTOHBTime} & \mem{\FASTtomcatHBMem} 	& \mem{\FASTtomcatFTOHBMem} \\
\bench{xalan} 	& \FASTxalanTotalThreads	& (\FASTxalanMaxLiveThreads)	& \loc{176}{175,784} & \et{\FASTxalanEvents}    & \nfpet{\xalanRECAPONoFPOtherTotal}{\xalanRECAPONoFPEventTotal}		& \per{\xalanRECAPOOneLockHeld} 	& \per{\xalanRECAPOTwoNestedLocks} 		& \per{\xalanRECAPOThreeNestedLocks} 	& \per{\xalanRECAPOExTotalCheck} 	& \per{\xalanRECAPOExTotalUpdate} & \roh{\FASTxalanHBTime}		& \roh{\FASTxalanFTOHBTime} & \mem{\FASTxalanHBMem}		& \mem{\FASTxalanFTOHBMem} \\
\end{tabular}

\later{
\jake{TODO: Test thread counts as RoadRunner does not correctly represent
thread count values using the -noTidGC option (maxActive field should dec at join.\\
Another possibility to represent the data is to provide total thread counts
without maximum active thread counts or provide a range of maximum active
thread counts.
\mike{Addressed by removing incorrect maximum active thread counts.}}
}

\caption{Run-time characteristics of the evaluated programs. NSEAs are \emph{non-same-epoch accesses}.
The last two major columns report run time and memory usage for FastTrack-based \HB analyses, relative to uninstrumented execution.}
\label{tab:characteristics:bench-stats}
\end{table*}

\subsection{Methodology}

Our evaluation uses the DaCapo
benchmarks, version 9.12-bach, which are real, widely used concurrent programs
that have been harnessed for evaluating performance~\cite{dacapo-benchmarks-conf}.
While the DaCapo suite is not expressly intended for evaluating data race detection,
the programs do contain data races.

RoadRunner bundles a version
of the DaCapo benchmarks, modified to work with RoadRunner,
that executes workloads similar to the default workloads. RoadRunner
does not currently support \bench{eclipse}, \bench{tradebeans}, or
\bench{tradesoap}, and \bench{fop} is single threaded, so our evaluation excludes those programs.

The experiments run on a quiet system with an Intel Xeon E5-2683 14-core
processor with hyperthreading disabled and 256 GB of main memory running Linux 3.10.0. 
We run the implementations with the HotSpot 1.8.0 JVM and let it choose and adjust the heap size on the fly.

Each reported performance result, race count, or frequency statistic for an evaluated program
is the arithmetic mean of 10 trials.
We measure execution time as wall-clock time within the benchmarked harness of the evaluated program,
and memory usage as the maximum resident set size during execution according to the GNU \code{time} program.
We measure time, memory, and races within the same runs,
and frequency statistics in separate statistics-collecting runs.

\iftoggle{techReport}{%
Appendices~\ref{appendix:detailed-performance-results}--\ref{appendix:smarttrack-stats} provide
}{%
Our extended arXiv paper provides
}%
detailed performance results,
predictable race coverage results,
vindication results,
95\% confidence intervals for all results,
and frequency statistics for \RE algorithm cases\iftoggle{techReport}{}{~\cite{smarttrack-extended-arxiv}}.


\subsection{Run-Time Characteristics}

Table~\ref{tab:characteristics:bench-stats} shows run-time characteristics relevant to the analyses.
The \col{\#Thr} column shows the total number of threads created.
\later{and, in parentheses, the maximum number of active threads at any time.\footnote{RoadRunner
detects terminated threads lazily, leading to nondeterminism in its accounting of maximum active threads.}}%
\col{Events} are the total executed program events (\col{All})
and non-same-epoch accesses (\col{NSEAs}).

The \col{Locks held at NSEAs} columns
report percentages of non-same-epoch accesses holding at least one, two, or three locks, respectively.
These counts are important because non-\RE predictive analyses 
perform substantial work per held lock at non-same-epoch accesses.
While all programs generally benefit from epoch and ownership optimizations,
only programs that perform many accesses holding one or more locks 
benefit substantially from CCS optimizations.
Notably, \bench{h2}, \bench{luindex}, and \bench{xalan} have the
highest average locks held per access.
Unsurprisingly, these programs have the highest
\FTO-based predictive analysis overhead
and benefit the most from \RE's optimizations (Section~\ref{subsec:perf}).

The \col{``Ancillary'' metadata} columns report percentages of non-same-epoch accesses that 
detect non-null ancillary metadata at a \col{Check}
(lines~\ref{line:RE-VC:rdExtraAtWrStart} and \ref{line:RE-VC:wrExtraAtRdStart} in Algorithm~\ref{alg:RE-VC})
and that \col{Use} ancillary metadata to add critical section ordering 
(lines~\ref{line:RE-VC:rdExtraAtWrUpdate} and \ref{line:RE-VC:wrExtraAtRdUpdate} in Algorithm~\ref{alg:RE-VC}).
Ancillary metadata is rarely if ever \emph{used},
but some programs perform a significant number of \emph{checks},
which can degrade performance.

\subsection{Comparing Baselines}
\label{sec:eval:fasttrack}

\iftoggle{twoColumnText}{
}{
\begin{table}
\newcommand{\rzero}{0\xspace}
\newcommand{\roh}[1]{\ifthenelse{\equal{#1}{\rzero}}{0}{#1$\;\!\times$}} 
\newcommand{\rna}{N/A}
\newcommand{\memna}{N/A} 
\newcommand{\st}[1]{(#1~s)} 
\newcommand{\dt}[1]{#1~s} 
\newcommand{\mem}[1]{\ifthenelse{\equal{#1}{\memna}}{\rna}{#1$\;\!\times$}}
\newcommand{\baseMem}[1]{#1~MB} 
\newcommand{\et}[1]{#1M} 
\newcommand{\enfp}[1]{(#1M)} 
\newcommand{\base}[1]{#1~s} 
\input{result-macros/PIP_slowTool_noCoresSet}
\input{result-macros/PIP_fastTool_extraOpt2Quiet}
\small
\centering
\newcommand{\ics}{\;}
\begin{tabular}{@{}l|HHHHHHcc|rr|rr@{}}
& \mc{2}{Z}{} & \mc{2}{Z}{} & Base & \mc{3}{c|}{\HB} & \mc{2}{c|}{Unopt-\DC} & \mc{2}{c@{}}{Unopt-\CAPO} \\
Program & \mc{2}{Z}{Events} & \mc{2}{Z}{\#Thr} & time & RR\FTTwoAbbrv & \FTTwoAbbrv & \FTO & w/ G & w/o G & w/ G & w/o G \\\hline 
\bench{avrora} 	& \et{\FASTavroraEvents}   & \enfp{\FASTavroraNoFPEvents}	& \FASTavroraTotalThreads 	& (\FASTavroraMaxLiveThreads)	& \base{\FASTavroraBaseTime}   & \roh{\FASTavroraFTTime}   & \roh{\FASTavroraHBTime}	& \roh{\FASTavroraFTOHBTime}	& \roh{\SLOWavroraDCExcTime}	& \roh{\SLOWavroraDCnoGExcTime}	& \roh{\SLOWavroraCAPOExcTime}	& \roh{\SLOWavroraCAPOnoGExcTime}\\
\bench{batik} 	& \et{\FASTbatikEvents}    & \enfp{\FASTbatikNoFPEvents}	& \FASTbatikTotalThreads 	& (\FASTbatikMaxLiveThreads)	& \base{\FASTbatikBaseTime}    & \roh{\FASTbatikFTTime}    & \roh{\FASTbatikHBTime}   	& \roh{\FASTbatikFTOHBTime}		& \roh{\SLOWbatikDCExcTime}	& \roh{\SLOWbatikDCnoGExcTime}		& \roh{\SLOWbatikCAPOExcTime}	& \roh{\SLOWbatikCAPOnoGExcTime} \\
\bench{h2} 		& \et{\FASThtwoEvents}     & \enfp{\FASThtwoNoFPEvents}	    & \FASThtwoTotalThreads 	& (\FASThtwoMaxLiveThreads)	    & \base{\FASThtwoBaseTime}     & \roh{\FASThtwoFTTime}     & \roh{\FASThtwoHBTime}    	& \roh{\FASThtwoFTOHBTime}		& \roh{\SLOWhtwoDCExcTime}		& \roh{\SLOWhtwoDCnoGExcTime}		& \roh{\SLOWhtwoCAPOExcTime}	& \roh{\SLOWhtwoCAPOnoGExcTime}\\
\bench{jython} 	& \et{\FASTjythonEvents}   & \enfp{\FASTjythonNoFPEvents}	& \FASTjythonTotalThreads	& (\FASTjythonMaxLiveThreads)	& \base{\FASTjythonBaseTime}   & \roh{\FASTjythonFTTime}   & \roh{\FASTjythonHBTime} 	& \roh{\FASTjythonFTOHBTime}	& \roh{\SLOWjythonDCExcTime}	& \roh{\SLOWjythonDCnoGExcTime}	& \roh{\SLOWjythonCAPOExcTime}	& \roh{\SLOWjythonCAPOnoGExcTime} \\
\bench{luindex} & \et{\FASTluindexEvents}  & \enfp{\FASTluindexNoFPEvents}	& \FASTluindexTotalThreads	& (\FASTluindexMaxLiveThreads)	& \base{\FASTluindexBaseTime}  & \roh{\FASTluindexFTTime}  & \roh{\FASTluindexHBTime} 	& \roh{\FASTluindexFTOHBTime}	& \roh{\SLOWluindexDCExcTime}	& \roh{\SLOWluindexDCnoGExcTime}	& \roh{\SLOWluindexCAPOExcTime}	& \roh{\SLOWluindexCAPOnoGExcTime} \\
\bench{lusearch}& \et{\FASTlusearchEvents} & \enfp{\FASTlusearchNoFPEvents}	& \FASTlusearchTotalThreads	& (\FASTlusearchMaxLiveThreads)	& \base{\FASTlusearchBaseTime} & \roh{\FASTlusearchFTTime} & \roh{\FASTlusearchHBTime}	& \roh{\FASTlusearchFTOHBTime}	& \roh{\SLOWlusearchDCExcTime}	& \roh{\SLOWlusearchDCnoGExcTime}	& \roh{\SLOWlusearchCAPOExcTime}& \roh{\SLOWlusearchCAPOnoGExcTime} \\
\bench{pmd}		& \et{\FASTpmdEvents}      & \enfp{\FASTpmdNoFPEvents}      & \FASTpmdTotalThreads	    & (\FASTpmdMaxLiveThreads)	    & \base{\FASTpmdBaseTime}      & \roh{\FASTpmdFTTime}      & \roh{\FASTpmdHBTime}		& \roh{\FASTpmdFTOHBTime}		& \roh{\SLOWpmdDCExcTime}		& \roh{\SLOWpmdDCnoGExcTime}		& \roh{\SLOWpmdCAPOExcTime}		& \roh{\SLOWpmdCAPOnoGExcTime} \\
\bench{sunflow} & \et{\FASTsunflowEvents}  & \enfp{\FASTsunflowNoFPEvents}	& \FASTsunflowTotalThreads	& (\FASTsunflowMaxLiveThreads)	& \base{\FASTsunflowBaseTime}  & \roh{\FASTsunflowFTTime}  & \roh{\FASTsunflowHBTime}	& \roh{\FASTsunflowFTOHBTime}	& \roh{\SLOWsunflowDCExcTime}	& \roh{\SLOWsunflowDCnoGExcTime}	& \roh{\SLOWsunflowCAPOExcTime}	& \roh{\SLOWsunflowCAPOnoGExcTime} \\
\bench{tomcat} 	& \et{\FASTtomcatEvents}   & \enfp{\FASTtomcatNoFPEvents}	& \FASTtomcatTotalThreads	& (\FASTtomcatMaxLiveThreads)	& \base{\FASTtomcatBaseTime}   & \roh{\FASTtomcatFTTime}   & \roh{\FASTtomcatHBTime} 	& \roh{\FASTtomcatFTOHBTime}	& \roh{\SLOWtomcatDCExcTime}	& \roh{\SLOWtomcatDCnoGExcTime}	& \roh{\SLOWtomcatCAPOExcTime}	& \roh{\SLOWtomcatCAPOnoGExcTime} \\
\bench{xalan} 	& \et{\FASTxalanEvents}    & \enfp{\FASTxalanNoFPEvents}	& \FASTxalanTotalThreads	& (\FASTxalanMaxLiveThreads)	& \base{\FASTxalanBaseTime}    & \roh{\FASTxalanFTTime}    & \roh{\FASTxalanHBTime}		& \roh{\FASTxalanFTOHBTime}		& \roh{\SLOWxalanDCExcTime}	& \roh{\SLOWxalanDCnoGExcTime} 	& \roh{\SLOWxalanCAPOExcTime}	& \roh{\SLOWxalanCAPOnoGExcTime}\\\hline
geomean	& &	& &	& & \roh{\FASTFTTimeGeoMean} & \roh{\FASTHBTimeGeoMean} & \roh{\FASTFTOHBTimeGeoMean} & \roh{\SLOWDCExcTimeGeoMean} & \roh{\SLOWDCnoGExcTimeGeoMean} & \roh{\SLOWCAPOExcTimeGeoMean} & \roh{\SLOWCAPOnoGExcTimeGeoMean} \\
\mc{1}{c|}{} & \mc{12}{c}{Run time} \\
\end{tabular}%
\hfill\mbox{\vline height 3cm depth 2.5cm}\hfill\null
\begin{tabular}{@{}HHHHHHHcc|rr|rr@{}}
& \mc{2}{Z}{} & \mc{2}{Z}{} & Base & \mc{3}{c|}{\HB} & \mc{2}{c|}{Unopt-\DC} & \mc{2}{c@{}}{Unopt-\CAPO} \\
Program & \mc{2}{Z}{Events} & \mc{2}{Z}{\#Thr} & memory & RR\FTTwoAbbrv & \FTTwoAbbrv & \FTO & w/ G & w/o G & w/ G & w/o G \\\hline 
\bench{avrora} 	& \et{\FASTavroraEvents}   & \enfp{\FASTavroraNoFPEvents}	& \FASTavroraTotalThreads 	& (\FASTavroraMaxLiveThreads)	& \baseMem{\FASTavroraBaseMem}   & \mem{\FASTavroraFTMem}   & \mem{\FASTavroraHBMem}	& \mem{\FASTavroraFTOHBMem}	& \mem{\SLOWavroraDCExcMem}	& \mem{\SLOWavroraDCnoGExcMem}	& \mem{\SLOWavroraCAPOExcMem}	& \mem{\SLOWavroraCAPOnoGExcMem}\\
\bench{batik} 	& \et{\FASTbatikEvents}    & \enfp{\FASTbatikNoFPEvents}	& \FASTbatikTotalThreads 	& (\FASTbatikMaxLiveThreads)	& \baseMem{\FASTbatikBaseMem}    & \mem{\FASTbatikFTMem}    & \mem{\FASTbatikHBMem}   	& \mem{\FASTbatikFTOHBMem}		& \mem{\SLOWbatikDCExcMem}	& \mem{\SLOWbatikDCnoGExcMem}		& \mem{\SLOWbatikCAPOExcMem}	& \mem{\SLOWbatikCAPOnoGExcMem} \\
\bench{h2} 		& \et{\FASThtwoEvents}     & \enfp{\FASThtwoNoFPEvents}	    & \FASThtwoTotalThreads 	& (\FASThtwoMaxLiveThreads)	    & \baseMem{\FASThtwoBaseMem}     & \mem{\FASThtwoFTMem}     & \mem{\FASThtwoHBMem}    	& \mem{\FASThtwoFTOHBMem}		& \mem{\SLOWhtwoDCExcMem}		& \mem{\SLOWhtwoDCnoGExcMem}		& \mem{\SLOWhtwoCAPOExcMem}	& \mem{\SLOWhtwoCAPOnoGExcMem}\\
\bench{jython} 	& \et{\FASTjythonEvents}   & \enfp{\FASTjythonNoFPEvents}	& \FASTjythonTotalThreads	& (\FASTjythonMaxLiveThreads)	& \baseMem{\FASTjythonBaseMem}   & \mem{\FASTjythonFTMem}   & \mem{\FASTjythonHBMem} 	& \mem{\FASTjythonFTOHBMem}	& \mem{\SLOWjythonDCExcMem}	& \mem{\SLOWjythonDCnoGExcMem}	& \mem{\SLOWjythonCAPOExcMem}	& \mem{\SLOWjythonCAPOnoGExcMem} \\
\bench{luindex} & \et{\FASTluindexEvents}  & \enfp{\FASTluindexNoFPEvents}	& \FASTluindexTotalThreads	& (\FASTluindexMaxLiveThreads)	& \baseMem{\FASTluindexBaseMem}  & \mem{\FASTluindexFTMem}  & \mem{\FASTluindexHBMem} 	& \mem{\FASTluindexFTOHBMem}	& \mem{\SLOWluindexDCExcMem}	& \mem{\SLOWluindexDCnoGExcMem}	& \mem{\SLOWluindexCAPOExcMem}	& \mem{\SLOWluindexCAPOnoGExcMem} \\
\bench{lusearch}& \et{\FASTlusearchEvents} & \enfp{\FASTlusearchNoFPEvents}	& \FASTlusearchTotalThreads	& (\FASTlusearchMaxLiveThreads)	& \baseMem{\FASTlusearchBaseMem} & \mem{\FASTlusearchFTMem} & \mem{\FASTlusearchHBMem}	& \mem{\FASTlusearchFTOHBMem}	& \mem{\SLOWlusearchDCExcMem}	& \mem{\SLOWlusearchDCnoGExcMem}	& \mem{\SLOWlusearchCAPOExcMem}& \mem{\SLOWlusearchCAPOnoGExcMem} \\
\bench{pmd}		& \et{\FASTpmdEvents}      & \enfp{\FASTpmdNoFPEvents}      & \FASTpmdTotalThreads	    & (\FASTpmdMaxLiveThreads)	    & \baseMem{\FASTpmdBaseMem}      & \mem{\FASTpmdFTMem}      & \mem{\FASTpmdHBMem}		& \mem{\FASTpmdFTOHBMem}		& \mem{\SLOWpmdDCExcMem}		& \mem{\SLOWpmdDCnoGExcMem}		& \mem{\SLOWpmdCAPOExcMem}		& \mem{\SLOWpmdCAPOnoGExcMem} \\
\bench{sunflow} & \et{\FASTsunflowEvents}  & \enfp{\FASTsunflowNoFPEvents}	& \FASTsunflowTotalThreads	& (\FASTsunflowMaxLiveThreads)	& \baseMem{\FASTsunflowBaseMem}  & \mem{\FASTsunflowFTMem}  & \mem{\FASTsunflowHBMem}	& \mem{\FASTsunflowFTOHBMem}	& \mem{\SLOWsunflowDCExcMem}	& \mem{\SLOWsunflowDCnoGExcMem}	& \mem{\SLOWsunflowCAPOExcMem}	& \mem{\SLOWsunflowCAPOnoGExcMem} \\
\bench{tomcat} 	& \et{\FASTtomcatEvents}   & \enfp{\FASTtomcatNoFPEvents}	& \FASTtomcatTotalThreads	& (\FASTtomcatMaxLiveThreads)	& \baseMem{\FASTtomcatBaseMem}   & \mem{\FASTtomcatFTMem}   & \mem{\FASTtomcatHBMem} 	& \mem{\FASTtomcatFTOHBMem}	& \mem{\SLOWtomcatDCExcMem}	& \mem{\SLOWtomcatDCnoGExcMem}	& \mem{\SLOWtomcatCAPOExcMem}	& \mem{\SLOWtomcatCAPOnoGExcMem} \\
\bench{xalan} 	& \et{\FASTxalanEvents}    & \enfp{\FASTxalanNoFPEvents}	& \FASTxalanTotalThreads	& (\FASTxalanMaxLiveThreads)	& \baseMem{\FASTxalanBaseMem}    & \mem{\FASTxalanFTMem}    & \mem{\FASTxalanHBMem}		& \mem{\FASTxalanFTOHBMem}		& \mem{\SLOWxalanDCExcMem}	& \mem{\SLOWxalanDCnoGExcMem} 	& \mem{\SLOWxalanCAPOExcMem}	& \mem{\SLOWxalanCAPOnoGExcMem}\\\hline
geomean	& &	& &	& & \mem{\FASTFTMemGeoMean} & \mem{\FASTHBMemGeoMean} & \mem{\FASTFTOHBMemGeoMean} & \mem{\SLOWDCExcMemGeoMean} & \mem{\SLOWDCnoGExcMemGeoMean} & \mem{\SLOWCAPOExcMemGeoMean} & \mem{\SLOWCAPOnoGExcMemGeoMean} \\
\mc{13}{c}{Memory usage} \\
\end{tabular}

\caption{Run time and memory usage for FastTrack-based \HB analyses
and for unoptimized \DC and \CAPO analyses, relative to uninstrumented execution.}
\label{tab:performance:fasttrack}
\end{table}
}

\iftoggle{twoColumnText}{
The rightmost columns of Table~\ref{tab:characteristics:bench-stats}}{
Table~\ref{tab:performance:fasttrack}}
show results that help determine whether we are using valid baselines compared with prior work.
\iftoggle{twoColumnText}{%
\emph{Run time}}{%
The left side of the table}
reports slowdowns relative to uninstrumented (unanalyzed) execution,
and
\iftoggle{twoColumnText}{%
\emph{Memory usage}}{%
the right side} 
reports memory used relative to uninstrumented execution.

\paragraph{FastTrack comparison.}

The
\iftoggle{twoColumnText}{%
\col{Run time} and \col{Memory usage}
}{%
\col{\HB}}%
columns report the performance of two variants of the FastTrack algorithm.
\col{\FTTwoAbbrv} is our implementation of the \FTTwo algorithm~\cite{fasttrack2},
based closely on RoadRunner's implementation of \FTTwo,
which is the default FastTrack tool in RoadRunner. 
The main difference between \FTTwoAbbrv and RoadRunner's \FTTwo
lies in how they handle detected races.
RoadRunner's \FTTwo does not update last-access metadata at read (but not write) events that detect a race (for unknown reasons);
it does not perform analysis on future accesses to a variable after it detects a race on the variable; and
it limits the number of races it counts by class field and array type.
In contrast, our \FTTwoAbbrv updates last-access metadata after every event even if it detects a race;
it does not stop performing analysis on any events; and
it counts every race.

\col{\FTO} is our implementation of \HBO analysis, implemented in the same RoadRunner tool as \col{\FTTwoAbbrv}.
Overall \HBO performs quite similarly to \FTTwoAbbrv.
The rest of the paper's results compare against \HBO as the representative from the FastTrack family of optimized \HB analyses.

\iftoggle{twoColumnText}{
\notes{The rest of the results evaluate whether our optimizations bridge the performance gap
between prior work's unoptimized predictive analyses and optimized non-predictive (\HB) analysis.}
}{
\paragraph{Unoptimized analyses.}

Table~\ref{tab:performance:fasttrack}'s \col{Unopt-$\ast$} columns compare the performance of
unoptimized \DC and \CAPO analyses, with and without support for vindication.
To support verifying that \DC-races (or \CAPO-races) are (true) predictable races,
\DC (\CAPO) analysis can build a constraint graph $G$ during the analysis.
\col{Unopt-\DC w/$G$} represents the cost incurred by prior work to detect \DC-races and be able to check them after execution.
It also represents the cost that would be incurred by a replayed execution that builds $G$
in order to verify \DC-races
detected by a recorded run that used \REDC analysis or some other \DC analysis that does \emph{not} build $G$ (Section~\ref{subsec:record-replay}).
Likewise, \col{Unopt-\CAPO w/$G$} shows the cost of a replayed execution checking \CAPO-races.

\col{Unopt-\DC w/o $G$} represents the cost incurred by prior work to detect \DC-races without being able to check them,
which is still useful because few if any \DC-races are false positives in practice, and a second replayed run can optionally check \DC-races.
Likewise, \col{Unopt-\CAPO w/o $G$} shows the cost of detecting \CAPO-races without being able to check them.
The rest of the results compare our optimized analyses against unoptimized analyses that do \emph{not} build a constraint graph (\col{Unopt-$\ast$ w/o $G$}).

The results show that the costs of unoptimized predictive analyses are high, whether or not they build a constraint graph,
compared with existing optimized non-predictive (\HB) analyses. The rest of the results evaluate whether our optimizations help to bridge this
performance gap.
}

\subsection{Run-Time and Memory Performance}
\label{subsec:perf}

This section evaluates the performance of our optimized analyses,
compared with competing approaches from prior work.
\later{IF FIGURE EXCLUDED: Table~\ref{tab:performance:geoMean} and
Figure~\ref{fig:hbNorm:allDaCapo}
present two different views of the paper's main results:
run time and memory usage of the 11 analyses from Table~\ref{tab:configs}.
Table~\ref{tab:performance:geoMean} provides the cost of analyzing a program with each analysis and
Figure~\ref{fig:hbNorm:allDaCapo} directly evaluates the performance gap between predictive analysis
and non-predictive optimized \FTO-\HB analysis.
\jake{Do you mean if we keep it in the Appendix? I think that's fine.
\mike{Yeah}}
}%
Table~\ref{tab:performance:geoMean}
presents the paper's main results:
run time and memory usage of the 11 analyses from Table~\ref{tab:configs}.
\iftoggle{techReport}{%
Appendix~\ref{appendix:detailed-performance-results}
presents the performance results normalized to \HBO
and shows results for each program.
}{%
\notes{Our extended arXiv paper
presents the performance results normalized to \HBO
and shows results for each program~\cite{smarttrack-extended-arxiv}.}%
}%

The table reports relative run time and memory usage across all programs.
For example, a cell in column \col{\RE-} and row \col{\DC} shows slowdown or memory usage of \REDC analysis
relative to uninstrumented execution.
\notes{Note that we do not evaluate vindication when comparing unoptimized to optimized 
analyses since the column \col{Unopt-} represents \col{Unopt-$\ast$ w/o G}.
\mike{Seems like readers won't be wondering about this (the paper hasn't talked much about vindication).}}%
\later{Figure~\ref{fig:hbNorm:allDaCapo}
shows the same results, but normalized to \FTO-\HB execution (and omitting Unopt-\HB),
with separate results for each program.
For example, \bench{avrora} shows three groups of bars corresponding to \WCP, \DC, and \CAPO analyses
where each group shows three bars corresponding to the optimization level applied to each analysis.
Appendix~\ref{appendix:confidence-interval-results} presents the same results in a third view,
Tables~\ref{tab:performance:allDaCapo:CI} and \ref{tab:memory:allDaCapo:CI}
provides each separate program result for Table~\ref{tab:performance:geoMean} with 95\% confidence interavls.}%

\begin{table}[t]
\newcommand{\rzero}{0\xspace}
\newcommand{\roh}[1]{\ifthenelse{\equal{#1}{\rna}}{\rna}{#1$\;\!\times$}} 
\newcommand{\rna}{N/A}
\newcommand{\memna}{N/A} 
\newcommand{\st}[1]{(#1~s)} 
\newcommand{\dt}[1]{#1~s} 
\newcommand{\mem}[1]{\ifthenelse{\equal{#1}{\memna}}{\memna}{#1$\;\!\times$}} 
\newcommand{\et}[1]{#1M} 
\newcommand{\enfp}[1]{(#1M)} 
\newcommand{\base}[1]{#1~s} 
\input{result-macros/PIP_slowTool_noCoresSet}
\input{result-macros/PIP_fastTool_extraOpt2Quiet}
\small
\centering
\begin{tabular}{@{}l|Hrr@{\;\;}r@{}}
        & w/ G & Unopt- & \FTO- & \footnotesize \RE- \\\hline
\HB		& \rna							& \roh{\SLOWHBTimeGeoMean}			& \roh{\FASTFTOHBTimeGeoMean}	& \rna \\
\WCP	& \rna							& \roh{\SLOWWCPTimeGeoMean}			& \roh{\FASTFTOWCPTimeGeoMean}	& \roh{\FASTREWCPTimeGeoMean} \\
\DC		& \roh{\SLOWDCExcTimeGeoMean}	& \roh{\SLOWDCnoGExcTimeGeoMean}	& \roh{\FASTFTODCTimeGeoMean}	& \roh{\FASTREDCTimeGeoMean} \\
\CAPO	& \roh{\SLOWCAPOExcTimeGeoMean}	& \roh{\SLOWCAPOnoGExcTimeGeoMean}	& \roh{\FASTFTOCAPOTimeGeoMean}	& \roh{\FASTRECAPOTimeGeoMean} \\
\mc{2}{H}{} & \mc{3}{c}{Run time} \\
\end{tabular}
\hfill
\begin{tabular}{@{}HHrr@{\;\;}r@{}}
        & w/ G & Unopt- & \FTO- & \footnotesize \RE- \\\hline
\HB		& \rna							& \mem{\SLOWHBMemGeoMean}			& \mem{\FASTFTOHBMemGeoMean}	& \rna \\
\WCP	& \rna							& \mem{\SLOWWCPMemGeoMean}			& \mem{\FASTFTOWCPMemGeoMean}	& \mem{\FASTREWCPMemGeoMean} \\
\DC		& \mem{\SLOWDCExcMemGeoMean}	& \mem{\SLOWDCnoGExcMemGeoMean}	& \mem{\FASTFTODCMemGeoMean}	& \mem{\FASTREDCMemGeoMean} \\
\CAPO	& \mem{\SLOWCAPOExcMemGeoMean}	& \mem{\SLOWCAPOnoGExcMemGeoMean}	& \mem{\FASTFTOCAPOMemGeoMean}	& \mem{\FASTRECAPOMemGeoMean} \\
\mc{2}{@{}H}{} & \mc{3}{c}{Memory usage} \\
\end{tabular}


\caption{Geometric mean of run time and memory usage across the evaluated programs.}
\label{tab:performance:geoMean}
\end{table}

\iftoggle{twoColumnText}{
\iftoggle{includeRaceResults}{
\begin{figure*}[t]
\centering
\subfloat[Run time]{
\includegraphics[width=0.9\textwidth,height=0.45\textwidth]{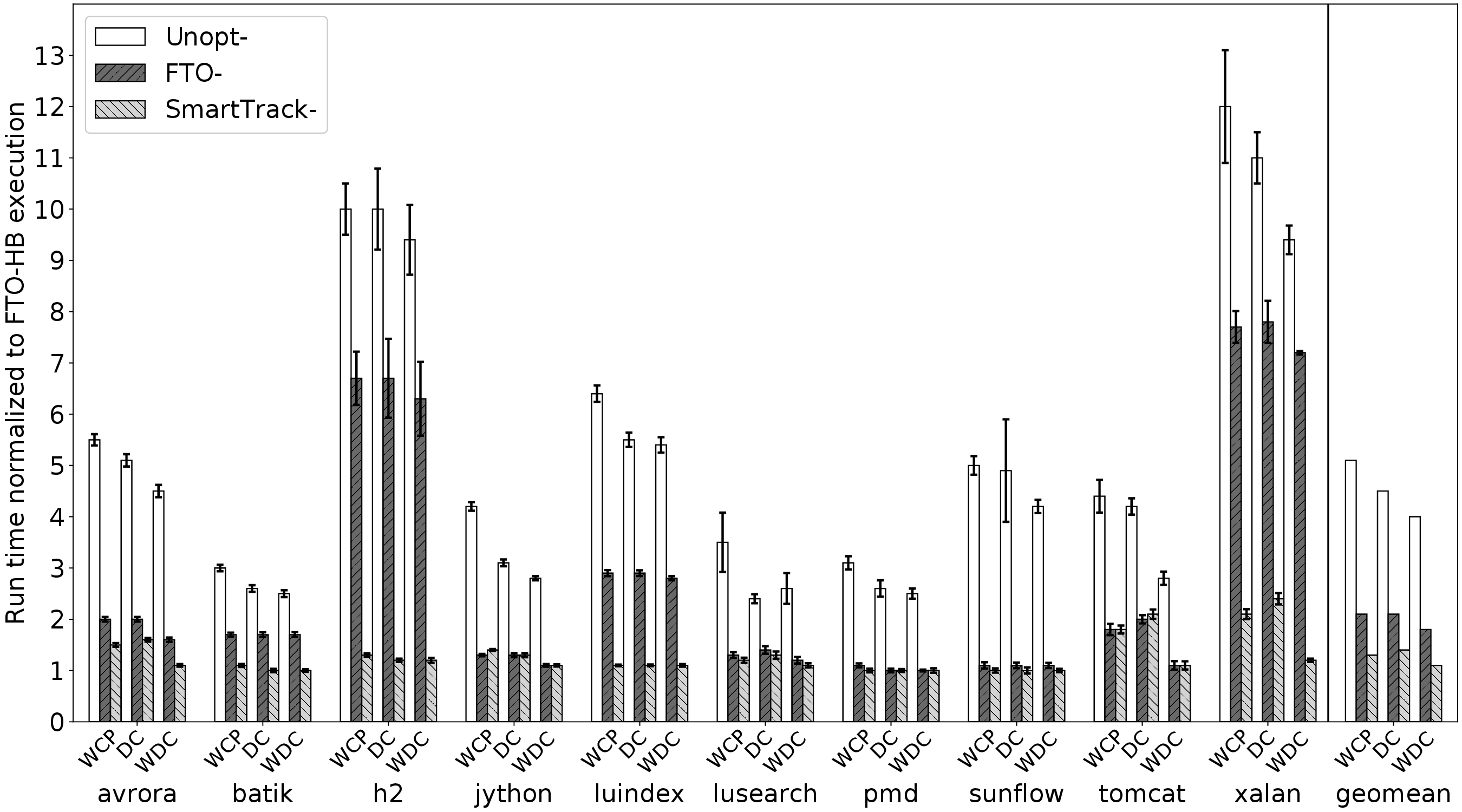}
\label{fig:runtime-hbNorm:allDaCapo}
}
\hfill
\centering
\subfloat[Memory usage]{
\includegraphics[width=0.9\textwidth,height=0.45\textwidth]{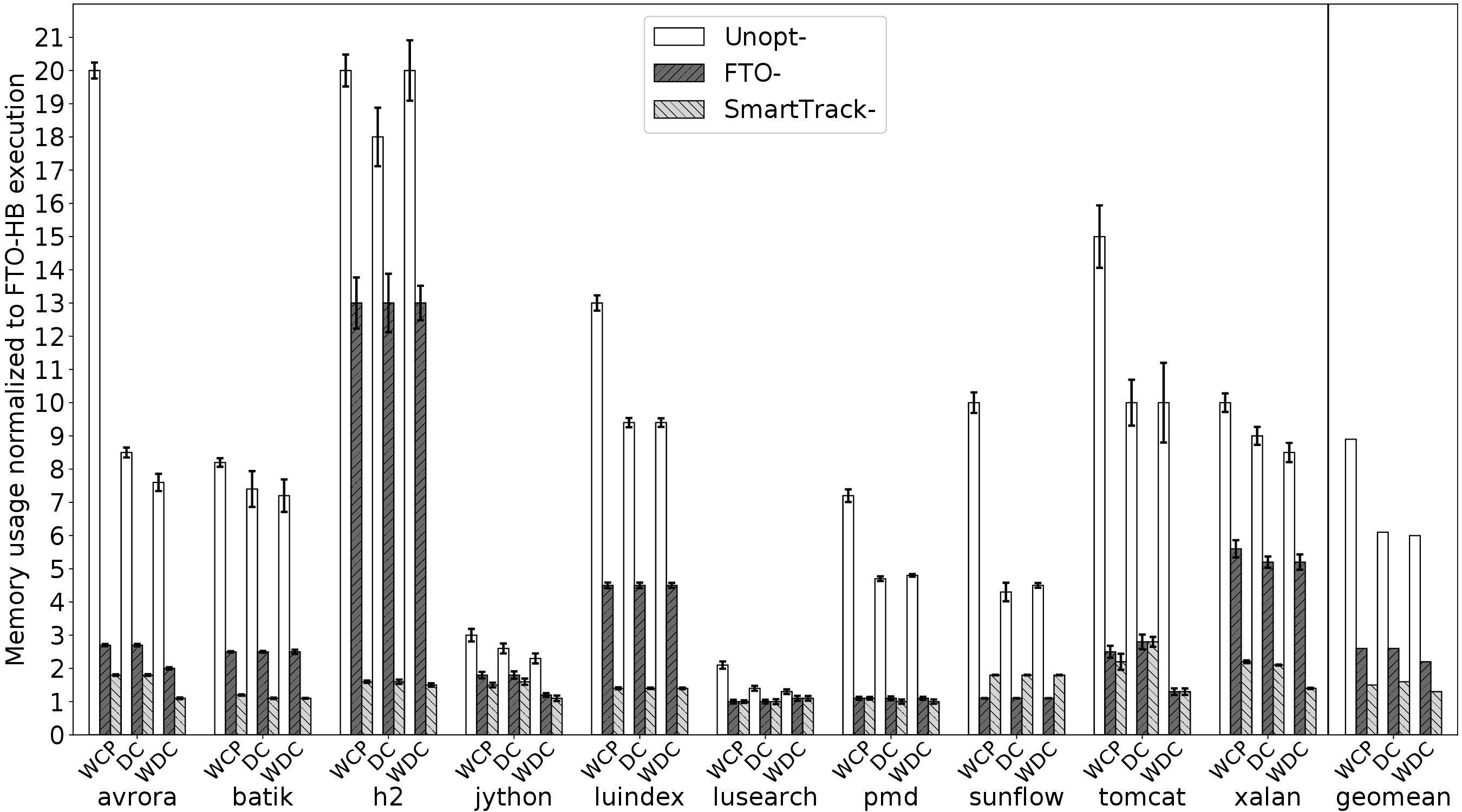}
\label{fig:memory-hbNorm:allDaCapo}
}
\caption{Run time and memory usage of various analyses, relative to \FTO-\HB, for each evaluated program.
The ranges are 95\% confidence intervals.}
\label{fig:hbNorm:allDaCapo}
\end{figure*}
}{}

}{

\begin{figure}[t]
\centering
\subfloat[Run time]{
\includegraphics[width=1\textwidth]{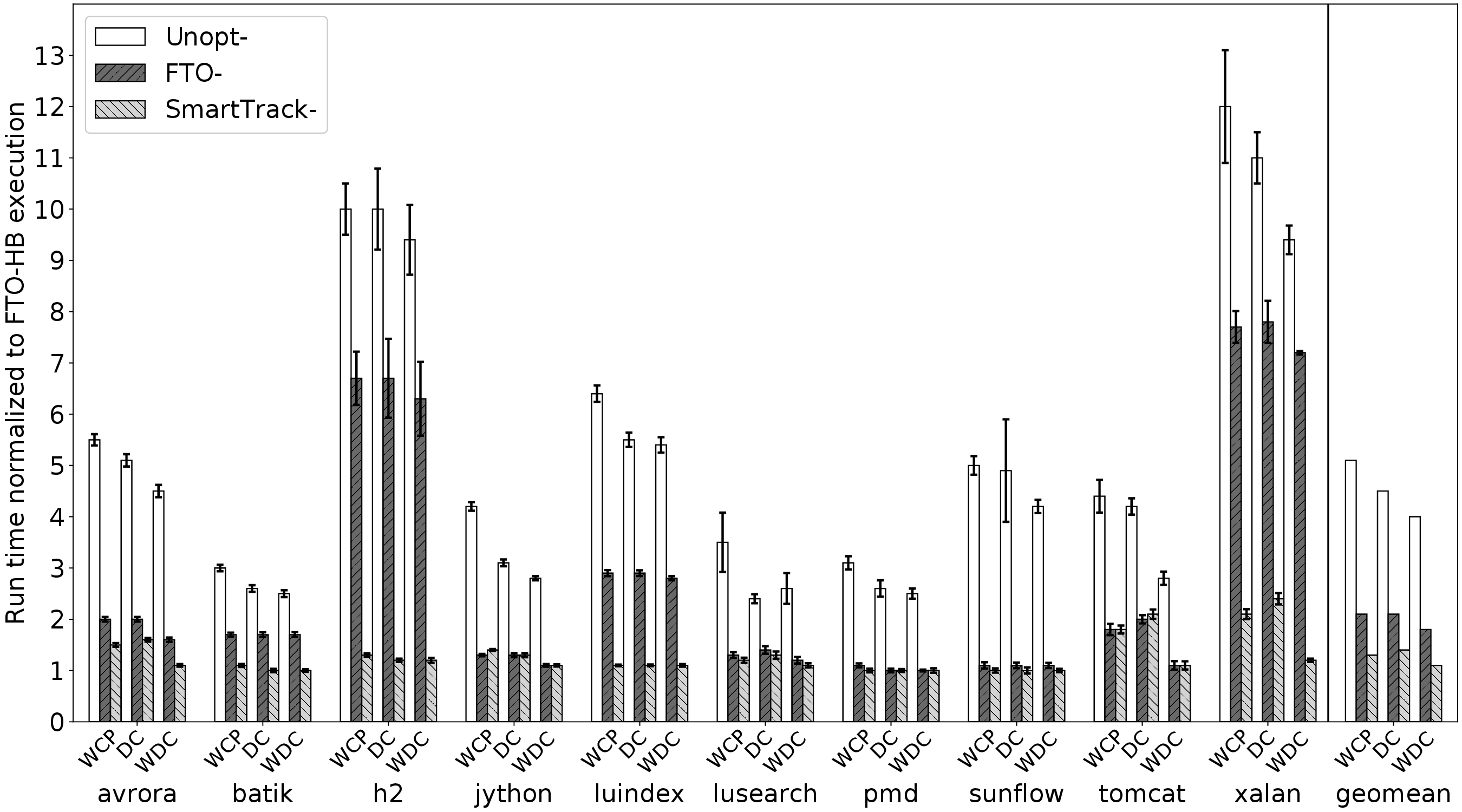}
\label{fig:runtime-hbNorm:allDaCapo}
}
\hfill
\centering
\subfloat[Memory usage]{
\includegraphics[width=1\textwidth]{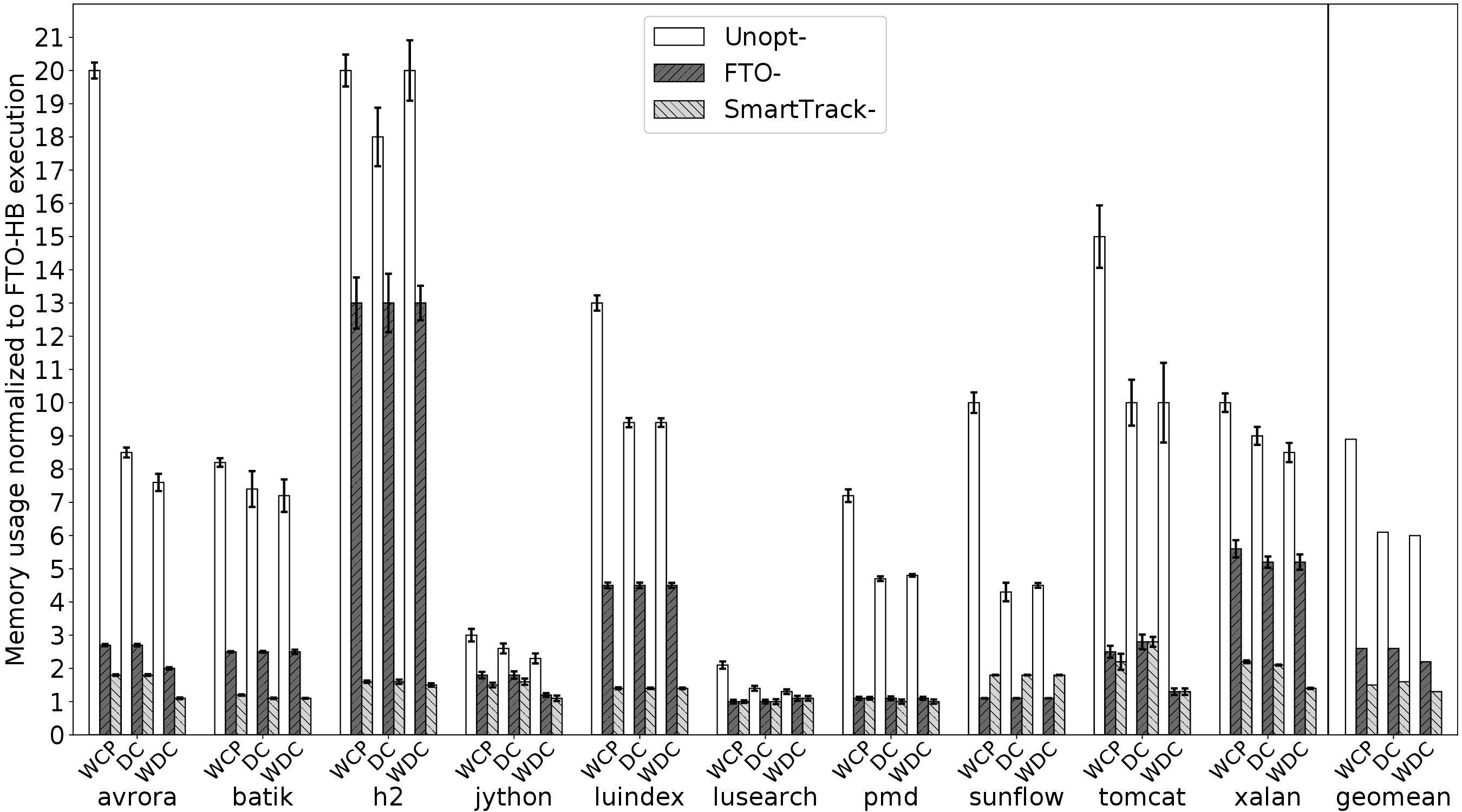}
\label{fig:memory-hbNorm:allDaCapo}
}
\caption{Run time and memory usage of various analyses, relative to \FTO-\HB, for each evaluated program.
The ranges are 95\% confidence intervals.}
\label{fig:hbNorm:allDaCapo}
\end{figure}
}


The main takeaway
is that \RE's optimizations
are effective at improving the performance of all three predictive analyses substantially,
achieving performance (notably run-time overhead) close to state-of-the-art \HB analysis (\HBO).
On average across the programs, the \FTO optimizations applied to predictive analyses result 
in a \factor{2.2--2.6} speedup and \factor{2.7--3.6} memory usage reduction
over unoptimized analyses (Unopt-$\ast$),
although the \FTO-based predictive analyses are still about twice as slow as \HBO on average.
\RE's CCS optimizations provide a \factor{1.5--1.7} average speedup and \factor{1.6--1.8} memory usage reduction over \FTO-$\ast$ analyses,
showing that CCS optimizations eliminate most of the remaining costs \FTO-based predictive analyses 
incur compared with \HBO.
\notes{\FTO-\WCP (which tracks CCSs and the \HB relation)
incurs a \factor{2.1} slowdown and \factor{2.7} increased memory usage compared with \FTO-\HB,
showing the relative cost of computing CCS.
\jake{I think these factors should be removed since the cost of computing CCS is not represented by comparing \FTO-\WCP and \FTO-\HB.
The factors more show the cost of computing \WCP in addition to \HB.
Comparing \FTO-\HB and \FTO-\CAPO is a better comparison but
more shows the difference between conditional and unconditional cost of ordering critical sections.}}
Overall, \RE optimizations yield \factor{3.3--4.1} average speedups and \factor{4.2--6.3} memory usage reductions over unoptimized analyses,
closing the performance gap compared with \HBO.
Both \FTO and CCS optimizations contribute proportionate improvements
to achieve predictive analysis with performance close to that of state-of-the-art \HB analysis.


\HB analysis generally outperforms predictive analyses at each optimization level because it is the most straightforward analysis,
eschewing the cost of computing CCSs.
Unopt-\WCP performs worse than Unopt-\DC due to the additional cost of computing \HB (needed to compute \WCP).
\WCPO and \REWCP reduce this analysis cost significantly.
At the same time, \DC \Rule{b} is somewhat more complex to compute than \WCP \Rule{b} (Section~\ref{sec:background:predictive-analyses}).
These two effects cancel out on average, leading to little or no average performance difference
between \WCPO and \DCO and between \REWCP and \REDC.
\CAPO analysis eliminates computing \Rule{b}, achieving better performance than \DC analysis at all optimization levels.


\RE thus enables three kinds of predictive analysis,
each offering a different coverage--soundness tradeoff,
with performance approaching that of \HB analysis.

%
%

\subsection{Predictable Race Coverage}
\label{eval:subsection:race-coverage}

\later{
\begin{table}[t]
\newcommand{\rzero}{0\xspace}
\newcommand{\roh}[1]{\ifthenelse{\equal{#1}{\rna}}{\rna}{#1$\;\!\times$}} 
\newcommand{\rna}{N/A}
\newcommand{\memna}{N/A} 
\newcommand{\st}[1]{(#1~s)} 
\newcommand{\dt}[1]{#1~s} 
\newcommand{\mem}[1]{\ifthenelse{\equal{#1}{\memna}}{\memna}{#1$\;\!Mb$}} 
\newcommand{\et}[1]{#1M} 
\newcommand{\enfp}[1]{(#1M)} 
\newcommand{\base}[1]{#1~s} 
\newcommand{\sdr}[2]{#1{\;}(#2)}
\input{result-macros/PIP_slowTool_noCoresSet}
\input{result-macros/PIP_fastTool_extraOpt2Quiet}
\small
\centering
\begin{tabular}{@{}l|Hccc@{}}
        & Unopt w/G & Unopt- & \FTO- & \RE- \\\hline
\HB		& \rna												& \sdr{\SLOWHBTotal}{\SLOWHBDynamicTotal}					& \sdr{\FASTFTOHBTotal}{\FASTFTOHBDynamicTotal}		& \rna \\
\WCP	& \rna												& \sdr{\SLOWWCPTotal}{\SLOWWCPDynamicTotal}					& \sdr{\FASTFTOWCPTotal}{\FASTFTOWCPDynamicTotal}	& \sdr{\FASTREWCPTotal}{\FASTREWCPDynamicTotal} \\
\DC		& \sdr{\SLOWDCExcTotal}{\SLOWDCExcDynamicTotal}	& \sdr{\SLOWDCnoGExcTotal}{\SLOWDCnoGExcDynamicTotal}		& \sdr{\FASTFTODCTotal}{\FASTFTODCDynamicTotal}		& \sdr{\FASTREDCTotal}{\FASTREDCDynamicTotal} \\
\CAPO	& \sdr{\SLOWCAPOExcTotal}{\SLOWCAPOExcDynamicTotal}	& \sdr{\SLOWCAPOnoGExcTotal}{\SLOWCAPOnoGExcDynamicTotal}	& \sdr{\FASTFTOCAPOTotal}{\FASTFTOCAPODynamicTotal}	& \sdr{\FASTRECAPOTotal}{\FASTRECAPODynamicTotal} \\
\end{tabular}

\caption{Total average races reported by various analyses.
In each cell, the first value is statically distinct races
(\ie, distinct program locations) and the second value, in parentheses, is total dynamic races.}
\label{tab:race:Total}
\end{table}
Table~\ref{tab:race:Total}...
}

\iftoggle{includeRaceResults}{
\begin{table*}[t]
\newcommand{\rzero}{0\xspace}
\newcommand{\roh}[1]{\ifthenelse{\equal{#1}{\rna}}{\rna}{#1$\;\!\times$}} 
\newcommand{\rna}{N/A}
\newcommand{\memna}{N/A} 
\newcommand{\st}[1]{(#1~s)} 
\newcommand{\dt}[1]{#1~s} 
\newcommand{\mem}[1]{\ifthenelse{\equal{#1}{\memna}}{\memna}{#1$\;Mb$}} 
\newcommand{\et}[1]{#1M} 
\newcommand{\enfp}[1]{(#1M)} 
\newcommand{\base}[1]{#1~s} 
\newcommand{\sdr}[2]{#1{\;}(#2)}
\input{result-macros/PIP_slowTool_noCoresSet}
\input{result-macros/PIP_fastTool_extraOpt2Quiet}
\smaller
\centering
\begin{tabular}{@{}l|Hrrr@{}}
        & w/ G & Unopt- & \FTO- & \REAbbrv- \\\hline
\HB		& \rna													& \sdr{\SLOWavroraHB}{\SLOWavroraHBDynamic}					& \sdr{\FASTavroraFTOHB}{\FASTavroraFTOHBDynamic}		& \rna \\
\WCP	& \rna													& \sdr{\SLOWavroraWCP}{\SLOWavroraWCPDynamic}				& \sdr{\FASTavroraFTOWCP}{\FASTavroraFTOWCPDynamic}		& \sdr{\FASTavroraREWCP}{\FASTavroraREWCPDynamic} \\
\DC		& \sdr{\SLOWavroraDCExc}{\SLOWavroraDCExcDynamic}		& \sdr{\SLOWavroraDCnoGExc}{\SLOWavroraDCnoGExcDynamic}	& \sdr{\FASTavroraFTODC}{\FASTavroraFTODCDynamic}		& \sdr{\FASTavroraREDC}{\FASTavroraREDCDynamic} \\
\CAPO	& \sdr{\SLOWavroraCAPOExc}{\SLOWavroraCAPOExcDynamic}	& \sdr{\SLOWavroraCAPOnoGExc}{\SLOWavroraCAPOnoGExcDynamic}	& \sdr{\FASTavroraFTOCAPO}{\FASTavroraFTOCAPODynamic}	& \sdr{\FASTavroraRECAPO}{\FASTavroraRECAPODynamic} \\
\mc{1}{c}{} & \mc{4}{c}{\bench{avrora}} \\
\end{tabular}
\hfill
\begin{tabular}{@{}HHrrr@{}}
        & w/ G & Unopt- & \FTO- & \REAbbrv- \\\hline
\HB		& \rna													& \sdr{\SLOWhtwoHB}{\SLOWhtwoHBDynamic}					& \sdr{\FASThtwoFTOHB}{\FASThtwoFTOHBDynamic}		& \rna \\
\WCP	& \rna													& \sdr{\SLOWhtwoWCP}{\SLOWhtwoWCPDynamic}				& \sdr{\FASThtwoFTOWCP}{\FASThtwoFTOWCPDynamic}		& \sdr{\FASThtwoREWCP}{\FASThtwoREWCPDynamic} \\
\DC		& \sdr{\SLOWhtwoDCExc}{\SLOWhtwoDCExcDynamic}		& \sdr{\SLOWhtwoDCnoGExc}{\SLOWhtwoDCnoGExcDynamic}	& \sdr{\FASThtwoFTODC}{\FASThtwoFTODCDynamic}		& \sdr{\FASThtwoREDC}{\FASThtwoREDCDynamic} \\
\CAPO	& \sdr{\SLOWhtwoCAPOExc}{\SLOWhtwoCAPOExcDynamic}	& \sdr{\SLOWhtwoCAPOnoGExc}{\SLOWhtwoCAPOnoGExcDynamic}	& \sdr{\FASThtwoFTOCAPO}{\FASThtwoFTOCAPODynamic}	& \sdr{\FASThtwoRECAPO}{\FASThtwoRECAPODynamic} \\
		& \mc{4}{c}{\bench{h2}} \\
\end{tabular}
\hfill
\begin{tabular}{@{}HHrrr@{}}
        & w/ G & Unopt- & \FTO- & \REAbbrv- \\\hline
\HB		& \rna													& \sdr{\SLOWjythonHB}{\SLOWjythonHBDynamic}					& \sdr{\FASTjythonFTOHB}{\FASTjythonFTOHBDynamic}		& \rna \\
\WCP	& \rna													& \sdr{\SLOWjythonWCP}{\SLOWjythonWCPDynamic}				& \sdr{\FASTjythonFTOWCP}{\FASTjythonFTOWCPDynamic}		& \sdr{\FASTjythonREWCP}{\FASTjythonREWCPDynamic} \\
\DC		& \sdr{\SLOWjythonDCExc}{\SLOWjythonDCExcDynamic}		& \sdr{\SLOWjythonDCnoGExc}{\SLOWjythonDCnoGExcDynamic}	& \sdr{\FASTjythonFTODC}{\FASTjythonFTODCDynamic}		& \sdr{\FASTjythonREDC}{\FASTjythonREDCDynamic} \\
\CAPO	& \sdr{\SLOWjythonCAPOExc}{\SLOWjythonCAPOExcDynamic}	& \sdr{\SLOWjythonCAPOnoGExc}{\SLOWjythonCAPOnoGExcDynamic}	& \sdr{\FASTjythonFTOCAPO}{\FASTjythonFTOCAPODynamic}	& \sdr{\FASTjythonRECAPO}{\FASTjythonRECAPODynamic} \\
		& \mc{4}{c}{\bench{jython}} \\
\end{tabular}
\medskip\\
\begin{tabular}{@{}l|Hrrr@{}}
        & w/ G & Unopt- & \FTO- & \REAbbrv- \\\hline
\HB		& \rna													& \sdr{\SLOWluindexHB}{\SLOWluindexHBDynamic}					& \sdr{\FASTluindexFTOHB}{\FASTluindexFTOHBDynamic}		& \rna \\
\WCP	& \rna													& \sdr{\SLOWluindexWCP}{\SLOWluindexWCPDynamic}				& \sdr{\FASTluindexFTOWCP}{\FASTluindexFTOWCPDynamic}		& \sdr{\FASTluindexREWCP}{\FASTluindexREWCPDynamic} \\
\DC		& \sdr{\SLOWluindexDCExc}{\SLOWluindexDCExcDynamic}		& \sdr{\SLOWluindexDCnoGExc}{\SLOWluindexDCnoGExcDynamic}	& \sdr{\FASTluindexFTODC}{\FASTluindexFTODCDynamic}		& \sdr{\FASTluindexREDC}{\FASTluindexREDCDynamic} \\
\CAPO	& \sdr{\SLOWluindexCAPOExc}{\SLOWluindexCAPOExcDynamic}	& \sdr{\SLOWluindexCAPOnoGExc}{\SLOWluindexCAPOnoGExcDynamic}	& \sdr{\FASTluindexFTOCAPO}{\FASTluindexFTOCAPODynamic}	& \sdr{\FASTluindexRECAPO}{\FASTluindexRECAPODynamic} \\
\mc{1}{c}{} & \mc{4}{c}{\bench{luindex}} \\
\end{tabular}
\hfill
\begin{tabular}{@{}HHrrr@{}}
        & w/ G & Unopt- & \FTO- & \REAbbrv- \\\hline
\HB		& \rna													& \sdr{\SLOWpmdHB}{\SLOWpmdHBDynamic}					& \sdr{\FASTpmdFTOHB}{\FASTpmdFTOHBDynamic}		& \rna \\
\WCP	& \rna													& \sdr{\SLOWpmdWCP}{\SLOWpmdWCPDynamic}				& \sdr{\FASTpmdFTOWCP}{\FASTpmdFTOWCPDynamic}		& \sdr{\FASTpmdREWCP}{\FASTpmdREWCPDynamic} \\
\DC		& \sdr{\SLOWpmdDCExc}{\SLOWpmdDCExcDynamic}		& \sdr{\SLOWpmdDCnoGExc}{\SLOWpmdDCnoGExcDynamic}	& \sdr{\FASTpmdFTODC}{\FASTpmdFTODCDynamic}		& \sdr{\FASTpmdREDC}{\FASTpmdREDCDynamic} \\
\CAPO	& \sdr{\SLOWpmdCAPOExc}{\SLOWpmdCAPOExcDynamic}	& \sdr{\SLOWpmdCAPOnoGExc}{\SLOWpmdCAPOnoGExcDynamic}	& \sdr{\FASTpmdFTOCAPO}{\FASTpmdFTOCAPODynamic}	& \sdr{\FASTpmdRECAPO}{\FASTpmdRECAPODynamic} \\
		& \mc{4}{c}{\bench{pmd}} \\
\end{tabular}
\hfill
\begin{tabular}{@{}HHrrr@{}}
        & w/ G & Unopt- & \FTO- & \REAbbrv- \\\hline
\HB		& \rna													& \sdr{\SLOWsunflowHB}{\SLOWsunflowHBDynamic}					& \sdr{\FASTsunflowFTOHB}{\FASTsunflowFTOHBDynamic}		& \rna \\
\WCP	& \rna													& \sdr{\SLOWsunflowWCP}{\SLOWsunflowWCPDynamic}				& \sdr{\FASTsunflowFTOWCP}{\FASTsunflowFTOWCPDynamic}		& \sdr{\FASTsunflowREWCP}{\FASTsunflowREWCPDynamic} \\
\DC		& \sdr{\SLOWsunflowDCExc}{\SLOWsunflowDCExcDynamic}		& \sdr{\SLOWsunflowDCnoGExc}{\SLOWsunflowDCnoGExcDynamic}	& \sdr{\FASTsunflowFTODC}{\FASTsunflowFTODCDynamic}		& \sdr{\FASTsunflowREDC}{\FASTsunflowREDCDynamic} \\
\CAPO	& \sdr{\SLOWsunflowCAPOExc}{\SLOWsunflowCAPOExcDynamic}	& \sdr{\SLOWsunflowCAPOnoGExc}{\SLOWsunflowCAPOnoGExcDynamic}	& \sdr{\FASTsunflowFTOCAPO}{\FASTsunflowFTOCAPODynamic}	& \sdr{\FASTsunflowRECAPO}{\FASTsunflowRECAPODynamic} \\
		& \mc{4}{c}{\bench{sunflow}} \\
\end{tabular}
\medskip\\
\begin{tabular}{@{}l|Hrrr@{}}
        & w/ G & Unopt- & \FTO- & \REAbbrv- \\\hline
\HB		& \rna													& \sdr{\SLOWtomcatHB}{\SLOWtomcatHBDynamic}					& \sdr{\FASTtomcatFTOHB}{\FASTtomcatFTOHBDynamic}		& \rna \\
\WCP	& \rna													& \sdr{\SLOWtomcatWCP}{\SLOWtomcatWCPDynamic}				& \sdr{\FASTtomcatFTOWCP}{\FASTtomcatFTOWCPDynamic}		& \sdr{\FASTtomcatREWCP}{\FASTtomcatREWCPDynamic} \\
\DC		& \sdr{\SLOWtomcatDCExc}{\SLOWtomcatDCExcDynamic}		& \sdr{\SLOWtomcatDCnoGExc}{\SLOWtomcatDCnoGExcDynamic}	& \sdr{\FASTtomcatFTODC}{\FASTtomcatFTODCDynamic}		& \sdr{\FASTtomcatREDC}{\FASTtomcatREDCDynamic} \\
\CAPO	& \sdr{\SLOWtomcatCAPOExc}{\SLOWtomcatCAPOExcDynamic}	& \sdr{\SLOWtomcatCAPOnoGExc}{\SLOWtomcatCAPOnoGExcDynamic}	& \sdr{\FASTtomcatFTOCAPO}{\FASTtomcatFTOCAPODynamic}	& \sdr{\FASTtomcatRECAPO}{\FASTtomcatRECAPODynamic} \\
\mc{1}{c}{} & \mc{4}{c}{\bench{tomcat}} \\
\end{tabular}
\hfill
\begin{tabular}{@{}HHrrr@{}}
        & w/ G & Unopt- & \FTO- & \REAbbrv- \\\hline
\HB		& \rna													& \sdr{\SLOWxalanHB}{\SLOWxalanHBDynamic}					& \sdr{\FASTxalanFTOHB}{\FASTxalanFTOHBDynamic}		& \rna \\
\WCP	& \rna													& \sdr{\SLOWxalanWCP}{\SLOWxalanWCPDynamic}				& \sdr{\FASTxalanFTOWCP}{\FASTxalanFTOWCPDynamic}		& \sdr{\FASTxalanREWCP}{\FASTxalanREWCPDynamic} \\
\DC		& \sdr{\SLOWxalanDCExc}{\SLOWxalanDCExcDynamic}		& \sdr{\SLOWxalanDCnoGExc}{\SLOWxalanDCnoGExcDynamic}	& \sdr{\FASTxalanFTODC}{\FASTxalanFTODCDynamic}		& \sdr{\FASTxalanREDC}{\FASTxalanREDCDynamic} \\
\CAPO	& \sdr{\SLOWxalanCAPOExc}{\SLOWxalanCAPOExcDynamic}	& \sdr{\SLOWxalanCAPOnoGExc}{\SLOWxalanCAPOnoGExcDynamic}	& \sdr{\FASTxalanFTOCAPO}{\FASTxalanFTOCAPODynamic}	& \sdr{\FASTxalanRECAPO}{\FASTxalanRECAPODynamic} \\
		& \mc{4}{c}{\bench{xalan}} \\
\end{tabular}
\caption{Average races reported by various analyses for each evaluated program
(excluding \bench{batik} and \bench{lusearch}, for which all analyses report no races).
In each cell, the first value is statically distinct races
(\ie, distinct program locations) and the second value, in parentheses, is total dynamic races.
As the text explains, significant differences between the algorithms (\col{Unopt-}, \col{\FTO-}, \col{\RE-})
are attributable to how the algorithms behave after detecting the first race.}
\label{tab:race:allDaCapo}
\end{table*}
}{}

\iftoggle{includeRaceResults}{


Although our evaluation focuses on the performance of our optimizations,
and prior work has established that \WCP and \DC analyses detect more races than \HB analysis~\cite{wcp,vindicator},
we have also evaluated how many races each analysis detects.

%
%

Table~\ref{tab:race:allDaCapo} reports how many races
each analysis finds.
For each cell, the second value (in parentheses) is total dynamic races reported,
and the first value is \emph{statically distinct} races.
Two dynamic races detected at the same static program location
are the same statically distinct race.

\paragraph{Comparing relations.}

In general, the results confirm that weaker relations find more races than stronger relations.
However, although the analyses get progressively more powerful from top to bottom
(\eg, every \DC-race is a \CAPO-race),
this relationship does not always hold empirically for two reasons.
First, run-to-run variation naturally affects repeatability.
Appendix~\ref{appendix:confidence-interval-results} (Table~\ref{tab:race:allDaCapo:CI})
provides 95\% confidence intervals for these results,
showing that many of the differences involve overlapping confidence intervals.
Second, analyses have different performance characteristics that may affect
an execution's timing and memory access interleaving, leading to different races occurring.

Run-to-run variation can explain the unintuitive race results across relations for \bench{tomcat}
(\ie, for each algorithm, weaker relations do not always report more races),
as the confidence intervals in Table~\ref{tab:race:allDaCapo:CI} in Appendix~\ref{appendix:confidence-interval-results} show.
The table reports one anomalous result for \bench{jython}:
\WCPO reports fewer races than expected;
we would expect the race counts to fall between the race counts of \HBO and \DCO.
This result is statistically significant (Table~\ref{tab:race:allDaCapo:CI}).
We are still investigating the cause of this anomaly.

Prior work has shown the relative effectiveness of \WCP and \DC analyses
by performing \HB, \WCP, and \DC analyses on the same observed trace~\cite{vindicator,wcp}.
(Our results often report many more races, especially dynamic races, than prior work's results that used
the RoadRunner Vindicator implementation and the DaCapo benchmarks~\cite{vindicator}.
These differences occur because the prior work used default RoadRunner behavior
that stops performing analysis for a field after 100 dynamic races detected on the field,
whereas our analyses disable that behavior.)

The results do show that despite using a weaker relation than \DC analysis,
\CAPO analysis does not on average report more races than \DC analysis,
which suggests that \CAPO analysis's optimization does not lead to false races in practice.
In separate experiments with the Unopt-\{\DC, \CAPO{}\} w/$G$ analyses,
\textsc{VindicateRace} successfully vindicates every \DC- and \CAPO-race detected across 10 trials,
confirming every dynamic \DC- and \CAPO-race is a true predictable race.

\paragraph{Comparing optimizations.}

For each relation, the different algorithms (\col{Unopt-}, \col{\FTO-}, \col{\RE-}) often report comparable race counts,
but sometimes the counts differ significantly.
These differences---particularly between unoptimized (\col{Unopt-}) and optimized (\col{FTO-}, \col{ST-}) race counts---occur
because of run-to-run variation and performance characteristics, but primarily for a third reason:
the different optimization levels have different behavior after they detect the first race,
affecting race counts by using different metadata (\eg, epochs vs.\ vector clocks)
to update racing accesses and detect future races
(Section~\ref{subsec:impl}).

Thus for each relation, the differences between the \emph{algorithms} (\col{Unopt-}, \col{\FTO-}, \col{\RE-})
are not a reflection of race detection effectiveness across optimizations.
Any extra races detected by one algorithm
are likely to be related to each other (\eg, extra races involve accesses
to the same data structure as accesses in races reported by all algorithms,
or extra races may be dependent on races reported by all algorithms),
and thus not be of much use to programmers.
Rather, the race differences serve to show that the proposed optimizations and our implementations of them
lead to reasonable race detection results.
\notes{We have looked at individual static program locations reported by the analyses,
acting as a sanity check and providing confidence in the correctness of our implementations.}%

\subsection{Summary}

As the results show, prior work's \WCP and \DC analyses are expensive,
particularly for programs that frequently access variables in critical sections.
The \RE-optimized \WCP and \DC analyses improve run time and memory usage by several times on average,
achieving performance comparable to \HB analysis.

\RE's optimizations are effective across predictive analyses
and demonstrates practical use during in-house testing as a possible substitute to \HB analysis.
Sound \WCP analysis detects fewer races than other predictive analyses and, in its unoptimized form, has the highest overhead.
\REWCP provides performance on par with \HB analysis and other predictive analyses.
At the other end of the coverage--soundness tradeoff,
\CAPO analysis has the potential to detect the most false races (although in practice it detects only true predictable races),
and it has the lowest overhead among predictive analyses.
We emphasize that \ST-\DC and \ST-\CAPO analyses do \emph{not} perform vindication. 
They instead report \DC- and \CAPO-races without vindication; 
or they can use record \& replay techniques to replay an execution 
and collect enough information to perform vindication.

Evaluating analysis effectiveness across multiple executions is admittedly beyond the scope of the paper. 
A key challenge is the plethora of different options for choosing multiple executions (\cf~Section~\ref{Sec:related}).
Simply re-executing a program natively is likely to expose a similar set of HB-races, 
while an advanced race-exposing technique could potentially expose HB-races effectively, 
but likely at a higher cost.
Despite this, the results show that predictive analyses can be practical data race
detectors that are competitive with standard highly optimized \HB data race detectors.
}{}

\iftoggle{includeRaceResults}{}{
Although our evaluation focuses on the performance of our optimizations,
and prior work has established that \WCP and \DC analyses detect more races than \HB analysis~\cite{wcp,vindicator},
we have also evaluated how many races each analysis detects.
\iftoggle{techReport}{%
Appendix~\ref{appendix:race-coverage} presents full results, which we summarize here.
}{%
\notes{Our extended arXiv paper presents full results~\cite{smarttrack-extended-arxiv}, which we summarize here.}%
}%

In general, the results confirm that weaker relations find more races than stronger relations
(except \CAPO analysis does not report more races than \DC analysis).
In addition, for each relation, the different optimizations (Unopt-, \FTO-, and \RE-)
generally report comparable race counts.
The differences that exist across optimizations are attributable to
run-to-run variation (as reported confidence intervals show)
and differences in how the optimized analyses detect races after the first race (Section~\ref{subsec:impl}).
Thus the race count differences do not serve to compare race detection effectiveness across optimizations,
but rather to verify that the proposed optimizations and our implementations of them
lead to reasonable race detection results.

In experiments with configurations of Unopt-\DC and Unopt-\CAPO that build constraint graphs and perform vindication,
every detected \DC- and \CAPO-race was successfully vindicated (results not shown).
We cross-referenced the static races detected by unoptimized and \RE-based analyses
in order to confirm that every race reported by the \RE-based
analyses was a true race.

\subsection{Results Summary}

As the results show, prior work's \WCP and \DC analyses are costly,
especially when accesses in critical sections are frequent.
The \RE-optimized \WCP and \DC analyses improve run time and memory usage by several times on average,
achieving performance comparable to \HB analysis.

\RE's optimizations are effective across predictive analyses.
Sound \WCP analysis detects fewer races than other predictive analyses and, in its unoptimized form, has the highest overhead.
\REWCP provides performance on par with \HB analysis and other predictive analyses.
At the other end of the coverage--soundness tradeoff,
\CAPO has the most potential for false positives---although in practice it detects only true races---and
it has the lowest overhead among predictive analyses.
\RECAPO provides the best performance of any predictive analysis, nearly matching the performance of optimized \HB analysis (\HBO).
The coverage--soundness tradeoff provides flexibility to choose different analyses
depending on a programmer's tolerance for the possibility of false races
(although deploying with record \& replay allows vindicating reported \DC- or \CAPO-races) and
the empirically observed differences among the analyses for the programmer's application.

Overall, the results show that predictive analyses can be practical data race
detectors that are competitive with standard highly optimized \HB data race detectors.
}

%
%
%
%

%% file: result-macros/PIP_fastTool_extraOpt2Stat.tex
\newcommand{\FASTavroraEvents}{1,400}
\newcommand{\FASTavroraNoFPEvents}{5.7}
\newcommand{\avroraHBEventTotal}{1,400}
\newcommand{\avroraHBNoFPEventTotal}{150}
\newcommand{\avroraHBNoFPAccessTotal}{150}
\newcommand{\avroraHBNoFPOtherTotal}{3.6}
\newcommand{\avroraHBReadTotal}{69.4}
\newcommand{\avroraHBWriteTotal}{28.3}
\newcommand{\avroraHBNoFPAccessInCS}{30.2}
\newcommand{\avroraHBNoFPAccessOutCS}{47.7}
\newcommand{\avroraHBAcqRelTotal}{51.4}
\newcommand{\avroraHBOtherTotal}{11.3}
\newcommand{\avroraHBNoFPReadTotal}{110}
\newcommand{\avroraHBReadInCS}{6.05}
\newcommand{\avroraHBReadOutCS}{94}
\newcommand{\avroraHBReadSameEp}{0.0242}
\newcommand{\avroraHBReadSharedSameEp}{\cna}
\newcommand{\avroraHBReadExclusive}{96.5}
\newcommand{\avroraHBReadOwned}{\cna}
\newcommand{\avroraHBReadShare}{0.0206}
\newcommand{\avroraHBReadShared}{3.51}
\newcommand{\avroraHBReadSharedOwned}{\cna}
\newcommand{\avroraHBNoFPHonestWriteTotal}{44}
\newcommand{\avroraHBWriteInCS}{6.21}
\newcommand{\avroraHBWriteOutCS}{96.2}
\newcommand{\avroraHBNoFPWriteTotal}{44}
\newcommand{\avroraHBWriteSameEp}{2.4}
\newcommand{\avroraHBWriteExclusive}{99.5}
\newcommand{\avroraHBWriteOwned}{\cna}
\newcommand{\avroraHBWriteShared}{0.476}
\newcommand{\avroraHBNoFPOtherEventTotal}{3563253}
\newcommand{\avroraHBAcqRelOtherTotal}{82.0}
\newcommand{\avroraHBNoAcqRelOtherTotal}{641918}
\newcommand{\avroraHBFork}{9.35E-4}
\newcommand{\avroraHBJoin}{9.35E-4}
\newcommand{\avroraHBPreWait}{50.0}
\newcommand{\avroraHBPostWait}{50.0}
\newcommand{\avroraHBVolatileTotal}{0.0}
\newcommand{\avroraHBClassInit}{0.0195}
\newcommand{\avroraHBClassAccess}{0.0288}
\newcommand{\avroraHBRaceTotal}{624316}
\newcommand{\avroraHBWrRdRace}{34.9}
\newcommand{\avroraHBWrWrRace}{32.5}
\newcommand{\avroraHBRdWrRace}{0.0767}
\newcommand{\avroraHBRdShWrRace}{32.5}
\newcommand{\avroraHBHoldLocksTotal}{9.2}
\newcommand{\avroraHBOneLockHeld}{5.69}
\newcommand{\avroraHBTwoNestedLocks}{<0.1}
\newcommand{\avroraHBThreeNestedLocks}{\cna}
\newcommand{\avroraHBFourNestedLocks}{\cna}
\newcommand{\avroraHBFiveNestedLocks}{\cna}
\newcommand{\avroraHBSixNestedLocks}{\cna}
\newcommand{\avroraHBSevenNestedLocks}{\cna}
\newcommand{\avroraHBEightNestedLocks}{\cna}
\newcommand{\avroraHBNineNestedLocks}{\cna}
\newcommand{\avroraHBTenNestedLocks}{\cna}
\newcommand{\avroraHBHundredNestedLocks}{\cna}
\newcommand{\avroraHBExWrSet}{\ena}
\newcommand{\avroraHBExWrCheck}{\ena}
\newcommand{\avroraHBExWrUpdate}{\ena}
\newcommand{\avroraHBExRdCheck}{\ena}
\newcommand{\avroraHBExRdUpdate}{\ena}
\newcommand{\avroraHBExTotalCheck}{\ena}
\newcommand{\avroraHBExTotalUpdate}{\ena}
\newcommand{\avroraFTOHBEventTotal}{1,400}
\newcommand{\avroraFTOHBNoFPEventTotal}{140}
\newcommand{\avroraFTOHBNoFPAccessTotal}{140}
\newcommand{\avroraFTOHBNoFPOtherTotal}{3.6}
\newcommand{\avroraFTOHBReadTotal}{66.6}
\newcommand{\avroraFTOHBWriteTotal}{30.9}
\newcommand{\avroraFTOHBNoFPAccessInCS}{28.7}
\newcommand{\avroraFTOHBNoFPAccessOutCS}{48.5}
\newcommand{\avroraFTOHBAcqRelTotal}{52.7}
\newcommand{\avroraFTOHBOtherTotal}{11.9}
\newcommand{\avroraFTOHBNoFPReadTotal}{94}
\newcommand{\avroraFTOHBReadInCS}{6.63}
\newcommand{\avroraFTOHBReadOutCS}{93.4}
\newcommand{\avroraFTOHBReadSameEp}{0.0276}
\newcommand{\avroraFTOHBReadSharedSameEp}{\cna}
\newcommand{\avroraFTOHBReadExclusive}{5.38}
\newcommand{\avroraFTOHBReadOwned}{90.5}
\newcommand{\avroraFTOHBReadShare}{0.217}
\newcommand{\avroraFTOHBReadShared}{0.14}
\newcommand{\avroraFTOHBReadSharedOwned}{3.81}
\newcommand{\avroraFTOHBNoFPHonestWriteTotal}{44}
\newcommand{\avroraFTOHBWriteInCS}{6.21}
\newcommand{\avroraFTOHBWriteOutCS}{96.2}
\newcommand{\avroraFTOHBNoFPWriteTotal}{44}
\newcommand{\avroraFTOHBWriteSameEp}{2.4}
\newcommand{\avroraFTOHBWriteExclusive}{0.372}
\newcommand{\avroraFTOHBWriteOwned}{99.2}
\newcommand{\avroraFTOHBWriteShared}{0.417}
\newcommand{\avroraFTOHBNoFPOtherEventTotal}{3581909}
\newcommand{\avroraFTOHBAcqRelOtherTotal}{81.6}
\newcommand{\avroraFTOHBNoAcqRelOtherTotal}{658970}
\newcommand{\avroraFTOHBFork}{9.11E-4}
\newcommand{\avroraFTOHBJoin}{9.11E-4}
\newcommand{\avroraFTOHBPreWait}{50.0}
\newcommand{\avroraFTOHBPostWait}{50.0}
\newcommand{\avroraFTOHBVolatileTotal}{0.0}
\newcommand{\avroraFTOHBClassInit}{0.019}
\newcommand{\avroraFTOHBClassAccess}{0.0281}
\newcommand{\avroraFTOHBRaceTotal}{449288}
\newcommand{\avroraFTOHBWrRdRace}{55.0}
\newcommand{\avroraFTOHBWrWrRace}{0.0}
\newcommand{\avroraFTOHBRdWrRace}{4.98}
\newcommand{\avroraFTOHBRdShWrRace}{40.0}
\newcommand{\avroraFTOHBHoldLocksTotal}{8.4}
\newcommand{\avroraFTOHBOneLockHeld}{5.76}
\newcommand{\avroraFTOHBTwoNestedLocks}{<0.1}
\newcommand{\avroraFTOHBThreeNestedLocks}{\cna}
\newcommand{\avroraFTOHBFourNestedLocks}{\cna}
\newcommand{\avroraFTOHBFiveNestedLocks}{\cna}
\newcommand{\avroraFTOHBSixNestedLocks}{\cna}
\newcommand{\avroraFTOHBSevenNestedLocks}{\cna}
\newcommand{\avroraFTOHBEightNestedLocks}{\cna}
\newcommand{\avroraFTOHBNineNestedLocks}{\cna}
\newcommand{\avroraFTOHBTenNestedLocks}{\cna}
\newcommand{\avroraFTOHBHundredNestedLocks}{\cna}
\newcommand{\avroraFTOHBExWrSet}{\ena}
\newcommand{\avroraFTOHBExWrCheck}{\ena}
\newcommand{\avroraFTOHBExWrUpdate}{\ena}
\newcommand{\avroraFTOHBExRdCheck}{\ena}
\newcommand{\avroraFTOHBExRdUpdate}{\ena}
\newcommand{\avroraFTOHBExTotalCheck}{\ena}
\newcommand{\avroraFTOHBExTotalUpdate}{\ena}
\newcommand{\avroraFTOWCPEventTotal}{1,400}
\newcommand{\avroraFTOWCPNoFPEventTotal}{140}
\newcommand{\avroraFTOWCPNoFPAccessTotal}{140}
\newcommand{\avroraFTOWCPNoFPOtherTotal}{3.7}
\newcommand{\avroraFTOWCPReadTotal}{66.6}
\newcommand{\avroraFTOWCPWriteTotal}{30.8}
\newcommand{\avroraFTOWCPNoFPAccessInCS}{28.7}
\newcommand{\avroraFTOWCPNoFPAccessOutCS}{49.0}
\newcommand{\avroraFTOWCPAcqRelTotal}{52.0}
\newcommand{\avroraFTOWCPOtherTotal}{13.3}
\newcommand{\avroraFTOWCPNoFPReadTotal}{94}
\newcommand{\avroraFTOWCPReadInCS}{6.72}
\newcommand{\avroraFTOWCPReadOutCS}{93.3}
\newcommand{\avroraFTOWCPReadSameEp}{0.0275}
\newcommand{\avroraFTOWCPReadSharedSameEp}{\cna}
\newcommand{\avroraFTOWCPReadExclusive}{2.86}
\newcommand{\avroraFTOWCPReadOwned}{90.3}
\newcommand{\avroraFTOWCPReadShare}{0.533}
\newcommand{\avroraFTOWCPReadShared}{0.449}
\newcommand{\avroraFTOWCPReadSharedOwned}{5.83}
\newcommand{\avroraFTOWCPNoFPHonestWriteTotal}{44}
\newcommand{\avroraFTOWCPWriteInCS}{6.17}
\newcommand{\avroraFTOWCPWriteOutCS}{96.3}
\newcommand{\avroraFTOWCPNoFPWriteTotal}{44}
\newcommand{\avroraFTOWCPWriteSameEp}{2.44}
\newcommand{\avroraFTOWCPWriteExclusive}{0.312}
\newcommand{\avroraFTOWCPWriteOwned}{98.9}
\newcommand{\avroraFTOWCPWriteShared}{0.782}
\newcommand{\avroraFTOWCPNoFPOtherEventTotal}{3712313}
\newcommand{\avroraFTOWCPAcqRelOtherTotal}{79.7}
\newcommand{\avroraFTOWCPNoAcqRelOtherTotal}{754895}
\newcommand{\avroraFTOWCPFork}{7.95E-4}
\newcommand{\avroraFTOWCPJoin}{7.95E-4}
\newcommand{\avroraFTOWCPPreWait}{50.0}
\newcommand{\avroraFTOWCPPostWait}{50.0}
\newcommand{\avroraFTOWCPVolatileTotal}{0.0}
\newcommand{\avroraFTOWCPClassInit}{0.0166}
\newcommand{\avroraFTOWCPClassAccess}{0.0245}
\newcommand{\avroraFTOWCPRaceTotal}{439867}
\newcommand{\avroraFTOWCPWrRdRace}{53.9}
\newcommand{\avroraFTOWCPWrWrRace}{0.0}
\newcommand{\avroraFTOWCPRdWrRace}{3.57}
\newcommand{\avroraFTOWCPRdShWrRace}{42.6}
\newcommand{\avroraFTOWCPHoldLocksTotal}{8.5}
\newcommand{\avroraFTOWCPOneLockHeld}{5.82}
\newcommand{\avroraFTOWCPTwoNestedLocks}{<0.1}
\newcommand{\avroraFTOWCPThreeNestedLocks}{\cna}
\newcommand{\avroraFTOWCPFourNestedLocks}{\cna}
\newcommand{\avroraFTOWCPFiveNestedLocks}{\cna}
\newcommand{\avroraFTOWCPSixNestedLocks}{\cna}
\newcommand{\avroraFTOWCPSevenNestedLocks}{\cna}
\newcommand{\avroraFTOWCPEightNestedLocks}{\cna}
\newcommand{\avroraFTOWCPNineNestedLocks}{\cna}
\newcommand{\avroraFTOWCPTenNestedLocks}{\cna}
\newcommand{\avroraFTOWCPHundredNestedLocks}{\cna}
\newcommand{\avroraFTOWCPExWrSet}{\ena}
\newcommand{\avroraFTOWCPExWrCheck}{\ena}
\newcommand{\avroraFTOWCPExWrUpdate}{\ena}
\newcommand{\avroraFTOWCPExRdCheck}{\ena}
\newcommand{\avroraFTOWCPExRdUpdate}{\ena}
\newcommand{\avroraFTOWCPExTotalCheck}{\ena}
\newcommand{\avroraFTOWCPExTotalUpdate}{\ena}
\newcommand{\avroraREWCPEventTotal}{1,400}
\newcommand{\avroraREWCPNoFPEventTotal}{140}
\newcommand{\avroraREWCPNoFPAccessTotal}{140}
\newcommand{\avroraREWCPNoFPOtherTotal}{3.6}
\newcommand{\avroraREWCPReadTotal}{66.6}
\newcommand{\avroraREWCPWriteTotal}{30.8}
\newcommand{\avroraREWCPNoFPAccessInCS}{33.1}
\newcommand{\avroraREWCPNoFPAccessOutCS}{45.8}
\newcommand{\avroraREWCPAcqRelTotal}{51.4}
\newcommand{\avroraREWCPOtherTotal}{12.1}
\newcommand{\avroraREWCPNoFPReadTotal}{94}
\newcommand{\avroraREWCPReadInCS}{6.66}
\newcommand{\avroraREWCPReadOutCS}{93.4}
\newcommand{\avroraREWCPReadSameEp}{0.0275}
\newcommand{\avroraREWCPReadSharedSameEp}{\cna}
\newcommand{\avroraREWCPReadExclusive}{2.21}
\newcommand{\avroraREWCPReadOwned}{42.3}
\newcommand{\avroraREWCPReadShare}{0.848}
\newcommand{\avroraREWCPReadShared}{0.744}
\newcommand{\avroraREWCPReadSharedOwned}{53.9}
\newcommand{\avroraREWCPNoFPHonestWriteTotal}{44}
\newcommand{\avroraREWCPWriteInCS}{6.18}
\newcommand{\avroraREWCPWriteOutCS}{96.2}
\newcommand{\avroraREWCPNoFPWriteTotal}{44}
\newcommand{\avroraREWCPWriteSameEp}{2.42}
\newcommand{\avroraREWCPWriteExclusive}{0.339}
\newcommand{\avroraREWCPWriteOwned}{98.4}
\newcommand{\avroraREWCPWriteShared}{1.3}
\newcommand{\avroraREWCPNoFPOtherEventTotal}{3634229}
\newcommand{\avroraREWCPAcqRelOtherTotal}{80.9}
\newcommand{\avroraREWCPNoAcqRelOtherTotal}{693507}
\newcommand{\avroraREWCPFork}{8.65E-4}
\newcommand{\avroraREWCPJoin}{8.65E-4}
\newcommand{\avroraREWCPPreWait}{50.0}
\newcommand{\avroraREWCPPostWait}{50.0}
\newcommand{\avroraREWCPVolatileTotal}{0.0}
\newcommand{\avroraREWCPClassInit}{0.018}
\newcommand{\avroraREWCPClassAccess}{0.0267}
\newcommand{\avroraREWCPRaceTotal}{441847}
\newcommand{\avroraREWCPWrRdRace}{54.1}
\newcommand{\avroraREWCPWrWrRace}{0.0}
\newcommand{\avroraREWCPRdWrRace}{3.9}
\newcommand{\avroraREWCPRdShWrRace}{42.0}
\newcommand{\avroraREWCPHoldLocksTotal}{8.5}
\newcommand{\avroraREWCPOneLockHeld}{5.87}
\newcommand{\avroraREWCPTwoNestedLocks}{<0.1}
\newcommand{\avroraREWCPThreeNestedLocks}{\cna}
\newcommand{\avroraREWCPFourNestedLocks}{\cna}
\newcommand{\avroraREWCPFiveNestedLocks}{\cna}
\newcommand{\avroraREWCPSixNestedLocks}{\cna}
\newcommand{\avroraREWCPSevenNestedLocks}{\cna}
\newcommand{\avroraREWCPEightNestedLocks}{\cna}
\newcommand{\avroraREWCPNineNestedLocks}{\cna}
\newcommand{\avroraREWCPTenNestedLocks}{\cna}
\newcommand{\avroraREWCPHundredNestedLocks}{\cna}
\newcommand{\avroraREWCPExWrSet}{421}
\newcommand{\avroraREWCPExWrCheck}{5429273}
\newcommand{\avroraREWCPExWrUpdate}{\ena}
\newcommand{\avroraREWCPExRdCheck}{4057349}
\newcommand{\avroraREWCPExRdUpdate}{\ena}
\newcommand{\avroraREWCPExTotalCheck}{9486622}
\newcommand{\avroraREWCPExTotalUpdate}{\ena}
\newcommand{\avroraFTODCEventTotal}{1,400}
\newcommand{\avroraFTODCNoFPEventTotal}{140}
\newcommand{\avroraFTODCNoFPAccessTotal}{140}
\newcommand{\avroraFTODCNoFPOtherTotal}{3.7}
\newcommand{\avroraFTODCReadTotal}{66.6}
\newcommand{\avroraFTODCWriteTotal}{30.8}
\newcommand{\avroraFTODCNoFPAccessInCS}{28.8}
\newcommand{\avroraFTODCNoFPAccessOutCS}{48.9}
\newcommand{\avroraFTODCAcqRelTotal}{52.1}
\newcommand{\avroraFTODCOtherTotal}{13.2}
\newcommand{\avroraFTODCNoFPReadTotal}{94}
\newcommand{\avroraFTODCReadInCS}{6.71}
\newcommand{\avroraFTODCReadOutCS}{93.3}
\newcommand{\avroraFTODCReadSameEp}{0.0275}
\newcommand{\avroraFTODCReadSharedSameEp}{\cna}
\newcommand{\avroraFTODCReadExclusive}{2.6}
\newcommand{\avroraFTODCReadOwned}{90.3}
\newcommand{\avroraFTODCReadShare}{0.669}
\newcommand{\avroraFTODCReadShared}{0.558}
\newcommand{\avroraFTODCReadSharedOwned}{5.83}
\newcommand{\avroraFTODCNoFPHonestWriteTotal}{44}
\newcommand{\avroraFTODCWriteInCS}{6.16}
\newcommand{\avroraFTODCWriteOutCS}{96.3}
\newcommand{\avroraFTODCNoFPWriteTotal}{44}
\newcommand{\avroraFTODCWriteSameEp}{2.43}
\newcommand{\avroraFTODCWriteExclusive}{0.307}
\newcommand{\avroraFTODCWriteOwned}{98.8}
\newcommand{\avroraFTODCWriteShared}{0.94}
\newcommand{\avroraFTODCNoFPOtherEventTotal}{3700058}
\newcommand{\avroraFTODCAcqRelOtherTotal}{79.7}
\newcommand{\avroraFTODCNoAcqRelOtherTotal}{750009}
\newcommand{\avroraFTODCFork}{8.0E-4}
\newcommand{\avroraFTODCJoin}{8.0E-4}
\newcommand{\avroraFTODCPreWait}{50.0}
\newcommand{\avroraFTODCPostWait}{50.0}
\newcommand{\avroraFTODCVolatileTotal}{0.0}
\newcommand{\avroraFTODCClassInit}{0.0167}
\newcommand{\avroraFTODCClassAccess}{0.0247}
\newcommand{\avroraFTODCRaceTotal}{437875}
\newcommand{\avroraFTODCWrRdRace}{53.6}
\newcommand{\avroraFTODCWrWrRace}{0.0}
\newcommand{\avroraFTODCRdWrRace}{3.28}
\newcommand{\avroraFTODCRdShWrRace}{43.1}
\newcommand{\avroraFTODCHoldLocksTotal}{8.4}
\newcommand{\avroraFTODCOneLockHeld}{5.81}
\newcommand{\avroraFTODCTwoNestedLocks}{<0.1}
\newcommand{\avroraFTODCThreeNestedLocks}{\cna}
\newcommand{\avroraFTODCFourNestedLocks}{\cna}
\newcommand{\avroraFTODCFiveNestedLocks}{\cna}
\newcommand{\avroraFTODCSixNestedLocks}{\cna}
\newcommand{\avroraFTODCSevenNestedLocks}{\cna}
\newcommand{\avroraFTODCEightNestedLocks}{\cna}
\newcommand{\avroraFTODCNineNestedLocks}{\cna}
\newcommand{\avroraFTODCTenNestedLocks}{\cna}
\newcommand{\avroraFTODCHundredNestedLocks}{\cna}
\newcommand{\avroraFTODCExWrSet}{\ena}
\newcommand{\avroraFTODCExWrCheck}{\ena}
\newcommand{\avroraFTODCExWrUpdate}{\ena}
\newcommand{\avroraFTODCExRdCheck}{\ena}
\newcommand{\avroraFTODCExRdUpdate}{\ena}
\newcommand{\avroraFTODCExTotalCheck}{\ena}
\newcommand{\avroraFTODCExTotalUpdate}{\ena}
\newcommand{\avroraREDCEventTotal}{1,400}
\newcommand{\avroraREDCNoFPEventTotal}{140}
\newcommand{\avroraREDCNoFPAccessTotal}{140}
\newcommand{\avroraREDCNoFPOtherTotal}{3.7}
\newcommand{\avroraREDCReadTotal}{66.6}
\newcommand{\avroraREDCWriteTotal}{30.8}
\newcommand{\avroraREDCNoFPAccessInCS}{34.7}
\newcommand{\avroraREDCNoFPAccessOutCS}{44.7}
\newcommand{\avroraREDCAcqRelTotal}{50.7}
\newcommand{\avroraREDCOtherTotal}{12.4}
\newcommand{\avroraREDCNoFPReadTotal}{94}
\newcommand{\avroraREDCReadInCS}{6.67}
\newcommand{\avroraREDCReadOutCS}{93.4}
\newcommand{\avroraREDCReadSameEp}{0.0275}
\newcommand{\avroraREDCReadSharedSameEp}{\cna}
\newcommand{\avroraREDCReadExclusive}{1.74}
\newcommand{\avroraREDCReadOwned}{42.3}
\newcommand{\avroraREDCReadShare}{1.09}
\newcommand{\avroraREDCReadShared}{0.926}
\newcommand{\avroraREDCReadSharedOwned}{53.9}
\newcommand{\avroraREDCNoFPHonestWriteTotal}{44}
\newcommand{\avroraREDCWriteInCS}{6.16}
\newcommand{\avroraREDCWriteOutCS}{96.3}
\newcommand{\avroraREDCNoFPWriteTotal}{44}
\newcommand{\avroraREDCWriteSameEp}{2.42}
\newcommand{\avroraREDCWriteExclusive}{0.323}
\newcommand{\avroraREDCWriteOwned}{98}
\newcommand{\avroraREDCWriteShared}{1.71}
\newcommand{\avroraREDCNoFPOtherEventTotal}{3653557}
\newcommand{\avroraREDCAcqRelOtherTotal}{80.4}
\newcommand{\avroraREDCNoAcqRelOtherTotal}{715953}
\newcommand{\avroraREDCFork}{8.38E-4}
\newcommand{\avroraREDCJoin}{8.38E-4}
\newcommand{\avroraREDCPreWait}{50.0}
\newcommand{\avroraREDCPostWait}{50.0}
\newcommand{\avroraREDCVolatileTotal}{0.0}
\newcommand{\avroraREDCClassInit}{0.0175}
\newcommand{\avroraREDCClassAccess}{0.0258}
\newcommand{\avroraREDCRaceTotal}{439883}
\newcommand{\avroraREDCWrRdRace}{53.8}
\newcommand{\avroraREDCWrWrRace}{0.0}
\newcommand{\avroraREDCRdWrRace}{3.51}
\newcommand{\avroraREDCRdShWrRace}{42.7}
\newcommand{\avroraREDCHoldLocksTotal}{8.6}
\newcommand{\avroraREDCOneLockHeld}{5.90}
\newcommand{\avroraREDCTwoNestedLocks}{<0.1}
\newcommand{\avroraREDCThreeNestedLocks}{\cna}
\newcommand{\avroraREDCFourNestedLocks}{\cna}
\newcommand{\avroraREDCFiveNestedLocks}{\cna}
\newcommand{\avroraREDCSixNestedLocks}{\cna}
\newcommand{\avroraREDCSevenNestedLocks}{\cna}
\newcommand{\avroraREDCEightNestedLocks}{\cna}
\newcommand{\avroraREDCNineNestedLocks}{\cna}
\newcommand{\avroraREDCTenNestedLocks}{\cna}
\newcommand{\avroraREDCHundredNestedLocks}{\cna}
\newcommand{\avroraREDCExWrSet}{421}
\newcommand{\avroraREDCExWrCheck}{5384222}
\newcommand{\avroraREDCExWrUpdate}{\ena}
\newcommand{\avroraREDCExRdCheck}{4018577}
\newcommand{\avroraREDCExRdUpdate}{\ena}
\newcommand{\avroraREDCExTotalCheck}{9402799}
\newcommand{\avroraREDCExTotalUpdate}{\ena}
\newcommand{\avroraFTOCAPOEventTotal}{1,400}
\newcommand{\avroraFTOCAPONoFPEventTotal}{140}
\newcommand{\avroraFTOCAPONoFPAccessTotal}{140}
\newcommand{\avroraFTOCAPONoFPOtherTotal}{3.7}
\newcommand{\avroraFTOCAPOReadTotal}{66.6}
\newcommand{\avroraFTOCAPOWriteTotal}{30.8}
\newcommand{\avroraFTOCAPONoFPAccessInCS}{28.6}
\newcommand{\avroraFTOCAPONoFPAccessOutCS}{48.9}
\newcommand{\avroraFTOCAPOAcqRelTotal}{52.1}
\newcommand{\avroraFTOCAPOOtherTotal}{13.0}
\newcommand{\avroraFTOCAPONoFPReadTotal}{94}
\newcommand{\avroraFTOCAPOReadInCS}{6.7}
\newcommand{\avroraFTOCAPOReadOutCS}{93.3}
\newcommand{\avroraFTOCAPOReadSameEp}{0.0275}
\newcommand{\avroraFTOCAPOReadSharedSameEp}{\cna}
\newcommand{\avroraFTOCAPOReadExclusive}{2.58}
\newcommand{\avroraFTOCAPOReadOwned}{90.3}
\newcommand{\avroraFTOCAPOReadShare}{0.686}
\newcommand{\avroraFTOCAPOReadShared}{0.578}
\newcommand{\avroraFTOCAPOReadSharedOwned}{5.82}
\newcommand{\avroraFTOCAPONoFPHonestWriteTotal}{44}
\newcommand{\avroraFTOCAPOWriteInCS}{6.17}
\newcommand{\avroraFTOCAPOWriteOutCS}{96.3}
\newcommand{\avroraFTOCAPONoFPWriteTotal}{44}
\newcommand{\avroraFTOCAPOWriteSameEp}{2.43}
\newcommand{\avroraFTOCAPOWriteExclusive}{0.332}
\newcommand{\avroraFTOCAPOWriteOwned}{98.7}
\newcommand{\avroraFTOCAPOWriteShared}{0.936}
\newcommand{\avroraFTOCAPONoFPOtherEventTotal}{3682500}
\newcommand{\avroraFTOCAPOAcqRelOtherTotal}{80.0}
\newcommand{\avroraFTOCAPONoAcqRelOtherTotal}{735199}
\newcommand{\avroraFTOCAPOFork}{8.16E-4}
\newcommand{\avroraFTOCAPOJoin}{8.16E-4}
\newcommand{\avroraFTOCAPOPreWait}{50.0}
\newcommand{\avroraFTOCAPOPostWait}{50.0}
\newcommand{\avroraFTOCAPOVolatileTotal}{0.0}
\newcommand{\avroraFTOCAPOClassInit}{0.017}
\newcommand{\avroraFTOCAPOClassAccess}{0.0252}
\newcommand{\avroraFTOCAPORaceTotal}{446027}
\newcommand{\avroraFTOCAPOWrRdRace}{54.6}
\newcommand{\avroraFTOCAPOWrWrRace}{0.0}
\newcommand{\avroraFTOCAPORdWrRace}{4.31}
\newcommand{\avroraFTOCAPORdShWrRace}{41.1}
\newcommand{\avroraFTOCAPOHoldLocksTotal}{8.4}
\newcommand{\avroraFTOCAPOOneLockHeld}{5.80}
\newcommand{\avroraFTOCAPOTwoNestedLocks}{<0.1}
\newcommand{\avroraFTOCAPOThreeNestedLocks}{\cna}
\newcommand{\avroraFTOCAPOFourNestedLocks}{\cna}
\newcommand{\avroraFTOCAPOFiveNestedLocks}{\cna}
\newcommand{\avroraFTOCAPOSixNestedLocks}{\cna}
\newcommand{\avroraFTOCAPOSevenNestedLocks}{\cna}
\newcommand{\avroraFTOCAPOEightNestedLocks}{\cna}
\newcommand{\avroraFTOCAPONineNestedLocks}{\cna}
\newcommand{\avroraFTOCAPOTenNestedLocks}{\cna}
\newcommand{\avroraFTOCAPOHundredNestedLocks}{\cna}
\newcommand{\avroraFTOCAPOExWrSet}{\ena}
\newcommand{\avroraFTOCAPOExWrCheck}{\ena}
\newcommand{\avroraFTOCAPOExWrUpdate}{\ena}
\newcommand{\avroraFTOCAPOExRdCheck}{\ena}
\newcommand{\avroraFTOCAPOExRdUpdate}{\ena}
\newcommand{\avroraFTOCAPOExTotalCheck}{\ena}
\newcommand{\avroraFTOCAPOExTotalUpdate}{\ena}
\newcommand{\avroraRECAPOEventTotal}{1,400}
\newcommand{\avroraRECAPONoFPEventTotal}{140}
\newcommand{\avroraRECAPONoFPAccessTotal}{140}
\newcommand{\avroraRECAPONoFPOtherTotal}{3.6}
\newcommand{\avroraRECAPOReadTotal}{66.6}
\newcommand{\avroraRECAPOWriteTotal}{30.9}
\newcommand{\avroraRECAPONoFPAccessInCS}{34.6}
\newcommand{\avroraRECAPONoFPAccessOutCS}{44.6}
\newcommand{\avroraRECAPOAcqRelTotal}{51.1}
\newcommand{\avroraRECAPOOtherTotal}{11.4}
\newcommand{\avroraRECAPONoFPReadTotal}{94}
\newcommand{\avroraRECAPOReadInCS}{6.63}
\newcommand{\avroraRECAPOReadOutCS}{93.4}
\newcommand{\avroraRECAPOReadSameEp}{0.0276}
\newcommand{\avroraRECAPOReadSharedSameEp}{\cna}
\newcommand{\avroraRECAPOReadExclusive}{1.8}
\newcommand{\avroraRECAPOReadOwned}{42.2}
\newcommand{\avroraRECAPOReadShare}{1.1}
\newcommand{\avroraRECAPOReadShared}{0.94}
\newcommand{\avroraRECAPOReadSharedOwned}{53.9}
\newcommand{\avroraRECAPONoFPHonestWriteTotal}{44}
\newcommand{\avroraRECAPOWriteInCS}{6.21}
\newcommand{\avroraRECAPOWriteOutCS}{96.2}
\newcommand{\avroraRECAPONoFPWriteTotal}{44}
\newcommand{\avroraRECAPOWriteSameEp}{2.4}
\newcommand{\avroraRECAPOWriteExclusive}{0.37}
\newcommand{\avroraRECAPOWriteOwned}{98}
\newcommand{\avroraRECAPOWriteShared}{1.7}
\newcommand{\avroraRECAPONoFPOtherEventTotal}{3576757}
\newcommand{\avroraRECAPOAcqRelOtherTotal}{81.8}
\newcommand{\avroraRECAPONoAcqRelOtherTotal}{651189}
\newcommand{\avroraRECAPOFork}{9.21E-4}
\newcommand{\avroraRECAPOJoin}{9.21E-4}
\newcommand{\avroraRECAPOPreWait}{50.0}
\newcommand{\avroraRECAPOPostWait}{50.0}
\newcommand{\avroraRECAPOVolatileTotal}{0.0}
\newcommand{\avroraRECAPOClassInit}{0.0192}
\newcommand{\avroraRECAPOClassAccess}{0.0284}
\newcommand{\avroraRECAPORaceTotal}{447573}
\newcommand{\avroraRECAPOWrRdRace}{54.8}
\newcommand{\avroraRECAPOWrWrRace}{0.0}
\newcommand{\avroraRECAPORdWrRace}{4.55}
\newcommand{\avroraRECAPORdShWrRace}{40.7}
\newcommand{\avroraRECAPOHoldLocksTotal}{8.5}
\newcommand{\avroraRECAPOOneLockHeld}{5.9}
\newcommand{\avroraRECAPOTwoNestedLocks}{<0.1}
\newcommand{\avroraRECAPOThreeNestedLocks}{\cna}
\newcommand{\avroraRECAPOFourNestedLocks}{\cna}
\newcommand{\avroraRECAPOFiveNestedLocks}{\cna}
\newcommand{\avroraRECAPOSixNestedLocks}{\cna}
\newcommand{\avroraRECAPOSevenNestedLocks}{\cna}
\newcommand{\avroraRECAPOEightNestedLocks}{\cna}
\newcommand{\avroraRECAPONineNestedLocks}{\cna}
\newcommand{\avroraRECAPOTenNestedLocks}{\cna}
\newcommand{\avroraRECAPOHundredNestedLocks}{\cna}
\newcommand{\avroraRECAPOExWrSet}{421}
\newcommand{\avroraRECAPOExWrCheck}{5411195}
\newcommand{\avroraRECAPOExWrUpdate}{\ena}
\newcommand{\avroraRECAPOExRdCheck}{4052423}
\newcommand{\avroraRECAPOExRdUpdate}{\ena}
\newcommand{\avroraRECAPOExTotalCheck}{6.8}
\newcommand{\avroraRECAPOExTotalUpdate}{\ena}
\newcommand{\FASTavroraMaxLiveThreads}{7}
\newcommand{\FASTavroraTotalThreads}{7}
\newcommand{\FASTavroraBaseTime}{2.5}
\newcommand{\FASTavroraBaseTimeCI}{14}
\newcommand{\FASTavroraEmptyTime}{\rna}
\newcommand{\FASTavroraEmptyTimeCI}{\rna}
\newcommand{\FASTavroraEmptyTimeCIMIN}{\rna}
\newcommand{\FASTavroraEmptyTimeCIMAX}{\rna}
\newcommand{\FASTavroraFTTime}{6.5}
\newcommand{\FASTavroraFTTimeCI}{0.14}
\newcommand{\FASTavroraHBTime}{5.3}
\newcommand{\FASTavroraHBTimeCI}{0.09}
\newcommand{\FASTavroraFTOHBTime}{5.4}
\newcommand{\FASTavroraFTOHBTimeCI}{0.039}
\newcommand{\FASTavroraWCPTime}{\rna}
\newcommand{\FASTavroraWCPTimeCI}{\rna}
\newcommand{\FASTavroraWCPTimeCIMIN}{\rna}
\newcommand{\FASTavroraWCPTimeCIMAX}{\rna}
\newcommand{\FASTavroraFTOWCPTime}{9.0}
\newcommand{\FASTavroraFTOWCPTimeCI}{0.14}
\newcommand{\FASTavroraREWCPTime}{7.3}
\newcommand{\FASTavroraREWCPTimeCI}{0.13}
\newcommand{\FASTavroraDCTime}{\rna}
\newcommand{\FASTavroraDCTimeCI}{\rna}
\newcommand{\FASTavroraDCTimeCIMIN}{\rna}
\newcommand{\FASTavroraDCTimeCIMAX}{\rna}
\newcommand{\FASTavroraFTODCTime}{9.5}
\newcommand{\FASTavroraFTODCTimeCI}{0.11}
\newcommand{\FASTavroraREDCTime}{7.8}
\newcommand{\FASTavroraREDCTimeCI}{0.13}
\newcommand{\FASTavroraCAPOTime}{\rna}
\newcommand{\FASTavroraCAPOTimeCI}{\rna}
\newcommand{\FASTavroraCAPOTimeCIMIN}{\rna}
\newcommand{\FASTavroraCAPOTimeCIMAX}{\rna}
\newcommand{\FASTavroraFTOCAPOTime}{7.8}
\newcommand{\FASTavroraFTOCAPOTimeCI}{0.085}
\newcommand{\FASTavroraRECAPOTime}{6.0}
\newcommand{\FASTavroraRECAPOTimeCI}{0.095}
\newcommand{\FASTavroraAGGCAPOTime}{\rna}
\newcommand{\FASTavroraAGGCAPOTimeCI}{\rna}
\newcommand{\FASTavroraAGGCAPOTimeCIMIN}{\rna}
\newcommand{\FASTavroraAGGCAPOTimeCIMAX}{\rna}
\newcommand{\FASTavroraStaticTime}{\rzero}
\newcommand{\FASTavroraDynamicTime}{\rzero}
\newcommand{\FASTavroraBaseMem}{180}
\newcommand{\FASTavroraBaseMemCI}{6.4}
\newcommand{\FASTavroraFTMem}{22}
\newcommand{\FASTavroraFTMemCI}{2.6}
\newcommand{\FASTavroraHBMem}{13}
\newcommand{\FASTavroraHBMemCI}{0.45}
\newcommand{\FASTavroraFTOHBMem}{13}
\newcommand{\FASTavroraFTOHBMemCI}{0.44}
\newcommand{\FASTavroraWCPMem}{\memna}
\newcommand{\FASTavroraWCPMemCI}{\memna}
\newcommand{\FASTavroraWCPMemCIMIN}{\memna}
\newcommand{\FASTavroraWCPMemCIMAX}{\memna}
\newcommand{\FASTavroraFTOWCPMem}{28}
\newcommand{\FASTavroraFTOWCPMemCI}{0.97}
\newcommand{\FASTavroraREWCPMem}{20}
\newcommand{\FASTavroraREWCPMemCI}{0.96}
\newcommand{\FASTavroraDCMem}{\memna}
\newcommand{\FASTavroraDCMemCI}{\memna}
\newcommand{\FASTavroraDCMemCIMIN}{\memna}
\newcommand{\FASTavroraDCMemCIMAX}{\memna}
\newcommand{\FASTavroraFTODCMem}{28}
\newcommand{\FASTavroraFTODCMemCI}{0.94}
\newcommand{\FASTavroraREDCMem}{21}
\newcommand{\FASTavroraREDCMemCI}{0.59}
\newcommand{\FASTavroraCAPOMem}{\memna}
\newcommand{\FASTavroraCAPOMemCI}{\memna}
\newcommand{\FASTavroraCAPOMemCIMIN}{\memna}
\newcommand{\FASTavroraCAPOMemCIMAX}{\memna}
\newcommand{\FASTavroraFTOCAPOMem}{25}
\newcommand{\FASTavroraFTOCAPOMemCI}{0.69}
\newcommand{\FASTavroraRECAPOMem}{14}
\newcommand{\FASTavroraRECAPOMemCI}{0.42}
\newcommand{\FASTavroraAGGCAPOMem}{\memna}
\newcommand{\FASTavroraAGGCAPOMemCI}{\memna}
\newcommand{\FASTavroraAGGCAPOMemCIMIN}{\memna}
\newcommand{\FASTavroraAGGCAPOMemCIMAX}{\memna}
\newcommand{\FASTavroraEventsCI}{17,167}
\newcommand{\FASTavroraEventsCIMIN}{1,437,241,646}
\newcommand{\FASTavroraEventsCIMAX}{1,437,275,980}
\newcommand{\FASTavroraNoFPEventsCI}{4,638}
\newcommand{\FASTavroraNoFPEventsCIMIN}{5,716,126}
\newcommand{\FASTavroraNoFPEventsCIMAX}{5,725,402}
\newcommand{\FASTavroraFT}{3}
\newcommand{\FASTavroraFTCI}{0.0}
\newcommand{\FASTavroraFTCIMIN}{3}
\newcommand{\FASTavroraFTCIMAX}{3}
\newcommand{\FASTavroraFTDynamic}{756,300}
\newcommand{\FASTavroraFTDynamicCI}{2,317}
\newcommand{\FASTavroraFTDynamicCIMIN}{753,983}
\newcommand{\FASTavroraFTDynamicCIMAX}{758,617}
\newcommand{\FASTavroraHB}{6}
\newcommand{\FASTavroraHBCI}{0.0}
\newcommand{\FASTavroraHBCIMIN}{6}
\newcommand{\FASTavroraHBCIMAX}{6}
\newcommand{\FASTavroraHBDynamic}{421,626}
\newcommand{\FASTavroraHBDynamicCI}{287}
\newcommand{\FASTavroraHBDynamicCIMIN}{421,339}
\newcommand{\FASTavroraHBDynamicCIMAX}{421,913}
\newcommand{\FASTavroraFTOHB}{6}
\newcommand{\FASTavroraFTOHBCI}{0.0}
\newcommand{\FASTavroraFTOHBCIMIN}{6}
\newcommand{\FASTavroraFTOHBCIMAX}{6}
\newcommand{\FASTavroraFTOHBDynamic}{449,288}
\newcommand{\FASTavroraFTOHBDynamicCI}{772}
\newcommand{\FASTavroraFTOHBDynamicCIMIN}{448,516}
\newcommand{\FASTavroraFTOHBDynamicCIMAX}{450,060}
\newcommand{\FASTavroraWCP}{\rna}
\newcommand{\FASTavroraWCPCI}{\rna}
\newcommand{\FASTavroraWCPCIMIN}{\rna}
\newcommand{\FASTavroraWCPCIMAX}{\rna}
\newcommand{\FASTavroraWCPDynamic}{\rna}
\newcommand{\FASTavroraWCPDynamicCI}{\rna}
\newcommand{\FASTavroraWCPDynamicCIMIN}{\rna}
\newcommand{\FASTavroraWCPDynamicCIMAX}{\rna}
\newcommand{\FASTavroraFTOWCP}{6}
\newcommand{\FASTavroraFTOWCPCI}{0.0}
\newcommand{\FASTavroraFTOWCPCIMIN}{6}
\newcommand{\FASTavroraFTOWCPCIMAX}{6}
\newcommand{\FASTavroraFTOWCPDynamic}{439,867}
\newcommand{\FASTavroraFTOWCPDynamicCI}{2,131}
\newcommand{\FASTavroraFTOWCPDynamicCIMIN}{437,736}
\newcommand{\FASTavroraFTOWCPDynamicCIMAX}{441,998}
\newcommand{\FASTavroraREWCP}{6}
\newcommand{\FASTavroraREWCPCI}{0.0}
\newcommand{\FASTavroraREWCPCIMIN}{6}
\newcommand{\FASTavroraREWCPCIMAX}{6}
\newcommand{\FASTavroraREWCPDynamic}{441,847}
\newcommand{\FASTavroraREWCPDynamicCI}{1,700}
\newcommand{\FASTavroraREWCPDynamicCIMIN}{440,147}
\newcommand{\FASTavroraREWCPDynamicCIMAX}{443,547}
\newcommand{\FASTavroraDC}{\rna}
\newcommand{\FASTavroraDCCI}{\rna}
\newcommand{\FASTavroraDCCIMIN}{\rna}
\newcommand{\FASTavroraDCCIMAX}{\rna}
\newcommand{\FASTavroraDCDynamic}{\rna}
\newcommand{\FASTavroraDCDynamicCI}{\rna}
\newcommand{\FASTavroraDCDynamicCIMIN}{\rna}
\newcommand{\FASTavroraDCDynamicCIMAX}{\rna}
\newcommand{\FASTavroraFTODC}{6}
\newcommand{\FASTavroraFTODCCI}{0.0}
\newcommand{\FASTavroraFTODCCIMIN}{6}
\newcommand{\FASTavroraFTODCCIMAX}{6}
\newcommand{\FASTavroraFTODCDynamic}{437,875}
\newcommand{\FASTavroraFTODCDynamicCI}{1,587}
\newcommand{\FASTavroraFTODCDynamicCIMIN}{436,288}
\newcommand{\FASTavroraFTODCDynamicCIMAX}{439,462}
\newcommand{\FASTavroraREDC}{6}
\newcommand{\FASTavroraREDCCI}{0.0}
\newcommand{\FASTavroraREDCCIMIN}{6}
\newcommand{\FASTavroraREDCCIMAX}{6}
\newcommand{\FASTavroraREDCDynamic}{439,883}
\newcommand{\FASTavroraREDCDynamicCI}{1,280}
\newcommand{\FASTavroraREDCDynamicCIMIN}{438,603}
\newcommand{\FASTavroraREDCDynamicCIMAX}{441,163}
\newcommand{\FASTavroraCAPO}{\rna}
\newcommand{\FASTavroraCAPOCI}{\rna}
\newcommand{\FASTavroraCAPOCIMIN}{\rna}
\newcommand{\FASTavroraCAPOCIMAX}{\rna}
\newcommand{\FASTavroraCAPODynamic}{\rna}
\newcommand{\FASTavroraCAPODynamicCI}{\rna}
\newcommand{\FASTavroraCAPODynamicCIMIN}{\rna}
\newcommand{\FASTavroraCAPODynamicCIMAX}{\rna}
\newcommand{\FASTavroraFTOCAPO}{6}
\newcommand{\FASTavroraFTOCAPOCI}{0.0}
\newcommand{\FASTavroraFTOCAPOCIMIN}{6}
\newcommand{\FASTavroraFTOCAPOCIMAX}{6}
\newcommand{\FASTavroraFTOCAPODynamic}{446,027}
\newcommand{\FASTavroraFTOCAPODynamicCI}{866}
\newcommand{\FASTavroraFTOCAPODynamicCIMIN}{445,161}
\newcommand{\FASTavroraFTOCAPODynamicCIMAX}{446,893}
\newcommand{\FASTavroraRECAPO}{6}
\newcommand{\FASTavroraRECAPOCI}{0.0}
\newcommand{\FASTavroraRECAPOCIMIN}{6}
\newcommand{\FASTavroraRECAPOCIMAX}{6}
\newcommand{\FASTavroraRECAPODynamic}{447,573}
\newcommand{\FASTavroraRECAPODynamicCI}{1,209}
\newcommand{\FASTavroraRECAPODynamicCIMIN}{446,364}
\newcommand{\FASTavroraRECAPODynamicCIMAX}{448,782}
\newcommand{\FASTavroraAGGCAPO}{\rna}
\newcommand{\FASTavroraAGGCAPOCI}{\rna}
\newcommand{\FASTavroraAGGCAPOCIMIN}{\rna}
\newcommand{\FASTavroraAGGCAPOCIMAX}{\rna}
\newcommand{\FASTavroraAGGCAPODynamic}{\rna}
\newcommand{\FASTavroraAGGCAPODynamicCI}{\rna}
\newcommand{\FASTavroraAGGCAPODynamicCIMIN}{\rna}
\newcommand{\FASTavroraAGGCAPODynamicCIMAX}{\rna}
\newcommand{\FASTbatikEvents}{160}
\newcommand{\FASTbatikNoFPEvents}{5.1}
\newcommand{\batikHBEventTotal}{160}
\newcommand{\batikHBNoFPEventTotal}{6.5}
\newcommand{\batikHBNoFPAccessTotal}{6.3}
\newcommand{\batikHBNoFPOtherTotal}{0.22}
\newcommand{\batikHBReadTotal}{60.4}
\newcommand{\batikHBWriteTotal}{36.2}
\newcommand{\batikHBNoFPAccessInCS}{76.2}
\newcommand{\batikHBNoFPAccessOutCS}{23.8}
\newcommand{\batikHBAcqRelTotal}{2.05}
\newcommand{\batikHBOtherTotal}{2.27}
\newcommand{\batikHBNoFPReadTotal}{3.9}
\newcommand{\batikHBReadInCS}{91.8}
\newcommand{\batikHBReadOutCS}{77.1}
\newcommand{\batikHBReadSameEp}{68.9}
\newcommand{\batikHBReadSharedSameEp}{\cna}
\newcommand{\batikHBReadExclusive}{100}
\newcommand{\batikHBReadOwned}{\cna}
\newcommand{\batikHBReadShare}{\cna}
\newcommand{\batikHBReadShared}{\cna}
\newcommand{\batikHBReadSharedOwned}{\cna}
\newcommand{\batikHBNoFPHonestWriteTotal}{2.4}
\newcommand{\batikHBWriteInCS}{134}
\newcommand{\batikHBWriteOutCS}{59.2}
\newcommand{\batikHBNoFPWriteTotal}{2.4}
\newcommand{\batikHBWriteSameEp}{93.3}
\newcommand{\batikHBWriteExclusive}{100}
\newcommand{\batikHBWriteOwned}{\cna}
\newcommand{\batikHBWriteShared}{\cna}
\newcommand{\batikHBNoFPOtherEventTotal}{221213}
\newcommand{\batikHBAcqRelOtherTotal}{47.4}
\newcommand{\batikHBNoAcqRelOtherTotal}{116273}
\newcommand{\batikHBFork}{0.00516}
\newcommand{\batikHBJoin}{0.0}
\newcommand{\batikHBPreWait}{0.0086}
\newcommand{\batikHBPostWait}{0.0086}
\newcommand{\batikHBVolatileTotal}{99.7}
\newcommand{\batikHBClassInit}{0.134}
\newcommand{\batikHBClassAccess}{0.0989}
\newcommand{\batikHBRaceTotal}{\rna}
\newcommand{\batikHBWrRdRace}{\rna}
\newcommand{\batikHBWrWrRace}{\rna}
\newcommand{\batikHBRdWrRace}{\rna}
\newcommand{\batikHBRdShWrRace}{\rna}
\newcommand{\batikHBHoldLocksTotal}{6.8}
\newcommand{\batikHBOneLockHeld}{107.}
\newcommand{\batikHBTwoNestedLocks}{<0.1}
\newcommand{\batikHBThreeNestedLocks}{<0.1}
\newcommand{\batikHBFourNestedLocks}{\cna}
\newcommand{\batikHBFiveNestedLocks}{\cna}
\newcommand{\batikHBSixNestedLocks}{\cna}
\newcommand{\batikHBSevenNestedLocks}{\cna}
\newcommand{\batikHBEightNestedLocks}{\cna}
\newcommand{\batikHBNineNestedLocks}{\cna}
\newcommand{\batikHBTenNestedLocks}{\cna}
\newcommand{\batikHBHundredNestedLocks}{\cna}
\newcommand{\batikHBExWrSet}{\ena}
\newcommand{\batikHBExWrCheck}{\ena}
\newcommand{\batikHBExWrUpdate}{\ena}
\newcommand{\batikHBExRdCheck}{\ena}
\newcommand{\batikHBExRdUpdate}{\ena}
\newcommand{\batikHBExTotalCheck}{\ena}
\newcommand{\batikHBExTotalUpdate}{\ena}
\newcommand{\batikFTOHBEventTotal}{160}
\newcommand{\batikFTOHBNoFPEventTotal}{5.8}
\newcommand{\batikFTOHBNoFPAccessTotal}{5.6}
\newcommand{\batikFTOHBNoFPOtherTotal}{0.22}
\newcommand{\batikFTOHBReadTotal}{55.7}
\newcommand{\batikFTOHBWriteTotal}{40.5}
\newcommand{\batikFTOHBNoFPAccessInCS}{76.2}
\newcommand{\batikFTOHBNoFPAccessOutCS}{23.8}
\newcommand{\batikFTOHBAcqRelTotal}{2.05}
\newcommand{\batikFTOHBOtherTotal}{2.27}
\newcommand{\batikFTOHBNoFPReadTotal}{3.2}
\newcommand{\batikFTOHBReadInCS}{97.6}
\newcommand{\batikFTOHBReadOutCS}{86.2}
\newcommand{\batikFTOHBReadSameEp}{83.8}
\newcommand{\batikFTOHBReadSharedSameEp}{\cna}
\newcommand{\batikFTOHBReadExclusive}{0.00708}
\newcommand{\batikFTOHBReadOwned}{100}
\newcommand{\batikFTOHBReadShare}{\cna}
\newcommand{\batikFTOHBReadShared}{\cna}
\newcommand{\batikFTOHBReadSharedOwned}{\cna}
\newcommand{\batikFTOHBNoFPHonestWriteTotal}{2.4}
\newcommand{\batikFTOHBWriteInCS}{134}
\newcommand{\batikFTOHBWriteOutCS}{59.2}
\newcommand{\batikFTOHBNoFPWriteTotal}{2.4}
\newcommand{\batikFTOHBWriteSameEp}{93.3}
\newcommand{\batikFTOHBWriteExclusive}{\cna}
\newcommand{\batikFTOHBWriteOwned}{100}
\newcommand{\batikFTOHBWriteShared}{\cna}
\newcommand{\batikFTOHBNoFPOtherEventTotal}{221213}
\newcommand{\batikFTOHBAcqRelOtherTotal}{47.4}
\newcommand{\batikFTOHBNoAcqRelOtherTotal}{116273}
\newcommand{\batikFTOHBFork}{0.00516}
\newcommand{\batikFTOHBJoin}{0.0}
\newcommand{\batikFTOHBPreWait}{0.0086}
\newcommand{\batikFTOHBPostWait}{0.0086}
\newcommand{\batikFTOHBVolatileTotal}{99.7}
\newcommand{\batikFTOHBClassInit}{0.134}
\newcommand{\batikFTOHBClassAccess}{0.0989}
\newcommand{\batikFTOHBRaceTotal}{\rna}
\newcommand{\batikFTOHBWrRdRace}{\rna}
\newcommand{\batikFTOHBWrWrRace}{\rna}
\newcommand{\batikFTOHBRdWrRace}{\rna}
\newcommand{\batikFTOHBRdShWrRace}{\rna}
\newcommand{\batikFTOHBHoldLocksTotal}{2.6}
\newcommand{\batikFTOHBOneLockHeld}{46.0}
\newcommand{\batikFTOHBTwoNestedLocks}{<0.1}
\newcommand{\batikFTOHBThreeNestedLocks}{<0.1}
\newcommand{\batikFTOHBFourNestedLocks}{\cna}
\newcommand{\batikFTOHBFiveNestedLocks}{\cna}
\newcommand{\batikFTOHBSixNestedLocks}{\cna}
\newcommand{\batikFTOHBSevenNestedLocks}{\cna}
\newcommand{\batikFTOHBEightNestedLocks}{\cna}
\newcommand{\batikFTOHBNineNestedLocks}{\cna}
\newcommand{\batikFTOHBTenNestedLocks}{\cna}
\newcommand{\batikFTOHBHundredNestedLocks}{\cna}
\newcommand{\batikFTOHBExWrSet}{\ena}
\newcommand{\batikFTOHBExWrCheck}{\ena}
\newcommand{\batikFTOHBExWrUpdate}{\ena}
\newcommand{\batikFTOHBExRdCheck}{\ena}
\newcommand{\batikFTOHBExRdUpdate}{\ena}
\newcommand{\batikFTOHBExTotalCheck}{\ena}
\newcommand{\batikFTOHBExTotalUpdate}{\ena}
\newcommand{\batikFTOWCPEventTotal}{160}
\newcommand{\batikFTOWCPNoFPEventTotal}{5.8}
\newcommand{\batikFTOWCPNoFPAccessTotal}{5.6}
\newcommand{\batikFTOWCPNoFPOtherTotal}{0.22}
\newcommand{\batikFTOWCPReadTotal}{55.6}
\newcommand{\batikFTOWCPWriteTotal}{40.6}
\newcommand{\batikFTOWCPNoFPAccessInCS}{76.2}
\newcommand{\batikFTOWCPNoFPAccessOutCS}{23.8}
\newcommand{\batikFTOWCPAcqRelTotal}{2.05}
\newcommand{\batikFTOWCPOtherTotal}{2.27}
\newcommand{\batikFTOWCPNoFPReadTotal}{3.2}
\newcommand{\batikFTOWCPReadInCS}{97.6}
\newcommand{\batikFTOWCPReadOutCS}{86.1}
\newcommand{\batikFTOWCPReadSameEp}{83.8}
\newcommand{\batikFTOWCPReadSharedSameEp}{\cna}
\newcommand{\batikFTOWCPReadExclusive}{0.0069}
\newcommand{\batikFTOWCPReadOwned}{100}
\newcommand{\batikFTOWCPReadShare}{<0.001}
\newcommand{\batikFTOWCPReadShared}{\cna}
\newcommand{\batikFTOWCPReadSharedOwned}{\cna}
\newcommand{\batikFTOWCPNoFPHonestWriteTotal}{2.4}
\newcommand{\batikFTOWCPWriteInCS}{134}
\newcommand{\batikFTOWCPWriteOutCS}{59}
\newcommand{\batikFTOWCPNoFPWriteTotal}{2.4}
\newcommand{\batikFTOWCPWriteSameEp}{93}
\newcommand{\batikFTOWCPWriteExclusive}{\cna}
\newcommand{\batikFTOWCPWriteOwned}{100}
\newcommand{\batikFTOWCPWriteShared}{<0.001}
\newcommand{\batikFTOWCPNoFPOtherEventTotal}{221213}
\newcommand{\batikFTOWCPAcqRelOtherTotal}{47.4}
\newcommand{\batikFTOWCPNoAcqRelOtherTotal}{116273}
\newcommand{\batikFTOWCPFork}{0.00516}
\newcommand{\batikFTOWCPJoin}{0.0}
\newcommand{\batikFTOWCPPreWait}{0.0086}
\newcommand{\batikFTOWCPPostWait}{0.0086}
\newcommand{\batikFTOWCPVolatileTotal}{99.7}
\newcommand{\batikFTOWCPClassInit}{0.134}
\newcommand{\batikFTOWCPClassAccess}{0.0989}
\newcommand{\batikFTOWCPRaceTotal}{\rna}
\newcommand{\batikFTOWCPWrRdRace}{\rna}
\newcommand{\batikFTOWCPWrWrRace}{\rna}
\newcommand{\batikFTOWCPRdWrRace}{\rna}
\newcommand{\batikFTOWCPRdShWrRace}{\rna}
\newcommand{\batikFTOWCPHoldLocksTotal}{2.6}
\newcommand{\batikFTOWCPOneLockHeld}{46.1}
\newcommand{\batikFTOWCPTwoNestedLocks}{<0.1}
\newcommand{\batikFTOWCPThreeNestedLocks}{<0.1}
\newcommand{\batikFTOWCPFourNestedLocks}{\cna}
\newcommand{\batikFTOWCPFiveNestedLocks}{\cna}
\newcommand{\batikFTOWCPSixNestedLocks}{\cna}
\newcommand{\batikFTOWCPSevenNestedLocks}{\cna}
\newcommand{\batikFTOWCPEightNestedLocks}{\cna}
\newcommand{\batikFTOWCPNineNestedLocks}{\cna}
\newcommand{\batikFTOWCPTenNestedLocks}{\cna}
\newcommand{\batikFTOWCPHundredNestedLocks}{\cna}
\newcommand{\batikFTOWCPExWrSet}{\ena}
\newcommand{\batikFTOWCPExWrCheck}{\ena}
\newcommand{\batikFTOWCPExWrUpdate}{\ena}
\newcommand{\batikFTOWCPExRdCheck}{\ena}
\newcommand{\batikFTOWCPExRdUpdate}{\ena}
\newcommand{\batikFTOWCPExTotalCheck}{\ena}
\newcommand{\batikFTOWCPExTotalUpdate}{\ena}
\newcommand{\batikREWCPEventTotal}{160}
\newcommand{\batikREWCPNoFPEventTotal}{5.8}
\newcommand{\batikREWCPNoFPAccessTotal}{5.6}
\newcommand{\batikREWCPNoFPOtherTotal}{0.22}
\newcommand{\batikREWCPReadTotal}{55.6}
\newcommand{\batikREWCPWriteTotal}{40.6}
\newcommand{\batikREWCPNoFPAccessInCS}{76.2}
\newcommand{\batikREWCPNoFPAccessOutCS}{23.8}
\newcommand{\batikREWCPAcqRelTotal}{2.05}
\newcommand{\batikREWCPOtherTotal}{2.27}
\newcommand{\batikREWCPNoFPReadTotal}{3.2}
\newcommand{\batikREWCPReadInCS}{97.6}
\newcommand{\batikREWCPReadOutCS}{86.1}
\newcommand{\batikREWCPReadSameEp}{83.8}
\newcommand{\batikREWCPReadSharedSameEp}{\cna}
\newcommand{\batikREWCPReadExclusive}{0.0069}
\newcommand{\batikREWCPReadOwned}{100}
\newcommand{\batikREWCPReadShare}{<0.001}
\newcommand{\batikREWCPReadShared}{\cna}
\newcommand{\batikREWCPReadSharedOwned}{\cna}
\newcommand{\batikREWCPNoFPHonestWriteTotal}{2.4}
\newcommand{\batikREWCPWriteInCS}{134}
\newcommand{\batikREWCPWriteOutCS}{59}
\newcommand{\batikREWCPNoFPWriteTotal}{2.4}
\newcommand{\batikREWCPWriteSameEp}{93}
\newcommand{\batikREWCPWriteExclusive}{\cna}
\newcommand{\batikREWCPWriteOwned}{100}
\newcommand{\batikREWCPWriteShared}{<0.001}
\newcommand{\batikREWCPNoFPOtherEventTotal}{221213}
\newcommand{\batikREWCPAcqRelOtherTotal}{47.4}
\newcommand{\batikREWCPNoAcqRelOtherTotal}{116273}
\newcommand{\batikREWCPFork}{0.00516}
\newcommand{\batikREWCPJoin}{0.0}
\newcommand{\batikREWCPPreWait}{0.0086}
\newcommand{\batikREWCPPostWait}{0.0086}
\newcommand{\batikREWCPVolatileTotal}{99.7}
\newcommand{\batikREWCPClassInit}{0.134}
\newcommand{\batikREWCPClassAccess}{0.0989}
\newcommand{\batikREWCPRaceTotal}{\rna}
\newcommand{\batikREWCPWrRdRace}{\rna}
\newcommand{\batikREWCPWrWrRace}{\rna}
\newcommand{\batikREWCPRdWrRace}{\rna}
\newcommand{\batikREWCPRdShWrRace}{\rna}
\newcommand{\batikREWCPHoldLocksTotal}{2.6}
\newcommand{\batikREWCPOneLockHeld}{46.1}
\newcommand{\batikREWCPTwoNestedLocks}{<0.1}
\newcommand{\batikREWCPThreeNestedLocks}{<0.1}
\newcommand{\batikREWCPFourNestedLocks}{\cna}
\newcommand{\batikREWCPFiveNestedLocks}{\cna}
\newcommand{\batikREWCPSixNestedLocks}{\cna}
\newcommand{\batikREWCPSevenNestedLocks}{\cna}
\newcommand{\batikREWCPEightNestedLocks}{\cna}
\newcommand{\batikREWCPNineNestedLocks}{\cna}
\newcommand{\batikREWCPTenNestedLocks}{\cna}
\newcommand{\batikREWCPHundredNestedLocks}{\cna}
\newcommand{\batikREWCPExWrSet}{\ena}
\newcommand{\batikREWCPExWrCheck}{\ena}
\newcommand{\batikREWCPExWrUpdate}{\ena}
\newcommand{\batikREWCPExRdCheck}{\ena}
\newcommand{\batikREWCPExRdUpdate}{\ena}
\newcommand{\batikREWCPExTotalCheck}{\ena}
\newcommand{\batikREWCPExTotalUpdate}{\ena}
\newcommand{\batikFTODCEventTotal}{160}
\newcommand{\batikFTODCNoFPEventTotal}{5.8}
\newcommand{\batikFTODCNoFPAccessTotal}{5.6}
\newcommand{\batikFTODCNoFPOtherTotal}{0.22}
\newcommand{\batikFTODCReadTotal}{55.6}
\newcommand{\batikFTODCWriteTotal}{40.6}
\newcommand{\batikFTODCNoFPAccessInCS}{76.2}
\newcommand{\batikFTODCNoFPAccessOutCS}{23.8}
\newcommand{\batikFTODCAcqRelTotal}{2.05}
\newcommand{\batikFTODCOtherTotal}{2.27}
\newcommand{\batikFTODCNoFPReadTotal}{3.2}
\newcommand{\batikFTODCReadInCS}{97.6}
\newcommand{\batikFTODCReadOutCS}{86.1}
\newcommand{\batikFTODCReadSameEp}{83.8}
\newcommand{\batikFTODCReadSharedSameEp}{\cna}
\newcommand{\batikFTODCReadExclusive}{0.0069}
\newcommand{\batikFTODCReadOwned}{100}
\newcommand{\batikFTODCReadShare}{<0.001}
\newcommand{\batikFTODCReadShared}{\cna}
\newcommand{\batikFTODCReadSharedOwned}{\cna}
\newcommand{\batikFTODCNoFPHonestWriteTotal}{2.4}
\newcommand{\batikFTODCWriteInCS}{134}
\newcommand{\batikFTODCWriteOutCS}{59}
\newcommand{\batikFTODCNoFPWriteTotal}{2.4}
\newcommand{\batikFTODCWriteSameEp}{93}
\newcommand{\batikFTODCWriteExclusive}{\cna}
\newcommand{\batikFTODCWriteOwned}{100}
\newcommand{\batikFTODCWriteShared}{<0.001}
\newcommand{\batikFTODCNoFPOtherEventTotal}{221213}
\newcommand{\batikFTODCAcqRelOtherTotal}{47.4}
\newcommand{\batikFTODCNoAcqRelOtherTotal}{116273}
\newcommand{\batikFTODCFork}{0.00516}
\newcommand{\batikFTODCJoin}{0.0}
\newcommand{\batikFTODCPreWait}{0.0086}
\newcommand{\batikFTODCPostWait}{0.0086}
\newcommand{\batikFTODCVolatileTotal}{99.7}
\newcommand{\batikFTODCClassInit}{0.134}
\newcommand{\batikFTODCClassAccess}{0.0989}
\newcommand{\batikFTODCRaceTotal}{\rna}
\newcommand{\batikFTODCWrRdRace}{\rna}
\newcommand{\batikFTODCWrWrRace}{\rna}
\newcommand{\batikFTODCRdWrRace}{\rna}
\newcommand{\batikFTODCRdShWrRace}{\rna}
\newcommand{\batikFTODCHoldLocksTotal}{2.6}
\newcommand{\batikFTODCOneLockHeld}{46.1}
\newcommand{\batikFTODCTwoNestedLocks}{<0.1}
\newcommand{\batikFTODCThreeNestedLocks}{<0.1}
\newcommand{\batikFTODCFourNestedLocks}{\cna}
\newcommand{\batikFTODCFiveNestedLocks}{\cna}
\newcommand{\batikFTODCSixNestedLocks}{\cna}
\newcommand{\batikFTODCSevenNestedLocks}{\cna}
\newcommand{\batikFTODCEightNestedLocks}{\cna}
\newcommand{\batikFTODCNineNestedLocks}{\cna}
\newcommand{\batikFTODCTenNestedLocks}{\cna}
\newcommand{\batikFTODCHundredNestedLocks}{\cna}
\newcommand{\batikFTODCExWrSet}{\ena}
\newcommand{\batikFTODCExWrCheck}{\ena}
\newcommand{\batikFTODCExWrUpdate}{\ena}
\newcommand{\batikFTODCExRdCheck}{\ena}
\newcommand{\batikFTODCExRdUpdate}{\ena}
\newcommand{\batikFTODCExTotalCheck}{\ena}
\newcommand{\batikFTODCExTotalUpdate}{\ena}
\newcommand{\batikREDCEventTotal}{160}
\newcommand{\batikREDCNoFPEventTotal}{5.8}
\newcommand{\batikREDCNoFPAccessTotal}{5.6}
\newcommand{\batikREDCNoFPOtherTotal}{0.22}
\newcommand{\batikREDCReadTotal}{55.6}
\newcommand{\batikREDCWriteTotal}{40.6}
\newcommand{\batikREDCNoFPAccessInCS}{76.2}
\newcommand{\batikREDCNoFPAccessOutCS}{23.8}
\newcommand{\batikREDCAcqRelTotal}{2.05}
\newcommand{\batikREDCOtherTotal}{2.27}
\newcommand{\batikREDCNoFPReadTotal}{3.2}
\newcommand{\batikREDCReadInCS}{97.6}
\newcommand{\batikREDCReadOutCS}{86.1}
\newcommand{\batikREDCReadSameEp}{83.8}
\newcommand{\batikREDCReadSharedSameEp}{\cna}
\newcommand{\batikREDCReadExclusive}{0.0069}
\newcommand{\batikREDCReadOwned}{100}
\newcommand{\batikREDCReadShare}{<0.001}
\newcommand{\batikREDCReadShared}{\cna}
\newcommand{\batikREDCReadSharedOwned}{\cna}
\newcommand{\batikREDCNoFPHonestWriteTotal}{2.4}
\newcommand{\batikREDCWriteInCS}{134}
\newcommand{\batikREDCWriteOutCS}{59}
\newcommand{\batikREDCNoFPWriteTotal}{2.4}
\newcommand{\batikREDCWriteSameEp}{93}
\newcommand{\batikREDCWriteExclusive}{\cna}
\newcommand{\batikREDCWriteOwned}{100}
\newcommand{\batikREDCWriteShared}{<0.001}
\newcommand{\batikREDCNoFPOtherEventTotal}{221213}
\newcommand{\batikREDCAcqRelOtherTotal}{47.4}
\newcommand{\batikREDCNoAcqRelOtherTotal}{116273}
\newcommand{\batikREDCFork}{0.00516}
\newcommand{\batikREDCJoin}{0.0}
\newcommand{\batikREDCPreWait}{0.0086}
\newcommand{\batikREDCPostWait}{0.0086}
\newcommand{\batikREDCVolatileTotal}{99.7}
\newcommand{\batikREDCClassInit}{0.134}
\newcommand{\batikREDCClassAccess}{0.0989}
\newcommand{\batikREDCRaceTotal}{\rna}
\newcommand{\batikREDCWrRdRace}{\rna}
\newcommand{\batikREDCWrWrRace}{\rna}
\newcommand{\batikREDCRdWrRace}{\rna}
\newcommand{\batikREDCRdShWrRace}{\rna}
\newcommand{\batikREDCHoldLocksTotal}{2.6}
\newcommand{\batikREDCOneLockHeld}{46.1}
\newcommand{\batikREDCTwoNestedLocks}{<0.1}
\newcommand{\batikREDCThreeNestedLocks}{<0.1}
\newcommand{\batikREDCFourNestedLocks}{\cna}
\newcommand{\batikREDCFiveNestedLocks}{\cna}
\newcommand{\batikREDCSixNestedLocks}{\cna}
\newcommand{\batikREDCSevenNestedLocks}{\cna}
\newcommand{\batikREDCEightNestedLocks}{\cna}
\newcommand{\batikREDCNineNestedLocks}{\cna}
\newcommand{\batikREDCTenNestedLocks}{\cna}
\newcommand{\batikREDCHundredNestedLocks}{\cna}
\newcommand{\batikREDCExWrSet}{\ena}
\newcommand{\batikREDCExWrCheck}{\ena}
\newcommand{\batikREDCExWrUpdate}{\ena}
\newcommand{\batikREDCExRdCheck}{\ena}
\newcommand{\batikREDCExRdUpdate}{\ena}
\newcommand{\batikREDCExTotalCheck}{\ena}
\newcommand{\batikREDCExTotalUpdate}{\ena}
\newcommand{\batikFTOCAPOEventTotal}{160}
\newcommand{\batikFTOCAPONoFPEventTotal}{5.8}
\newcommand{\batikFTOCAPONoFPAccessTotal}{5.6}
\newcommand{\batikFTOCAPONoFPOtherTotal}{0.22}
\newcommand{\batikFTOCAPOReadTotal}{55.6}
\newcommand{\batikFTOCAPOWriteTotal}{40.6}
\newcommand{\batikFTOCAPONoFPAccessInCS}{76.2}
\newcommand{\batikFTOCAPONoFPAccessOutCS}{23.8}
\newcommand{\batikFTOCAPOAcqRelTotal}{2.05}
\newcommand{\batikFTOCAPOOtherTotal}{2.27}
\newcommand{\batikFTOCAPONoFPReadTotal}{3.2}
\newcommand{\batikFTOCAPOReadInCS}{97.6}
\newcommand{\batikFTOCAPOReadOutCS}{86.1}
\newcommand{\batikFTOCAPOReadSameEp}{83.8}
\newcommand{\batikFTOCAPOReadSharedSameEp}{\cna}
\newcommand{\batikFTOCAPOReadExclusive}{0.0069}
\newcommand{\batikFTOCAPOReadOwned}{100}
\newcommand{\batikFTOCAPOReadShare}{<0.001}
\newcommand{\batikFTOCAPOReadShared}{\cna}
\newcommand{\batikFTOCAPOReadSharedOwned}{\cna}
\newcommand{\batikFTOCAPONoFPHonestWriteTotal}{2.4}
\newcommand{\batikFTOCAPOWriteInCS}{134}
\newcommand{\batikFTOCAPOWriteOutCS}{59}
\newcommand{\batikFTOCAPONoFPWriteTotal}{2.4}
\newcommand{\batikFTOCAPOWriteSameEp}{93}
\newcommand{\batikFTOCAPOWriteExclusive}{\cna}
\newcommand{\batikFTOCAPOWriteOwned}{100}
\newcommand{\batikFTOCAPOWriteShared}{<0.001}
\newcommand{\batikFTOCAPONoFPOtherEventTotal}{221213}
\newcommand{\batikFTOCAPOAcqRelOtherTotal}{47.4}
\newcommand{\batikFTOCAPONoAcqRelOtherTotal}{116273}
\newcommand{\batikFTOCAPOFork}{0.00516}
\newcommand{\batikFTOCAPOJoin}{0.0}
\newcommand{\batikFTOCAPOPreWait}{0.0086}
\newcommand{\batikFTOCAPOPostWait}{0.0086}
\newcommand{\batikFTOCAPOVolatileTotal}{99.7}
\newcommand{\batikFTOCAPOClassInit}{0.134}
\newcommand{\batikFTOCAPOClassAccess}{0.0989}
\newcommand{\batikFTOCAPORaceTotal}{\rna}
\newcommand{\batikFTOCAPOWrRdRace}{\rna}
\newcommand{\batikFTOCAPOWrWrRace}{\rna}
\newcommand{\batikFTOCAPORdWrRace}{\rna}
\newcommand{\batikFTOCAPORdShWrRace}{\rna}
\newcommand{\batikFTOCAPOHoldLocksTotal}{2.6}
\newcommand{\batikFTOCAPOOneLockHeld}{46.1}
\newcommand{\batikFTOCAPOTwoNestedLocks}{<0.1}
\newcommand{\batikFTOCAPOThreeNestedLocks}{<0.1}
\newcommand{\batikFTOCAPOFourNestedLocks}{\cna}
\newcommand{\batikFTOCAPOFiveNestedLocks}{\cna}
\newcommand{\batikFTOCAPOSixNestedLocks}{\cna}
\newcommand{\batikFTOCAPOSevenNestedLocks}{\cna}
\newcommand{\batikFTOCAPOEightNestedLocks}{\cna}
\newcommand{\batikFTOCAPONineNestedLocks}{\cna}
\newcommand{\batikFTOCAPOTenNestedLocks}{\cna}
\newcommand{\batikFTOCAPOHundredNestedLocks}{\cna}
\newcommand{\batikFTOCAPOExWrSet}{\ena}
\newcommand{\batikFTOCAPOExWrCheck}{\ena}
\newcommand{\batikFTOCAPOExWrUpdate}{\ena}
\newcommand{\batikFTOCAPOExRdCheck}{\ena}
\newcommand{\batikFTOCAPOExRdUpdate}{\ena}
\newcommand{\batikFTOCAPOExTotalCheck}{\ena}
\newcommand{\batikFTOCAPOExTotalUpdate}{\ena}
\newcommand{\batikRECAPOEventTotal}{160}
\newcommand{\batikRECAPONoFPEventTotal}{5.8}
\newcommand{\batikRECAPONoFPAccessTotal}{5.6}
\newcommand{\batikRECAPONoFPOtherTotal}{0.22}
\newcommand{\batikRECAPOReadTotal}{55.6}
\newcommand{\batikRECAPOWriteTotal}{40.6}
\newcommand{\batikRECAPONoFPAccessInCS}{76.2}
\newcommand{\batikRECAPONoFPAccessOutCS}{23.8}
\newcommand{\batikRECAPOAcqRelTotal}{2.05}
\newcommand{\batikRECAPOOtherTotal}{2.27}
\newcommand{\batikRECAPONoFPReadTotal}{3.2}
\newcommand{\batikRECAPOReadInCS}{97.6}
\newcommand{\batikRECAPOReadOutCS}{86.1}
\newcommand{\batikRECAPOReadSameEp}{83.8}
\newcommand{\batikRECAPOReadSharedSameEp}{\cna}
\newcommand{\batikRECAPOReadExclusive}{0.0069}
\newcommand{\batikRECAPOReadOwned}{100}
\newcommand{\batikRECAPOReadShare}{<0.001}
\newcommand{\batikRECAPOReadShared}{\cna}
\newcommand{\batikRECAPOReadSharedOwned}{\cna}
\newcommand{\batikRECAPONoFPHonestWriteTotal}{2.4}
\newcommand{\batikRECAPOWriteInCS}{134}
\newcommand{\batikRECAPOWriteOutCS}{59}
\newcommand{\batikRECAPONoFPWriteTotal}{2.4}
\newcommand{\batikRECAPOWriteSameEp}{93}
\newcommand{\batikRECAPOWriteExclusive}{\cna}
\newcommand{\batikRECAPOWriteOwned}{100}
\newcommand{\batikRECAPOWriteShared}{<0.001}
\newcommand{\batikRECAPONoFPOtherEventTotal}{221213}
\newcommand{\batikRECAPOAcqRelOtherTotal}{47.4}
\newcommand{\batikRECAPONoAcqRelOtherTotal}{116273}
\newcommand{\batikRECAPOFork}{0.00516}
\newcommand{\batikRECAPOJoin}{0.0}
\newcommand{\batikRECAPOPreWait}{0.0086}
\newcommand{\batikRECAPOPostWait}{0.0086}
\newcommand{\batikRECAPOVolatileTotal}{99.7}
\newcommand{\batikRECAPOClassInit}{0.134}
\newcommand{\batikRECAPOClassAccess}{0.0989}
\newcommand{\batikRECAPORaceTotal}{\rna}
\newcommand{\batikRECAPOWrRdRace}{\rna}
\newcommand{\batikRECAPOWrWrRace}{\rna}
\newcommand{\batikRECAPORdWrRace}{\rna}
\newcommand{\batikRECAPORdShWrRace}{\rna}
\newcommand{\batikRECAPOHoldLocksTotal}{2.6}
\newcommand{\batikRECAPOOneLockHeld}{46.1}
\newcommand{\batikRECAPOTwoNestedLocks}{<0.1}
\newcommand{\batikRECAPOThreeNestedLocks}{<0.1}
\newcommand{\batikRECAPOFourNestedLocks}{\cna}
\newcommand{\batikRECAPOFiveNestedLocks}{\cna}
\newcommand{\batikRECAPOSixNestedLocks}{\cna}
\newcommand{\batikRECAPOSevenNestedLocks}{\cna}
\newcommand{\batikRECAPOEightNestedLocks}{\cna}
\newcommand{\batikRECAPONineNestedLocks}{\cna}
\newcommand{\batikRECAPOTenNestedLocks}{\cna}
\newcommand{\batikRECAPOHundredNestedLocks}{\cna}
\newcommand{\batikRECAPOExWrSet}{\ena}
\newcommand{\batikRECAPOExWrCheck}{\ena}
\newcommand{\batikRECAPOExWrUpdate}{\ena}
\newcommand{\batikRECAPOExRdCheck}{\ena}
\newcommand{\batikRECAPOExRdUpdate}{\ena}
\newcommand{\batikRECAPOExTotalCheck}{\ena}
\newcommand{\batikRECAPOExTotalUpdate}{\ena}
\newcommand{\FASTbatikMaxLiveThreads}{7}
\newcommand{\FASTbatikTotalThreads}{7}
\newcommand{\FASTbatikBaseTime}{2.6}
\newcommand{\FASTbatikBaseTimeCI}{34}
\newcommand{\FASTbatikEmptyTime}{\rna}
\newcommand{\FASTbatikEmptyTimeCI}{\rna}
\newcommand{\FASTbatikEmptyTimeCIMIN}{\rna}
\newcommand{\FASTbatikEmptyTimeCIMAX}{\rna}
\newcommand{\FASTbatikFTTime}{4.1}
\newcommand{\FASTbatikFTTimeCI}{0.049}
\newcommand{\FASTbatikHBTime}{4.2}
\newcommand{\FASTbatikHBTimeCI}{0.071}
\newcommand{\FASTbatikFTOHBTime}{4.2}
\newcommand{\FASTbatikFTOHBTimeCI}{0.038}
\newcommand{\FASTbatikWCPTime}{\rna}
\newcommand{\FASTbatikWCPTimeCI}{\rna}
\newcommand{\FASTbatikWCPTimeCIMIN}{\rna}
\newcommand{\FASTbatikWCPTimeCIMAX}{\rna}
\newcommand{\FASTbatikFTOWCPTime}{7.0}
\newcommand{\FASTbatikFTOWCPTimeCI}{0.072}
\newcommand{\FASTbatikREWCPTime}{4.3}
\newcommand{\FASTbatikREWCPTimeCI}{0.062}
\newcommand{\FASTbatikDCTime}{\rna}
\newcommand{\FASTbatikDCTimeCI}{\rna}
\newcommand{\FASTbatikDCTimeCIMIN}{\rna}
\newcommand{\FASTbatikDCTimeCIMAX}{\rna}
\newcommand{\FASTbatikFTODCTime}{7.0}
\newcommand{\FASTbatikFTODCTimeCI}{0.074}
\newcommand{\FASTbatikREDCTime}{4.3}
\newcommand{\FASTbatikREDCTimeCI}{0.032}
\newcommand{\FASTbatikCAPOTime}{\rna}
\newcommand{\FASTbatikCAPOTimeCI}{\rna}
\newcommand{\FASTbatikCAPOTimeCIMIN}{\rna}
\newcommand{\FASTbatikCAPOTimeCIMAX}{\rna}
\newcommand{\FASTbatikFTOCAPOTime}{7.0}
\newcommand{\FASTbatikFTOCAPOTimeCI}{0.074}
\newcommand{\FASTbatikRECAPOTime}{4.3}
\newcommand{\FASTbatikRECAPOTimeCI}{0.046}
\newcommand{\FASTbatikAGGCAPOTime}{\rna}
\newcommand{\FASTbatikAGGCAPOTimeCI}{\rna}
\newcommand{\FASTbatikAGGCAPOTimeCIMIN}{\rna}
\newcommand{\FASTbatikAGGCAPOTimeCIMAX}{\rna}
\newcommand{\FASTbatikStaticTime}{\rzero}
\newcommand{\FASTbatikDynamicTime}{\rzero}
\newcommand{\FASTbatikBaseMem}{240}
\newcommand{\FASTbatikBaseMemCI}{0.89}
\newcommand{\FASTbatikFTMem}{4.9}
\newcommand{\FASTbatikFTMemCI}{0.036}
\newcommand{\FASTbatikHBMem}{4.9}
\newcommand{\FASTbatikHBMemCI}{0.043}
\newcommand{\FASTbatikFTOHBMem}{4.9}
\newcommand{\FASTbatikFTOHBMemCI}{0.072}
\newcommand{\FASTbatikWCPMem}{\memna}
\newcommand{\FASTbatikWCPMemCI}{\memna}
\newcommand{\FASTbatikWCPMemCIMIN}{\memna}
\newcommand{\FASTbatikWCPMemCIMAX}{\memna}
\newcommand{\FASTbatikFTOWCPMem}{13}
\newcommand{\FASTbatikFTOWCPMemCI}{0.22}
\newcommand{\FASTbatikREWCPMem}{5.7}
\newcommand{\FASTbatikREWCPMemCI}{0.055}
\newcommand{\FASTbatikDCMem}{\memna}
\newcommand{\FASTbatikDCMemCI}{\memna}
\newcommand{\FASTbatikDCMemCIMIN}{\memna}
\newcommand{\FASTbatikDCMemCIMAX}{\memna}
\newcommand{\FASTbatikFTODCMem}{13}
\newcommand{\FASTbatikFTODCMemCI}{0.21}
\newcommand{\FASTbatikREDCMem}{5.7}
\newcommand{\FASTbatikREDCMemCI}{0.042}
\newcommand{\FASTbatikCAPOMem}{\memna}
\newcommand{\FASTbatikCAPOMemCI}{\memna}
\newcommand{\FASTbatikCAPOMemCIMIN}{\memna}
\newcommand{\FASTbatikCAPOMemCIMAX}{\memna}
\newcommand{\FASTbatikFTOCAPOMem}{13}
\newcommand{\FASTbatikFTOCAPOMemCI}{0.085}
\newcommand{\FASTbatikRECAPOMem}{5.7}
\newcommand{\FASTbatikRECAPOMemCI}{0.028}
\newcommand{\FASTbatikAGGCAPOMem}{\memna}
\newcommand{\FASTbatikAGGCAPOMemCI}{\memna}
\newcommand{\FASTbatikAGGCAPOMemCIMIN}{\memna}
\newcommand{\FASTbatikAGGCAPOMemCIMAX}{\memna}
\newcommand{\FASTbatikEventsCI}{0}
\newcommand{\FASTbatikEventsCIMIN}{155,322,726}
\newcommand{\FASTbatikEventsCIMAX}{155,322,726}
\newcommand{\FASTbatikNoFPEventsCI}{0}
\newcommand{\FASTbatikNoFPEventsCIMIN}{5,127,824}
\newcommand{\FASTbatikNoFPEventsCIMAX}{5,127,824}
\newcommand{\FASTbatikFT}{0}
\newcommand{\FASTbatikFTCI}{0.0}
\newcommand{\FASTbatikFTCIMIN}{0}
\newcommand{\FASTbatikFTCIMAX}{0}
\newcommand{\FASTbatikFTDynamic}{0}
\newcommand{\FASTbatikFTDynamicCI}{0.0}
\newcommand{\FASTbatikFTDynamicCIMIN}{0}
\newcommand{\FASTbatikFTDynamicCIMAX}{0}
\newcommand{\FASTbatikHB}{0}
\newcommand{\FASTbatikHBCI}{0.0}
\newcommand{\FASTbatikHBCIMIN}{0}
\newcommand{\FASTbatikHBCIMAX}{0}
\newcommand{\FASTbatikHBDynamic}{0}
\newcommand{\FASTbatikHBDynamicCI}{0.0}
\newcommand{\FASTbatikHBDynamicCIMIN}{0}
\newcommand{\FASTbatikHBDynamicCIMAX}{0}
\newcommand{\FASTbatikFTOHB}{0}
\newcommand{\FASTbatikFTOHBCI}{0.0}
\newcommand{\FASTbatikFTOHBCIMIN}{0}
\newcommand{\FASTbatikFTOHBCIMAX}{0}
\newcommand{\FASTbatikFTOHBDynamic}{0}
\newcommand{\FASTbatikFTOHBDynamicCI}{0.0}
\newcommand{\FASTbatikFTOHBDynamicCIMIN}{0}
\newcommand{\FASTbatikFTOHBDynamicCIMAX}{0}
\newcommand{\FASTbatikWCP}{\rna}
\newcommand{\FASTbatikWCPCI}{\rna}
\newcommand{\FASTbatikWCPCIMIN}{\rna}
\newcommand{\FASTbatikWCPCIMAX}{\rna}
\newcommand{\FASTbatikWCPDynamic}{\rna}
\newcommand{\FASTbatikWCPDynamicCI}{\rna}
\newcommand{\FASTbatikWCPDynamicCIMIN}{\rna}
\newcommand{\FASTbatikWCPDynamicCIMAX}{\rna}
\newcommand{\FASTbatikFTOWCP}{0}
\newcommand{\FASTbatikFTOWCPCI}{0.0}
\newcommand{\FASTbatikFTOWCPCIMIN}{0}
\newcommand{\FASTbatikFTOWCPCIMAX}{0}
\newcommand{\FASTbatikFTOWCPDynamic}{0}
\newcommand{\FASTbatikFTOWCPDynamicCI}{0.0}
\newcommand{\FASTbatikFTOWCPDynamicCIMIN}{0}
\newcommand{\FASTbatikFTOWCPDynamicCIMAX}{0}
\newcommand{\FASTbatikREWCP}{0}
\newcommand{\FASTbatikREWCPCI}{0.0}
\newcommand{\FASTbatikREWCPCIMIN}{0}
\newcommand{\FASTbatikREWCPCIMAX}{0}
\newcommand{\FASTbatikREWCPDynamic}{0}
\newcommand{\FASTbatikREWCPDynamicCI}{0.0}
\newcommand{\FASTbatikREWCPDynamicCIMIN}{0}
\newcommand{\FASTbatikREWCPDynamicCIMAX}{0}
\newcommand{\FASTbatikDC}{\rna}
\newcommand{\FASTbatikDCCI}{\rna}
\newcommand{\FASTbatikDCCIMIN}{\rna}
\newcommand{\FASTbatikDCCIMAX}{\rna}
\newcommand{\FASTbatikDCDynamic}{\rna}
\newcommand{\FASTbatikDCDynamicCI}{\rna}
\newcommand{\FASTbatikDCDynamicCIMIN}{\rna}
\newcommand{\FASTbatikDCDynamicCIMAX}{\rna}
\newcommand{\FASTbatikFTODC}{0}
\newcommand{\FASTbatikFTODCCI}{0.0}
\newcommand{\FASTbatikFTODCCIMIN}{0}
\newcommand{\FASTbatikFTODCCIMAX}{0}
\newcommand{\FASTbatikFTODCDynamic}{0}
\newcommand{\FASTbatikFTODCDynamicCI}{0.0}
\newcommand{\FASTbatikFTODCDynamicCIMIN}{0}
\newcommand{\FASTbatikFTODCDynamicCIMAX}{0}
\newcommand{\FASTbatikREDC}{0}
\newcommand{\FASTbatikREDCCI}{0.0}
\newcommand{\FASTbatikREDCCIMIN}{0}
\newcommand{\FASTbatikREDCCIMAX}{0}
\newcommand{\FASTbatikREDCDynamic}{0}
\newcommand{\FASTbatikREDCDynamicCI}{0.0}
\newcommand{\FASTbatikREDCDynamicCIMIN}{0}
\newcommand{\FASTbatikREDCDynamicCIMAX}{0}
\newcommand{\FASTbatikCAPO}{\rna}
\newcommand{\FASTbatikCAPOCI}{\rna}
\newcommand{\FASTbatikCAPOCIMIN}{\rna}
\newcommand{\FASTbatikCAPOCIMAX}{\rna}
\newcommand{\FASTbatikCAPODynamic}{\rna}
\newcommand{\FASTbatikCAPODynamicCI}{\rna}
\newcommand{\FASTbatikCAPODynamicCIMIN}{\rna}
\newcommand{\FASTbatikCAPODynamicCIMAX}{\rna}
\newcommand{\FASTbatikFTOCAPO}{0}
\newcommand{\FASTbatikFTOCAPOCI}{0.0}
\newcommand{\FASTbatikFTOCAPOCIMIN}{0}
\newcommand{\FASTbatikFTOCAPOCIMAX}{0}
\newcommand{\FASTbatikFTOCAPODynamic}{0}
\newcommand{\FASTbatikFTOCAPODynamicCI}{0.0}
\newcommand{\FASTbatikFTOCAPODynamicCIMIN}{0}
\newcommand{\FASTbatikFTOCAPODynamicCIMAX}{0}
\newcommand{\FASTbatikRECAPO}{0}
\newcommand{\FASTbatikRECAPOCI}{0.0}
\newcommand{\FASTbatikRECAPOCIMIN}{0}
\newcommand{\FASTbatikRECAPOCIMAX}{0}
\newcommand{\FASTbatikRECAPODynamic}{0}
\newcommand{\FASTbatikRECAPODynamicCI}{0.0}
\newcommand{\FASTbatikRECAPODynamicCIMIN}{0}
\newcommand{\FASTbatikRECAPODynamicCIMAX}{0}
\newcommand{\FASTbatikAGGCAPO}{\rna}
\newcommand{\FASTbatikAGGCAPOCI}{\rna}
\newcommand{\FASTbatikAGGCAPOCIMIN}{\rna}
\newcommand{\FASTbatikAGGCAPOCIMAX}{\rna}
\newcommand{\FASTbatikAGGCAPODynamic}{\rna}
\newcommand{\FASTbatikAGGCAPODynamicCI}{\rna}
\newcommand{\FASTbatikAGGCAPODynamicCIMIN}{\rna}
\newcommand{\FASTbatikAGGCAPODynamicCIMAX}{\rna}
\newcommand{\FASThtwoEvents}{3,800}
\newcommand{\FASThtwoNoFPEvents}{120}
\newcommand{\htwoHBEventTotal}{3,800}
\newcommand{\htwoHBNoFPEventTotal}{340}
\newcommand{\htwoHBNoFPAccessTotal}{340}
\newcommand{\htwoHBNoFPOtherTotal}{5.1}
\newcommand{\htwoHBReadTotal}{84.9}
\newcommand{\htwoHBWriteTotal}{13.6}
\newcommand{\htwoHBNoFPAccessInCS}{96.9}
\newcommand{\htwoHBNoFPAccessOutCS}{3.02}
\newcommand{\htwoHBAcqRelTotal}{3.17}
\newcommand{\htwoHBOtherTotal}{1.13}
\newcommand{\htwoHBNoFPReadTotal}{290}
\newcommand{\htwoHBReadInCS}{94.8}
\newcommand{\htwoHBReadOutCS}{15.2}
\newcommand{\htwoHBReadSameEp}{9.99}
\newcommand{\htwoHBReadSharedSameEp}{\cna}
\newcommand{\htwoHBReadExclusive}{98.7}
\newcommand{\htwoHBReadOwned}{\cna}
\newcommand{\htwoHBReadShare}{<0.001}
\newcommand{\htwoHBReadShared}{1.26}
\newcommand{\htwoHBReadSharedOwned}{\cna}
\newcommand{\htwoHBNoFPHonestWriteTotal}{46}
\newcommand{\htwoHBWriteInCS}{258}
\newcommand{\htwoHBWriteOutCS}{24.3}
\newcommand{\htwoHBNoFPWriteTotal}{46}
\newcommand{\htwoHBWriteSameEp}{182}
\newcommand{\htwoHBWriteExclusive}{99.8}
\newcommand{\htwoHBWriteOwned}{\cna}
\newcommand{\htwoHBWriteShared}{0.24}
\newcommand{\htwoHBNoFPOtherEventTotal}{5100560}
\newcommand{\htwoHBAcqRelOtherTotal}{73.7}
\newcommand{\htwoHBNoAcqRelOtherTotal}{1339822}
\newcommand{\htwoHBFork}{0.00112}
\newcommand{\htwoHBJoin}{0.00112}
\newcommand{\htwoHBPreWait}{0.0537}
\newcommand{\htwoHBPostWait}{0.0537}
\newcommand{\htwoHBVolatileTotal}{99.9}
\newcommand{\htwoHBClassInit}{0.00269}
\newcommand{\htwoHBClassAccess}{0.0214}
\newcommand{\htwoHBRaceTotal}{127277}
\newcommand{\htwoHBWrRdRace}{32.8}
\newcommand{\htwoHBWrWrRace}{31.5}
\newcommand{\htwoHBRdWrRace}{0.0}
\newcommand{\htwoHBRdShWrRace}{35.8}
\newcommand{\htwoHBHoldLocksTotal}{390}
\newcommand{\htwoHBOneLockHeld}{117.}
\newcommand{\htwoHBTwoNestedLocks}{115.}
\newcommand{\htwoHBThreeNestedLocks}{0.14}
\newcommand{\htwoHBFourNestedLocks}{<0.1}
\newcommand{\htwoHBFiveNestedLocks}{\cna}
\newcommand{\htwoHBSixNestedLocks}{\cna}
\newcommand{\htwoHBSevenNestedLocks}{\cna}
\newcommand{\htwoHBEightNestedLocks}{\cna}
\newcommand{\htwoHBNineNestedLocks}{\cna}
\newcommand{\htwoHBTenNestedLocks}{\cna}
\newcommand{\htwoHBHundredNestedLocks}{\cna}
\newcommand{\htwoHBExWrSet}{\ena}
\newcommand{\htwoHBExWrCheck}{\ena}
\newcommand{\htwoHBExWrUpdate}{\ena}
\newcommand{\htwoHBExRdCheck}{\ena}
\newcommand{\htwoHBExRdUpdate}{\ena}
\newcommand{\htwoHBExTotalCheck}{\ena}
\newcommand{\htwoHBExTotalUpdate}{\ena}
\newcommand{\htwoFTOHBEventTotal}{3,800}
\newcommand{\htwoFTOHBNoFPEventTotal}{300}
\newcommand{\htwoFTOHBNoFPAccessTotal}{300}
\newcommand{\htwoFTOHBNoFPOtherTotal}{5.1}
\newcommand{\htwoFTOHBReadTotal}{82.9}
\newcommand{\htwoFTOHBWriteTotal}{15.4}
\newcommand{\htwoFTOHBNoFPAccessInCS}{96.9}
\newcommand{\htwoFTOHBNoFPAccessOutCS}{3.01}
\newcommand{\htwoFTOHBAcqRelTotal}{3.17}
\newcommand{\htwoFTOHBOtherTotal}{1.13}
\newcommand{\htwoFTOHBNoFPReadTotal}{250}
\newcommand{\htwoFTOHBReadInCS}{94.2}
\newcommand{\htwoFTOHBReadOutCS}{17.3}
\newcommand{\htwoFTOHBReadSameEp}{11.6}
\newcommand{\htwoFTOHBReadSharedSameEp}{\cna}
\newcommand{\htwoFTOHBReadExclusive}{14.3}
\newcommand{\htwoFTOHBReadOwned}{84.3}
\newcommand{\htwoFTOHBReadShare}{0.0157}
\newcommand{\htwoFTOHBReadShared}{0.00631}
\newcommand{\htwoFTOHBReadSharedOwned}{1.39}
\newcommand{\htwoFTOHBNoFPHonestWriteTotal}{46}
\newcommand{\htwoFTOHBWriteInCS}{258}
\newcommand{\htwoFTOHBWriteOutCS}{24.3}
\newcommand{\htwoFTOHBNoFPWriteTotal}{46}
\newcommand{\htwoFTOHBWriteSameEp}{182}
\newcommand{\htwoFTOHBWriteExclusive}{0.299}
\newcommand{\htwoFTOHBWriteOwned}{99.6}
\newcommand{\htwoFTOHBWriteShared}{0.0834}
\newcommand{\htwoFTOHBNoFPOtherEventTotal}{5100480}
\newcommand{\htwoFTOHBAcqRelOtherTotal}{73.7}
\newcommand{\htwoFTOHBNoAcqRelOtherTotal}{1339765}
\newcommand{\htwoFTOHBFork}{0.00112}
\newcommand{\htwoFTOHBJoin}{0.00112}
\newcommand{\htwoFTOHBPreWait}{0.0519}
\newcommand{\htwoFTOHBPostWait}{0.0519}
\newcommand{\htwoFTOHBVolatileTotal}{99.9}
\newcommand{\htwoFTOHBClassInit}{0.00269}
\newcommand{\htwoFTOHBClassAccess}{0.0214}
\newcommand{\htwoFTOHBRaceTotal}{76525}
\newcommand{\htwoFTOHBWrRdRace}{49.9}
\newcommand{\htwoFTOHBWrWrRace}{0.0}
\newcommand{\htwoFTOHBRdWrRace}{0.397}
\newcommand{\htwoFTOHBRdShWrRace}{49.7}
\newcommand{\htwoFTOHBHoldLocksTotal}{250}
\newcommand{\htwoFTOHBOneLockHeld}{82.6}
\newcommand{\htwoFTOHBTwoNestedLocks}{80.6}
\newcommand{\htwoFTOHBThreeNestedLocks}{0.17}
\newcommand{\htwoFTOHBFourNestedLocks}{<0.1}
\newcommand{\htwoFTOHBFiveNestedLocks}{\cna}
\newcommand{\htwoFTOHBSixNestedLocks}{\cna}
\newcommand{\htwoFTOHBSevenNestedLocks}{\cna}
\newcommand{\htwoFTOHBEightNestedLocks}{\cna}
\newcommand{\htwoFTOHBNineNestedLocks}{\cna}
\newcommand{\htwoFTOHBTenNestedLocks}{\cna}
\newcommand{\htwoFTOHBHundredNestedLocks}{\cna}
\newcommand{\htwoFTOHBExWrSet}{\ena}
\newcommand{\htwoFTOHBExWrCheck}{\ena}
\newcommand{\htwoFTOHBExWrUpdate}{\ena}
\newcommand{\htwoFTOHBExRdCheck}{\ena}
\newcommand{\htwoFTOHBExRdUpdate}{\ena}
\newcommand{\htwoFTOHBExTotalCheck}{\ena}
\newcommand{\htwoFTOHBExTotalUpdate}{\ena}
\newcommand{\htwoFTOWCPEventTotal}{3,800}
\newcommand{\htwoFTOWCPNoFPEventTotal}{310}
\newcommand{\htwoFTOWCPNoFPAccessTotal}{300}
\newcommand{\htwoFTOWCPNoFPOtherTotal}{5.1}
\newcommand{\htwoFTOWCPReadTotal}{83.1}
\newcommand{\htwoFTOWCPWriteTotal}{15.2}
\newcommand{\htwoFTOWCPNoFPAccessInCS}{96.9}
\newcommand{\htwoFTOWCPNoFPAccessOutCS}{3.02}
\newcommand{\htwoFTOWCPAcqRelTotal}{3.17}
\newcommand{\htwoFTOWCPOtherTotal}{1.13}
\newcommand{\htwoFTOWCPNoFPReadTotal}{250}
\newcommand{\htwoFTOWCPReadInCS}{94.3}
\newcommand{\htwoFTOWCPReadOutCS}{17.1}
\newcommand{\htwoFTOWCPReadSameEp}{11.4}
\newcommand{\htwoFTOWCPReadSharedSameEp}{\cna}
\newcommand{\htwoFTOWCPReadExclusive}{11.7}
\newcommand{\htwoFTOWCPReadOwned}{83.5}
\newcommand{\htwoFTOWCPReadShare}{0.0482}
\newcommand{\htwoFTOWCPReadShared}{0.181}
\newcommand{\htwoFTOWCPReadSharedOwned}{4.53}
\newcommand{\htwoFTOWCPNoFPHonestWriteTotal}{46}
\newcommand{\htwoFTOWCPWriteInCS}{258}
\newcommand{\htwoFTOWCPWriteOutCS}{24.3}
\newcommand{\htwoFTOWCPNoFPWriteTotal}{46}
\newcommand{\htwoFTOWCPWriteSameEp}{182}
\newcommand{\htwoFTOWCPWriteExclusive}{0.306}
\newcommand{\htwoFTOWCPWriteOwned}{99.5}
\newcommand{\htwoFTOWCPWriteShared}{0.172}
\newcommand{\htwoFTOWCPNoFPOtherEventTotal}{5103979}
\newcommand{\htwoFTOWCPAcqRelOtherTotal}{73.7}
\newcommand{\htwoFTOWCPNoAcqRelOtherTotal}{1342259}
\newcommand{\htwoFTOWCPFork}{0.00112}
\newcommand{\htwoFTOWCPJoin}{0.00112}
\newcommand{\htwoFTOWCPPreWait}{0.128}
\newcommand{\htwoFTOWCPPostWait}{0.128}
\newcommand{\htwoFTOWCPVolatileTotal}{99.7}
\newcommand{\htwoFTOWCPClassInit}{0.00268}
\newcommand{\htwoFTOWCPClassAccess}{0.0214}
\newcommand{\htwoFTOWCPRaceTotal}{86043}
\newcommand{\htwoFTOWCPWrRdRace}{49.9}
\newcommand{\htwoFTOWCPWrWrRace}{0.0}
\newcommand{\htwoFTOWCPRdWrRace}{0.449}
\newcommand{\htwoFTOWCPRdShWrRace}{49.7}
\newcommand{\htwoFTOWCPHoldLocksTotal}{250}
\newcommand{\htwoFTOWCPOneLockHeld}{82.8}
\newcommand{\htwoFTOWCPTwoNestedLocks}{80.1}
\newcommand{\htwoFTOWCPThreeNestedLocks}{0.17}
\newcommand{\htwoFTOWCPFourNestedLocks}{<0.1}
\newcommand{\htwoFTOWCPFiveNestedLocks}{\cna}
\newcommand{\htwoFTOWCPSixNestedLocks}{\cna}
\newcommand{\htwoFTOWCPSevenNestedLocks}{\cna}
\newcommand{\htwoFTOWCPEightNestedLocks}{\cna}
\newcommand{\htwoFTOWCPNineNestedLocks}{\cna}
\newcommand{\htwoFTOWCPTenNestedLocks}{\cna}
\newcommand{\htwoFTOWCPHundredNestedLocks}{\cna}
\newcommand{\htwoFTOWCPExWrSet}{\ena}
\newcommand{\htwoFTOWCPExWrCheck}{\ena}
\newcommand{\htwoFTOWCPExWrUpdate}{\ena}
\newcommand{\htwoFTOWCPExRdCheck}{\ena}
\newcommand{\htwoFTOWCPExRdUpdate}{\ena}
\newcommand{\htwoFTOWCPExTotalCheck}{\ena}
\newcommand{\htwoFTOWCPExTotalUpdate}{\ena}
\newcommand{\htwoREWCPEventTotal}{3,800}
\newcommand{\htwoREWCPNoFPEventTotal}{300}
\newcommand{\htwoREWCPNoFPAccessTotal}{300}
\newcommand{\htwoREWCPNoFPOtherTotal}{5.1}
\newcommand{\htwoREWCPReadTotal}{83.1}
\newcommand{\htwoREWCPWriteTotal}{15.2}
\newcommand{\htwoREWCPNoFPAccessInCS}{96.9}
\newcommand{\htwoREWCPNoFPAccessOutCS}{3.01}
\newcommand{\htwoREWCPAcqRelTotal}{3.17}
\newcommand{\htwoREWCPOtherTotal}{1.13}
\newcommand{\htwoREWCPNoFPReadTotal}{250}
\newcommand{\htwoREWCPReadInCS}{94.3}
\newcommand{\htwoREWCPReadOutCS}{17.1}
\newcommand{\htwoREWCPReadSameEp}{11.5}
\newcommand{\htwoREWCPReadSharedSameEp}{\cna}
\newcommand{\htwoREWCPReadExclusive}{9.95}
\newcommand{\htwoREWCPReadOwned}{83.1}
\newcommand{\htwoREWCPReadShare}{0.202}
\newcommand{\htwoREWCPReadShared}{0.389}
\newcommand{\htwoREWCPReadSharedOwned}{6.39}
\newcommand{\htwoREWCPNoFPHonestWriteTotal}{46}
\newcommand{\htwoREWCPWriteInCS}{258}
\newcommand{\htwoREWCPWriteOutCS}{24.2}
\newcommand{\htwoREWCPNoFPWriteTotal}{46}
\newcommand{\htwoREWCPWriteSameEp}{182}
\newcommand{\htwoREWCPWriteExclusive}{0.29}
\newcommand{\htwoREWCPWriteOwned}{98.9}
\newcommand{\htwoREWCPWriteShared}{0.808}
\newcommand{\htwoREWCPNoFPOtherEventTotal}{5100890}
\newcommand{\htwoREWCPAcqRelOtherTotal}{73.7}
\newcommand{\htwoREWCPNoAcqRelOtherTotal}{1340060}
\newcommand{\htwoREWCPFork}{0.00112}
\newcommand{\htwoREWCPJoin}{0.00112}
\newcommand{\htwoREWCPPreWait}{0.0616}
\newcommand{\htwoREWCPPostWait}{0.0616}
\newcommand{\htwoREWCPVolatileTotal}{99.9}
\newcommand{\htwoREWCPClassInit}{0.00269}
\newcommand{\htwoREWCPClassAccess}{0.0214}
\newcommand{\htwoREWCPRaceTotal}{82217}
\newcommand{\htwoREWCPWrRdRace}{49.9}
\newcommand{\htwoREWCPWrWrRace}{0.0}
\newcommand{\htwoREWCPRdWrRace}{0.167}
\newcommand{\htwoREWCPRdShWrRace}{49.9}
\newcommand{\htwoREWCPHoldLocksTotal}{250}
\newcommand{\htwoREWCPOneLockHeld}{82.8}
\newcommand{\htwoREWCPTwoNestedLocks}{80.1}
\newcommand{\htwoREWCPThreeNestedLocks}{0.17}
\newcommand{\htwoREWCPFourNestedLocks}{<0.1}
\newcommand{\htwoREWCPFiveNestedLocks}{\cna}
\newcommand{\htwoREWCPSixNestedLocks}{\cna}
\newcommand{\htwoREWCPSevenNestedLocks}{\cna}
\newcommand{\htwoREWCPEightNestedLocks}{\cna}
\newcommand{\htwoREWCPNineNestedLocks}{\cna}
\newcommand{\htwoREWCPTenNestedLocks}{\cna}
\newcommand{\htwoREWCPHundredNestedLocks}{\cna}
\newcommand{\htwoREWCPExWrSet}{288965}
\newcommand{\htwoREWCPExWrCheck}{523644}
\newcommand{\htwoREWCPExWrUpdate}{\ena}
\newcommand{\htwoREWCPExRdCheck}{836967}
\newcommand{\htwoREWCPExRdUpdate}{\ena}
\newcommand{\htwoREWCPExTotalCheck}{1360610}
\newcommand{\htwoREWCPExTotalUpdate}{\ena}
\newcommand{\htwoFTODCEventTotal}{3,800}
\newcommand{\htwoFTODCNoFPEventTotal}{310}
\newcommand{\htwoFTODCNoFPAccessTotal}{300}
\newcommand{\htwoFTODCNoFPOtherTotal}{5.1}
\newcommand{\htwoFTODCReadTotal}{83.1}
\newcommand{\htwoFTODCWriteTotal}{15.2}
\newcommand{\htwoFTODCNoFPAccessInCS}{96.9}
\newcommand{\htwoFTODCNoFPAccessOutCS}{3.02}
\newcommand{\htwoFTODCAcqRelTotal}{3.17}
\newcommand{\htwoFTODCOtherTotal}{1.13}
\newcommand{\htwoFTODCNoFPReadTotal}{250}
\newcommand{\htwoFTODCReadInCS}{94.3}
\newcommand{\htwoFTODCReadOutCS}{17.1}
\newcommand{\htwoFTODCReadSameEp}{11.4}
\newcommand{\htwoFTODCReadSharedSameEp}{\cna}
\newcommand{\htwoFTODCReadExclusive}{11.1}
\newcommand{\htwoFTODCReadOwned}{83.4}
\newcommand{\htwoFTODCReadShare}{0.0764}
\newcommand{\htwoFTODCReadShared}{0.305}
\newcommand{\htwoFTODCReadSharedOwned}{5.17}
\newcommand{\htwoFTODCNoFPHonestWriteTotal}{46}
\newcommand{\htwoFTODCWriteInCS}{258}
\newcommand{\htwoFTODCWriteOutCS}{24.3}
\newcommand{\htwoFTODCNoFPWriteTotal}{46}
\newcommand{\htwoFTODCWriteSameEp}{182}
\newcommand{\htwoFTODCWriteExclusive}{0.303}
\newcommand{\htwoFTODCWriteOwned}{99.5}
\newcommand{\htwoFTODCWriteShared}{0.237}
\newcommand{\htwoFTODCNoFPOtherEventTotal}{5103256}
\newcommand{\htwoFTODCAcqRelOtherTotal}{73.7}
\newcommand{\htwoFTODCNoAcqRelOtherTotal}{1342080}
\newcommand{\htwoFTODCFork}{0.00112}
\newcommand{\htwoFTODCJoin}{0.00112}
\newcommand{\htwoFTODCPreWait}{0.127}
\newcommand{\htwoFTODCPostWait}{0.127}
\newcommand{\htwoFTODCVolatileTotal}{99.7}
\newcommand{\htwoFTODCClassInit}{0.00268}
\newcommand{\htwoFTODCClassAccess}{0.0214}
\newcommand{\htwoFTODCRaceTotal}{86742}
\newcommand{\htwoFTODCWrRdRace}{49.8}
\newcommand{\htwoFTODCWrWrRace}{0.0}
\newcommand{\htwoFTODCRdWrRace}{0.416}
\newcommand{\htwoFTODCRdShWrRace}{49.8}
\newcommand{\htwoFTODCHoldLocksTotal}{250}
\newcommand{\htwoFTODCOneLockHeld}{82.8}
\newcommand{\htwoFTODCTwoNestedLocks}{80.1}
\newcommand{\htwoFTODCThreeNestedLocks}{0.17}
\newcommand{\htwoFTODCFourNestedLocks}{<0.1}
\newcommand{\htwoFTODCFiveNestedLocks}{\cna}
\newcommand{\htwoFTODCSixNestedLocks}{\cna}
\newcommand{\htwoFTODCSevenNestedLocks}{\cna}
\newcommand{\htwoFTODCEightNestedLocks}{\cna}
\newcommand{\htwoFTODCNineNestedLocks}{\cna}
\newcommand{\htwoFTODCTenNestedLocks}{\cna}
\newcommand{\htwoFTODCHundredNestedLocks}{\cna}
\newcommand{\htwoFTODCExWrSet}{\ena}
\newcommand{\htwoFTODCExWrCheck}{\ena}
\newcommand{\htwoFTODCExWrUpdate}{\ena}
\newcommand{\htwoFTODCExRdCheck}{\ena}
\newcommand{\htwoFTODCExRdUpdate}{\ena}
\newcommand{\htwoFTODCExTotalCheck}{\ena}
\newcommand{\htwoFTODCExTotalUpdate}{\ena}
\newcommand{\htwoREDCEventTotal}{3,800}
\newcommand{\htwoREDCNoFPEventTotal}{300}
\newcommand{\htwoREDCNoFPAccessTotal}{300}
\newcommand{\htwoREDCNoFPOtherTotal}{5.1}
\newcommand{\htwoREDCReadTotal}{83.1}
\newcommand{\htwoREDCWriteTotal}{15.2}
\newcommand{\htwoREDCNoFPAccessInCS}{96.9}
\newcommand{\htwoREDCNoFPAccessOutCS}{3.01}
\newcommand{\htwoREDCAcqRelTotal}{3.17}
\newcommand{\htwoREDCOtherTotal}{1.13}
\newcommand{\htwoREDCNoFPReadTotal}{250}
\newcommand{\htwoREDCReadInCS}{94.3}
\newcommand{\htwoREDCReadOutCS}{17.1}
\newcommand{\htwoREDCReadSameEp}{11.5}
\newcommand{\htwoREDCReadSharedSameEp}{\cna}
\newcommand{\htwoREDCReadExclusive}{7.46}
\newcommand{\htwoREDCReadOwned}{82.6}
\newcommand{\htwoREDCReadShare}{0.26}
\newcommand{\htwoREDCReadShared}{0.877}
\newcommand{\htwoREDCReadSharedOwned}{8.77}
\newcommand{\htwoREDCNoFPHonestWriteTotal}{46}
\newcommand{\htwoREDCWriteInCS}{258}
\newcommand{\htwoREDCWriteOutCS}{24.2}
\newcommand{\htwoREDCNoFPWriteTotal}{46}
\newcommand{\htwoREDCWriteSameEp}{182}
\newcommand{\htwoREDCWriteExclusive}{0.287}
\newcommand{\htwoREDCWriteOwned}{98.8}
\newcommand{\htwoREDCWriteShared}{0.886}
\newcommand{\htwoREDCNoFPOtherEventTotal}{5100793}
\newcommand{\htwoREDCAcqRelOtherTotal}{73.7}
\newcommand{\htwoREDCNoAcqRelOtherTotal}{1340096}
\newcommand{\htwoREDCFork}{0.00112}
\newcommand{\htwoREDCJoin}{0.00112}
\newcommand{\htwoREDCPreWait}{0.0639}
\newcommand{\htwoREDCPostWait}{0.0639}
\newcommand{\htwoREDCVolatileTotal}{99.8}
\newcommand{\htwoREDCClassInit}{0.00269}
\newcommand{\htwoREDCClassAccess}{0.0214}
\newcommand{\htwoREDCRaceTotal}{83243}
\newcommand{\htwoREDCWrRdRace}{49.8}
\newcommand{\htwoREDCWrWrRace}{0.0}
\newcommand{\htwoREDCRdWrRace}{0.157}
\newcommand{\htwoREDCRdShWrRace}{50.0}
\newcommand{\htwoREDCHoldLocksTotal}{250}
\newcommand{\htwoREDCOneLockHeld}{82.8}
\newcommand{\htwoREDCTwoNestedLocks}{80.1}
\newcommand{\htwoREDCThreeNestedLocks}{0.17}
\newcommand{\htwoREDCFourNestedLocks}{<0.1}
\newcommand{\htwoREDCFiveNestedLocks}{\cna}
\newcommand{\htwoREDCSixNestedLocks}{\cna}
\newcommand{\htwoREDCSevenNestedLocks}{\cna}
\newcommand{\htwoREDCEightNestedLocks}{\cna}
\newcommand{\htwoREDCNineNestedLocks}{\cna}
\newcommand{\htwoREDCTenNestedLocks}{\cna}
\newcommand{\htwoREDCHundredNestedLocks}{\cna}
\newcommand{\htwoREDCExWrSet}{315750}
\newcommand{\htwoREDCExWrCheck}{545958}
\newcommand{\htwoREDCExWrUpdate}{0}
\newcommand{\htwoREDCExRdCheck}{850572}
\newcommand{\htwoREDCExRdUpdate}{0}
\newcommand{\htwoREDCExTotalCheck}{1396530}
\newcommand{\htwoREDCExTotalUpdate}{0}
\newcommand{\htwoFTOCAPOEventTotal}{3,800}
\newcommand{\htwoFTOCAPONoFPEventTotal}{300}
\newcommand{\htwoFTOCAPONoFPAccessTotal}{300}
\newcommand{\htwoFTOCAPONoFPOtherTotal}{5.1}
\newcommand{\htwoFTOCAPOReadTotal}{83.1}
\newcommand{\htwoFTOCAPOWriteTotal}{15.2}
\newcommand{\htwoFTOCAPONoFPAccessInCS}{96.9}
\newcommand{\htwoFTOCAPONoFPAccessOutCS}{3.02}
\newcommand{\htwoFTOCAPOAcqRelTotal}{3.17}
\newcommand{\htwoFTOCAPOOtherTotal}{1.13}
\newcommand{\htwoFTOCAPONoFPReadTotal}{250}
\newcommand{\htwoFTOCAPOReadInCS}{94.3}
\newcommand{\htwoFTOCAPOReadOutCS}{17.1}
\newcommand{\htwoFTOCAPOReadSameEp}{11.4}
\newcommand{\htwoFTOCAPOReadSharedSameEp}{\cna}
\newcommand{\htwoFTOCAPOReadExclusive}{10.9}
\newcommand{\htwoFTOCAPOReadOwned}{83.3}
\newcommand{\htwoFTOCAPOReadShare}{0.0799}
\newcommand{\htwoFTOCAPOReadShared}{0.352}
\newcommand{\htwoFTOCAPOReadSharedOwned}{5.35}
\newcommand{\htwoFTOCAPONoFPHonestWriteTotal}{46}
\newcommand{\htwoFTOCAPOWriteInCS}{258}
\newcommand{\htwoFTOCAPOWriteOutCS}{24.3}
\newcommand{\htwoFTOCAPONoFPWriteTotal}{46}
\newcommand{\htwoFTOCAPOWriteSameEp}{182}
\newcommand{\htwoFTOCAPOWriteExclusive}{0.303}
\newcommand{\htwoFTOCAPOWriteOwned}{99.5}
\newcommand{\htwoFTOCAPOWriteShared}{0.238}
\newcommand{\htwoFTOCAPONoFPOtherEventTotal}{5102764}
\newcommand{\htwoFTOCAPOAcqRelOtherTotal}{73.7}
\newcommand{\htwoFTOCAPONoAcqRelOtherTotal}{1342004}
\newcommand{\htwoFTOCAPOFork}{0.00112}
\newcommand{\htwoFTOCAPOJoin}{0.00112}
\newcommand{\htwoFTOCAPOPreWait}{0.126}
\newcommand{\htwoFTOCAPOPostWait}{0.126}
\newcommand{\htwoFTOCAPOVolatileTotal}{99.7}
\newcommand{\htwoFTOCAPOClassInit}{0.00268}
\newcommand{\htwoFTOCAPOClassAccess}{0.0213}
\newcommand{\htwoFTOCAPORaceTotal}{86839}
\newcommand{\htwoFTOCAPOWrRdRace}{49.8}
\newcommand{\htwoFTOCAPOWrWrRace}{0.0}
\newcommand{\htwoFTOCAPORdWrRace}{0.58}
\newcommand{\htwoFTOCAPORdShWrRace}{49.6}
\newcommand{\htwoFTOCAPOHoldLocksTotal}{250}
\newcommand{\htwoFTOCAPOOneLockHeld}{82.8}
\newcommand{\htwoFTOCAPOTwoNestedLocks}{80.1}
\newcommand{\htwoFTOCAPOThreeNestedLocks}{0.17}
\newcommand{\htwoFTOCAPOFourNestedLocks}{<0.1}
\newcommand{\htwoFTOCAPOFiveNestedLocks}{\cna}
\newcommand{\htwoFTOCAPOSixNestedLocks}{\cna}
\newcommand{\htwoFTOCAPOSevenNestedLocks}{\cna}
\newcommand{\htwoFTOCAPOEightNestedLocks}{\cna}
\newcommand{\htwoFTOCAPONineNestedLocks}{\cna}
\newcommand{\htwoFTOCAPOTenNestedLocks}{\cna}
\newcommand{\htwoFTOCAPOHundredNestedLocks}{\cna}
\newcommand{\htwoFTOCAPOExWrSet}{\ena}
\newcommand{\htwoFTOCAPOExWrCheck}{\ena}
\newcommand{\htwoFTOCAPOExWrUpdate}{\ena}
\newcommand{\htwoFTOCAPOExRdCheck}{\ena}
\newcommand{\htwoFTOCAPOExRdUpdate}{\ena}
\newcommand{\htwoFTOCAPOExTotalCheck}{\ena}
\newcommand{\htwoFTOCAPOExTotalUpdate}{\ena}
\newcommand{\htwoRECAPOEventTotal}{3,800}
\newcommand{\htwoRECAPONoFPEventTotal}{300}
\newcommand{\htwoRECAPONoFPAccessTotal}{300}
\newcommand{\htwoRECAPONoFPOtherTotal}{5.1}
\newcommand{\htwoRECAPOReadTotal}{83.1}
\newcommand{\htwoRECAPOWriteTotal}{15.2}
\newcommand{\htwoRECAPONoFPAccessInCS}{96.9}
\newcommand{\htwoRECAPONoFPAccessOutCS}{3.01}
\newcommand{\htwoRECAPOAcqRelTotal}{3.17}
\newcommand{\htwoRECAPOOtherTotal}{1.13}
\newcommand{\htwoRECAPONoFPReadTotal}{250}
\newcommand{\htwoRECAPOReadInCS}{94.3}
\newcommand{\htwoRECAPOReadOutCS}{17.1}
\newcommand{\htwoRECAPOReadSameEp}{11.5}
\newcommand{\htwoRECAPOReadSharedSameEp}{\cna}
\newcommand{\htwoRECAPOReadExclusive}{7.6}
\newcommand{\htwoRECAPOReadOwned}{82.7}
\newcommand{\htwoRECAPOReadShare}{0.25}
\newcommand{\htwoRECAPOReadShared}{0.85}
\newcommand{\htwoRECAPOReadSharedOwned}{8.7}
\newcommand{\htwoRECAPONoFPHonestWriteTotal}{46}
\newcommand{\htwoRECAPOWriteInCS}{258}
\newcommand{\htwoRECAPOWriteOutCS}{24.2}
\newcommand{\htwoRECAPONoFPWriteTotal}{46}
\newcommand{\htwoRECAPOWriteSameEp}{182}
\newcommand{\htwoRECAPOWriteExclusive}{0.28}
\newcommand{\htwoRECAPOWriteOwned}{98.9}
\newcommand{\htwoRECAPOWriteShared}{0.85}
\newcommand{\htwoRECAPONoFPOtherEventTotal}{5100781}
\newcommand{\htwoRECAPOAcqRelOtherTotal}{73.7}
\newcommand{\htwoRECAPONoAcqRelOtherTotal}{1339963}
\newcommand{\htwoRECAPOFork}{0.00112}
\newcommand{\htwoRECAPOJoin}{0.00112}
\newcommand{\htwoRECAPOPreWait}{0.0584}
\newcommand{\htwoRECAPOPostWait}{0.0584}
\newcommand{\htwoRECAPOVolatileTotal}{99.9}
\newcommand{\htwoRECAPOClassInit}{0.00269}
\newcommand{\htwoRECAPOClassAccess}{0.0214}
\newcommand{\htwoRECAPORaceTotal}{81677}
\newcommand{\htwoRECAPOWrRdRace}{49.9}
\newcommand{\htwoRECAPOWrWrRace}{0.0}
\newcommand{\htwoRECAPORdWrRace}{0.362}
\newcommand{\htwoRECAPORdShWrRace}{49.8}
\newcommand{\htwoRECAPOHoldLocksTotal}{250}
\newcommand{\htwoRECAPOOneLockHeld}{82.8}
\newcommand{\htwoRECAPOTwoNestedLocks}{80.1}
\newcommand{\htwoRECAPOThreeNestedLocks}{0.17}
\newcommand{\htwoRECAPOFourNestedLocks}{<0.1}
\newcommand{\htwoRECAPOFiveNestedLocks}{\cna}
\newcommand{\htwoRECAPOSixNestedLocks}{\cna}
\newcommand{\htwoRECAPOSevenNestedLocks}{\cna}
\newcommand{\htwoRECAPOEightNestedLocks}{\cna}
\newcommand{\htwoRECAPONineNestedLocks}{\cna}
\newcommand{\htwoRECAPOTenNestedLocks}{\cna}
\newcommand{\htwoRECAPOHundredNestedLocks}{\cna}
\newcommand{\htwoRECAPOExWrSet}{303342}
\newcommand{\htwoRECAPOExWrCheck}{537103}
\newcommand{\htwoRECAPOExWrUpdate}{25}
\newcommand{\htwoRECAPOExRdCheck}{844158}
\newcommand{\htwoRECAPOExRdUpdate}{24}
\newcommand{\htwoRECAPOExTotalCheck}{0.46}
\newcommand{\htwoRECAPOExTotalUpdate}{<0.001}
\newcommand{\FASThtwoMaxLiveThreads}{15}
\newcommand{\FASThtwoTotalThreads}{16}
\newcommand{\FASThtwoBaseTime}{4.7}
\newcommand{\FASThtwoBaseTimeCI}{87}
\newcommand{\FASThtwoEmptyTime}{\rna}
\newcommand{\FASThtwoEmptyTimeCI}{\rna}
\newcommand{\FASThtwoEmptyTimeCIMIN}{\rna}
\newcommand{\FASThtwoEmptyTimeCIMAX}{\rna}
\newcommand{\FASThtwoFTTime}{9.4}
\newcommand{\FASThtwoFTTimeCI}{0.48}
\newcommand{\FASThtwoHBTime}{9.5}
\newcommand{\FASThtwoHBTimeCI}{0.12}
\newcommand{\FASThtwoFTOHBTime}{9.3}
\newcommand{\FASThtwoFTOHBTimeCI}{0.13}
\newcommand{\FASThtwoWCPTime}{\rna}
\newcommand{\FASThtwoWCPTimeCI}{\rna}
\newcommand{\FASThtwoWCPTimeCIMIN}{\rna}
\newcommand{\FASThtwoWCPTimeCIMAX}{\rna}
\newcommand{\FASThtwoFTOWCPTime}{57}
\newcommand{\FASThtwoFTOWCPTimeCI}{3.8}
\newcommand{\FASThtwoREWCPTime}{12}
\newcommand{\FASThtwoREWCPTimeCI}{0.44}
\newcommand{\FASThtwoDCTime}{\rna}
\newcommand{\FASThtwoDCTimeCI}{\rna}
\newcommand{\FASThtwoDCTimeCIMIN}{\rna}
\newcommand{\FASThtwoDCTimeCIMAX}{\rna}
\newcommand{\FASThtwoFTODCTime}{61}
\newcommand{\FASThtwoFTODCTimeCI}{6.8}
\newcommand{\FASThtwoREDCTime}{12}
\newcommand{\FASThtwoREDCTimeCI}{0.17}
\newcommand{\FASThtwoCAPOTime}{\rna}
\newcommand{\FASThtwoCAPOTimeCI}{\rna}
\newcommand{\FASThtwoCAPOTimeCIMIN}{\rna}
\newcommand{\FASThtwoCAPOTimeCIMAX}{\rna}
\newcommand{\FASThtwoFTOCAPOTime}{60}
\newcommand{\FASThtwoFTOCAPOTimeCI}{6.5}
\newcommand{\FASThtwoRECAPOTime}{11}
\newcommand{\FASThtwoRECAPOTimeCI}{0.19}
\newcommand{\FASThtwoAGGCAPOTime}{\rna}
\newcommand{\FASThtwoAGGCAPOTimeCI}{\rna}
\newcommand{\FASThtwoAGGCAPOTimeCIMIN}{\rna}
\newcommand{\FASThtwoAGGCAPOTimeCIMAX}{\rna}
\newcommand{\FASThtwoStaticTime}{\rzero}
\newcommand{\FASThtwoDynamicTime}{\rzero}
\newcommand{\FASThtwoBaseMem}{1,800}
\newcommand{\FASThtwoBaseMemCI}{42.0}
\newcommand{\FASThtwoFTMem}{3.0}
\newcommand{\FASThtwoFTMemCI}{0.059}
\newcommand{\FASThtwoHBMem}{3.0}
\newcommand{\FASThtwoHBMemCI}{0.11}
\newcommand{\FASThtwoFTOHBMem}{3.0}
\newcommand{\FASThtwoFTOHBMemCI}{0.078}
\newcommand{\FASThtwoWCPMem}{\memna}
\newcommand{\FASThtwoWCPMemCI}{\memna}
\newcommand{\FASThtwoWCPMemCIMIN}{\memna}
\newcommand{\FASThtwoWCPMemCIMAX}{\memna}
\newcommand{\FASThtwoFTOWCPMem}{37}
\newcommand{\FASThtwoFTOWCPMemCI}{1.7}
\newcommand{\FASThtwoREWCPMem}{4.9}
\newcommand{\FASThtwoREWCPMemCI}{0.19}
\newcommand{\FASThtwoDCMem}{\memna}
\newcommand{\FASThtwoDCMemCI}{\memna}
\newcommand{\FASThtwoDCMemCIMIN}{\memna}
\newcommand{\FASThtwoDCMemCIMAX}{\memna}
\newcommand{\FASThtwoFTODCMem}{41}
\newcommand{\FASThtwoFTODCMemCI}{2.0}
\newcommand{\FASThtwoREDCMem}{4.8}
\newcommand{\FASThtwoREDCMemCI}{0.23}
\newcommand{\FASThtwoCAPOMem}{\memna}
\newcommand{\FASThtwoCAPOMemCI}{\memna}
\newcommand{\FASThtwoCAPOMemCIMIN}{\memna}
\newcommand{\FASThtwoCAPOMemCIMAX}{\memna}
\newcommand{\FASThtwoFTOCAPOMem}{38}
\newcommand{\FASThtwoFTOCAPOMemCI}{2.5}
\newcommand{\FASThtwoRECAPOMem}{4.3}
\newcommand{\FASThtwoRECAPOMemCI}{0.17}
\newcommand{\FASThtwoAGGCAPOMem}{\memna}
\newcommand{\FASThtwoAGGCAPOMemCI}{\memna}
\newcommand{\FASThtwoAGGCAPOMemCIMIN}{\memna}
\newcommand{\FASThtwoAGGCAPOMemCIMAX}{\memna}
\newcommand{\FASThtwoEventsCI}{1,223,292}
\newcommand{\FASThtwoEventsCIMIN}{3,810,420,016}
\newcommand{\FASThtwoEventsCIMAX}{3,812,866,600}
\newcommand{\FASThtwoNoFPEventsCI}{3,615}
\newcommand{\FASThtwoNoFPEventsCIMIN}{118,683,440}
\newcommand{\FASThtwoNoFPEventsCIMAX}{118,690,670}
\newcommand{\FASThtwoFT}{11}
\newcommand{\FASThtwoFTCI}{0.0}
\newcommand{\FASThtwoFTCIMIN}{11}
\newcommand{\FASThtwoFTCIMAX}{11}
\newcommand{\FASThtwoFTDynamic}{106,816}
\newcommand{\FASThtwoFTDynamicCI}{1,318}
\newcommand{\FASThtwoFTDynamicCIMIN}{105,498}
\newcommand{\FASThtwoFTDynamicCIMAX}{108,134}
\newcommand{\FASThtwoHB}{13}
\newcommand{\FASThtwoHBCI}{0.0}
\newcommand{\FASThtwoHBCIMIN}{13}
\newcommand{\FASThtwoHBCIMAX}{13}
\newcommand{\FASThtwoHBDynamic}{87,248}
\newcommand{\FASThtwoHBDynamicCI}{262}
\newcommand{\FASThtwoHBDynamicCIMIN}{86,986}
\newcommand{\FASThtwoHBDynamicCIMAX}{87,510}
\newcommand{\FASThtwoFTOHB}{13}
\newcommand{\FASThtwoFTOHBCI}{0.0}
\newcommand{\FASThtwoFTOHBCIMIN}{13}
\newcommand{\FASThtwoFTOHBCIMAX}{13}
\newcommand{\FASThtwoFTOHBDynamic}{76,525}
\newcommand{\FASThtwoFTOHBDynamicCI}{1,728}
\newcommand{\FASThtwoFTOHBDynamicCIMIN}{74,797}
\newcommand{\FASThtwoFTOHBDynamicCIMAX}{78,253}
\newcommand{\FASThtwoWCP}{\rna}
\newcommand{\FASThtwoWCPCI}{\rna}
\newcommand{\FASThtwoWCPCIMIN}{\rna}
\newcommand{\FASThtwoWCPCIMAX}{\rna}
\newcommand{\FASThtwoWCPDynamic}{\rna}
\newcommand{\FASThtwoWCPDynamicCI}{\rna}
\newcommand{\FASThtwoWCPDynamicCIMIN}{\rna}
\newcommand{\FASThtwoWCPDynamicCIMAX}{\rna}
\newcommand{\FASThtwoFTOWCP}{13}
\newcommand{\FASThtwoFTOWCPCI}{0.2}
\newcommand{\FASThtwoFTOWCPCIMIN}{13}
\newcommand{\FASThtwoFTOWCPCIMAX}{13}
\newcommand{\FASThtwoFTOWCPDynamic}{86,043}
\newcommand{\FASThtwoFTOWCPDynamicCI}{307}
\newcommand{\FASThtwoFTOWCPDynamicCIMIN}{85,736}
\newcommand{\FASThtwoFTOWCPDynamicCIMAX}{86,350}
\newcommand{\FASThtwoREWCP}{13}
\newcommand{\FASThtwoREWCPCI}{0.0}
\newcommand{\FASThtwoREWCPCIMIN}{13}
\newcommand{\FASThtwoREWCPCIMAX}{13}
\newcommand{\FASThtwoREWCPDynamic}{82,217}
\newcommand{\FASThtwoREWCPDynamicCI}{562}
\newcommand{\FASThtwoREWCPDynamicCIMIN}{81,655}
\newcommand{\FASThtwoREWCPDynamicCIMAX}{82,779}
\newcommand{\FASThtwoDC}{\rna}
\newcommand{\FASThtwoDCCI}{\rna}
\newcommand{\FASThtwoDCCIMIN}{\rna}
\newcommand{\FASThtwoDCCIMAX}{\rna}
\newcommand{\FASThtwoDCDynamic}{\rna}
\newcommand{\FASThtwoDCDynamicCI}{\rna}
\newcommand{\FASThtwoDCDynamicCIMIN}{\rna}
\newcommand{\FASThtwoDCDynamicCIMAX}{\rna}
\newcommand{\FASThtwoFTODC}{13}
\newcommand{\FASThtwoFTODCCI}{0.0}
\newcommand{\FASThtwoFTODCCIMIN}{13}
\newcommand{\FASThtwoFTODCCIMAX}{13}
\newcommand{\FASThtwoFTODCDynamic}{86,742}
\newcommand{\FASThtwoFTODCDynamicCI}{173}
\newcommand{\FASThtwoFTODCDynamicCIMIN}{86,569}
\newcommand{\FASThtwoFTODCDynamicCIMAX}{86,915}
\newcommand{\FASThtwoREDC}{13}
\newcommand{\FASThtwoREDCCI}{0.0}
\newcommand{\FASThtwoREDCCIMIN}{13}
\newcommand{\FASThtwoREDCCIMAX}{13}
\newcommand{\FASThtwoREDCDynamic}{83,243}
\newcommand{\FASThtwoREDCDynamicCI}{464}
\newcommand{\FASThtwoREDCDynamicCIMIN}{82,779}
\newcommand{\FASThtwoREDCDynamicCIMAX}{83,707}
\newcommand{\FASThtwoCAPO}{\rna}
\newcommand{\FASThtwoCAPOCI}{\rna}
\newcommand{\FASThtwoCAPOCIMIN}{\rna}
\newcommand{\FASThtwoCAPOCIMAX}{\rna}
\newcommand{\FASThtwoCAPODynamic}{\rna}
\newcommand{\FASThtwoCAPODynamicCI}{\rna}
\newcommand{\FASThtwoCAPODynamicCIMIN}{\rna}
\newcommand{\FASThtwoCAPODynamicCIMAX}{\rna}
\newcommand{\FASThtwoFTOCAPO}{13}
\newcommand{\FASThtwoFTOCAPOCI}{0.0}
\newcommand{\FASThtwoFTOCAPOCIMIN}{13}
\newcommand{\FASThtwoFTOCAPOCIMAX}{13}
\newcommand{\FASThtwoFTOCAPODynamic}{86,839}
\newcommand{\FASThtwoFTOCAPODynamicCI}{207}
\newcommand{\FASThtwoFTOCAPODynamicCIMIN}{86,632}
\newcommand{\FASThtwoFTOCAPODynamicCIMAX}{87,046}
\newcommand{\FASThtwoRECAPO}{13}
\newcommand{\FASThtwoRECAPOCI}{0.0}
\newcommand{\FASThtwoRECAPOCIMIN}{13}
\newcommand{\FASThtwoRECAPOCIMAX}{13}
\newcommand{\FASThtwoRECAPODynamic}{81,677}
\newcommand{\FASThtwoRECAPODynamicCI}{859}
\newcommand{\FASThtwoRECAPODynamicCIMIN}{80,818}
\newcommand{\FASThtwoRECAPODynamicCIMAX}{82,536}
\newcommand{\FASThtwoAGGCAPO}{\rna}
\newcommand{\FASThtwoAGGCAPOCI}{\rna}
\newcommand{\FASThtwoAGGCAPOCIMIN}{\rna}
\newcommand{\FASThtwoAGGCAPOCIMAX}{\rna}
\newcommand{\FASThtwoAGGCAPODynamic}{\rna}
\newcommand{\FASThtwoAGGCAPODynamicCI}{\rna}
\newcommand{\FASThtwoAGGCAPODynamicCIMIN}{\rna}
\newcommand{\FASThtwoAGGCAPODynamicCIMAX}{\rna}
\newcommand{\FASTjythonEvents}{730}
\newcommand{\FASTjythonNoFPEvents}{130}
\newcommand{\jythonHBEventTotal}{730}
\newcommand{\jythonHBNoFPEventTotal}{190}
\newcommand{\jythonHBNoFPAccessTotal}{170}
\newcommand{\jythonHBNoFPOtherTotal}{27}
\newcommand{\jythonHBReadTotal}{71.5}
\newcommand{\jythonHBWriteTotal}{14.7}
\newcommand{\jythonHBNoFPAccessInCS}{5.42}
\newcommand{\jythonHBNoFPAccessOutCS}{94.6}
\newcommand{\jythonHBAcqRelTotal}{17.2}
\newcommand{\jythonHBOtherTotal}{3.6}
\newcommand{\jythonHBNoFPReadTotal}{140}
\newcommand{\jythonHBReadInCS}{4.18}
\newcommand{\jythonHBReadOutCS}{106}
\newcommand{\jythonHBReadSameEp}{10.3}
\newcommand{\jythonHBReadSharedSameEp}{\cna}
\newcommand{\jythonHBReadExclusive}{97.7}
\newcommand{\jythonHBReadOwned}{\cna}
\newcommand{\jythonHBReadShare}{<0.001}
\newcommand{\jythonHBReadShared}{2.25}
\newcommand{\jythonHBReadSharedOwned}{\cna}
\newcommand{\jythonHBNoFPHonestWriteTotal}{28}
\newcommand{\jythonHBWriteInCS}{19.1}
\newcommand{\jythonHBWriteOutCS}{388}
\newcommand{\jythonHBNoFPWriteTotal}{28}
\newcommand{\jythonHBWriteSameEp}{307}
\newcommand{\jythonHBWriteExclusive}{100}
\newcommand{\jythonHBWriteOwned}{\cna}
\newcommand{\jythonHBWriteShared}{<0.001}
\newcommand{\jythonHBNoFPOtherEventTotal}{26523557}
\newcommand{\jythonHBAcqRelOtherTotal}{82.7}
\newcommand{\jythonHBNoAcqRelOtherTotal}{4594891}
\newcommand{\jythonHBFork}{0.0}
\newcommand{\jythonHBJoin}{0.0}
\newcommand{\jythonHBPreWait}{0.0}
\newcommand{\jythonHBPostWait}{0.0}
\newcommand{\jythonHBVolatileTotal}{100.0}
\newcommand{\jythonHBClassInit}{0.00372}
\newcommand{\jythonHBClassAccess}{0.00303}
\newcommand{\jythonHBRaceTotal}{30}
\newcommand{\jythonHBWrRdRace}{80.0}
\newcommand{\jythonHBWrWrRace}{10.0}
\newcommand{\jythonHBRdWrRace}{6.67}
\newcommand{\jythonHBRdShWrRace}{3.33}
\newcommand{\jythonHBHoldLocksTotal}{11}
\newcommand{\jythonHBOneLockHeld}{6.73}
\newcommand{\jythonHBTwoNestedLocks}{0.26}
\newcommand{\jythonHBThreeNestedLocks}{<0.1}
\newcommand{\jythonHBFourNestedLocks}{\cna}
\newcommand{\jythonHBFiveNestedLocks}{\cna}
\newcommand{\jythonHBSixNestedLocks}{\cna}
\newcommand{\jythonHBSevenNestedLocks}{\cna}
\newcommand{\jythonHBEightNestedLocks}{\cna}
\newcommand{\jythonHBNineNestedLocks}{\cna}
\newcommand{\jythonHBTenNestedLocks}{\cna}
\newcommand{\jythonHBHundredNestedLocks}{\cna}
\newcommand{\jythonHBExWrSet}{\ena}
\newcommand{\jythonHBExWrCheck}{\ena}
\newcommand{\jythonHBExWrUpdate}{\ena}
\newcommand{\jythonHBExRdCheck}{\ena}
\newcommand{\jythonHBExRdUpdate}{\ena}
\newcommand{\jythonHBExTotalCheck}{\ena}
\newcommand{\jythonHBExTotalUpdate}{\ena}
\newcommand{\jythonFTOHBEventTotal}{730}
\newcommand{\jythonFTOHBNoFPEventTotal}{170}
\newcommand{\jythonFTOHBNoFPAccessTotal}{140}
\newcommand{\jythonFTOHBNoFPOtherTotal}{27}
\newcommand{\jythonFTOHBReadTotal}{67.3}
\newcommand{\jythonFTOHBWriteTotal}{16.9}
\newcommand{\jythonFTOHBNoFPAccessInCS}{5.42}
\newcommand{\jythonFTOHBNoFPAccessOutCS}{94.6}
\newcommand{\jythonFTOHBAcqRelTotal}{17.2}
\newcommand{\jythonFTOHBOtherTotal}{3.6}
\newcommand{\jythonFTOHBNoFPReadTotal}{110}
\newcommand{\jythonFTOHBReadInCS}{4.84}
\newcommand{\jythonFTOHBReadOutCS}{108}
\newcommand{\jythonFTOHBReadSameEp}{12.5}
\newcommand{\jythonFTOHBReadSharedSameEp}{\cna}
\newcommand{\jythonFTOHBReadExclusive}{<0.001}
\newcommand{\jythonFTOHBReadOwned}{97.3}
\newcommand{\jythonFTOHBReadShare}{<0.001}
\newcommand{\jythonFTOHBReadShared}{\cna}
\newcommand{\jythonFTOHBReadSharedOwned}{2.74}
\newcommand{\jythonFTOHBNoFPHonestWriteTotal}{28}
\newcommand{\jythonFTOHBWriteInCS}{19.1}
\newcommand{\jythonFTOHBWriteOutCS}{388}
\newcommand{\jythonFTOHBNoFPWriteTotal}{28}
\newcommand{\jythonFTOHBWriteSameEp}{307}
\newcommand{\jythonFTOHBWriteExclusive}{<0.001}
\newcommand{\jythonFTOHBWriteOwned}{100}
\newcommand{\jythonFTOHBWriteShared}{<0.001}
\newcommand{\jythonFTOHBNoFPOtherEventTotal}{26523556}
\newcommand{\jythonFTOHBAcqRelOtherTotal}{82.7}
\newcommand{\jythonFTOHBNoAcqRelOtherTotal}{4594890}
\newcommand{\jythonFTOHBFork}{0.0}
\newcommand{\jythonFTOHBJoin}{0.0}
\newcommand{\jythonFTOHBPreWait}{0.0}
\newcommand{\jythonFTOHBPostWait}{0.0}
\newcommand{\jythonFTOHBVolatileTotal}{100.0}
\newcommand{\jythonFTOHBClassInit}{0.00372}
\newcommand{\jythonFTOHBClassAccess}{0.00303}
\newcommand{\jythonFTOHBRaceTotal}{27}
\newcommand{\jythonFTOHBWrRdRace}{88.9}
\newcommand{\jythonFTOHBWrWrRace}{0.0}
\newcommand{\jythonFTOHBRdWrRace}{7.41}
\newcommand{\jythonFTOHBRdShWrRace}{3.7}
\newcommand{\jythonFTOHBHoldLocksTotal}{5.4}
\newcommand{\jythonFTOHBOneLockHeld}{3.81}
\newcommand{\jythonFTOHBTwoNestedLocks}{0.23}
\newcommand{\jythonFTOHBThreeNestedLocks}{<0.1}
\newcommand{\jythonFTOHBFourNestedLocks}{\cna}
\newcommand{\jythonFTOHBFiveNestedLocks}{\cna}
\newcommand{\jythonFTOHBSixNestedLocks}{\cna}
\newcommand{\jythonFTOHBSevenNestedLocks}{\cna}
\newcommand{\jythonFTOHBEightNestedLocks}{\cna}
\newcommand{\jythonFTOHBNineNestedLocks}{\cna}
\newcommand{\jythonFTOHBTenNestedLocks}{\cna}
\newcommand{\jythonFTOHBHundredNestedLocks}{\cna}
\newcommand{\jythonFTOHBExWrSet}{\ena}
\newcommand{\jythonFTOHBExWrCheck}{\ena}
\newcommand{\jythonFTOHBExWrUpdate}{\ena}
\newcommand{\jythonFTOHBExRdCheck}{\ena}
\newcommand{\jythonFTOHBExRdUpdate}{\ena}
\newcommand{\jythonFTOHBExTotalCheck}{\ena}
\newcommand{\jythonFTOHBExTotalUpdate}{\ena}
\newcommand{\jythonFTOWCPEventTotal}{730}
\newcommand{\jythonFTOWCPNoFPEventTotal}{170}
\newcommand{\jythonFTOWCPNoFPAccessTotal}{140}
\newcommand{\jythonFTOWCPNoFPOtherTotal}{27}
\newcommand{\jythonFTOWCPReadTotal}{67.3}
\newcommand{\jythonFTOWCPWriteTotal}{16.9}
\newcommand{\jythonFTOWCPNoFPAccessInCS}{5.42}
\newcommand{\jythonFTOWCPNoFPAccessOutCS}{94.6}
\newcommand{\jythonFTOWCPAcqRelTotal}{17.2}
\newcommand{\jythonFTOWCPOtherTotal}{3.6}
\newcommand{\jythonFTOWCPNoFPReadTotal}{110}
\newcommand{\jythonFTOWCPReadInCS}{4.84}
\newcommand{\jythonFTOWCPReadOutCS}{108}
\newcommand{\jythonFTOWCPReadSameEp}{12.5}
\newcommand{\jythonFTOWCPReadSharedSameEp}{\cna}
\newcommand{\jythonFTOWCPReadExclusive}{<0.001}
\newcommand{\jythonFTOWCPReadOwned}{95.4}
\newcommand{\jythonFTOWCPReadShare}{<0.001}
\newcommand{\jythonFTOWCPReadShared}{\cna}
\newcommand{\jythonFTOWCPReadSharedOwned}{4.61}
\newcommand{\jythonFTOWCPNoFPHonestWriteTotal}{28}
\newcommand{\jythonFTOWCPWriteInCS}{19.1}
\newcommand{\jythonFTOWCPWriteOutCS}{388}
\newcommand{\jythonFTOWCPNoFPWriteTotal}{28}
\newcommand{\jythonFTOWCPWriteSameEp}{307}
\newcommand{\jythonFTOWCPWriteExclusive}{<0.001}
\newcommand{\jythonFTOWCPWriteOwned}{100}
\newcommand{\jythonFTOWCPWriteShared}{<0.001}
\newcommand{\jythonFTOWCPNoFPOtherEventTotal}{26523557}
\newcommand{\jythonFTOWCPAcqRelOtherTotal}{82.7}
\newcommand{\jythonFTOWCPNoAcqRelOtherTotal}{4594891}
\newcommand{\jythonFTOWCPFork}{0.0}
\newcommand{\jythonFTOWCPJoin}{0.0}
\newcommand{\jythonFTOWCPPreWait}{0.0}
\newcommand{\jythonFTOWCPPostWait}{0.0}
\newcommand{\jythonFTOWCPVolatileTotal}{100.0}
\newcommand{\jythonFTOWCPClassInit}{0.00372}
\newcommand{\jythonFTOWCPClassAccess}{0.00303}
\newcommand{\jythonFTOWCPRaceTotal}{20}
\newcommand{\jythonFTOWCPWrRdRace}{85.0}
\newcommand{\jythonFTOWCPWrWrRace}{0.0}
\newcommand{\jythonFTOWCPRdWrRace}{10.0}
\newcommand{\jythonFTOWCPRdShWrRace}{5.0}
\newcommand{\jythonFTOWCPHoldLocksTotal}{5.4}
\newcommand{\jythonFTOWCPOneLockHeld}{3.8}
\newcommand{\jythonFTOWCPTwoNestedLocks}{0.23}
\newcommand{\jythonFTOWCPThreeNestedLocks}{<0.1}
\newcommand{\jythonFTOWCPFourNestedLocks}{\cna}
\newcommand{\jythonFTOWCPFiveNestedLocks}{\cna}
\newcommand{\jythonFTOWCPSixNestedLocks}{\cna}
\newcommand{\jythonFTOWCPSevenNestedLocks}{\cna}
\newcommand{\jythonFTOWCPEightNestedLocks}{\cna}
\newcommand{\jythonFTOWCPNineNestedLocks}{\cna}
\newcommand{\jythonFTOWCPTenNestedLocks}{\cna}
\newcommand{\jythonFTOWCPHundredNestedLocks}{\cna}
\newcommand{\jythonFTOWCPExWrSet}{\ena}
\newcommand{\jythonFTOWCPExWrCheck}{\ena}
\newcommand{\jythonFTOWCPExWrUpdate}{\ena}
\newcommand{\jythonFTOWCPExRdCheck}{\ena}
\newcommand{\jythonFTOWCPExRdUpdate}{\ena}
\newcommand{\jythonFTOWCPExTotalCheck}{\ena}
\newcommand{\jythonFTOWCPExTotalUpdate}{\ena}
\newcommand{\jythonREWCPEventTotal}{730}
\newcommand{\jythonREWCPNoFPEventTotal}{170}
\newcommand{\jythonREWCPNoFPAccessTotal}{140}
\newcommand{\jythonREWCPNoFPOtherTotal}{27}
\newcommand{\jythonREWCPReadTotal}{67.3}
\newcommand{\jythonREWCPWriteTotal}{16.9}
\newcommand{\jythonREWCPNoFPAccessInCS}{5.42}
\newcommand{\jythonREWCPNoFPAccessOutCS}{94.6}
\newcommand{\jythonREWCPAcqRelTotal}{17.2}
\newcommand{\jythonREWCPOtherTotal}{3.6}
\newcommand{\jythonREWCPNoFPReadTotal}{110}
\newcommand{\jythonREWCPReadInCS}{4.84}
\newcommand{\jythonREWCPReadOutCS}{108}
\newcommand{\jythonREWCPReadSameEp}{12.5}
\newcommand{\jythonREWCPReadSharedSameEp}{\cna}
\newcommand{\jythonREWCPReadExclusive}{<0.001}
\newcommand{\jythonREWCPReadOwned}{95.4}
\newcommand{\jythonREWCPReadShare}{<0.001}
\newcommand{\jythonREWCPReadShared}{\cna}
\newcommand{\jythonREWCPReadSharedOwned}{4.6}
\newcommand{\jythonREWCPNoFPHonestWriteTotal}{28}
\newcommand{\jythonREWCPWriteInCS}{19.1}
\newcommand{\jythonREWCPWriteOutCS}{388}
\newcommand{\jythonREWCPNoFPWriteTotal}{28}
\newcommand{\jythonREWCPWriteSameEp}{307}
\newcommand{\jythonREWCPWriteExclusive}{<0.001}
\newcommand{\jythonREWCPWriteOwned}{100}
\newcommand{\jythonREWCPWriteShared}{<0.001}
\newcommand{\jythonREWCPNoFPOtherEventTotal}{26523556}
\newcommand{\jythonREWCPAcqRelOtherTotal}{82.7}
\newcommand{\jythonREWCPNoAcqRelOtherTotal}{4594890}
\newcommand{\jythonREWCPFork}{0.0}
\newcommand{\jythonREWCPJoin}{0.0}
\newcommand{\jythonREWCPPreWait}{0.0}
\newcommand{\jythonREWCPPostWait}{0.0}
\newcommand{\jythonREWCPVolatileTotal}{100.0}
\newcommand{\jythonREWCPClassInit}{0.00372}
\newcommand{\jythonREWCPClassAccess}{0.00303}
\newcommand{\jythonREWCPRaceTotal}{26}
\newcommand{\jythonREWCPWrRdRace}{88.5}
\newcommand{\jythonREWCPWrWrRace}{0.0}
\newcommand{\jythonREWCPRdWrRace}{7.69}
\newcommand{\jythonREWCPRdShWrRace}{3.85}
\newcommand{\jythonREWCPHoldLocksTotal}{5.4}
\newcommand{\jythonREWCPOneLockHeld}{3.82}
\newcommand{\jythonREWCPTwoNestedLocks}{0.23}
\newcommand{\jythonREWCPThreeNestedLocks}{<0.1}
\newcommand{\jythonREWCPFourNestedLocks}{\cna}
\newcommand{\jythonREWCPFiveNestedLocks}{\cna}
\newcommand{\jythonREWCPSixNestedLocks}{\cna}
\newcommand{\jythonREWCPSevenNestedLocks}{\cna}
\newcommand{\jythonREWCPEightNestedLocks}{\cna}
\newcommand{\jythonREWCPNineNestedLocks}{\cna}
\newcommand{\jythonREWCPTenNestedLocks}{\cna}
\newcommand{\jythonREWCPHundredNestedLocks}{\cna}
\newcommand{\jythonREWCPExWrSet}{\ena}
\newcommand{\jythonREWCPExWrCheck}{\ena}
\newcommand{\jythonREWCPExWrUpdate}{\ena}
\newcommand{\jythonREWCPExRdCheck}{\ena}
\newcommand{\jythonREWCPExRdUpdate}{\ena}
\newcommand{\jythonREWCPExTotalCheck}{\ena}
\newcommand{\jythonREWCPExTotalUpdate}{\ena}
\newcommand{\jythonFTODCEventTotal}{730}
\newcommand{\jythonFTODCNoFPEventTotal}{170}
\newcommand{\jythonFTODCNoFPAccessTotal}{140}
\newcommand{\jythonFTODCNoFPOtherTotal}{27}
\newcommand{\jythonFTODCReadTotal}{67.3}
\newcommand{\jythonFTODCWriteTotal}{16.9}
\newcommand{\jythonFTODCNoFPAccessInCS}{5.42}
\newcommand{\jythonFTODCNoFPAccessOutCS}{94.6}
\newcommand{\jythonFTODCAcqRelTotal}{17.2}
\newcommand{\jythonFTODCOtherTotal}{3.6}
\newcommand{\jythonFTODCNoFPReadTotal}{110}
\newcommand{\jythonFTODCReadInCS}{4.84}
\newcommand{\jythonFTODCReadOutCS}{108}
\newcommand{\jythonFTODCReadSameEp}{12.5}
\newcommand{\jythonFTODCReadSharedSameEp}{\cna}
\newcommand{\jythonFTODCReadExclusive}{\cna}
\newcommand{\jythonFTODCReadOwned}{95.4}
\newcommand{\jythonFTODCReadShare}{<0.001}
\newcommand{\jythonFTODCReadShared}{\cna}
\newcommand{\jythonFTODCReadSharedOwned}{4.63}
\newcommand{\jythonFTODCNoFPHonestWriteTotal}{28}
\newcommand{\jythonFTODCWriteInCS}{19.1}
\newcommand{\jythonFTODCWriteOutCS}{388}
\newcommand{\jythonFTODCNoFPWriteTotal}{28}
\newcommand{\jythonFTODCWriteSameEp}{307}
\newcommand{\jythonFTODCWriteExclusive}{<0.001}
\newcommand{\jythonFTODCWriteOwned}{100}
\newcommand{\jythonFTODCWriteShared}{<0.001}
\newcommand{\jythonFTODCNoFPOtherEventTotal}{26523555}
\newcommand{\jythonFTODCAcqRelOtherTotal}{82.7}
\newcommand{\jythonFTODCNoAcqRelOtherTotal}{4594889}
\newcommand{\jythonFTODCFork}{0.0}
\newcommand{\jythonFTODCJoin}{0.0}
\newcommand{\jythonFTODCPreWait}{0.0}
\newcommand{\jythonFTODCPostWait}{0.0}
\newcommand{\jythonFTODCVolatileTotal}{100.0}
\newcommand{\jythonFTODCClassInit}{0.00372}
\newcommand{\jythonFTODCClassAccess}{0.00303}
\newcommand{\jythonFTODCRaceTotal}{29}
\newcommand{\jythonFTODCWrRdRace}{79.3}
\newcommand{\jythonFTODCWrWrRace}{0.0}
\newcommand{\jythonFTODCRdWrRace}{6.9}
\newcommand{\jythonFTODCRdShWrRace}{13.8}
\newcommand{\jythonFTODCHoldLocksTotal}{5.4}
\newcommand{\jythonFTODCOneLockHeld}{3.82}
\newcommand{\jythonFTODCTwoNestedLocks}{0.23}
\newcommand{\jythonFTODCThreeNestedLocks}{<0.1}
\newcommand{\jythonFTODCFourNestedLocks}{\cna}
\newcommand{\jythonFTODCFiveNestedLocks}{\cna}
\newcommand{\jythonFTODCSixNestedLocks}{\cna}
\newcommand{\jythonFTODCSevenNestedLocks}{\cna}
\newcommand{\jythonFTODCEightNestedLocks}{\cna}
\newcommand{\jythonFTODCNineNestedLocks}{\cna}
\newcommand{\jythonFTODCTenNestedLocks}{\cna}
\newcommand{\jythonFTODCHundredNestedLocks}{\cna}
\newcommand{\jythonFTODCExWrSet}{\ena}
\newcommand{\jythonFTODCExWrCheck}{\ena}
\newcommand{\jythonFTODCExWrUpdate}{\ena}
\newcommand{\jythonFTODCExRdCheck}{\ena}
\newcommand{\jythonFTODCExRdUpdate}{\ena}
\newcommand{\jythonFTODCExTotalCheck}{\ena}
\newcommand{\jythonFTODCExTotalUpdate}{\ena}
\newcommand{\jythonREDCEventTotal}{730}
\newcommand{\jythonREDCNoFPEventTotal}{170}
\newcommand{\jythonREDCNoFPAccessTotal}{140}
\newcommand{\jythonREDCNoFPOtherTotal}{27}
\newcommand{\jythonREDCReadTotal}{67.3}
\newcommand{\jythonREDCWriteTotal}{16.9}
\newcommand{\jythonREDCNoFPAccessInCS}{5.42}
\newcommand{\jythonREDCNoFPAccessOutCS}{94.6}
\newcommand{\jythonREDCAcqRelTotal}{17.2}
\newcommand{\jythonREDCOtherTotal}{3.6}
\newcommand{\jythonREDCNoFPReadTotal}{110}
\newcommand{\jythonREDCReadInCS}{4.84}
\newcommand{\jythonREDCReadOutCS}{108}
\newcommand{\jythonREDCReadSameEp}{12.5}
\newcommand{\jythonREDCReadSharedSameEp}{\cna}
\newcommand{\jythonREDCReadExclusive}{\cna}
\newcommand{\jythonREDCReadOwned}{95.4}
\newcommand{\jythonREDCReadShare}{<0.001}
\newcommand{\jythonREDCReadShared}{\cna}
\newcommand{\jythonREDCReadSharedOwned}{4.58}
\newcommand{\jythonREDCNoFPHonestWriteTotal}{28}
\newcommand{\jythonREDCWriteInCS}{19.1}
\newcommand{\jythonREDCWriteOutCS}{388}
\newcommand{\jythonREDCNoFPWriteTotal}{28}
\newcommand{\jythonREDCWriteSameEp}{307}
\newcommand{\jythonREDCWriteExclusive}{<0.001}
\newcommand{\jythonREDCWriteOwned}{100}
\newcommand{\jythonREDCWriteShared}{<0.001}
\newcommand{\jythonREDCNoFPOtherEventTotal}{26523556}
\newcommand{\jythonREDCAcqRelOtherTotal}{82.7}
\newcommand{\jythonREDCNoAcqRelOtherTotal}{4594890}
\newcommand{\jythonREDCFork}{0.0}
\newcommand{\jythonREDCJoin}{0.0}
\newcommand{\jythonREDCPreWait}{0.0}
\newcommand{\jythonREDCPostWait}{0.0}
\newcommand{\jythonREDCVolatileTotal}{100.0}
\newcommand{\jythonREDCClassInit}{0.00372}
\newcommand{\jythonREDCClassAccess}{0.00303}
\newcommand{\jythonREDCRaceTotal}{33}
\newcommand{\jythonREDCWrRdRace}{81.8}
\newcommand{\jythonREDCWrWrRace}{0.0}
\newcommand{\jythonREDCRdWrRace}{6.06}
\newcommand{\jythonREDCRdShWrRace}{12.1}
\newcommand{\jythonREDCHoldLocksTotal}{5.4}
\newcommand{\jythonREDCOneLockHeld}{3.82}
\newcommand{\jythonREDCTwoNestedLocks}{0.23}
\newcommand{\jythonREDCThreeNestedLocks}{<0.1}
\newcommand{\jythonREDCFourNestedLocks}{\cna}
\newcommand{\jythonREDCFiveNestedLocks}{\cna}
\newcommand{\jythonREDCSixNestedLocks}{\cna}
\newcommand{\jythonREDCSevenNestedLocks}{\cna}
\newcommand{\jythonREDCEightNestedLocks}{\cna}
\newcommand{\jythonREDCNineNestedLocks}{\cna}
\newcommand{\jythonREDCTenNestedLocks}{\cna}
\newcommand{\jythonREDCHundredNestedLocks}{\cna}
\newcommand{\jythonREDCExWrSet}{\ena}
\newcommand{\jythonREDCExWrCheck}{\ena}
\newcommand{\jythonREDCExWrUpdate}{\ena}
\newcommand{\jythonREDCExRdCheck}{\ena}
\newcommand{\jythonREDCExRdUpdate}{\ena}
\newcommand{\jythonREDCExTotalCheck}{\ena}
\newcommand{\jythonREDCExTotalUpdate}{\ena}
\newcommand{\jythonFTOCAPOEventTotal}{730}
\newcommand{\jythonFTOCAPONoFPEventTotal}{170}
\newcommand{\jythonFTOCAPONoFPAccessTotal}{140}
\newcommand{\jythonFTOCAPONoFPOtherTotal}{27}
\newcommand{\jythonFTOCAPOReadTotal}{67.3}
\newcommand{\jythonFTOCAPOWriteTotal}{16.9}
\newcommand{\jythonFTOCAPONoFPAccessInCS}{5.42}
\newcommand{\jythonFTOCAPONoFPAccessOutCS}{94.6}
\newcommand{\jythonFTOCAPOAcqRelTotal}{17.2}
\newcommand{\jythonFTOCAPOOtherTotal}{3.6}
\newcommand{\jythonFTOCAPONoFPReadTotal}{110}
\newcommand{\jythonFTOCAPOReadInCS}{4.84}
\newcommand{\jythonFTOCAPOReadOutCS}{108}
\newcommand{\jythonFTOCAPOReadSameEp}{12.5}
\newcommand{\jythonFTOCAPOReadSharedSameEp}{\cna}
\newcommand{\jythonFTOCAPOReadExclusive}{\cna}
\newcommand{\jythonFTOCAPOReadOwned}{95.3}
\newcommand{\jythonFTOCAPOReadShare}{<0.001}
\newcommand{\jythonFTOCAPOReadShared}{\cna}
\newcommand{\jythonFTOCAPOReadSharedOwned}{4.67}
\newcommand{\jythonFTOCAPONoFPHonestWriteTotal}{28}
\newcommand{\jythonFTOCAPOWriteInCS}{19.1}
\newcommand{\jythonFTOCAPOWriteOutCS}{388}
\newcommand{\jythonFTOCAPONoFPWriteTotal}{28}
\newcommand{\jythonFTOCAPOWriteSameEp}{307}
\newcommand{\jythonFTOCAPOWriteExclusive}{<0.001}
\newcommand{\jythonFTOCAPOWriteOwned}{100}
\newcommand{\jythonFTOCAPOWriteShared}{<0.001}
\newcommand{\jythonFTOCAPONoFPOtherEventTotal}{26523557}
\newcommand{\jythonFTOCAPOAcqRelOtherTotal}{82.7}
\newcommand{\jythonFTOCAPONoAcqRelOtherTotal}{4594891}
\newcommand{\jythonFTOCAPOFork}{0.0}
\newcommand{\jythonFTOCAPOJoin}{0.0}
\newcommand{\jythonFTOCAPOPreWait}{0.0}
\newcommand{\jythonFTOCAPOPostWait}{0.0}
\newcommand{\jythonFTOCAPOVolatileTotal}{100.0}
\newcommand{\jythonFTOCAPOClassInit}{0.00372}
\newcommand{\jythonFTOCAPOClassAccess}{0.00303}
\newcommand{\jythonFTOCAPORaceTotal}{31}
\newcommand{\jythonFTOCAPOWrRdRace}{80.6}
\newcommand{\jythonFTOCAPOWrWrRace}{0.0}
\newcommand{\jythonFTOCAPORdWrRace}{6.45}
\newcommand{\jythonFTOCAPORdShWrRace}{12.9}
\newcommand{\jythonFTOCAPOHoldLocksTotal}{5.4}
\newcommand{\jythonFTOCAPOOneLockHeld}{3.82}
\newcommand{\jythonFTOCAPOTwoNestedLocks}{0.23}
\newcommand{\jythonFTOCAPOThreeNestedLocks}{<0.1}
\newcommand{\jythonFTOCAPOFourNestedLocks}{\cna}
\newcommand{\jythonFTOCAPOFiveNestedLocks}{\cna}
\newcommand{\jythonFTOCAPOSixNestedLocks}{\cna}
\newcommand{\jythonFTOCAPOSevenNestedLocks}{\cna}
\newcommand{\jythonFTOCAPOEightNestedLocks}{\cna}
\newcommand{\jythonFTOCAPONineNestedLocks}{\cna}
\newcommand{\jythonFTOCAPOTenNestedLocks}{\cna}
\newcommand{\jythonFTOCAPOHundredNestedLocks}{\cna}
\newcommand{\jythonFTOCAPOExWrSet}{\ena}
\newcommand{\jythonFTOCAPOExWrCheck}{\ena}
\newcommand{\jythonFTOCAPOExWrUpdate}{\ena}
\newcommand{\jythonFTOCAPOExRdCheck}{\ena}
\newcommand{\jythonFTOCAPOExRdUpdate}{\ena}
\newcommand{\jythonFTOCAPOExTotalCheck}{\ena}
\newcommand{\jythonFTOCAPOExTotalUpdate}{\ena}
\newcommand{\jythonRECAPOEventTotal}{730}
\newcommand{\jythonRECAPONoFPEventTotal}{170}
\newcommand{\jythonRECAPONoFPAccessTotal}{140}
\newcommand{\jythonRECAPONoFPOtherTotal}{27}
\newcommand{\jythonRECAPOReadTotal}{67.3}
\newcommand{\jythonRECAPOWriteTotal}{16.9}
\newcommand{\jythonRECAPONoFPAccessInCS}{5.42}
\newcommand{\jythonRECAPONoFPAccessOutCS}{94.6}
\newcommand{\jythonRECAPOAcqRelTotal}{17.2}
\newcommand{\jythonRECAPOOtherTotal}{3.6}
\newcommand{\jythonRECAPONoFPReadTotal}{110}
\newcommand{\jythonRECAPOReadInCS}{4.84}
\newcommand{\jythonRECAPOReadOutCS}{108}
\newcommand{\jythonRECAPOReadSameEp}{12.5}
\newcommand{\jythonRECAPOReadSharedSameEp}{\cna}
\newcommand{\jythonRECAPOReadExclusive}{\cna}
\newcommand{\jythonRECAPOReadOwned}{95.2}
\newcommand{\jythonRECAPOReadShare}{<0.001}
\newcommand{\jythonRECAPOReadShared}{\cna}
\newcommand{\jythonRECAPOReadSharedOwned}{4.8}
\newcommand{\jythonRECAPONoFPHonestWriteTotal}{28}
\newcommand{\jythonRECAPOWriteInCS}{19.1}
\newcommand{\jythonRECAPOWriteOutCS}{388}
\newcommand{\jythonRECAPONoFPWriteTotal}{28}
\newcommand{\jythonRECAPOWriteSameEp}{307}
\newcommand{\jythonRECAPOWriteExclusive}{<0.001}
\newcommand{\jythonRECAPOWriteOwned}{100}
\newcommand{\jythonRECAPOWriteShared}{<0.001}
\newcommand{\jythonRECAPONoFPOtherEventTotal}{26523557}
\newcommand{\jythonRECAPOAcqRelOtherTotal}{82.7}
\newcommand{\jythonRECAPONoAcqRelOtherTotal}{4594891}
\newcommand{\jythonRECAPOFork}{0.0}
\newcommand{\jythonRECAPOJoin}{0.0}
\newcommand{\jythonRECAPOPreWait}{0.0}
\newcommand{\jythonRECAPOPostWait}{0.0}
\newcommand{\jythonRECAPOVolatileTotal}{100.0}
\newcommand{\jythonRECAPOClassInit}{0.00372}
\newcommand{\jythonRECAPOClassAccess}{0.00303}
\newcommand{\jythonRECAPORaceTotal}{30}
\newcommand{\jythonRECAPOWrRdRace}{80.0}
\newcommand{\jythonRECAPOWrWrRace}{0.0}
\newcommand{\jythonRECAPORdWrRace}{6.67}
\newcommand{\jythonRECAPORdShWrRace}{13.3}
\newcommand{\jythonRECAPOHoldLocksTotal}{5.4}
\newcommand{\jythonRECAPOOneLockHeld}{3.8}
\newcommand{\jythonRECAPOTwoNestedLocks}{0.23}
\newcommand{\jythonRECAPOThreeNestedLocks}{<0.1}
\newcommand{\jythonRECAPOFourNestedLocks}{\cna}
\newcommand{\jythonRECAPOFiveNestedLocks}{\cna}
\newcommand{\jythonRECAPOSixNestedLocks}{\cna}
\newcommand{\jythonRECAPOSevenNestedLocks}{\cna}
\newcommand{\jythonRECAPOEightNestedLocks}{\cna}
\newcommand{\jythonRECAPONineNestedLocks}{\cna}
\newcommand{\jythonRECAPOTenNestedLocks}{\cna}
\newcommand{\jythonRECAPOHundredNestedLocks}{\cna}
\newcommand{\jythonRECAPOExWrSet}{\ena}
\newcommand{\jythonRECAPOExWrCheck}{\ena}
\newcommand{\jythonRECAPOExWrUpdate}{\ena}
\newcommand{\jythonRECAPOExRdCheck}{\ena}
\newcommand{\jythonRECAPOExRdUpdate}{\ena}
\newcommand{\jythonRECAPOExTotalCheck}{\ena}
\newcommand{\jythonRECAPOExTotalUpdate}{\ena}
\newcommand{\FASTjythonMaxLiveThreads}{2}
\newcommand{\FASTjythonTotalThreads}{2}
\newcommand{\FASTjythonBaseTime}{3.9}
\newcommand{\FASTjythonBaseTimeCI}{110}
\newcommand{\FASTjythonEmptyTime}{\rna}
\newcommand{\FASTjythonEmptyTimeCI}{\rna}
\newcommand{\FASTjythonEmptyTimeCIMIN}{\rna}
\newcommand{\FASTjythonEmptyTimeCIMAX}{\rna}
\newcommand{\FASTjythonFTTime}{7.8}
\newcommand{\FASTjythonFTTimeCI}{0.2}
\newcommand{\FASTjythonHBTime}{8.3}
\newcommand{\FASTjythonHBTimeCI}{0.28}
\newcommand{\FASTjythonFTOHBTime}{8.3}
\newcommand{\FASTjythonFTOHBTimeCI}{0.32}
\newcommand{\FASTjythonWCPTime}{\rna}
\newcommand{\FASTjythonWCPTimeCI}{\rna}
\newcommand{\FASTjythonWCPTimeCIMIN}{\rna}
\newcommand{\FASTjythonWCPTimeCIMAX}{\rna}
\newcommand{\FASTjythonFTOWCPTime}{11}
\newcommand{\FASTjythonFTOWCPTimeCI}{0.33}
\newcommand{\FASTjythonREWCPTime}{11}
\newcommand{\FASTjythonREWCPTimeCI}{0.38}
\newcommand{\FASTjythonDCTime}{\rna}
\newcommand{\FASTjythonDCTimeCI}{\rna}
\newcommand{\FASTjythonDCTimeCIMIN}{\rna}
\newcommand{\FASTjythonDCTimeCIMAX}{\rna}
\newcommand{\FASTjythonFTODCTime}{11}
\newcommand{\FASTjythonFTODCTimeCI}{0.38}
\newcommand{\FASTjythonREDCTime}{11}
\newcommand{\FASTjythonREDCTimeCI}{0.40}
\newcommand{\FASTjythonCAPOTime}{\rna}
\newcommand{\FASTjythonCAPOTimeCI}{\rna}
\newcommand{\FASTjythonCAPOTimeCIMIN}{\rna}
\newcommand{\FASTjythonCAPOTimeCIMAX}{\rna}
\newcommand{\FASTjythonFTOCAPOTime}{8.7}
\newcommand{\FASTjythonFTOCAPOTimeCI}{0.23}
\newcommand{\FASTjythonRECAPOTime}{8.8}
\newcommand{\FASTjythonRECAPOTimeCI}{0.26}
\newcommand{\FASTjythonAGGCAPOTime}{\rna}
\newcommand{\FASTjythonAGGCAPOTimeCI}{\rna}
\newcommand{\FASTjythonAGGCAPOTimeCIMIN}{\rna}
\newcommand{\FASTjythonAGGCAPOTimeCIMAX}{\rna}
\newcommand{\FASTjythonStaticTime}{\rzero}
\newcommand{\FASTjythonDynamicTime}{\rzero}
\newcommand{\FASTjythonBaseMem}{750}
\newcommand{\FASTjythonBaseMemCI}{2.1}
\newcommand{\FASTjythonFTMem}{6.0}
\newcommand{\FASTjythonFTMemCI}{0.42}
\newcommand{\FASTjythonHBMem}{7.0}
\newcommand{\FASTjythonHBMemCI}{0.12}
\newcommand{\FASTjythonFTOHBMem}{7.0}
\newcommand{\FASTjythonFTOHBMemCI}{0.1}
\newcommand{\FASTjythonWCPMem}{\memna}
\newcommand{\FASTjythonWCPMemCI}{\memna}
\newcommand{\FASTjythonWCPMemCIMIN}{\memna}
\newcommand{\FASTjythonWCPMemCIMAX}{\memna}
\newcommand{\FASTjythonFTOWCPMem}{13}
\newcommand{\FASTjythonFTOWCPMemCI}{0.57}
\newcommand{\FASTjythonREWCPMem}{12}
\newcommand{\FASTjythonREWCPMemCI}{0.75}
\newcommand{\FASTjythonDCMem}{\memna}
\newcommand{\FASTjythonDCMemCI}{\memna}
\newcommand{\FASTjythonDCMemCIMIN}{\memna}
\newcommand{\FASTjythonDCMemCIMAX}{\memna}
\newcommand{\FASTjythonFTODCMem}{13}
\newcommand{\FASTjythonFTODCMemCI}{0.37}
\newcommand{\FASTjythonREDCMem}{12}
\newcommand{\FASTjythonREDCMemCI}{0.72}
\newcommand{\FASTjythonCAPOMem}{\memna}
\newcommand{\FASTjythonCAPOMemCI}{\memna}
\newcommand{\FASTjythonCAPOMemCIMIN}{\memna}
\newcommand{\FASTjythonCAPOMemCIMAX}{\memna}
\newcommand{\FASTjythonFTOCAPOMem}{8.9}
\newcommand{\FASTjythonFTOCAPOMemCI}{0.14}
\newcommand{\FASTjythonRECAPOMem}{8.0}
\newcommand{\FASTjythonRECAPOMemCI}{0.30}
\newcommand{\FASTjythonAGGCAPOMem}{\memna}
\newcommand{\FASTjythonAGGCAPOMemCI}{\memna}
\newcommand{\FASTjythonAGGCAPOMemCIMIN}{\memna}
\newcommand{\FASTjythonAGGCAPOMemCIMAX}{\memna}
\newcommand{\FASTjythonEventsCI}{221}
\newcommand{\FASTjythonEventsCIMIN}{728,376,622}
\newcommand{\FASTjythonEventsCIMAX}{728,377,064}
\newcommand{\FASTjythonNoFPEventsCI}{7}
\newcommand{\FASTjythonNoFPEventsCIMIN}{127,785,453}
\newcommand{\FASTjythonNoFPEventsCIMAX}{127,785,467}
\newcommand{\FASTjythonFT}{22}
\newcommand{\FASTjythonFTCI}{1.2}
\newcommand{\FASTjythonFTCIMIN}{21}
\newcommand{\FASTjythonFTCIMAX}{23}
\newcommand{\FASTjythonFTDynamic}{47}
\newcommand{\FASTjythonFTDynamicCI}{1.1}
\newcommand{\FASTjythonFTDynamicCIMIN}{46}
\newcommand{\FASTjythonFTDynamicCIMAX}{48}
\newcommand{\FASTjythonHB}{24}
\newcommand{\FASTjythonHBCI}{1.3}
\newcommand{\FASTjythonHBCIMIN}{23}
\newcommand{\FASTjythonHBCIMAX}{25}
\newcommand{\FASTjythonHBDynamic}{27}
\newcommand{\FASTjythonHBDynamicCI}{1.4}
\newcommand{\FASTjythonHBDynamicCIMIN}{26}
\newcommand{\FASTjythonHBDynamicCIMAX}{28}
\newcommand{\FASTjythonFTOHB}{24}
\newcommand{\FASTjythonFTOHBCI}{1.3}
\newcommand{\FASTjythonFTOHBCIMIN}{23}
\newcommand{\FASTjythonFTOHBCIMAX}{25}
\newcommand{\FASTjythonFTOHBDynamic}{27}
\newcommand{\FASTjythonFTOHBDynamicCI}{1.4}
\newcommand{\FASTjythonFTOHBDynamicCIMIN}{26}
\newcommand{\FASTjythonFTOHBDynamicCIMAX}{28}
\newcommand{\FASTjythonWCP}{\rna}
\newcommand{\FASTjythonWCPCI}{\rna}
\newcommand{\FASTjythonWCPCIMIN}{\rna}
\newcommand{\FASTjythonWCPCIMAX}{\rna}
\newcommand{\FASTjythonWCPDynamic}{\rna}
\newcommand{\FASTjythonWCPDynamicCI}{\rna}
\newcommand{\FASTjythonWCPDynamicCIMIN}{\rna}
\newcommand{\FASTjythonWCPDynamicCIMAX}{\rna}
\newcommand{\FASTjythonFTOWCP}{19}
\newcommand{\FASTjythonFTOWCPCI}{0.64}
\newcommand{\FASTjythonFTOWCPCIMIN}{18}
\newcommand{\FASTjythonFTOWCPCIMAX}{20}
\newcommand{\FASTjythonFTOWCPDynamic}{20}
\newcommand{\FASTjythonFTOWCPDynamicCI}{1.0}
\newcommand{\FASTjythonFTOWCPDynamicCIMIN}{19}
\newcommand{\FASTjythonFTOWCPDynamicCIMAX}{21}
\newcommand{\FASTjythonREWCP}{24}
\newcommand{\FASTjythonREWCPCI}{1.1}
\newcommand{\FASTjythonREWCPCIMIN}{23}
\newcommand{\FASTjythonREWCPCIMAX}{25}
\newcommand{\FASTjythonREWCPDynamic}{26}
\newcommand{\FASTjythonREWCPDynamicCI}{1.6}
\newcommand{\FASTjythonREWCPDynamicCIMIN}{24}
\newcommand{\FASTjythonREWCPDynamicCIMAX}{28}
\newcommand{\FASTjythonDC}{\rna}
\newcommand{\FASTjythonDCCI}{\rna}
\newcommand{\FASTjythonDCCIMIN}{\rna}
\newcommand{\FASTjythonDCCIMAX}{\rna}
\newcommand{\FASTjythonDCDynamic}{\rna}
\newcommand{\FASTjythonDCDynamicCI}{\rna}
\newcommand{\FASTjythonDCDynamicCIMIN}{\rna}
\newcommand{\FASTjythonDCDynamicCIMAX}{\rna}
\newcommand{\FASTjythonFTODC}{27}
\newcommand{\FASTjythonFTODCCI}{0.0}
\newcommand{\FASTjythonFTODCCIMIN}{27}
\newcommand{\FASTjythonFTODCCIMAX}{27}
\newcommand{\FASTjythonFTODCDynamic}{29}
\newcommand{\FASTjythonFTODCDynamicCI}{0.32}
\newcommand{\FASTjythonFTODCDynamicCIMIN}{29}
\newcommand{\FASTjythonFTODCDynamicCIMAX}{29}
\newcommand{\FASTjythonREDC}{30}
\newcommand{\FASTjythonREDCCI}{1.0}
\newcommand{\FASTjythonREDCCIMIN}{29}
\newcommand{\FASTjythonREDCCIMAX}{31}
\newcommand{\FASTjythonREDCDynamic}{33}
\newcommand{\FASTjythonREDCDynamicCI}{1.0}
\newcommand{\FASTjythonREDCDynamicCIMIN}{32}
\newcommand{\FASTjythonREDCDynamicCIMAX}{34}
\newcommand{\FASTjythonCAPO}{\rna}
\newcommand{\FASTjythonCAPOCI}{\rna}
\newcommand{\FASTjythonCAPOCIMIN}{\rna}
\newcommand{\FASTjythonCAPOCIMAX}{\rna}
\newcommand{\FASTjythonCAPODynamic}{\rna}
\newcommand{\FASTjythonCAPODynamicCI}{\rna}
\newcommand{\FASTjythonCAPODynamicCIMIN}{\rna}
\newcommand{\FASTjythonCAPODynamicCIMAX}{\rna}
\newcommand{\FASTjythonFTOCAPO}{28}
\newcommand{\FASTjythonFTOCAPOCI}{1.2}
\newcommand{\FASTjythonFTOCAPOCIMIN}{27}
\newcommand{\FASTjythonFTOCAPOCIMAX}{29}
\newcommand{\FASTjythonFTOCAPODynamic}{31}
\newcommand{\FASTjythonFTOCAPODynamicCI}{1.2}
\newcommand{\FASTjythonFTOCAPODynamicCIMIN}{30}
\newcommand{\FASTjythonFTOCAPODynamicCIMAX}{32}
\newcommand{\FASTjythonRECAPO}{27}
\newcommand{\FASTjythonRECAPOCI}{0.0}
\newcommand{\FASTjythonRECAPOCIMIN}{27}
\newcommand{\FASTjythonRECAPOCIMAX}{27}
\newcommand{\FASTjythonRECAPODynamic}{30}
\newcommand{\FASTjythonRECAPODynamicCI}{0.0}
\newcommand{\FASTjythonRECAPODynamicCIMIN}{30}
\newcommand{\FASTjythonRECAPODynamicCIMAX}{30}
\newcommand{\FASTjythonAGGCAPO}{\rna}
\newcommand{\FASTjythonAGGCAPOCI}{\rna}
\newcommand{\FASTjythonAGGCAPOCIMIN}{\rna}
\newcommand{\FASTjythonAGGCAPOCIMAX}{\rna}
\newcommand{\FASTjythonAGGCAPODynamic}{\rna}
\newcommand{\FASTjythonAGGCAPODynamicCI}{\rna}
\newcommand{\FASTjythonAGGCAPODynamicCIMIN}{\rna}
\newcommand{\FASTjythonAGGCAPODynamicCIMAX}{\rna}
\newcommand{\FASTluindexEvents}{400}
\newcommand{\FASTluindexNoFPEvents}{2.9}
\newcommand{\luindexHBEventTotal}{400}
\newcommand{\luindexHBNoFPEventTotal}{43}
\newcommand{\luindexHBNoFPAccessTotal}{42}
\newcommand{\luindexHBNoFPOtherTotal}{0.42}
\newcommand{\luindexHBReadTotal}{67.6}
\newcommand{\luindexHBWriteTotal}{31.4}
\newcommand{\luindexHBNoFPAccessInCS}{8.08}
\newcommand{\luindexHBNoFPAccessOutCS}{91.9}
\newcommand{\luindexHBAcqRelTotal}{14.2}
\newcommand{\luindexHBOtherTotal}{0.0197}
\newcommand{\luindexHBNoFPReadTotal}{29}
\newcommand{\luindexHBReadInCS}{23.8}
\newcommand{\luindexHBReadOutCS}{81.3}
\newcommand{\luindexHBReadSameEp}{5.1}
\newcommand{\luindexHBReadSharedSameEp}{\cna}
\newcommand{\luindexHBReadExclusive}{100}
\newcommand{\luindexHBReadOwned}{\cna}
\newcommand{\luindexHBReadShare}{<0.001}
\newcommand{\luindexHBReadShared}{<0.001}
\newcommand{\luindexHBReadSharedOwned}{\cna}
\newcommand{\luindexHBNoFPHonestWriteTotal}{13}
\newcommand{\luindexHBWriteInCS}{32.1}
\newcommand{\luindexHBWriteOutCS}{75.5}
\newcommand{\luindexHBNoFPWriteTotal}{13}
\newcommand{\luindexHBWriteSameEp}{7.57}
\newcommand{\luindexHBWriteExclusive}{100}
\newcommand{\luindexHBWriteOwned}{\cna}
\newcommand{\luindexHBWriteShared}{<0.001}
\newcommand{\luindexHBNoFPOtherEventTotal}{415314}
\newcommand{\luindexHBAcqRelOtherTotal}{99.9}
\newcommand{\luindexHBNoAcqRelOtherTotal}{574}
\newcommand{\luindexHBFork}{0.174}
\newcommand{\luindexHBJoin}{0.0}
\newcommand{\luindexHBPreWait}{0.348}
\newcommand{\luindexHBPostWait}{0.348}
\newcommand{\luindexHBVolatileTotal}{77.2}
\newcommand{\luindexHBClassInit}{11.5}
\newcommand{\luindexHBClassAccess}{10.5}
\newcommand{\luindexHBRaceTotal}{1}
\newcommand{\luindexHBWrRdRace}{0.0}
\newcommand{\luindexHBWrWrRace}{0.0}
\newcommand{\luindexHBRdWrRace}{0.0}
\newcommand{\luindexHBRdShWrRace}{100.0}
\newcommand{\luindexHBHoldLocksTotal}{11}
\newcommand{\luindexHBOneLockHeld}{26.3}
\newcommand{\luindexHBTwoNestedLocks}{26.0}
\newcommand{\luindexHBThreeNestedLocks}{25.9}
\newcommand{\luindexHBFourNestedLocks}{0.69}
\newcommand{\luindexHBFiveNestedLocks}{<0.1}
\newcommand{\luindexHBSixNestedLocks}{<0.1}
\newcommand{\luindexHBSevenNestedLocks}{\cna}
\newcommand{\luindexHBEightNestedLocks}{\cna}
\newcommand{\luindexHBNineNestedLocks}{\cna}
\newcommand{\luindexHBTenNestedLocks}{\cna}
\newcommand{\luindexHBHundredNestedLocks}{\cna}
\newcommand{\luindexHBExWrSet}{\ena}
\newcommand{\luindexHBExWrCheck}{\ena}
\newcommand{\luindexHBExWrUpdate}{\ena}
\newcommand{\luindexHBExRdCheck}{\ena}
\newcommand{\luindexHBExRdUpdate}{\ena}
\newcommand{\luindexHBExTotalCheck}{\ena}
\newcommand{\luindexHBExTotalUpdate}{\ena}
\newcommand{\luindexFTOHBEventTotal}{400}
\newcommand{\luindexFTOHBNoFPEventTotal}{41}
\newcommand{\luindexFTOHBNoFPAccessTotal}{40}
\newcommand{\luindexFTOHBNoFPOtherTotal}{0.42}
\newcommand{\luindexFTOHBReadTotal}{66.0}
\newcommand{\luindexFTOHBWriteTotal}{33.0}
\newcommand{\luindexFTOHBNoFPAccessInCS}{8.08}
\newcommand{\luindexFTOHBNoFPAccessOutCS}{91.9}
\newcommand{\luindexFTOHBAcqRelTotal}{14.2}
\newcommand{\luindexFTOHBOtherTotal}{0.0197}
\newcommand{\luindexFTOHBNoFPReadTotal}{27}
\newcommand{\luindexFTOHBReadInCS}{23.4}
\newcommand{\luindexFTOHBReadOutCS}{82}
\newcommand{\luindexFTOHBReadSameEp}{5.48}
\newcommand{\luindexFTOHBReadSharedSameEp}{\cna}
\newcommand{\luindexFTOHBReadExclusive}{<0.001}
\newcommand{\luindexFTOHBReadOwned}{100}
\newcommand{\luindexFTOHBReadShare}{<0.001}
\newcommand{\luindexFTOHBReadShared}{\cna}
\newcommand{\luindexFTOHBReadSharedOwned}{<0.001}
\newcommand{\luindexFTOHBNoFPHonestWriteTotal}{13}
\newcommand{\luindexFTOHBWriteInCS}{32.1}
\newcommand{\luindexFTOHBWriteOutCS}{75.5}
\newcommand{\luindexFTOHBNoFPWriteTotal}{13}
\newcommand{\luindexFTOHBWriteSameEp}{7.57}
\newcommand{\luindexFTOHBWriteExclusive}{<0.001}
\newcommand{\luindexFTOHBWriteOwned}{100}
\newcommand{\luindexFTOHBWriteShared}{<0.001}
\newcommand{\luindexFTOHBNoFPOtherEventTotal}{415314}
\newcommand{\luindexFTOHBAcqRelOtherTotal}{99.9}
\newcommand{\luindexFTOHBNoAcqRelOtherTotal}{574}
\newcommand{\luindexFTOHBFork}{0.174}
\newcommand{\luindexFTOHBJoin}{0.0}
\newcommand{\luindexFTOHBPreWait}{0.348}
\newcommand{\luindexFTOHBPostWait}{0.348}
\newcommand{\luindexFTOHBVolatileTotal}{77.2}
\newcommand{\luindexFTOHBClassInit}{11.5}
\newcommand{\luindexFTOHBClassAccess}{10.5}
\newcommand{\luindexFTOHBRaceTotal}{1}
\newcommand{\luindexFTOHBWrRdRace}{0.0}
\newcommand{\luindexFTOHBWrWrRace}{0.0}
\newcommand{\luindexFTOHBRdWrRace}{0.0}
\newcommand{\luindexFTOHBRdShWrRace}{100.0}
\newcommand{\luindexFTOHBHoldLocksTotal}{10}
\newcommand{\luindexFTOHBOneLockHeld}{25.8}
\newcommand{\luindexFTOHBTwoNestedLocks}{25.4}
\newcommand{\luindexFTOHBThreeNestedLocks}{25.3}
\newcommand{\luindexFTOHBFourNestedLocks}{0.61}
\newcommand{\luindexFTOHBFiveNestedLocks}{<0.1}
\newcommand{\luindexFTOHBSixNestedLocks}{<0.1}
\newcommand{\luindexFTOHBSevenNestedLocks}{\cna}
\newcommand{\luindexFTOHBEightNestedLocks}{\cna}
\newcommand{\luindexFTOHBNineNestedLocks}{\cna}
\newcommand{\luindexFTOHBTenNestedLocks}{\cna}
\newcommand{\luindexFTOHBHundredNestedLocks}{\cna}
\newcommand{\luindexFTOHBExWrSet}{\ena}
\newcommand{\luindexFTOHBExWrCheck}{\ena}
\newcommand{\luindexFTOHBExWrUpdate}{\ena}
\newcommand{\luindexFTOHBExRdCheck}{\ena}
\newcommand{\luindexFTOHBExRdUpdate}{\ena}
\newcommand{\luindexFTOHBExTotalCheck}{\ena}
\newcommand{\luindexFTOHBExTotalUpdate}{\ena}
\newcommand{\luindexFTOWCPEventTotal}{400}
\newcommand{\luindexFTOWCPNoFPEventTotal}{41}
\newcommand{\luindexFTOWCPNoFPAccessTotal}{40}
\newcommand{\luindexFTOWCPNoFPOtherTotal}{0.42}
\newcommand{\luindexFTOWCPReadTotal}{66.0}
\newcommand{\luindexFTOWCPWriteTotal}{33.0}
\newcommand{\luindexFTOWCPNoFPAccessInCS}{8.08}
\newcommand{\luindexFTOWCPNoFPAccessOutCS}{91.9}
\newcommand{\luindexFTOWCPAcqRelTotal}{14.2}
\newcommand{\luindexFTOWCPOtherTotal}{0.02}
\newcommand{\luindexFTOWCPNoFPReadTotal}{27}
\newcommand{\luindexFTOWCPReadInCS}{23.5}
\newcommand{\luindexFTOWCPReadOutCS}{82}
\newcommand{\luindexFTOWCPReadSameEp}{5.48}
\newcommand{\luindexFTOWCPReadSharedSameEp}{\cna}
\newcommand{\luindexFTOWCPReadExclusive}{<0.001}
\newcommand{\luindexFTOWCPReadOwned}{100}
\newcommand{\luindexFTOWCPReadShare}{<0.001}
\newcommand{\luindexFTOWCPReadShared}{\cna}
\newcommand{\luindexFTOWCPReadSharedOwned}{<0.001}
\newcommand{\luindexFTOWCPNoFPHonestWriteTotal}{13}
\newcommand{\luindexFTOWCPWriteInCS}{32.1}
\newcommand{\luindexFTOWCPWriteOutCS}{75.5}
\newcommand{\luindexFTOWCPNoFPWriteTotal}{13}
\newcommand{\luindexFTOWCPWriteSameEp}{7.57}
\newcommand{\luindexFTOWCPWriteExclusive}{<0.001}
\newcommand{\luindexFTOWCPWriteOwned}{100}
\newcommand{\luindexFTOWCPWriteShared}{<0.001}
\newcommand{\luindexFTOWCPNoFPOtherEventTotal}{415322}
\newcommand{\luindexFTOWCPAcqRelOtherTotal}{99.9}
\newcommand{\luindexFTOWCPNoAcqRelOtherTotal}{582}
\newcommand{\luindexFTOWCPFork}{0.172}
\newcommand{\luindexFTOWCPJoin}{0.0}
\newcommand{\luindexFTOWCPPreWait}{0.344}
\newcommand{\luindexFTOWCPPostWait}{0.344}
\newcommand{\luindexFTOWCPVolatileTotal}{77.5}
\newcommand{\luindexFTOWCPClassInit}{11.3}
\newcommand{\luindexFTOWCPClassAccess}{10.3}
\newcommand{\luindexFTOWCPRaceTotal}{1}
\newcommand{\luindexFTOWCPWrRdRace}{0.0}
\newcommand{\luindexFTOWCPWrWrRace}{0.0}
\newcommand{\luindexFTOWCPRdWrRace}{0.0}
\newcommand{\luindexFTOWCPRdShWrRace}{100.0}
\newcommand{\luindexFTOWCPHoldLocksTotal}{10}
\newcommand{\luindexFTOWCPOneLockHeld}{25.8}
\newcommand{\luindexFTOWCPTwoNestedLocks}{25.4}
\newcommand{\luindexFTOWCPThreeNestedLocks}{25.3}
\newcommand{\luindexFTOWCPFourNestedLocks}{0.61}
\newcommand{\luindexFTOWCPFiveNestedLocks}{<0.1}
\newcommand{\luindexFTOWCPSixNestedLocks}{<0.1}
\newcommand{\luindexFTOWCPSevenNestedLocks}{\cna}
\newcommand{\luindexFTOWCPEightNestedLocks}{\cna}
\newcommand{\luindexFTOWCPNineNestedLocks}{\cna}
\newcommand{\luindexFTOWCPTenNestedLocks}{\cna}
\newcommand{\luindexFTOWCPHundredNestedLocks}{\cna}
\newcommand{\luindexFTOWCPExWrSet}{\ena}
\newcommand{\luindexFTOWCPExWrCheck}{\ena}
\newcommand{\luindexFTOWCPExWrUpdate}{\ena}
\newcommand{\luindexFTOWCPExRdCheck}{\ena}
\newcommand{\luindexFTOWCPExRdUpdate}{\ena}
\newcommand{\luindexFTOWCPExTotalCheck}{\ena}
\newcommand{\luindexFTOWCPExTotalUpdate}{\ena}
\newcommand{\luindexREWCPEventTotal}{400}
\newcommand{\luindexREWCPNoFPEventTotal}{41}
\newcommand{\luindexREWCPNoFPAccessTotal}{40}
\newcommand{\luindexREWCPNoFPOtherTotal}{0.42}
\newcommand{\luindexREWCPReadTotal}{66.0}
\newcommand{\luindexREWCPWriteTotal}{33.0}
\newcommand{\luindexREWCPNoFPAccessInCS}{8.08}
\newcommand{\luindexREWCPNoFPAccessOutCS}{91.9}
\newcommand{\luindexREWCPAcqRelTotal}{14.2}
\newcommand{\luindexREWCPOtherTotal}{0.0197}
\newcommand{\luindexREWCPNoFPReadTotal}{27}
\newcommand{\luindexREWCPReadInCS}{23.5}
\newcommand{\luindexREWCPReadOutCS}{82}
\newcommand{\luindexREWCPReadSameEp}{5.48}
\newcommand{\luindexREWCPReadSharedSameEp}{\cna}
\newcommand{\luindexREWCPReadExclusive}{<0.001}
\newcommand{\luindexREWCPReadOwned}{100}
\newcommand{\luindexREWCPReadShare}{<0.001}
\newcommand{\luindexREWCPReadShared}{\cna}
\newcommand{\luindexREWCPReadSharedOwned}{<0.001}
\newcommand{\luindexREWCPNoFPHonestWriteTotal}{13}
\newcommand{\luindexREWCPWriteInCS}{32.1}
\newcommand{\luindexREWCPWriteOutCS}{75.5}
\newcommand{\luindexREWCPNoFPWriteTotal}{13}
\newcommand{\luindexREWCPWriteSameEp}{7.57}
\newcommand{\luindexREWCPWriteExclusive}{<0.001}
\newcommand{\luindexREWCPWriteOwned}{100}
\newcommand{\luindexREWCPWriteShared}{<0.001}
\newcommand{\luindexREWCPNoFPOtherEventTotal}{415315}
\newcommand{\luindexREWCPAcqRelOtherTotal}{99.9}
\newcommand{\luindexREWCPNoAcqRelOtherTotal}{575}
\newcommand{\luindexREWCPFork}{0.174}
\newcommand{\luindexREWCPJoin}{0.0}
\newcommand{\luindexREWCPPreWait}{0.348}
\newcommand{\luindexREWCPPostWait}{0.348}
\newcommand{\luindexREWCPVolatileTotal}{77.2}
\newcommand{\luindexREWCPClassInit}{11.5}
\newcommand{\luindexREWCPClassAccess}{10.4}
\newcommand{\luindexREWCPRaceTotal}{1}
\newcommand{\luindexREWCPWrRdRace}{0.0}
\newcommand{\luindexREWCPWrWrRace}{0.0}
\newcommand{\luindexREWCPRdWrRace}{0.0}
\newcommand{\luindexREWCPRdShWrRace}{100.0}
\newcommand{\luindexREWCPHoldLocksTotal}{10}
\newcommand{\luindexREWCPOneLockHeld}{25.8}
\newcommand{\luindexREWCPTwoNestedLocks}{25.4}
\newcommand{\luindexREWCPThreeNestedLocks}{25.3}
\newcommand{\luindexREWCPFourNestedLocks}{0.61}
\newcommand{\luindexREWCPFiveNestedLocks}{<0.1}
\newcommand{\luindexREWCPSixNestedLocks}{<0.1}
\newcommand{\luindexREWCPSevenNestedLocks}{\cna}
\newcommand{\luindexREWCPEightNestedLocks}{\cna}
\newcommand{\luindexREWCPNineNestedLocks}{\cna}
\newcommand{\luindexREWCPTenNestedLocks}{\cna}
\newcommand{\luindexREWCPHundredNestedLocks}{\cna}
\newcommand{\luindexREWCPExWrSet}{\ena}
\newcommand{\luindexREWCPExWrCheck}{\ena}
\newcommand{\luindexREWCPExWrUpdate}{\ena}
\newcommand{\luindexREWCPExRdCheck}{\ena}
\newcommand{\luindexREWCPExRdUpdate}{\ena}
\newcommand{\luindexREWCPExTotalCheck}{\ena}
\newcommand{\luindexREWCPExTotalUpdate}{\ena}
\newcommand{\luindexFTODCEventTotal}{400}
\newcommand{\luindexFTODCNoFPEventTotal}{41}
\newcommand{\luindexFTODCNoFPAccessTotal}{40}
\newcommand{\luindexFTODCNoFPOtherTotal}{0.42}
\newcommand{\luindexFTODCReadTotal}{66.0}
\newcommand{\luindexFTODCWriteTotal}{33.0}
\newcommand{\luindexFTODCNoFPAccessInCS}{8.08}
\newcommand{\luindexFTODCNoFPAccessOutCS}{91.9}
\newcommand{\luindexFTODCAcqRelTotal}{14.2}
\newcommand{\luindexFTODCOtherTotal}{0.02}
\newcommand{\luindexFTODCNoFPReadTotal}{27}
\newcommand{\luindexFTODCReadInCS}{23.5}
\newcommand{\luindexFTODCReadOutCS}{82}
\newcommand{\luindexFTODCReadSameEp}{5.48}
\newcommand{\luindexFTODCReadSharedSameEp}{\cna}
\newcommand{\luindexFTODCReadExclusive}{<0.001}
\newcommand{\luindexFTODCReadOwned}{100}
\newcommand{\luindexFTODCReadShare}{<0.001}
\newcommand{\luindexFTODCReadShared}{\cna}
\newcommand{\luindexFTODCReadSharedOwned}{<0.001}
\newcommand{\luindexFTODCNoFPHonestWriteTotal}{13}
\newcommand{\luindexFTODCWriteInCS}{32.1}
\newcommand{\luindexFTODCWriteOutCS}{75.5}
\newcommand{\luindexFTODCNoFPWriteTotal}{13}
\newcommand{\luindexFTODCWriteSameEp}{7.57}
\newcommand{\luindexFTODCWriteExclusive}{<0.001}
\newcommand{\luindexFTODCWriteOwned}{100}
\newcommand{\luindexFTODCWriteShared}{<0.001}
\newcommand{\luindexFTODCNoFPOtherEventTotal}{415322}
\newcommand{\luindexFTODCAcqRelOtherTotal}{99.9}
\newcommand{\luindexFTODCNoAcqRelOtherTotal}{582}
\newcommand{\luindexFTODCFork}{0.172}
\newcommand{\luindexFTODCJoin}{0.0}
\newcommand{\luindexFTODCPreWait}{0.344}
\newcommand{\luindexFTODCPostWait}{0.344}
\newcommand{\luindexFTODCVolatileTotal}{77.5}
\newcommand{\luindexFTODCClassInit}{11.3}
\newcommand{\luindexFTODCClassAccess}{10.3}
\newcommand{\luindexFTODCRaceTotal}{1}
\newcommand{\luindexFTODCWrRdRace}{0.0}
\newcommand{\luindexFTODCWrWrRace}{0.0}
\newcommand{\luindexFTODCRdWrRace}{0.0}
\newcommand{\luindexFTODCRdShWrRace}{100.0}
\newcommand{\luindexFTODCHoldLocksTotal}{10}
\newcommand{\luindexFTODCOneLockHeld}{25.8}
\newcommand{\luindexFTODCTwoNestedLocks}{25.4}
\newcommand{\luindexFTODCThreeNestedLocks}{25.3}
\newcommand{\luindexFTODCFourNestedLocks}{0.61}
\newcommand{\luindexFTODCFiveNestedLocks}{<0.1}
\newcommand{\luindexFTODCSixNestedLocks}{<0.1}
\newcommand{\luindexFTODCSevenNestedLocks}{\cna}
\newcommand{\luindexFTODCEightNestedLocks}{\cna}
\newcommand{\luindexFTODCNineNestedLocks}{\cna}
\newcommand{\luindexFTODCTenNestedLocks}{\cna}
\newcommand{\luindexFTODCHundredNestedLocks}{\cna}
\newcommand{\luindexFTODCExWrSet}{\ena}
\newcommand{\luindexFTODCExWrCheck}{\ena}
\newcommand{\luindexFTODCExWrUpdate}{\ena}
\newcommand{\luindexFTODCExRdCheck}{\ena}
\newcommand{\luindexFTODCExRdUpdate}{\ena}
\newcommand{\luindexFTODCExTotalCheck}{\ena}
\newcommand{\luindexFTODCExTotalUpdate}{\ena}
\newcommand{\luindexREDCEventTotal}{400}
\newcommand{\luindexREDCNoFPEventTotal}{41}
\newcommand{\luindexREDCNoFPAccessTotal}{40}
\newcommand{\luindexREDCNoFPOtherTotal}{0.42}
\newcommand{\luindexREDCReadTotal}{66.0}
\newcommand{\luindexREDCWriteTotal}{33.0}
\newcommand{\luindexREDCNoFPAccessInCS}{8.08}
\newcommand{\luindexREDCNoFPAccessOutCS}{91.9}
\newcommand{\luindexREDCAcqRelTotal}{14.2}
\newcommand{\luindexREDCOtherTotal}{0.0198}
\newcommand{\luindexREDCNoFPReadTotal}{27}
\newcommand{\luindexREDCReadInCS}{23.5}
\newcommand{\luindexREDCReadOutCS}{82}
\newcommand{\luindexREDCReadSameEp}{5.48}
\newcommand{\luindexREDCReadSharedSameEp}{\cna}
\newcommand{\luindexREDCReadExclusive}{<0.001}
\newcommand{\luindexREDCReadOwned}{100}
\newcommand{\luindexREDCReadShare}{<0.001}
\newcommand{\luindexREDCReadShared}{\cna}
\newcommand{\luindexREDCReadSharedOwned}{<0.001}
\newcommand{\luindexREDCNoFPHonestWriteTotal}{13}
\newcommand{\luindexREDCWriteInCS}{32.1}
\newcommand{\luindexREDCWriteOutCS}{75.5}
\newcommand{\luindexREDCNoFPWriteTotal}{13}
\newcommand{\luindexREDCWriteSameEp}{7.57}
\newcommand{\luindexREDCWriteExclusive}{<0.001}
\newcommand{\luindexREDCWriteOwned}{100}
\newcommand{\luindexREDCWriteShared}{<0.001}
\newcommand{\luindexREDCNoFPOtherEventTotal}{415316}
\newcommand{\luindexREDCAcqRelOtherTotal}{99.9}
\newcommand{\luindexREDCNoAcqRelOtherTotal}{576}
\newcommand{\luindexREDCFork}{0.174}
\newcommand{\luindexREDCJoin}{0.0}
\newcommand{\luindexREDCPreWait}{0.347}
\newcommand{\luindexREDCPostWait}{0.347}
\newcommand{\luindexREDCVolatileTotal}{77.3}
\newcommand{\luindexREDCClassInit}{11.5}
\newcommand{\luindexREDCClassAccess}{10.4}
\newcommand{\luindexREDCRaceTotal}{1}
\newcommand{\luindexREDCWrRdRace}{0.0}
\newcommand{\luindexREDCWrWrRace}{0.0}
\newcommand{\luindexREDCRdWrRace}{0.0}
\newcommand{\luindexREDCRdShWrRace}{100.0}
\newcommand{\luindexREDCHoldLocksTotal}{10}
\newcommand{\luindexREDCOneLockHeld}{25.8}
\newcommand{\luindexREDCTwoNestedLocks}{25.4}
\newcommand{\luindexREDCThreeNestedLocks}{25.3}
\newcommand{\luindexREDCFourNestedLocks}{0.61}
\newcommand{\luindexREDCFiveNestedLocks}{<0.1}
\newcommand{\luindexREDCSixNestedLocks}{<0.1}
\newcommand{\luindexREDCSevenNestedLocks}{\cna}
\newcommand{\luindexREDCEightNestedLocks}{\cna}
\newcommand{\luindexREDCNineNestedLocks}{\cna}
\newcommand{\luindexREDCTenNestedLocks}{\cna}
\newcommand{\luindexREDCHundredNestedLocks}{\cna}
\newcommand{\luindexREDCExWrSet}{\ena}
\newcommand{\luindexREDCExWrCheck}{\ena}
\newcommand{\luindexREDCExWrUpdate}{\ena}
\newcommand{\luindexREDCExRdCheck}{\ena}
\newcommand{\luindexREDCExRdUpdate}{\ena}
\newcommand{\luindexREDCExTotalCheck}{\ena}
\newcommand{\luindexREDCExTotalUpdate}{\ena}
\newcommand{\luindexFTOCAPOEventTotal}{400}
\newcommand{\luindexFTOCAPONoFPEventTotal}{41}
\newcommand{\luindexFTOCAPONoFPAccessTotal}{40}
\newcommand{\luindexFTOCAPONoFPOtherTotal}{0.42}
\newcommand{\luindexFTOCAPOReadTotal}{66.0}
\newcommand{\luindexFTOCAPOWriteTotal}{33.0}
\newcommand{\luindexFTOCAPONoFPAccessInCS}{8.08}
\newcommand{\luindexFTOCAPONoFPAccessOutCS}{91.9}
\newcommand{\luindexFTOCAPOAcqRelTotal}{14.2}
\newcommand{\luindexFTOCAPOOtherTotal}{0.02}
\newcommand{\luindexFTOCAPONoFPReadTotal}{27}
\newcommand{\luindexFTOCAPOReadInCS}{23.5}
\newcommand{\luindexFTOCAPOReadOutCS}{82}
\newcommand{\luindexFTOCAPOReadSameEp}{5.48}
\newcommand{\luindexFTOCAPOReadSharedSameEp}{\cna}
\newcommand{\luindexFTOCAPOReadExclusive}{<0.001}
\newcommand{\luindexFTOCAPOReadOwned}{100}
\newcommand{\luindexFTOCAPOReadShare}{<0.001}
\newcommand{\luindexFTOCAPOReadShared}{\cna}
\newcommand{\luindexFTOCAPOReadSharedOwned}{<0.001}
\newcommand{\luindexFTOCAPONoFPHonestWriteTotal}{13}
\newcommand{\luindexFTOCAPOWriteInCS}{32.1}
\newcommand{\luindexFTOCAPOWriteOutCS}{75.5}
\newcommand{\luindexFTOCAPONoFPWriteTotal}{13}
\newcommand{\luindexFTOCAPOWriteSameEp}{7.57}
\newcommand{\luindexFTOCAPOWriteExclusive}{<0.001}
\newcommand{\luindexFTOCAPOWriteOwned}{100}
\newcommand{\luindexFTOCAPOWriteShared}{<0.001}
\newcommand{\luindexFTOCAPONoFPOtherEventTotal}{415322}
\newcommand{\luindexFTOCAPOAcqRelOtherTotal}{99.9}
\newcommand{\luindexFTOCAPONoAcqRelOtherTotal}{582}
\newcommand{\luindexFTOCAPOFork}{0.172}
\newcommand{\luindexFTOCAPOJoin}{0.0}
\newcommand{\luindexFTOCAPOPreWait}{0.344}
\newcommand{\luindexFTOCAPOPostWait}{0.344}
\newcommand{\luindexFTOCAPOVolatileTotal}{77.5}
\newcommand{\luindexFTOCAPOClassInit}{11.3}
\newcommand{\luindexFTOCAPOClassAccess}{10.3}
\newcommand{\luindexFTOCAPORaceTotal}{1}
\newcommand{\luindexFTOCAPOWrRdRace}{0.0}
\newcommand{\luindexFTOCAPOWrWrRace}{0.0}
\newcommand{\luindexFTOCAPORdWrRace}{0.0}
\newcommand{\luindexFTOCAPORdShWrRace}{100.0}
\newcommand{\luindexFTOCAPOHoldLocksTotal}{10}
\newcommand{\luindexFTOCAPOOneLockHeld}{25.8}
\newcommand{\luindexFTOCAPOTwoNestedLocks}{25.4}
\newcommand{\luindexFTOCAPOThreeNestedLocks}{25.3}
\newcommand{\luindexFTOCAPOFourNestedLocks}{0.61}
\newcommand{\luindexFTOCAPOFiveNestedLocks}{<0.1}
\newcommand{\luindexFTOCAPOSixNestedLocks}{<0.1}
\newcommand{\luindexFTOCAPOSevenNestedLocks}{\cna}
\newcommand{\luindexFTOCAPOEightNestedLocks}{\cna}
\newcommand{\luindexFTOCAPONineNestedLocks}{\cna}
\newcommand{\luindexFTOCAPOTenNestedLocks}{\cna}
\newcommand{\luindexFTOCAPOHundredNestedLocks}{\cna}
\newcommand{\luindexFTOCAPOExWrSet}{\ena}
\newcommand{\luindexFTOCAPOExWrCheck}{\ena}
\newcommand{\luindexFTOCAPOExWrUpdate}{\ena}
\newcommand{\luindexFTOCAPOExRdCheck}{\ena}
\newcommand{\luindexFTOCAPOExRdUpdate}{\ena}
\newcommand{\luindexFTOCAPOExTotalCheck}{\ena}
\newcommand{\luindexFTOCAPOExTotalUpdate}{\ena}
\newcommand{\luindexRECAPOEventTotal}{400}
\newcommand{\luindexRECAPONoFPEventTotal}{41}
\newcommand{\luindexRECAPONoFPAccessTotal}{40}
\newcommand{\luindexRECAPONoFPOtherTotal}{0.42}
\newcommand{\luindexRECAPOReadTotal}{66.0}
\newcommand{\luindexRECAPOWriteTotal}{33.0}
\newcommand{\luindexRECAPONoFPAccessInCS}{8.08}
\newcommand{\luindexRECAPONoFPAccessOutCS}{91.9}
\newcommand{\luindexRECAPOAcqRelTotal}{14.2}
\newcommand{\luindexRECAPOOtherTotal}{0.0198}
\newcommand{\luindexRECAPONoFPReadTotal}{27}
\newcommand{\luindexRECAPOReadInCS}{23.5}
\newcommand{\luindexRECAPOReadOutCS}{82}
\newcommand{\luindexRECAPOReadSameEp}{5.48}
\newcommand{\luindexRECAPOReadSharedSameEp}{\cna}
\newcommand{\luindexRECAPOReadExclusive}{<0.001}
\newcommand{\luindexRECAPOReadOwned}{100}
\newcommand{\luindexRECAPOReadShare}{<0.001}
\newcommand{\luindexRECAPOReadShared}{\cna}
\newcommand{\luindexRECAPOReadSharedOwned}{<0.001}
\newcommand{\luindexRECAPONoFPHonestWriteTotal}{13}
\newcommand{\luindexRECAPOWriteInCS}{32.1}
\newcommand{\luindexRECAPOWriteOutCS}{75.5}
\newcommand{\luindexRECAPONoFPWriteTotal}{13}
\newcommand{\luindexRECAPOWriteSameEp}{7.57}
\newcommand{\luindexRECAPOWriteExclusive}{<0.001}
\newcommand{\luindexRECAPOWriteOwned}{100}
\newcommand{\luindexRECAPOWriteShared}{<0.001}
\newcommand{\luindexRECAPONoFPOtherEventTotal}{415318}
\newcommand{\luindexRECAPOAcqRelOtherTotal}{99.9}
\newcommand{\luindexRECAPONoAcqRelOtherTotal}{578}
\newcommand{\luindexRECAPOFork}{0.173}
\newcommand{\luindexRECAPOJoin}{0.0}
\newcommand{\luindexRECAPOPreWait}{0.346}
\newcommand{\luindexRECAPOPostWait}{0.346}
\newcommand{\luindexRECAPOVolatileTotal}{77.3}
\newcommand{\luindexRECAPOClassInit}{11.4}
\newcommand{\luindexRECAPOClassAccess}{10.4}
\newcommand{\luindexRECAPORaceTotal}{1}
\newcommand{\luindexRECAPOWrRdRace}{0.0}
\newcommand{\luindexRECAPOWrWrRace}{0.0}
\newcommand{\luindexRECAPORdWrRace}{0.0}
\newcommand{\luindexRECAPORdShWrRace}{100.0}
\newcommand{\luindexRECAPOHoldLocksTotal}{10}
\newcommand{\luindexRECAPOOneLockHeld}{25.8}
\newcommand{\luindexRECAPOTwoNestedLocks}{25.4}
\newcommand{\luindexRECAPOThreeNestedLocks}{25.3}
\newcommand{\luindexRECAPOFourNestedLocks}{0.61}
\newcommand{\luindexRECAPOFiveNestedLocks}{<0.1}
\newcommand{\luindexRECAPOSixNestedLocks}{<0.1}
\newcommand{\luindexRECAPOSevenNestedLocks}{\cna}
\newcommand{\luindexRECAPOEightNestedLocks}{\cna}
\newcommand{\luindexRECAPONineNestedLocks}{\cna}
\newcommand{\luindexRECAPOTenNestedLocks}{\cna}
\newcommand{\luindexRECAPOHundredNestedLocks}{\cna}
\newcommand{\luindexRECAPOExWrSet}{\ena}
\newcommand{\luindexRECAPOExWrCheck}{\ena}
\newcommand{\luindexRECAPOExWrUpdate}{\ena}
\newcommand{\luindexRECAPOExRdCheck}{\ena}
\newcommand{\luindexRECAPOExRdUpdate}{\ena}
\newcommand{\luindexRECAPOExTotalCheck}{\ena}
\newcommand{\luindexRECAPOExTotalUpdate}{\ena}
\newcommand{\FASTluindexMaxLiveThreads}{3}
\newcommand{\FASTluindexTotalThreads}{3}
\newcommand{\FASTluindexBaseTime}{1.2}
\newcommand{\FASTluindexBaseTimeCI}{32}
\newcommand{\FASTluindexEmptyTime}{\rna}
\newcommand{\FASTluindexEmptyTimeCI}{\rna}
\newcommand{\FASTluindexEmptyTimeCIMIN}{\rna}
\newcommand{\FASTluindexEmptyTimeCIMAX}{\rna}
\newcommand{\FASTluindexFTTime}{7.5}
\newcommand{\FASTluindexFTTimeCI}{0.12}
\newcommand{\FASTluindexHBTime}{7.9}
\newcommand{\FASTluindexHBTimeCI}{0.16}
\newcommand{\FASTluindexFTOHBTime}{8.0}
\newcommand{\FASTluindexFTOHBTimeCI}{0.2}
\newcommand{\FASTluindexWCPTime}{\rna}
\newcommand{\FASTluindexWCPTimeCI}{\rna}
\newcommand{\FASTluindexWCPTimeCIMIN}{\rna}
\newcommand{\FASTluindexWCPTimeCIMAX}{\rna}
\newcommand{\FASTluindexFTOWCPTime}{22}
\newcommand{\FASTluindexFTOWCPTimeCI}{0.45}
\newcommand{\FASTluindexREWCPTime}{8.8}
\newcommand{\FASTluindexREWCPTimeCI}{0.18}
\newcommand{\FASTluindexDCTime}{\rna}
\newcommand{\FASTluindexDCTimeCI}{\rna}
\newcommand{\FASTluindexDCTimeCIMIN}{\rna}
\newcommand{\FASTluindexDCTimeCIMAX}{\rna}
\newcommand{\FASTluindexFTODCTime}{21}
\newcommand{\FASTluindexFTODCTimeCI}{0.44}
\newcommand{\FASTluindexREDCTime}{8.6}
\newcommand{\FASTluindexREDCTimeCI}{0.18}
\newcommand{\FASTluindexCAPOTime}{\rna}
\newcommand{\FASTluindexCAPOTimeCI}{\rna}
\newcommand{\FASTluindexCAPOTimeCIMIN}{\rna}
\newcommand{\FASTluindexCAPOTimeCIMAX}{\rna}
\newcommand{\FASTluindexFTOCAPOTime}{21}
\newcommand{\FASTluindexFTOCAPOTimeCI}{0.30}
\newcommand{\FASTluindexRECAPOTime}{8.5}
\newcommand{\FASTluindexRECAPOTimeCI}{0.14}
\newcommand{\FASTluindexAGGCAPOTime}{\rna}
\newcommand{\FASTluindexAGGCAPOTimeCI}{\rna}
\newcommand{\FASTluindexAGGCAPOTimeCIMIN}{\rna}
\newcommand{\FASTluindexAGGCAPOTimeCIMAX}{\rna}
\newcommand{\FASTluindexStaticTime}{\rzero}
\newcommand{\FASTluindexDynamicTime}{\rzero}
\newcommand{\FASTluindexBaseMem}{150}
\newcommand{\FASTluindexBaseMemCI}{1.6}
\newcommand{\FASTluindexFTMem}{4.3}
\newcommand{\FASTluindexFTMemCI}{0.064}
\newcommand{\FASTluindexHBMem}{4.3}
\newcommand{\FASTluindexHBMemCI}{0.078}
\newcommand{\FASTluindexFTOHBMem}{4.3}
\newcommand{\FASTluindexFTOHBMemCI}{0.076}
\newcommand{\FASTluindexWCPMem}{\memna}
\newcommand{\FASTluindexWCPMemCI}{\memna}
\newcommand{\FASTluindexWCPMemCIMIN}{\memna}
\newcommand{\FASTluindexWCPMemCIMAX}{\memna}
\newcommand{\FASTluindexFTOWCPMem}{20}
\newcommand{\FASTluindexFTOWCPMemCI}{0.28}
\newcommand{\FASTluindexREWCPMem}{6.1}
\newcommand{\FASTluindexREWCPMemCI}{0.039}
\newcommand{\FASTluindexDCMem}{\memna}
\newcommand{\FASTluindexDCMemCI}{\memna}
\newcommand{\FASTluindexDCMemCIMIN}{\memna}
\newcommand{\FASTluindexDCMemCIMAX}{\memna}
\newcommand{\FASTluindexFTODCMem}{20}
\newcommand{\FASTluindexFTODCMemCI}{0.23}
\newcommand{\FASTluindexREDCMem}{6.1}
\newcommand{\FASTluindexREDCMemCI}{0.076}
\newcommand{\FASTluindexCAPOMem}{\memna}
\newcommand{\FASTluindexCAPOMemCI}{\memna}
\newcommand{\FASTluindexCAPOMemCIMIN}{\memna}
\newcommand{\FASTluindexCAPOMemCIMAX}{\memna}
\newcommand{\FASTluindexFTOCAPOMem}{20}
\newcommand{\FASTluindexFTOCAPOMemCI}{0.32}
\newcommand{\FASTluindexRECAPOMem}{6.1}
\newcommand{\FASTluindexRECAPOMemCI}{0.096}
\newcommand{\FASTluindexAGGCAPOMem}{\memna}
\newcommand{\FASTluindexAGGCAPOMemCI}{\memna}
\newcommand{\FASTluindexAGGCAPOMemCIMIN}{\memna}
\newcommand{\FASTluindexAGGCAPOMemCIMAX}{\memna}
\newcommand{\FASTluindexEventsCI}{3}
\newcommand{\FASTluindexEventsCIMIN}{396,268,365}
\newcommand{\FASTluindexEventsCIMAX}{396,268,371}
\newcommand{\FASTluindexNoFPEventsCI}{2}
\newcommand{\FASTluindexNoFPEventsCIMIN}{2,915,286}
\newcommand{\FASTluindexNoFPEventsCIMAX}{2,915,290}
\newcommand{\FASTluindexFT}{1}
\newcommand{\FASTluindexFTCI}{0.0}
\newcommand{\FASTluindexFTCIMIN}{1}
\newcommand{\FASTluindexFTCIMAX}{1}
\newcommand{\FASTluindexFTDynamic}{1}
\newcommand{\FASTluindexFTDynamicCI}{0.0}
\newcommand{\FASTluindexFTDynamicCIMIN}{1}
\newcommand{\FASTluindexFTDynamicCIMAX}{1}
\newcommand{\FASTluindexHB}{1}
\newcommand{\FASTluindexHBCI}{0.0}
\newcommand{\FASTluindexHBCIMIN}{1}
\newcommand{\FASTluindexHBCIMAX}{1}
\newcommand{\FASTluindexHBDynamic}{1}
\newcommand{\FASTluindexHBDynamicCI}{0.0}
\newcommand{\FASTluindexHBDynamicCIMIN}{1}
\newcommand{\FASTluindexHBDynamicCIMAX}{1}
\newcommand{\FASTluindexFTOHB}{1}
\newcommand{\FASTluindexFTOHBCI}{0.0}
\newcommand{\FASTluindexFTOHBCIMIN}{1}
\newcommand{\FASTluindexFTOHBCIMAX}{1}
\newcommand{\FASTluindexFTOHBDynamic}{1}
\newcommand{\FASTluindexFTOHBDynamicCI}{0.0}
\newcommand{\FASTluindexFTOHBDynamicCIMIN}{1}
\newcommand{\FASTluindexFTOHBDynamicCIMAX}{1}
\newcommand{\FASTluindexWCP}{\rna}
\newcommand{\FASTluindexWCPCI}{\rna}
\newcommand{\FASTluindexWCPCIMIN}{\rna}
\newcommand{\FASTluindexWCPCIMAX}{\rna}
\newcommand{\FASTluindexWCPDynamic}{\rna}
\newcommand{\FASTluindexWCPDynamicCI}{\rna}
\newcommand{\FASTluindexWCPDynamicCIMIN}{\rna}
\newcommand{\FASTluindexWCPDynamicCIMAX}{\rna}
\newcommand{\FASTluindexFTOWCP}{1}
\newcommand{\FASTluindexFTOWCPCI}{0.0}
\newcommand{\FASTluindexFTOWCPCIMIN}{1}
\newcommand{\FASTluindexFTOWCPCIMAX}{1}
\newcommand{\FASTluindexFTOWCPDynamic}{1}
\newcommand{\FASTluindexFTOWCPDynamicCI}{0.0}
\newcommand{\FASTluindexFTOWCPDynamicCIMIN}{1}
\newcommand{\FASTluindexFTOWCPDynamicCIMAX}{1}
\newcommand{\FASTluindexREWCP}{1}
\newcommand{\FASTluindexREWCPCI}{0.0}
\newcommand{\FASTluindexREWCPCIMIN}{1}
\newcommand{\FASTluindexREWCPCIMAX}{1}
\newcommand{\FASTluindexREWCPDynamic}{1}
\newcommand{\FASTluindexREWCPDynamicCI}{0.0}
\newcommand{\FASTluindexREWCPDynamicCIMIN}{1}
\newcommand{\FASTluindexREWCPDynamicCIMAX}{1}
\newcommand{\FASTluindexDC}{\rna}
\newcommand{\FASTluindexDCCI}{\rna}
\newcommand{\FASTluindexDCCIMIN}{\rna}
\newcommand{\FASTluindexDCCIMAX}{\rna}
\newcommand{\FASTluindexDCDynamic}{\rna}
\newcommand{\FASTluindexDCDynamicCI}{\rna}
\newcommand{\FASTluindexDCDynamicCIMIN}{\rna}
\newcommand{\FASTluindexDCDynamicCIMAX}{\rna}
\newcommand{\FASTluindexFTODC}{1}
\newcommand{\FASTluindexFTODCCI}{0.0}
\newcommand{\FASTluindexFTODCCIMIN}{1}
\newcommand{\FASTluindexFTODCCIMAX}{1}
\newcommand{\FASTluindexFTODCDynamic}{1}
\newcommand{\FASTluindexFTODCDynamicCI}{0.0}
\newcommand{\FASTluindexFTODCDynamicCIMIN}{1}
\newcommand{\FASTluindexFTODCDynamicCIMAX}{1}
\newcommand{\FASTluindexREDC}{1}
\newcommand{\FASTluindexREDCCI}{0.0}
\newcommand{\FASTluindexREDCCIMIN}{1}
\newcommand{\FASTluindexREDCCIMAX}{1}
\newcommand{\FASTluindexREDCDynamic}{1}
\newcommand{\FASTluindexREDCDynamicCI}{0.0}
\newcommand{\FASTluindexREDCDynamicCIMIN}{1}
\newcommand{\FASTluindexREDCDynamicCIMAX}{1}
\newcommand{\FASTluindexCAPO}{\rna}
\newcommand{\FASTluindexCAPOCI}{\rna}
\newcommand{\FASTluindexCAPOCIMIN}{\rna}
\newcommand{\FASTluindexCAPOCIMAX}{\rna}
\newcommand{\FASTluindexCAPODynamic}{\rna}
\newcommand{\FASTluindexCAPODynamicCI}{\rna}
\newcommand{\FASTluindexCAPODynamicCIMIN}{\rna}
\newcommand{\FASTluindexCAPODynamicCIMAX}{\rna}
\newcommand{\FASTluindexFTOCAPO}{1}
\newcommand{\FASTluindexFTOCAPOCI}{0.0}
\newcommand{\FASTluindexFTOCAPOCIMIN}{1}
\newcommand{\FASTluindexFTOCAPOCIMAX}{1}
\newcommand{\FASTluindexFTOCAPODynamic}{1}
\newcommand{\FASTluindexFTOCAPODynamicCI}{0.0}
\newcommand{\FASTluindexFTOCAPODynamicCIMIN}{1}
\newcommand{\FASTluindexFTOCAPODynamicCIMAX}{1}
\newcommand{\FASTluindexRECAPO}{1}
\newcommand{\FASTluindexRECAPOCI}{0.0}
\newcommand{\FASTluindexRECAPOCIMIN}{1}
\newcommand{\FASTluindexRECAPOCIMAX}{1}
\newcommand{\FASTluindexRECAPODynamic}{1}
\newcommand{\FASTluindexRECAPODynamicCI}{0.0}
\newcommand{\FASTluindexRECAPODynamicCIMIN}{1}
\newcommand{\FASTluindexRECAPODynamicCIMAX}{1}
\newcommand{\FASTluindexAGGCAPO}{\rna}
\newcommand{\FASTluindexAGGCAPOCI}{\rna}
\newcommand{\FASTluindexAGGCAPOCIMIN}{\rna}
\newcommand{\FASTluindexAGGCAPOCIMAX}{\rna}
\newcommand{\FASTluindexAGGCAPODynamic}{\rna}
\newcommand{\FASTluindexAGGCAPODynamicCI}{\rna}
\newcommand{\FASTluindexAGGCAPODynamicCIMIN}{\rna}
\newcommand{\FASTluindexAGGCAPODynamicCIMAX}{\rna}
\newcommand{\FASTlusearchEvents}{1,400}
\newcommand{\FASTlusearchNoFPEvents}{43}
\newcommand{\lusearchHBEventTotal}{1,400}
\newcommand{\lusearchHBNoFPEventTotal}{150}
\newcommand{\lusearchHBNoFPAccessTotal}{150}
\newcommand{\lusearchHBNoFPOtherTotal}{3.8}
\newcommand{\lusearchHBReadTotal}{79.4}
\newcommand{\lusearchHBWriteTotal}{18.1}
\newcommand{\lusearchHBNoFPAccessInCS}{4.89}
\newcommand{\lusearchHBNoFPAccessOutCS}{95.1}
\newcommand{\lusearchHBAcqRelTotal}{6.23}
\newcommand{\lusearchHBOtherTotal}{2.72}
\newcommand{\lusearchHBNoFPReadTotal}{120}
\newcommand{\lusearchHBReadInCS}{3.36}
\newcommand{\lusearchHBReadOutCS}{103}
\newcommand{\lusearchHBReadSameEp}{6.47}
\newcommand{\lusearchHBReadSharedSameEp}{\cna}
\newcommand{\lusearchHBReadExclusive}{96.6}
\newcommand{\lusearchHBReadOwned}{\cna}
\newcommand{\lusearchHBReadShare}{<0.001}
\newcommand{\lusearchHBReadShared}{3.41}
\newcommand{\lusearchHBReadSharedOwned}{\cna}
\newcommand{\lusearchHBNoFPHonestWriteTotal}{28}
\newcommand{\lusearchHBWriteInCS}{10}
\newcommand{\lusearchHBWriteOutCS}{200}
\newcommand{\lusearchHBNoFPWriteTotal}{28}
\newcommand{\lusearchHBWriteSameEp}{110}
\newcommand{\lusearchHBWriteExclusive}{100}
\newcommand{\lusearchHBWriteOwned}{\cna}
\newcommand{\lusearchHBWriteShared}{\cna}
\newcommand{\lusearchHBNoFPOtherEventTotal}{3816275}
\newcommand{\lusearchHBAcqRelOtherTotal}{69.6}
\newcommand{\lusearchHBNoAcqRelOtherTotal}{1158339}
\newcommand{\lusearchHBFork}{0.00121}
\newcommand{\lusearchHBJoin}{0.0}
\newcommand{\lusearchHBPreWait}{0.00518}
\newcommand{\lusearchHBPostWait}{0.00518}
\newcommand{\lusearchHBVolatileTotal}{100.0}
\newcommand{\lusearchHBClassInit}{0.00311}
\newcommand{\lusearchHBClassAccess}{0.0293}
\newcommand{\lusearchHBRaceTotal}{\rna}
\newcommand{\lusearchHBWrRdRace}{\rna}
\newcommand{\lusearchHBWrWrRace}{\rna}
\newcommand{\lusearchHBRdWrRace}{\rna}
\newcommand{\lusearchHBRdShWrRace}{\rna}
\newcommand{\lusearchHBHoldLocksTotal}{6.9}
\newcommand{\lusearchHBOneLockHeld}{4.60}
\newcommand{\lusearchHBTwoNestedLocks}{0.26}
\newcommand{\lusearchHBThreeNestedLocks}{<0.1}
\newcommand{\lusearchHBFourNestedLocks}{\cna}
\newcommand{\lusearchHBFiveNestedLocks}{\cna}
\newcommand{\lusearchHBSixNestedLocks}{\cna}
\newcommand{\lusearchHBSevenNestedLocks}{\cna}
\newcommand{\lusearchHBEightNestedLocks}{\cna}
\newcommand{\lusearchHBNineNestedLocks}{\cna}
\newcommand{\lusearchHBTenNestedLocks}{\cna}
\newcommand{\lusearchHBHundredNestedLocks}{\cna}
\newcommand{\lusearchHBExWrSet}{\ena}
\newcommand{\lusearchHBExWrCheck}{\ena}
\newcommand{\lusearchHBExWrUpdate}{\ena}
\newcommand{\lusearchHBExRdCheck}{\ena}
\newcommand{\lusearchHBExRdUpdate}{\ena}
\newcommand{\lusearchHBExTotalCheck}{\ena}
\newcommand{\lusearchHBExTotalUpdate}{\ena}
\newcommand{\lusearchFTOHBEventTotal}{1,400}
\newcommand{\lusearchFTOHBNoFPEventTotal}{140}
\newcommand{\lusearchFTOHBNoFPAccessTotal}{130}
\newcommand{\lusearchFTOHBNoFPOtherTotal}{3.8}
\newcommand{\lusearchFTOHBReadTotal}{76.9}
\newcommand{\lusearchFTOHBWriteTotal}{20.3}
\newcommand{\lusearchFTOHBNoFPAccessInCS}{4.89}
\newcommand{\lusearchFTOHBNoFPAccessOutCS}{95.1}
\newcommand{\lusearchFTOHBAcqRelTotal}{6.23}
\newcommand{\lusearchFTOHBOtherTotal}{2.72}
\newcommand{\lusearchFTOHBNoFPReadTotal}{110}
\newcommand{\lusearchFTOHBReadInCS}{3.53}
\newcommand{\lusearchFTOHBReadOutCS}{104}
\newcommand{\lusearchFTOHBReadSameEp}{7.49}
\newcommand{\lusearchFTOHBReadSharedSameEp}{\cna}
\newcommand{\lusearchFTOHBReadExclusive}{<0.001}
\newcommand{\lusearchFTOHBReadOwned}{96.1}
\newcommand{\lusearchFTOHBReadShare}{<0.001}
\newcommand{\lusearchFTOHBReadShared}{0.0011}
\newcommand{\lusearchFTOHBReadSharedOwned}{3.9}
\newcommand{\lusearchFTOHBNoFPHonestWriteTotal}{28}
\newcommand{\lusearchFTOHBWriteInCS}{10}
\newcommand{\lusearchFTOHBWriteOutCS}{200}
\newcommand{\lusearchFTOHBNoFPWriteTotal}{28}
\newcommand{\lusearchFTOHBWriteSameEp}{110}
\newcommand{\lusearchFTOHBWriteExclusive}{\cna}
\newcommand{\lusearchFTOHBWriteOwned}{100}
\newcommand{\lusearchFTOHBWriteShared}{\cna}
\newcommand{\lusearchFTOHBNoFPOtherEventTotal}{3816271}
\newcommand{\lusearchFTOHBAcqRelOtherTotal}{69.6}
\newcommand{\lusearchFTOHBNoAcqRelOtherTotal}{1158335}
\newcommand{\lusearchFTOHBFork}{0.00121}
\newcommand{\lusearchFTOHBJoin}{0.0}
\newcommand{\lusearchFTOHBPreWait}{0.00527}
\newcommand{\lusearchFTOHBPostWait}{0.00527}
\newcommand{\lusearchFTOHBVolatileTotal}{100.0}
\newcommand{\lusearchFTOHBClassInit}{0.00311}
\newcommand{\lusearchFTOHBClassAccess}{0.0293}
\newcommand{\lusearchFTOHBRaceTotal}{\rna}
\newcommand{\lusearchFTOHBWrRdRace}{\rna}
\newcommand{\lusearchFTOHBWrWrRace}{\rna}
\newcommand{\lusearchFTOHBRdWrRace}{\rna}
\newcommand{\lusearchFTOHBRdShWrRace}{\rna}
\newcommand{\lusearchFTOHBHoldLocksTotal}{4.7}
\newcommand{\lusearchFTOHBOneLockHeld}{3.47}
\newcommand{\lusearchFTOHBTwoNestedLocks}{0.29}
\newcommand{\lusearchFTOHBThreeNestedLocks}{<0.1}
\newcommand{\lusearchFTOHBFourNestedLocks}{\cna}
\newcommand{\lusearchFTOHBFiveNestedLocks}{\cna}
\newcommand{\lusearchFTOHBSixNestedLocks}{\cna}
\newcommand{\lusearchFTOHBSevenNestedLocks}{\cna}
\newcommand{\lusearchFTOHBEightNestedLocks}{\cna}
\newcommand{\lusearchFTOHBNineNestedLocks}{\cna}
\newcommand{\lusearchFTOHBTenNestedLocks}{\cna}
\newcommand{\lusearchFTOHBHundredNestedLocks}{\cna}
\newcommand{\lusearchFTOHBExWrSet}{\ena}
\newcommand{\lusearchFTOHBExWrCheck}{\ena}
\newcommand{\lusearchFTOHBExWrUpdate}{\ena}
\newcommand{\lusearchFTOHBExRdCheck}{\ena}
\newcommand{\lusearchFTOHBExRdUpdate}{\ena}
\newcommand{\lusearchFTOHBExTotalCheck}{\ena}
\newcommand{\lusearchFTOHBExTotalUpdate}{\ena}
\newcommand{\lusearchFTOWCPEventTotal}{1,400}
\newcommand{\lusearchFTOWCPNoFPEventTotal}{140}
\newcommand{\lusearchFTOWCPNoFPAccessTotal}{130}
\newcommand{\lusearchFTOWCPNoFPOtherTotal}{3.8}
\newcommand{\lusearchFTOWCPReadTotal}{77.0}
\newcommand{\lusearchFTOWCPWriteTotal}{20.2}
\newcommand{\lusearchFTOWCPNoFPAccessInCS}{4.89}
\newcommand{\lusearchFTOWCPNoFPAccessOutCS}{95.1}
\newcommand{\lusearchFTOWCPAcqRelTotal}{6.23}
\newcommand{\lusearchFTOWCPOtherTotal}{2.72}
\newcommand{\lusearchFTOWCPNoFPReadTotal}{110}
\newcommand{\lusearchFTOWCPReadInCS}{3.9}
\newcommand{\lusearchFTOWCPReadOutCS}{104}
\newcommand{\lusearchFTOWCPReadSameEp}{7.45}
\newcommand{\lusearchFTOWCPReadSharedSameEp}{\cna}
\newcommand{\lusearchFTOWCPReadExclusive}{<0.001}
\newcommand{\lusearchFTOWCPReadOwned}{96.1}
\newcommand{\lusearchFTOWCPReadShare}{<0.001}
\newcommand{\lusearchFTOWCPReadShared}{0.0011}
\newcommand{\lusearchFTOWCPReadSharedOwned}{3.9}
\newcommand{\lusearchFTOWCPNoFPHonestWriteTotal}{28}
\newcommand{\lusearchFTOWCPWriteInCS}{10}
\newcommand{\lusearchFTOWCPWriteOutCS}{200}
\newcommand{\lusearchFTOWCPNoFPWriteTotal}{28}
\newcommand{\lusearchFTOWCPWriteSameEp}{110}
\newcommand{\lusearchFTOWCPWriteExclusive}{\cna}
\newcommand{\lusearchFTOWCPWriteOwned}{100}
\newcommand{\lusearchFTOWCPWriteShared}{<0.001}
\newcommand{\lusearchFTOWCPNoFPOtherEventTotal}{3816274}
\newcommand{\lusearchFTOWCPAcqRelOtherTotal}{69.6}
\newcommand{\lusearchFTOWCPNoAcqRelOtherTotal}{1158338}
\newcommand{\lusearchFTOWCPFork}{0.00121}
\newcommand{\lusearchFTOWCPJoin}{0.0}
\newcommand{\lusearchFTOWCPPreWait}{0.00527}
\newcommand{\lusearchFTOWCPPostWait}{0.00527}
\newcommand{\lusearchFTOWCPVolatileTotal}{100.0}
\newcommand{\lusearchFTOWCPClassInit}{0.00311}
\newcommand{\lusearchFTOWCPClassAccess}{0.0293}
\newcommand{\lusearchFTOWCPRaceTotal}{\rna}
\newcommand{\lusearchFTOWCPWrRdRace}{\rna}
\newcommand{\lusearchFTOWCPWrWrRace}{\rna}
\newcommand{\lusearchFTOWCPRdWrRace}{\rna}
\newcommand{\lusearchFTOWCPRdShWrRace}{\rna}
\newcommand{\lusearchFTOWCPHoldLocksTotal}{5.1}
\newcommand{\lusearchFTOWCPOneLockHeld}{3.8}
\newcommand{\lusearchFTOWCPTwoNestedLocks}{0.39}
\newcommand{\lusearchFTOWCPThreeNestedLocks}{<0.1}
\newcommand{\lusearchFTOWCPFourNestedLocks}{\cna}
\newcommand{\lusearchFTOWCPFiveNestedLocks}{\cna}
\newcommand{\lusearchFTOWCPSixNestedLocks}{\cna}
\newcommand{\lusearchFTOWCPSevenNestedLocks}{\cna}
\newcommand{\lusearchFTOWCPEightNestedLocks}{\cna}
\newcommand{\lusearchFTOWCPNineNestedLocks}{\cna}
\newcommand{\lusearchFTOWCPTenNestedLocks}{\cna}
\newcommand{\lusearchFTOWCPHundredNestedLocks}{\cna}
\newcommand{\lusearchFTOWCPExWrSet}{\ena}
\newcommand{\lusearchFTOWCPExWrCheck}{\ena}
\newcommand{\lusearchFTOWCPExWrUpdate}{\ena}
\newcommand{\lusearchFTOWCPExRdCheck}{\ena}
\newcommand{\lusearchFTOWCPExRdUpdate}{\ena}
\newcommand{\lusearchFTOWCPExTotalCheck}{\ena}
\newcommand{\lusearchFTOWCPExTotalUpdate}{\ena}
\newcommand{\lusearchREWCPEventTotal}{1,400}
\newcommand{\lusearchREWCPNoFPEventTotal}{140}
\newcommand{\lusearchREWCPNoFPAccessTotal}{130}
\newcommand{\lusearchREWCPNoFPOtherTotal}{3.8}
\newcommand{\lusearchREWCPReadTotal}{77.0}
\newcommand{\lusearchREWCPWriteTotal}{20.2}
\newcommand{\lusearchREWCPNoFPAccessInCS}{4.89}
\newcommand{\lusearchREWCPNoFPAccessOutCS}{95.1}
\newcommand{\lusearchREWCPAcqRelTotal}{6.23}
\newcommand{\lusearchREWCPOtherTotal}{2.72}
\newcommand{\lusearchREWCPNoFPReadTotal}{110}
\newcommand{\lusearchREWCPReadInCS}{3.9}
\newcommand{\lusearchREWCPReadOutCS}{104}
\newcommand{\lusearchREWCPReadSameEp}{7.45}
\newcommand{\lusearchREWCPReadSharedSameEp}{\cna}
\newcommand{\lusearchREWCPReadExclusive}{<0.001}
\newcommand{\lusearchREWCPReadOwned}{96.1}
\newcommand{\lusearchREWCPReadShare}{<0.001}
\newcommand{\lusearchREWCPReadShared}{0.0011}
\newcommand{\lusearchREWCPReadSharedOwned}{3.9}
\newcommand{\lusearchREWCPNoFPHonestWriteTotal}{28}
\newcommand{\lusearchREWCPWriteInCS}{10}
\newcommand{\lusearchREWCPWriteOutCS}{200}
\newcommand{\lusearchREWCPNoFPWriteTotal}{28}
\newcommand{\lusearchREWCPWriteSameEp}{110}
\newcommand{\lusearchREWCPWriteExclusive}{\cna}
\newcommand{\lusearchREWCPWriteOwned}{100}
\newcommand{\lusearchREWCPWriteShared}{<0.001}
\newcommand{\lusearchREWCPNoFPOtherEventTotal}{3816273}
\newcommand{\lusearchREWCPAcqRelOtherTotal}{69.6}
\newcommand{\lusearchREWCPNoAcqRelOtherTotal}{1158337}
\newcommand{\lusearchREWCPFork}{0.00121}
\newcommand{\lusearchREWCPJoin}{0.0}
\newcommand{\lusearchREWCPPreWait}{0.00518}
\newcommand{\lusearchREWCPPostWait}{0.00518}
\newcommand{\lusearchREWCPVolatileTotal}{100.0}
\newcommand{\lusearchREWCPClassInit}{0.00311}
\newcommand{\lusearchREWCPClassAccess}{0.0293}
\newcommand{\lusearchREWCPRaceTotal}{\rna}
\newcommand{\lusearchREWCPWrRdRace}{\rna}
\newcommand{\lusearchREWCPWrWrRace}{\rna}
\newcommand{\lusearchREWCPRdWrRace}{\rna}
\newcommand{\lusearchREWCPRdShWrRace}{\rna}
\newcommand{\lusearchREWCPHoldLocksTotal}{5.1}
\newcommand{\lusearchREWCPOneLockHeld}{3.79}
\newcommand{\lusearchREWCPTwoNestedLocks}{0.39}
\newcommand{\lusearchREWCPThreeNestedLocks}{<0.1}
\newcommand{\lusearchREWCPFourNestedLocks}{\cna}
\newcommand{\lusearchREWCPFiveNestedLocks}{\cna}
\newcommand{\lusearchREWCPSixNestedLocks}{\cna}
\newcommand{\lusearchREWCPSevenNestedLocks}{\cna}
\newcommand{\lusearchREWCPEightNestedLocks}{\cna}
\newcommand{\lusearchREWCPNineNestedLocks}{\cna}
\newcommand{\lusearchREWCPTenNestedLocks}{\cna}
\newcommand{\lusearchREWCPHundredNestedLocks}{\cna}
\newcommand{\lusearchREWCPExWrSet}{15}
\newcommand{\lusearchREWCPExWrCheck}{27}
\newcommand{\lusearchREWCPExWrUpdate}{\ena}
\newcommand{\lusearchREWCPExRdCheck}{27}
\newcommand{\lusearchREWCPExRdUpdate}{\ena}
\newcommand{\lusearchREWCPExTotalCheck}{54}
\newcommand{\lusearchREWCPExTotalUpdate}{\ena}
\newcommand{\lusearchFTODCEventTotal}{1,400}
\newcommand{\lusearchFTODCNoFPEventTotal}{140}
\newcommand{\lusearchFTODCNoFPAccessTotal}{130}
\newcommand{\lusearchFTODCNoFPOtherTotal}{3.8}
\newcommand{\lusearchFTODCReadTotal}{77.0}
\newcommand{\lusearchFTODCWriteTotal}{20.2}
\newcommand{\lusearchFTODCNoFPAccessInCS}{4.89}
\newcommand{\lusearchFTODCNoFPAccessOutCS}{95.1}
\newcommand{\lusearchFTODCAcqRelTotal}{6.23}
\newcommand{\lusearchFTODCOtherTotal}{2.72}
\newcommand{\lusearchFTODCNoFPReadTotal}{110}
\newcommand{\lusearchFTODCReadInCS}{3.9}
\newcommand{\lusearchFTODCReadOutCS}{104}
\newcommand{\lusearchFTODCReadSameEp}{7.45}
\newcommand{\lusearchFTODCReadSharedSameEp}{\cna}
\newcommand{\lusearchFTODCReadExclusive}{<0.001}
\newcommand{\lusearchFTODCReadOwned}{96.1}
\newcommand{\lusearchFTODCReadShare}{<0.001}
\newcommand{\lusearchFTODCReadShared}{0.0011}
\newcommand{\lusearchFTODCReadSharedOwned}{3.9}
\newcommand{\lusearchFTODCNoFPHonestWriteTotal}{28}
\newcommand{\lusearchFTODCWriteInCS}{10}
\newcommand{\lusearchFTODCWriteOutCS}{200}
\newcommand{\lusearchFTODCNoFPWriteTotal}{28}
\newcommand{\lusearchFTODCWriteSameEp}{110}
\newcommand{\lusearchFTODCWriteExclusive}{\cna}
\newcommand{\lusearchFTODCWriteOwned}{100}
\newcommand{\lusearchFTODCWriteShared}{<0.001}
\newcommand{\lusearchFTODCNoFPOtherEventTotal}{3816253}
\newcommand{\lusearchFTODCAcqRelOtherTotal}{69.6}
\newcommand{\lusearchFTODCNoAcqRelOtherTotal}{1158317}
\newcommand{\lusearchFTODCFork}{0.00121}
\newcommand{\lusearchFTODCJoin}{0.0}
\newcommand{\lusearchFTODCPreWait}{0.00535}
\newcommand{\lusearchFTODCPostWait}{0.00535}
\newcommand{\lusearchFTODCVolatileTotal}{100.0}
\newcommand{\lusearchFTODCClassInit}{0.00311}
\newcommand{\lusearchFTODCClassAccess}{0.0293}
\newcommand{\lusearchFTODCRaceTotal}{\rna}
\newcommand{\lusearchFTODCWrRdRace}{\rna}
\newcommand{\lusearchFTODCWrWrRace}{\rna}
\newcommand{\lusearchFTODCRdWrRace}{\rna}
\newcommand{\lusearchFTODCRdShWrRace}{\rna}
\newcommand{\lusearchFTODCHoldLocksTotal}{5.1}
\newcommand{\lusearchFTODCOneLockHeld}{3.79}
\newcommand{\lusearchFTODCTwoNestedLocks}{0.39}
\newcommand{\lusearchFTODCThreeNestedLocks}{<0.1}
\newcommand{\lusearchFTODCFourNestedLocks}{\cna}
\newcommand{\lusearchFTODCFiveNestedLocks}{\cna}
\newcommand{\lusearchFTODCSixNestedLocks}{\cna}
\newcommand{\lusearchFTODCSevenNestedLocks}{\cna}
\newcommand{\lusearchFTODCEightNestedLocks}{\cna}
\newcommand{\lusearchFTODCNineNestedLocks}{\cna}
\newcommand{\lusearchFTODCTenNestedLocks}{\cna}
\newcommand{\lusearchFTODCHundredNestedLocks}{\cna}
\newcommand{\lusearchFTODCExWrSet}{\ena}
\newcommand{\lusearchFTODCExWrCheck}{\ena}
\newcommand{\lusearchFTODCExWrUpdate}{\ena}
\newcommand{\lusearchFTODCExRdCheck}{\ena}
\newcommand{\lusearchFTODCExRdUpdate}{\ena}
\newcommand{\lusearchFTODCExTotalCheck}{\ena}
\newcommand{\lusearchFTODCExTotalUpdate}{\ena}
\newcommand{\lusearchREDCEventTotal}{1,400}
\newcommand{\lusearchREDCNoFPEventTotal}{140}
\newcommand{\lusearchREDCNoFPAccessTotal}{130}
\newcommand{\lusearchREDCNoFPOtherTotal}{3.8}
\newcommand{\lusearchREDCReadTotal}{77.0}
\newcommand{\lusearchREDCWriteTotal}{20.2}
\newcommand{\lusearchREDCNoFPAccessInCS}{4.89}
\newcommand{\lusearchREDCNoFPAccessOutCS}{95.1}
\newcommand{\lusearchREDCAcqRelTotal}{6.23}
\newcommand{\lusearchREDCOtherTotal}{2.72}
\newcommand{\lusearchREDCNoFPReadTotal}{110}
\newcommand{\lusearchREDCReadInCS}{3.9}
\newcommand{\lusearchREDCReadOutCS}{104}
\newcommand{\lusearchREDCReadSameEp}{7.45}
\newcommand{\lusearchREDCReadSharedSameEp}{\cna}
\newcommand{\lusearchREDCReadExclusive}{<0.001}
\newcommand{\lusearchREDCReadOwned}{96.1}
\newcommand{\lusearchREDCReadShare}{<0.001}
\newcommand{\lusearchREDCReadShared}{0.0011}
\newcommand{\lusearchREDCReadSharedOwned}{3.9}
\newcommand{\lusearchREDCNoFPHonestWriteTotal}{28}
\newcommand{\lusearchREDCWriteInCS}{10}
\newcommand{\lusearchREDCWriteOutCS}{200}
\newcommand{\lusearchREDCNoFPWriteTotal}{28}
\newcommand{\lusearchREDCWriteSameEp}{110}
\newcommand{\lusearchREDCWriteExclusive}{\cna}
\newcommand{\lusearchREDCWriteOwned}{100}
\newcommand{\lusearchREDCWriteShared}{<0.001}
\newcommand{\lusearchREDCNoFPOtherEventTotal}{3816258}
\newcommand{\lusearchREDCAcqRelOtherTotal}{69.6}
\newcommand{\lusearchREDCNoAcqRelOtherTotal}{1158322}
\newcommand{\lusearchREDCFork}{0.00121}
\newcommand{\lusearchREDCJoin}{0.0}
\newcommand{\lusearchREDCPreWait}{0.00535}
\newcommand{\lusearchREDCPostWait}{0.00535}
\newcommand{\lusearchREDCVolatileTotal}{100.0}
\newcommand{\lusearchREDCClassInit}{0.00311}
\newcommand{\lusearchREDCClassAccess}{0.0293}
\newcommand{\lusearchREDCRaceTotal}{\rna}
\newcommand{\lusearchREDCWrRdRace}{\rna}
\newcommand{\lusearchREDCWrWrRace}{\rna}
\newcommand{\lusearchREDCRdWrRace}{\rna}
\newcommand{\lusearchREDCRdShWrRace}{\rna}
\newcommand{\lusearchREDCHoldLocksTotal}{5.1}
\newcommand{\lusearchREDCOneLockHeld}{3.79}
\newcommand{\lusearchREDCTwoNestedLocks}{0.39}
\newcommand{\lusearchREDCThreeNestedLocks}{<0.1}
\newcommand{\lusearchREDCFourNestedLocks}{\cna}
\newcommand{\lusearchREDCFiveNestedLocks}{\cna}
\newcommand{\lusearchREDCSixNestedLocks}{\cna}
\newcommand{\lusearchREDCSevenNestedLocks}{\cna}
\newcommand{\lusearchREDCEightNestedLocks}{\cna}
\newcommand{\lusearchREDCNineNestedLocks}{\cna}
\newcommand{\lusearchREDCTenNestedLocks}{\cna}
\newcommand{\lusearchREDCHundredNestedLocks}{\cna}
\newcommand{\lusearchREDCExWrSet}{16}
\newcommand{\lusearchREDCExWrCheck}{26}
\newcommand{\lusearchREDCExWrUpdate}{\ena}
\newcommand{\lusearchREDCExRdCheck}{26}
\newcommand{\lusearchREDCExRdUpdate}{\ena}
\newcommand{\lusearchREDCExTotalCheck}{52}
\newcommand{\lusearchREDCExTotalUpdate}{\ena}
\newcommand{\lusearchFTOCAPOEventTotal}{1,400}
\newcommand{\lusearchFTOCAPONoFPEventTotal}{140}
\newcommand{\lusearchFTOCAPONoFPAccessTotal}{130}
\newcommand{\lusearchFTOCAPONoFPOtherTotal}{3.8}
\newcommand{\lusearchFTOCAPOReadTotal}{77.0}
\newcommand{\lusearchFTOCAPOWriteTotal}{20.2}
\newcommand{\lusearchFTOCAPONoFPAccessInCS}{4.89}
\newcommand{\lusearchFTOCAPONoFPAccessOutCS}{95.1}
\newcommand{\lusearchFTOCAPOAcqRelTotal}{6.23}
\newcommand{\lusearchFTOCAPOOtherTotal}{2.72}
\newcommand{\lusearchFTOCAPONoFPReadTotal}{110}
\newcommand{\lusearchFTOCAPOReadInCS}{3.9}
\newcommand{\lusearchFTOCAPOReadOutCS}{104}
\newcommand{\lusearchFTOCAPOReadSameEp}{7.45}
\newcommand{\lusearchFTOCAPOReadSharedSameEp}{\cna}
\newcommand{\lusearchFTOCAPOReadExclusive}{<0.001}
\newcommand{\lusearchFTOCAPOReadOwned}{96.1}
\newcommand{\lusearchFTOCAPOReadShare}{<0.001}
\newcommand{\lusearchFTOCAPOReadShared}{0.0011}
\newcommand{\lusearchFTOCAPOReadSharedOwned}{3.9}
\newcommand{\lusearchFTOCAPONoFPHonestWriteTotal}{28}
\newcommand{\lusearchFTOCAPOWriteInCS}{10}
\newcommand{\lusearchFTOCAPOWriteOutCS}{200}
\newcommand{\lusearchFTOCAPONoFPWriteTotal}{28}
\newcommand{\lusearchFTOCAPOWriteSameEp}{110}
\newcommand{\lusearchFTOCAPOWriteExclusive}{\cna}
\newcommand{\lusearchFTOCAPOWriteOwned}{100}
\newcommand{\lusearchFTOCAPOWriteShared}{<0.001}
\newcommand{\lusearchFTOCAPONoFPOtherEventTotal}{3816284}
\newcommand{\lusearchFTOCAPOAcqRelOtherTotal}{69.6}
\newcommand{\lusearchFTOCAPONoAcqRelOtherTotal}{1158348}
\newcommand{\lusearchFTOCAPOFork}{0.00121}
\newcommand{\lusearchFTOCAPOJoin}{0.0}
\newcommand{\lusearchFTOCAPOPreWait}{0.00518}
\newcommand{\lusearchFTOCAPOPostWait}{0.00518}
\newcommand{\lusearchFTOCAPOVolatileTotal}{100.0}
\newcommand{\lusearchFTOCAPOClassInit}{0.00311}
\newcommand{\lusearchFTOCAPOClassAccess}{0.0293}
\newcommand{\lusearchFTOCAPORaceTotal}{\rna}
\newcommand{\lusearchFTOCAPOWrRdRace}{\rna}
\newcommand{\lusearchFTOCAPOWrWrRace}{\rna}
\newcommand{\lusearchFTOCAPORdWrRace}{\rna}
\newcommand{\lusearchFTOCAPORdShWrRace}{\rna}
\newcommand{\lusearchFTOCAPOHoldLocksTotal}{5.1}
\newcommand{\lusearchFTOCAPOOneLockHeld}{3.79}
\newcommand{\lusearchFTOCAPOTwoNestedLocks}{0.39}
\newcommand{\lusearchFTOCAPOThreeNestedLocks}{<0.1}
\newcommand{\lusearchFTOCAPOFourNestedLocks}{\cna}
\newcommand{\lusearchFTOCAPOFiveNestedLocks}{\cna}
\newcommand{\lusearchFTOCAPOSixNestedLocks}{\cna}
\newcommand{\lusearchFTOCAPOSevenNestedLocks}{\cna}
\newcommand{\lusearchFTOCAPOEightNestedLocks}{\cna}
\newcommand{\lusearchFTOCAPONineNestedLocks}{\cna}
\newcommand{\lusearchFTOCAPOTenNestedLocks}{\cna}
\newcommand{\lusearchFTOCAPOHundredNestedLocks}{\cna}
\newcommand{\lusearchFTOCAPOExWrSet}{\ena}
\newcommand{\lusearchFTOCAPOExWrCheck}{\ena}
\newcommand{\lusearchFTOCAPOExWrUpdate}{\ena}
\newcommand{\lusearchFTOCAPOExRdCheck}{\ena}
\newcommand{\lusearchFTOCAPOExRdUpdate}{\ena}
\newcommand{\lusearchFTOCAPOExTotalCheck}{\ena}
\newcommand{\lusearchFTOCAPOExTotalUpdate}{\ena}
\newcommand{\lusearchRECAPOEventTotal}{1,400}
\newcommand{\lusearchRECAPONoFPEventTotal}{140}
\newcommand{\lusearchRECAPONoFPAccessTotal}{130}
\newcommand{\lusearchRECAPONoFPOtherTotal}{3.8}
\newcommand{\lusearchRECAPOReadTotal}{77.0}
\newcommand{\lusearchRECAPOWriteTotal}{20.2}
\newcommand{\lusearchRECAPONoFPAccessInCS}{4.89}
\newcommand{\lusearchRECAPONoFPAccessOutCS}{95.1}
\newcommand{\lusearchRECAPOAcqRelTotal}{6.23}
\newcommand{\lusearchRECAPOOtherTotal}{2.72}
\newcommand{\lusearchRECAPONoFPReadTotal}{110}
\newcommand{\lusearchRECAPOReadInCS}{3.9}
\newcommand{\lusearchRECAPOReadOutCS}{104}
\newcommand{\lusearchRECAPOReadSameEp}{7.45}
\newcommand{\lusearchRECAPOReadSharedSameEp}{\cna}
\newcommand{\lusearchRECAPOReadExclusive}{<0.001}
\newcommand{\lusearchRECAPOReadOwned}{96.1}
\newcommand{\lusearchRECAPOReadShare}{<0.001}
\newcommand{\lusearchRECAPOReadShared}{0.0011}
\newcommand{\lusearchRECAPOReadSharedOwned}{3.9}
\newcommand{\lusearchRECAPONoFPHonestWriteTotal}{28}
\newcommand{\lusearchRECAPOWriteInCS}{10}
\newcommand{\lusearchRECAPOWriteOutCS}{200}
\newcommand{\lusearchRECAPONoFPWriteTotal}{28}
\newcommand{\lusearchRECAPOWriteSameEp}{110}
\newcommand{\lusearchRECAPOWriteExclusive}{\cna}
\newcommand{\lusearchRECAPOWriteOwned}{100}
\newcommand{\lusearchRECAPOWriteShared}{<0.001}
\newcommand{\lusearchRECAPONoFPOtherEventTotal}{3816281}
\newcommand{\lusearchRECAPOAcqRelOtherTotal}{69.6}
\newcommand{\lusearchRECAPONoAcqRelOtherTotal}{1158345}
\newcommand{\lusearchRECAPOFork}{0.00121}
\newcommand{\lusearchRECAPOJoin}{0.0}
\newcommand{\lusearchRECAPOPreWait}{0.00509}
\newcommand{\lusearchRECAPOPostWait}{0.00509}
\newcommand{\lusearchRECAPOVolatileTotal}{100.0}
\newcommand{\lusearchRECAPOClassInit}{0.00311}
\newcommand{\lusearchRECAPOClassAccess}{0.0293}
\newcommand{\lusearchRECAPORaceTotal}{\rna}
\newcommand{\lusearchRECAPOWrRdRace}{\rna}
\newcommand{\lusearchRECAPOWrWrRace}{\rna}
\newcommand{\lusearchRECAPORdWrRace}{\rna}
\newcommand{\lusearchRECAPORdShWrRace}{\rna}
\newcommand{\lusearchRECAPOHoldLocksTotal}{5.1}
\newcommand{\lusearchRECAPOOneLockHeld}{3.8}
\newcommand{\lusearchRECAPOTwoNestedLocks}{0.39}
\newcommand{\lusearchRECAPOThreeNestedLocks}{<0.1}
\newcommand{\lusearchRECAPOFourNestedLocks}{\cna}
\newcommand{\lusearchRECAPOFiveNestedLocks}{\cna}
\newcommand{\lusearchRECAPOSixNestedLocks}{\cna}
\newcommand{\lusearchRECAPOSevenNestedLocks}{\cna}
\newcommand{\lusearchRECAPOEightNestedLocks}{\cna}
\newcommand{\lusearchRECAPONineNestedLocks}{\cna}
\newcommand{\lusearchRECAPOTenNestedLocks}{\cna}
\newcommand{\lusearchRECAPOHundredNestedLocks}{\cna}
\newcommand{\lusearchRECAPOExWrSet}{14}
\newcommand{\lusearchRECAPOExWrCheck}{26}
\newcommand{\lusearchRECAPOExWrUpdate}{\ena}
\newcommand{\lusearchRECAPOExRdCheck}{26}
\newcommand{\lusearchRECAPOExRdUpdate}{\ena}
\newcommand{\lusearchRECAPOExTotalCheck}{<0.001}
\newcommand{\lusearchRECAPOExTotalUpdate}{\ena}
\newcommand{\FASTlusearchMaxLiveThreads}{15}
\newcommand{\FASTlusearchTotalThreads}{16}
\newcommand{\FASTlusearchBaseTime}{1.0}
\newcommand{\FASTlusearchBaseTimeCI}{110}
\newcommand{\FASTlusearchEmptyTime}{\rna}
\newcommand{\FASTlusearchEmptyTimeCI}{\rna}
\newcommand{\FASTlusearchEmptyTimeCIMIN}{\rna}
\newcommand{\FASTlusearchEmptyTimeCIMAX}{\rna}
\newcommand{\FASTlusearchFTTime}{11}
\newcommand{\FASTlusearchFTTimeCI}{1.4}
\newcommand{\FASTlusearchHBTime}{11}
\newcommand{\FASTlusearchHBTimeCI}{1.2}
\newcommand{\FASTlusearchFTOHBTime}{12}
\newcommand{\FASTlusearchFTOHBTimeCI}{1.7}
\newcommand{\FASTlusearchWCPTime}{\rna}
\newcommand{\FASTlusearchWCPTimeCI}{\rna}
\newcommand{\FASTlusearchWCPTimeCIMIN}{\rna}
\newcommand{\FASTlusearchWCPTimeCIMAX}{\rna}
\newcommand{\FASTlusearchFTOWCPTime}{14}
\newcommand{\FASTlusearchFTOWCPTimeCI}{1.9}
\newcommand{\FASTlusearchREWCPTime}{13}
\newcommand{\FASTlusearchREWCPTimeCI}{1.4}
\newcommand{\FASTlusearchDCTime}{\rna}
\newcommand{\FASTlusearchDCTimeCI}{\rna}
\newcommand{\FASTlusearchDCTimeCIMIN}{\rna}
\newcommand{\FASTlusearchDCTimeCIMAX}{\rna}
\newcommand{\FASTlusearchFTODCTime}{13}
\newcommand{\FASTlusearchFTODCTimeCI}{1.9}
\newcommand{\FASTlusearchREDCTime}{13}
\newcommand{\FASTlusearchREDCTimeCI}{1.6}
\newcommand{\FASTlusearchCAPOTime}{\rna}
\newcommand{\FASTlusearchCAPOTimeCI}{\rna}
\newcommand{\FASTlusearchCAPOTimeCIMIN}{\rna}
\newcommand{\FASTlusearchCAPOTimeCIMAX}{\rna}
\newcommand{\FASTlusearchFTOCAPOTime}{12}
\newcommand{\FASTlusearchFTOCAPOTimeCI}{1.6}
\newcommand{\FASTlusearchRECAPOTime}{13}
\newcommand{\FASTlusearchRECAPOTimeCI}{2.3}
\newcommand{\FASTlusearchAGGCAPOTime}{\rna}
\newcommand{\FASTlusearchAGGCAPOTimeCI}{\rna}
\newcommand{\FASTlusearchAGGCAPOTimeCIMIN}{\rna}
\newcommand{\FASTlusearchAGGCAPOTimeCIMAX}{\rna}
\newcommand{\FASTlusearchStaticTime}{\rzero}
\newcommand{\FASTlusearchDynamicTime}{\rzero}
\newcommand{\FASTlusearchBaseMem}{1,500}
\newcommand{\FASTlusearchBaseMemCI}{120.0}
\newcommand{\FASTlusearchFTMem}{9.7}
\newcommand{\FASTlusearchFTMemCI}{0.79}
\newcommand{\FASTlusearchHBMem}{9.7}
\newcommand{\FASTlusearchHBMemCI}{1.0}
\newcommand{\FASTlusearchFTOHBMem}{10}
\newcommand{\FASTlusearchFTOHBMemCI}{1.0}
\newcommand{\FASTlusearchWCPMem}{\memna}
\newcommand{\FASTlusearchWCPMemCI}{\memna}
\newcommand{\FASTlusearchWCPMemCIMIN}{\memna}
\newcommand{\FASTlusearchWCPMemCIMAX}{\memna}
\newcommand{\FASTlusearchFTOWCPMem}{11}
\newcommand{\FASTlusearchFTOWCPMemCI}{0.9}
\newcommand{\FASTlusearchREWCPMem}{9.7}
\newcommand{\FASTlusearchREWCPMemCI}{1.0}
\newcommand{\FASTlusearchDCMem}{\memna}
\newcommand{\FASTlusearchDCMemCI}{\memna}
\newcommand{\FASTlusearchDCMemCIMIN}{\memna}
\newcommand{\FASTlusearchDCMemCIMAX}{\memna}
\newcommand{\FASTlusearchFTODCMem}{10}
\newcommand{\FASTlusearchFTODCMemCI}{1.0}
\newcommand{\FASTlusearchREDCMem}{10}
\newcommand{\FASTlusearchREDCMemCI}{0.91}
\newcommand{\FASTlusearchCAPOMem}{\memna}
\newcommand{\FASTlusearchCAPOMemCI}{\memna}
\newcommand{\FASTlusearchCAPOMemCIMIN}{\memna}
\newcommand{\FASTlusearchCAPOMemCIMAX}{\memna}
\newcommand{\FASTlusearchFTOCAPOMem}{10}
\newcommand{\FASTlusearchFTOCAPOMemCI}{1.1}
\newcommand{\FASTlusearchRECAPOMem}{10}
\newcommand{\FASTlusearchRECAPOMemCI}{0.93}
\newcommand{\FASTlusearchAGGCAPOMem}{\memna}
\newcommand{\FASTlusearchAGGCAPOMemCI}{\memna}
\newcommand{\FASTlusearchAGGCAPOMemCIMIN}{\memna}
\newcommand{\FASTlusearchAGGCAPOMemCIMAX}{\memna}
\newcommand{\FASTlusearchEventsCI}{41}
\newcommand{\FASTlusearchEventsCIMIN}{1,443,227,375}
\newcommand{\FASTlusearchEventsCIMAX}{1,443,227,457}
\newcommand{\FASTlusearchNoFPEventsCI}{41}
\newcommand{\FASTlusearchNoFPEventsCIMIN}{42,639,863}
\newcommand{\FASTlusearchNoFPEventsCIMAX}{42,639,945}
\newcommand{\FASTlusearchFT}{0}
\newcommand{\FASTlusearchFTCI}{0.0}
\newcommand{\FASTlusearchFTCIMIN}{0}
\newcommand{\FASTlusearchFTCIMAX}{0}
\newcommand{\FASTlusearchFTDynamic}{0}
\newcommand{\FASTlusearchFTDynamicCI}{0.0}
\newcommand{\FASTlusearchFTDynamicCIMIN}{0}
\newcommand{\FASTlusearchFTDynamicCIMAX}{0}
\newcommand{\FASTlusearchHB}{0}
\newcommand{\FASTlusearchHBCI}{0.0}
\newcommand{\FASTlusearchHBCIMIN}{0}
\newcommand{\FASTlusearchHBCIMAX}{0}
\newcommand{\FASTlusearchHBDynamic}{0}
\newcommand{\FASTlusearchHBDynamicCI}{0.0}
\newcommand{\FASTlusearchHBDynamicCIMIN}{0}
\newcommand{\FASTlusearchHBDynamicCIMAX}{0}
\newcommand{\FASTlusearchFTOHB}{0}
\newcommand{\FASTlusearchFTOHBCI}{0.0}
\newcommand{\FASTlusearchFTOHBCIMIN}{0}
\newcommand{\FASTlusearchFTOHBCIMAX}{0}
\newcommand{\FASTlusearchFTOHBDynamic}{0}
\newcommand{\FASTlusearchFTOHBDynamicCI}{0.0}
\newcommand{\FASTlusearchFTOHBDynamicCIMIN}{0}
\newcommand{\FASTlusearchFTOHBDynamicCIMAX}{0}
\newcommand{\FASTlusearchWCP}{\rna}
\newcommand{\FASTlusearchWCPCI}{\rna}
\newcommand{\FASTlusearchWCPCIMIN}{\rna}
\newcommand{\FASTlusearchWCPCIMAX}{\rna}
\newcommand{\FASTlusearchWCPDynamic}{\rna}
\newcommand{\FASTlusearchWCPDynamicCI}{\rna}
\newcommand{\FASTlusearchWCPDynamicCIMIN}{\rna}
\newcommand{\FASTlusearchWCPDynamicCIMAX}{\rna}
\newcommand{\FASTlusearchFTOWCP}{0}
\newcommand{\FASTlusearchFTOWCPCI}{0.0}
\newcommand{\FASTlusearchFTOWCPCIMIN}{0}
\newcommand{\FASTlusearchFTOWCPCIMAX}{0}
\newcommand{\FASTlusearchFTOWCPDynamic}{0}
\newcommand{\FASTlusearchFTOWCPDynamicCI}{0.0}
\newcommand{\FASTlusearchFTOWCPDynamicCIMIN}{0}
\newcommand{\FASTlusearchFTOWCPDynamicCIMAX}{0}
\newcommand{\FASTlusearchREWCP}{0}
\newcommand{\FASTlusearchREWCPCI}{0.0}
\newcommand{\FASTlusearchREWCPCIMIN}{0}
\newcommand{\FASTlusearchREWCPCIMAX}{0}
\newcommand{\FASTlusearchREWCPDynamic}{0}
\newcommand{\FASTlusearchREWCPDynamicCI}{0.0}
\newcommand{\FASTlusearchREWCPDynamicCIMIN}{0}
\newcommand{\FASTlusearchREWCPDynamicCIMAX}{0}
\newcommand{\FASTlusearchDC}{\rna}
\newcommand{\FASTlusearchDCCI}{\rna}
\newcommand{\FASTlusearchDCCIMIN}{\rna}
\newcommand{\FASTlusearchDCCIMAX}{\rna}
\newcommand{\FASTlusearchDCDynamic}{\rna}
\newcommand{\FASTlusearchDCDynamicCI}{\rna}
\newcommand{\FASTlusearchDCDynamicCIMIN}{\rna}
\newcommand{\FASTlusearchDCDynamicCIMAX}{\rna}
\newcommand{\FASTlusearchFTODC}{0}
\newcommand{\FASTlusearchFTODCCI}{0.0}
\newcommand{\FASTlusearchFTODCCIMIN}{0}
\newcommand{\FASTlusearchFTODCCIMAX}{0}
\newcommand{\FASTlusearchFTODCDynamic}{0}
\newcommand{\FASTlusearchFTODCDynamicCI}{0.0}
\newcommand{\FASTlusearchFTODCDynamicCIMIN}{0}
\newcommand{\FASTlusearchFTODCDynamicCIMAX}{0}
\newcommand{\FASTlusearchREDC}{0}
\newcommand{\FASTlusearchREDCCI}{0.0}
\newcommand{\FASTlusearchREDCCIMIN}{0}
\newcommand{\FASTlusearchREDCCIMAX}{0}
\newcommand{\FASTlusearchREDCDynamic}{0}
\newcommand{\FASTlusearchREDCDynamicCI}{0.0}
\newcommand{\FASTlusearchREDCDynamicCIMIN}{0}
\newcommand{\FASTlusearchREDCDynamicCIMAX}{0}
\newcommand{\FASTlusearchCAPO}{\rna}
\newcommand{\FASTlusearchCAPOCI}{\rna}
\newcommand{\FASTlusearchCAPOCIMIN}{\rna}
\newcommand{\FASTlusearchCAPOCIMAX}{\rna}
\newcommand{\FASTlusearchCAPODynamic}{\rna}
\newcommand{\FASTlusearchCAPODynamicCI}{\rna}
\newcommand{\FASTlusearchCAPODynamicCIMIN}{\rna}
\newcommand{\FASTlusearchCAPODynamicCIMAX}{\rna}
\newcommand{\FASTlusearchFTOCAPO}{0}
\newcommand{\FASTlusearchFTOCAPOCI}{0.0}
\newcommand{\FASTlusearchFTOCAPOCIMIN}{0}
\newcommand{\FASTlusearchFTOCAPOCIMAX}{0}
\newcommand{\FASTlusearchFTOCAPODynamic}{0}
\newcommand{\FASTlusearchFTOCAPODynamicCI}{0.0}
\newcommand{\FASTlusearchFTOCAPODynamicCIMIN}{0}
\newcommand{\FASTlusearchFTOCAPODynamicCIMAX}{0}
\newcommand{\FASTlusearchRECAPO}{0}
\newcommand{\FASTlusearchRECAPOCI}{0.0}
\newcommand{\FASTlusearchRECAPOCIMIN}{0}
\newcommand{\FASTlusearchRECAPOCIMAX}{0}
\newcommand{\FASTlusearchRECAPODynamic}{0}
\newcommand{\FASTlusearchRECAPODynamicCI}{0.0}
\newcommand{\FASTlusearchRECAPODynamicCIMIN}{0}
\newcommand{\FASTlusearchRECAPODynamicCIMAX}{0}
\newcommand{\FASTlusearchAGGCAPO}{\rna}
\newcommand{\FASTlusearchAGGCAPOCI}{\rna}
\newcommand{\FASTlusearchAGGCAPOCIMIN}{\rna}
\newcommand{\FASTlusearchAGGCAPOCIMAX}{\rna}
\newcommand{\FASTlusearchAGGCAPODynamic}{\rna}
\newcommand{\FASTlusearchAGGCAPODynamicCI}{\rna}
\newcommand{\FASTlusearchAGGCAPODynamicCIMIN}{\rna}
\newcommand{\FASTlusearchAGGCAPODynamicCIMAX}{\rna}
\newcommand{\FASTpmdEvents}{210}
\newcommand{\FASTpmdNoFPEvents}{13}
\newcommand{\pmdHBEventTotal}{210}
\newcommand{\pmdHBNoFPEventTotal}{13}
\newcommand{\pmdHBNoFPAccessTotal}{13}
\newcommand{\pmdHBNoFPOtherTotal}{<0.1}
\newcommand{\pmdHBReadTotal}{95.6}
\newcommand{\pmdHBWriteTotal}{4.27}
\newcommand{\pmdHBNoFPAccessInCS}{10.3}
\newcommand{\pmdHBNoFPAccessOutCS}{89.7}
\newcommand{\pmdHBAcqRelTotal}{0.134}
\newcommand{\pmdHBOtherTotal}{0.0055}
\newcommand{\pmdHBNoFPReadTotal}{12}
\newcommand{\pmdHBReadInCS}{11}
\newcommand{\pmdHBReadOutCS}{110}
\newcommand{\pmdHBReadSameEp}{21.2}
\newcommand{\pmdHBReadSharedSameEp}{\cna}
\newcommand{\pmdHBReadExclusive}{99}
\newcommand{\pmdHBReadOwned}{\cna}
\newcommand{\pmdHBReadShare}{0.0163}
\newcommand{\pmdHBReadShared}{0.948}
\newcommand{\pmdHBReadSharedOwned}{\cna}
\newcommand{\pmdHBNoFPHonestWriteTotal}{0.54}
\newcommand{\pmdHBWriteInCS}{26.4}
\newcommand{\pmdHBWriteOutCS}{2000}
\newcommand{\pmdHBNoFPWriteTotal}{0.54}
\newcommand{\pmdHBWriteSameEp}{1930}
\newcommand{\pmdHBWriteExclusive}{100}
\newcommand{\pmdHBWriteOwned}{\cna}
\newcommand{\pmdHBWriteShared}{<0.001}
\newcommand{\pmdHBNoFPOtherEventTotal}{18323}
\newcommand{\pmdHBAcqRelOtherTotal}{96.1}
\newcommand{\pmdHBNoAcqRelOtherTotal}{721}
\newcommand{\pmdHBFork}{0.0}
\newcommand{\pmdHBJoin}{0.0}
\newcommand{\pmdHBPreWait}{0.0}
\newcommand{\pmdHBPostWait}{0.0}
\newcommand{\pmdHBVolatileTotal}{0.0}
\newcommand{\pmdHBClassInit}{7.91}
\newcommand{\pmdHBClassAccess}{92.1}
\newcommand{\pmdHBRaceTotal}{1851}
\newcommand{\pmdHBWrRdRace}{99.0}
\newcommand{\pmdHBWrWrRace}{0.864}
\newcommand{\pmdHBRdWrRace}{0.0}
\newcommand{\pmdHBRdShWrRace}{0.108}
\newcommand{\pmdHBHoldLocksTotal}{1.5}
\newcommand{\pmdHBOneLockHeld}{11.6}
\newcommand{\pmdHBTwoNestedLocks}{\cna}
\newcommand{\pmdHBThreeNestedLocks}{\cna}
\newcommand{\pmdHBFourNestedLocks}{\cna}
\newcommand{\pmdHBFiveNestedLocks}{\cna}
\newcommand{\pmdHBSixNestedLocks}{\cna}
\newcommand{\pmdHBSevenNestedLocks}{\cna}
\newcommand{\pmdHBEightNestedLocks}{\cna}
\newcommand{\pmdHBNineNestedLocks}{\cna}
\newcommand{\pmdHBTenNestedLocks}{\cna}
\newcommand{\pmdHBHundredNestedLocks}{\cna}
\newcommand{\pmdHBExWrSet}{\ena}
\newcommand{\pmdHBExWrCheck}{\ena}
\newcommand{\pmdHBExWrUpdate}{\ena}
\newcommand{\pmdHBExRdCheck}{\ena}
\newcommand{\pmdHBExRdUpdate}{\ena}
\newcommand{\pmdHBExTotalCheck}{\ena}
\newcommand{\pmdHBExTotalUpdate}{\ena}
\newcommand{\pmdFTOHBEventTotal}{210}
\newcommand{\pmdFTOHBNoFPEventTotal}{8.0}
\newcommand{\pmdFTOHBNoFPAccessTotal}{7.9}
\newcommand{\pmdFTOHBNoFPOtherTotal}{<0.1}
\newcommand{\pmdFTOHBReadTotal}{92.9}
\newcommand{\pmdFTOHBWriteTotal}{6.85}
\newcommand{\pmdFTOHBNoFPAccessInCS}{10.3}
\newcommand{\pmdFTOHBNoFPAccessOutCS}{89.7}
\newcommand{\pmdFTOHBAcqRelTotal}{0.134}
\newcommand{\pmdFTOHBOtherTotal}{0.00552}
\newcommand{\pmdFTOHBNoFPReadTotal}{7.4}
\newcommand{\pmdFTOHBReadInCS}{17.5}
\newcommand{\pmdFTOHBReadOutCS}{117}
\newcommand{\pmdFTOHBReadSameEp}{34.9}
\newcommand{\pmdFTOHBReadSharedSameEp}{\cna}
\newcommand{\pmdFTOHBReadExclusive}{0.313}
\newcommand{\pmdFTOHBReadOwned}{98.1}
\newcommand{\pmdFTOHBReadShare}{0.0333}
\newcommand{\pmdFTOHBReadShared}{0.144}
\newcommand{\pmdFTOHBReadSharedOwned}{1.42}
\newcommand{\pmdFTOHBNoFPHonestWriteTotal}{0.54}
\newcommand{\pmdFTOHBWriteInCS}{26.4}
\newcommand{\pmdFTOHBWriteOutCS}{2000}
\newcommand{\pmdFTOHBNoFPWriteTotal}{0.54}
\newcommand{\pmdFTOHBWriteSameEp}{1930}
\newcommand{\pmdFTOHBWriteExclusive}{0.00257}
\newcommand{\pmdFTOHBWriteOwned}{100}
\newcommand{\pmdFTOHBWriteShared}{<0.001}
\newcommand{\pmdFTOHBNoFPOtherEventTotal}{18325}
\newcommand{\pmdFTOHBAcqRelOtherTotal}{96.1}
\newcommand{\pmdFTOHBNoAcqRelOtherTotal}{723}
\newcommand{\pmdFTOHBFork}{0.0}
\newcommand{\pmdFTOHBJoin}{0.0}
\newcommand{\pmdFTOHBPreWait}{0.0}
\newcommand{\pmdFTOHBPostWait}{0.0}
\newcommand{\pmdFTOHBVolatileTotal}{0.0}
\newcommand{\pmdFTOHBClassInit}{7.88}
\newcommand{\pmdFTOHBClassAccess}{92.1}
\newcommand{\pmdFTOHBRaceTotal}{1728}
\newcommand{\pmdFTOHBWrRdRace}{99.1}
\newcommand{\pmdFTOHBWrWrRace}{0.0}
\newcommand{\pmdFTOHBRdWrRace}{0.81}
\newcommand{\pmdFTOHBRdShWrRace}{0.0}
\newcommand{\pmdFTOHBHoldLocksTotal}{0.093}
\newcommand{\pmdFTOHBOneLockHeld}{1.14}
\newcommand{\pmdFTOHBTwoNestedLocks}{\cna}
\newcommand{\pmdFTOHBThreeNestedLocks}{\cna}
\newcommand{\pmdFTOHBFourNestedLocks}{\cna}
\newcommand{\pmdFTOHBFiveNestedLocks}{\cna}
\newcommand{\pmdFTOHBSixNestedLocks}{\cna}
\newcommand{\pmdFTOHBSevenNestedLocks}{\cna}
\newcommand{\pmdFTOHBEightNestedLocks}{\cna}
\newcommand{\pmdFTOHBNineNestedLocks}{\cna}
\newcommand{\pmdFTOHBTenNestedLocks}{\cna}
\newcommand{\pmdFTOHBHundredNestedLocks}{\cna}
\newcommand{\pmdFTOHBExWrSet}{\ena}
\newcommand{\pmdFTOHBExWrCheck}{\ena}
\newcommand{\pmdFTOHBExWrUpdate}{\ena}
\newcommand{\pmdFTOHBExRdCheck}{\ena}
\newcommand{\pmdFTOHBExRdUpdate}{\ena}
\newcommand{\pmdFTOHBExTotalCheck}{\ena}
\newcommand{\pmdFTOHBExTotalUpdate}{\ena}
\newcommand{\pmdFTOWCPEventTotal}{210}
\newcommand{\pmdFTOWCPNoFPEventTotal}{8.0}
\newcommand{\pmdFTOWCPNoFPAccessTotal}{7.9}
\newcommand{\pmdFTOWCPNoFPOtherTotal}{<0.1}
\newcommand{\pmdFTOWCPReadTotal}{92.9}
\newcommand{\pmdFTOWCPWriteTotal}{6.85}
\newcommand{\pmdFTOWCPNoFPAccessInCS}{10.3}
\newcommand{\pmdFTOWCPNoFPAccessOutCS}{89.7}
\newcommand{\pmdFTOWCPAcqRelTotal}{0.134}
\newcommand{\pmdFTOWCPOtherTotal}{0.00552}
\newcommand{\pmdFTOWCPNoFPReadTotal}{7.4}
\newcommand{\pmdFTOWCPReadInCS}{17.5}
\newcommand{\pmdFTOWCPReadOutCS}{117}
\newcommand{\pmdFTOWCPReadSameEp}{34.9}
\newcommand{\pmdFTOWCPReadSharedSameEp}{\cna}
\newcommand{\pmdFTOWCPReadExclusive}{0.254}
\newcommand{\pmdFTOWCPReadOwned}{98}
\newcommand{\pmdFTOWCPReadShare}{0.034}
\newcommand{\pmdFTOWCPReadShared}{0.146}
\newcommand{\pmdFTOWCPReadSharedOwned}{1.55}
\newcommand{\pmdFTOWCPNoFPHonestWriteTotal}{0.54}
\newcommand{\pmdFTOWCPWriteInCS}{26.4}
\newcommand{\pmdFTOWCPWriteOutCS}{2000}
\newcommand{\pmdFTOWCPNoFPWriteTotal}{0.54}
\newcommand{\pmdFTOWCPWriteSameEp}{1930}
\newcommand{\pmdFTOWCPWriteExclusive}{0.00257}
\newcommand{\pmdFTOWCPWriteOwned}{100}
\newcommand{\pmdFTOWCPWriteShared}{<0.001}
\newcommand{\pmdFTOWCPNoFPOtherEventTotal}{18325}
\newcommand{\pmdFTOWCPAcqRelOtherTotal}{96.1}
\newcommand{\pmdFTOWCPNoAcqRelOtherTotal}{723}
\newcommand{\pmdFTOWCPFork}{0.0}
\newcommand{\pmdFTOWCPJoin}{0.0}
\newcommand{\pmdFTOWCPPreWait}{0.0}
\newcommand{\pmdFTOWCPPostWait}{0.0}
\newcommand{\pmdFTOWCPVolatileTotal}{0.0}
\newcommand{\pmdFTOWCPClassInit}{7.88}
\newcommand{\pmdFTOWCPClassAccess}{92.1}
\newcommand{\pmdFTOWCPRaceTotal}{1789}
\newcommand{\pmdFTOWCPWrRdRace}{99.2}
\newcommand{\pmdFTOWCPWrWrRace}{0.0}
\newcommand{\pmdFTOWCPRdWrRace}{0.783}
\newcommand{\pmdFTOWCPRdShWrRace}{0.0}
\newcommand{\pmdFTOWCPHoldLocksTotal}{0.093}
\newcommand{\pmdFTOWCPOneLockHeld}{1.14}
\newcommand{\pmdFTOWCPTwoNestedLocks}{\cna}
\newcommand{\pmdFTOWCPThreeNestedLocks}{\cna}
\newcommand{\pmdFTOWCPFourNestedLocks}{\cna}
\newcommand{\pmdFTOWCPFiveNestedLocks}{\cna}
\newcommand{\pmdFTOWCPSixNestedLocks}{\cna}
\newcommand{\pmdFTOWCPSevenNestedLocks}{\cna}
\newcommand{\pmdFTOWCPEightNestedLocks}{\cna}
\newcommand{\pmdFTOWCPNineNestedLocks}{\cna}
\newcommand{\pmdFTOWCPTenNestedLocks}{\cna}
\newcommand{\pmdFTOWCPHundredNestedLocks}{\cna}
\newcommand{\pmdFTOWCPExWrSet}{\ena}
\newcommand{\pmdFTOWCPExWrCheck}{\ena}
\newcommand{\pmdFTOWCPExWrUpdate}{\ena}
\newcommand{\pmdFTOWCPExRdCheck}{\ena}
\newcommand{\pmdFTOWCPExRdUpdate}{\ena}
\newcommand{\pmdFTOWCPExTotalCheck}{\ena}
\newcommand{\pmdFTOWCPExTotalUpdate}{\ena}
\newcommand{\pmdREWCPEventTotal}{210}
\newcommand{\pmdREWCPNoFPEventTotal}{8.0}
\newcommand{\pmdREWCPNoFPAccessTotal}{7.9}
\newcommand{\pmdREWCPNoFPOtherTotal}{<0.1}
\newcommand{\pmdREWCPReadTotal}{92.9}
\newcommand{\pmdREWCPWriteTotal}{6.85}
\newcommand{\pmdREWCPNoFPAccessInCS}{10.3}
\newcommand{\pmdREWCPNoFPAccessOutCS}{89.7}
\newcommand{\pmdREWCPAcqRelTotal}{0.134}
\newcommand{\pmdREWCPOtherTotal}{0.00551}
\newcommand{\pmdREWCPNoFPReadTotal}{7.4}
\newcommand{\pmdREWCPReadInCS}{17.5}
\newcommand{\pmdREWCPReadOutCS}{117}
\newcommand{\pmdREWCPReadSameEp}{34.9}
\newcommand{\pmdREWCPReadSharedSameEp}{\cna}
\newcommand{\pmdREWCPReadExclusive}{0.222}
\newcommand{\pmdREWCPReadOwned}{98}
\newcommand{\pmdREWCPReadShare}{0.0679}
\newcommand{\pmdREWCPReadShared}{0.147}
\newcommand{\pmdREWCPReadSharedOwned}{1.55}
\newcommand{\pmdREWCPNoFPHonestWriteTotal}{0.54}
\newcommand{\pmdREWCPWriteInCS}{26.4}
\newcommand{\pmdREWCPWriteOutCS}{2000}
\newcommand{\pmdREWCPNoFPWriteTotal}{0.54}
\newcommand{\pmdREWCPWriteSameEp}{1930}
\newcommand{\pmdREWCPWriteExclusive}{0.00257}
\newcommand{\pmdREWCPWriteOwned}{99.5}
\newcommand{\pmdREWCPWriteShared}{0.46}
\newcommand{\pmdREWCPNoFPOtherEventTotal}{18324}
\newcommand{\pmdREWCPAcqRelOtherTotal}{96.1}
\newcommand{\pmdREWCPNoAcqRelOtherTotal}{722}
\newcommand{\pmdREWCPFork}{0.0}
\newcommand{\pmdREWCPJoin}{0.0}
\newcommand{\pmdREWCPPreWait}{0.0}
\newcommand{\pmdREWCPPostWait}{0.0}
\newcommand{\pmdREWCPVolatileTotal}{0.0}
\newcommand{\pmdREWCPClassInit}{7.89}
\newcommand{\pmdREWCPClassAccess}{92.1}
\newcommand{\pmdREWCPRaceTotal}{1752}
\newcommand{\pmdREWCPWrRdRace}{99.1}
\newcommand{\pmdREWCPWrWrRace}{0.0}
\newcommand{\pmdREWCPRdWrRace}{0.799}
\newcommand{\pmdREWCPRdShWrRace}{0.0571}
\newcommand{\pmdREWCPHoldLocksTotal}{0.093}
\newcommand{\pmdREWCPOneLockHeld}{1.14}
\newcommand{\pmdREWCPTwoNestedLocks}{\cna}
\newcommand{\pmdREWCPThreeNestedLocks}{\cna}
\newcommand{\pmdREWCPFourNestedLocks}{\cna}
\newcommand{\pmdREWCPFiveNestedLocks}{\cna}
\newcommand{\pmdREWCPSixNestedLocks}{\cna}
\newcommand{\pmdREWCPSevenNestedLocks}{\cna}
\newcommand{\pmdREWCPEightNestedLocks}{\cna}
\newcommand{\pmdREWCPNineNestedLocks}{\cna}
\newcommand{\pmdREWCPTenNestedLocks}{\cna}
\newcommand{\pmdREWCPHundredNestedLocks}{\cna}
\newcommand{\pmdREWCPExWrSet}{\ena}
\newcommand{\pmdREWCPExWrCheck}{\ena}
\newcommand{\pmdREWCPExWrUpdate}{\ena}
\newcommand{\pmdREWCPExRdCheck}{\ena}
\newcommand{\pmdREWCPExRdUpdate}{\ena}
\newcommand{\pmdREWCPExTotalCheck}{\ena}
\newcommand{\pmdREWCPExTotalUpdate}{\ena}
\newcommand{\pmdFTODCEventTotal}{210}
\newcommand{\pmdFTODCNoFPEventTotal}{8.0}
\newcommand{\pmdFTODCNoFPAccessTotal}{7.9}
\newcommand{\pmdFTODCNoFPOtherTotal}{<0.1}
\newcommand{\pmdFTODCReadTotal}{92.9}
\newcommand{\pmdFTODCWriteTotal}{6.84}
\newcommand{\pmdFTODCNoFPAccessInCS}{10.3}
\newcommand{\pmdFTODCNoFPAccessOutCS}{89.7}
\newcommand{\pmdFTODCAcqRelTotal}{0.134}
\newcommand{\pmdFTODCOtherTotal}{0.00551}
\newcommand{\pmdFTODCNoFPReadTotal}{7.4}
\newcommand{\pmdFTODCReadInCS}{17.5}
\newcommand{\pmdFTODCReadOutCS}{117}
\newcommand{\pmdFTODCReadSameEp}{34.9}
\newcommand{\pmdFTODCReadSharedSameEp}{\cna}
\newcommand{\pmdFTODCReadExclusive}{0.229}
\newcommand{\pmdFTODCReadOwned}{98}
\newcommand{\pmdFTODCReadShare}{0.0524}
\newcommand{\pmdFTODCReadShared}{0.151}
\newcommand{\pmdFTODCReadSharedOwned}{1.6}
\newcommand{\pmdFTODCNoFPHonestWriteTotal}{0.54}
\newcommand{\pmdFTODCWriteInCS}{26.4}
\newcommand{\pmdFTODCWriteOutCS}{2000}
\newcommand{\pmdFTODCNoFPWriteTotal}{0.54}
\newcommand{\pmdFTODCWriteSameEp}{1930}
\newcommand{\pmdFTODCWriteExclusive}{0.00257}
\newcommand{\pmdFTODCWriteOwned}{100}
\newcommand{\pmdFTODCWriteShared}{<0.001}
\newcommand{\pmdFTODCNoFPOtherEventTotal}{18324}
\newcommand{\pmdFTODCAcqRelOtherTotal}{96.1}
\newcommand{\pmdFTODCNoAcqRelOtherTotal}{722}
\newcommand{\pmdFTODCFork}{0.0}
\newcommand{\pmdFTODCJoin}{0.0}
\newcommand{\pmdFTODCPreWait}{0.0}
\newcommand{\pmdFTODCPostWait}{0.0}
\newcommand{\pmdFTODCVolatileTotal}{0.0}
\newcommand{\pmdFTODCClassInit}{7.89}
\newcommand{\pmdFTODCClassAccess}{92.1}
\newcommand{\pmdFTODCRaceTotal}{3709}
\newcommand{\pmdFTODCWrRdRace}{99.6}
\newcommand{\pmdFTODCWrWrRace}{0.0}
\newcommand{\pmdFTODCRdWrRace}{0.377}
\newcommand{\pmdFTODCRdShWrRace}{0.027}
\newcommand{\pmdFTODCHoldLocksTotal}{0.093}
\newcommand{\pmdFTODCOneLockHeld}{1.12}
\newcommand{\pmdFTODCTwoNestedLocks}{\cna}
\newcommand{\pmdFTODCThreeNestedLocks}{\cna}
\newcommand{\pmdFTODCFourNestedLocks}{\cna}
\newcommand{\pmdFTODCFiveNestedLocks}{\cna}
\newcommand{\pmdFTODCSixNestedLocks}{\cna}
\newcommand{\pmdFTODCSevenNestedLocks}{\cna}
\newcommand{\pmdFTODCEightNestedLocks}{\cna}
\newcommand{\pmdFTODCNineNestedLocks}{\cna}
\newcommand{\pmdFTODCTenNestedLocks}{\cna}
\newcommand{\pmdFTODCHundredNestedLocks}{\cna}
\newcommand{\pmdFTODCExWrSet}{\ena}
\newcommand{\pmdFTODCExWrCheck}{\ena}
\newcommand{\pmdFTODCExWrUpdate}{\ena}
\newcommand{\pmdFTODCExRdCheck}{\ena}
\newcommand{\pmdFTODCExRdUpdate}{\ena}
\newcommand{\pmdFTODCExTotalCheck}{\ena}
\newcommand{\pmdFTODCExTotalUpdate}{\ena}
\newcommand{\pmdREDCEventTotal}{210}
\newcommand{\pmdREDCNoFPEventTotal}{8.0}
\newcommand{\pmdREDCNoFPAccessTotal}{7.9}
\newcommand{\pmdREDCNoFPOtherTotal}{<0.1}
\newcommand{\pmdREDCReadTotal}{92.9}
\newcommand{\pmdREDCWriteTotal}{6.84}
\newcommand{\pmdREDCNoFPAccessInCS}{10.3}
\newcommand{\pmdREDCNoFPAccessOutCS}{89.7}
\newcommand{\pmdREDCAcqRelTotal}{0.134}
\newcommand{\pmdREDCOtherTotal}{0.00549}
\newcommand{\pmdREDCNoFPReadTotal}{7.4}
\newcommand{\pmdREDCReadInCS}{17.5}
\newcommand{\pmdREDCReadOutCS}{117}
\newcommand{\pmdREDCReadSameEp}{34.9}
\newcommand{\pmdREDCReadSharedSameEp}{\cna}
\newcommand{\pmdREDCReadExclusive}{0.12}
\newcommand{\pmdREDCReadOwned}{98}
\newcommand{\pmdREDCReadShare}{0.16}
\newcommand{\pmdREDCReadShared}{0.15}
\newcommand{\pmdREDCReadSharedOwned}{1.6}
\newcommand{\pmdREDCNoFPHonestWriteTotal}{0.54}
\newcommand{\pmdREDCWriteInCS}{26.4}
\newcommand{\pmdREDCWriteOutCS}{2000}
\newcommand{\pmdREDCNoFPWriteTotal}{0.54}
\newcommand{\pmdREDCWriteSameEp}{1930}
\newcommand{\pmdREDCWriteExclusive}{0.00257}
\newcommand{\pmdREDCWriteOwned}{98.5}
\newcommand{\pmdREDCWriteShared}{1.47}
\newcommand{\pmdREDCNoFPOtherEventTotal}{18321}
\newcommand{\pmdREDCAcqRelOtherTotal}{96.1}
\newcommand{\pmdREDCNoAcqRelOtherTotal}{719}
\newcommand{\pmdREDCFork}{0.0}
\newcommand{\pmdREDCJoin}{0.0}
\newcommand{\pmdREDCPreWait}{0.0}
\newcommand{\pmdREDCPostWait}{0.0}
\newcommand{\pmdREDCVolatileTotal}{0.0}
\newcommand{\pmdREDCClassInit}{7.93}
\newcommand{\pmdREDCClassAccess}{92.1}
\newcommand{\pmdREDCRaceTotal}{3627}
\newcommand{\pmdREDCWrRdRace}{99.6}
\newcommand{\pmdREDCWrWrRace}{0.0}
\newcommand{\pmdREDCRdWrRace}{0.386}
\newcommand{\pmdREDCRdShWrRace}{0.0}
\newcommand{\pmdREDCHoldLocksTotal}{0.093}
\newcommand{\pmdREDCOneLockHeld}{1.12}
\newcommand{\pmdREDCTwoNestedLocks}{\cna}
\newcommand{\pmdREDCThreeNestedLocks}{\cna}
\newcommand{\pmdREDCFourNestedLocks}{\cna}
\newcommand{\pmdREDCFiveNestedLocks}{\cna}
\newcommand{\pmdREDCSixNestedLocks}{\cna}
\newcommand{\pmdREDCSevenNestedLocks}{\cna}
\newcommand{\pmdREDCEightNestedLocks}{\cna}
\newcommand{\pmdREDCNineNestedLocks}{\cna}
\newcommand{\pmdREDCTenNestedLocks}{\cna}
\newcommand{\pmdREDCHundredNestedLocks}{\cna}
\newcommand{\pmdREDCExWrSet}{\ena}
\newcommand{\pmdREDCExWrCheck}{\ena}
\newcommand{\pmdREDCExWrUpdate}{\ena}
\newcommand{\pmdREDCExRdCheck}{\ena}
\newcommand{\pmdREDCExRdUpdate}{\ena}
\newcommand{\pmdREDCExTotalCheck}{\ena}
\newcommand{\pmdREDCExTotalUpdate}{\ena}
\newcommand{\pmdFTOCAPOEventTotal}{210}
\newcommand{\pmdFTOCAPONoFPEventTotal}{8.0}
\newcommand{\pmdFTOCAPONoFPAccessTotal}{7.9}
\newcommand{\pmdFTOCAPONoFPOtherTotal}{<0.1}
\newcommand{\pmdFTOCAPOReadTotal}{92.9}
\newcommand{\pmdFTOCAPOWriteTotal}{6.85}
\newcommand{\pmdFTOCAPONoFPAccessInCS}{10.3}
\newcommand{\pmdFTOCAPONoFPAccessOutCS}{89.7}
\newcommand{\pmdFTOCAPOAcqRelTotal}{0.134}
\newcommand{\pmdFTOCAPOOtherTotal}{0.00548}
\newcommand{\pmdFTOCAPONoFPReadTotal}{7.4}
\newcommand{\pmdFTOCAPOReadInCS}{17.5}
\newcommand{\pmdFTOCAPOReadOutCS}{117}
\newcommand{\pmdFTOCAPOReadSameEp}{34.9}
\newcommand{\pmdFTOCAPOReadSharedSameEp}{\cna}
\newcommand{\pmdFTOCAPOReadExclusive}{0.226}
\newcommand{\pmdFTOCAPOReadOwned}{98}
\newcommand{\pmdFTOCAPOReadShare}{0.0523}
\newcommand{\pmdFTOCAPOReadShared}{0.15}
\newcommand{\pmdFTOCAPOReadSharedOwned}{1.6}
\newcommand{\pmdFTOCAPONoFPHonestWriteTotal}{0.54}
\newcommand{\pmdFTOCAPOWriteInCS}{26.4}
\newcommand{\pmdFTOCAPOWriteOutCS}{2000}
\newcommand{\pmdFTOCAPONoFPWriteTotal}{0.54}
\newcommand{\pmdFTOCAPOWriteSameEp}{1930}
\newcommand{\pmdFTOCAPOWriteExclusive}{0.00275}
\newcommand{\pmdFTOCAPOWriteOwned}{100}
\newcommand{\pmdFTOCAPOWriteShared}{<0.001}
\newcommand{\pmdFTOCAPONoFPOtherEventTotal}{18320}
\newcommand{\pmdFTOCAPOAcqRelOtherTotal}{96.1}
\newcommand{\pmdFTOCAPONoAcqRelOtherTotal}{718}
\newcommand{\pmdFTOCAPOFork}{0.0}
\newcommand{\pmdFTOCAPOJoin}{0.0}
\newcommand{\pmdFTOCAPOPreWait}{0.0}
\newcommand{\pmdFTOCAPOPostWait}{0.0}
\newcommand{\pmdFTOCAPOVolatileTotal}{0.0}
\newcommand{\pmdFTOCAPOClassInit}{7.94}
\newcommand{\pmdFTOCAPOClassAccess}{92.1}
\newcommand{\pmdFTOCAPORaceTotal}{3910}
\newcommand{\pmdFTOCAPOWrRdRace}{99.6}
\newcommand{\pmdFTOCAPOWrWrRace}{0.0}
\newcommand{\pmdFTOCAPORdWrRace}{0.384}
\newcommand{\pmdFTOCAPORdShWrRace}{0.0256}
\newcommand{\pmdFTOCAPOHoldLocksTotal}{0.093}
\newcommand{\pmdFTOCAPOOneLockHeld}{1.12}
\newcommand{\pmdFTOCAPOTwoNestedLocks}{\cna}
\newcommand{\pmdFTOCAPOThreeNestedLocks}{\cna}
\newcommand{\pmdFTOCAPOFourNestedLocks}{\cna}
\newcommand{\pmdFTOCAPOFiveNestedLocks}{\cna}
\newcommand{\pmdFTOCAPOSixNestedLocks}{\cna}
\newcommand{\pmdFTOCAPOSevenNestedLocks}{\cna}
\newcommand{\pmdFTOCAPOEightNestedLocks}{\cna}
\newcommand{\pmdFTOCAPONineNestedLocks}{\cna}
\newcommand{\pmdFTOCAPOTenNestedLocks}{\cna}
\newcommand{\pmdFTOCAPOHundredNestedLocks}{\cna}
\newcommand{\pmdFTOCAPOExWrSet}{\ena}
\newcommand{\pmdFTOCAPOExWrCheck}{\ena}
\newcommand{\pmdFTOCAPOExWrUpdate}{\ena}
\newcommand{\pmdFTOCAPOExRdCheck}{\ena}
\newcommand{\pmdFTOCAPOExRdUpdate}{\ena}
\newcommand{\pmdFTOCAPOExTotalCheck}{\ena}
\newcommand{\pmdFTOCAPOExTotalUpdate}{\ena}
\newcommand{\pmdRECAPOEventTotal}{210}
\newcommand{\pmdRECAPONoFPEventTotal}{8.0}
\newcommand{\pmdRECAPONoFPAccessTotal}{7.9}
\newcommand{\pmdRECAPONoFPOtherTotal}{<0.1}
\newcommand{\pmdRECAPOReadTotal}{92.9}
\newcommand{\pmdRECAPOWriteTotal}{6.84}
\newcommand{\pmdRECAPONoFPAccessInCS}{10.3}
\newcommand{\pmdRECAPONoFPAccessOutCS}{89.7}
\newcommand{\pmdRECAPOAcqRelTotal}{0.134}
\newcommand{\pmdRECAPOOtherTotal}{0.0055}
\newcommand{\pmdRECAPONoFPReadTotal}{7.4}
\newcommand{\pmdRECAPOReadInCS}{17.5}
\newcommand{\pmdRECAPOReadOutCS}{117}
\newcommand{\pmdRECAPOReadSameEp}{34.9}
\newcommand{\pmdRECAPOReadSharedSameEp}{\cna}
\newcommand{\pmdRECAPOReadExclusive}{0.12}
\newcommand{\pmdRECAPOReadOwned}{98}
\newcommand{\pmdRECAPOReadShare}{0.16}
\newcommand{\pmdRECAPOReadShared}{0.15}
\newcommand{\pmdRECAPOReadSharedOwned}{1.6}
\newcommand{\pmdRECAPONoFPHonestWriteTotal}{0.54}
\newcommand{\pmdRECAPOWriteInCS}{26.4}
\newcommand{\pmdRECAPOWriteOutCS}{2000}
\newcommand{\pmdRECAPONoFPWriteTotal}{0.54}
\newcommand{\pmdRECAPOWriteSameEp}{1930}
\newcommand{\pmdRECAPOWriteExclusive}{0.0026}
\newcommand{\pmdRECAPOWriteOwned}{98.5}
\newcommand{\pmdRECAPOWriteShared}{1.5}
\newcommand{\pmdRECAPONoFPOtherEventTotal}{18323}
\newcommand{\pmdRECAPOAcqRelOtherTotal}{96.1}
\newcommand{\pmdRECAPONoAcqRelOtherTotal}{721}
\newcommand{\pmdRECAPOFork}{0.0}
\newcommand{\pmdRECAPOJoin}{0.0}
\newcommand{\pmdRECAPOPreWait}{0.0}
\newcommand{\pmdRECAPOPostWait}{0.0}
\newcommand{\pmdRECAPOVolatileTotal}{0.0}
\newcommand{\pmdRECAPOClassInit}{7.91}
\newcommand{\pmdRECAPOClassAccess}{92.1}
\newcommand{\pmdRECAPORaceTotal}{3774}
\newcommand{\pmdRECAPOWrRdRace}{99.6}
\newcommand{\pmdRECAPOWrWrRace}{0.0}
\newcommand{\pmdRECAPORdWrRace}{0.371}
\newcommand{\pmdRECAPORdShWrRace}{0.0265}
\newcommand{\pmdRECAPOHoldLocksTotal}{0.093}
\newcommand{\pmdRECAPOOneLockHeld}{1.1}
\newcommand{\pmdRECAPOTwoNestedLocks}{\cna}
\newcommand{\pmdRECAPOThreeNestedLocks}{\cna}
\newcommand{\pmdRECAPOFourNestedLocks}{\cna}
\newcommand{\pmdRECAPOFiveNestedLocks}{\cna}
\newcommand{\pmdRECAPOSixNestedLocks}{\cna}
\newcommand{\pmdRECAPOSevenNestedLocks}{\cna}
\newcommand{\pmdRECAPOEightNestedLocks}{\cna}
\newcommand{\pmdRECAPONineNestedLocks}{\cna}
\newcommand{\pmdRECAPOTenNestedLocks}{\cna}
\newcommand{\pmdRECAPOHundredNestedLocks}{\cna}
\newcommand{\pmdRECAPOExWrSet}{\ena}
\newcommand{\pmdRECAPOExWrCheck}{\ena}
\newcommand{\pmdRECAPOExWrUpdate}{\ena}
\newcommand{\pmdRECAPOExRdCheck}{\ena}
\newcommand{\pmdRECAPOExRdUpdate}{\ena}
\newcommand{\pmdRECAPOExTotalCheck}{\ena}
\newcommand{\pmdRECAPOExTotalUpdate}{\ena}
\newcommand{\FASTpmdMaxLiveThreads}{15}
\newcommand{\FASTpmdTotalThreads}{15}
\newcommand{\FASTpmdBaseTime}{1.2}
\newcommand{\FASTpmdBaseTimeCI}{18}
\newcommand{\FASTpmdEmptyTime}{\rna}
\newcommand{\FASTpmdEmptyTimeCI}{\rna}
\newcommand{\FASTpmdEmptyTimeCIMIN}{\rna}
\newcommand{\FASTpmdEmptyTimeCIMAX}{\rna}
\newcommand{\FASTpmdFTTime}{6.7}
\newcommand{\FASTpmdFTTimeCI}{0.17}
\newcommand{\FASTpmdHBTime}{6.5}
\newcommand{\FASTpmdHBTimeCI}{0.11}
\newcommand{\FASTpmdFTOHBTime}{6.6}
\newcommand{\FASTpmdFTOHBTimeCI}{0.19}
\newcommand{\FASTpmdWCPTime}{\rna}
\newcommand{\FASTpmdWCPTimeCI}{\rna}
\newcommand{\FASTpmdWCPTimeCIMIN}{\rna}
\newcommand{\FASTpmdWCPTimeCIMAX}{\rna}
\newcommand{\FASTpmdFTOWCPTime}{6.9}
\newcommand{\FASTpmdFTOWCPTimeCI}{0.12}
\newcommand{\FASTpmdREWCPTime}{6.9}
\newcommand{\FASTpmdREWCPTimeCI}{0.21}
\newcommand{\FASTpmdDCTime}{\rna}
\newcommand{\FASTpmdDCTimeCI}{\rna}
\newcommand{\FASTpmdDCTimeCIMIN}{\rna}
\newcommand{\FASTpmdDCTimeCIMAX}{\rna}
\newcommand{\FASTpmdFTODCTime}{6.9}
\newcommand{\FASTpmdFTODCTimeCI}{0.21}
\newcommand{\FASTpmdREDCTime}{7.0}
\newcommand{\FASTpmdREDCTimeCI}{0.27}
\newcommand{\FASTpmdCAPOTime}{\rna}
\newcommand{\FASTpmdCAPOTimeCI}{\rna}
\newcommand{\FASTpmdCAPOTimeCIMIN}{\rna}
\newcommand{\FASTpmdCAPOTimeCIMAX}{\rna}
\newcommand{\FASTpmdFTOCAPOTime}{6.9}
\newcommand{\FASTpmdFTOCAPOTimeCI}{0.23}
\newcommand{\FASTpmdRECAPOTime}{6.7}
\newcommand{\FASTpmdRECAPOTimeCI}{0.17}
\newcommand{\FASTpmdAGGCAPOTime}{\rna}
\newcommand{\FASTpmdAGGCAPOTimeCI}{\rna}
\newcommand{\FASTpmdAGGCAPOTimeCIMIN}{\rna}
\newcommand{\FASTpmdAGGCAPOTimeCIMAX}{\rna}
\newcommand{\FASTpmdStaticTime}{\rzero}
\newcommand{\FASTpmdDynamicTime}{\rzero}
\newcommand{\FASTpmdBaseMem}{630}
\newcommand{\FASTpmdBaseMemCI}{12.0}
\newcommand{\FASTpmdFTMem}{2.9}
\newcommand{\FASTpmdFTMemCI}{0.045}
\newcommand{\FASTpmdHBMem}{2.9}
\newcommand{\FASTpmdHBMemCI}{0.13}
\newcommand{\FASTpmdFTOHBMem}{2.7}
\newcommand{\FASTpmdFTOHBMemCI}{0.16}
\newcommand{\FASTpmdWCPMem}{\memna}
\newcommand{\FASTpmdWCPMemCI}{\memna}
\newcommand{\FASTpmdWCPMemCIMIN}{\memna}
\newcommand{\FASTpmdWCPMemCIMAX}{\memna}
\newcommand{\FASTpmdFTOWCPMem}{3.0}
\newcommand{\FASTpmdFTOWCPMemCI}{0.15}
\newcommand{\FASTpmdREWCPMem}{3.0}
\newcommand{\FASTpmdREWCPMemCI}{0.15}
\newcommand{\FASTpmdDCMem}{\memna}
\newcommand{\FASTpmdDCMemCI}{\memna}
\newcommand{\FASTpmdDCMemCIMIN}{\memna}
\newcommand{\FASTpmdDCMemCIMAX}{\memna}
\newcommand{\FASTpmdFTODCMem}{3.0}
\newcommand{\FASTpmdFTODCMemCI}{0.094}
\newcommand{\FASTpmdREDCMem}{3.0}
\newcommand{\FASTpmdREDCMemCI}{0.13}
\newcommand{\FASTpmdCAPOMem}{\memna}
\newcommand{\FASTpmdCAPOMemCI}{\memna}
\newcommand{\FASTpmdCAPOMemCIMIN}{\memna}
\newcommand{\FASTpmdCAPOMemCIMAX}{\memna}
\newcommand{\FASTpmdFTOCAPOMem}{3.0}
\newcommand{\FASTpmdFTOCAPOMemCI}{0.046}
\newcommand{\FASTpmdRECAPOMem}{2.9}
\newcommand{\FASTpmdRECAPOMemCI}{0.13}
\newcommand{\FASTpmdAGGCAPOMem}{\memna}
\newcommand{\FASTpmdAGGCAPOMemCI}{\memna}
\newcommand{\FASTpmdAGGCAPOMemCIMIN}{\memna}
\newcommand{\FASTpmdAGGCAPOMemCIMAX}{\memna}
\newcommand{\FASTpmdEventsCI}{1,092}
\newcommand{\FASTpmdEventsCIMIN}{207,312,897}
\newcommand{\FASTpmdEventsCIMAX}{207,315,081}
\newcommand{\FASTpmdNoFPEventsCI}{920}
\newcommand{\FASTpmdNoFPEventsCIMIN}{13,102,231}
\newcommand{\FASTpmdNoFPEventsCIMAX}{13,104,071}
\newcommand{\FASTpmdFT}{17}
\newcommand{\FASTpmdFTCI}{0.64}
\newcommand{\FASTpmdFTCIMIN}{16}
\newcommand{\FASTpmdFTCIMAX}{18}
\newcommand{\FASTpmdFTDynamic}{3,986}
\newcommand{\FASTpmdFTDynamicCI}{385}
\newcommand{\FASTpmdFTDynamicCIMIN}{3,601}
\newcommand{\FASTpmdFTDynamicCIMAX}{4,371}
\newcommand{\FASTpmdHB}{18}
\newcommand{\FASTpmdHBCI}{0.57}
\newcommand{\FASTpmdHBCIMIN}{17}
\newcommand{\FASTpmdHBCIMAX}{19}
\newcommand{\FASTpmdHBDynamic}{1,849}
\newcommand{\FASTpmdHBDynamicCI}{206}
\newcommand{\FASTpmdHBDynamicCIMIN}{1,643}
\newcommand{\FASTpmdHBDynamicCIMAX}{2,055}
\newcommand{\FASTpmdFTOHB}{18}
\newcommand{\FASTpmdFTOHBCI}{0.2}
\newcommand{\FASTpmdFTOHBCIMIN}{18}
\newcommand{\FASTpmdFTOHBCIMAX}{18}
\newcommand{\FASTpmdFTOHBDynamic}{1,728}
\newcommand{\FASTpmdFTOHBDynamicCI}{106}
\newcommand{\FASTpmdFTOHBDynamicCIMIN}{1,622}
\newcommand{\FASTpmdFTOHBDynamicCIMAX}{1,834}
\newcommand{\FASTpmdWCP}{\rna}
\newcommand{\FASTpmdWCPCI}{\rna}
\newcommand{\FASTpmdWCPCIMIN}{\rna}
\newcommand{\FASTpmdWCPCIMAX}{\rna}
\newcommand{\FASTpmdWCPDynamic}{\rna}
\newcommand{\FASTpmdWCPDynamicCI}{\rna}
\newcommand{\FASTpmdWCPDynamicCIMIN}{\rna}
\newcommand{\FASTpmdWCPDynamicCIMAX}{\rna}
\newcommand{\FASTpmdFTOWCP}{18}
\newcommand{\FASTpmdFTOWCPCI}{0.2}
\newcommand{\FASTpmdFTOWCPCIMIN}{18}
\newcommand{\FASTpmdFTOWCPCIMAX}{18}
\newcommand{\FASTpmdFTOWCPDynamic}{1,789}
\newcommand{\FASTpmdFTOWCPDynamicCI}{89}
\newcommand{\FASTpmdFTOWCPDynamicCIMIN}{1,700}
\newcommand{\FASTpmdFTOWCPDynamicCIMAX}{1,878}
\newcommand{\FASTpmdREWCP}{18}
\newcommand{\FASTpmdREWCPCI}{0.30}
\newcommand{\FASTpmdREWCPCIMIN}{18}
\newcommand{\FASTpmdREWCPCIMAX}{18}
\newcommand{\FASTpmdREWCPDynamic}{1,752}
\newcommand{\FASTpmdREWCPDynamicCI}{78}
\newcommand{\FASTpmdREWCPDynamicCIMIN}{1,674}
\newcommand{\FASTpmdREWCPDynamicCIMAX}{1,830}
\newcommand{\FASTpmdDC}{\rna}
\newcommand{\FASTpmdDCCI}{\rna}
\newcommand{\FASTpmdDCCIMIN}{\rna}
\newcommand{\FASTpmdDCCIMAX}{\rna}
\newcommand{\FASTpmdDCDynamic}{\rna}
\newcommand{\FASTpmdDCDynamicCI}{\rna}
\newcommand{\FASTpmdDCDynamicCIMIN}{\rna}
\newcommand{\FASTpmdDCDynamicCIMAX}{\rna}
\newcommand{\FASTpmdFTODC}{18}
\newcommand{\FASTpmdFTODCCI}{0.42}
\newcommand{\FASTpmdFTODCCIMIN}{18}
\newcommand{\FASTpmdFTODCCIMAX}{18}
\newcommand{\FASTpmdFTODCDynamic}{3,709}
\newcommand{\FASTpmdFTODCDynamicCI}{185}
\newcommand{\FASTpmdFTODCDynamicCIMIN}{3,524}
\newcommand{\FASTpmdFTODCDynamicCIMAX}{3,894}
\newcommand{\FASTpmdREDC}{18}
\newcommand{\FASTpmdREDCCI}{0.2}
\newcommand{\FASTpmdREDCCIMIN}{18}
\newcommand{\FASTpmdREDCCIMAX}{18}
\newcommand{\FASTpmdREDCDynamic}{3,627}
\newcommand{\FASTpmdREDCDynamicCI}{174}
\newcommand{\FASTpmdREDCDynamicCIMIN}{3,453}
\newcommand{\FASTpmdREDCDynamicCIMAX}{3,801}
\newcommand{\FASTpmdCAPO}{\rna}
\newcommand{\FASTpmdCAPOCI}{\rna}
\newcommand{\FASTpmdCAPOCIMIN}{\rna}
\newcommand{\FASTpmdCAPOCIMAX}{\rna}
\newcommand{\FASTpmdCAPODynamic}{\rna}
\newcommand{\FASTpmdCAPODynamicCI}{\rna}
\newcommand{\FASTpmdCAPODynamicCIMIN}{\rna}
\newcommand{\FASTpmdCAPODynamicCIMAX}{\rna}
\newcommand{\FASTpmdFTOCAPO}{18}
\newcommand{\FASTpmdFTOCAPOCI}{0.32}
\newcommand{\FASTpmdFTOCAPOCIMIN}{18}
\newcommand{\FASTpmdFTOCAPOCIMAX}{18}
\newcommand{\FASTpmdFTOCAPODynamic}{3,910}
\newcommand{\FASTpmdFTOCAPODynamicCI}{206}
\newcommand{\FASTpmdFTOCAPODynamicCIMIN}{3,704}
\newcommand{\FASTpmdFTOCAPODynamicCIMAX}{4,116}
\newcommand{\FASTpmdRECAPO}{19}
\newcommand{\FASTpmdRECAPOCI}{0.44}
\newcommand{\FASTpmdRECAPOCIMIN}{19}
\newcommand{\FASTpmdRECAPOCIMAX}{19}
\newcommand{\FASTpmdRECAPODynamic}{3,774}
\newcommand{\FASTpmdRECAPODynamicCI}{198}
\newcommand{\FASTpmdRECAPODynamicCIMIN}{3,576}
\newcommand{\FASTpmdRECAPODynamicCIMAX}{3,972}
\newcommand{\FASTpmdAGGCAPO}{\rna}
\newcommand{\FASTpmdAGGCAPOCI}{\rna}
\newcommand{\FASTpmdAGGCAPOCIMIN}{\rna}
\newcommand{\FASTpmdAGGCAPOCIMAX}{\rna}
\newcommand{\FASTpmdAGGCAPODynamic}{\rna}
\newcommand{\FASTpmdAGGCAPODynamicCI}{\rna}
\newcommand{\FASTpmdAGGCAPODynamicCIMIN}{\rna}
\newcommand{\FASTpmdAGGCAPODynamicCIMAX}{\rna}
\newcommand{\FASTsunflowEvents}{9,700}
\newcommand{\FASTsunflowNoFPEvents}{400}
\newcommand{\sunflowHBEventTotal}{9,700}
\newcommand{\sunflowHBNoFPEventTotal}{370}
\newcommand{\sunflowHBNoFPAccessTotal}{370}
\newcommand{\sunflowHBNoFPOtherTotal}{<0.1}
\newcommand{\sunflowHBReadTotal}{99.7}
\newcommand{\sunflowHBWriteTotal}{0.26}
\newcommand{\sunflowHBNoFPAccessInCS}{0.229}
\newcommand{\sunflowHBNoFPAccessOutCS}{99.8}
\newcommand{\sunflowHBAcqRelTotal}{4.05E-4}
\newcommand{\sunflowHBOtherTotal}{5.27E-5}
\newcommand{\sunflowHBNoFPReadTotal}{370}
\newcommand{\sunflowHBReadInCS}{0.235}
\newcommand{\sunflowHBReadOutCS}{100}
\newcommand{\sunflowHBReadSameEp}{0.277}
\newcommand{\sunflowHBReadSharedSameEp}{<0.001}
\newcommand{\sunflowHBReadExclusive}{99.4}
\newcommand{\sunflowHBReadOwned}{\cna}
\newcommand{\sunflowHBReadShare}{0.0223}
\newcommand{\sunflowHBReadShared}{0.595}
\newcommand{\sunflowHBReadSharedOwned}{\cna}
\newcommand{\sunflowHBNoFPHonestWriteTotal}{0.96}
\newcommand{\sunflowHBWriteInCS}{95.1}
\newcommand{\sunflowHBWriteOutCS}{41400}
\newcommand{\sunflowHBNoFPWriteTotal}{0.96}
\newcommand{\sunflowHBWriteSameEp}{41400}
\newcommand{\sunflowHBWriteExclusive}{100}
\newcommand{\sunflowHBWriteOwned}{\cna}
\newcommand{\sunflowHBWriteShared}{<0.001}
\newcommand{\sunflowHBNoFPOtherEventTotal}{1816}
\newcommand{\sunflowHBAcqRelOtherTotal}{88.5}
\newcommand{\sunflowHBNoAcqRelOtherTotal}{209}
\newcommand{\sunflowHBFork}{13.4}
\newcommand{\sunflowHBJoin}{13.4}
\newcommand{\sunflowHBPreWait}{0.0}
\newcommand{\sunflowHBPostWait}{0.0}
\newcommand{\sunflowHBVolatileTotal}{0.0}
\newcommand{\sunflowHBClassInit}{9.09}
\newcommand{\sunflowHBClassAccess}{64.1}
\newcommand{\sunflowHBRaceTotal}{55}
\newcommand{\sunflowHBWrRdRace}{92.7}
\newcommand{\sunflowHBWrWrRace}{0.0}
\newcommand{\sunflowHBRdWrRace}{0.0}
\newcommand{\sunflowHBRdShWrRace}{7.27}
\newcommand{\sunflowHBHoldLocksTotal}{1.8}
\newcommand{\sunflowHBOneLockHeld}{0.48}
\newcommand{\sunflowHBTwoNestedLocks}{<0.1}
\newcommand{\sunflowHBThreeNestedLocks}{\cna}
\newcommand{\sunflowHBFourNestedLocks}{\cna}
\newcommand{\sunflowHBFiveNestedLocks}{\cna}
\newcommand{\sunflowHBSixNestedLocks}{\cna}
\newcommand{\sunflowHBSevenNestedLocks}{\cna}
\newcommand{\sunflowHBEightNestedLocks}{\cna}
\newcommand{\sunflowHBNineNestedLocks}{\cna}
\newcommand{\sunflowHBTenNestedLocks}{\cna}
\newcommand{\sunflowHBHundredNestedLocks}{\cna}
\newcommand{\sunflowHBExWrSet}{\ena}
\newcommand{\sunflowHBExWrCheck}{\ena}
\newcommand{\sunflowHBExWrUpdate}{\ena}
\newcommand{\sunflowHBExRdCheck}{\ena}
\newcommand{\sunflowHBExRdUpdate}{\ena}
\newcommand{\sunflowHBExTotalCheck}{\ena}
\newcommand{\sunflowHBExTotalUpdate}{\ena}
\newcommand{\sunflowFTOHBEventTotal}{9,700}
\newcommand{\sunflowFTOHBNoFPEventTotal}{3.5}
\newcommand{\sunflowFTOHBNoFPAccessTotal}{3.5}
\newcommand{\sunflowFTOHBNoFPOtherTotal}{<0.1}
\newcommand{\sunflowFTOHBReadTotal}{72.3}
\newcommand{\sunflowFTOHBWriteTotal}{27.6}
\newcommand{\sunflowFTOHBNoFPAccessInCS}{0.229}
\newcommand{\sunflowFTOHBNoFPAccessOutCS}{99.8}
\newcommand{\sunflowFTOHBAcqRelTotal}{4.05E-4}
\newcommand{\sunflowFTOHBOtherTotal}{5.27E-5}
\newcommand{\sunflowFTOHBNoFPReadTotal}{2.5}
\newcommand{\sunflowFTOHBReadInCS}{1.09}
\newcommand{\sunflowFTOHBReadOutCS}{139}
\newcommand{\sunflowFTOHBReadSameEp}{40.5}
\newcommand{\sunflowFTOHBReadSharedSameEp}{<0.001}
\newcommand{\sunflowFTOHBReadExclusive}{5.76}
\newcommand{\sunflowFTOHBReadOwned}{3.96}
\newcommand{\sunflowFTOHBReadShare}{3.27}
\newcommand{\sunflowFTOHBReadShared}{30.9}
\newcommand{\sunflowFTOHBReadSharedOwned}{56.1}
\newcommand{\sunflowFTOHBNoFPHonestWriteTotal}{0.96}
\newcommand{\sunflowFTOHBWriteInCS}{95.1}
\newcommand{\sunflowFTOHBWriteOutCS}{41400}
\newcommand{\sunflowFTOHBNoFPWriteTotal}{0.96}
\newcommand{\sunflowFTOHBWriteSameEp}{41400}
\newcommand{\sunflowFTOHBWriteExclusive}{<0.001}
\newcommand{\sunflowFTOHBWriteOwned}{100}
\newcommand{\sunflowFTOHBWriteShared}{<0.001}
\newcommand{\sunflowFTOHBNoFPOtherEventTotal}{1817}
\newcommand{\sunflowFTOHBAcqRelOtherTotal}{88.5}
\newcommand{\sunflowFTOHBNoAcqRelOtherTotal}{209}
\newcommand{\sunflowFTOHBFork}{13.4}
\newcommand{\sunflowFTOHBJoin}{13.4}
\newcommand{\sunflowFTOHBPreWait}{0.0}
\newcommand{\sunflowFTOHBPostWait}{0.0}
\newcommand{\sunflowFTOHBVolatileTotal}{0.0}
\newcommand{\sunflowFTOHBClassInit}{9.09}
\newcommand{\sunflowFTOHBClassAccess}{64.1}
\newcommand{\sunflowFTOHBRaceTotal}{51}
\newcommand{\sunflowFTOHBWrRdRace}{90.2}
\newcommand{\sunflowFTOHBWrWrRace}{0.0}
\newcommand{\sunflowFTOHBRdWrRace}{0.0}
\newcommand{\sunflowFTOHBRdShWrRace}{9.8}
\newcommand{\sunflowFTOHBHoldLocksTotal}{0.027}
\newcommand{\sunflowFTOHBOneLockHeld}{0.79}
\newcommand{\sunflowFTOHBTwoNestedLocks}{<0.1}
\newcommand{\sunflowFTOHBThreeNestedLocks}{\cna}
\newcommand{\sunflowFTOHBFourNestedLocks}{\cna}
\newcommand{\sunflowFTOHBFiveNestedLocks}{\cna}
\newcommand{\sunflowFTOHBSixNestedLocks}{\cna}
\newcommand{\sunflowFTOHBSevenNestedLocks}{\cna}
\newcommand{\sunflowFTOHBEightNestedLocks}{\cna}
\newcommand{\sunflowFTOHBNineNestedLocks}{\cna}
\newcommand{\sunflowFTOHBTenNestedLocks}{\cna}
\newcommand{\sunflowFTOHBHundredNestedLocks}{\cna}
\newcommand{\sunflowFTOHBExWrSet}{\ena}
\newcommand{\sunflowFTOHBExWrCheck}{\ena}
\newcommand{\sunflowFTOHBExWrUpdate}{\ena}
\newcommand{\sunflowFTOHBExRdCheck}{\ena}
\newcommand{\sunflowFTOHBExRdUpdate}{\ena}
\newcommand{\sunflowFTOHBExTotalCheck}{\ena}
\newcommand{\sunflowFTOHBExTotalUpdate}{\ena}
\newcommand{\sunflowFTOWCPEventTotal}{9,700}
\newcommand{\sunflowFTOWCPNoFPEventTotal}{3.5}
\newcommand{\sunflowFTOWCPNoFPAccessTotal}{3.5}
\newcommand{\sunflowFTOWCPNoFPOtherTotal}{<0.1}
\newcommand{\sunflowFTOWCPReadTotal}{72.3}
\newcommand{\sunflowFTOWCPWriteTotal}{27.6}
\newcommand{\sunflowFTOWCPNoFPAccessInCS}{0.229}
\newcommand{\sunflowFTOWCPNoFPAccessOutCS}{99.8}
\newcommand{\sunflowFTOWCPAcqRelTotal}{4.05E-4}
\newcommand{\sunflowFTOWCPOtherTotal}{5.27E-5}
\newcommand{\sunflowFTOWCPNoFPReadTotal}{2.5}
\newcommand{\sunflowFTOWCPReadInCS}{1.09}
\newcommand{\sunflowFTOWCPReadOutCS}{139}
\newcommand{\sunflowFTOWCPReadSameEp}{40.5}
\newcommand{\sunflowFTOWCPReadSharedSameEp}{<0.001}
\newcommand{\sunflowFTOWCPReadExclusive}{5.91}
\newcommand{\sunflowFTOWCPReadOwned}{3.78}
\newcommand{\sunflowFTOWCPReadShare}{3.27}
\newcommand{\sunflowFTOWCPReadShared}{31.2}
\newcommand{\sunflowFTOWCPReadSharedOwned}{55.9}
\newcommand{\sunflowFTOWCPNoFPHonestWriteTotal}{0.96}
\newcommand{\sunflowFTOWCPWriteInCS}{95.1}
\newcommand{\sunflowFTOWCPWriteOutCS}{41400}
\newcommand{\sunflowFTOWCPNoFPWriteTotal}{0.96}
\newcommand{\sunflowFTOWCPWriteSameEp}{41400}
\newcommand{\sunflowFTOWCPWriteExclusive}{<0.001}
\newcommand{\sunflowFTOWCPWriteOwned}{100}
\newcommand{\sunflowFTOWCPWriteShared}{<0.001}
\newcommand{\sunflowFTOWCPNoFPOtherEventTotal}{1818}
\newcommand{\sunflowFTOWCPAcqRelOtherTotal}{88.5}
\newcommand{\sunflowFTOWCPNoAcqRelOtherTotal}{209}
\newcommand{\sunflowFTOWCPFork}{13.4}
\newcommand{\sunflowFTOWCPJoin}{13.4}
\newcommand{\sunflowFTOWCPPreWait}{0.0}
\newcommand{\sunflowFTOWCPPostWait}{0.0}
\newcommand{\sunflowFTOWCPVolatileTotal}{0.0}
\newcommand{\sunflowFTOWCPClassInit}{9.09}
\newcommand{\sunflowFTOWCPClassAccess}{64.1}
\newcommand{\sunflowFTOWCPRaceTotal}{217}
\newcommand{\sunflowFTOWCPWrRdRace}{97.7}
\newcommand{\sunflowFTOWCPWrWrRace}{0.0}
\newcommand{\sunflowFTOWCPRdWrRace}{0.0}
\newcommand{\sunflowFTOWCPRdShWrRace}{2.3}
\newcommand{\sunflowFTOWCPHoldLocksTotal}{0.027}
\newcommand{\sunflowFTOWCPOneLockHeld}{0.78}
\newcommand{\sunflowFTOWCPTwoNestedLocks}{<0.1}
\newcommand{\sunflowFTOWCPThreeNestedLocks}{\cna}
\newcommand{\sunflowFTOWCPFourNestedLocks}{\cna}
\newcommand{\sunflowFTOWCPFiveNestedLocks}{\cna}
\newcommand{\sunflowFTOWCPSixNestedLocks}{\cna}
\newcommand{\sunflowFTOWCPSevenNestedLocks}{\cna}
\newcommand{\sunflowFTOWCPEightNestedLocks}{\cna}
\newcommand{\sunflowFTOWCPNineNestedLocks}{\cna}
\newcommand{\sunflowFTOWCPTenNestedLocks}{\cna}
\newcommand{\sunflowFTOWCPHundredNestedLocks}{\cna}
\newcommand{\sunflowFTOWCPExWrSet}{\ena}
\newcommand{\sunflowFTOWCPExWrCheck}{\ena}
\newcommand{\sunflowFTOWCPExWrUpdate}{\ena}
\newcommand{\sunflowFTOWCPExRdCheck}{\ena}
\newcommand{\sunflowFTOWCPExRdUpdate}{\ena}
\newcommand{\sunflowFTOWCPExTotalCheck}{\ena}
\newcommand{\sunflowFTOWCPExTotalUpdate}{\ena}
\newcommand{\sunflowREWCPEventTotal}{9,700}
\newcommand{\sunflowREWCPNoFPEventTotal}{3.5}
\newcommand{\sunflowREWCPNoFPAccessTotal}{3.5}
\newcommand{\sunflowREWCPNoFPOtherTotal}{<0.1}
\newcommand{\sunflowREWCPReadTotal}{72.3}
\newcommand{\sunflowREWCPWriteTotal}{27.6}
\newcommand{\sunflowREWCPNoFPAccessInCS}{0.229}
\newcommand{\sunflowREWCPNoFPAccessOutCS}{99.8}
\newcommand{\sunflowREWCPAcqRelTotal}{4.05E-4}
\newcommand{\sunflowREWCPOtherTotal}{5.27E-5}
\newcommand{\sunflowREWCPNoFPReadTotal}{2.5}
\newcommand{\sunflowREWCPReadInCS}{1.09}
\newcommand{\sunflowREWCPReadOutCS}{139}
\newcommand{\sunflowREWCPReadSameEp}{40.5}
\newcommand{\sunflowREWCPReadSharedSameEp}{<0.001}
\newcommand{\sunflowREWCPReadExclusive}{5.73}
\newcommand{\sunflowREWCPReadOwned}{3.9}
\newcommand{\sunflowREWCPReadShare}{3.27}
\newcommand{\sunflowREWCPReadShared}{31.2}
\newcommand{\sunflowREWCPReadSharedOwned}{55.8}
\newcommand{\sunflowREWCPNoFPHonestWriteTotal}{0.96}
\newcommand{\sunflowREWCPWriteInCS}{95.1}
\newcommand{\sunflowREWCPWriteOutCS}{41400}
\newcommand{\sunflowREWCPNoFPWriteTotal}{0.96}
\newcommand{\sunflowREWCPWriteSameEp}{41400}
\newcommand{\sunflowREWCPWriteExclusive}{<0.001}
\newcommand{\sunflowREWCPWriteOwned}{100}
\newcommand{\sunflowREWCPWriteShared}{0.0162}
\newcommand{\sunflowREWCPNoFPOtherEventTotal}{1816}
\newcommand{\sunflowREWCPAcqRelOtherTotal}{88.5}
\newcommand{\sunflowREWCPNoAcqRelOtherTotal}{209}
\newcommand{\sunflowREWCPFork}{13.4}
\newcommand{\sunflowREWCPJoin}{13.4}
\newcommand{\sunflowREWCPPreWait}{0.0}
\newcommand{\sunflowREWCPPostWait}{0.0}
\newcommand{\sunflowREWCPVolatileTotal}{0.0}
\newcommand{\sunflowREWCPClassInit}{9.09}
\newcommand{\sunflowREWCPClassAccess}{64.1}
\newcommand{\sunflowREWCPRaceTotal}{248}
\newcommand{\sunflowREWCPWrRdRace}{98.4}
\newcommand{\sunflowREWCPWrWrRace}{0.0}
\newcommand{\sunflowREWCPRdWrRace}{0.0}
\newcommand{\sunflowREWCPRdShWrRace}{1.61}
\newcommand{\sunflowREWCPHoldLocksTotal}{0.027}
\newcommand{\sunflowREWCPOneLockHeld}{0.78}
\newcommand{\sunflowREWCPTwoNestedLocks}{<0.1}
\newcommand{\sunflowREWCPThreeNestedLocks}{\cna}
\newcommand{\sunflowREWCPFourNestedLocks}{\cna}
\newcommand{\sunflowREWCPFiveNestedLocks}{\cna}
\newcommand{\sunflowREWCPSixNestedLocks}{\cna}
\newcommand{\sunflowREWCPSevenNestedLocks}{\cna}
\newcommand{\sunflowREWCPEightNestedLocks}{\cna}
\newcommand{\sunflowREWCPNineNestedLocks}{\cna}
\newcommand{\sunflowREWCPTenNestedLocks}{\cna}
\newcommand{\sunflowREWCPHundredNestedLocks}{\cna}
\newcommand{\sunflowREWCPExWrSet}{\ena}
\newcommand{\sunflowREWCPExWrCheck}{\ena}
\newcommand{\sunflowREWCPExWrUpdate}{\ena}
\newcommand{\sunflowREWCPExRdCheck}{\ena}
\newcommand{\sunflowREWCPExRdUpdate}{\ena}
\newcommand{\sunflowREWCPExTotalCheck}{\ena}
\newcommand{\sunflowREWCPExTotalUpdate}{\ena}
\newcommand{\sunflowFTODCEventTotal}{9,700}
\newcommand{\sunflowFTODCNoFPEventTotal}{3.5}
\newcommand{\sunflowFTODCNoFPAccessTotal}{3.5}
\newcommand{\sunflowFTODCNoFPOtherTotal}{<0.1}
\newcommand{\sunflowFTODCReadTotal}{72.3}
\newcommand{\sunflowFTODCWriteTotal}{27.6}
\newcommand{\sunflowFTODCNoFPAccessInCS}{0.229}
\newcommand{\sunflowFTODCNoFPAccessOutCS}{99.8}
\newcommand{\sunflowFTODCAcqRelTotal}{4.06E-4}
\newcommand{\sunflowFTODCOtherTotal}{5.27E-5}
\newcommand{\sunflowFTODCNoFPReadTotal}{2.5}
\newcommand{\sunflowFTODCReadInCS}{1.09}
\newcommand{\sunflowFTODCReadOutCS}{139}
\newcommand{\sunflowFTODCReadSameEp}{40.5}
\newcommand{\sunflowFTODCReadSharedSameEp}{<0.001}
\newcommand{\sunflowFTODCReadExclusive}{5.81}
\newcommand{\sunflowFTODCReadOwned}{3.86}
\newcommand{\sunflowFTODCReadShare}{3.27}
\newcommand{\sunflowFTODCReadShared}{31.1}
\newcommand{\sunflowFTODCReadSharedOwned}{56}
\newcommand{\sunflowFTODCNoFPHonestWriteTotal}{0.96}
\newcommand{\sunflowFTODCWriteInCS}{95.1}
\newcommand{\sunflowFTODCWriteOutCS}{41400}
\newcommand{\sunflowFTODCNoFPWriteTotal}{0.96}
\newcommand{\sunflowFTODCWriteSameEp}{41400}
\newcommand{\sunflowFTODCWriteExclusive}{<0.001}
\newcommand{\sunflowFTODCWriteOwned}{100}
\newcommand{\sunflowFTODCWriteShared}{<0.001}
\newcommand{\sunflowFTODCNoFPOtherEventTotal}{1819}
\newcommand{\sunflowFTODCAcqRelOtherTotal}{88.5}
\newcommand{\sunflowFTODCNoAcqRelOtherTotal}{209}
\newcommand{\sunflowFTODCFork}{13.4}
\newcommand{\sunflowFTODCJoin}{13.4}
\newcommand{\sunflowFTODCPreWait}{0.0}
\newcommand{\sunflowFTODCPostWait}{0.0}
\newcommand{\sunflowFTODCVolatileTotal}{0.0}
\newcommand{\sunflowFTODCClassInit}{9.09}
\newcommand{\sunflowFTODCClassAccess}{64.1}
\newcommand{\sunflowFTODCRaceTotal}{397}
\newcommand{\sunflowFTODCWrRdRace}{98.7}
\newcommand{\sunflowFTODCWrWrRace}{0.0}
\newcommand{\sunflowFTODCRdWrRace}{0.0}
\newcommand{\sunflowFTODCRdShWrRace}{1.26}
\newcommand{\sunflowFTODCHoldLocksTotal}{0.027}
\newcommand{\sunflowFTODCOneLockHeld}{0.78}
\newcommand{\sunflowFTODCTwoNestedLocks}{<0.1}
\newcommand{\sunflowFTODCThreeNestedLocks}{\cna}
\newcommand{\sunflowFTODCFourNestedLocks}{\cna}
\newcommand{\sunflowFTODCFiveNestedLocks}{\cna}
\newcommand{\sunflowFTODCSixNestedLocks}{\cna}
\newcommand{\sunflowFTODCSevenNestedLocks}{\cna}
\newcommand{\sunflowFTODCEightNestedLocks}{\cna}
\newcommand{\sunflowFTODCNineNestedLocks}{\cna}
\newcommand{\sunflowFTODCTenNestedLocks}{\cna}
\newcommand{\sunflowFTODCHundredNestedLocks}{\cna}
\newcommand{\sunflowFTODCExWrSet}{\ena}
\newcommand{\sunflowFTODCExWrCheck}{\ena}
\newcommand{\sunflowFTODCExWrUpdate}{\ena}
\newcommand{\sunflowFTODCExRdCheck}{\ena}
\newcommand{\sunflowFTODCExRdUpdate}{\ena}
\newcommand{\sunflowFTODCExTotalCheck}{\ena}
\newcommand{\sunflowFTODCExTotalUpdate}{\ena}
\newcommand{\sunflowREDCEventTotal}{9,700}
\newcommand{\sunflowREDCNoFPEventTotal}{3.5}
\newcommand{\sunflowREDCNoFPAccessTotal}{3.5}
\newcommand{\sunflowREDCNoFPOtherTotal}{<0.1}
\newcommand{\sunflowREDCReadTotal}{72.3}
\newcommand{\sunflowREDCWriteTotal}{27.6}
\newcommand{\sunflowREDCNoFPAccessInCS}{0.229}
\newcommand{\sunflowREDCNoFPAccessOutCS}{99.8}
\newcommand{\sunflowREDCAcqRelTotal}{4.05E-4}
\newcommand{\sunflowREDCOtherTotal}{5.27E-5}
\newcommand{\sunflowREDCNoFPReadTotal}{2.5}
\newcommand{\sunflowREDCReadInCS}{1.09}
\newcommand{\sunflowREDCReadOutCS}{139}
\newcommand{\sunflowREDCReadSameEp}{40.5}
\newcommand{\sunflowREDCReadSharedSameEp}{<0.001}
\newcommand{\sunflowREDCReadExclusive}{5.73}
\newcommand{\sunflowREDCReadOwned}{3.9}
\newcommand{\sunflowREDCReadShare}{3.28}
\newcommand{\sunflowREDCReadShared}{31.1}
\newcommand{\sunflowREDCReadSharedOwned}{56}
\newcommand{\sunflowREDCNoFPHonestWriteTotal}{0.96}
\newcommand{\sunflowREDCWriteInCS}{95.1}
\newcommand{\sunflowREDCWriteOutCS}{41400}
\newcommand{\sunflowREDCNoFPWriteTotal}{0.96}
\newcommand{\sunflowREDCWriteSameEp}{41400}
\newcommand{\sunflowREDCWriteExclusive}{<0.001}
\newcommand{\sunflowREDCWriteOwned}{100}
\newcommand{\sunflowREDCWriteShared}{0.0162}
\newcommand{\sunflowREDCNoFPOtherEventTotal}{1817}
\newcommand{\sunflowREDCAcqRelOtherTotal}{88.5}
\newcommand{\sunflowREDCNoAcqRelOtherTotal}{209}
\newcommand{\sunflowREDCFork}{13.4}
\newcommand{\sunflowREDCJoin}{13.4}
\newcommand{\sunflowREDCPreWait}{0.0}
\newcommand{\sunflowREDCPostWait}{0.0}
\newcommand{\sunflowREDCVolatileTotal}{0.0}
\newcommand{\sunflowREDCClassInit}{9.09}
\newcommand{\sunflowREDCClassAccess}{64.1}
\newcommand{\sunflowREDCRaceTotal}{412}
\newcommand{\sunflowREDCWrRdRace}{98.8}
\newcommand{\sunflowREDCWrWrRace}{0.0}
\newcommand{\sunflowREDCRdWrRace}{0.0}
\newcommand{\sunflowREDCRdShWrRace}{1.21}
\newcommand{\sunflowREDCHoldLocksTotal}{0.027}
\newcommand{\sunflowREDCOneLockHeld}{0.78}
\newcommand{\sunflowREDCTwoNestedLocks}{<0.1}
\newcommand{\sunflowREDCThreeNestedLocks}{\cna}
\newcommand{\sunflowREDCFourNestedLocks}{\cna}
\newcommand{\sunflowREDCFiveNestedLocks}{\cna}
\newcommand{\sunflowREDCSixNestedLocks}{\cna}
\newcommand{\sunflowREDCSevenNestedLocks}{\cna}
\newcommand{\sunflowREDCEightNestedLocks}{\cna}
\newcommand{\sunflowREDCNineNestedLocks}{\cna}
\newcommand{\sunflowREDCTenNestedLocks}{\cna}
\newcommand{\sunflowREDCHundredNestedLocks}{\cna}
\newcommand{\sunflowREDCExWrSet}{\ena}
\newcommand{\sunflowREDCExWrCheck}{\ena}
\newcommand{\sunflowREDCExWrUpdate}{\ena}
\newcommand{\sunflowREDCExRdCheck}{\ena}
\newcommand{\sunflowREDCExRdUpdate}{\ena}
\newcommand{\sunflowREDCExTotalCheck}{\ena}
\newcommand{\sunflowREDCExTotalUpdate}{\ena}
\newcommand{\sunflowFTOCAPOEventTotal}{9,700}
\newcommand{\sunflowFTOCAPONoFPEventTotal}{3.5}
\newcommand{\sunflowFTOCAPONoFPAccessTotal}{3.5}
\newcommand{\sunflowFTOCAPONoFPOtherTotal}{<0.1}
\newcommand{\sunflowFTOCAPOReadTotal}{72.3}
\newcommand{\sunflowFTOCAPOWriteTotal}{27.6}
\newcommand{\sunflowFTOCAPONoFPAccessInCS}{0.229}
\newcommand{\sunflowFTOCAPONoFPAccessOutCS}{99.8}
\newcommand{\sunflowFTOCAPOAcqRelTotal}{4.05E-4}
\newcommand{\sunflowFTOCAPOOtherTotal}{5.27E-5}
\newcommand{\sunflowFTOCAPONoFPReadTotal}{2.5}
\newcommand{\sunflowFTOCAPOReadInCS}{1.09}
\newcommand{\sunflowFTOCAPOReadOutCS}{139}
\newcommand{\sunflowFTOCAPOReadSameEp}{40.5}
\newcommand{\sunflowFTOCAPOReadSharedSameEp}{\cna}
\newcommand{\sunflowFTOCAPOReadExclusive}{5.9}
\newcommand{\sunflowFTOCAPOReadOwned}{3.78}
\newcommand{\sunflowFTOCAPOReadShare}{3.27}
\newcommand{\sunflowFTOCAPOReadShared}{31.1}
\newcommand{\sunflowFTOCAPOReadSharedOwned}{55.9}
\newcommand{\sunflowFTOCAPONoFPHonestWriteTotal}{0.96}
\newcommand{\sunflowFTOCAPOWriteInCS}{95.1}
\newcommand{\sunflowFTOCAPOWriteOutCS}{41400}
\newcommand{\sunflowFTOCAPONoFPWriteTotal}{0.96}
\newcommand{\sunflowFTOCAPOWriteSameEp}{41400}
\newcommand{\sunflowFTOCAPOWriteExclusive}{<0.001}
\newcommand{\sunflowFTOCAPOWriteOwned}{100}
\newcommand{\sunflowFTOCAPOWriteShared}{<0.001}
\newcommand{\sunflowFTOCAPONoFPOtherEventTotal}{1818}
\newcommand{\sunflowFTOCAPOAcqRelOtherTotal}{88.5}
\newcommand{\sunflowFTOCAPONoAcqRelOtherTotal}{209}
\newcommand{\sunflowFTOCAPOFork}{13.4}
\newcommand{\sunflowFTOCAPOJoin}{13.4}
\newcommand{\sunflowFTOCAPOPreWait}{0.0}
\newcommand{\sunflowFTOCAPOPostWait}{0.0}
\newcommand{\sunflowFTOCAPOVolatileTotal}{0.0}
\newcommand{\sunflowFTOCAPOClassInit}{9.09}
\newcommand{\sunflowFTOCAPOClassAccess}{64.1}
\newcommand{\sunflowFTOCAPORaceTotal}{397}
\newcommand{\sunflowFTOCAPOWrRdRace}{99.0}
\newcommand{\sunflowFTOCAPOWrWrRace}{0.0}
\newcommand{\sunflowFTOCAPORdWrRace}{0.0}
\newcommand{\sunflowFTOCAPORdShWrRace}{1.26}
\newcommand{\sunflowFTOCAPOHoldLocksTotal}{0.027}
\newcommand{\sunflowFTOCAPOOneLockHeld}{0.78}
\newcommand{\sunflowFTOCAPOTwoNestedLocks}{<0.1}
\newcommand{\sunflowFTOCAPOThreeNestedLocks}{\cna}
\newcommand{\sunflowFTOCAPOFourNestedLocks}{\cna}
\newcommand{\sunflowFTOCAPOFiveNestedLocks}{\cna}
\newcommand{\sunflowFTOCAPOSixNestedLocks}{\cna}
\newcommand{\sunflowFTOCAPOSevenNestedLocks}{\cna}
\newcommand{\sunflowFTOCAPOEightNestedLocks}{\cna}
\newcommand{\sunflowFTOCAPONineNestedLocks}{\cna}
\newcommand{\sunflowFTOCAPOTenNestedLocks}{\cna}
\newcommand{\sunflowFTOCAPOHundredNestedLocks}{\cna}
\newcommand{\sunflowFTOCAPOExWrSet}{\ena}
\newcommand{\sunflowFTOCAPOExWrCheck}{\ena}
\newcommand{\sunflowFTOCAPOExWrUpdate}{\ena}
\newcommand{\sunflowFTOCAPOExRdCheck}{\ena}
\newcommand{\sunflowFTOCAPOExRdUpdate}{\ena}
\newcommand{\sunflowFTOCAPOExTotalCheck}{\ena}
\newcommand{\sunflowFTOCAPOExTotalUpdate}{\ena}
\newcommand{\sunflowRECAPOEventTotal}{9,700}
\newcommand{\sunflowRECAPONoFPEventTotal}{3.5}
\newcommand{\sunflowRECAPONoFPAccessTotal}{3.5}
\newcommand{\sunflowRECAPONoFPOtherTotal}{<0.1}
\newcommand{\sunflowRECAPOReadTotal}{72.3}
\newcommand{\sunflowRECAPOWriteTotal}{27.6}
\newcommand{\sunflowRECAPONoFPAccessInCS}{0.229}
\newcommand{\sunflowRECAPONoFPAccessOutCS}{99.8}
\newcommand{\sunflowRECAPOAcqRelTotal}{4.05E-4}
\newcommand{\sunflowRECAPOOtherTotal}{5.27E-5}
\newcommand{\sunflowRECAPONoFPReadTotal}{2.5}
\newcommand{\sunflowRECAPOReadInCS}{1.09}
\newcommand{\sunflowRECAPOReadOutCS}{139}
\newcommand{\sunflowRECAPOReadSameEp}{40.5}
\newcommand{\sunflowRECAPOReadSharedSameEp}{<0.001}
\newcommand{\sunflowRECAPOReadExclusive}{5.7}
\newcommand{\sunflowRECAPOReadOwned}{3.9}
\newcommand{\sunflowRECAPOReadShare}{3.3}
\newcommand{\sunflowRECAPOReadShared}{31}
\newcommand{\sunflowRECAPOReadSharedOwned}{56}
\newcommand{\sunflowRECAPONoFPHonestWriteTotal}{0.96}
\newcommand{\sunflowRECAPOWriteInCS}{95.1}
\newcommand{\sunflowRECAPOWriteOutCS}{41400}
\newcommand{\sunflowRECAPONoFPWriteTotal}{0.96}
\newcommand{\sunflowRECAPOWriteSameEp}{41400}
\newcommand{\sunflowRECAPOWriteExclusive}{<0.001}
\newcommand{\sunflowRECAPOWriteOwned}{100}
\newcommand{\sunflowRECAPOWriteShared}{0.016}
\newcommand{\sunflowRECAPONoFPOtherEventTotal}{1817}
\newcommand{\sunflowRECAPOAcqRelOtherTotal}{88.5}
\newcommand{\sunflowRECAPONoAcqRelOtherTotal}{209}
\newcommand{\sunflowRECAPOFork}{13.4}
\newcommand{\sunflowRECAPOJoin}{13.4}
\newcommand{\sunflowRECAPOPreWait}{0.0}
\newcommand{\sunflowRECAPOPostWait}{0.0}
\newcommand{\sunflowRECAPOVolatileTotal}{0.0}
\newcommand{\sunflowRECAPOClassInit}{9.09}
\newcommand{\sunflowRECAPOClassAccess}{64.1}
\newcommand{\sunflowRECAPORaceTotal}{405}
\newcommand{\sunflowRECAPOWrRdRace}{99.0}
\newcommand{\sunflowRECAPOWrWrRace}{0.0}
\newcommand{\sunflowRECAPORdWrRace}{0.0}
\newcommand{\sunflowRECAPORdShWrRace}{1.23}
\newcommand{\sunflowRECAPOHoldLocksTotal}{0.027}
\newcommand{\sunflowRECAPOOneLockHeld}{0.78}
\newcommand{\sunflowRECAPOTwoNestedLocks}{<0.1}
\newcommand{\sunflowRECAPOThreeNestedLocks}{\cna}
\newcommand{\sunflowRECAPOFourNestedLocks}{\cna}
\newcommand{\sunflowRECAPOFiveNestedLocks}{\cna}
\newcommand{\sunflowRECAPOSixNestedLocks}{\cna}
\newcommand{\sunflowRECAPOSevenNestedLocks}{\cna}
\newcommand{\sunflowRECAPOEightNestedLocks}{\cna}
\newcommand{\sunflowRECAPONineNestedLocks}{\cna}
\newcommand{\sunflowRECAPOTenNestedLocks}{\cna}
\newcommand{\sunflowRECAPOHundredNestedLocks}{\cna}
\newcommand{\sunflowRECAPOExWrSet}{\ena}
\newcommand{\sunflowRECAPOExWrCheck}{\ena}
\newcommand{\sunflowRECAPOExWrUpdate}{\ena}
\newcommand{\sunflowRECAPOExRdCheck}{\ena}
\newcommand{\sunflowRECAPOExRdUpdate}{\ena}
\newcommand{\sunflowRECAPOExTotalCheck}{\ena}
\newcommand{\sunflowRECAPOExTotalUpdate}{\ena}
\newcommand{\FASTsunflowMaxLiveThreads}{26}
\newcommand{\FASTsunflowTotalThreads}{29}
\newcommand{\FASTsunflowBaseTime}{1.5}
\newcommand{\FASTsunflowBaseTimeCI}{110}
\newcommand{\FASTsunflowEmptyTime}{\rna}
\newcommand{\FASTsunflowEmptyTimeCI}{\rna}
\newcommand{\FASTsunflowEmptyTimeCIMIN}{\rna}
\newcommand{\FASTsunflowEmptyTimeCIMAX}{\rna}
\newcommand{\FASTsunflowFTTime}{16}
\newcommand{\FASTsunflowFTTimeCI}{0.98}
\newcommand{\FASTsunflowHBTime}{17}
\newcommand{\FASTsunflowHBTimeCI}{1.3}
\newcommand{\FASTsunflowFTOHBTime}{17}
\newcommand{\FASTsunflowFTOHBTimeCI}{1.2}
\newcommand{\FASTsunflowWCPTime}{\rna}
\newcommand{\FASTsunflowWCPTimeCI}{\rna}
\newcommand{\FASTsunflowWCPTimeCIMIN}{\rna}
\newcommand{\FASTsunflowWCPTimeCIMAX}{\rna}
\newcommand{\FASTsunflowFTOWCPTime}{18}
\newcommand{\FASTsunflowFTOWCPTimeCI}{1.3}
\newcommand{\FASTsunflowREWCPTime}{17}
\newcommand{\FASTsunflowREWCPTimeCI}{1.1}
\newcommand{\FASTsunflowDCTime}{\rna}
\newcommand{\FASTsunflowDCTimeCI}{\rna}
\newcommand{\FASTsunflowDCTimeCIMIN}{\rna}
\newcommand{\FASTsunflowDCTimeCIMAX}{\rna}
\newcommand{\FASTsunflowFTODCTime}{18}
\newcommand{\FASTsunflowFTODCTimeCI}{1.5}
\newcommand{\FASTsunflowREDCTime}{17}
\newcommand{\FASTsunflowREDCTimeCI}{1.3}
\newcommand{\FASTsunflowCAPOTime}{\rna}
\newcommand{\FASTsunflowCAPOTimeCI}{\rna}
\newcommand{\FASTsunflowCAPOTimeCIMIN}{\rna}
\newcommand{\FASTsunflowCAPOTimeCIMAX}{\rna}
\newcommand{\FASTsunflowFTOCAPOTime}{18}
\newcommand{\FASTsunflowFTOCAPOTimeCI}{1.2}
\newcommand{\FASTsunflowRECAPOTime}{17}
\newcommand{\FASTsunflowRECAPOTimeCI}{1.2}
\newcommand{\FASTsunflowAGGCAPOTime}{\rna}
\newcommand{\FASTsunflowAGGCAPOTimeCI}{\rna}
\newcommand{\FASTsunflowAGGCAPOTimeCIMIN}{\rna}
\newcommand{\FASTsunflowAGGCAPOTimeCIMAX}{\rna}
\newcommand{\FASTsunflowStaticTime}{\rzero}
\newcommand{\FASTsunflowDynamicTime}{\rzero}
\newcommand{\FASTsunflowBaseMem}{630}
\newcommand{\FASTsunflowBaseMemCI}{2.8}
\newcommand{\FASTsunflowFTMem}{8.4}
\newcommand{\FASTsunflowFTMemCI}{0.054}
\newcommand{\FASTsunflowHBMem}{8.4}
\newcommand{\FASTsunflowHBMemCI}{0.081}
\newcommand{\FASTsunflowFTOHBMem}{8.4}
\newcommand{\FASTsunflowFTOHBMemCI}{0.04}
\newcommand{\FASTsunflowWCPMem}{\memna}
\newcommand{\FASTsunflowWCPMemCI}{\memna}
\newcommand{\FASTsunflowWCPMemCIMIN}{\memna}
\newcommand{\FASTsunflowWCPMemCIMAX}{\memna}
\newcommand{\FASTsunflowFTOWCPMem}{9.0}
\newcommand{\FASTsunflowFTOWCPMemCI}{0.061}
\newcommand{\FASTsunflowREWCPMem}{15}
\newcommand{\FASTsunflowREWCPMemCI}{0.091}
\newcommand{\FASTsunflowDCMem}{\memna}
\newcommand{\FASTsunflowDCMemCI}{\memna}
\newcommand{\FASTsunflowDCMemCIMIN}{\memna}
\newcommand{\FASTsunflowDCMemCIMAX}{\memna}
\newcommand{\FASTsunflowFTODCMem}{9.0}
\newcommand{\FASTsunflowFTODCMemCI}{0.056}
\newcommand{\FASTsunflowREDCMem}{15}
\newcommand{\FASTsunflowREDCMemCI}{0.084}
\newcommand{\FASTsunflowCAPOMem}{\memna}
\newcommand{\FASTsunflowCAPOMemCI}{\memna}
\newcommand{\FASTsunflowCAPOMemCIMIN}{\memna}
\newcommand{\FASTsunflowCAPOMemCIMAX}{\memna}
\newcommand{\FASTsunflowFTOCAPOMem}{9.0}
\newcommand{\FASTsunflowFTOCAPOMemCI}{0.066}
\newcommand{\FASTsunflowRECAPOMem}{15}
\newcommand{\FASTsunflowRECAPOMemCI}{0.081}
\newcommand{\FASTsunflowAGGCAPOMem}{\memna}
\newcommand{\FASTsunflowAGGCAPOMemCI}{\memna}
\newcommand{\FASTsunflowAGGCAPOMemCIMIN}{\memna}
\newcommand{\FASTsunflowAGGCAPOMemCIMAX}{\memna}
\newcommand{\FASTsunflowEventsCI}{4}
\newcommand{\FASTsunflowEventsCIMIN}{9,681,323,545}
\newcommand{\FASTsunflowEventsCIMAX}{9,681,323,553}
\newcommand{\FASTsunflowNoFPEventsCI}{11}
\newcommand{\FASTsunflowNoFPEventsCIMIN}{396,882,620}
\newcommand{\FASTsunflowNoFPEventsCIMAX}{396,882,642}
\newcommand{\FASTsunflowFT}{5}
\newcommand{\FASTsunflowFTCI}{0.0}
\newcommand{\FASTsunflowFTCIMIN}{5}
\newcommand{\FASTsunflowFTCIMAX}{5}
\newcommand{\FASTsunflowFTDynamic}{119}
\newcommand{\FASTsunflowFTDynamicCI}{11}
\newcommand{\FASTsunflowFTDynamicCIMIN}{108}
\newcommand{\FASTsunflowFTDynamicCIMAX}{130}
\newcommand{\FASTsunflowHB}{6}
\newcommand{\FASTsunflowHBCI}{0.0}
\newcommand{\FASTsunflowHBCIMIN}{6}
\newcommand{\FASTsunflowHBCIMAX}{6}
\newcommand{\FASTsunflowHBDynamic}{55}
\newcommand{\FASTsunflowHBDynamicCI}{4.2}
\newcommand{\FASTsunflowHBDynamicCIMIN}{51}
\newcommand{\FASTsunflowHBDynamicCIMAX}{59}
\newcommand{\FASTsunflowFTOHB}{6}
\newcommand{\FASTsunflowFTOHBCI}{0.0}
\newcommand{\FASTsunflowFTOHBCIMIN}{6}
\newcommand{\FASTsunflowFTOHBCIMAX}{6}
\newcommand{\FASTsunflowFTOHBDynamic}{51}
\newcommand{\FASTsunflowFTOHBDynamicCI}{1.5}
\newcommand{\FASTsunflowFTOHBDynamicCIMIN}{49}
\newcommand{\FASTsunflowFTOHBDynamicCIMAX}{53}
\newcommand{\FASTsunflowWCP}{\rna}
\newcommand{\FASTsunflowWCPCI}{\rna}
\newcommand{\FASTsunflowWCPCIMIN}{\rna}
\newcommand{\FASTsunflowWCPCIMAX}{\rna}
\newcommand{\FASTsunflowWCPDynamic}{\rna}
\newcommand{\FASTsunflowWCPDynamicCI}{\rna}
\newcommand{\FASTsunflowWCPDynamicCIMIN}{\rna}
\newcommand{\FASTsunflowWCPDynamicCIMAX}{\rna}
\newcommand{\FASTsunflowFTOWCP}{18}
\newcommand{\FASTsunflowFTOWCPCI}{0.0}
\newcommand{\FASTsunflowFTOWCPCIMIN}{18}
\newcommand{\FASTsunflowFTOWCPCIMAX}{18}
\newcommand{\FASTsunflowFTOWCPDynamic}{217}
\newcommand{\FASTsunflowFTOWCPDynamicCI}{8.7}
\newcommand{\FASTsunflowFTOWCPDynamicCIMIN}{208}
\newcommand{\FASTsunflowFTOWCPDynamicCIMAX}{226}
\newcommand{\FASTsunflowREWCP}{19}
\newcommand{\FASTsunflowREWCPCI}{0.0}
\newcommand{\FASTsunflowREWCPCIMIN}{19}
\newcommand{\FASTsunflowREWCPCIMAX}{19}
\newcommand{\FASTsunflowREWCPDynamic}{248}
\newcommand{\FASTsunflowREWCPDynamicCI}{4.8}
\newcommand{\FASTsunflowREWCPDynamicCIMIN}{243}
\newcommand{\FASTsunflowREWCPDynamicCIMAX}{253}
\newcommand{\FASTsunflowDC}{\rna}
\newcommand{\FASTsunflowDCCI}{\rna}
\newcommand{\FASTsunflowDCCIMIN}{\rna}
\newcommand{\FASTsunflowDCCIMAX}{\rna}
\newcommand{\FASTsunflowDCDynamic}{\rna}
\newcommand{\FASTsunflowDCDynamicCI}{\rna}
\newcommand{\FASTsunflowDCDynamicCIMIN}{\rna}
\newcommand{\FASTsunflowDCDynamicCIMAX}{\rna}
\newcommand{\FASTsunflowFTODC}{19}
\newcommand{\FASTsunflowFTODCCI}{0.0}
\newcommand{\FASTsunflowFTODCCIMIN}{19}
\newcommand{\FASTsunflowFTODCCIMAX}{19}
\newcommand{\FASTsunflowFTODCDynamic}{397}
\newcommand{\FASTsunflowFTODCDynamicCI}{14}
\newcommand{\FASTsunflowFTODCDynamicCIMIN}{383}
\newcommand{\FASTsunflowFTODCDynamicCIMAX}{411}
\newcommand{\FASTsunflowREDC}{19}
\newcommand{\FASTsunflowREDCCI}{0.0}
\newcommand{\FASTsunflowREDCCIMIN}{19}
\newcommand{\FASTsunflowREDCCIMAX}{19}
\newcommand{\FASTsunflowREDCDynamic}{412}
\newcommand{\FASTsunflowREDCDynamicCI}{7.9}
\newcommand{\FASTsunflowREDCDynamicCIMIN}{404}
\newcommand{\FASTsunflowREDCDynamicCIMAX}{420}
\newcommand{\FASTsunflowCAPO}{\rna}
\newcommand{\FASTsunflowCAPOCI}{\rna}
\newcommand{\FASTsunflowCAPOCIMIN}{\rna}
\newcommand{\FASTsunflowCAPOCIMAX}{\rna}
\newcommand{\FASTsunflowCAPODynamic}{\rna}
\newcommand{\FASTsunflowCAPODynamicCI}{\rna}
\newcommand{\FASTsunflowCAPODynamicCIMIN}{\rna}
\newcommand{\FASTsunflowCAPODynamicCIMAX}{\rna}
\newcommand{\FASTsunflowFTOCAPO}{19}
\newcommand{\FASTsunflowFTOCAPOCI}{0.0}
\newcommand{\FASTsunflowFTOCAPOCIMIN}{19}
\newcommand{\FASTsunflowFTOCAPOCIMAX}{19}
\newcommand{\FASTsunflowFTOCAPODynamic}{397}
\newcommand{\FASTsunflowFTOCAPODynamicCI}{19}
\newcommand{\FASTsunflowFTOCAPODynamicCIMIN}{378}
\newcommand{\FASTsunflowFTOCAPODynamicCIMAX}{416}
\newcommand{\FASTsunflowRECAPO}{19}
\newcommand{\FASTsunflowRECAPOCI}{0.0}
\newcommand{\FASTsunflowRECAPOCIMIN}{19}
\newcommand{\FASTsunflowRECAPOCIMAX}{19}
\newcommand{\FASTsunflowRECAPODynamic}{405}
\newcommand{\FASTsunflowRECAPODynamicCI}{5.4}
\newcommand{\FASTsunflowRECAPODynamicCIMIN}{400}
\newcommand{\FASTsunflowRECAPODynamicCIMAX}{410}
\newcommand{\FASTsunflowAGGCAPO}{\rna}
\newcommand{\FASTsunflowAGGCAPOCI}{\rna}
\newcommand{\FASTsunflowAGGCAPOCIMIN}{\rna}
\newcommand{\FASTsunflowAGGCAPOCIMAX}{\rna}
\newcommand{\FASTsunflowAGGCAPODynamic}{\rna}
\newcommand{\FASTsunflowAGGCAPODynamicCI}{\rna}
\newcommand{\FASTsunflowAGGCAPODynamicCIMIN}{\rna}
\newcommand{\FASTsunflowAGGCAPODynamicCIMAX}{\rna}
\newcommand{\FASTtomcatEvents}{44}
\newcommand{\FASTtomcatNoFPEvents}{11}
\newcommand{\tomcatHBEventTotal}{44}
\newcommand{\tomcatHBNoFPEventTotal}{10}
\newcommand{\tomcatHBNoFPAccessTotal}{9.6}
\newcommand{\tomcatHBNoFPOtherTotal}{0.77}
\newcommand{\tomcatHBReadTotal}{54.9}
\newcommand{\tomcatHBWriteTotal}{37.6}
\newcommand{\tomcatHBNoFPAccessInCS}{2.84}
\newcommand{\tomcatHBNoFPAccessOutCS}{85.8}
\newcommand{\tomcatHBAcqRelTotal}{6.45}
\newcommand{\tomcatHBOtherTotal}{0.576}
\newcommand{\tomcatHBNoFPReadTotal}{5.7}
\newcommand{\tomcatHBReadInCS}{21.9}
\newcommand{\tomcatHBReadOutCS}{163}
\newcommand{\tomcatHBReadSameEp}{84.9}
\newcommand{\tomcatHBReadSharedSameEp}{<0.001}
\newcommand{\tomcatHBReadExclusive}{37.7}
\newcommand{\tomcatHBReadOwned}{\cna}
\newcommand{\tomcatHBReadShare}{0.308}
\newcommand{\tomcatHBReadShared}{61.9}
\newcommand{\tomcatHBReadSharedOwned}{\cna}
\newcommand{\tomcatHBNoFPHonestWriteTotal}{3.9}
\newcommand{\tomcatHBWriteInCS}{5.3}
\newcommand{\tomcatHBWriteOutCS}{114}
\newcommand{\tomcatHBNoFPWriteTotal}{3.9}
\newcommand{\tomcatHBWriteSameEp}{19.7}
\newcommand{\tomcatHBWriteExclusive}{65.7}
\newcommand{\tomcatHBWriteOwned}{\cna}
\newcommand{\tomcatHBWriteShared}{34.3}
\newcommand{\tomcatHBNoFPOtherEventTotal}{774756}
\newcommand{\tomcatHBAcqRelOtherTotal}{91.8}
\newcommand{\tomcatHBNoAcqRelOtherTotal}{63451}
\newcommand{\tomcatHBFork}{0.0583}
\newcommand{\tomcatHBJoin}{0.0236}
\newcommand{\tomcatHBPreWait}{0.011}
\newcommand{\tomcatHBPostWait}{0.00946}
\newcommand{\tomcatHBVolatileTotal}{96.9}
\newcommand{\tomcatHBClassInit}{0.389}
\newcommand{\tomcatHBClassAccess}{2.63}
\newcommand{\tomcatHBRaceTotal}{2575062}
\newcommand{\tomcatHBWrRdRace}{16.7}
\newcommand{\tomcatHBWrWrRace}{61.9}
\newcommand{\tomcatHBRdWrRace}{1.21}
\newcommand{\tomcatHBRdShWrRace}{20.1}
\newcommand{\tomcatHBHoldLocksTotal}{1.5}
\newcommand{\tomcatHBOneLockHeld}{<0.1}
\newcommand{\tomcatHBTwoNestedLocks}{<0.1}
\newcommand{\tomcatHBThreeNestedLocks}{<0.1}
\newcommand{\tomcatHBFourNestedLocks}{<0.1}
\newcommand{\tomcatHBFiveNestedLocks}{<0.1}
\newcommand{\tomcatHBSixNestedLocks}{<0.1}
\newcommand{\tomcatHBSevenNestedLocks}{<0.1}
\newcommand{\tomcatHBEightNestedLocks}{<0.1}
\newcommand{\tomcatHBNineNestedLocks}{<0.1}
\newcommand{\tomcatHBTenNestedLocks}{<0.1}
\newcommand{\tomcatHBHundredNestedLocks}{\cna}
\newcommand{\tomcatHBExWrSet}{\ena}
\newcommand{\tomcatHBExWrCheck}{\ena}
\newcommand{\tomcatHBExWrUpdate}{\ena}
\newcommand{\tomcatHBExRdCheck}{\ena}
\newcommand{\tomcatHBExRdUpdate}{\ena}
\newcommand{\tomcatHBExTotalCheck}{\ena}
\newcommand{\tomcatHBExTotalUpdate}{\ena}
\newcommand{\tomcatFTOHBEventTotal}{43}
\newcommand{\tomcatFTOHBNoFPEventTotal}{9.6}
\newcommand{\tomcatFTOHBNoFPAccessTotal}{8.8}
\newcommand{\tomcatFTOHBNoFPOtherTotal}{0.77}
\newcommand{\tomcatFTOHBReadTotal}{51.4}
\newcommand{\tomcatFTOHBWriteTotal}{40.5}
\newcommand{\tomcatFTOHBNoFPAccessInCS}{2.34}
\newcommand{\tomcatFTOHBNoFPAccessOutCS}{77.2}
\newcommand{\tomcatFTOHBAcqRelTotal}{6.95}
\newcommand{\tomcatFTOHBOtherTotal}{0.62}
\newcommand{\tomcatFTOHBNoFPReadTotal}{4.9}
\newcommand{\tomcatFTOHBReadInCS}{24.8}
\newcommand{\tomcatFTOHBReadOutCS}{173}
\newcommand{\tomcatFTOHBReadSameEp}{97.8}
\newcommand{\tomcatFTOHBReadSharedSameEp}{<0.001}
\newcommand{\tomcatFTOHBReadExclusive}{6.47}
\newcommand{\tomcatFTOHBReadOwned}{38.7}
\newcommand{\tomcatFTOHBReadShare}{7.05}
\newcommand{\tomcatFTOHBReadShared}{3.14}
\newcommand{\tomcatFTOHBReadSharedOwned}{44.6}
\newcommand{\tomcatFTOHBNoFPHonestWriteTotal}{3.9}
\newcommand{\tomcatFTOHBWriteInCS}{5.31}
\newcommand{\tomcatFTOHBWriteOutCS}{114}
\newcommand{\tomcatFTOHBNoFPWriteTotal}{3.9}
\newcommand{\tomcatFTOHBWriteSameEp}{19.7}
\newcommand{\tomcatFTOHBWriteExclusive}{51.2}
\newcommand{\tomcatFTOHBWriteOwned}{40.1}
\newcommand{\tomcatFTOHBWriteShared}{8.73}
\newcommand{\tomcatFTOHBNoFPOtherEventTotal}{774349}
\newcommand{\tomcatFTOHBAcqRelOtherTotal}{91.8}
\newcommand{\tomcatFTOHBNoAcqRelOtherTotal}{63464}
\newcommand{\tomcatFTOHBFork}{0.0583}
\newcommand{\tomcatFTOHBJoin}{0.0236}
\newcommand{\tomcatFTOHBPreWait}{0.011}
\newcommand{\tomcatFTOHBPostWait}{0.00945}
\newcommand{\tomcatFTOHBVolatileTotal}{96.9}
\newcommand{\tomcatFTOHBClassInit}{0.389}
\newcommand{\tomcatFTOHBClassAccess}{2.63}
\newcommand{\tomcatFTOHBRaceTotal}{1933879}
\newcommand{\tomcatFTOHBWrRdRace}{17.5}
\newcommand{\tomcatFTOHBWrWrRace}{0.0}
\newcommand{\tomcatFTOHBRdWrRace}{66.4}
\newcommand{\tomcatFTOHBRdShWrRace}{16.1}
\newcommand{\tomcatFTOHBHoldLocksTotal}{1.2}
\newcommand{\tomcatFTOHBOneLockHeld}{<0.1}
\newcommand{\tomcatFTOHBTwoNestedLocks}{<0.1}
\newcommand{\tomcatFTOHBThreeNestedLocks}{<0.1}
\newcommand{\tomcatFTOHBFourNestedLocks}{<0.1}
\newcommand{\tomcatFTOHBFiveNestedLocks}{<0.1}
\newcommand{\tomcatFTOHBSixNestedLocks}{<0.1}
\newcommand{\tomcatFTOHBSevenNestedLocks}{<0.1}
\newcommand{\tomcatFTOHBEightNestedLocks}{<0.1}
\newcommand{\tomcatFTOHBNineNestedLocks}{<0.1}
\newcommand{\tomcatFTOHBTenNestedLocks}{<0.1}
\newcommand{\tomcatFTOHBHundredNestedLocks}{\cna}
\newcommand{\tomcatFTOHBExWrSet}{\ena}
\newcommand{\tomcatFTOHBExWrCheck}{\ena}
\newcommand{\tomcatFTOHBExWrUpdate}{\ena}
\newcommand{\tomcatFTOHBExRdCheck}{\ena}
\newcommand{\tomcatFTOHBExRdUpdate}{\ena}
\newcommand{\tomcatFTOHBExTotalCheck}{\ena}
\newcommand{\tomcatFTOHBExTotalUpdate}{\ena}
\newcommand{\tomcatFTOWCPEventTotal}{43}
\newcommand{\tomcatFTOWCPNoFPEventTotal}{9.7}
\newcommand{\tomcatFTOWCPNoFPAccessTotal}{8.9}
\newcommand{\tomcatFTOWCPNoFPOtherTotal}{0.77}
\newcommand{\tomcatFTOWCPReadTotal}{51.6}
\newcommand{\tomcatFTOWCPWriteTotal}{40.3}
\newcommand{\tomcatFTOWCPNoFPAccessInCS}{2.38}
\newcommand{\tomcatFTOWCPNoFPAccessOutCS}{77.7}
\newcommand{\tomcatFTOWCPAcqRelTotal}{7.07}
\newcommand{\tomcatFTOWCPOtherTotal}{0.63}
\newcommand{\tomcatFTOWCPNoFPReadTotal}{5.0}
\newcommand{\tomcatFTOWCPReadInCS}{25.5}
\newcommand{\tomcatFTOWCPReadOutCS}{171}
\newcommand{\tomcatFTOWCPReadSameEp}{96.8}
\newcommand{\tomcatFTOWCPReadSharedSameEp}{<0.001}
\newcommand{\tomcatFTOWCPReadExclusive}{5.89}
\newcommand{\tomcatFTOWCPReadOwned}{36.9}
\newcommand{\tomcatFTOWCPReadShare}{6.73}
\newcommand{\tomcatFTOWCPReadShared}{2.98}
\newcommand{\tomcatFTOWCPReadSharedOwned}{47.5}
\newcommand{\tomcatFTOWCPNoFPHonestWriteTotal}{3.9}
\newcommand{\tomcatFTOWCPWriteInCS}{5.33}
\newcommand{\tomcatFTOWCPWriteOutCS}{114}
\newcommand{\tomcatFTOWCPNoFPWriteTotal}{3.9}
\newcommand{\tomcatFTOWCPWriteSameEp}{19.6}
\newcommand{\tomcatFTOWCPWriteExclusive}{50.1}
\newcommand{\tomcatFTOWCPWriteOwned}{41.5}
\newcommand{\tomcatFTOWCPWriteShared}{8.4}
\newcommand{\tomcatFTOWCPNoFPOtherEventTotal}{774972}
\newcommand{\tomcatFTOWCPAcqRelOtherTotal}{91.8}
\newcommand{\tomcatFTOWCPNoAcqRelOtherTotal}{63433}
\newcommand{\tomcatFTOWCPFork}{0.0583}
\newcommand{\tomcatFTOWCPJoin}{0.0236}
\newcommand{\tomcatFTOWCPPreWait}{0.011}
\newcommand{\tomcatFTOWCPPostWait}{0.00946}
\newcommand{\tomcatFTOWCPVolatileTotal}{96.9}
\newcommand{\tomcatFTOWCPClassInit}{0.389}
\newcommand{\tomcatFTOWCPClassAccess}{2.58}
\newcommand{\tomcatFTOWCPRaceTotal}{1853595}
\newcommand{\tomcatFTOWCPWrRdRace}{17.5}
\newcommand{\tomcatFTOWCPWrWrRace}{0.0}
\newcommand{\tomcatFTOWCPRdWrRace}{66.4}
\newcommand{\tomcatFTOWCPRdShWrRace}{16.0}
\newcommand{\tomcatFTOWCPHoldLocksTotal}{1.3}
\newcommand{\tomcatFTOWCPOneLockHeld}{<0.1}
\newcommand{\tomcatFTOWCPTwoNestedLocks}{<0.1}
\newcommand{\tomcatFTOWCPThreeNestedLocks}{<0.1}
\newcommand{\tomcatFTOWCPFourNestedLocks}{<0.1}
\newcommand{\tomcatFTOWCPFiveNestedLocks}{<0.1}
\newcommand{\tomcatFTOWCPSixNestedLocks}{<0.1}
\newcommand{\tomcatFTOWCPSevenNestedLocks}{<0.1}
\newcommand{\tomcatFTOWCPEightNestedLocks}{<0.1}
\newcommand{\tomcatFTOWCPNineNestedLocks}{<0.1}
\newcommand{\tomcatFTOWCPTenNestedLocks}{<0.1}
\newcommand{\tomcatFTOWCPHundredNestedLocks}{\cna}
\newcommand{\tomcatFTOWCPExWrSet}{\ena}
\newcommand{\tomcatFTOWCPExWrCheck}{\ena}
\newcommand{\tomcatFTOWCPExWrUpdate}{\ena}
\newcommand{\tomcatFTOWCPExRdCheck}{\ena}
\newcommand{\tomcatFTOWCPExRdUpdate}{\ena}
\newcommand{\tomcatFTOWCPExTotalCheck}{\ena}
\newcommand{\tomcatFTOWCPExTotalUpdate}{\ena}
\newcommand{\tomcatREWCPEventTotal}{45}
\newcommand{\tomcatREWCPNoFPEventTotal}{9.7}
\newcommand{\tomcatREWCPNoFPAccessTotal}{8.9}
\newcommand{\tomcatREWCPNoFPOtherTotal}{0.78}
\newcommand{\tomcatREWCPReadTotal}{51.8}
\newcommand{\tomcatREWCPWriteTotal}{40.2}
\newcommand{\tomcatREWCPNoFPAccessInCS}{7.6}
\newcommand{\tomcatREWCPNoFPAccessOutCS}{73.3}
\newcommand{\tomcatREWCPAcqRelTotal}{6.66}
\newcommand{\tomcatREWCPOtherTotal}{0.595}
\newcommand{\tomcatREWCPNoFPReadTotal}{5.0}
\newcommand{\tomcatREWCPReadInCS}{30.3}
\newcommand{\tomcatREWCPReadOutCS}{170}
\newcommand{\tomcatREWCPReadSameEp}{100}
\newcommand{\tomcatREWCPReadSharedSameEp}{<0.001}
\newcommand{\tomcatREWCPReadExclusive}{5.65}
\newcommand{\tomcatREWCPReadOwned}{36.8}
\newcommand{\tomcatREWCPReadShare}{6.94}
\newcommand{\tomcatREWCPReadShared}{3.07}
\newcommand{\tomcatREWCPReadSharedOwned}{47.5}
\newcommand{\tomcatREWCPNoFPHonestWriteTotal}{3.9}
\newcommand{\tomcatREWCPWriteInCS}{13.7}
\newcommand{\tomcatREWCPWriteOutCS}{114}
\newcommand{\tomcatREWCPNoFPWriteTotal}{3.9}
\newcommand{\tomcatREWCPWriteSameEp}{27.8}
\newcommand{\tomcatREWCPWriteExclusive}{50.5}
\newcommand{\tomcatREWCPWriteOwned}{40.8}
\newcommand{\tomcatREWCPWriteShared}{8.68}
\newcommand{\tomcatREWCPNoFPOtherEventTotal}{775516}
\newcommand{\tomcatREWCPAcqRelOtherTotal}{91.8}
\newcommand{\tomcatREWCPNoAcqRelOtherTotal}{63573}
\newcommand{\tomcatREWCPFork}{0.0582}
\newcommand{\tomcatREWCPJoin}{0.0236}
\newcommand{\tomcatREWCPPreWait}{0.011}
\newcommand{\tomcatREWCPPostWait}{0.00944}
\newcommand{\tomcatREWCPVolatileTotal}{96.8}
\newcommand{\tomcatREWCPClassInit}{0.434}
\newcommand{\tomcatREWCPClassAccess}{2.66}
\newcommand{\tomcatREWCPRaceTotal}{1893870}
\newcommand{\tomcatREWCPWrRdRace}{17.6}
\newcommand{\tomcatREWCPWrWrRace}{0.0}
\newcommand{\tomcatREWCPRdWrRace}{66.4}
\newcommand{\tomcatREWCPRdShWrRace}{16.1}
\newcommand{\tomcatREWCPHoldLocksTotal}{1.3}
\newcommand{\tomcatREWCPOneLockHeld}{<0.1}
\newcommand{\tomcatREWCPTwoNestedLocks}{<0.1}
\newcommand{\tomcatREWCPThreeNestedLocks}{<0.1}
\newcommand{\tomcatREWCPFourNestedLocks}{<0.1}
\newcommand{\tomcatREWCPFiveNestedLocks}{<0.1}
\newcommand{\tomcatREWCPSixNestedLocks}{<0.1}
\newcommand{\tomcatREWCPSevenNestedLocks}{<0.1}
\newcommand{\tomcatREWCPEightNestedLocks}{<0.1}
\newcommand{\tomcatREWCPNineNestedLocks}{<0.1}
\newcommand{\tomcatREWCPTenNestedLocks}{<0.1}
\newcommand{\tomcatREWCPHundredNestedLocks}{\cna}
\newcommand{\tomcatREWCPExWrSet}{4838}
\newcommand{\tomcatREWCPExWrCheck}{8415}
\newcommand{\tomcatREWCPExWrUpdate}{3}
\newcommand{\tomcatREWCPExRdCheck}{4664}
\newcommand{\tomcatREWCPExRdUpdate}{1}
\newcommand{\tomcatREWCPExTotalCheck}{13079}
\newcommand{\tomcatREWCPExTotalUpdate}{4}
\newcommand{\tomcatFTODCEventTotal}{43}
\newcommand{\tomcatFTODCNoFPEventTotal}{9.6}
\newcommand{\tomcatFTODCNoFPAccessTotal}{8.9}
\newcommand{\tomcatFTODCNoFPOtherTotal}{0.77}
\newcommand{\tomcatFTODCReadTotal}{51.6}
\newcommand{\tomcatFTODCWriteTotal}{40.3}
\newcommand{\tomcatFTODCNoFPAccessInCS}{2.39}
\newcommand{\tomcatFTODCNoFPAccessOutCS}{77.7}
\newcommand{\tomcatFTODCAcqRelTotal}{7.06}
\newcommand{\tomcatFTODCOtherTotal}{0.63}
\newcommand{\tomcatFTODCNoFPReadTotal}{5.0}
\newcommand{\tomcatFTODCReadInCS}{25.5}
\newcommand{\tomcatFTODCReadOutCS}{171}
\newcommand{\tomcatFTODCReadSameEp}{96.9}
\newcommand{\tomcatFTODCReadSharedSameEp}{<0.001}
\newcommand{\tomcatFTODCReadExclusive}{5.98}
\newcommand{\tomcatFTODCReadOwned}{36.8}
\newcommand{\tomcatFTODCReadShare}{6.73}
\newcommand{\tomcatFTODCReadShared}{2.99}
\newcommand{\tomcatFTODCReadSharedOwned}{47.5}
\newcommand{\tomcatFTODCNoFPHonestWriteTotal}{3.9}
\newcommand{\tomcatFTODCWriteInCS}{5.34}
\newcommand{\tomcatFTODCWriteOutCS}{114}
\newcommand{\tomcatFTODCNoFPWriteTotal}{3.9}
\newcommand{\tomcatFTODCWriteSameEp}{19.7}
\newcommand{\tomcatFTODCWriteExclusive}{50.3}
\newcommand{\tomcatFTODCWriteOwned}{41.3}
\newcommand{\tomcatFTODCWriteShared}{8.39}
\newcommand{\tomcatFTODCNoFPOtherEventTotal}{774361}
\newcommand{\tomcatFTODCAcqRelOtherTotal}{91.8}
\newcommand{\tomcatFTODCNoAcqRelOtherTotal}{63441}
\newcommand{\tomcatFTODCFork}{0.0583}
\newcommand{\tomcatFTODCJoin}{0.0236}
\newcommand{\tomcatFTODCPreWait}{0.011}
\newcommand{\tomcatFTODCPostWait}{0.00946}
\newcommand{\tomcatFTODCVolatileTotal}{96.9}
\newcommand{\tomcatFTODCClassInit}{0.389}
\newcommand{\tomcatFTODCClassAccess}{2.57}
\newcommand{\tomcatFTODCRaceTotal}{1852393}
\newcommand{\tomcatFTODCWrRdRace}{17.5}
\newcommand{\tomcatFTODCWrWrRace}{0.0}
\newcommand{\tomcatFTODCRdWrRace}{66.5}
\newcommand{\tomcatFTODCRdShWrRace}{16.0}
\newcommand{\tomcatFTODCHoldLocksTotal}{1.3}
\newcommand{\tomcatFTODCOneLockHeld}{<0.1}
\newcommand{\tomcatFTODCTwoNestedLocks}{<0.1}
\newcommand{\tomcatFTODCThreeNestedLocks}{<0.1}
\newcommand{\tomcatFTODCFourNestedLocks}{<0.1}
\newcommand{\tomcatFTODCFiveNestedLocks}{<0.1}
\newcommand{\tomcatFTODCSixNestedLocks}{<0.1}
\newcommand{\tomcatFTODCSevenNestedLocks}{<0.1}
\newcommand{\tomcatFTODCEightNestedLocks}{<0.1}
\newcommand{\tomcatFTODCNineNestedLocks}{<0.1}
\newcommand{\tomcatFTODCTenNestedLocks}{<0.1}
\newcommand{\tomcatFTODCHundredNestedLocks}{\cna}
\newcommand{\tomcatFTODCExWrSet}{\ena}
\newcommand{\tomcatFTODCExWrCheck}{\ena}
\newcommand{\tomcatFTODCExWrUpdate}{\ena}
\newcommand{\tomcatFTODCExRdCheck}{\ena}
\newcommand{\tomcatFTODCExRdUpdate}{\ena}
\newcommand{\tomcatFTODCExTotalCheck}{\ena}
\newcommand{\tomcatFTODCExTotalUpdate}{\ena}
\newcommand{\tomcatREDCEventTotal}{44}
\newcommand{\tomcatREDCNoFPEventTotal}{9.7}
\newcommand{\tomcatREDCNoFPAccessTotal}{8.9}
\newcommand{\tomcatREDCNoFPOtherTotal}{0.78}
\newcommand{\tomcatREDCReadTotal}{51.7}
\newcommand{\tomcatREDCWriteTotal}{40.3}
\newcommand{\tomcatREDCNoFPAccessInCS}{5.96}
\newcommand{\tomcatREDCNoFPAccessOutCS}{74.8}
\newcommand{\tomcatREDCAcqRelTotal}{6.81}
\newcommand{\tomcatREDCOtherTotal}{0.609}
\newcommand{\tomcatREDCNoFPReadTotal}{5.0}
\newcommand{\tomcatREDCReadInCS}{28.9}
\newcommand{\tomcatREDCReadOutCS}{171}
\newcommand{\tomcatREDCReadSameEp}{99.5}
\newcommand{\tomcatREDCReadSharedSameEp}{<0.001}
\newcommand{\tomcatREDCReadExclusive}{5.74}
\newcommand{\tomcatREDCReadOwned}{36.7}
\newcommand{\tomcatREDCReadShare}{6.87}
\newcommand{\tomcatREDCReadShared}{3.03}
\newcommand{\tomcatREDCReadSharedOwned}{47.7}
\newcommand{\tomcatREDCNoFPHonestWriteTotal}{3.9}
\newcommand{\tomcatREDCWriteInCS}{10.7}
\newcommand{\tomcatREDCWriteOutCS}{114}
\newcommand{\tomcatREDCNoFPWriteTotal}{3.9}
\newcommand{\tomcatREDCWriteSameEp}{24.9}
\newcommand{\tomcatREDCWriteExclusive}{50.3}
\newcommand{\tomcatREDCWriteOwned}{41.1}
\newcommand{\tomcatREDCWriteShared}{8.58}
\newcommand{\tomcatREDCNoFPOtherEventTotal}{775257}
\newcommand{\tomcatREDCAcqRelOtherTotal}{91.8}
\newcommand{\tomcatREDCNoAcqRelOtherTotal}{63629}
\newcommand{\tomcatREDCFork}{0.0581}
\newcommand{\tomcatREDCJoin}{0.0236}
\newcommand{\tomcatREDCPreWait}{0.011}
\newcommand{\tomcatREDCPostWait}{0.00943}
\newcommand{\tomcatREDCVolatileTotal}{96.9}
\newcommand{\tomcatREDCClassInit}{0.418}
\newcommand{\tomcatREDCClassAccess}{2.62}
\newcommand{\tomcatREDCRaceTotal}{1866329}
\newcommand{\tomcatREDCWrRdRace}{17.6}
\newcommand{\tomcatREDCWrWrRace}{0.0}
\newcommand{\tomcatREDCRdWrRace}{66.4}
\newcommand{\tomcatREDCRdShWrRace}{16.0}
\newcommand{\tomcatREDCHoldLocksTotal}{1.3}
\newcommand{\tomcatREDCOneLockHeld}{<0.1}
\newcommand{\tomcatREDCTwoNestedLocks}{<0.1}
\newcommand{\tomcatREDCThreeNestedLocks}{<0.1}
\newcommand{\tomcatREDCFourNestedLocks}{<0.1}
\newcommand{\tomcatREDCFiveNestedLocks}{<0.1}
\newcommand{\tomcatREDCSixNestedLocks}{<0.1}
\newcommand{\tomcatREDCSevenNestedLocks}{<0.1}
\newcommand{\tomcatREDCEightNestedLocks}{<0.1}
\newcommand{\tomcatREDCNineNestedLocks}{<0.1}
\newcommand{\tomcatREDCTenNestedLocks}{<0.1}
\newcommand{\tomcatREDCHundredNestedLocks}{\cna}
\newcommand{\tomcatREDCExWrSet}{4937}
\newcommand{\tomcatREDCExWrCheck}{8442}
\newcommand{\tomcatREDCExWrUpdate}{2}
\newcommand{\tomcatREDCExRdCheck}{4599}
\newcommand{\tomcatREDCExRdUpdate}{4}
\newcommand{\tomcatREDCExTotalCheck}{13041}
\newcommand{\tomcatREDCExTotalUpdate}{5}
\newcommand{\tomcatFTOCAPOEventTotal}{43}
\newcommand{\tomcatFTOCAPONoFPEventTotal}{9.6}
\newcommand{\tomcatFTOCAPONoFPAccessTotal}{8.9}
\newcommand{\tomcatFTOCAPONoFPOtherTotal}{0.77}
\newcommand{\tomcatFTOCAPOReadTotal}{51.6}
\newcommand{\tomcatFTOCAPOWriteTotal}{40.3}
\newcommand{\tomcatFTOCAPONoFPAccessInCS}{2.37}
\newcommand{\tomcatFTOCAPONoFPAccessOutCS}{77.6}
\newcommand{\tomcatFTOCAPOAcqRelTotal}{7.03}
\newcommand{\tomcatFTOCAPOOtherTotal}{0.627}
\newcommand{\tomcatFTOCAPONoFPReadTotal}{5.0}
\newcommand{\tomcatFTOCAPOReadInCS}{25.5}
\newcommand{\tomcatFTOCAPOReadOutCS}{171}
\newcommand{\tomcatFTOCAPOReadSameEp}{96.9}
\newcommand{\tomcatFTOCAPOReadSharedSameEp}{<0.001}
\newcommand{\tomcatFTOCAPOReadExclusive}{5.87}
\newcommand{\tomcatFTOCAPOReadOwned}{36.8}
\newcommand{\tomcatFTOCAPOReadShare}{6.8}
\newcommand{\tomcatFTOCAPOReadShared}{3.02}
\newcommand{\tomcatFTOCAPOReadSharedOwned}{47.5}
\newcommand{\tomcatFTOCAPONoFPHonestWriteTotal}{3.9}
\newcommand{\tomcatFTOCAPOWriteInCS}{5.32}
\newcommand{\tomcatFTOCAPOWriteOutCS}{114}
\newcommand{\tomcatFTOCAPONoFPWriteTotal}{3.9}
\newcommand{\tomcatFTOCAPOWriteSameEp}{19.6}
\newcommand{\tomcatFTOCAPOWriteExclusive}{50.5}
\newcommand{\tomcatFTOCAPOWriteOwned}{41}
\newcommand{\tomcatFTOCAPOWriteShared}{8.48}
\newcommand{\tomcatFTOCAPONoFPOtherEventTotal}{773986}
\newcommand{\tomcatFTOCAPOAcqRelOtherTotal}{91.8}
\newcommand{\tomcatFTOCAPONoAcqRelOtherTotal}{63375}
\newcommand{\tomcatFTOCAPOFork}{0.0584}
\newcommand{\tomcatFTOCAPOJoin}{0.0237}
\newcommand{\tomcatFTOCAPOPreWait}{0.011}
\newcommand{\tomcatFTOCAPOPostWait}{0.00947}
\newcommand{\tomcatFTOCAPOVolatileTotal}{96.9}
\newcommand{\tomcatFTOCAPOClassInit}{0.39}
\newcommand{\tomcatFTOCAPOClassAccess}{2.59}
\newcommand{\tomcatFTOCAPORaceTotal}{1870245}
\newcommand{\tomcatFTOCAPOWrRdRace}{17.6}
\newcommand{\tomcatFTOCAPOWrWrRace}{0.0}
\newcommand{\tomcatFTOCAPORdWrRace}{66.4}
\newcommand{\tomcatFTOCAPORdShWrRace}{16.0}
\newcommand{\tomcatFTOCAPOHoldLocksTotal}{1.3}
\newcommand{\tomcatFTOCAPOOneLockHeld}{<0.1}
\newcommand{\tomcatFTOCAPOTwoNestedLocks}{<0.1}
\newcommand{\tomcatFTOCAPOThreeNestedLocks}{<0.1}
\newcommand{\tomcatFTOCAPOFourNestedLocks}{<0.1}
\newcommand{\tomcatFTOCAPOFiveNestedLocks}{<0.1}
\newcommand{\tomcatFTOCAPOSixNestedLocks}{<0.1}
\newcommand{\tomcatFTOCAPOSevenNestedLocks}{<0.1}
\newcommand{\tomcatFTOCAPOEightNestedLocks}{<0.1}
\newcommand{\tomcatFTOCAPONineNestedLocks}{<0.1}
\newcommand{\tomcatFTOCAPOTenNestedLocks}{<0.1}
\newcommand{\tomcatFTOCAPOHundredNestedLocks}{\cna}
\newcommand{\tomcatFTOCAPOExWrSet}{\ena}
\newcommand{\tomcatFTOCAPOExWrCheck}{\ena}
\newcommand{\tomcatFTOCAPOExWrUpdate}{\ena}
\newcommand{\tomcatFTOCAPOExRdCheck}{\ena}
\newcommand{\tomcatFTOCAPOExRdUpdate}{\ena}
\newcommand{\tomcatFTOCAPOExTotalCheck}{\ena}
\newcommand{\tomcatFTOCAPOExTotalUpdate}{\ena}
\newcommand{\tomcatRECAPOEventTotal}{44}
\newcommand{\tomcatRECAPONoFPEventTotal}{9.7}
\newcommand{\tomcatRECAPONoFPAccessTotal}{8.9}
\newcommand{\tomcatRECAPONoFPOtherTotal}{0.78}
\newcommand{\tomcatRECAPOReadTotal}{51.7}
\newcommand{\tomcatRECAPOWriteTotal}{40.3}
\newcommand{\tomcatRECAPONoFPAccessInCS}{5.13}
\newcommand{\tomcatRECAPONoFPAccessOutCS}{74.8}
\newcommand{\tomcatRECAPOAcqRelTotal}{6.73}
\newcommand{\tomcatRECAPOOtherTotal}{0.601}
\newcommand{\tomcatRECAPONoFPReadTotal}{5.0}
\newcommand{\tomcatRECAPOReadInCS}{28.2}
\newcommand{\tomcatRECAPOReadOutCS}{171}
\newcommand{\tomcatRECAPOReadSameEp}{98.9}
\newcommand{\tomcatRECAPOReadSharedSameEp}{<0.001}
\newcommand{\tomcatRECAPOReadExclusive}{5.6}
\newcommand{\tomcatRECAPOReadOwned}{36.5}
\newcommand{\tomcatRECAPOReadShare}{7.2}
\newcommand{\tomcatRECAPOReadShared}{3.2}
\newcommand{\tomcatRECAPOReadSharedOwned}{47.6}
\newcommand{\tomcatRECAPONoFPHonestWriteTotal}{3.9}
\newcommand{\tomcatRECAPOWriteInCS}{9.61}
\newcommand{\tomcatRECAPOWriteOutCS}{114}
\newcommand{\tomcatRECAPONoFPWriteTotal}{3.9}
\newcommand{\tomcatRECAPOWriteSameEp}{23.8}
\newcommand{\tomcatRECAPOWriteExclusive}{51.4}
\newcommand{\tomcatRECAPOWriteOwned}{39.6}
\newcommand{\tomcatRECAPOWriteShared}{9.0}
\newcommand{\tomcatRECAPONoFPOtherEventTotal}{775057}
\newcommand{\tomcatRECAPOAcqRelOtherTotal}{91.8}
\newcommand{\tomcatRECAPONoAcqRelOtherTotal}{63571}
\newcommand{\tomcatRECAPOFork}{0.0582}
\newcommand{\tomcatRECAPOJoin}{0.0236}
\newcommand{\tomcatRECAPOPreWait}{0.011}
\newcommand{\tomcatRECAPOPostWait}{0.00944}
\newcommand{\tomcatRECAPOVolatileTotal}{96.8}
\newcommand{\tomcatRECAPOClassInit}{0.417}
\newcommand{\tomcatRECAPOClassAccess}{2.68}
\newcommand{\tomcatRECAPORaceTotal}{1965501}
\newcommand{\tomcatRECAPOWrRdRace}{17.5}
\newcommand{\tomcatRECAPOWrWrRace}{0.0}
\newcommand{\tomcatRECAPORdWrRace}{66.4}
\newcommand{\tomcatRECAPORdShWrRace}{16.1}
\newcommand{\tomcatRECAPOHoldLocksTotal}{1.3}
\newcommand{\tomcatRECAPOOneLockHeld}{13.1}
\newcommand{\tomcatRECAPOTwoNestedLocks}{8.0}
\newcommand{\tomcatRECAPOThreeNestedLocks}{3.9}
\newcommand{\tomcatRECAPOFourNestedLocks}{<0.1}
\newcommand{\tomcatRECAPOFiveNestedLocks}{<0.1}
\newcommand{\tomcatRECAPOSixNestedLocks}{<0.1}
\newcommand{\tomcatRECAPOSevenNestedLocks}{<0.1}
\newcommand{\tomcatRECAPOEightNestedLocks}{<0.1}
\newcommand{\tomcatRECAPONineNestedLocks}{<0.1}
\newcommand{\tomcatRECAPOTenNestedLocks}{<0.1}
\newcommand{\tomcatRECAPOHundredNestedLocks}{\cna}
\newcommand{\tomcatRECAPOExWrSet}{4976}
\newcommand{\tomcatRECAPOExWrCheck}{8292}
\newcommand{\tomcatRECAPOExWrUpdate}{2}
\newcommand{\tomcatRECAPOExRdCheck}{4543}
\newcommand{\tomcatRECAPOExRdUpdate}{2}
\newcommand{\tomcatRECAPOExTotalCheck}{0.13}
\newcommand{\tomcatRECAPOExTotalUpdate}{<0.001}
\newcommand{\FASTtomcatMaxLiveThreads}{55}
\newcommand{\FASTtomcatTotalThreads}{55}
\newcommand{\FASTtomcatBaseTime}{0.89}
\newcommand{\FASTtomcatBaseTimeCI}{5.3}
\newcommand{\FASTtomcatEmptyTime}{\rna}
\newcommand{\FASTtomcatEmptyTimeCI}{\rna}
\newcommand{\FASTtomcatEmptyTimeCIMIN}{\rna}
\newcommand{\FASTtomcatEmptyTimeCIMAX}{\rna}
\newcommand{\FASTtomcatFTTime}{31}
\newcommand{\FASTtomcatFTTimeCI}{1.1}
\newcommand{\FASTtomcatHBTime}{20}
\newcommand{\FASTtomcatHBTimeCI}{0.72}
\newcommand{\FASTtomcatFTOHBTime}{19}
\newcommand{\FASTtomcatFTOHBTimeCI}{0.58}
\newcommand{\FASTtomcatWCPTime}{\rna}
\newcommand{\FASTtomcatWCPTimeCI}{\rna}
\newcommand{\FASTtomcatWCPTimeCIMIN}{\rna}
\newcommand{\FASTtomcatWCPTimeCIMAX}{\rna}
\newcommand{\FASTtomcatFTOWCPTime}{21}
\newcommand{\FASTtomcatFTOWCPTimeCI}{0.74}
\newcommand{\FASTtomcatREWCPTime}{24}
\newcommand{\FASTtomcatREWCPTimeCI}{1.4}
\newcommand{\FASTtomcatDCTime}{\rna}
\newcommand{\FASTtomcatDCTimeCI}{\rna}
\newcommand{\FASTtomcatDCTimeCIMIN}{\rna}
\newcommand{\FASTtomcatDCTimeCIMAX}{\rna}
\newcommand{\FASTtomcatFTODCTime}{23}
\newcommand{\FASTtomcatFTODCTimeCI}{0.6}
\newcommand{\FASTtomcatREDCTime}{24}
\newcommand{\FASTtomcatREDCTimeCI}{0.88}
\newcommand{\FASTtomcatCAPOTime}{\rna}
\newcommand{\FASTtomcatCAPOTimeCI}{\rna}
\newcommand{\FASTtomcatCAPOTimeCIMIN}{\rna}
\newcommand{\FASTtomcatCAPOTimeCIMAX}{\rna}
\newcommand{\FASTtomcatFTOCAPOTime}{19}
\newcommand{\FASTtomcatFTOCAPOTimeCI}{0.6}
\newcommand{\FASTtomcatRECAPOTime}{21}
\newcommand{\FASTtomcatRECAPOTimeCI}{1.3}
\newcommand{\FASTtomcatAGGCAPOTime}{\rna}
\newcommand{\FASTtomcatAGGCAPOTimeCI}{\rna}
\newcommand{\FASTtomcatAGGCAPOTimeCIMIN}{\rna}
\newcommand{\FASTtomcatAGGCAPOTimeCIMAX}{\rna}
\newcommand{\FASTtomcatStaticTime}{\rzero}
\newcommand{\FASTtomcatDynamicTime}{\rzero}
\newcommand{\FASTtomcatBaseMem}{600}
\newcommand{\FASTtomcatBaseMemCI}{3.9}
\newcommand{\FASTtomcatFTMem}{78}
\newcommand{\FASTtomcatFTMemCI}{5.0}
\newcommand{\FASTtomcatHBMem}{55}
\newcommand{\FASTtomcatHBMemCI}{4.5}
\newcommand{\FASTtomcatFTOHBMem}{61}
\newcommand{\FASTtomcatFTOHBMemCI}{10.0}
\newcommand{\FASTtomcatWCPMem}{\memna}
\newcommand{\FASTtomcatWCPMemCI}{\memna}
\newcommand{\FASTtomcatWCPMemCIMIN}{\memna}
\newcommand{\FASTtomcatWCPMemCIMAX}{\memna}
\newcommand{\FASTtomcatFTOWCPMem}{70}
\newcommand{\FASTtomcatFTOWCPMemCI}{4.8}
\newcommand{\FASTtomcatREWCPMem}{71}
\newcommand{\FASTtomcatREWCPMemCI}{6.8}
\newcommand{\FASTtomcatDCMem}{\memna}
\newcommand{\FASTtomcatDCMemCI}{\memna}
\newcommand{\FASTtomcatDCMemCIMIN}{\memna}
\newcommand{\FASTtomcatDCMemCIMAX}{\memna}
\newcommand{\FASTtomcatFTODCMem}{70}
\newcommand{\FASTtomcatFTODCMemCI}{4.5}
\newcommand{\FASTtomcatREDCMem}{73}
\newcommand{\FASTtomcatREDCMemCI}{7.8}
\newcommand{\FASTtomcatCAPOMem}{\memna}
\newcommand{\FASTtomcatCAPOMemCI}{\memna}
\newcommand{\FASTtomcatCAPOMemCIMIN}{\memna}
\newcommand{\FASTtomcatCAPOMemCIMAX}{\memna}
\newcommand{\FASTtomcatFTOCAPOMem}{70}
\newcommand{\FASTtomcatFTOCAPOMemCI}{4.3}
\newcommand{\FASTtomcatRECAPOMem}{70}
\newcommand{\FASTtomcatRECAPOMemCI}{9.1}
\newcommand{\FASTtomcatAGGCAPOMem}{\memna}
\newcommand{\FASTtomcatAGGCAPOMemCI}{\memna}
\newcommand{\FASTtomcatAGGCAPOMemCIMIN}{\memna}
\newcommand{\FASTtomcatAGGCAPOMemCIMAX}{\memna}
\newcommand{\FASTtomcatEventsCI}{1,008,837}
\newcommand{\FASTtomcatEventsCIMIN}{43,147,719}
\newcommand{\FASTtomcatEventsCIMAX}{45,165,393}
\newcommand{\FASTtomcatNoFPEventsCI}{292,915}
\newcommand{\FASTtomcatNoFPEventsCIMIN}{10,285,256}
\newcommand{\FASTtomcatNoFPEventsCIMAX}{10,871,086}
\newcommand{\FASTtomcatFT}{425}
\newcommand{\FASTtomcatFTCI}{3.0}
\newcommand{\FASTtomcatFTCIMIN}{422}
\newcommand{\FASTtomcatFTCIMAX}{428}
\newcommand{\FASTtomcatFTDynamic}{3,600,666}
\newcommand{\FASTtomcatFTDynamicCI}{115,168}
\newcommand{\FASTtomcatFTDynamicCIMIN}{3,485,498}
\newcommand{\FASTtomcatFTDynamicCIMAX}{3,715,834}
\newcommand{\FASTtomcatHB}{640}
\newcommand{\FASTtomcatHBCI}{4.3}
\newcommand{\FASTtomcatHBCIMIN}{636}
\newcommand{\FASTtomcatHBCIMAX}{644}
\newcommand{\FASTtomcatHBDynamic}{2,079,269}
\newcommand{\FASTtomcatHBDynamicCI}{100,157}
\newcommand{\FASTtomcatHBDynamicCIMIN}{1,979,112}
\newcommand{\FASTtomcatHBDynamicCIMAX}{2,179,426}
\newcommand{\FASTtomcatFTOHB}{597}
\newcommand{\FASTtomcatFTOHBCI}{2.8}
\newcommand{\FASTtomcatFTOHBCIMIN}{594}
\newcommand{\FASTtomcatFTOHBCIMAX}{600}
\newcommand{\FASTtomcatFTOHBDynamic}{1,933,879}
\newcommand{\FASTtomcatFTOHBDynamicCI}{71,170}
\newcommand{\FASTtomcatFTOHBDynamicCIMIN}{1,862,709}
\newcommand{\FASTtomcatFTOHBDynamicCIMAX}{2,005,049}
\newcommand{\FASTtomcatWCP}{\rna}
\newcommand{\FASTtomcatWCPCI}{\rna}
\newcommand{\FASTtomcatWCPCIMIN}{\rna}
\newcommand{\FASTtomcatWCPCIMAX}{\rna}
\newcommand{\FASTtomcatWCPDynamic}{\rna}
\newcommand{\FASTtomcatWCPDynamicCI}{\rna}
\newcommand{\FASTtomcatWCPDynamicCIMIN}{\rna}
\newcommand{\FASTtomcatWCPDynamicCIMAX}{\rna}
\newcommand{\FASTtomcatFTOWCP}{600}
\newcommand{\FASTtomcatFTOWCPCI}{3.9}
\newcommand{\FASTtomcatFTOWCPCIMIN}{596}
\newcommand{\FASTtomcatFTOWCPCIMAX}{604}
\newcommand{\FASTtomcatFTOWCPDynamic}{1,853,595}
\newcommand{\FASTtomcatFTOWCPDynamicCI}{50,376}
\newcommand{\FASTtomcatFTOWCPDynamicCIMIN}{1,803,219}
\newcommand{\FASTtomcatFTOWCPDynamicCIMAX}{1,903,971}
\newcommand{\FASTtomcatREWCP}{606}
\newcommand{\FASTtomcatREWCPCI}{5.5}
\newcommand{\FASTtomcatREWCPCIMIN}{601}
\newcommand{\FASTtomcatREWCPCIMAX}{611}
\newcommand{\FASTtomcatREWCPDynamic}{1,893,870}
\newcommand{\FASTtomcatREWCPDynamicCI}{83,509}
\newcommand{\FASTtomcatREWCPDynamicCIMIN}{1,810,361}
\newcommand{\FASTtomcatREWCPDynamicCIMAX}{1,977,379}
\newcommand{\FASTtomcatDC}{\rna}
\newcommand{\FASTtomcatDCCI}{\rna}
\newcommand{\FASTtomcatDCCIMIN}{\rna}
\newcommand{\FASTtomcatDCCIMAX}{\rna}
\newcommand{\FASTtomcatDCDynamic}{\rna}
\newcommand{\FASTtomcatDCDynamicCI}{\rna}
\newcommand{\FASTtomcatDCDynamicCIMIN}{\rna}
\newcommand{\FASTtomcatDCDynamicCIMAX}{\rna}
\newcommand{\FASTtomcatFTODC}{608}
\newcommand{\FASTtomcatFTODCCI}{5.1}
\newcommand{\FASTtomcatFTODCCIMIN}{603}
\newcommand{\FASTtomcatFTODCCIMAX}{613}
\newcommand{\FASTtomcatFTODCDynamic}{1,852,366}
\newcommand{\FASTtomcatFTODCDynamicCI}{53,266}
\newcommand{\FASTtomcatFTODCDynamicCIMIN}{1,799,100}
\newcommand{\FASTtomcatFTODCDynamicCIMAX}{1,905,632}
\newcommand{\FASTtomcatREDC}{607}
\newcommand{\FASTtomcatREDCCI}{4.7}
\newcommand{\FASTtomcatREDCCIMIN}{602}
\newcommand{\FASTtomcatREDCCIMAX}{612}
\newcommand{\FASTtomcatREDCDynamic}{1,866,329}
\newcommand{\FASTtomcatREDCDynamicCI}{60,779}
\newcommand{\FASTtomcatREDCDynamicCIMIN}{1,805,550}
\newcommand{\FASTtomcatREDCDynamicCIMAX}{1,927,108}
\newcommand{\FASTtomcatCAPO}{\rna}
\newcommand{\FASTtomcatCAPOCI}{\rna}
\newcommand{\FASTtomcatCAPOCIMIN}{\rna}
\newcommand{\FASTtomcatCAPOCIMAX}{\rna}
\newcommand{\FASTtomcatCAPODynamic}{\rna}
\newcommand{\FASTtomcatCAPODynamicCI}{\rna}
\newcommand{\FASTtomcatCAPODynamicCIMIN}{\rna}
\newcommand{\FASTtomcatCAPODynamicCIMAX}{\rna}
\newcommand{\FASTtomcatFTOCAPO}{608}
\newcommand{\FASTtomcatFTOCAPOCI}{3.7}
\newcommand{\FASTtomcatFTOCAPOCIMIN}{604}
\newcommand{\FASTtomcatFTOCAPOCIMAX}{612}
\newcommand{\FASTtomcatFTOCAPODynamic}{1,870,245}
\newcommand{\FASTtomcatFTOCAPODynamicCI}{65,103}
\newcommand{\FASTtomcatFTOCAPODynamicCIMIN}{1,805,142}
\newcommand{\FASTtomcatFTOCAPODynamicCIMAX}{1,935,348}
\newcommand{\FASTtomcatRECAPO}{608}
\newcommand{\FASTtomcatRECAPOCI}{7.3}
\newcommand{\FASTtomcatRECAPOCIMIN}{601}
\newcommand{\FASTtomcatRECAPOCIMAX}{615}
\newcommand{\FASTtomcatRECAPODynamic}{1,965,501}
\newcommand{\FASTtomcatRECAPODynamicCI}{55,639}
\newcommand{\FASTtomcatRECAPODynamicCIMIN}{1,909,862}
\newcommand{\FASTtomcatRECAPODynamicCIMAX}{2,021,140}
\newcommand{\FASTtomcatAGGCAPO}{\rna}
\newcommand{\FASTtomcatAGGCAPOCI}{\rna}
\newcommand{\FASTtomcatAGGCAPOCIMIN}{\rna}
\newcommand{\FASTtomcatAGGCAPOCIMAX}{\rna}
\newcommand{\FASTtomcatAGGCAPODynamic}{\rna}
\newcommand{\FASTtomcatAGGCAPODynamicCI}{\rna}
\newcommand{\FASTtomcatAGGCAPODynamicCIMIN}{\rna}
\newcommand{\FASTtomcatAGGCAPODynamicCIMAX}{\rna}
\newcommand{\FASTxalanEvents}{630}
\newcommand{\FASTxalanNoFPEvents}{48}
\newcommand{\xalanHBEventTotal}{630}
\newcommand{\xalanHBNoFPEventTotal}{240}
\newcommand{\xalanHBNoFPAccessTotal}{230}
\newcommand{\xalanHBNoFPOtherTotal}{8.9}
\newcommand{\xalanHBReadTotal}{79.5}
\newcommand{\xalanHBWriteTotal}{16.7}
\newcommand{\xalanHBNoFPAccessInCS}{98.5}
\newcommand{\xalanHBNoFPAccessOutCS}{1.49}
\newcommand{\xalanHBAcqRelTotal}{22.1}
\newcommand{\xalanHBOtherTotal}{7.66E-4}
\newcommand{\xalanHBNoFPReadTotal}{190}
\newcommand{\xalanHBReadInCS}{110}
\newcommand{\xalanHBReadOutCS}{0.0808}
\newcommand{\xalanHBReadSameEp}{10}
\newcommand{\xalanHBReadSharedSameEp}{\cna}
\newcommand{\xalanHBReadExclusive}{85}
\newcommand{\xalanHBReadOwned}{\cna}
\newcommand{\xalanHBReadShare}{0.00197}
\newcommand{\xalanHBReadShared}{15}
\newcommand{\xalanHBReadSharedOwned}{\cna}
\newcommand{\xalanHBNoFPHonestWriteTotal}{40}
\newcommand{\xalanHBWriteInCS}{130}
\newcommand{\xalanHBWriteOutCS}{1.16}
\newcommand{\xalanHBNoFPWriteTotal}{40}
\newcommand{\xalanHBWriteSameEp}{31}
\newcommand{\xalanHBWriteExclusive}{100}
\newcommand{\xalanHBWriteOwned}{\cna}
\newcommand{\xalanHBWriteShared}{<0.001}
\newcommand{\xalanHBNoFPOtherEventTotal}{8928581}
\newcommand{\xalanHBAcqRelOtherTotal}{100.0}
\newcommand{\xalanHBNoAcqRelOtherTotal}{309}
\newcommand{\xalanHBFork}{4.53}
\newcommand{\xalanHBJoin}{4.53}
\newcommand{\xalanHBPreWait}{4.53}
\newcommand{\xalanHBPostWait}{4.53}
\newcommand{\xalanHBVolatileTotal}{0.0}
\newcommand{\xalanHBClassInit}{10.4}
\newcommand{\xalanHBClassAccess}{71.5}
\newcommand{\xalanHBRaceTotal}{2521}
\newcommand{\xalanHBWrRdRace}{98.7}
\newcommand{\xalanHBWrWrRace}{0.595}
\newcommand{\xalanHBRdWrRace}{0.0}
\newcommand{\xalanHBRdShWrRace}{0.674}
\newcommand{\xalanHBHoldLocksTotal}{260}
\newcommand{\xalanHBOneLockHeld}{113.}
\newcommand{\xalanHBTwoNestedLocks}{113.}
\newcommand{\xalanHBThreeNestedLocks}{1.17}
\newcommand{\xalanHBFourNestedLocks}{<0.1}
\newcommand{\xalanHBFiveNestedLocks}{\cna}
\newcommand{\xalanHBSixNestedLocks}{\cna}
\newcommand{\xalanHBSevenNestedLocks}{\cna}
\newcommand{\xalanHBEightNestedLocks}{\cna}
\newcommand{\xalanHBNineNestedLocks}{\cna}
\newcommand{\xalanHBTenNestedLocks}{\cna}
\newcommand{\xalanHBHundredNestedLocks}{\cna}
\newcommand{\xalanHBExWrSet}{\ena}
\newcommand{\xalanHBExWrCheck}{\ena}
\newcommand{\xalanHBExWrUpdate}{\ena}
\newcommand{\xalanHBExRdCheck}{\ena}
\newcommand{\xalanHBExRdUpdate}{\ena}
\newcommand{\xalanHBExTotalCheck}{\ena}
\newcommand{\xalanHBExTotalUpdate}{\ena}
\newcommand{\xalanFTOHBEventTotal}{630}
\newcommand{\xalanFTOHBNoFPEventTotal}{230}
\newcommand{\xalanFTOHBNoFPAccessTotal}{220}
\newcommand{\xalanFTOHBNoFPOtherTotal}{8.9}
\newcommand{\xalanFTOHBReadTotal}{79.1}
\newcommand{\xalanFTOHBWriteTotal}{17.1}
\newcommand{\xalanFTOHBNoFPAccessInCS}{98.5}
\newcommand{\xalanFTOHBNoFPAccessOutCS}{1.49}
\newcommand{\xalanFTOHBAcqRelTotal}{22.1}
\newcommand{\xalanFTOHBOtherTotal}{7.63E-4}
\newcommand{\xalanFTOHBNoFPReadTotal}{180}
\newcommand{\xalanFTOHBReadInCS}{110}
\newcommand{\xalanFTOHBReadOutCS}{0.0477}
\newcommand{\xalanFTOHBReadSameEp}{10.3}
\newcommand{\xalanFTOHBReadSharedSameEp}{<0.001}
\newcommand{\xalanFTOHBReadExclusive}{1.66}
\newcommand{\xalanFTOHBReadOwned}{82.9}
\newcommand{\xalanFTOHBReadShare}{0.00197}
\newcommand{\xalanFTOHBReadShared}{0.0234}
\newcommand{\xalanFTOHBReadSharedOwned}{15.4}
\newcommand{\xalanFTOHBNoFPHonestWriteTotal}{40}
\newcommand{\xalanFTOHBWriteInCS}{130}
\newcommand{\xalanFTOHBWriteOutCS}{1.16}
\newcommand{\xalanFTOHBNoFPWriteTotal}{40}
\newcommand{\xalanFTOHBWriteSameEp}{31}
\newcommand{\xalanFTOHBWriteExclusive}{10.6}
\newcommand{\xalanFTOHBWriteOwned}{89.4}
\newcommand{\xalanFTOHBWriteShared}{<0.001}
\newcommand{\xalanFTOHBNoFPOtherEventTotal}{8928580}
\newcommand{\xalanFTOHBAcqRelOtherTotal}{100.0}
\newcommand{\xalanFTOHBNoAcqRelOtherTotal}{308}
\newcommand{\xalanFTOHBFork}{4.55}
\newcommand{\xalanFTOHBJoin}{4.55}
\newcommand{\xalanFTOHBPreWait}{4.55}
\newcommand{\xalanFTOHBPostWait}{4.55}
\newcommand{\xalanFTOHBVolatileTotal}{0.0}
\newcommand{\xalanFTOHBClassInit}{10.4}
\newcommand{\xalanFTOHBClassAccess}{71.4}
\newcommand{\xalanFTOHBRaceTotal}{2544}
\newcommand{\xalanFTOHBWrRdRace}{99.4}
\newcommand{\xalanFTOHBWrWrRace}{0.0}
\newcommand{\xalanFTOHBRdWrRace}{0.511}
\newcommand{\xalanFTOHBRdShWrRace}{0.0786}
\newcommand{\xalanFTOHBHoldLocksTotal}{220}
\newcommand{\xalanFTOHBOneLockHeld}{99.9}
\newcommand{\xalanFTOHBTwoNestedLocks}{99.7}
\newcommand{\xalanFTOHBThreeNestedLocks}{1.10}
\newcommand{\xalanFTOHBFourNestedLocks}{\cna}
\newcommand{\xalanFTOHBFiveNestedLocks}{\cna}
\newcommand{\xalanFTOHBSixNestedLocks}{\cna}
\newcommand{\xalanFTOHBSevenNestedLocks}{\cna}
\newcommand{\xalanFTOHBEightNestedLocks}{\cna}
\newcommand{\xalanFTOHBNineNestedLocks}{\cna}
\newcommand{\xalanFTOHBTenNestedLocks}{\cna}
\newcommand{\xalanFTOHBHundredNestedLocks}{\cna}
\newcommand{\xalanFTOHBExWrSet}{\ena}
\newcommand{\xalanFTOHBExWrCheck}{\ena}
\newcommand{\xalanFTOHBExWrUpdate}{\ena}
\newcommand{\xalanFTOHBExRdCheck}{\ena}
\newcommand{\xalanFTOHBExRdUpdate}{\ena}
\newcommand{\xalanFTOHBExTotalCheck}{\ena}
\newcommand{\xalanFTOHBExTotalUpdate}{\ena}
\newcommand{\xalanFTOWCPEventTotal}{630}
\newcommand{\xalanFTOWCPNoFPEventTotal}{240}
\newcommand{\xalanFTOWCPNoFPAccessTotal}{230}
\newcommand{\xalanFTOWCPNoFPOtherTotal}{8.9}
\newcommand{\xalanFTOWCPReadTotal}{79.4}
\newcommand{\xalanFTOWCPWriteTotal}{16.8}
\newcommand{\xalanFTOWCPNoFPAccessInCS}{89.4}
\newcommand{\xalanFTOWCPNoFPAccessOutCS}{1.21}
\newcommand{\xalanFTOWCPAcqRelTotal}{18.8}
\newcommand{\xalanFTOWCPOtherTotal}{6.43E-4}
\newcommand{\xalanFTOWCPNoFPReadTotal}{190}
\newcommand{\xalanFTOWCPReadInCS}{110}
\newcommand{\xalanFTOWCPReadOutCS}{0.0467}
\newcommand{\xalanFTOWCPReadSameEp}{10.1}
\newcommand{\xalanFTOWCPReadSharedSameEp}{\cna}
\newcommand{\xalanFTOWCPReadExclusive}{0.0399}
\newcommand{\xalanFTOWCPReadOwned}{82.5}
\newcommand{\xalanFTOWCPReadShare}{0.0184}
\newcommand{\xalanFTOWCPReadShared}{0.0358}
\newcommand{\xalanFTOWCPReadSharedOwned}{17.4}
\newcommand{\xalanFTOWCPNoFPHonestWriteTotal}{40}
\newcommand{\xalanFTOWCPWriteInCS}{130}
\newcommand{\xalanFTOWCPWriteOutCS}{1.15}
\newcommand{\xalanFTOWCPNoFPWriteTotal}{40}
\newcommand{\xalanFTOWCPWriteSameEp}{31}
\newcommand{\xalanFTOWCPWriteExclusive}{10.8}
\newcommand{\xalanFTOWCPWriteOwned}{89.2}
\newcommand{\xalanFTOWCPWriteShared}{0.0726}
\newcommand{\xalanFTOWCPNoFPOtherEventTotal}{8928578}
\newcommand{\xalanFTOWCPAcqRelOtherTotal}{100.0}
\newcommand{\xalanFTOWCPNoAcqRelOtherTotal}{306}
\newcommand{\xalanFTOWCPFork}{4.58}
\newcommand{\xalanFTOWCPJoin}{4.58}
\newcommand{\xalanFTOWCPPreWait}{4.58}
\newcommand{\xalanFTOWCPPostWait}{4.58}
\newcommand{\xalanFTOWCPVolatileTotal}{0.0}
\newcommand{\xalanFTOWCPClassInit}{10.5}
\newcommand{\xalanFTOWCPClassAccess}{71.2}
\newcommand{\xalanFTOWCPRaceTotal}{3616942}
\newcommand{\xalanFTOWCPWrRdRace}{0.877}
\newcommand{\xalanFTOWCPWrWrRace}{0.0}
\newcommand{\xalanFTOWCPRdWrRace}{98.7}
\newcommand{\xalanFTOWCPRdShWrRace}{0.40}
\newcommand{\xalanFTOWCPHoldLocksTotal}{230}
\newcommand{\xalanFTOWCPOneLockHeld}{99.9}
\newcommand{\xalanFTOWCPTwoNestedLocks}{99.7}
\newcommand{\xalanFTOWCPThreeNestedLocks}{1.27}
\newcommand{\xalanFTOWCPFourNestedLocks}{\cna}
\newcommand{\xalanFTOWCPFiveNestedLocks}{\cna}
\newcommand{\xalanFTOWCPSixNestedLocks}{\cna}
\newcommand{\xalanFTOWCPSevenNestedLocks}{\cna}
\newcommand{\xalanFTOWCPEightNestedLocks}{\cna}
\newcommand{\xalanFTOWCPNineNestedLocks}{\cna}
\newcommand{\xalanFTOWCPTenNestedLocks}{\cna}
\newcommand{\xalanFTOWCPHundredNestedLocks}{\cna}
\newcommand{\xalanFTOWCPExWrSet}{\ena}
\newcommand{\xalanFTOWCPExWrCheck}{\ena}
\newcommand{\xalanFTOWCPExWrUpdate}{\ena}
\newcommand{\xalanFTOWCPExRdCheck}{\ena}
\newcommand{\xalanFTOWCPExRdUpdate}{\ena}
\newcommand{\xalanFTOWCPExTotalCheck}{\ena}
\newcommand{\xalanFTOWCPExTotalUpdate}{\ena}
\newcommand{\xalanREWCPEventTotal}{630}
\newcommand{\xalanREWCPNoFPEventTotal}{240}
\newcommand{\xalanREWCPNoFPAccessTotal}{230}
\newcommand{\xalanREWCPNoFPOtherTotal}{8.9}
\newcommand{\xalanREWCPReadTotal}{79.4}
\newcommand{\xalanREWCPWriteTotal}{16.8}
\newcommand{\xalanREWCPNoFPAccessInCS}{89.5}
\newcommand{\xalanREWCPNoFPAccessOutCS}{1.21}
\newcommand{\xalanREWCPAcqRelTotal}{18.8}
\newcommand{\xalanREWCPOtherTotal}{6.45E-4}
\newcommand{\xalanREWCPNoFPReadTotal}{190}
\newcommand{\xalanREWCPReadInCS}{110}
\newcommand{\xalanREWCPReadOutCS}{0.0467}
\newcommand{\xalanREWCPReadSameEp}{10.1}
\newcommand{\xalanREWCPReadSharedSameEp}{\cna}
\newcommand{\xalanREWCPReadExclusive}{0.0177}
\newcommand{\xalanREWCPReadOwned}{82.4}
\newcommand{\xalanREWCPReadShare}{0.0199}
\newcommand{\xalanREWCPReadShared}{0.0376}
\newcommand{\xalanREWCPReadSharedOwned}{17.5}
\newcommand{\xalanREWCPNoFPHonestWriteTotal}{40}
\newcommand{\xalanREWCPWriteInCS}{130}
\newcommand{\xalanREWCPWriteOutCS}{1.15}
\newcommand{\xalanREWCPNoFPWriteTotal}{40}
\newcommand{\xalanREWCPWriteSameEp}{31}
\newcommand{\xalanREWCPWriteExclusive}{10.8}
\newcommand{\xalanREWCPWriteOwned}{89.1}
\newcommand{\xalanREWCPWriteShared}{0.0784}
\newcommand{\xalanREWCPNoFPOtherEventTotal}{8928579}
\newcommand{\xalanREWCPAcqRelOtherTotal}{100.0}
\newcommand{\xalanREWCPNoAcqRelOtherTotal}{307}
\newcommand{\xalanREWCPFork}{4.56}
\newcommand{\xalanREWCPJoin}{4.56}
\newcommand{\xalanREWCPPreWait}{4.56}
\newcommand{\xalanREWCPPostWait}{4.56}
\newcommand{\xalanREWCPVolatileTotal}{0.0}
\newcommand{\xalanREWCPClassInit}{10.4}
\newcommand{\xalanREWCPClassAccess}{71.7}
\newcommand{\xalanREWCPRaceTotal}{3605462}
\newcommand{\xalanREWCPWrRdRace}{0.868}
\newcommand{\xalanREWCPWrWrRace}{0.0}
\newcommand{\xalanREWCPRdWrRace}{98.7}
\newcommand{\xalanREWCPRdShWrRace}{0.395}
\newcommand{\xalanREWCPHoldLocksTotal}{230}
\newcommand{\xalanREWCPOneLockHeld}{99.9}
\newcommand{\xalanREWCPTwoNestedLocks}{99.7}
\newcommand{\xalanREWCPThreeNestedLocks}{1.29}
\newcommand{\xalanREWCPFourNestedLocks}{\cna}
\newcommand{\xalanREWCPFiveNestedLocks}{\cna}
\newcommand{\xalanREWCPSixNestedLocks}{\cna}
\newcommand{\xalanREWCPSevenNestedLocks}{\cna}
\newcommand{\xalanREWCPEightNestedLocks}{\cna}
\newcommand{\xalanREWCPNineNestedLocks}{\cna}
\newcommand{\xalanREWCPTenNestedLocks}{\cna}
\newcommand{\xalanREWCPHundredNestedLocks}{\cna}
\newcommand{\xalanREWCPExWrSet}{4188293}
\newcommand{\xalanREWCPExWrCheck}{10624224}
\newcommand{\xalanREWCPExWrUpdate}{\ena}
\newcommand{\xalanREWCPExRdCheck}{4971766}
\newcommand{\xalanREWCPExRdUpdate}{\ena}
\newcommand{\xalanREWCPExTotalCheck}{15595990}
\newcommand{\xalanREWCPExTotalUpdate}{\ena}
\newcommand{\xalanFTODCEventTotal}{630}
\newcommand{\xalanFTODCNoFPEventTotal}{240}
\newcommand{\xalanFTODCNoFPAccessTotal}{230}
\newcommand{\xalanFTODCNoFPOtherTotal}{8.9}
\newcommand{\xalanFTODCReadTotal}{79.4}
\newcommand{\xalanFTODCWriteTotal}{16.8}
\newcommand{\xalanFTODCNoFPAccessInCS}{88.6}
\newcommand{\xalanFTODCNoFPAccessOutCS}{1.18}
\newcommand{\xalanFTODCAcqRelTotal}{18.5}
\newcommand{\xalanFTODCOtherTotal}{6.32E-4}
\newcommand{\xalanFTODCNoFPReadTotal}{190}
\newcommand{\xalanFTODCReadInCS}{110}
\newcommand{\xalanFTODCReadOutCS}{0.0467}
\newcommand{\xalanFTODCReadSameEp}{10.1}
\newcommand{\xalanFTODCReadSharedSameEp}{\cna}
\newcommand{\xalanFTODCReadExclusive}{0.0201}
\newcommand{\xalanFTODCReadOwned}{82.4}
\newcommand{\xalanFTODCReadShare}{0.0196}
\newcommand{\xalanFTODCReadShared}{0.0374}
\newcommand{\xalanFTODCReadSharedOwned}{17.6}
\newcommand{\xalanFTODCNoFPHonestWriteTotal}{40}
\newcommand{\xalanFTODCWriteInCS}{130}
\newcommand{\xalanFTODCWriteOutCS}{1.15}
\newcommand{\xalanFTODCNoFPWriteTotal}{40}
\newcommand{\xalanFTODCWriteSameEp}{31}
\newcommand{\xalanFTODCWriteExclusive}{10.8}
\newcommand{\xalanFTODCWriteOwned}{89.1}
\newcommand{\xalanFTODCWriteShared}{0.0778}
\newcommand{\xalanFTODCNoFPOtherEventTotal}{8928578}
\newcommand{\xalanFTODCAcqRelOtherTotal}{100.0}
\newcommand{\xalanFTODCNoAcqRelOtherTotal}{306}
\newcommand{\xalanFTODCFork}{4.58}
\newcommand{\xalanFTODCJoin}{4.58}
\newcommand{\xalanFTODCPreWait}{4.58}
\newcommand{\xalanFTODCPostWait}{4.58}
\newcommand{\xalanFTODCVolatileTotal}{0.0}
\newcommand{\xalanFTODCClassInit}{10.5}
\newcommand{\xalanFTODCClassAccess}{71.2}
\newcommand{\xalanFTODCRaceTotal}{4027027}
\newcommand{\xalanFTODCWrRdRace}{0.855}
\newcommand{\xalanFTODCWrWrRace}{0.0}
\newcommand{\xalanFTODCRdWrRace}{98.8}
\newcommand{\xalanFTODCRdShWrRace}{0.385}
\newcommand{\xalanFTODCHoldLocksTotal}{230}
\newcommand{\xalanFTODCOneLockHeld}{99.9}
\newcommand{\xalanFTODCTwoNestedLocks}{99.7}
\newcommand{\xalanFTODCThreeNestedLocks}{1.10}
\newcommand{\xalanFTODCFourNestedLocks}{\cna}
\newcommand{\xalanFTODCFiveNestedLocks}{\cna}
\newcommand{\xalanFTODCSixNestedLocks}{\cna}
\newcommand{\xalanFTODCSevenNestedLocks}{\cna}
\newcommand{\xalanFTODCEightNestedLocks}{\cna}
\newcommand{\xalanFTODCNineNestedLocks}{\cna}
\newcommand{\xalanFTODCTenNestedLocks}{\cna}
\newcommand{\xalanFTODCHundredNestedLocks}{\cna}
\newcommand{\xalanFTODCExWrSet}{\ena}
\newcommand{\xalanFTODCExWrCheck}{\ena}
\newcommand{\xalanFTODCExWrUpdate}{\ena}
\newcommand{\xalanFTODCExRdCheck}{\ena}
\newcommand{\xalanFTODCExRdUpdate}{\ena}
\newcommand{\xalanFTODCExTotalCheck}{\ena}
\newcommand{\xalanFTODCExTotalUpdate}{\ena}
\newcommand{\xalanREDCEventTotal}{630}
\newcommand{\xalanREDCNoFPEventTotal}{240}
\newcommand{\xalanREDCNoFPAccessTotal}{230}
\newcommand{\xalanREDCNoFPOtherTotal}{8.9}
\newcommand{\xalanREDCReadTotal}{79.4}
\newcommand{\xalanREDCWriteTotal}{16.8}
\newcommand{\xalanREDCNoFPAccessInCS}{88.6}
\newcommand{\xalanREDCNoFPAccessOutCS}{1.18}
\newcommand{\xalanREDCAcqRelTotal}{18.4}
\newcommand{\xalanREDCOtherTotal}{6.3E-4}
\newcommand{\xalanREDCNoFPReadTotal}{190}
\newcommand{\xalanREDCReadInCS}{110}
\newcommand{\xalanREDCReadOutCS}{0.0467}
\newcommand{\xalanREDCReadSameEp}{10.1}
\newcommand{\xalanREDCReadSharedSameEp}{\cna}
\newcommand{\xalanREDCReadExclusive}{0.018}
\newcommand{\xalanREDCReadOwned}{82.4}
\newcommand{\xalanREDCReadShare}{0.0199}
\newcommand{\xalanREDCReadShared}{0.0379}
\newcommand{\xalanREDCReadSharedOwned}{17.5}
\newcommand{\xalanREDCNoFPHonestWriteTotal}{40}
\newcommand{\xalanREDCWriteInCS}{130}
\newcommand{\xalanREDCWriteOutCS}{1.15}
\newcommand{\xalanREDCNoFPWriteTotal}{40}
\newcommand{\xalanREDCWriteSameEp}{31}
\newcommand{\xalanREDCWriteExclusive}{10.8}
\newcommand{\xalanREDCWriteOwned}{89.1}
\newcommand{\xalanREDCWriteShared}{0.0787}
\newcommand{\xalanREDCNoFPOtherEventTotal}{8928577}
\newcommand{\xalanREDCAcqRelOtherTotal}{100.0}
\newcommand{\xalanREDCNoAcqRelOtherTotal}{305}
\newcommand{\xalanREDCFork}{4.59}
\newcommand{\xalanREDCJoin}{4.59}
\newcommand{\xalanREDCPreWait}{4.59}
\newcommand{\xalanREDCPostWait}{4.59}
\newcommand{\xalanREDCVolatileTotal}{0.0}
\newcommand{\xalanREDCClassInit}{10.5}
\newcommand{\xalanREDCClassAccess}{71.5}
\newcommand{\xalanREDCRaceTotal}{4022878}
\newcommand{\xalanREDCWrRdRace}{0.858}
\newcommand{\xalanREDCWrWrRace}{0.0}
\newcommand{\xalanREDCRdWrRace}{98.8}
\newcommand{\xalanREDCRdShWrRace}{0.386}
\newcommand{\xalanREDCHoldLocksTotal}{230}
\newcommand{\xalanREDCOneLockHeld}{99.9}
\newcommand{\xalanREDCTwoNestedLocks}{99.7}
\newcommand{\xalanREDCThreeNestedLocks}{1.10}
\newcommand{\xalanREDCFourNestedLocks}{\cna}
\newcommand{\xalanREDCFiveNestedLocks}{\cna}
\newcommand{\xalanREDCSixNestedLocks}{\cna}
\newcommand{\xalanREDCSevenNestedLocks}{\cna}
\newcommand{\xalanREDCEightNestedLocks}{\cna}
\newcommand{\xalanREDCNineNestedLocks}{\cna}
\newcommand{\xalanREDCTenNestedLocks}{\cna}
\newcommand{\xalanREDCHundredNestedLocks}{\cna}
\newcommand{\xalanREDCExWrSet}{4195402}
\newcommand{\xalanREDCExWrCheck}{10624975}
\newcommand{\xalanREDCExWrUpdate}{\ena}
\newcommand{\xalanREDCExRdCheck}{4973988}
\newcommand{\xalanREDCExRdUpdate}{\ena}
\newcommand{\xalanREDCExTotalCheck}{15598964}
\newcommand{\xalanREDCExTotalUpdate}{\ena}
\newcommand{\xalanFTOCAPOEventTotal}{630}
\newcommand{\xalanFTOCAPONoFPEventTotal}{240}
\newcommand{\xalanFTOCAPONoFPAccessTotal}{230}
\newcommand{\xalanFTOCAPONoFPOtherTotal}{8.9}
\newcommand{\xalanFTOCAPOReadTotal}{79.4}
\newcommand{\xalanFTOCAPOWriteTotal}{16.8}
\newcommand{\xalanFTOCAPONoFPAccessInCS}{88.7}
\newcommand{\xalanFTOCAPONoFPAccessOutCS}{1.18}
\newcommand{\xalanFTOCAPOAcqRelTotal}{18.5}
\newcommand{\xalanFTOCAPOOtherTotal}{6.37E-4}
\newcommand{\xalanFTOCAPONoFPReadTotal}{190}
\newcommand{\xalanFTOCAPOReadInCS}{110}
\newcommand{\xalanFTOCAPOReadOutCS}{0.0467}
\newcommand{\xalanFTOCAPOReadSameEp}{10.1}
\newcommand{\xalanFTOCAPOReadSharedSameEp}{<0.001}
\newcommand{\xalanFTOCAPOReadExclusive}{0.0193}
\newcommand{\xalanFTOCAPOReadOwned}{82.4}
\newcommand{\xalanFTOCAPOReadShare}{0.0195}
\newcommand{\xalanFTOCAPOReadShared}{0.0373}
\newcommand{\xalanFTOCAPOReadSharedOwned}{17.6}
\newcommand{\xalanFTOCAPONoFPHonestWriteTotal}{40}
\newcommand{\xalanFTOCAPOWriteInCS}{130}
\newcommand{\xalanFTOCAPOWriteOutCS}{1.15}
\newcommand{\xalanFTOCAPONoFPWriteTotal}{40}
\newcommand{\xalanFTOCAPOWriteSameEp}{31}
\newcommand{\xalanFTOCAPOWriteExclusive}{10.7}
\newcommand{\xalanFTOCAPOWriteOwned}{89.2}
\newcommand{\xalanFTOCAPOWriteShared}{0.0773}
\newcommand{\xalanFTOCAPONoFPOtherEventTotal}{8928580}
\newcommand{\xalanFTOCAPOAcqRelOtherTotal}{100.0}
\newcommand{\xalanFTOCAPONoAcqRelOtherTotal}{308}
\newcommand{\xalanFTOCAPOFork}{4.55}
\newcommand{\xalanFTOCAPOJoin}{4.55}
\newcommand{\xalanFTOCAPOPreWait}{4.55}
\newcommand{\xalanFTOCAPOPostWait}{4.55}
\newcommand{\xalanFTOCAPOVolatileTotal}{0.0}
\newcommand{\xalanFTOCAPOClassInit}{10.4}
\newcommand{\xalanFTOCAPOClassAccess}{71.4}
\newcommand{\xalanFTOCAPORaceTotal}{4006697}
\newcommand{\xalanFTOCAPOWrRdRace}{0.852}
\newcommand{\xalanFTOCAPOWrWrRace}{0.0}
\newcommand{\xalanFTOCAPORdWrRace}{98.8}
\newcommand{\xalanFTOCAPORdShWrRace}{0.385}
\newcommand{\xalanFTOCAPOHoldLocksTotal}{230}
\newcommand{\xalanFTOCAPOOneLockHeld}{99.9}
\newcommand{\xalanFTOCAPOTwoNestedLocks}{99.7}
\newcommand{\xalanFTOCAPOThreeNestedLocks}{1.10}
\newcommand{\xalanFTOCAPOFourNestedLocks}{\cna}
\newcommand{\xalanFTOCAPOFiveNestedLocks}{\cna}
\newcommand{\xalanFTOCAPOSixNestedLocks}{\cna}
\newcommand{\xalanFTOCAPOSevenNestedLocks}{\cna}
\newcommand{\xalanFTOCAPOEightNestedLocks}{\cna}
\newcommand{\xalanFTOCAPONineNestedLocks}{\cna}
\newcommand{\xalanFTOCAPOTenNestedLocks}{\cna}
\newcommand{\xalanFTOCAPOHundredNestedLocks}{\cna}
\newcommand{\xalanFTOCAPOExWrSet}{\ena}
\newcommand{\xalanFTOCAPOExWrCheck}{\ena}
\newcommand{\xalanFTOCAPOExWrUpdate}{\ena}
\newcommand{\xalanFTOCAPOExRdCheck}{\ena}
\newcommand{\xalanFTOCAPOExRdUpdate}{\ena}
\newcommand{\xalanFTOCAPOExTotalCheck}{\ena}
\newcommand{\xalanFTOCAPOExTotalUpdate}{\ena}
\newcommand{\xalanRECAPOEventTotal}{630}
\newcommand{\xalanRECAPONoFPEventTotal}{240}
\newcommand{\xalanRECAPONoFPAccessTotal}{230}
\newcommand{\xalanRECAPONoFPOtherTotal}{8.9}
\newcommand{\xalanRECAPOReadTotal}{79.4}
\newcommand{\xalanRECAPOWriteTotal}{16.8}
\newcommand{\xalanRECAPONoFPAccessInCS}{88.6}
\newcommand{\xalanRECAPONoFPAccessOutCS}{1.18}
\newcommand{\xalanRECAPOAcqRelTotal}{18.4}
\newcommand{\xalanRECAPOOtherTotal}{6.36E-4}
\newcommand{\xalanRECAPONoFPReadTotal}{190}
\newcommand{\xalanRECAPOReadInCS}{110}
\newcommand{\xalanRECAPOReadOutCS}{0.0467}
\newcommand{\xalanRECAPOReadSameEp}{10.1}
\newcommand{\xalanRECAPOReadSharedSameEp}{\cna}
\newcommand{\xalanRECAPOReadExclusive}{0.012}
\newcommand{\xalanRECAPOReadOwned}{82.3}
\newcommand{\xalanRECAPOReadShare}{0.020}
\newcommand{\xalanRECAPOReadShared}{0.039}
\newcommand{\xalanRECAPOReadSharedOwned}{17.6}
\newcommand{\xalanRECAPONoFPHonestWriteTotal}{40}
\newcommand{\xalanRECAPOWriteInCS}{130}
\newcommand{\xalanRECAPOWriteOutCS}{1.15}
\newcommand{\xalanRECAPONoFPWriteTotal}{40}
\newcommand{\xalanRECAPOWriteSameEp}{31}
\newcommand{\xalanRECAPOWriteExclusive}{10.8}
\newcommand{\xalanRECAPOWriteOwned}{89.1}
\newcommand{\xalanRECAPOWriteShared}{0.079}
\newcommand{\xalanRECAPONoFPOtherEventTotal}{8928580}
\newcommand{\xalanRECAPOAcqRelOtherTotal}{100.0}
\newcommand{\xalanRECAPONoAcqRelOtherTotal}{308}
\newcommand{\xalanRECAPOFork}{4.55}
\newcommand{\xalanRECAPOJoin}{4.55}
\newcommand{\xalanRECAPOPreWait}{4.55}
\newcommand{\xalanRECAPOPostWait}{4.55}
\newcommand{\xalanRECAPOVolatileTotal}{0.0}
\newcommand{\xalanRECAPOClassInit}{10.4}
\newcommand{\xalanRECAPOClassAccess}{71.4}
\newcommand{\xalanRECAPORaceTotal}{4025914}
\newcommand{\xalanRECAPOWrRdRace}{0.85}
\newcommand{\xalanRECAPOWrWrRace}{0.0}
\newcommand{\xalanRECAPORdWrRace}{98.8}
\newcommand{\xalanRECAPORdShWrRace}{0.385}
\newcommand{\xalanRECAPOHoldLocksTotal}{230}
\newcommand{\xalanRECAPOOneLockHeld}{99.9}
\newcommand{\xalanRECAPOTwoNestedLocks}{99.7}
\newcommand{\xalanRECAPOThreeNestedLocks}{1.1}
\newcommand{\xalanRECAPOFourNestedLocks}{\cna}
\newcommand{\xalanRECAPOFiveNestedLocks}{\cna}
\newcommand{\xalanRECAPOSixNestedLocks}{\cna}
\newcommand{\xalanRECAPOSevenNestedLocks}{\cna}
\newcommand{\xalanRECAPOEightNestedLocks}{\cna}
\newcommand{\xalanRECAPONineNestedLocks}{\cna}
\newcommand{\xalanRECAPOTenNestedLocks}{\cna}
\newcommand{\xalanRECAPOHundredNestedLocks}{\cna}
\newcommand{\xalanRECAPOExWrSet}{4203486}
\newcommand{\xalanRECAPOExWrCheck}{10642976}
\newcommand{\xalanRECAPOExWrUpdate}{\ena}
\newcommand{\xalanRECAPOExRdCheck}{4976903}
\newcommand{\xalanRECAPOExRdUpdate}{\ena}
\newcommand{\xalanRECAPOExTotalCheck}{6.5}
\newcommand{\xalanRECAPOExTotalUpdate}{\ena}
\newcommand{\FASTxalanMaxLiveThreads}{15}
\newcommand{\FASTxalanTotalThreads}{15}
\newcommand{\FASTxalanBaseTime}{2.1}
\newcommand{\FASTxalanBaseTimeCI}{130}
\newcommand{\FASTxalanEmptyTime}{\rna}
\newcommand{\FASTxalanEmptyTimeCI}{\rna}
\newcommand{\FASTxalanEmptyTimeCIMIN}{\rna}
\newcommand{\FASTxalanEmptyTimeCIMAX}{\rna}
\newcommand{\FASTxalanFTTime}{4.0}
\newcommand{\FASTxalanFTTimeCI}{0.26}
\newcommand{\FASTxalanHBTime}{4.1}
\newcommand{\FASTxalanHBTimeCI}{0.29}
\newcommand{\FASTxalanFTOHBTime}{4.4}
\newcommand{\FASTxalanFTOHBTimeCI}{0.30}
\newcommand{\FASTxalanWCPTime}{\rna}
\newcommand{\FASTxalanWCPTimeCI}{\rna}
\newcommand{\FASTxalanWCPTimeCIMIN}{\rna}
\newcommand{\FASTxalanWCPTimeCIMAX}{\rna}
\newcommand{\FASTxalanFTOWCPTime}{36}
\newcommand{\FASTxalanFTOWCPTimeCI}{3.0}
\newcommand{\FASTxalanREWCPTime}{12}
\newcommand{\FASTxalanREWCPTimeCI}{0.79}
\newcommand{\FASTxalanDCTime}{\rna}
\newcommand{\FASTxalanDCTimeCI}{\rna}
\newcommand{\FASTxalanDCTimeCIMIN}{\rna}
\newcommand{\FASTxalanDCTimeCIMAX}{\rna}
\newcommand{\FASTxalanFTODCTime}{34}
\newcommand{\FASTxalanFTODCTimeCI}{2.1}
\newcommand{\FASTxalanREDCTime}{14}
\newcommand{\FASTxalanREDCTimeCI}{1.0}
\newcommand{\FASTxalanCAPOTime}{\rna}
\newcommand{\FASTxalanCAPOTimeCI}{\rna}
\newcommand{\FASTxalanCAPOTimeCIMIN}{\rna}
\newcommand{\FASTxalanCAPOTimeCIMAX}{\rna}
\newcommand{\FASTxalanFTOCAPOTime}{31}
\newcommand{\FASTxalanFTOCAPOTimeCI}{2.6}
\newcommand{\FASTxalanRECAPOTime}{10}
\newcommand{\FASTxalanRECAPOTimeCI}{0.63}
\newcommand{\FASTxalanAGGCAPOTime}{\rna}
\newcommand{\FASTxalanAGGCAPOTimeCI}{\rna}
\newcommand{\FASTxalanAGGCAPOTimeCIMIN}{\rna}
\newcommand{\FASTxalanAGGCAPOTimeCIMAX}{\rna}
\newcommand{\FASTxalanStaticTime}{\rzero}
\newcommand{\FASTxalanDynamicTime}{\rzero}
\newcommand{\FASTxalanBaseMem}{690}
\newcommand{\FASTxalanBaseMemCI}{5.3}
\newcommand{\FASTxalanFTMem}{6.3}
\newcommand{\FASTxalanFTMemCI}{0.054}
\newcommand{\FASTxalanHBMem}{6.3}
\newcommand{\FASTxalanHBMemCI}{0.054}
\newcommand{\FASTxalanFTOHBMem}{6.3}
\newcommand{\FASTxalanFTOHBMemCI}{0.051}
\newcommand{\FASTxalanWCPMem}{\memna}
\newcommand{\FASTxalanWCPMemCI}{\memna}
\newcommand{\FASTxalanWCPMemCIMIN}{\memna}
\newcommand{\FASTxalanWCPMemCIMAX}{\memna}
\newcommand{\FASTxalanFTOWCPMem}{56}
\newcommand{\FASTxalanFTOWCPMemCI}{1.8}
\newcommand{\FASTxalanREWCPMem}{39}
\newcommand{\FASTxalanREWCPMemCI}{0.7}
\newcommand{\FASTxalanDCMem}{\memna}
\newcommand{\FASTxalanDCMemCI}{\memna}
\newcommand{\FASTxalanDCMemCIMIN}{\memna}
\newcommand{\FASTxalanDCMemCIMAX}{\memna}
\newcommand{\FASTxalanFTODCMem}{58}
\newcommand{\FASTxalanFTODCMemCI}{1.1}
\newcommand{\FASTxalanREDCMem}{45}
\newcommand{\FASTxalanREDCMemCI}{1.1}
\newcommand{\FASTxalanCAPOMem}{\memna}
\newcommand{\FASTxalanCAPOMemCI}{\memna}
\newcommand{\FASTxalanCAPOMemCIMIN}{\memna}
\newcommand{\FASTxalanCAPOMemCIMAX}{\memna}
\newcommand{\FASTxalanFTOCAPOMem}{59}
\newcommand{\FASTxalanFTOCAPOMemCI}{0.87}
\newcommand{\FASTxalanRECAPOMem}{53}
\newcommand{\FASTxalanRECAPOMemCI}{0.62}
\newcommand{\FASTxalanAGGCAPOMem}{\memna}
\newcommand{\FASTxalanAGGCAPOMemCI}{\memna}
\newcommand{\FASTxalanAGGCAPOMemCIMIN}{\memna}
\newcommand{\FASTxalanAGGCAPOMemCIMAX}{\memna}
\newcommand{\FASTxalanEventsCI}{10,162}
\newcommand{\FASTxalanEventsCIMIN}{630,377,983}
\newcommand{\FASTxalanEventsCIMAX}{630,398,307}
\newcommand{\FASTxalanNoFPEventsCI}{20,343}
\newcommand{\FASTxalanNoFPEventsCIMIN}{48,384,550}
\newcommand{\FASTxalanNoFPEventsCIMAX}{48,425,236}
\newcommand{\FASTxalanFT}{8}
\newcommand{\FASTxalanFTCI}{0.59}
\newcommand{\FASTxalanFTCIMIN}{7}
\newcommand{\FASTxalanFTCIMAX}{9}
\newcommand{\FASTxalanFTDynamic}{2,822}
\newcommand{\FASTxalanFTDynamicCI}{29}
\newcommand{\FASTxalanFTDynamicCIMIN}{2,793}
\newcommand{\FASTxalanFTDynamicCIMAX}{2,851}
\newcommand{\FASTxalanHB}{8}
\newcommand{\FASTxalanHBCI}{0.0}
\newcommand{\FASTxalanHBCIMIN}{8}
\newcommand{\FASTxalanHBCIMAX}{8}
\newcommand{\FASTxalanHBDynamic}{2,506}
\newcommand{\FASTxalanHBDynamicCI}{130}
\newcommand{\FASTxalanHBDynamicCIMIN}{2,376}
\newcommand{\FASTxalanHBDynamicCIMAX}{2,636}
\newcommand{\FASTxalanFTOHB}{8}
\newcommand{\FASTxalanFTOHBCI}{0.82}
\newcommand{\FASTxalanFTOHBCIMIN}{7}
\newcommand{\FASTxalanFTOHBCIMAX}{9}
\newcommand{\FASTxalanFTOHBDynamic}{2,544}
\newcommand{\FASTxalanFTOHBDynamicCI}{73}
\newcommand{\FASTxalanFTOHBDynamicCIMIN}{2,471}
\newcommand{\FASTxalanFTOHBDynamicCIMAX}{2,617}
\newcommand{\FASTxalanWCP}{\rna}
\newcommand{\FASTxalanWCPCI}{\rna}
\newcommand{\FASTxalanWCPCIMIN}{\rna}
\newcommand{\FASTxalanWCPCIMAX}{\rna}
\newcommand{\FASTxalanWCPDynamic}{\rna}
\newcommand{\FASTxalanWCPDynamicCI}{\rna}
\newcommand{\FASTxalanWCPDynamicCIMIN}{\rna}
\newcommand{\FASTxalanWCPDynamicCIMAX}{\rna}
\newcommand{\FASTxalanFTOWCP}{42}
\newcommand{\FASTxalanFTOWCPCI}{1.1}
\newcommand{\FASTxalanFTOWCPCIMIN}{41}
\newcommand{\FASTxalanFTOWCPCIMAX}{43}
\newcommand{\FASTxalanFTOWCPDynamic}{3,616,942}
\newcommand{\FASTxalanFTOWCPDynamicCI}{9,924}
\newcommand{\FASTxalanFTOWCPDynamicCIMIN}{3,607,018}
\newcommand{\FASTxalanFTOWCPDynamicCIMAX}{3,626,866}
\newcommand{\FASTxalanREWCP}{49}
\newcommand{\FASTxalanREWCPCI}{1.0}
\newcommand{\FASTxalanREWCPCIMIN}{48}
\newcommand{\FASTxalanREWCPCIMAX}{50}
\newcommand{\FASTxalanREWCPDynamic}{3,605,462}
\newcommand{\FASTxalanREWCPDynamicCI}{13,016}
\newcommand{\FASTxalanREWCPDynamicCIMIN}{3,592,446}
\newcommand{\FASTxalanREWCPDynamicCIMAX}{3,618,478}
\newcommand{\FASTxalanDC}{\rna}
\newcommand{\FASTxalanDCCI}{\rna}
\newcommand{\FASTxalanDCCIMIN}{\rna}
\newcommand{\FASTxalanDCCIMAX}{\rna}
\newcommand{\FASTxalanDCDynamic}{\rna}
\newcommand{\FASTxalanDCDynamicCI}{\rna}
\newcommand{\FASTxalanDCDynamicCIMIN}{\rna}
\newcommand{\FASTxalanDCDynamicCIMAX}{\rna}
\newcommand{\FASTxalanFTODC}{54}
\newcommand{\FASTxalanFTODCCI}{0.32}
\newcommand{\FASTxalanFTODCCIMIN}{54}
\newcommand{\FASTxalanFTODCCIMAX}{54}
\newcommand{\FASTxalanFTODCDynamic}{4,027,027}
\newcommand{\FASTxalanFTODCDynamicCI}{9,173}
\newcommand{\FASTxalanFTODCDynamicCIMIN}{4,017,854}
\newcommand{\FASTxalanFTODCDynamicCIMAX}{4,036,200}
\newcommand{\FASTxalanREDC}{53}
\newcommand{\FASTxalanREDCCI}{0.74}
\newcommand{\FASTxalanREDCCIMIN}{52}
\newcommand{\FASTxalanREDCCIMAX}{54}
\newcommand{\FASTxalanREDCDynamic}{4,022,878}
\newcommand{\FASTxalanREDCDynamicCI}{14,505}
\newcommand{\FASTxalanREDCDynamicCIMIN}{4,008,373}
\newcommand{\FASTxalanREDCDynamicCIMAX}{4,037,383}
\newcommand{\FASTxalanCAPO}{\rna}
\newcommand{\FASTxalanCAPOCI}{\rna}
\newcommand{\FASTxalanCAPOCIMIN}{\rna}
\newcommand{\FASTxalanCAPOCIMAX}{\rna}
\newcommand{\FASTxalanCAPODynamic}{\rna}
\newcommand{\FASTxalanCAPODynamicCI}{\rna}
\newcommand{\FASTxalanCAPODynamicCIMIN}{\rna}
\newcommand{\FASTxalanCAPODynamicCIMAX}{\rna}
\newcommand{\FASTxalanFTOCAPO}{53}
\newcommand{\FASTxalanFTOCAPOCI}{0.42}
\newcommand{\FASTxalanFTOCAPOCIMIN}{53}
\newcommand{\FASTxalanFTOCAPOCIMAX}{53}
\newcommand{\FASTxalanFTOCAPODynamic}{4,006,697}
\newcommand{\FASTxalanFTOCAPODynamicCI}{10,411}
\newcommand{\FASTxalanFTOCAPODynamicCIMIN}{3,996,286}
\newcommand{\FASTxalanFTOCAPODynamicCIMAX}{4,017,108}
\newcommand{\FASTxalanRECAPO}{54}
\newcommand{\FASTxalanRECAPOCI}{0.59}
\newcommand{\FASTxalanRECAPOCIMIN}{53}
\newcommand{\FASTxalanRECAPOCIMAX}{55}
\newcommand{\FASTxalanRECAPODynamic}{4,025,914}
\newcommand{\FASTxalanRECAPODynamicCI}{10,175}
\newcommand{\FASTxalanRECAPODynamicCIMIN}{4,015,739}
\newcommand{\FASTxalanRECAPODynamicCIMAX}{4,036,089}
\newcommand{\FASTxalanAGGCAPO}{\rna}
\newcommand{\FASTxalanAGGCAPOCI}{\rna}
\newcommand{\FASTxalanAGGCAPOCIMIN}{\rna}
\newcommand{\FASTxalanAGGCAPOCIMAX}{\rna}
\newcommand{\FASTxalanAGGCAPODynamic}{\rna}
\newcommand{\FASTxalanAGGCAPODynamicCI}{\rna}
\newcommand{\FASTxalanAGGCAPODynamicCIMIN}{\rna}
\newcommand{\FASTxalanAGGCAPODynamicCIMAX}{\rna}
\newcommand{\FASTBaseTimeGeoMean}{1900}
\newcommand{\FASTEmptyTimeGeoMean}{\rna}
\newcommand{\FASTFTTimeGeoMean}{8.5}
\newcommand{\FASTHBTimeGeoMean}{8.2}
\newcommand{\FASTFTOHBTimeGeoMean}{8.3}
\newcommand{\FASTWCPTimeGeoMean}{\rna}
\newcommand{\FASTFTOWCPTimeGeoMean}{16}
\newcommand{\FASTREWCPTimeGeoMean}{10}
\newcommand{\FASTDCTimeGeoMean}{\rna}
\newcommand{\FASTFTODCTimeGeoMean}{16}
\newcommand{\FASTREDCTimeGeoMean}{11}
\newcommand{\FASTCAPOTimeGeoMean}{\rna}
\newcommand{\FASTFTOCAPOTimeGeoMean}{15}
\newcommand{\FASTRECAPOTimeGeoMean}{9.6}
\newcommand{\FASTAGGCAPOTimeGeoMean}{\rna}
\newcommand{\FASTBaseMemGeoMean}{540}
\newcommand{\FASTEmptyMemGeoMean}{0.0}
\newcommand{\FASTFTMemGeoMean}{7.9}
\newcommand{\FASTHBMemGeoMean}{7.4}
\newcommand{\FASTFTOHBMemGeoMean}{7.5}
\newcommand{\FASTWCPMemGeoMean}{\memna}
\newcommand{\FASTFTOWCPMemGeoMean}{18}
\newcommand{\FASTREWCPMemGeoMean}{12}
\newcommand{\FASTDCMemGeoMean}{\memna}
\newcommand{\FASTFTODCMemGeoMean}{18}
\newcommand{\FASTREDCMemGeoMean}{12}
\newcommand{\FASTCAPOMemGeoMean}{\memna}
\newcommand{\FASTFTOCAPOMemGeoMean}{17}
\newcommand{\FASTRECAPOMemGeoMean}{11}
\newcommand{\FASTAGGCAPOMemGeoMean}{\memna}
\newcommand{\FASTWCPDynamicTotal}{0}
\newcommand{\FASTFTODCTotal}{746}
\newcommand{\FASTCAPOTotal}{0}
\newcommand{\FASTWCPTotal}{0}
\newcommand{\FASTAGGCAPODynamicTotal}{0}
\newcommand{\FASTRECAPOTotal}{747}
\newcommand{\FASTFTTotal}{492}
\newcommand{\FASTHBDynamicTotal}{2,592,581}
\newcommand{\FASTREDCTotal}{747}
\newcommand{\FASTREWCPTotal}{736}
\newcommand{\FASTFTOWCPDynamicTotal}{5,998,474}
\newcommand{\FASTRECAPODynamicTotal}{6,524,875}
\newcommand{\FASTFTOWCPTotal}{717}
\newcommand{\FASTAGGCAPOTotal}{0}
\newcommand{\FASTCAPODynamicTotal}{0}
\newcommand{\FASTFTOCAPOTotal}{746}
\newcommand{\FASTFTODCDynamicTotal}{6,408,146}
\newcommand{\FASTFTDynamicTotal}{4,470,757}
\newcommand{\FASTREWCPDynamicTotal}{6,025,423}
\newcommand{\FASTREDCDynamicTotal}{6,416,406}
\newcommand{\FASTFTOCAPODynamicTotal}{6,414,147}
\newcommand{\FASTFTOHBDynamicTotal}{2,464,043}
\newcommand{\FASTHBTotal}{716}
\newcommand{\FASTFTOHBTotal}{673}
\newcommand{\FASTDCDynamicTotal}{0}
\newcommand{\FASTDCTotal}{0}

%% file: relatedwork.tex
\section{Related Work}
\label{Sec:related}

This section considers
prior work \emph{other than} happens-before (HB) and partial-order-based predictive analyses 
discussed in Section~\ref{sec:background}~\cite{fasttrack,fasttrack2,fib,causally-precedes,dighr,pavlogiannis-2019,raptor,wcp,vindicator,google-tsan-v1,google-tsan-v2,intel-inspector,multirace,goldilocks-pldi-2007}.


Our recent work introduces two partial-order-based analyses,
\emph{strong-dependently-precedes (SDP)} and \emph{weak-dependently-pre\-cedes (WDP)} analyses,
that have more precise notions of dependence than \WCP and \DC analyses,
respectively~\cite{depaware}.
SDP and WDP do not generally order write--write conflicting critical sections,
making it challenging to apply epoch and CCS optimizations to these analyses.
\notes{Applying \STFull to WDP analysis presents similar challenges;
furthermore, the benefits of optimizing WDP analysis are limited because it report false races regularly in practice and thus
relies on costly construction of a constraint graph during analysis to support vindication.}%

An alternative to partial-order-based predictive analysis is \emph{SMT-based approaches},
which encode reordering constraints as SMT constraints~\cite{rvpredict-pldi-2014,said-nfm-2011,ipa,rdit-oopsla-2016,jpredictor,maximal-causal-models}.
However, the number of constraints and the solving time scale superlinearly with trace length,
so prior work analyzes bounded windows of execution, typically missing races that are more than a few thousand events apart.
Prior work shows that a predictable race's accesses may be millions of events apart~\cite{vindicator,depaware}.

An alternative to \HB analysis is \emph{lockset analysis}, which detects races that violate a
locking discipline, but inherently reports false
races~\cite{dinning-schonberg,ocallahan-hybrid-racedet-2003,eraser,object-racedet-2001,choi-racedet-2002,racedet-escape-nishiyama-2004}.
Hybrid lockset--HB lockset analyses typically incur the disadvantages of at least one
kind of analysis~\cite{ocallahan-hybrid-racedet-2003,racetrack,multirace}.

A sound, non-predictive alternative to \HB analysis is analyses that detect or infer simultaneous conflicting regions or
accesses~\cite{frost,valor-oopsla-2015,ifrit,datacollider,racefuzzer,racechaser-caper-cc-2017}.

Dynamic race detection analyses can target \emph{production runs} by trading race coverage for
performance~\cite{literace,pacer-2010,racemob,datacollider,racechaser-caper-cc-2017,racez,prorace}
or using custom hardware~\cite{radish,zhou-hard,lard,clean-isca-2015,parsnip}.

\emph{Static analysis} can detect all data races in
all possible executions of a
program~\cite{naik-static-racedet-2006,naik-static-racedet-2007,locksmith,racerx,relay-2007},
but for real programs, it reports thousands of false races~\cite{racechaser-caper-cc-2017,chimera}.

\emph{RacerD} and \emph{RacerDX} are recent static race detectors that find few false races in practice~\cite{racerd,racerdx}.
RacerDX provably reports no false races under a set of well-defined assumptions~\cite{racerdx}.
However, these assumptions are not realistic; for example, RacerDX reports false races for well-synchronized programs
that violate a locking discipline~\cite{racerdx-personal}.
The RacerDX evaluation uses a few of the same programs as our evaluation, but the results are incomparable
because the papers use different methodology for counting distinct races.
\notes{while our evaluation counts races as distinct static code locations at which races are detected,
the RacerDX evaluation counts races as pairs of distinct ``access paths''~\cite{racerdx}.}%

\emph{Schedule exploration} approaches execute programs multiple times
using either systematic exploration (often called \emph{model checking})
or using heuristics~\cite{mcr,mcr-s,chess,randomized-scheduler,racageddon,racefuzzer,drfinder,racedet-model-checking-2004}.
Schedule exploration is complementary with predictive analysis,
which seeks to find more races in a given schedule.

%% file: conclusion.tex
\section{Conclusion}
\label{sec:conclusion}

This paper's contributions---notably SmartTrack's novel conflicting critical section (CCS) optimizations---enable
predictive race detectors to perform nearly as well as state-of-the-art non-predictive race detectors.
\STFull's optimizations are applicable to existing predictive analyses and to this paper's new \CAPO analysis,
offering compelling new options in the performance--detection space.
This work substantially improves the performance of predictive race detection analyses,
making a case for predictive analysis to be the prevailing approach for detecting data races.


%% file: acks.tex
\begin{acks}

Thanks to Yufan Xu for early help with this project;
Steve Freund for help with RoadRunner; and
Ilya Sergey and Peter O'Hearn
for discussions about RacerDX.
Thanks to the anonymous reviewers and the paper's shepherd, Grigore Ro{\c{s}}u,
for many suggestions that helped improve the paper.

This material is based upon work supported by the National Science Foundation
under Grants CAREER-1253703, CCF-1421612, and XPS-1629126.

\end{acks}

%% file: correctness.tex
\section{Correctness}

\mike{This section is just notes, not intended for the paper or extended version.
We don't plan to prove correctness.
We might try to provide a stronger correctness argument, but it probably wouldn't look like this.}

It seems like the main thing to prove is that \RE is a
sound and complete predictive race detector.
For example, prove that \REDC is a sound and complete \DC-race detector.

\begin{theorem}[Soundness of \REDC]
If \REDC reports a race, then the execution trace has a \DC-race.
\end{theorem}

\begin{proof}
Suppose \REDC reports a race for an execution trace \tr, but \tr has no \DC-race.

\REDC must report at line~\ref{??} in \sc{MultiCheck}.
\dots
\end{proof}

\begin{theorem}[Completeness of \REDC]
If an execution trace has a \DC-race, \REDC reports a race.
\end{theorem}

\begin{proof}
Suppose an execution trace \tr has a \DC-race, but \REDC reports no race for \tr.
\dots
\end{proof}

The ancillary metadata seems like one of the biggest challenges in the proofs.
More generally, we may need to prove that every step of the analysis maintains
invariants such as representing \DC order up to that point (like in the Raptor~\cite{raptor} proofs).
At each point in the trace, the analysis metdata (including ancillary metadata)
represents \DC ordering for each variable to the current point in the trace.
At a high level it's proof by induction on the elements in the trace.

Or: William Mansky and Joe Devietti had a paper about this.
There's also work work by Steve and Cormac, I think.
One of those is called VerifiedFT?
Did one or both of those papers use mechnical theorem proving (Coq)?
I chatted with William and Joe at PLDI 2019 and they mentioned they might be interested
in proving \RE correct.
Not sure why they needed to prove FastTrack correct, considering that
the FastTrack paper had a correctness proof? But that proof was with respect
to operational semantics, and it was not extremely rigorous, so maybe the newer verification
work is with respect to and it's highly rigorous since it's with respect
to an algorithm or implementation and it's a machine-checked/-generated proof?

%% file: ci-results.tex
\iftoggle{twoColumnText}{
\iftoggle{includeRaceResults}{}{
\section{Detailed Performance Results}
\label{appendix:detailed-performance-results}

Figure~\ref{fig:hbNorm:allDaCapo} presents a different view of the paper's main results than
Table~\ref{tab:performance:geoMean},
focusing on the performance gap between predictive and non-predictive analyses.
The figure shows predictive analysis run times,
normalized to \FTO-\HB, for each evaluated program.
For each program, the three groups of bars correspond to \WCP, \DC, and \CAPO analyses, respectively.
In a group of bars, the three bars correspond to the optimization level applied to each analysis.
\notes{Appendix~\ref{appendix:confidence-interval-results} presents the same results in a third view,
Tables~\ref{tab:performance:allDaCapo:CI} and \ref{tab:memory:allDaCapo:CI}
provides each separate program result for Table~\ref{tab:performance:geoMean} with 95\% confidence interavls.}%

\begin{figure*}[t]
\centering
\subfloat[Run time]{
\includegraphics[width=0.9\textwidth,height=0.45\textwidth]{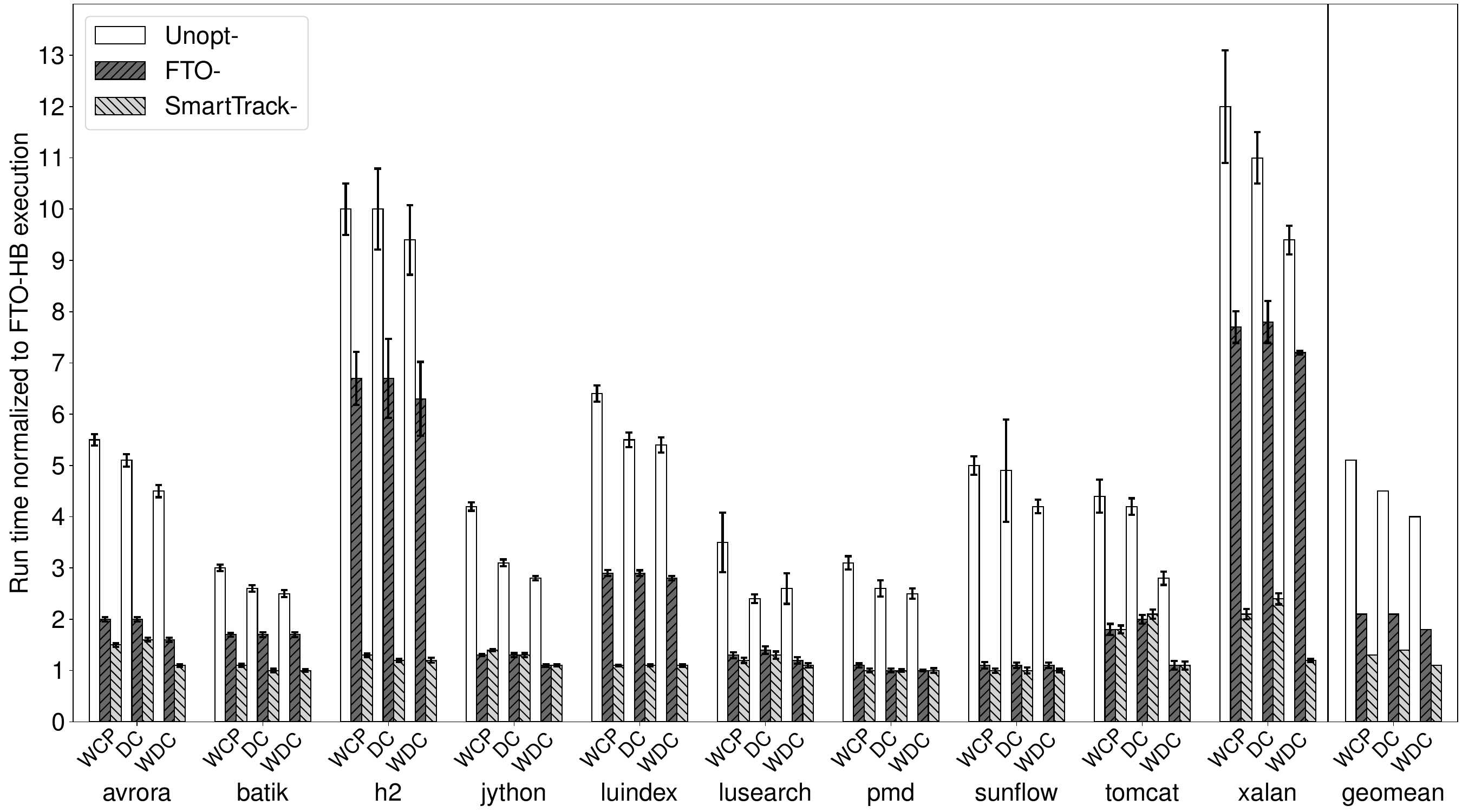}
\label{fig:runtime-hbNorm:allDaCapo}
}
\hfill
\centering
\subfloat[Memory usage]{
\includegraphics[width=0.9\textwidth,height=0.45\textwidth]{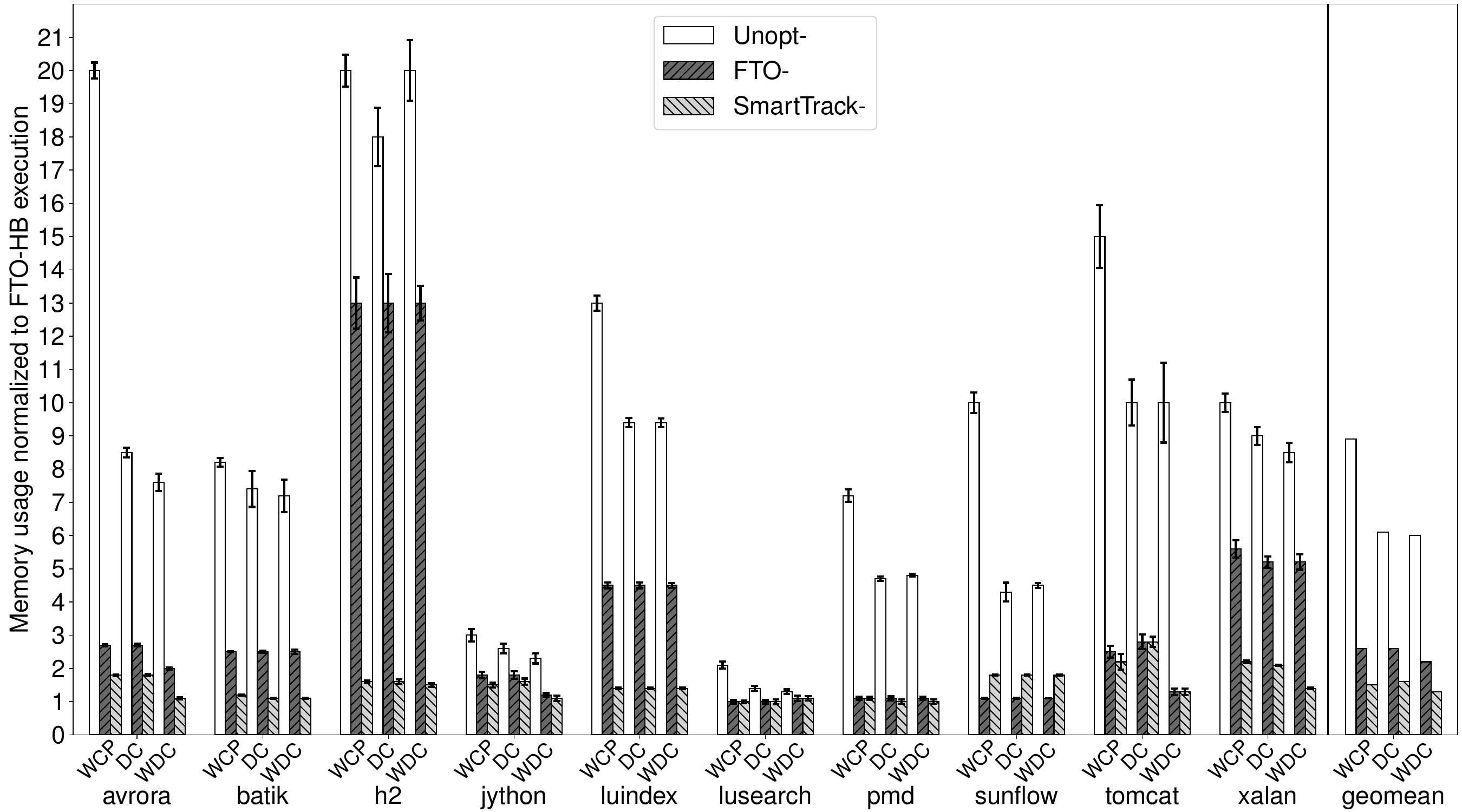}
\label{fig:memory-hbNorm:allDaCapo}
}
\caption{Run time and memory usage of various analyses, relative to \FTO-\HB, for each evaluated program.
The intervals are 95\% confidence intervals.}
\label{fig:hbNorm:allDaCapo}
\end{figure*}
}}{}

\iftoggle{includeRaceResults}{}{%
\section{Predictable Race Coverage}
\label{appendix:race-coverage}

Table~\ref{tab:race:allDaCapo} reports how many races
each analysis finds.
For each table cell, the second value (in parentheses) is total dynamic races reported,
and the first value is \emph{statically distinct} races.
Two dynamic races detected at the same static program location
are the same statically distinct race.

Although the analyses get progressively more powerful from top to bottom
(\eg, every \DC-race is a \CAPO-race),
this relationship does not always hold empirically for two reasons.
First, run-to-run variation naturally affects repeatability.
Appendix~\ref{appendix:confidence-interval-results} provides 95\% confidence intervals for these results,
showing that many of the differences involve overlapping confidence intervals.
Second, analyses have different performance characteristics that may affect the evaluated programs'
memory interleaving behavior, leading to different races occurring.
The table reports one anomalous result for \bench{jython} that we have been unable to diagnose:
\WCPO reports fewer races than expected;
we would expect the race counts to fall between the race counts of \HBO and \DCO.
This result is statistically significant (Table~\ref{tab:race:allDaCapo:CI} in Appendix~\ref{appendix:confidence-interval-results}).

For each relation, the different algorithms (\col{Unopt-}, \col{\FTO-}, \col{\REAbbrv-}) often report comparable race counts,
but sometimes the counts differ significantly, especially the counts of statically distinct races.
These differences occur for the above reasons plus a third reason:
the different optimization levels have different behavior after they detect the first race,
affecting race counts by using different metadata (\eg, epochs vs.\ vector clocks) 
to update racing accesses and detect future races
(Section~\ref{subsec:impl}).
These differences have the most impact on counts of statically distinct races.
%

For each relation, the differences between the \emph{algorithms} (\col{Unopt-}, \col{\FTO-}, \col{\REAbbrv-})
are not a reflection of race detection effectiveness.
The extra races detected by one algorithm, even if statistically significant,
are likely to be related to each other---involving accesses to the same data structure
as other reported races, or being dependent on other reported races---and thus not be of much use to programmers.
Rather, the race differences serve to show that the proposed optimizations and our implementations of them
lead to reasonable race detection results.
\notes{We have looked at individual static program locations reported by the analyses,
acting as a sanity check and providing confidence in the correctness of our implementations.}%

Likewise, the differences across relations do not serve mainly to demonstrate the effectiveness
of weaker predictive analyses.
Prior work has shown the relative effectiveness of \WCP and \DC analyses
by performing \HB, \WCP, and \DC analyses on the same observed trace~\cite{vindicator,wcp}.
(Our results often report many more races, especially dynamic races, than our prior work's results that used
the RoadRunner Vindicator implementation and the DaCapo benchmarks~\cite{vindicator}.
These differences occur because our prior work used default RoadRunner behavior
that stops performing analysis for a field after 100 dynamic races detected on the field,
whereas this paper's analyses disable that behavior.)

The results \emph{do} show that despite using a weaker relation than \DC analysis,
\CAPO analysis does not on average report more races than \DC analysis,
which suggests that \CAPO analysis's optimization does not lead to false races in practice.
In separate experiments that ran vindication with \DC and \CAPO analyses,
every \DC- and \CAPO-race detected across 10 trials was successfully vindicated.

\iftoggle{twoColumnText}{
\begin{table*}[t]
\newcommand{\rzero}{0\xspace}
\newcommand{\roh}[1]{\ifthenelse{\equal{#1}{\rna}}{\rna}{#1$\;\!\times$}} 
\newcommand{\rna}{N/A}
\newcommand{\memna}{N/A} 
\newcommand{\st}[1]{(#1~s)} 
\newcommand{\dt}[1]{#1~s} 
\newcommand{\mem}[1]{\ifthenelse{\equal{#1}{\memna}}{\memna}{#1$\;Mb$}} 
\newcommand{\et}[1]{#1M} 
\newcommand{\enfp}[1]{(#1M)} 
\newcommand{\base}[1]{#1~s} 
\newcommand{\sdr}[2]{#1{\;}(#2)}
\input{result-macros/PIP_slowTool_noWriteOrReadRaceEdge}

\input{result-macros/PIP_fastTool_noWriteOrReadRaceEdge_accessMetaUpdate}

\small
\smaller
\centering
\begin{tabular}{@{}l|Hr@{\;\;}r@{\;\;}r@{}}
        & w/ G & Unopt- & \FTO- & \REAbbrv- \\\hline
\HB		& \rna													& \sdr{\SLOWavroraHB}{\SLOWavroraHBDynamic}					& \sdr{\FASTavroraFTOHB}{\FASTavroraFTOHBDynamic}		& \rna \\
\WCP	& \rna													& \sdr{\SLOWavroraWCP}{\SLOWavroraWCPDynamic}				& \sdr{\FASTavroraFTOWCP}{\FASTavroraFTOWCPDynamic}		& \sdr{\FASTavroraREWCP}{\FASTavroraREWCPDynamic} \\
\DC		& \sdr{\SLOWavroraDCExc}{\SLOWavroraDCExcDynamic}		& \sdr{\SLOWavroraDCnoGExc}{\SLOWavroraDCnoGExcDynamic}	& \sdr{\FASTavroraFTODC}{\FASTavroraFTODCDynamic}		& \sdr{\FASTavroraREDC}{\FASTavroraREDCDynamic} \\
\CAPO	& \sdr{\SLOWavroraCAPOExc}{\SLOWavroraCAPOExcDynamic}	& \sdr{\SLOWavroraCAPOnoGExc}{\SLOWavroraCAPOnoGExcDynamic}	& \sdr{\FASTavroraFTOCAPO}{\FASTavroraFTOCAPODynamic}	& \sdr{\FASTavroraRECAPO}{\FASTavroraRECAPODynamic} \\
\mc{1}{c}{} & \mc{4}{c}{\bench{avrora}} \\
\end{tabular}
\hfill
\begin{tabular}{@{}HHr@{\;\;}r@{\;\;}r@{}}
        & w/ G & Unopt- & \FTO- & \REAbbrv- \\\hline
\HB		& \rna													& \sdr{\SLOWhtwoHB}{\SLOWhtwoHBDynamic}					& \sdr{\FASThtwoFTOHB}{\FASThtwoFTOHBDynamic}		& \rna \\
\WCP	& \rna													& \sdr{\SLOWhtwoWCP}{\SLOWhtwoWCPDynamic}				& \sdr{\FASThtwoFTOWCP}{\FASThtwoFTOWCPDynamic}		& \sdr{\FASThtwoREWCP}{\FASThtwoREWCPDynamic} \\
\DC		& \sdr{\SLOWhtwoDCExc}{\SLOWhtwoDCExcDynamic}		& \sdr{\SLOWhtwoDCnoGExc}{\SLOWhtwoDCnoGExcDynamic}	& \sdr{\FASThtwoFTODC}{\FASThtwoFTODCDynamic}		& \sdr{\FASThtwoREDC}{\FASThtwoREDCDynamic} \\
\CAPO	& \sdr{\SLOWhtwoCAPOExc}{\SLOWhtwoCAPOExcDynamic}	& \sdr{\SLOWhtwoCAPOnoGExc}{\SLOWhtwoCAPOnoGExcDynamic}	& \sdr{\FASThtwoFTOCAPO}{\FASThtwoFTOCAPODynamic}	& \sdr{\FASThtwoRECAPO}{\FASThtwoRECAPODynamic} \\
		& \mc{4}{c}{\bench{h2}} \\
\end{tabular}
\hfill
\begin{tabular}{@{}HHr@{\;\;}r@{\;\;}r@{}}
        & w/ G & Unopt- & \FTO- & \REAbbrv- \\\hline
\HB		& \rna													& \sdr{\SLOWjythonHB}{\SLOWjythonHBDynamic}					& \sdr{\FASTjythonFTOHB}{\FASTjythonFTOHBDynamic}		& \rna \\
\WCP	& \rna													& \sdr{\SLOWjythonWCP}{\SLOWjythonWCPDynamic}				& \sdr{\FASTjythonFTOWCP}{\FASTjythonFTOWCPDynamic}		& \sdr{\FASTjythonREWCP}{\FASTjythonREWCPDynamic} \\
\DC		& \sdr{\SLOWjythonDCExc}{\SLOWjythonDCExcDynamic}		& \sdr{\SLOWjythonDCnoGExc}{\SLOWjythonDCnoGExcDynamic}	& \sdr{\FASTjythonFTODC}{\FASTjythonFTODCDynamic}		& \sdr{\FASTjythonREDC}{\FASTjythonREDCDynamic} \\
\CAPO	& \sdr{\SLOWjythonCAPOExc}{\SLOWjythonCAPOExcDynamic}	& \sdr{\SLOWjythonCAPOnoGExc}{\SLOWjythonCAPOnoGExcDynamic}	& \sdr{\FASTjythonFTOCAPO}{\FASTjythonFTOCAPODynamic}	& \sdr{\FASTjythonRECAPO}{\FASTjythonRECAPODynamic} \\
		& \mc{4}{c}{\bench{jython}} \\
\end{tabular}
\hfill
\begin{tabular}{@{}HHr@{\;\;}r@{\;\;}r@{}}
        & w/ G & Unopt- & \FTO- & \REAbbrv- \\\hline
\HB		& \rna													& \sdr{\SLOWluindexHB}{\SLOWluindexHBDynamic}					& \sdr{\FASTluindexFTOHB}{\FASTluindexFTOHBDynamic}		& \rna \\
\WCP	& \rna													& \sdr{\SLOWluindexWCP}{\SLOWluindexWCPDynamic}				& \sdr{\FASTluindexFTOWCP}{\FASTluindexFTOWCPDynamic}		& \sdr{\FASTluindexREWCP}{\FASTluindexREWCPDynamic} \\
\DC		& \sdr{\SLOWluindexDCExc}{\SLOWluindexDCExcDynamic}		& \sdr{\SLOWluindexDCnoGExc}{\SLOWluindexDCnoGExcDynamic}	& \sdr{\FASTluindexFTODC}{\FASTluindexFTODCDynamic}		& \sdr{\FASTluindexREDC}{\FASTluindexREDCDynamic} \\
\CAPO	& \sdr{\SLOWluindexCAPOExc}{\SLOWluindexCAPOExcDynamic}	& \sdr{\SLOWluindexCAPOnoGExc}{\SLOWluindexCAPOnoGExcDynamic}	& \sdr{\FASTluindexFTOCAPO}{\FASTluindexFTOCAPODynamic}	& \sdr{\FASTluindexRECAPO}{\FASTluindexRECAPODynamic} \\
		& \mc{4}{c}{\bench{luindex}} \\
\end{tabular}
\medskip\\
\begin{tabular}{@{}l|Hr@{\;}r@{\;}r@{}}
        & w/ G & Unopt- & \FTO- & \REAbbrv- \\\hline
\HB		& \rna													& \sdr{\SLOWpmdHB}{\SLOWpmdHBDynamic}					& \sdr{\FASTpmdFTOHB}{\FASTpmdFTOHBDynamic}		& \rna \\
\WCP	& \rna													& \sdr{\SLOWpmdWCP}{\SLOWpmdWCPDynamic}				& \sdr{\FASTpmdFTOWCP}{\FASTpmdFTOWCPDynamic}		& \sdr{\FASTpmdREWCP}{\FASTpmdREWCPDynamic} \\
\DC		& \sdr{\SLOWpmdDCExc}{\SLOWpmdDCExcDynamic}		& \sdr{\SLOWpmdDCnoGExc}{\SLOWpmdDCnoGExcDynamic}	& \sdr{\FASTpmdFTODC}{\FASTpmdFTODCDynamic}		& \sdr{\FASTpmdREDC}{\FASTpmdREDCDynamic} \\
\CAPO	& \sdr{\SLOWpmdCAPOExc}{\SLOWpmdCAPOExcDynamic}	& \sdr{\SLOWpmdCAPOnoGExc}{\SLOWpmdCAPOnoGExcDynamic}	& \sdr{\FASTpmdFTOCAPO}{\FASTpmdFTOCAPODynamic}	& \sdr{\FASTpmdRECAPO}{\FASTpmdRECAPODynamic} \\
\mc{1}{c}{} & \mc{4}{c}{\bench{pmd}} \\
\end{tabular}
\hfill
\begin{tabular}{@{}HHr@{\;}r@{\;}r@{}}
        & w/ G & Unopt- & \FTO- & \REAbbrv- \\\hline
\HB		& \rna													& \sdr{\SLOWsunflowHB}{\SLOWsunflowHBDynamic}					& \sdr{\FASTsunflowFTOHB}{\FASTsunflowFTOHBDynamic}		& \rna \\
\WCP	& \rna													& \sdr{\SLOWsunflowWCP}{\SLOWsunflowWCPDynamic}				& \sdr{\FASTsunflowFTOWCP}{\FASTsunflowFTOWCPDynamic}		& \sdr{\FASTsunflowREWCP}{\FASTsunflowREWCPDynamic} \\
\DC		& \sdr{\SLOWsunflowDCExc}{\SLOWsunflowDCExcDynamic}		& \sdr{\SLOWsunflowDCnoGExc}{\SLOWsunflowDCnoGExcDynamic}	& \sdr{\FASTsunflowFTODC}{\FASTsunflowFTODCDynamic}		& \sdr{\FASTsunflowREDC}{\FASTsunflowREDCDynamic} \\
\CAPO	& \sdr{\SLOWsunflowCAPOExc}{\SLOWsunflowCAPOExcDynamic}	& \sdr{\SLOWsunflowCAPOnoGExc}{\SLOWsunflowCAPOnoGExcDynamic}	& \sdr{\FASTsunflowFTOCAPO}{\FASTsunflowFTOCAPODynamic}	& \sdr{\FASTsunflowRECAPO}{\FASTsunflowRECAPODynamic} \\
		& \mc{4}{c}{\bench{sunflow}} \\
\end{tabular}
\hfill
\begin{tabular}{@{}HHr@{\;}r@{\;}r@{}}
        & w/ G & Unopt- & \FTO- & \REAbbrv- \\\hline
\HB		& \rna													& \sdr{\SLOWtomcatHB}{\SLOWtomcatHBDynamic}					& \sdr{\FASTtomcatFTOHB}{\FASTtomcatFTOHBDynamic}		& \rna \\
\WCP	& \rna													& \sdr{\SLOWtomcatWCP}{\SLOWtomcatWCPDynamic}				& \sdr{\FASTtomcatFTOWCP}{\FASTtomcatFTOWCPDynamic}		& \sdr{\FASTtomcatREWCP}{\FASTtomcatREWCPDynamic} \\
\DC		& \sdr{\SLOWtomcatDCExc}{\SLOWtomcatDCExcDynamic}		& \sdr{\SLOWtomcatDCnoGExc}{\SLOWtomcatDCnoGExcDynamic}	& \sdr{\FASTtomcatFTODC}{\FASTtomcatFTODCDynamic}		& \sdr{\FASTtomcatREDC}{\FASTtomcatREDCDynamic} \\
\CAPO	& \sdr{\SLOWtomcatCAPOExc}{\SLOWtomcatCAPOExcDynamic}	& \sdr{\SLOWtomcatCAPOnoGExc}{\SLOWtomcatCAPOnoGExcDynamic}	& \sdr{\FASTtomcatFTOCAPO}{\FASTtomcatFTOCAPODynamic}	& \sdr{\FASTtomcatRECAPO}{\FASTtomcatRECAPODynamic} \\
        & \mc{4}{c}{\bench{tomcat}} \\
\end{tabular}
\hfill
\begin{tabular}{@{}HHr@{\;}r@{\;}r@{}}
        & w/ G & Unopt- & \FTO- & \REAbbrv- \\\hline
\HB		& \rna													& \sdr{\SLOWxalanHB}{\SLOWxalanHBDynamic}					& \sdr{\FASTxalanFTOHB}{\FASTxalanFTOHBDynamic}		& \rna \\
\WCP	& \rna													& \sdr{\SLOWxalanWCP}{\SLOWxalanWCPDynamic}				& \sdr{\FASTxalanFTOWCP}{\FASTxalanFTOWCPDynamic}		& \sdr{\FASTxalanREWCP}{\FASTxalanREWCPDynamic} \\
\DC		& \sdr{\SLOWxalanDCExc}{\SLOWxalanDCExcDynamic}		& \sdr{\SLOWxalanDCnoGExc}{\SLOWxalanDCnoGExcDynamic}	& \sdr{\FASTxalanFTODC}{\FASTxalanFTODCDynamic}		& \sdr{\FASTxalanREDC}{\FASTxalanREDCDynamic} \\
\CAPO	& \sdr{\SLOWxalanCAPOExc}{\SLOWxalanCAPOExcDynamic}	& \sdr{\SLOWxalanCAPOnoGExc}{\SLOWxalanCAPOnoGExcDynamic}	& \sdr{\FASTxalanFTOCAPO}{\FASTxalanFTOCAPODynamic}	& \sdr{\FASTxalanRECAPO}{\FASTxalanRECAPODynamic} \\
& \mc{4}{c}{\bench{xalan}} \\
\end{tabular}
\jake{Remaining issue is that \HB finds more races than \FTO-\WCP for jython.}
\caption{Average races reported by various analyses for each evaluated program
(excluding \bench{batik} and \bench{lusearch}, for which all analyses report no races).
In each cell, the first value is statically distinct races
(\ie, distinct program locations) and the second value, in parentheses, is total dynamic races.}
\label{tab:race:allDaCapo}
\end{table*}
}{
\begin{table}[t]
\newcommand{\rzero}{0\xspace}
\newcommand{\roh}[1]{\ifthenelse{\equal{#1}{\rna}}{\rna}{#1$\;\!\times$}} 
\newcommand{\rna}{N/A}
\newcommand{\memna}{N/A} 
\newcommand{\st}[1]{(#1~s)} 
\newcommand{\dt}[1]{#1~s} 
\newcommand{\mem}[1]{\ifthenelse{\equal{#1}{\memna}}{\memna}{#1$\;Mb$}} 
\newcommand{\et}[1]{#1M} 
\newcommand{\enfp}[1]{(#1M)} 
\newcommand{\base}[1]{#1~s} 
\newcommand{\sdr}[2]{#1{\;}(#2)}

\input{result-macros/PIP_slowTool_noWriteOrReadRaceEdge}
\input{result-macros/PIP_fastTool_noWriteOrReadRaceEdge_accessMetaUpdate}
\small
\smaller
\centering
\begin{tabular}{@{}l|Hr@{\;\;}r@{\;\;}r@{}}
        & w/ G & Unopt- & \FTO- & \REAbbrv- \\\hline
\HB		& \rna													& \sdr{\SLOWavroraHB}{\SLOWavroraHBDynamic}					& \sdr{\FASTavroraFTOHB}{\FASTavroraFTOHBDynamic}		& \rna \\
\WCP	& \rna													& \sdr{\SLOWavroraWCP}{\SLOWavroraWCPDynamic}				& \sdr{\FASTavroraFTOWCP}{\FASTavroraFTOWCPDynamic}		& \sdr{\FASTavroraREWCP}{\FASTavroraREWCPDynamic} \\
\DC		& \sdr{\SLOWavroraDCExc}{\SLOWavroraDCExcDynamic}		& \sdr{\SLOWavroraDCnoGExc}{\SLOWavroraDCnoGExcDynamic}	& \sdr{\FASTavroraFTODC}{\FASTavroraFTODCDynamic}		& \sdr{\FASTavroraREDC}{\FASTavroraREDCDynamic} \\
\CAPO	& \sdr{\SLOWavroraCAPOExc}{\SLOWavroraCAPOExcDynamic}	& \sdr{\SLOWavroraCAPOnoGExc}{\SLOWavroraCAPOnoGExcDynamic}	& \sdr{\FASTavroraFTOCAPO}{\FASTavroraFTOCAPODynamic}	& \sdr{\FASTavroraRECAPO}{\FASTavroraRECAPODynamic} \\
\mc{1}{c}{} & \mc{4}{c}{\bench{avrora}} \\
\end{tabular}
\hfill
\begin{tabular}{@{}HHr@{\;\;}r@{\;\;}r@{}}
        & w/ G & Unopt- & \FTO- & \REAbbrv- \\\hline
\HB		& \rna													& \sdr{\SLOWhtwoHB}{\SLOWhtwoHBDynamic}					& \sdr{\FASThtwoFTOHB}{\FASThtwoFTOHBDynamic}		& \rna \\
\WCP	& \rna													& \sdr{\SLOWhtwoWCP}{\SLOWhtwoWCPDynamic}				& \sdr{\FASThtwoFTOWCP}{\FASThtwoFTOWCPDynamic}		& \sdr{\FASThtwoREWCP}{\FASThtwoREWCPDynamic} \\
\DC		& \sdr{\SLOWhtwoDCExc}{\SLOWhtwoDCExcDynamic}		& \sdr{\SLOWhtwoDCnoGExc}{\SLOWhtwoDCnoGExcDynamic}	& \sdr{\FASThtwoFTODC}{\FASThtwoFTODCDynamic}		& \sdr{\FASThtwoREDC}{\FASThtwoREDCDynamic} \\
\CAPO	& \sdr{\SLOWhtwoCAPOExc}{\SLOWhtwoCAPOExcDynamic}	& \sdr{\SLOWhtwoCAPOnoGExc}{\SLOWhtwoCAPOnoGExcDynamic}	& \sdr{\FASThtwoFTOCAPO}{\FASThtwoFTOCAPODynamic}	& \sdr{\FASThtwoRECAPO}{\FASThtwoRECAPODynamic} \\
		& \mc{4}{c}{\bench{h2}} \\
\end{tabular}
\hfill
\begin{tabular}{@{}HHr@{\;\;}r@{\;\;}r@{}}
        & w/ G & Unopt- & \FTO- & \REAbbrv- \\\hline
\HB		& \rna													& \sdr{\SLOWjythonHB}{\SLOWjythonHBDynamic}					& \sdr{\FASTjythonFTOHB}{\FASTjythonFTOHBDynamic}		& \rna \\
\WCP	& \rna													& \sdr{\SLOWjythonWCP}{\SLOWjythonWCPDynamic}				& \sdr{\FASTjythonFTOWCP}{\FASTjythonFTOWCPDynamic}		& \sdr{\FASTjythonREWCP}{\FASTjythonREWCPDynamic} \\
\DC		& \sdr{\SLOWjythonDCExc}{\SLOWjythonDCExcDynamic}		& \sdr{\SLOWjythonDCnoGExc}{\SLOWjythonDCnoGExcDynamic}	& \sdr{\FASTjythonFTODC}{\FASTjythonFTODCDynamic}		& \sdr{\FASTjythonREDC}{\FASTjythonREDCDynamic} \\
\CAPO	& \sdr{\SLOWjythonCAPOExc}{\SLOWjythonCAPOExcDynamic}	& \sdr{\SLOWjythonCAPOnoGExc}{\SLOWjythonCAPOnoGExcDynamic}	& \sdr{\FASTjythonFTOCAPO}{\FASTjythonFTOCAPODynamic}	& \sdr{\FASTjythonRECAPO}{\FASTjythonRECAPODynamic} \\
		& \mc{4}{c}{\bench{jython}} \\
\end{tabular}
\hfill
\begin{tabular}{@{}HHr@{\;\;}r@{\;\;}r@{}}
        & w/ G & Unopt- & \FTO- & \REAbbrv- \\\hline
\HB		& \rna													& \sdr{\SLOWluindexHB}{\SLOWluindexHBDynamic}					& \sdr{\FASTluindexFTOHB}{\FASTluindexFTOHBDynamic}		& \rna \\
\WCP	& \rna													& \sdr{\SLOWluindexWCP}{\SLOWluindexWCPDynamic}				& \sdr{\FASTluindexFTOWCP}{\FASTluindexFTOWCPDynamic}		& \sdr{\FASTluindexREWCP}{\FASTluindexREWCPDynamic} \\
\DC		& \sdr{\SLOWluindexDCExc}{\SLOWluindexDCExcDynamic}		& \sdr{\SLOWluindexDCnoGExc}{\SLOWluindexDCnoGExcDynamic}	& \sdr{\FASTluindexFTODC}{\FASTluindexFTODCDynamic}		& \sdr{\FASTluindexREDC}{\FASTluindexREDCDynamic} \\
\CAPO	& \sdr{\SLOWluindexCAPOExc}{\SLOWluindexCAPOExcDynamic}	& \sdr{\SLOWluindexCAPOnoGExc}{\SLOWluindexCAPOnoGExcDynamic}	& \sdr{\FASTluindexFTOCAPO}{\FASTluindexFTOCAPODynamic}	& \sdr{\FASTluindexRECAPO}{\FASTluindexRECAPODynamic} \\
		& \mc{4}{c}{\bench{luindex}} \\
\end{tabular}
\medskip\\
\begin{tabular}{@{}l|Hr@{\;}r@{\;}r@{}}
        & w/ G & Unopt- & \FTO- & \REAbbrv- \\\hline
\HB		& \rna													& \sdr{\SLOWpmdHB}{\SLOWpmdHBDynamic}					& \sdr{\FASTpmdFTOHB}{\FASTpmdFTOHBDynamic}		& \rna \\
\WCP	& \rna													& \sdr{\SLOWpmdWCP}{\SLOWpmdWCPDynamic}				& \sdr{\FASTpmdFTOWCP}{\FASTpmdFTOWCPDynamic}		& \sdr{\FASTpmdREWCP}{\FASTpmdREWCPDynamic} \\
\DC		& \sdr{\SLOWpmdDCExc}{\SLOWpmdDCExcDynamic}		& \sdr{\SLOWpmdDCnoGExc}{\SLOWpmdDCnoGExcDynamic}	& \sdr{\FASTpmdFTODC}{\FASTpmdFTODCDynamic}		& \sdr{\FASTpmdREDC}{\FASTpmdREDCDynamic} \\
\CAPO	& \sdr{\SLOWpmdCAPOExc}{\SLOWpmdCAPOExcDynamic}	& \sdr{\SLOWpmdCAPOnoGExc}{\SLOWpmdCAPOnoGExcDynamic}	& \sdr{\FASTpmdFTOCAPO}{\FASTpmdFTOCAPODynamic}	& \sdr{\FASTpmdRECAPO}{\FASTpmdRECAPODynamic} \\
\mc{1}{c}{} & \mc{4}{c}{\bench{pmd}} \\
\end{tabular}
\hfill
\begin{tabular}{@{}HHr@{\;}r@{\;}r@{}}
        & w/ G & Unopt- & \FTO- & \REAbbrv- \\\hline
\HB		& \rna													& \sdr{\SLOWsunflowHB}{\SLOWsunflowHBDynamic}					& \sdr{\FASTsunflowFTOHB}{\FASTsunflowFTOHBDynamic}		& \rna \\
\WCP	& \rna													& \sdr{\SLOWsunflowWCP}{\SLOWsunflowWCPDynamic}				& \sdr{\FASTsunflowFTOWCP}{\FASTsunflowFTOWCPDynamic}		& \sdr{\FASTsunflowREWCP}{\FASTsunflowREWCPDynamic} \\
\DC		& \sdr{\SLOWsunflowDCExc}{\SLOWsunflowDCExcDynamic}		& \sdr{\SLOWsunflowDCnoGExc}{\SLOWsunflowDCnoGExcDynamic}	& \sdr{\FASTsunflowFTODC}{\FASTsunflowFTODCDynamic}		& \sdr{\FASTsunflowREDC}{\FASTsunflowREDCDynamic} \\
\CAPO	& \sdr{\SLOWsunflowCAPOExc}{\SLOWsunflowCAPOExcDynamic}	& \sdr{\SLOWsunflowCAPOnoGExc}{\SLOWsunflowCAPOnoGExcDynamic}	& \sdr{\FASTsunflowFTOCAPO}{\FASTsunflowFTOCAPODynamic}	& \sdr{\FASTsunflowRECAPO}{\FASTsunflowRECAPODynamic} \\
		& \mc{4}{c}{\bench{sunflow}} \\
\end{tabular}
\hfill
\begin{tabular}{@{}HHr@{\;}r@{\;}r@{}}
        & w/ G & Unopt- & \FTO- & \REAbbrv- \\\hline
\HB		& \rna													& \sdr{\SLOWtomcatHB}{\SLOWtomcatHBDynamic}					& \sdr{\FASTtomcatFTOHB}{\FASTtomcatFTOHBDynamic}		& \rna \\
\WCP	& \rna													& \sdr{\SLOWtomcatWCP}{\SLOWtomcatWCPDynamic}				& \sdr{\FASTtomcatFTOWCP}{\FASTtomcatFTOWCPDynamic}		& \sdr{\FASTtomcatREWCP}{\FASTtomcatREWCPDynamic} \\
\DC		& \sdr{\SLOWtomcatDCExc}{\SLOWtomcatDCExcDynamic}		& \sdr{\SLOWtomcatDCnoGExc}{\SLOWtomcatDCnoGExcDynamic}	& \sdr{\FASTtomcatFTODC}{\FASTtomcatFTODCDynamic}		& \sdr{\FASTtomcatREDC}{\FASTtomcatREDCDynamic} \\
\CAPO	& \sdr{\SLOWtomcatCAPOExc}{\SLOWtomcatCAPOExcDynamic}	& \sdr{\SLOWtomcatCAPOnoGExc}{\SLOWtomcatCAPOnoGExcDynamic}	& \sdr{\FASTtomcatFTOCAPO}{\FASTtomcatFTOCAPODynamic}	& \sdr{\FASTtomcatRECAPO}{\FASTtomcatRECAPODynamic} \\
        & \mc{4}{c}{\bench{tomcat}} \\
\end{tabular}
\hfill
\begin{tabular}{@{}HHr@{\;}r@{\;}r@{}}
        & w/ G & Unopt- & \FTO- & \REAbbrv- \\\hline
\HB		& \rna													& \sdr{\SLOWxalanHB}{\SLOWxalanHBDynamic}					& \sdr{\FASTxalanFTOHB}{\FASTxalanFTOHBDynamic}		& \rna \\
\WCP	& \rna													& \sdr{\SLOWxalanWCP}{\SLOWxalanWCPDynamic}				& \sdr{\FASTxalanFTOWCP}{\FASTxalanFTOWCPDynamic}		& \sdr{\FASTxalanREWCP}{\FASTxalanREWCPDynamic} \\
\DC		& \sdr{\SLOWxalanDCExc}{\SLOWxalanDCExcDynamic}		& \sdr{\SLOWxalanDCnoGExc}{\SLOWxalanDCnoGExcDynamic}	& \sdr{\FASTxalanFTODC}{\FASTxalanFTODCDynamic}		& \sdr{\FASTxalanREDC}{\FASTxalanREDCDynamic} \\
\CAPO	& \sdr{\SLOWxalanCAPOExc}{\SLOWxalanCAPOExcDynamic}	& \sdr{\SLOWxalanCAPOnoGExc}{\SLOWxalanCAPOnoGExcDynamic}	& \sdr{\FASTxalanFTOCAPO}{\FASTxalanFTOCAPODynamic}	& \sdr{\FASTxalanRECAPO}{\FASTxalanRECAPODynamic} \\
& \mc{4}{c}{\bench{xalan}} \\
\end{tabular}
\notes{
\jake{Remaining issue is that \HB finds more races than \FTO-\WCP for jython.}
}
\caption{Average races reported by various analyses for each evaluated program
(excluding \bench{batik} and \bench{lusearch}, for which all analyses report no races).
In each cell, the first value is statically distinct races
(\ie, distinct program locations) and the second value, in parentheses, is total dynamic races.}
\label{tab:race:allDaCapo}
\end{table}
}
}

\section{Baselines with Confidence Intervals}
\label{appendix:baseline-confidence-interval-results}

Table~\ref{tab:performance:fasttrack:time-and-memoryCI}
shows the performance cost of several analyses.
The \emph{\HB} columns are the same as the rightmost columns
in Table~\ref{tab:characteristics:bench-stats}, but with 95\% confidence intervals.

The \col{Unopt-$\ast$} columns compare the performance of
unoptimized \DC and \CAPO analyses, with and without support for vindication.
The \col{w/$G$} configurations build a constraint graph during analysis
and perform vindication after the program completes, and \col{w/o $G$} configurations do not.
\col{Unopt-\DC w/$G$} represents the cost incurred by prior work to detect \DC-races and check them after execution.
It also represents the cost that would be incurred by a replayed execution that builds $G$
in order to verify \DC-races
detected in a recorded run that used \REDC analysis or some other \DC analysis that does \emph{not} build $G$ (Section~\ref{subsec:record-replay}).
Likewise, \col{Unopt-\CAPO w/$G$} shows the cost of a replayed execution checking \CAPO-races.

\col{Unopt-\DC w/o $G$} represents the cost incurred by prior work to detect \DC-races without checking them---a
realistic configuration because few if any \DC-races are false positives in practice, and a second replayed run can optionally check \DC-races.
Likewise, \col{Unopt-\CAPO w/o $G$} shows the cost of detecting \CAPO-races without checking them.

On average across the programs,
the results show that the costs of unoptimized predictive analyses are high, 
whether or not they build a constraint graph,
compared with existing optimized non-predictive (\HB) analyses.

\section{Main Results with Confidence Intervals}
\label{appendix:confidence-interval-results}

Tables~\ref{tab:performance:allDaCapo:CI} and \ref{tab:memory:allDaCapo:CI}
show the same performance results as Table~\ref{tab:performance:geoMean},
for each program separately and with 95\% confidence intervals.
Table~\ref{tab:race:allDaCapo:CI} 
shows the same race detection results as Table~\ref{tab:race:allDaCapo},
but with 95\% confidence intervals.

\iftoggle{twoColumnText}{
\begin{table*}[!b]
\begin{empty}
\newcommand{\rzero}{0\xspace}
\newcommand{\roh}[1]{\ifthenelse{\equal{#1}{\rzero}}{0}{#1$\;\!\times$}} 
\newcommand{\rna}{N/A}
\newcommand{\memna}{N/A} 
\newcommand{\st}[1]{(#1~s)} 
\newcommand{\dt}[1]{#1~s} 
\newcommand{\mem}[1]{\ifthenelse{\equal{#1}{\memna}}{\rna}{#1$\;\!\times$}}
\newcommand{\baseMem}[1]{#1~MB} 
\newcommand{\et}[1]{#1M} 
\newcommand{\enfp}[1]{(#1M)} 
\newcommand{\base}[1]{#1~s} 
\newcommand{\vci}[2]{\ifthenelse{\equal{#1}{\rna}}{\rna}{\roh{#1}{\;}\ci{#2}}}
\newcommand{\ci}[1]{\ensuremath{\pm}{\;}\roh{#1}}
\input{result-macros/PIP_slowTool_noCoresSet}
\input{result-macros/PIP_fastTool_extraOpt2Quiet}
\small
\centering
\newcommand{\ics}{\;}
\begin{tabular}{@{}l|HHHHHHll|l@{\ics}l|l@{\ics}l@{}}
& \mc{2}{Z}{} & \mc{2}{Z}{} & Base & \mc{3}{c|}{\HB} & \mc{2}{c|}{Unopt-\DC} & \mc{2}{c@{}}{Unopt-\CAPO} \\
Program & \mc{2}{Z}{Events} & \mc{2}{Z}{\#Thr} & time & RR\FTTwoAbbrv & \FTTwoAbbrv & \FTO & w/$G$ & w/o $G$ & w/$G$ & w/o $G$ \\\hline 
\bench{avrora} 	& \et{\FASTavroraEvents}{\FASTavroraEventsCI}   & \enfp{\FASTavroraNoFPEvents}{\FASTavroraNoFPEventsCI}	& \FASTavroraTotalThreads 	& (\FASTavroraMaxLiveThreads)	& \base{\FASTavroraBaseTime}   & \vci{\FASTavroraFTTime}{\FASTavroraFTTimeCI}   & \vci{\FASTavroraHBTime}{\FASTavroraHBTimeCI}	& \vci{\FASTavroraFTOHBTime}{\FASTavroraFTOHBTimeCI}	& \vci{\SLOWavroraDCExcTime}{\SLOWavroraDCExcTimeCI}	& \vci{\SLOWavroraDCnoGExcTime}{\SLOWavroraDCnoGExcTimeCI}	& \vci{\SLOWavroraCAPOExcTime}{\SLOWavroraCAPOExcTimeCI}	& \vci{\SLOWavroraCAPOnoGExcTime}{\SLOWavroraCAPOnoGExcTimeCI}\\
\bench{batik} 	& \et{\FASTbatikEvents}{\FASTbatikEventsCI}    & \enfp{\FASTbatikNoFPEvents}{\FASTbatikNoFPEventsCI}	& \FASTbatikTotalThreads 	& (\FASTbatikMaxLiveThreads)	& \base{\FASTbatikBaseTime}    & \vci{\FASTbatikFTTime}{\FASTbatikFTTimeCI}    & \vci{\FASTbatikHBTime}{\FASTbatikHBTimeCI}   	& \vci{\FASTbatikFTOHBTime}{\FASTbatikFTOHBTimeCI}		& \vci{\SLOWbatikDCExcTime}{\SLOWbatikDCExcTimeCI}	& \vci{\SLOWbatikDCnoGExcTime}{\SLOWbatikDCnoGExcTimeCI}		& \vci{\SLOWbatikCAPOExcTime}{\SLOWbatikCAPOExcTimeCI}	& \vci{\SLOWbatikCAPOnoGExcTime}{\SLOWbatikCAPOnoGExcTimeCI} \\
\bench{h2} 		& \et{\FASThtwoEvents}{\FASThtwoEventsCI}     & \enfp{\FASThtwoNoFPEvents}{\FASThtwoNoFPEventsCI}	    & \FASThtwoTotalThreads 	& (\FASThtwoMaxLiveThreads)	    & \base{\FASThtwoBaseTime}     & \vci{\FASThtwoFTTime}{\FASThtwoFTTimeCI}     & \vci{\FASThtwoHBTime}{\FASThtwoHBTimeCI}    	& \vci{\FASThtwoFTOHBTime}{\FASThtwoFTOHBTimeCI}		& \vci{\SLOWhtwoDCExcTime}{\SLOWhtwoDCExcTimeCI}		& \vci{\SLOWhtwoDCnoGExcTime}{\SLOWhtwoDCnoGExcTimeCI}		& \vci{\SLOWhtwoCAPOExcTime}{\SLOWhtwoCAPOExcTimeCI}	& \vci{\SLOWhtwoCAPOnoGExcTime}{\SLOWhtwoCAPOnoGExcTimeCI}\\
\bench{jython} 	& \et{\FASTjythonEvents}{\FASTjythonEventsCI}   & \enfp{\FASTjythonNoFPEvents}{\FASTjythonNoFPEventsCI}	& \FASTjythonTotalThreads	& (\FASTjythonMaxLiveThreads)	& \base{\FASTjythonBaseTime}   & \vci{\FASTjythonFTTime}{\FASTjythonFTTimeCI}   & \vci{\FASTjythonHBTime}{\FASTjythonHBTimeCI} 	& \vci{\FASTjythonFTOHBTime}{\FASTjythonFTOHBTimeCI}	& \vci{\SLOWjythonDCExcTime}{\SLOWjythonDCExcTimeCI}	& \vci{\SLOWjythonDCnoGExcTime}{\SLOWjythonDCnoGExcTimeCI}	& \vci{\SLOWjythonCAPOExcTime}{\SLOWjythonCAPOExcTimeCI}	& \vci{\SLOWjythonCAPOnoGExcTime}{\SLOWjythonCAPOnoGExcTimeCI} \\
\bench{luindex} & \et{\FASTluindexEvents}{\FASTluindexEventsCI}  & \enfp{\FASTluindexNoFPEvents}{\FASTluindexNoFPEventsCI}	& \FASTluindexTotalThreads	& (\FASTluindexMaxLiveThreads)	& \base{\FASTluindexBaseTime}  & \vci{\FASTluindexFTTime}{\FASTluindexFTTimeCI}  & \vci{\FASTluindexHBTime}{\FASTluindexHBTimeCI} 	& \vci{\FASTluindexFTOHBTime}{\FASTluindexFTOHBTimeCI}	& \vci{\SLOWluindexDCExcTime}{\SLOWluindexDCExcTimeCI}	& \vci{\SLOWluindexDCnoGExcTime}{\SLOWluindexDCnoGExcTimeCI}	& \vci{\SLOWluindexCAPOExcTime}{\SLOWluindexCAPOExcTimeCI}	& \vci{\SLOWluindexCAPOnoGExcTime}{\SLOWluindexCAPOnoGExcTimeCI} \\
\bench{lusearch}& \et{\FASTlusearchEvents}{\FASTlusearchEventsCI} & \enfp{\FASTlusearchNoFPEvents}{\FASTlusearchNoFPEventsCI}	& \FASTlusearchTotalThreads	& (\FASTlusearchMaxLiveThreads)	& \base{\FASTlusearchBaseTime} & \vci{\FASTlusearchFTTime}{\FASTlusearchFTTimeCI} & \vci{\FASTlusearchHBTime}{\FASTlusearchHBTimeCI}	& \vci{\FASTlusearchFTOHBTime}{\FASTlusearchFTOHBTimeCI}	& \vci{\SLOWlusearchDCExcTime}{\SLOWlusearchDCExcTimeCI}	& \vci{\SLOWlusearchDCnoGExcTime}{\SLOWlusearchDCnoGExcTimeCI}	& \vci{\SLOWlusearchCAPOExcTime}{\SLOWlusearchCAPOExcTimeCI}& \vci{\SLOWlusearchCAPOnoGExcTime}{\SLOWlusearchCAPOnoGExcTimeCI} \\
\bench{pmd}		& \et{\FASTpmdEvents}{\FASTpmdEventsCI}      & \enfp{\FASTpmdNoFPEvents}{\FASTpmdNoFPEventsCI}      & \FASTpmdTotalThreads	    & (\FASTpmdMaxLiveThreads)	    & \base{\FASTpmdBaseTime}      & \vci{\FASTpmdFTTime}{\FASTpmdFTTimeCI}      & \vci{\FASTpmdHBTime}{\FASTpmdHBTimeCI}		& \vci{\FASTpmdFTOHBTime}{\FASTpmdFTOHBTimeCI}		& \vci{\SLOWpmdDCExcTime}{\SLOWpmdDCExcTimeCI}		& \vci{\SLOWpmdDCnoGExcTime}{\SLOWpmdDCnoGExcTimeCI}		& \vci{\SLOWpmdCAPOExcTime}{\SLOWpmdCAPOExcTimeCI}		& \vci{\SLOWpmdCAPOnoGExcTime}{\SLOWpmdCAPOnoGExcTimeCI} \\
\bench{sunflow} & \et{\FASTsunflowEvents}{\FASTsunflowEventsCI}  & \enfp{\FASTsunflowNoFPEvents}{\FASTsunflowNoFPEventsCI}	& \FASTsunflowTotalThreads	& (\FASTsunflowMaxLiveThreads)	& \base{\FASTsunflowBaseTime}  & \vci{\FASTsunflowFTTime}{\FASTsunflowFTTimeCI}  & \vci{\FASTsunflowHBTime}{\FASTsunflowHBTimeCI}	& \vci{\FASTsunflowFTOHBTime}{\FASTsunflowFTOHBTimeCI}	& \vci{\SLOWsunflowDCExcTime}{\SLOWsunflowDCExcTimeCI}	& \vci{\SLOWsunflowDCnoGExcTime}{\SLOWsunflowDCnoGExcTimeCI}	& \vci{\SLOWsunflowCAPOExcTime}{\SLOWsunflowCAPOExcTimeCI}	& \vci{\SLOWsunflowCAPOnoGExcTime}{\SLOWsunflowCAPOnoGExcTimeCI} \\
\bench{tomcat} 	& \et{\FASTtomcatEvents}{\FASTtomcatEventsCI}   & \enfp{\FASTtomcatNoFPEvents}{\FASTtomcatNoFPEventsCI}	& \FASTtomcatTotalThreads	& (\FASTtomcatMaxLiveThreads)	& \base{\FASTtomcatBaseTime}   & \vci{\FASTtomcatFTTime}{\FASTtomcatFTTimeCI}   & \vci{\FASTtomcatHBTime}{\FASTtomcatHBTimeCI} 	& \vci{\FASTtomcatFTOHBTime}{\FASTtomcatFTOHBTimeCI}	& \vci{\SLOWtomcatDCExcTime}{\SLOWtomcatDCExcTimeCI}	& \vci{\SLOWtomcatDCnoGExcTime}{\SLOWtomcatDCnoGExcTimeCI}	& \vci{\SLOWtomcatCAPOExcTime}{\SLOWtomcatCAPOExcTimeCI}	& \vci{\SLOWtomcatCAPOnoGExcTime}{\SLOWtomcatCAPOnoGExcTimeCI} \\
\bench{xalan} 	& \et{\FASTxalanEvents}{\FASTxalanEventsCI}    & \enfp{\FASTxalanNoFPEvents}{\FASTxalanNoFPEventsCI}	& \FASTxalanTotalThreads	& (\FASTxalanMaxLiveThreads)	& \base{\FASTxalanBaseTime}    & \vci{\FASTxalanFTTime}{\FASTxalanFTTimeCI}    & \vci{\FASTxalanHBTime}{\FASTxalanHBTimeCI}		& \vci{\FASTxalanFTOHBTime}{\FASTxalanFTOHBTimeCI}		& \vci{\SLOWxalanDCExcTime}{\SLOWxalanDCExcTimeCI}	& \vci{\SLOWxalanDCnoGExcTime}{\SLOWxalanDCnoGExcTimeCI} 	& \vci{\SLOWxalanCAPOExcTime}{\SLOWxalanCAPOExcTimeCI}	& \vci{\SLOWxalanCAPOnoGExcTime}{\SLOWxalanCAPOnoGExcTimeCI}\\\hline
geomean	& &	& &	& & \roh{\FASTFTTimeGeoMean} & \roh{\FASTHBTimeGeoMean} & \roh{\FASTFTOHBTimeGeoMean} & \roh{\SLOWDCExcTimeGeoMean} & \roh{\SLOWDCnoGExcTimeGeoMean} & \roh{\SLOWCAPOExcTimeGeoMean} & \roh{\SLOWCAPOnoGExcTimeGeoMean} \\
\mc{1}{c}{} & \mc{12}{c}{Run time} \\
\end{tabular}
\end{empty}
\medskip
\begin{empty}
\newcommand{\rzero}{0\xspace}
\newcommand{\roh}[1]{\ifthenelse{\equal{#1}{\rzero}}{0}{#1$\;\!\times$}} 
\newcommand{\rna}{N/A}
\newcommand{\memna}{N/A} 
\newcommand{\st}[1]{(#1~s)} 
\newcommand{\dt}[1]{#1~s} 
\newcommand{\mem}[1]{\ifthenelse{\equal{#1}{\memna}}{\rna}{#1$\;\!\times$}}
\newcommand{\baseMem}[1]{#1~MB} 
\newcommand{\et}[1]{#1M} 
\newcommand{\enfp}[1]{(#1M)} 
\newcommand{\base}[1]{#1~s} 
\newcommand{\vci}[2]{\ifthenelse{\equal{#1}{\rna}}{\rna}{\roh{#1}{\;}\ci{#2}}}
\newcommand{\ci}[1]{\ensuremath{\pm}{\;}\roh{#1}}
\input{result-macros/PIP_slowTool_noCoresSet}
\input{result-macros/PIP_fastTool_extraOpt2Quiet}
\small
\centering
\newcommand{\ics}{\;\;}
\begin{tabular}{@{}l|HHHHHHll|l@{\ics}l|l@{\ics}l@{}}
& \mc{2}{Z}{} & \mc{2}{Z}{} & Base & \mc{3}{c|}{\HB} & \mc{2}{c|}{Unopt-\DC} & \mc{2}{c@{}}{Unopt-\CAPO} \\
Program & \mc{2}{Z}{Events} & \mc{2}{Z}{\#Thr} & memory & RR\FTTwoAbbrv & \FTTwoAbbrv & \FTO & w/$G$ & w/o $G$ & w/$G$ & w/o $G$ \\\hline 
\bench{avrora} 	& \et{\FASTavroraEvents}{\FASTavroraEventsCI}   & \enfp{\FASTavroraNoFPEvents}{\FASTavroraNoFPEventsCI}	& \FASTavroraTotalThreads 	& (\FASTavroraMaxLiveThreads)	& \baseMem{\FASTavroraBaseMem}   & \vci{\FASTavroraFTMem}{\FASTavroraFTMemCI}   & \vci{\FASTavroraHBMem}{\FASTavroraHBMemCI}	& \vci{\FASTavroraFTOHBMem}{\FASTavroraFTOHBMemCI}	& \vci{\SLOWavroraDCExcMem}{\SLOWavroraDCExcMemCI}	& \vci{\SLOWavroraDCnoGExcMem}{\SLOWavroraDCnoGExcMemCI}	& \vci{\SLOWavroraCAPOExcMem}{\SLOWavroraCAPOExcMemCI}	& \vci{\SLOWavroraCAPOnoGExcMem}{\SLOWavroraCAPOnoGExcMemCI}\\
\bench{batik} 	& \et{\FASTbatikEvents}{\FASTbatikEventsCI}    & \enfp{\FASTbatikNoFPEvents}{\FASTbatikNoFPEventsCI}	& \FASTbatikTotalThreads 	& (\FASTbatikMaxLiveThreads)	& \baseMem{\FASTbatikBaseMem}    & \vci{\FASTbatikFTMem}{\FASTbatikFTMemCI}    & \vci{\FASTbatikHBMem}{\FASTbatikHBMemCI}   	& \vci{\FASTbatikFTOHBMem}{\FASTbatikFTOHBMemCI}		& \vci{\SLOWbatikDCExcMem}{\SLOWbatikDCExcMemCI}	& \vci{\SLOWbatikDCnoGExcMem}{\SLOWbatikDCnoGExcMemCI}		& \vci{\SLOWbatikCAPOExcMem}{\SLOWbatikCAPOExcMemCI}	& \vci{\SLOWbatikCAPOnoGExcMem}{\SLOWbatikCAPOnoGExcMemCI} \\
\bench{h2} 		& \et{\FASThtwoEvents}{\FASThtwoEventsCI}     & \enfp{\FASThtwoNoFPEvents}{\FASThtwoNoFPEventsCI}	    & \FASThtwoTotalThreads 	& (\FASThtwoMaxLiveThreads)	    & \baseMem{\FASThtwoBaseMem}     & \vci{\FASThtwoFTMem}{\FASThtwoFTMemCI}     & \vci{\FASThtwoHBMem}{\FASThtwoHBMemCI}    	& \vci{\FASThtwoFTOHBMem}{\FASThtwoFTOHBMemCI}		& \vci{\SLOWhtwoDCExcMem}{\SLOWhtwoDCExcMemCI}		& \vci{\SLOWhtwoDCnoGExcMem}{\SLOWhtwoDCnoGExcMemCI}		& \vci{\SLOWhtwoCAPOExcMem}{\SLOWhtwoCAPOExcMemCI}	& \vci{\SLOWhtwoCAPOnoGExcMem}{\SLOWhtwoCAPOnoGExcMemCI}\\
\bench{jython} 	& \et{\FASTjythonEvents}{\FASTjythonEventsCI}   & \enfp{\FASTjythonNoFPEvents}{\FASTjythonNoFPEventsCI}	& \FASTjythonTotalThreads	& (\FASTjythonMaxLiveThreads)	& \baseMem{\FASTjythonBaseMem}   & \vci{\FASTjythonFTMem}{\FASTjythonFTMemCI}   & \vci{\FASTjythonHBMem}{\FASTjythonHBMemCI} 	& \vci{\FASTjythonFTOHBMem}{\FASTjythonFTOHBMemCI}	& \vci{\SLOWjythonDCExcMem}{\SLOWjythonDCExcMemCI}	& \vci{\SLOWjythonDCnoGExcMem}{\SLOWjythonDCnoGExcMemCI}	& \vci{\SLOWjythonCAPOExcMem}{\SLOWjythonCAPOExcMemCI}	& \vci{\SLOWjythonCAPOnoGExcMem}{\SLOWjythonCAPOnoGExcMemCI} \\
\bench{luindex} & \et{\FASTluindexEvents}{\FASTluindexEventsCI}  & \enfp{\FASTluindexNoFPEvents}{\FASTluindexNoFPEventsCI}	& \FASTluindexTotalThreads	& (\FASTluindexMaxLiveThreads)	& \baseMem{\FASTluindexBaseMem}  & \vci{\FASTluindexFTMem}{\FASTluindexFTMemCI}  & \vci{\FASTluindexHBMem}{\FASTluindexHBMemCI} 	& \vci{\FASTluindexFTOHBMem}{\FASTluindexFTOHBMemCI}	& \vci{\SLOWluindexDCExcMem}{\SLOWluindexDCExcMemCI}	& \vci{\SLOWluindexDCnoGExcMem}{\SLOWluindexDCnoGExcMemCI}	& \vci{\SLOWluindexCAPOExcMem}{\SLOWluindexCAPOExcMemCI}	& \vci{\SLOWluindexCAPOnoGExcMem}{\SLOWluindexCAPOnoGExcMemCI} \\
\bench{lusearch}& \et{\FASTlusearchEvents}{\FASTlusearchEventsCI} & \enfp{\FASTlusearchNoFPEvents}{\FASTlusearchNoFPEventsCI}	& \FASTlusearchTotalThreads	& (\FASTlusearchMaxLiveThreads)	& \baseMem{\FASTlusearchBaseMem} & \vci{\FASTlusearchFTMem}{\FASTlusearchFTMemCI} & \vci{\FASTlusearchHBMem}{\FASTlusearchHBMemCI}	& \vci{\FASTlusearchFTOHBMem}{\FASTlusearchFTOHBMemCI}	& \vci{\SLOWlusearchDCExcMem}{\SLOWlusearchDCExcMemCI}	& \vci{\SLOWlusearchDCnoGExcMem}{\SLOWlusearchDCnoGExcMemCI}	& \vci{\SLOWlusearchCAPOExcMem}{\SLOWlusearchCAPOExcMemCI}& \vci{\SLOWlusearchCAPOnoGExcMem}{\SLOWlusearchCAPOnoGExcMemCI} \\
\bench{pmd}		& \et{\FASTpmdEvents}{\FASTpmdEventsCI}      & \enfp{\FASTpmdNoFPEvents}{\FASTpmdNoFPEventsCI}      & \FASTpmdTotalThreads	    & (\FASTpmdMaxLiveThreads)	    & \baseMem{\FASTpmdBaseMem}      & \vci{\FASTpmdFTMem}{\FASTpmdFTMemCI}      & \vci{\FASTpmdHBMem}{\FASTpmdHBMemCI}		& \vci{\FASTpmdFTOHBMem}{\FASTpmdFTOHBMemCI}		& \vci{\SLOWpmdDCExcMem}{\SLOWpmdDCExcMemCI}		& \vci{\SLOWpmdDCnoGExcMem}{\SLOWpmdDCnoGExcMemCI}		& \vci{\SLOWpmdCAPOExcMem}{\SLOWpmdCAPOExcMemCI}		& \vci{\SLOWpmdCAPOnoGExcMem}{\SLOWpmdCAPOnoGExcMemCI} \\
\bench{sunflow} & \et{\FASTsunflowEvents}{\FASTsunflowEventsCI}  & \enfp{\FASTsunflowNoFPEvents}{\FASTsunflowNoFPEventsCI}	& \FASTsunflowTotalThreads	& (\FASTsunflowMaxLiveThreads)	& \baseMem{\FASTsunflowBaseMem}  & \vci{\FASTsunflowFTMem}{\FASTsunflowFTMemCI}  & \vci{\FASTsunflowHBMem}{\FASTsunflowHBMemCI}	& \vci{\FASTsunflowFTOHBMem}{\FASTsunflowFTOHBMemCI}	& \vci{\SLOWsunflowDCExcMem}{\SLOWsunflowDCExcMemCI}	& \vci{\SLOWsunflowDCnoGExcMem}{\SLOWsunflowDCnoGExcMemCI}	& \vci{\SLOWsunflowCAPOExcMem}{\SLOWsunflowCAPOExcMemCI}	& \vci{\SLOWsunflowCAPOnoGExcMem}{\SLOWsunflowCAPOnoGExcMemCI} \\
\bench{tomcat} 	& \et{\FASTtomcatEvents}{\FASTtomcatEventsCI}   & \enfp{\FASTtomcatNoFPEvents}{\FASTtomcatNoFPEventsCI}	& \FASTtomcatTotalThreads	& (\FASTtomcatMaxLiveThreads)	& \baseMem{\FASTtomcatBaseMem}   & \vci{\FASTtomcatFTMem}{\FASTtomcatFTMemCI}   & \vci{\FASTtomcatHBMem}{\FASTtomcatHBMemCI} 	& \vci{\FASTtomcatFTOHBMem}{\FASTtomcatFTOHBMemCI}	& \vci{\SLOWtomcatDCExcMem}{\SLOWtomcatDCExcMemCI}	& \vci{\SLOWtomcatDCnoGExcMem}{\SLOWtomcatDCnoGExcMemCI}	& \vci{\SLOWtomcatCAPOExcMem}{\SLOWtomcatCAPOExcMemCI}	& \vci{\SLOWtomcatCAPOnoGExcMem}{\SLOWtomcatCAPOnoGExcMemCI} \\
\bench{xalan} 	& \et{\FASTxalanEvents}{\FASTxalanEventsCI}    & \enfp{\FASTxalanNoFPEvents}{\FASTxalanNoFPEventsCI}	& \FASTxalanTotalThreads	& (\FASTxalanMaxLiveThreads)	& \baseMem{\FASTxalanBaseMem}    & \vci{\FASTxalanFTMem}{\FASTxalanFTMemCI}    & \vci{\FASTxalanHBMem}{\FASTxalanHBMemCI}		& \vci{\FASTxalanFTOHBMem}{\FASTxalanFTOHBMemCI}		& \vci{\SLOWxalanDCExcMem}{\SLOWxalanDCExcMemCI}	& \vci{\SLOWxalanDCnoGExcMem}{\SLOWxalanDCnoGExcMemCI} 	& \vci{\SLOWxalanCAPOExcMem}{\SLOWxalanCAPOExcMemCI}	& \vci{\SLOWxalanCAPOnoGExcMem}{\SLOWxalanCAPOnoGExcMemCI}\\\hline
geomean	& &	& &	& & \mem{\FASTFTMemGeoMean} & \mem{\FASTHBMemGeoMean} & \mem{\FASTFTOHBMemGeoMean} & \mem{\SLOWDCExcMemGeoMean} & \mem{\SLOWDCnoGExcMemGeoMean} & \mem{\SLOWCAPOExcMemGeoMean} & \mem{\SLOWCAPOnoGExcMemGeoMean} \\
\mc{13}{c}{Memory usage} \\
\end{tabular}

\caption{Run time and memory usage for FastTrack-based \HB analyses
and for unoptimized \DC and \CAPO analyses, relative to uninstrumented
execution, with 95\% confidence intervals.
The \col{w/$G$} configurations construct a constraint graph $G$ during analysis
and perform vindication.}
\label{tab:performance:fasttrack:time-and-memoryCI}
\end{empty}
\end{table*}
}{
\begin{table}[H]
\begin{empty}
\newcommand{\rzero}{0\xspace}
\newcommand{\roh}[1]{\ifthenelse{\equal{#1}{\rzero}}{0}{#1$\;\!\times$}} 
\newcommand{\rna}{N/A}
\newcommand{\memna}{N/A} 
\newcommand{\st}[1]{(#1~s)} 
\newcommand{\dt}[1]{#1~s} 
\newcommand{\mem}[1]{\ifthenelse{\equal{#1}{\memna}}{\rna}{#1$\;\!\times$}}
\newcommand{\baseMem}[1]{#1~MB} 
\newcommand{\et}[1]{#1M} 
\newcommand{\enfp}[1]{(#1M)} 
\newcommand{\base}[1]{#1~s} 
\newcommand{\vci}[2]{\ifthenelse{\equal{#1}{\rna}}{\rna}{\roh{#1}{\;}\ci{#2}}}
\newcommand{\ci}[1]{\ensuremath{\pm}{\;}\roh{#1}}
\input{result-macros/PIP_slowTool_noCoresSet}
\input{result-macros/PIP_fastTool_extraOpt2Quiet}
\small
\centering
\newcommand{\ics}{\;}
\begin{tabular}{@{}l|HHHHHHll|l@{\ics}l|l@{\ics}l@{}}
& \mc{2}{Z}{} & \mc{2}{Z}{} & Base & \mc{3}{c|}{\HB} & \mc{2}{c|}{Unopt-\DC} & \mc{2}{c@{}}{Unopt-\CAPO} \\
Program & \mc{2}{Z}{Events} & \mc{2}{Z}{\#Thr} & time & RR\FTTwoAbbrv & \FTTwoAbbrv & \FTO & w/ G & w/o G & w/ G & w/o G \\\hline 
\bench{avrora} 	& \et{\FASTavroraEvents}{\FASTavroraEventsCI}   & \enfp{\FASTavroraNoFPEvents}{\FASTavroraNoFPEventsCI}	& \FASTavroraTotalThreads 	& (\FASTavroraMaxLiveThreads)	& \base{\FASTavroraBaseTime}   & \vci{\FASTavroraFTTime}{\FASTavroraFTTimeCI}   & \vci{\FASTavroraHBTime}{\FASTavroraHBTimeCI}	& \vci{\FASTavroraFTOHBTime}{\FASTavroraFTOHBTimeCI}	& \vci{\SLOWavroraDCExcTime}{\SLOWavroraDCExcTimeCI}	& \vci{\SLOWavroraDCnoGExcTime}{\SLOWavroraDCnoGExcTimeCI}	& \vci{\SLOWavroraCAPOExcTime}{\SLOWavroraCAPOExcTimeCI}	& \vci{\SLOWavroraCAPOnoGExcTime}{\SLOWavroraCAPOnoGExcTimeCI}\\
\bench{batik} 	& \et{\FASTbatikEvents}{\FASTbatikEventsCI}    & \enfp{\FASTbatikNoFPEvents}{\FASTbatikNoFPEventsCI}	& \FASTbatikTotalThreads 	& (\FASTbatikMaxLiveThreads)	& \base{\FASTbatikBaseTime}    & \vci{\FASTbatikFTTime}{\FASTbatikFTTimeCI}    & \vci{\FASTbatikHBTime}{\FASTbatikHBTimeCI}   	& \vci{\FASTbatikFTOHBTime}{\FASTbatikFTOHBTimeCI}		& \vci{\SLOWbatikDCExcTime}{\SLOWbatikDCExcTimeCI}	& \vci{\SLOWbatikDCnoGExcTime}{\SLOWbatikDCnoGExcTimeCI}		& \vci{\SLOWbatikCAPOExcTime}{\SLOWbatikCAPOExcTimeCI}	& \vci{\SLOWbatikCAPOnoGExcTime}{\SLOWbatikCAPOnoGExcTimeCI} \\
\bench{h2} 		& \et{\FASThtwoEvents}{\FASThtwoEventsCI}     & \enfp{\FASThtwoNoFPEvents}{\FASThtwoNoFPEventsCI}	    & \FASThtwoTotalThreads 	& (\FASThtwoMaxLiveThreads)	    & \base{\FASThtwoBaseTime}     & \vci{\FASThtwoFTTime}{\FASThtwoFTTimeCI}     & \vci{\FASThtwoHBTime}{\FASThtwoHBTimeCI}    	& \vci{\FASThtwoFTOHBTime}{\FASThtwoFTOHBTimeCI}		& \vci{\SLOWhtwoDCExcTime}{\SLOWhtwoDCExcTimeCI}		& \vci{\SLOWhtwoDCnoGExcTime}{\SLOWhtwoDCnoGExcTimeCI}		& \vci{\SLOWhtwoCAPOExcTime}{\SLOWhtwoCAPOExcTimeCI}	& \vci{\SLOWhtwoCAPOnoGExcTime}{\SLOWhtwoCAPOnoGExcTimeCI}\\
\bench{jython} 	& \et{\FASTjythonEvents}{\FASTjythonEventsCI}   & \enfp{\FASTjythonNoFPEvents}{\FASTjythonNoFPEventsCI}	& \FASTjythonTotalThreads	& (\FASTjythonMaxLiveThreads)	& \base{\FASTjythonBaseTime}   & \vci{\FASTjythonFTTime}{\FASTjythonFTTimeCI}   & \vci{\FASTjythonHBTime}{\FASTjythonHBTimeCI} 	& \vci{\FASTjythonFTOHBTime}{\FASTjythonFTOHBTimeCI}	& \vci{\SLOWjythonDCExcTime}{\SLOWjythonDCExcTimeCI}	& \vci{\SLOWjythonDCnoGExcTime}{\SLOWjythonDCnoGExcTimeCI}	& \vci{\SLOWjythonCAPOExcTime}{\SLOWjythonCAPOExcTimeCI}	& \vci{\SLOWjythonCAPOnoGExcTime}{\SLOWjythonCAPOnoGExcTimeCI} \\
\bench{luindex} & \et{\FASTluindexEvents}{\FASTluindexEventsCI}  & \enfp{\FASTluindexNoFPEvents}{\FASTluindexNoFPEventsCI}	& \FASTluindexTotalThreads	& (\FASTluindexMaxLiveThreads)	& \base{\FASTluindexBaseTime}  & \vci{\FASTluindexFTTime}{\FASTluindexFTTimeCI}  & \vci{\FASTluindexHBTime}{\FASTluindexHBTimeCI} 	& \vci{\FASTluindexFTOHBTime}{\FASTluindexFTOHBTimeCI}	& \vci{\SLOWluindexDCExcTime}{\SLOWluindexDCExcTimeCI}	& \vci{\SLOWluindexDCnoGExcTime}{\SLOWluindexDCnoGExcTimeCI}	& \vci{\SLOWluindexCAPOExcTime}{\SLOWluindexCAPOExcTimeCI}	& \vci{\SLOWluindexCAPOnoGExcTime}{\SLOWluindexCAPOnoGExcTimeCI} \\
\bench{lusearch}& \et{\FASTlusearchEvents}{\FASTlusearchEventsCI} & \enfp{\FASTlusearchNoFPEvents}{\FASTlusearchNoFPEventsCI}	& \FASTlusearchTotalThreads	& (\FASTlusearchMaxLiveThreads)	& \base{\FASTlusearchBaseTime} & \vci{\FASTlusearchFTTime}{\FASTlusearchFTTimeCI} & \vci{\FASTlusearchHBTime}{\FASTlusearchHBTimeCI}	& \vci{\FASTlusearchFTOHBTime}{\FASTlusearchFTOHBTimeCI}	& \vci{\SLOWlusearchDCExcTime}{\SLOWlusearchDCExcTimeCI}	& \vci{\SLOWlusearchDCnoGExcTime}{\SLOWlusearchDCnoGExcTimeCI}	& \vci{\SLOWlusearchCAPOExcTime}{\SLOWlusearchCAPOExcTimeCI}& \vci{\SLOWlusearchCAPOnoGExcTime}{\SLOWlusearchCAPOnoGExcTimeCI} \\
\bench{pmd}		& \et{\FASTpmdEvents}{\FASTpmdEventsCI}      & \enfp{\FASTpmdNoFPEvents}{\FASTpmdNoFPEventsCI}      & \FASTpmdTotalThreads	    & (\FASTpmdMaxLiveThreads)	    & \base{\FASTpmdBaseTime}      & \vci{\FASTpmdFTTime}{\FASTpmdFTTimeCI}      & \vci{\FASTpmdHBTime}{\FASTpmdHBTimeCI}		& \vci{\FASTpmdFTOHBTime}{\FASTpmdFTOHBTimeCI}		& \vci{\SLOWpmdDCExcTime}{\SLOWpmdDCExcTimeCI}		& \vci{\SLOWpmdDCnoGExcTime}{\SLOWpmdDCnoGExcTimeCI}		& \vci{\SLOWpmdCAPOExcTime}{\SLOWpmdCAPOExcTimeCI}		& \vci{\SLOWpmdCAPOnoGExcTime}{\SLOWpmdCAPOnoGExcTimeCI} \\
\bench{sunflow} & \et{\FASTsunflowEvents}{\FASTsunflowEventsCI}  & \enfp{\FASTsunflowNoFPEvents}{\FASTsunflowNoFPEventsCI}	& \FASTsunflowTotalThreads	& (\FASTsunflowMaxLiveThreads)	& \base{\FASTsunflowBaseTime}  & \vci{\FASTsunflowFTTime}{\FASTsunflowFTTimeCI}  & \vci{\FASTsunflowHBTime}{\FASTsunflowHBTimeCI}	& \vci{\FASTsunflowFTOHBTime}{\FASTsunflowFTOHBTimeCI}	& \vci{\SLOWsunflowDCExcTime}{\SLOWsunflowDCExcTimeCI}	& \vci{\SLOWsunflowDCnoGExcTime}{\SLOWsunflowDCnoGExcTimeCI}	& \vci{\SLOWsunflowCAPOExcTime}{\SLOWsunflowCAPOExcTimeCI}	& \vci{\SLOWsunflowCAPOnoGExcTime}{\SLOWsunflowCAPOnoGExcTimeCI} \\
\bench{tomcat} 	& \et{\FASTtomcatEvents}{\FASTtomcatEventsCI}   & \enfp{\FASTtomcatNoFPEvents}{\FASTtomcatNoFPEventsCI}	& \FASTtomcatTotalThreads	& (\FASTtomcatMaxLiveThreads)	& \base{\FASTtomcatBaseTime}   & \vci{\FASTtomcatFTTime}{\FASTtomcatFTTimeCI}   & \vci{\FASTtomcatHBTime}{\FASTtomcatHBTimeCI} 	& \vci{\FASTtomcatFTOHBTime}{\FASTtomcatFTOHBTimeCI}	& \vci{\SLOWtomcatDCExcTime}{\SLOWtomcatDCExcTimeCI}	& \vci{\SLOWtomcatDCnoGExcTime}{\SLOWtomcatDCnoGExcTimeCI}	& \vci{\SLOWtomcatCAPOExcTime}{\SLOWtomcatCAPOExcTimeCI}	& \vci{\SLOWtomcatCAPOnoGExcTime}{\SLOWtomcatCAPOnoGExcTimeCI} \\
\bench{xalan} 	& \et{\FASTxalanEvents}{\FASTxalanEventsCI}    & \enfp{\FASTxalanNoFPEvents}{\FASTxalanNoFPEventsCI}	& \FASTxalanTotalThreads	& (\FASTxalanMaxLiveThreads)	& \base{\FASTxalanBaseTime}    & \vci{\FASTxalanFTTime}{\FASTxalanFTTimeCI}    & \vci{\FASTxalanHBTime}{\FASTxalanHBTimeCI}		& \vci{\FASTxalanFTOHBTime}{\FASTxalanFTOHBTimeCI}		& \vci{\SLOWxalanDCExcTime}{\SLOWxalanDCExcTimeCI}	& \vci{\SLOWxalanDCnoGExcTime}{\SLOWxalanDCnoGExcTimeCI} 	& \vci{\SLOWxalanCAPOExcTime}{\SLOWxalanCAPOExcTimeCI}	& \vci{\SLOWxalanCAPOnoGExcTime}{\SLOWxalanCAPOnoGExcTimeCI}\\\hline
geomean	& &	& &	& & \roh{\FASTFTTimeGeoMean} & \roh{\FASTHBTimeGeoMean} & \roh{\FASTFTOHBTimeGeoMean} & \roh{\SLOWDCExcTimeGeoMean} & \roh{\SLOWDCnoGExcTimeGeoMean} & \roh{\SLOWCAPOExcTimeGeoMean} & \roh{\SLOWCAPOnoGExcTimeGeoMean} \\
\mc{1}{c}{} & \mc{12}{c}{Run time} \\
\end{tabular}
\end{empty}
\medskip
\begin{empty}
\newcommand{\rzero}{0\xspace}
\newcommand{\roh}[1]{\ifthenelse{\equal{#1}{\rzero}}{0}{#1$\;\!\times$}} 
\newcommand{\rna}{N/A}
\newcommand{\memna}{N/A} 
\newcommand{\st}[1]{(#1~s)} 
\newcommand{\dt}[1]{#1~s} 
\newcommand{\mem}[1]{\ifthenelse{\equal{#1}{\memna}}{\rna}{#1$\;\!\times$}}
\newcommand{\baseMem}[1]{#1~MB} 
\newcommand{\et}[1]{#1M} 
\newcommand{\enfp}[1]{(#1M)} 
\newcommand{\base}[1]{#1~s} 
\newcommand{\vci}[2]{\ifthenelse{\equal{#1}{\rna}}{\rna}{\roh{#1}{\;}\ci{#2}}}
\newcommand{\ci}[1]{\ensuremath{\pm}{\;}\roh{#1}}
\input{result-macros/PIP_slowTool_noCoresSet}
\input{result-macros/PIP_fastTool_extraOpt2Quiet}
\small
\centering
\newcommand{\ics}{\;\;}
\begin{tabular}{@{}l|HHHHHHll|l@{\ics}l|l@{\ics}l@{}}
& \mc{2}{Z}{} & \mc{2}{Z}{} & Base & \mc{3}{c|}{\HB} & \mc{2}{c|}{Unopt-\DC} & \mc{2}{c@{}}{Unopt-\CAPO} \\
Program & \mc{2}{Z}{Events} & \mc{2}{Z}{\#Thr} & memory & RR\FTTwoAbbrv & \FTTwoAbbrv & \FTO & w/ G & w/o G & w/ G & w/o G \\\hline 
\bench{avrora} 	& \et{\FASTavroraEvents}{\FASTavroraEventsCI}   & \enfp{\FASTavroraNoFPEvents}{\FASTavroraNoFPEventsCI}	& \FASTavroraTotalThreads 	& (\FASTavroraMaxLiveThreads)	& \baseMem{\FASTavroraBaseMem}   & \vci{\FASTavroraFTMem}{\FASTavroraFTMemCI}   & \vci{\FASTavroraHBMem}{\FASTavroraHBMemCI}	& \vci{\FASTavroraFTOHBMem}{\FASTavroraFTOHBMemCI}	& \vci{\SLOWavroraDCExcMem}{\SLOWavroraDCExcMemCI}	& \vci{\SLOWavroraDCnoGExcMem}{\SLOWavroraDCnoGExcMemCI}	& \vci{\SLOWavroraCAPOExcMem}{\SLOWavroraCAPOExcMemCI}	& \vci{\SLOWavroraCAPOnoGExcMem}{\SLOWavroraCAPOnoGExcMemCI}\\
\bench{batik} 	& \et{\FASTbatikEvents}{\FASTbatikEventsCI}    & \enfp{\FASTbatikNoFPEvents}{\FASTbatikNoFPEventsCI}	& \FASTbatikTotalThreads 	& (\FASTbatikMaxLiveThreads)	& \baseMem{\FASTbatikBaseMem}    & \vci{\FASTbatikFTMem}{\FASTbatikFTMemCI}    & \vci{\FASTbatikHBMem}{\FASTbatikHBMemCI}   	& \vci{\FASTbatikFTOHBMem}{\FASTbatikFTOHBMemCI}		& \vci{\SLOWbatikDCExcMem}{\SLOWbatikDCExcMemCI}	& \vci{\SLOWbatikDCnoGExcMem}{\SLOWbatikDCnoGExcMemCI}		& \vci{\SLOWbatikCAPOExcMem}{\SLOWbatikCAPOExcMemCI}	& \vci{\SLOWbatikCAPOnoGExcMem}{\SLOWbatikCAPOnoGExcMemCI} \\
\bench{h2} 		& \et{\FASThtwoEvents}{\FASThtwoEventsCI}     & \enfp{\FASThtwoNoFPEvents}{\FASThtwoNoFPEventsCI}	    & \FASThtwoTotalThreads 	& (\FASThtwoMaxLiveThreads)	    & \baseMem{\FASThtwoBaseMem}     & \vci{\FASThtwoFTMem}{\FASThtwoFTMemCI}     & \vci{\FASThtwoHBMem}{\FASThtwoHBMemCI}    	& \vci{\FASThtwoFTOHBMem}{\FASThtwoFTOHBMemCI}		& \vci{\SLOWhtwoDCExcMem}{\SLOWhtwoDCExcMemCI}		& \vci{\SLOWhtwoDCnoGExcMem}{\SLOWhtwoDCnoGExcMemCI}		& \vci{\SLOWhtwoCAPOExcMem}{\SLOWhtwoCAPOExcMemCI}	& \vci{\SLOWhtwoCAPOnoGExcMem}{\SLOWhtwoCAPOnoGExcMemCI}\\
\bench{jython} 	& \et{\FASTjythonEvents}{\FASTjythonEventsCI}   & \enfp{\FASTjythonNoFPEvents}{\FASTjythonNoFPEventsCI}	& \FASTjythonTotalThreads	& (\FASTjythonMaxLiveThreads)	& \baseMem{\FASTjythonBaseMem}   & \vci{\FASTjythonFTMem}{\FASTjythonFTMemCI}   & \vci{\FASTjythonHBMem}{\FASTjythonHBMemCI} 	& \vci{\FASTjythonFTOHBMem}{\FASTjythonFTOHBMemCI}	& \vci{\SLOWjythonDCExcMem}{\SLOWjythonDCExcMemCI}	& \vci{\SLOWjythonDCnoGExcMem}{\SLOWjythonDCnoGExcMemCI}	& \vci{\SLOWjythonCAPOExcMem}{\SLOWjythonCAPOExcMemCI}	& \vci{\SLOWjythonCAPOnoGExcMem}{\SLOWjythonCAPOnoGExcMemCI} \\
\bench{luindex} & \et{\FASTluindexEvents}{\FASTluindexEventsCI}  & \enfp{\FASTluindexNoFPEvents}{\FASTluindexNoFPEventsCI}	& \FASTluindexTotalThreads	& (\FASTluindexMaxLiveThreads)	& \baseMem{\FASTluindexBaseMem}  & \vci{\FASTluindexFTMem}{\FASTluindexFTMemCI}  & \vci{\FASTluindexHBMem}{\FASTluindexHBMemCI} 	& \vci{\FASTluindexFTOHBMem}{\FASTluindexFTOHBMemCI}	& \vci{\SLOWluindexDCExcMem}{\SLOWluindexDCExcMemCI}	& \vci{\SLOWluindexDCnoGExcMem}{\SLOWluindexDCnoGExcMemCI}	& \vci{\SLOWluindexCAPOExcMem}{\SLOWluindexCAPOExcMemCI}	& \vci{\SLOWluindexCAPOnoGExcMem}{\SLOWluindexCAPOnoGExcMemCI} \\
\bench{lusearch}& \et{\FASTlusearchEvents}{\FASTlusearchEventsCI} & \enfp{\FASTlusearchNoFPEvents}{\FASTlusearchNoFPEventsCI}	& \FASTlusearchTotalThreads	& (\FASTlusearchMaxLiveThreads)	& \baseMem{\FASTlusearchBaseMem} & \vci{\FASTlusearchFTMem}{\FASTlusearchFTMemCI} & \vci{\FASTlusearchHBMem}{\FASTlusearchHBMemCI}	& \vci{\FASTlusearchFTOHBMem}{\FASTlusearchFTOHBMemCI}	& \vci{\SLOWlusearchDCExcMem}{\SLOWlusearchDCExcMemCI}	& \vci{\SLOWlusearchDCnoGExcMem}{\SLOWlusearchDCnoGExcMemCI}	& \vci{\SLOWlusearchCAPOExcMem}{\SLOWlusearchCAPOExcMemCI}& \vci{\SLOWlusearchCAPOnoGExcMem}{\SLOWlusearchCAPOnoGExcMemCI} \\
\bench{pmd}		& \et{\FASTpmdEvents}{\FASTpmdEventsCI}      & \enfp{\FASTpmdNoFPEvents}{\FASTpmdNoFPEventsCI}      & \FASTpmdTotalThreads	    & (\FASTpmdMaxLiveThreads)	    & \baseMem{\FASTpmdBaseMem}      & \vci{\FASTpmdFTMem}{\FASTpmdFTMemCI}      & \vci{\FASTpmdHBMem}{\FASTpmdHBMemCI}		& \vci{\FASTpmdFTOHBMem}{\FASTpmdFTOHBMemCI}		& \vci{\SLOWpmdDCExcMem}{\SLOWpmdDCExcMemCI}		& \vci{\SLOWpmdDCnoGExcMem}{\SLOWpmdDCnoGExcMemCI}		& \vci{\SLOWpmdCAPOExcMem}{\SLOWpmdCAPOExcMemCI}		& \vci{\SLOWpmdCAPOnoGExcMem}{\SLOWpmdCAPOnoGExcMemCI} \\
\bench{sunflow} & \et{\FASTsunflowEvents}{\FASTsunflowEventsCI}  & \enfp{\FASTsunflowNoFPEvents}{\FASTsunflowNoFPEventsCI}	& \FASTsunflowTotalThreads	& (\FASTsunflowMaxLiveThreads)	& \baseMem{\FASTsunflowBaseMem}  & \vci{\FASTsunflowFTMem}{\FASTsunflowFTMemCI}  & \vci{\FASTsunflowHBMem}{\FASTsunflowHBMemCI}	& \vci{\FASTsunflowFTOHBMem}{\FASTsunflowFTOHBMemCI}	& \vci{\SLOWsunflowDCExcMem}{\SLOWsunflowDCExcMemCI}	& \vci{\SLOWsunflowDCnoGExcMem}{\SLOWsunflowDCnoGExcMemCI}	& \vci{\SLOWsunflowCAPOExcMem}{\SLOWsunflowCAPOExcMemCI}	& \vci{\SLOWsunflowCAPOnoGExcMem}{\SLOWsunflowCAPOnoGExcMemCI} \\
\bench{tomcat} 	& \et{\FASTtomcatEvents}{\FASTtomcatEventsCI}   & \enfp{\FASTtomcatNoFPEvents}{\FASTtomcatNoFPEventsCI}	& \FASTtomcatTotalThreads	& (\FASTtomcatMaxLiveThreads)	& \baseMem{\FASTtomcatBaseMem}   & \vci{\FASTtomcatFTMem}{\FASTtomcatFTMemCI}   & \vci{\FASTtomcatHBMem}{\FASTtomcatHBMemCI} 	& \vci{\FASTtomcatFTOHBMem}{\FASTtomcatFTOHBMemCI}	& \vci{\SLOWtomcatDCExcMem}{\SLOWtomcatDCExcMemCI}	& \vci{\SLOWtomcatDCnoGExcMem}{\SLOWtomcatDCnoGExcMemCI}	& \vci{\SLOWtomcatCAPOExcMem}{\SLOWtomcatCAPOExcMemCI}	& \vci{\SLOWtomcatCAPOnoGExcMem}{\SLOWtomcatCAPOnoGExcMemCI} \\
\bench{xalan} 	& \et{\FASTxalanEvents}{\FASTxalanEventsCI}    & \enfp{\FASTxalanNoFPEvents}{\FASTxalanNoFPEventsCI}	& \FASTxalanTotalThreads	& (\FASTxalanMaxLiveThreads)	& \baseMem{\FASTxalanBaseMem}    & \vci{\FASTxalanFTMem}{\FASTxalanFTMemCI}    & \vci{\FASTxalanHBMem}{\FASTxalanHBMemCI}		& \vci{\FASTxalanFTOHBMem}{\FASTxalanFTOHBMemCI}		& \vci{\SLOWxalanDCExcMem}{\SLOWxalanDCExcMemCI}	& \vci{\SLOWxalanDCnoGExcMem}{\SLOWxalanDCnoGExcMemCI} 	& \vci{\SLOWxalanCAPOExcMem}{\SLOWxalanCAPOExcMemCI}	& \vci{\SLOWxalanCAPOnoGExcMem}{\SLOWxalanCAPOnoGExcMemCI}\\\hline
geomean	& &	& &	& & \mem{\FASTFTMemGeoMean} & \mem{\FASTHBMemGeoMean} & \mem{\FASTFTOHBMemGeoMean} & \mem{\SLOWDCExcMemGeoMean} & \mem{\SLOWDCnoGExcMemGeoMean} & \mem{\SLOWCAPOExcMemGeoMean} & \mem{\SLOWCAPOnoGExcMemGeoMean} \\
\mc{13}{c}{Memory usage} \\
\end{tabular}

\caption{Run time and memory usage for FastTrack-based \HB analyses
and for unoptimized \DC and \CAPO analyses, relative to uninstrumented
execution, with 95\% confidence intervals.}
\label{tab:performance:fasttrack:time-and-memoryCI}
\end{empty}
\end{table}
}

\iftoggle{twoColumnText}{
\begin{table*}[t]
\begin{empty}
\newcommand{\rzero}{0\xspace}
\newcommand{\roh}[1]{\ifthenelse{\equal{#1}{\rna}}{\rna}{#1$\;\!\times$}} 
\newcommand{\rna}{N/A}
\newcommand{\memna}{N/A} 
\newcommand{\st}[1]{(#1~s)} 
\newcommand{\dt}[1]{#1~s} 
\newcommand{\mem}[1]{#1} 
\newcommand{\et}[1]{#1M} 
\newcommand{\enfp}[1]{(#1M)} 
\newcommand{\base}[1]{#1~s} 
\newcommand{\vci}[2]{\ifthenelse{\equal{#1}{\rna}}{\rna}{\roh{#1}{\;}\ci{#2}}}
\newcommand{\ci}[1]{\ensuremath{\pm}{\;}\roh{#1}}
\input{result-macros/PIP_slowTool_noCoresSet}
\input{result-macros/PIP_fastTool_extraOpt2Quiet}
\smaller
\centering
\begin{tabular}{@{}l|Hl@{\;\;}l@{\;\;}l@{}}
        & w/G & Unopt- & \FTO- & \REAbbrv- \\\hline
\HB		& \rna							& \vci{\SLOWavroraHBTime}{\SLOWavroraHBTimeCI}			& \vci{\FASTavroraFTOHBTime}{\FASTavroraFTOHBTimeCI}	& \rna \\
\WCP	& \rna							& \vci{\SLOWavroraWCPTime}{\SLOWavroraWCPTimeCI}			& \vci{\FASTavroraFTOWCPTime}{\FASTavroraFTOWCPTimeCI}	& \vci{\FASTavroraREWCPTime}{\FASTavroraREWCPTimeCI} \\
\DC		& \vci{\SLOWavroraDCExcTime}{\SLOWavroraDCExcTimeCI}	& \vci{\SLOWavroraDCnoGExcTime}{\SLOWavroraDCnoGExcTimeCI}	& \vci{\FASTavroraFTODCTime}{\FASTavroraFTODCTimeCI}	& \vci{\FASTavroraREDCTime}{\FASTavroraREDCTimeCI} \\
\CAPO	& \vci{\SLOWavroraCAPOExcTime}{\SLOWavroraCAPOExcTimeCI}	& \vci{\SLOWavroraCAPOnoGExcTime}{\SLOWavroraCAPOnoGExcTimeCI}	& \vci{\FASTavroraFTOCAPOTime}{\FASTavroraFTOCAPOTimeCI}	& \vci{\FASTavroraRECAPOTime}{\FASTavroraRECAPOTimeCI} \\
\mc{1}{c}{} & \mc{4}{c}{\bench{avrora}} \\
\end{tabular}
\hfill
\begin{tabular}{@{}HHl@{\;\;}l@{\;\;}l@{}}
Relation& w/ G & Unopt- & \FTO- & \REAbbrv- \\\hline
\HB		& \rna							& \vci{\SLOWbatikHBTime}{\SLOWbatikHBTimeCI}			& \vci{\FASTbatikFTOHBTime}{\FASTbatikFTOHBTimeCI}	& \rna \\
\WCP	& \rna							& \vci{\SLOWbatikWCPTime}{\SLOWbatikWCPTimeCI}			& \vci{\FASTbatikFTOWCPTime}{\FASTbatikFTOWCPTimeCI}	& \vci{\FASTbatikREWCPTime}{\FASTbatikREWCPTimeCI} \\
\DC		& \vci{\SLOWbatikDCExcTime}{\SLOWbatikDCExcTimeCI}	& \vci{\SLOWbatikDCnoGExcTime}{\SLOWbatikDCnoGExcTimeCI}	& \vci{\FASTbatikFTODCTime}{\FASTbatikFTODCTimeCI}	& \vci{\FASTbatikREDCTime}{\FASTbatikREDCTimeCI} \\
\CAPO	& \vci{\SLOWbatikCAPOExcTime}{\SLOWbatikCAPOExcTimeCI}	& \vci{\SLOWbatikCAPOnoGExcTime}{\SLOWbatikCAPOnoGExcTimeCI}	& \vci{\FASTbatikFTOCAPOTime}{\FASTbatikFTOCAPOTimeCI}	& \vci{\FASTbatikRECAPOTime}{\FASTbatikRECAPOTimeCI} \\
		& \mc{4}{c}{\bench{batik}} \\
\end{tabular}
\hfill
\begin{tabular}{@{}HHl@{\;\;}l@{\;\;}l@{}}
		& w/ G & Unopt- & \FTO- & \REAbbrv- \\\hline
\HB		& \rna							& \vci{\SLOWhtwoHBTime}{\SLOWhtwoHBTimeCI}			& \vci{\FASThtwoFTOHBTime}{\FASThtwoFTOHBTimeCI}	& \rna \\
\WCP	& \rna							& \vci{\SLOWhtwoWCPTime}{\SLOWhtwoWCPTimeCI}			& \vci{\FASThtwoFTOWCPTime}{\FASThtwoFTOWCPTimeCI}	& \vci{\FASThtwoREWCPTime}{\FASThtwoREWCPTimeCI} \\
\DC		& \vci{\SLOWhtwoDCExcTime}{\SLOWhtwoDCExcTimeCI}	& \vci{\SLOWhtwoDCnoGExcTime}{\SLOWhtwoDCnoGExcTimeCI}	& \vci{\FASThtwoFTODCTime}{\FASThtwoFTODCTimeCI}	& \vci{\FASThtwoREDCTime}{\FASThtwoREDCTimeCI} \\
\CAPO	& \vci{\SLOWhtwoCAPOExcTime}{\SLOWhtwoCAPOExcTimeCI}	& \vci{\SLOWhtwoCAPOnoGExcTime}{\SLOWhtwoCAPOnoGExcTimeCI}	& \vci{\FASThtwoFTOCAPOTime}{\FASThtwoFTOCAPOTimeCI}	& \vci{\FASThtwoRECAPOTime}{\FASThtwoRECAPOTimeCI} \\
		& \mc{4}{c}{\bench{h2}} \\
\end{tabular}
\medskip\\
\begin{tabular}{@{}l|Hl@{\;\;}l@{\;\;}l@{}}
		& w/ G & Unopt- & \FTO- & \REAbbrv- \\\hline
\HB		& \rna							& \vci{\SLOWjythonHBTime}{\SLOWjythonHBTimeCI}			& \vci{\FASTjythonFTOHBTime}{\FASTjythonFTOHBTimeCI}	& \rna \\
\WCP	& \rna							& \vci{\SLOWjythonWCPTime}{\SLOWjythonWCPTimeCI}			& \vci{\FASTjythonFTOWCPTime}{\FASTjythonFTOWCPTimeCI}	& \vci{\FASTjythonREWCPTime}{\FASTjythonREWCPTimeCI} \\
\DC		& \vci{\SLOWjythonDCExcTime}{\SLOWjythonDCExcTimeCI}	& \vci{\SLOWjythonDCnoGExcTime}{\SLOWjythonDCnoGExcTimeCI}	& \vci{\FASTjythonFTODCTime}{\FASTjythonFTODCTimeCI}	& \vci{\FASTjythonREDCTime}{\FASTjythonREDCTimeCI} \\
\CAPO	& \vci{\SLOWjythonCAPOExcTime}{\SLOWjythonCAPOExcTimeCI}	& \vci{\SLOWjythonCAPOnoGExcTime}{\SLOWjythonCAPOnoGExcTimeCI}	& \vci{\FASTjythonFTOCAPOTime}{\FASTjythonFTOCAPOTimeCI}	& \vci{\FASTjythonRECAPOTime}{\FASTjythonRECAPOTimeCI} \\
\mc{1}{c}{} & \mc{4}{c}{\bench{jython}} \\
\end{tabular}
\hfill
\begin{tabular}{@{}HHl@{\;\;}l@{\;\;}l@{}}
		& w/ G & Unopt- & \FTO- & \REAbbrv- \\\hline
\HB		& \rna							& \vci{\SLOWluindexHBTime}{\SLOWluindexHBTimeCI}			& \vci{\FASTluindexFTOHBTime}{\FASTluindexFTOHBTimeCI}	& \rna \\
\WCP	& \rna							& \vci{\SLOWluindexWCPTime}{\SLOWluindexWCPTimeCI}			& \vci{\FASTluindexFTOWCPTime}{\FASTluindexFTOWCPTimeCI}	& \vci{\FASTluindexREWCPTime}{\FASTluindexREWCPTimeCI} \\
\DC		& \vci{\SLOWluindexDCExcTime}{\SLOWluindexDCExcTimeCI}	& \vci{\SLOWluindexDCnoGExcTime}{\SLOWluindexDCnoGExcTimeCI}	& \vci{\FASTluindexFTODCTime}{\FASTluindexFTODCTimeCI}	& \vci{\FASTluindexREDCTime}{\FASTluindexREDCTimeCI} \\
\CAPO	& \vci{\SLOWluindexCAPOExcTime}{\SLOWluindexCAPOExcTimeCI}	& \vci{\SLOWluindexCAPOnoGExcTime}{\SLOWluindexCAPOnoGExcTimeCI}	& \vci{\FASTluindexFTOCAPOTime}{\FASTluindexFTOCAPOTimeCI}	& \vci{\FASTluindexRECAPOTime}{\FASTluindexRECAPOTimeCI} \\
		& \mc{4}{c}{\bench{luindex}} \\
\end{tabular}
\hfill
\begin{tabular}{@{}HHl@{\;\;}l@{\;\;}l@{}}
        & w/ G & Unopt- & \FTO- & \REAbbrv- \\\hline
\HB		& \rna							& \vci{\SLOWlusearchHBTime}{\SLOWlusearchHBTimeCI}			& \vci{\FASTlusearchFTOHBTime}{\FASTlusearchFTOHBTimeCI}	& \rna \\
\WCP	& \rna							& \vci{\SLOWlusearchWCPTime}{\SLOWlusearchWCPTimeCI}			& \vci{\FASTlusearchFTOWCPTime}{\FASTlusearchFTOWCPTimeCI}	& \vci{\FASTlusearchREWCPTime}{\FASTlusearchREWCPTimeCI} \\
\DC		& \vci{\SLOWlusearchDCExcTime}{\SLOWlusearchDCExcTimeCI}	& \vci{\SLOWlusearchDCnoGExcTime}{\SLOWlusearchDCnoGExcTimeCI}	& \vci{\FASTlusearchFTODCTime}{\FASTlusearchFTODCTimeCI}	& \vci{\FASTlusearchREDCTime}{\FASTlusearchREDCTimeCI} \\
\CAPO	& \vci{\SLOWlusearchCAPOExcTime}{\SLOWlusearchCAPOExcTimeCI}	& \vci{\SLOWlusearchCAPOnoGExcTime}{\SLOWlusearchCAPOnoGExcTimeCI}	& \vci{\FASTlusearchFTOCAPOTime}{\FASTlusearchFTOCAPOTimeCI}	& \vci{\FASTlusearchRECAPOTime}{\FASTlusearchRECAPOTimeCI} \\
		& \mc{4}{c}{\bench{lusearch}} \\
\end{tabular}
\medskip\\
\begin{tabular}{@{}l|Hl@{\;\;}l@{\;\;}l@{}}
		& w/ G & Unopt- & \FTO- & \REAbbrv- \\\hline
\HB		& \rna							& \vci{\SLOWpmdHBTime}{\SLOWpmdHBTimeCI}			& \vci{\FASTpmdFTOHBTime}{\FASTpmdFTOHBTimeCI}	& \rna \\
\WCP	& \rna							& \vci{\SLOWpmdWCPTime}{\SLOWpmdWCPTimeCI}			& \vci{\FASTpmdFTOWCPTime}{\FASTpmdFTOWCPTimeCI}	& \vci{\FASTpmdREWCPTime}{\FASTpmdREWCPTimeCI} \\
\DC		& \vci{\SLOWpmdDCExcTime}{\SLOWpmdDCExcTimeCI}	& \vci{\SLOWpmdDCnoGExcTime}{\SLOWpmdDCnoGExcTimeCI}	& \vci{\FASTpmdFTODCTime}{\FASTpmdFTODCTimeCI}	& \vci{\FASTpmdREDCTime}{\FASTpmdREDCTimeCI} \\
\CAPO	& \vci{\SLOWpmdCAPOExcTime}{\SLOWpmdCAPOExcTimeCI}	& \vci{\SLOWpmdCAPOnoGExcTime}{\SLOWpmdCAPOnoGExcTimeCI}	& \vci{\FASTpmdFTOCAPOTime}{\FASTpmdFTOCAPOTimeCI}	& \vci{\FASTpmdRECAPOTime}{\FASTpmdRECAPOTimeCI} \\
\mc{1}{c}{} & \mc{4}{c}{\bench{pmd}} \\
\end{tabular}
\hfill
\begin{tabular}{@{}HHl@{\;\;}l@{\;\;}l@{}}
Relation& w/ G & Unopt- & \FTO- & \REAbbrv- \\\hline
\HB		& \rna							& \vci{\SLOWsunflowHBTime}{\SLOWsunflowHBTimeCI}			& \vci{\FASTsunflowFTOHBTime}{\FASTsunflowFTOHBTimeCI}	& \rna \\
\WCP	& \rna							& \vci{\SLOWsunflowWCPTime}{\SLOWsunflowWCPTimeCI}			& \vci{\FASTsunflowFTOWCPTime}{\FASTsunflowFTOWCPTimeCI}	& \vci{\FASTsunflowREWCPTime}{\FASTsunflowREWCPTimeCI} \\
\DC		& \vci{\SLOWsunflowDCExcTime}{\SLOWsunflowDCExcTimeCI}	& \vci{\SLOWsunflowDCnoGExcTime}{\SLOWsunflowDCnoGExcTimeCI}	& \vci{\FASTsunflowFTODCTime}{\FASTsunflowFTODCTimeCI}	& \vci{\FASTsunflowREDCTime}{\FASTsunflowREDCTimeCI} \\
\CAPO	& \vci{\SLOWsunflowCAPOExcTime}{\SLOWsunflowCAPOExcTimeCI}	& \vci{\SLOWsunflowCAPOnoGExcTime}{\SLOWsunflowCAPOnoGExcTimeCI}	& \vci{\FASTsunflowFTOCAPOTime}{\FASTsunflowFTOCAPOTimeCI}	& \vci{\FASTsunflowRECAPOTime}{\FASTsunflowRECAPOTimeCI} \\
		& \mc{4}{c}{\bench{sunflow}} \\
\end{tabular}
\hfill
\begin{tabular}{@{}HHl@{\;\;}l@{\;\;}l@{}}
		& w/ G & Unopt- & \FTO- & \REAbbrv- \\\hline
\HB		& \rna							& \vci{\SLOWtomcatHBTime}{\SLOWtomcatHBTimeCI}			& \vci{\FASTtomcatFTOHBTime}{\FASTtomcatFTOHBTimeCI}	& \rna \\
\WCP	& \rna							& \vci{\SLOWtomcatWCPTime}{\SLOWtomcatWCPTimeCI}			& \vci{\FASTtomcatFTOWCPTime}{\FASTtomcatFTOWCPTimeCI}	& \vci{\FASTtomcatREWCPTime}{\FASTtomcatREWCPTimeCI} \\
\DC		& \vci{\SLOWtomcatDCExcTime}{\SLOWtomcatDCExcTimeCI}	& \vci{\SLOWtomcatDCnoGExcTime}{\SLOWtomcatDCnoGExcTimeCI}	& \vci{\FASTtomcatFTODCTime}{\FASTtomcatFTODCTimeCI}	& \vci{\FASTtomcatREDCTime}{\FASTtomcatREDCTimeCI} \\
\CAPO	& \vci{\SLOWtomcatCAPOExcTime}{\SLOWtomcatCAPOExcTimeCI}	& \vci{\SLOWtomcatCAPOnoGExcTime}{\SLOWtomcatCAPOnoGExcTimeCI}	& \vci{\FASTtomcatFTOCAPOTime}{\FASTtomcatFTOCAPOTimeCI}	& \vci{\FASTtomcatRECAPOTime}{\FASTtomcatRECAPOTimeCI} \\
		& \mc{4}{c}{\bench{tomcat}} \\
\end{tabular}
\medskip\\
\begin{tabular}{@{}l|Hl@{\;\;}l@{\;\;}l@{}}
		& w/ G & Unopt- & \FTO- & \REAbbrv- \\\hline
\HB		& \rna							& \vci{\SLOWxalanHBTime}{\SLOWxalanHBTimeCI}			& \vci{\FASTxalanFTOHBTime}{\FASTxalanFTOHBTimeCI}	& \rna \\
\WCP	& \rna							& \vci{\SLOWxalanWCPTime}{\SLOWxalanWCPTimeCI}			& \vci{\FASTxalanFTOWCPTime}{\FASTxalanFTOWCPTimeCI}	& \vci{\FASTxalanREWCPTime}{\FASTxalanREWCPTimeCI} \\
\DC		& \vci{\SLOWxalanDCExcTime}{\SLOWxalanDCExcTimeCI}	& \vci{\SLOWxalanDCnoGExcTime}{\SLOWxalanDCnoGExcTimeCI}	& \vci{\FASTxalanFTODCTime}{\FASTxalanFTODCTimeCI}	& \vci{\FASTxalanREDCTime}{\FASTxalanREDCTimeCI} \\
\CAPO	& \vci{\SLOWxalanCAPOExcTime}{\SLOWxalanCAPOExcTimeCI}	& \vci{\SLOWxalanCAPOnoGExcTime}{\SLOWxalanCAPOnoGExcTimeCI}	& \vci{\FASTxalanFTOCAPOTime}{\FASTxalanFTOCAPOTimeCI}	& \vci{\FASTxalanRECAPOTime}{\FASTxalanRECAPOTimeCI} \\
\mc{1}{c}{} & \mc{4}{c}{\bench{xalan}} \\
\end{tabular}
\end{empty}
\caption{Run time, relative to uninstrumented execution, of various analyses for each evaluated program
with 95\% confidence intervals.}
\label{tab:performance:allDaCapo:CI}
\end{table*}

}{

\begin{table}[H]
\begin{empty}
\newcommand{\rzero}{0\xspace}
\newcommand{\roh}[1]{\ifthenelse{\equal{#1}{\rna}}{\rna}{#1$\;\!\times$}} 
\newcommand{\rna}{N/A}
\newcommand{\memna}{N/A} 
\newcommand{\st}[1]{(#1~s)} 
\newcommand{\dt}[1]{#1~s} 
\newcommand{\mem}[1]{#1} 
\newcommand{\et}[1]{#1M} 
\newcommand{\enfp}[1]{(#1M)} 
\newcommand{\base}[1]{#1~s} 
\newcommand{\vci}[2]{\ifthenelse{\equal{#1}{\rna}}{\rna}{\roh{#1}{\;}\ci{#2}}}
\newcommand{\ci}[1]{\ensuremath{\pm}{\;}\roh{#1}}
\input{result-macros/PIP_slowTool_noCoresSet}
\input{result-macros/PIP_fastTool_extraOpt2Quiet}
\small
\centering
\begin{tabular}{@{}l|Hlll@{}}
        & w/ G & Unopt- & \FTO- & \REAbbrv- \\\hline
\HB		& \rna							& \vci{\SLOWavroraHBTime}{\SLOWavroraHBTimeCI}			& \vci{\FASTavroraFTOHBTime}{\FASTavroraFTOHBTimeCI}	& \rna \\
\WCP	& \rna							& \vci{\SLOWavroraWCPTime}{\SLOWavroraWCPTimeCI}			& \vci{\FASTavroraFTOWCPTime}{\FASTavroraFTOWCPTimeCI}	& \vci{\FASTavroraREWCPTime}{\FASTavroraREWCPTimeCI} \\
\DC		& \vci{\SLOWavroraDCExcTime}{\SLOWavroraDCExcTimeCI}	& \vci{\SLOWavroraDCnoGExcTime}{\SLOWavroraDCnoGExcTimeCI}	& \vci{\FASTavroraFTODCTime}{\FASTavroraFTODCTimeCI}	& \vci{\FASTavroraREDCTime}{\FASTavroraREDCTimeCI} \\
\CAPO	& \vci{\SLOWavroraCAPOExcTime}{\SLOWavroraCAPOExcTimeCI}	& \vci{\SLOWavroraCAPOnoGExcTime}{\SLOWavroraCAPOnoGExcTimeCI}	& \vci{\FASTavroraFTOCAPOTime}{\FASTavroraFTOCAPOTimeCI}	& \vci{\FASTavroraRECAPOTime}{\FASTavroraRECAPOTimeCI} \\
\mc{1}{c}{} & \mc{4}{c}{\bench{avrora}} \\
\end{tabular}
\hfill
\begin{tabular}{@{}HHlll@{}}
Relation& w/ G & Unopt- & \FTO- & \REAbbrv- \\\hline
\HB		& \rna							& \vci{\SLOWbatikHBTime}{\SLOWbatikHBTimeCI}			& \vci{\FASTbatikFTOHBTime}{\FASTbatikFTOHBTimeCI}	& \rna \\
\WCP	& \rna							& \vci{\SLOWbatikWCPTime}{\SLOWbatikWCPTimeCI}			& \vci{\FASTbatikFTOWCPTime}{\FASTbatikFTOWCPTimeCI}	& \vci{\FASTbatikREWCPTime}{\FASTbatikREWCPTimeCI} \\
\DC		& \vci{\SLOWbatikDCExcTime}{\SLOWbatikDCExcTimeCI}	& \vci{\SLOWbatikDCnoGExcTime}{\SLOWbatikDCnoGExcTimeCI}	& \vci{\FASTbatikFTODCTime}{\FASTbatikFTODCTimeCI}	& \vci{\FASTbatikREDCTime}{\FASTbatikREDCTimeCI} \\
\CAPO	& \vci{\SLOWbatikCAPOExcTime}{\SLOWbatikCAPOExcTimeCI}	& \vci{\SLOWbatikCAPOnoGExcTime}{\SLOWbatikCAPOnoGExcTimeCI}	& \vci{\FASTbatikFTOCAPOTime}{\FASTbatikFTOCAPOTimeCI}	& \vci{\FASTbatikRECAPOTime}{\FASTbatikRECAPOTimeCI} \\
		& \mc{4}{c}{\bench{batik}} \\
\end{tabular}
\medskip\\
\begin{tabular}{@{}l|Hlll@{}}
		& w/ G & Unopt- & \FTO- & \REAbbrv- \\\hline
\HB		& \rna							& \vci{\SLOWhtwoHBTime}{\SLOWhtwoHBTimeCI}			& \vci{\FASThtwoFTOHBTime}{\FASThtwoFTOHBTimeCI}	& \rna \\
\WCP	& \rna							& \vci{\SLOWhtwoWCPTime}{\SLOWhtwoWCPTimeCI}			& \vci{\FASThtwoFTOWCPTime}{\FASThtwoFTOWCPTimeCI}	& \vci{\FASThtwoREWCPTime}{\FASThtwoREWCPTimeCI} \\
\DC		& \vci{\SLOWhtwoDCExcTime}{\SLOWhtwoDCExcTimeCI}	& \vci{\SLOWhtwoDCnoGExcTime}{\SLOWhtwoDCnoGExcTimeCI}	& \vci{\FASThtwoFTODCTime}{\FASThtwoFTODCTimeCI}	& \vci{\FASThtwoREDCTime}{\FASThtwoREDCTimeCI} \\
\CAPO	& \vci{\SLOWhtwoCAPOExcTime}{\SLOWhtwoCAPOExcTimeCI}	& \vci{\SLOWhtwoCAPOnoGExcTime}{\SLOWhtwoCAPOnoGExcTimeCI}	& \vci{\FASThtwoFTOCAPOTime}{\FASThtwoFTOCAPOTimeCI}	& \vci{\FASThtwoRECAPOTime}{\FASThtwoRECAPOTimeCI} \\
\mc{1}{c}{} & \mc{4}{c}{\bench{h2}} \\
\end{tabular}
\hfill
\begin{tabular}{@{}HHlll@{}}
		& w/ G & Unopt- & \FTO- & \REAbbrv- \\\hline
\HB		& \rna							& \vci{\SLOWjythonHBTime}{\SLOWjythonHBTimeCI}			& \vci{\FASTjythonFTOHBTime}{\FASTjythonFTOHBTimeCI}	& \rna \\
\WCP	& \rna							& \vci{\SLOWjythonWCPTime}{\SLOWjythonWCPTimeCI}			& \vci{\FASTjythonFTOWCPTime}{\FASTjythonFTOWCPTimeCI}	& \vci{\FASTjythonREWCPTime}{\FASTjythonREWCPTimeCI} \\
\DC		& \vci{\SLOWjythonDCExcTime}{\SLOWjythonDCExcTimeCI}	& \vci{\SLOWjythonDCnoGExcTime}{\SLOWjythonDCnoGExcTimeCI}	& \vci{\FASTjythonFTODCTime}{\FASTjythonFTODCTimeCI}	& \vci{\FASTjythonREDCTime}{\FASTjythonREDCTimeCI} \\
\CAPO	& \vci{\SLOWjythonCAPOExcTime}{\SLOWjythonCAPOExcTimeCI}	& \vci{\SLOWjythonCAPOnoGExcTime}{\SLOWjythonCAPOnoGExcTimeCI}	& \vci{\FASTjythonFTOCAPOTime}{\FASTjythonFTOCAPOTimeCI}	& \vci{\FASTjythonRECAPOTime}{\FASTjythonRECAPOTimeCI} \\
		& \mc{4}{c}{\bench{jython}} \\
\end{tabular}
\medskip\\
\begin{tabular}{@{}l|Hlll@{}}
		& w/ G & Unopt- & \FTO- & \REAbbrv- \\\hline
\HB		& \rna							& \vci{\SLOWluindexHBTime}{\SLOWluindexHBTimeCI}			& \vci{\FASTluindexFTOHBTime}{\FASTluindexFTOHBTimeCI}	& \rna \\
\WCP	& \rna							& \vci{\SLOWluindexWCPTime}{\SLOWluindexWCPTimeCI}			& \vci{\FASTluindexFTOWCPTime}{\FASTluindexFTOWCPTimeCI}	& \vci{\FASTluindexREWCPTime}{\FASTluindexREWCPTimeCI} \\
\DC		& \vci{\SLOWluindexDCExcTime}{\SLOWluindexDCExcTimeCI}	& \vci{\SLOWluindexDCnoGExcTime}{\SLOWluindexDCnoGExcTimeCI}	& \vci{\FASTluindexFTODCTime}{\FASTluindexFTODCTimeCI}	& \vci{\FASTluindexREDCTime}{\FASTluindexREDCTimeCI} \\
\CAPO	& \vci{\SLOWluindexCAPOExcTime}{\SLOWluindexCAPOExcTimeCI}	& \vci{\SLOWluindexCAPOnoGExcTime}{\SLOWluindexCAPOnoGExcTimeCI}	& \vci{\FASTluindexFTOCAPOTime}{\FASTluindexFTOCAPOTimeCI}	& \vci{\FASTluindexRECAPOTime}{\FASTluindexRECAPOTimeCI} \\
\mc{1}{c}{} & \mc{4}{c}{\bench{luindex}} \\
\end{tabular}
\hfill
\begin{tabular}{@{}HHlll@{}}
        & w/ G & Unopt- & \FTO- & \REAbbrv- \\\hline
\HB		& \rna							& \vci{\SLOWlusearchHBTime}{\SLOWlusearchHBTimeCI}			& \vci{\FASTlusearchFTOHBTime}{\FASTlusearchFTOHBTimeCI}	& \rna \\
\WCP	& \rna							& \vci{\SLOWlusearchWCPTime}{\SLOWlusearchWCPTimeCI}			& \vci{\FASTlusearchFTOWCPTime}{\FASTlusearchFTOWCPTimeCI}	& \vci{\FASTlusearchREWCPTime}{\FASTlusearchREWCPTimeCI} \\
\DC		& \vci{\SLOWlusearchDCExcTime}{\SLOWlusearchDCExcTimeCI}	& \vci{\SLOWlusearchDCnoGExcTime}{\SLOWlusearchDCnoGExcTimeCI}	& \vci{\FASTlusearchFTODCTime}{\FASTlusearchFTODCTimeCI}	& \vci{\FASTlusearchREDCTime}{\FASTlusearchREDCTimeCI} \\
\CAPO	& \vci{\SLOWlusearchCAPOExcTime}{\SLOWlusearchCAPOExcTimeCI}	& \vci{\SLOWlusearchCAPOnoGExcTime}{\SLOWlusearchCAPOnoGExcTimeCI}	& \vci{\FASTlusearchFTOCAPOTime}{\FASTlusearchFTOCAPOTimeCI}	& \vci{\FASTlusearchRECAPOTime}{\FASTlusearchRECAPOTimeCI} \\
		& \mc{4}{c}{\bench{lusearch}} \\
\end{tabular}
\medskip\\
\begin{tabular}{@{}l|Hlll@{}}
		& w/ G & Unopt- & \FTO- & \REAbbrv- \\\hline
\HB		& \rna							& \vci{\SLOWpmdHBTime}{\SLOWpmdHBTimeCI}			& \vci{\FASTpmdFTOHBTime}{\FASTpmdFTOHBTimeCI}	& \rna \\
\WCP	& \rna							& \vci{\SLOWpmdWCPTime}{\SLOWpmdWCPTimeCI}			& \vci{\FASTpmdFTOWCPTime}{\FASTpmdFTOWCPTimeCI}	& \vci{\FASTpmdREWCPTime}{\FASTpmdREWCPTimeCI} \\
\DC		& \vci{\SLOWpmdDCExcTime}{\SLOWpmdDCExcTimeCI}	& \vci{\SLOWpmdDCnoGExcTime}{\SLOWpmdDCnoGExcTimeCI}	& \vci{\FASTpmdFTODCTime}{\FASTpmdFTODCTimeCI}	& \vci{\FASTpmdREDCTime}{\FASTpmdREDCTimeCI} \\
\CAPO	& \vci{\SLOWpmdCAPOExcTime}{\SLOWpmdCAPOExcTimeCI}	& \vci{\SLOWpmdCAPOnoGExcTime}{\SLOWpmdCAPOnoGExcTimeCI}	& \vci{\FASTpmdFTOCAPOTime}{\FASTpmdFTOCAPOTimeCI}	& \vci{\FASTpmdRECAPOTime}{\FASTpmdRECAPOTimeCI} \\
\mc{1}{c}{} & \mc{4}{c}{\bench{pmd}} \\
\end{tabular}
\hfill
\begin{tabular}{@{}HHlll@{}}
Relation& w/ G & Unopt- & \FTO- & \REAbbrv- \\\hline
\HB		& \rna							& \vci{\SLOWsunflowHBTime}{\SLOWsunflowHBTimeCI}			& \vci{\FASTsunflowFTOHBTime}{\FASTsunflowFTOHBTimeCI}	& \rna \\
\WCP	& \rna							& \vci{\SLOWsunflowWCPTime}{\SLOWsunflowWCPTimeCI}			& \vci{\FASTsunflowFTOWCPTime}{\FASTsunflowFTOWCPTimeCI}	& \vci{\FASTsunflowREWCPTime}{\FASTsunflowREWCPTimeCI} \\
\DC		& \vci{\SLOWsunflowDCExcTime}{\SLOWsunflowDCExcTimeCI}	& \vci{\SLOWsunflowDCnoGExcTime}{\SLOWsunflowDCnoGExcTimeCI}	& \vci{\FASTsunflowFTODCTime}{\FASTsunflowFTODCTimeCI}	& \vci{\FASTsunflowREDCTime}{\FASTsunflowREDCTimeCI} \\
\CAPO	& \vci{\SLOWsunflowCAPOExcTime}{\SLOWsunflowCAPOExcTimeCI}	& \vci{\SLOWsunflowCAPOnoGExcTime}{\SLOWsunflowCAPOnoGExcTimeCI}	& \vci{\FASTsunflowFTOCAPOTime}{\FASTsunflowFTOCAPOTimeCI}	& \vci{\FASTsunflowRECAPOTime}{\FASTsunflowRECAPOTimeCI} \\
		& \mc{4}{c}{\bench{sunflow}} \\
\end{tabular}
\medskip\\
\begin{tabular}{@{}l|Hlll@{}}
		& w/ G & Unopt- & \FTO- & \REAbbrv- \\\hline
\HB		& \rna							& \vci{\SLOWtomcatHBTime}{\SLOWtomcatHBTimeCI}			& \vci{\FASTtomcatFTOHBTime}{\FASTtomcatFTOHBTimeCI}	& \rna \\
\WCP	& \rna							& \vci{\SLOWtomcatWCPTime}{\SLOWtomcatWCPTimeCI}			& \vci{\FASTtomcatFTOWCPTime}{\FASTtomcatFTOWCPTimeCI}	& \vci{\FASTtomcatREWCPTime}{\FASTtomcatREWCPTimeCI} \\
\DC		& \vci{\SLOWtomcatDCExcTime}{\SLOWtomcatDCExcTimeCI}	& \vci{\SLOWtomcatDCnoGExcTime}{\SLOWtomcatDCnoGExcTimeCI}	& \vci{\FASTtomcatFTODCTime}{\FASTtomcatFTODCTimeCI}	& \vci{\FASTtomcatREDCTime}{\FASTtomcatREDCTimeCI} \\
\CAPO	& \vci{\SLOWtomcatCAPOExcTime}{\SLOWtomcatCAPOExcTimeCI}	& \vci{\SLOWtomcatCAPOnoGExcTime}{\SLOWtomcatCAPOnoGExcTimeCI}	& \vci{\FASTtomcatFTOCAPOTime}{\FASTtomcatFTOCAPOTimeCI}	& \vci{\FASTtomcatRECAPOTime}{\FASTtomcatRECAPOTimeCI} \\
\mc{1}{c}{} & \mc{4}{c}{\bench{tomcat}} \\
\end{tabular}
\hfill
\begin{tabular}{@{}HHlll@{}}
		& w/ G & Unopt- & \FTO- & \REAbbrv- \\\hline
\HB		& \rna							& \vci{\SLOWxalanHBTime}{\SLOWxalanHBTimeCI}			& \vci{\FASTxalanFTOHBTime}{\FASTxalanFTOHBTimeCI}	& \rna \\
\WCP	& \rna							& \vci{\SLOWxalanWCPTime}{\SLOWxalanWCPTimeCI}			& \vci{\FASTxalanFTOWCPTime}{\FASTxalanFTOWCPTimeCI}	& \vci{\FASTxalanREWCPTime}{\FASTxalanREWCPTimeCI} \\
\DC		& \vci{\SLOWxalanDCExcTime}{\SLOWxalanDCExcTimeCI}	& \vci{\SLOWxalanDCnoGExcTime}{\SLOWxalanDCnoGExcTimeCI}	& \vci{\FASTxalanFTODCTime}{\FASTxalanFTODCTimeCI}	& \vci{\FASTxalanREDCTime}{\FASTxalanREDCTimeCI} \\
\CAPO	& \vci{\SLOWxalanCAPOExcTime}{\SLOWxalanCAPOExcTimeCI}	& \vci{\SLOWxalanCAPOnoGExcTime}{\SLOWxalanCAPOnoGExcTimeCI}	& \vci{\FASTxalanFTOCAPOTime}{\FASTxalanFTOCAPOTimeCI}	& \vci{\FASTxalanRECAPOTime}{\FASTxalanRECAPOTimeCI} \\
		& \mc{4}{c}{\bench{xalan}} \\
\end{tabular}
\end{empty}
\caption{Run time, relative to uninstrumented execution, of various analyses for each evaluated program
with 95\% confidence intervals.}
\label{tab:performance:allDaCapo:CI}
\end{table}
}

\iftoggle{twoColumnText}{
\begin{table*}[t]
\begin{empty}
\newcommand{\rzero}{0\xspace}
\newcommand{\roh}[1]{\ifthenelse{\equal{#1}{\rna}}{\rna}{#1$\;\!\times$}} 
\newcommand{\rna}{N/A}
\newcommand{\memna}{N/A} 
\newcommand{\st}[1]{(#1~s)} 
\newcommand{\dt}[1]{#1~s} 
\newcommand{\mem}[1]{\ifthenelse{\equal{#1}{\memna}}{\rna}{#1$\;\!\times$}}
\newcommand{\et}[1]{#1M} 
\newcommand{\enfp}[1]{(#1M)} 
\newcommand{\base}[1]{#1~s} 
\newcommand{\vci}[2]{\ifthenelse{\equal{#1}{\rna}}{\rna}{\roh{#1}{\;}\ci{#2}}}
\newcommand{\ci}[1]{\ensuremath{\pm}{\;}\roh{#1}}
\input{result-macros/PIP_slowTool_noCoresSet}
\input{result-macros/PIP_fastTool_extraOpt2Quiet}
\smaller
\centering
\begin{tabular}{@{}l|Hl@{\;\;}l@{\;\;}l@{}}
        & w/ G & Unopt- & \FTO- & \REAbbrv- \\\hline 
\HB		& \rna							& \vci{\SLOWavroraHBMem}{\SLOWavroraHBMemCI}		& \vci{\FASTavroraFTOHBMem}{\FASTavroraFTOHBMemCI}		& \rna \\
\WCP	& \rna							& \vci{\SLOWavroraWCPMem}{\SLOWavroraWCPMemCI}		& \vci{\FASTavroraFTOWCPMem}{\FASTavroraFTOWCPMemCI}	& \vci{\FASTavroraREWCPMem}{\FASTavroraREWCPMemCI} \\
\DC		& \vci{\SLOWavroraDCExcMem}{\SLOWavroraDCExcMemCI}	& \vci{\SLOWavroraDCnoGExcMem}{\SLOWavroraDCnoGExcMemCI}	& \vci{\FASTavroraFTODCMem}{\FASTavroraFTODCMemCI}		& \vci{\FASTavroraREDCMem}{\FASTavroraREDCMemCI} \\
\CAPO	& \vci{\SLOWavroraCAPOExcMem}{\SLOWavroraCAPOExcMemCI}	& \vci{\SLOWavroraCAPOnoGExcMem}{\SLOWavroraCAPOnoGExcMemCI}& \vci{\FASTavroraFTOCAPOMem}{\FASTavroraFTOCAPOMemCI}	& \vci{\FASTavroraRECAPOMem}{\FASTavroraRECAPOMemCI} \\
\mc{1}{c}{} & \mc{4}{c}{\bench{avrora}} \\
\end{tabular}
\hfill
\begin{tabular}{@{}HHl@{\;\;}l@{\;\;}l@{}}
Relation& w/ G & Unopt- & \FTO- & \REAbbrv- \\\hline 
\HB		& \rna							& \vci{\SLOWbatikHBMem}{\SLOWbatikHBMemCI}		& \vci{\FASTbatikFTOHBMem}{\FASTbatikFTOHBMemCI}		& \rna \\
\WCP	& \rna							& \vci{\SLOWbatikWCPMem}{\SLOWbatikWCPMemCI}		& \vci{\FASTbatikFTOWCPMem}{\FASTbatikFTOWCPMemCI}	& \vci{\FASTbatikREWCPMem}{\FASTbatikREWCPMemCI} \\
\DC		& \vci{\SLOWbatikDCExcMem}{\SLOWbatikDCExcMemCI}	& \vci{\SLOWbatikDCnoGExcMem}{\SLOWbatikDCnoGExcMemCI}	& \vci{\FASTbatikFTODCMem}{\FASTbatikFTODCMemCI}		& \vci{\FASTbatikREDCMem}{\FASTbatikREDCMemCI} \\
\CAPO	& \vci{\SLOWbatikCAPOExcMem}{\SLOWbatikCAPOExcMemCI}	& \vci{\SLOWbatikCAPOnoGExcMem}{\SLOWbatikCAPOnoGExcMemCI}& \vci{\FASTbatikFTOCAPOMem}{\FASTbatikFTOCAPOMemCI}	& \vci{\FASTbatikRECAPOMem}{\FASTbatikRECAPOMemCI} \\
		& \mc{4}{c}{\bench{batik}} \\
\end{tabular}
\hfill
\begin{tabular}{@{}HHl@{\;\;}l@{\;\;}l@{}}
		& w/ G & Unopt- & \FTO- & \REAbbrv- \\\hline 
\HB		& \rna							& \vci{\SLOWhtwoHBMem}{\SLOWhtwoHBMemCI}		& \vci{\FASThtwoFTOHBMem}{\FASThtwoFTOHBMemCI}		& \rna \\
\WCP	& \rna							& \vci{\SLOWhtwoWCPMem}{\SLOWhtwoWCPMemCI}		& \vci{\FASThtwoFTOWCPMem}{\FASThtwoFTOWCPMemCI}	& \vci{\FASThtwoREWCPMem}{\FASThtwoREWCPMemCI} \\
\DC		& \vci{\SLOWhtwoDCExcMem}{\SLOWhtwoDCExcMemCI}	& \vci{\SLOWhtwoDCnoGExcMem}{\SLOWhtwoDCnoGExcMemCI}	& \vci{\FASThtwoFTODCMem}{\FASThtwoFTODCMemCI}		& \vci{\FASThtwoREDCMem}{\FASThtwoREDCMemCI} \\
\CAPO	& \vci{\SLOWhtwoCAPOExcMem}{\SLOWhtwoCAPOExcMemCI}	& \vci{\SLOWhtwoCAPOnoGExcMem}{\SLOWhtwoCAPOnoGExcMemCI}& \vci{\FASThtwoFTOCAPOMem}{\FASThtwoFTOCAPOMemCI}	& \vci{\FASThtwoRECAPOMem}{\FASThtwoRECAPOMemCI} \\
		& \mc{4}{c}{\bench{h2}} \\
\end{tabular}
\medskip\\
\begin{tabular}{@{}l|Hl@{\;\;}l@{\;\;}l@{}}
		& w/ G & Unopt- & \FTO- & \REAbbrv- \\\hline 
\HB		& \rna							& \vci{\SLOWjythonHBMem}{\SLOWjythonHBMemCI}		& \vci{\FASTjythonFTOHBMem}{\FASTjythonFTOHBMemCI}		& \rna \\
\WCP	& \rna							& \vci{\SLOWjythonWCPMem}{\SLOWjythonWCPMemCI}		& \vci{\FASTjythonFTOWCPMem}{\FASTjythonFTOWCPMemCI}	& \vci{\FASTjythonREWCPMem}{\FASTjythonREWCPMemCI} \\
\DC		& \vci{\SLOWjythonDCExcMem}{\SLOWjythonDCExcMemCI}	& \vci{\SLOWjythonDCnoGExcMem}{\SLOWjythonDCnoGExcMemCI}	& \vci{\FASTjythonFTODCMem}{\FASTjythonFTODCMemCI}		& \vci{\FASTjythonREDCMem}{\FASTjythonREDCMemCI} \\
\CAPO	& \vci{\SLOWjythonCAPOExcMem}{\SLOWjythonCAPOExcMemCI}	& \vci{\SLOWjythonCAPOnoGExcMem}{\SLOWjythonCAPOnoGExcMemCI}& \vci{\FASTjythonFTOCAPOMem}{\FASTjythonFTOCAPOMemCI}	& \vci{\FASTjythonRECAPOMem}{\FASTjythonRECAPOMemCI} \\
		& \mc{4}{c}{\bench{jython}} \\
\end{tabular}
\hfill
\begin{tabular}{@{}HHl@{\;\;}l@{\;\;}l@{}}
		& w/ G & Unopt- & \FTO- & \REAbbrv- \\\hline 
\HB		& \rna							& \vci{\SLOWluindexHBMem}{\SLOWluindexHBMemCI}		& \vci{\FASTluindexFTOHBMem}{\FASTluindexFTOHBMemCI}		& \rna \\
\WCP	& \rna							& \vci{\SLOWluindexWCPMem}{\SLOWluindexWCPMemCI}		& \vci{\FASTluindexFTOWCPMem}{\FASTluindexFTOWCPMemCI}	& \vci{\FASTluindexREWCPMem}{\FASTluindexREWCPMemCI} \\
\DC		& \vci{\SLOWluindexDCExcMem}{\SLOWluindexDCExcMemCI}	& \vci{\SLOWluindexDCnoGExcMem}{\SLOWluindexDCnoGExcMemCI}	& \vci{\FASTluindexFTODCMem}{\FASTluindexFTODCMemCI}		& \vci{\FASTluindexREDCMem}{\FASTluindexREDCMemCI} \\
\CAPO	& \vci{\SLOWluindexCAPOExcMem}{\SLOWluindexCAPOExcMemCI}	& \vci{\SLOWluindexCAPOnoGExcMem}{\SLOWluindexCAPOnoGExcMemCI}& \vci{\FASTluindexFTOCAPOMem}{\FASTluindexFTOCAPOMemCI}	& \vci{\FASTluindexRECAPOMem}{\FASTluindexRECAPOMemCI} \\
		& \mc{4}{c}{\bench{luindex}} \\
\end{tabular}
\hfill
\begin{tabular}{@{}HHl@{\;\;}l@{\;\;}l@{}}
Relation& w/ G & Unopt- & \FTO- & \REAbbrv- \\\hline 
\HB		& \rna							& \vci{\SLOWlusearchHBMem}{\SLOWlusearchHBMemCI}		& \vci{\FASTlusearchFTOHBMem}{\FASTlusearchFTOHBMemCI}		& \rna \\
\WCP	& \rna							& \vci{\SLOWlusearchWCPMem}{\SLOWlusearchWCPMemCI}		& \vci{\FASTlusearchFTOWCPMem}{\FASTlusearchFTOWCPMemCI}	& \vci{\FASTlusearchREWCPMem}{\FASTlusearchREWCPMemCI} \\
\DC		& \vci{\SLOWlusearchDCExcMem}{\SLOWlusearchDCExcMemCI}	& \vci{\SLOWlusearchDCnoGExcMem}{\SLOWlusearchDCnoGExcMemCI}	& \vci{\FASTlusearchFTODCMem}{\FASTlusearchFTODCMemCI}		& \vci{\FASTlusearchREDCMem}{\FASTlusearchREDCMemCI} \\
\CAPO	& \vci{\SLOWlusearchCAPOExcMem}{\SLOWlusearchCAPOExcMemCI}	& \vci{\SLOWlusearchCAPOnoGExcMem}{\SLOWlusearchCAPOnoGExcMemCI}& \vci{\FASTlusearchFTOCAPOMem}{\FASTlusearchFTOCAPOMemCI}	& \vci{\FASTlusearchRECAPOMem}{\FASTlusearchRECAPOMemCI} \\
		& \mc{4}{c}{\bench{lusearch}} \\
\end{tabular}
\medskip\\
\begin{tabular}{@{}l|Hl@{\;\;}l@{\;\;}l@{}}
		& w/ G & Unopt- & \FTO- & \REAbbrv- \\\hline 
\HB		& \rna							& \vci{\SLOWpmdHBMem}{\SLOWpmdHBMemCI}		& \vci{\FASTpmdFTOHBMem}{\FASTpmdFTOHBMemCI}		& \rna \\
\WCP	& \rna							& \vci{\SLOWpmdWCPMem}{\SLOWpmdWCPMemCI}		& \vci{\FASTpmdFTOWCPMem}{\FASTpmdFTOWCPMemCI}	& \vci{\FASTpmdREWCPMem}{\FASTpmdREWCPMemCI} \\
\DC		& \vci{\SLOWpmdDCExcMem}{\SLOWpmdDCExcMemCI}	& \vci{\SLOWpmdDCnoGExcMem}{\SLOWpmdDCnoGExcMemCI}	& \vci{\FASTpmdFTODCMem}{\FASTpmdFTODCMemCI}		& \vci{\FASTpmdREDCMem}{\FASTpmdREDCMemCI} \\
\CAPO	& \vci{\SLOWpmdCAPOExcMem}{\SLOWpmdCAPOExcMemCI}	& \vci{\SLOWpmdCAPOnoGExcMem}{\SLOWpmdCAPOnoGExcMemCI}& \vci{\FASTpmdFTOCAPOMem}{\FASTpmdFTOCAPOMemCI}	& \vci{\FASTpmdRECAPOMem}{\FASTpmdRECAPOMemCI} \\
\mc{1}{c}{} & \mc{4}{c}{\bench{pmd}} \\
\end{tabular}
\hfill
\begin{tabular}{@{}HHl@{\;\;}l@{\;\;}l@{}}
Relation& w/ G & Unopt- & \FTO- & \REAbbrv- \\\hline 
\HB		& \rna							& \vci{\SLOWsunflowHBMem}{\SLOWsunflowHBMemCI}		& \vci{\FASTsunflowFTOHBMem}{\FASTsunflowFTOHBMemCI}		& \rna \\
\WCP	& \rna							& \vci{\SLOWsunflowWCPMem}{\SLOWsunflowWCPMemCI}		& \vci{\FASTsunflowFTOWCPMem}{\FASTsunflowFTOWCPMemCI}	& \vci{\FASTsunflowREWCPMem}{\FASTsunflowREWCPMemCI} \\
\DC		& \vci{\SLOWsunflowDCExcMem}{\SLOWsunflowDCExcMemCI}	& \vci{\SLOWsunflowDCnoGExcMem}{\SLOWsunflowDCnoGExcMemCI}	& \vci{\FASTsunflowFTODCMem}{\FASTsunflowFTODCMemCI}		& \vci{\FASTsunflowREDCMem}{\FASTsunflowREDCMemCI} \\
\CAPO	& \vci{\SLOWsunflowCAPOExcMem}{\SLOWsunflowCAPOExcMemCI}	& \vci{\SLOWsunflowCAPOnoGExcMem}{\SLOWsunflowCAPOnoGExcMemCI}& \vci{\FASTsunflowFTOCAPOMem}{\FASTsunflowFTOCAPOMemCI}	& \vci{\FASTsunflowRECAPOMem}{\FASTsunflowRECAPOMemCI} \\
		& \mc{4}{c}{\bench{sunflow}} \\
\end{tabular}
\hfill
\begin{tabular}{@{}HHl@{\;\;}l@{\;\;}l@{}}
		& w/ G & Unopt- & \FTO- & \REAbbrv- \\\hline 
\HB		& \rna							& \vci{\SLOWtomcatHBMem}{\SLOWtomcatHBMemCI}		& \vci{\FASTtomcatFTOHBMem}{\FASTtomcatFTOHBMemCI}		& \rna \\
\WCP	& \rna							& \vci{\SLOWtomcatWCPMem}{\SLOWtomcatWCPMemCI}		& \vci{\FASTtomcatFTOWCPMem}{\FASTtomcatFTOWCPMemCI}	& \vci{\FASTtomcatREWCPMem}{\FASTtomcatREWCPMemCI} \\
\DC		& \vci{\SLOWtomcatDCExcMem}{\SLOWtomcatDCExcMemCI}	& \vci{\SLOWtomcatDCnoGExcMem}{\SLOWtomcatDCnoGExcMemCI}	& \vci{\FASTtomcatFTODCMem}{\FASTtomcatFTODCMemCI}		& \vci{\FASTtomcatREDCMem}{\FASTtomcatREDCMemCI} \\
\CAPO	& \vci{\SLOWtomcatCAPOExcMem}{\SLOWtomcatCAPOExcMemCI}	& \vci{\SLOWtomcatCAPOnoGExcMem}{\SLOWtomcatCAPOnoGExcMemCI}& \vci{\FASTtomcatFTOCAPOMem}{\FASTtomcatFTOCAPOMemCI}	& \vci{\FASTtomcatRECAPOMem}{\FASTtomcatRECAPOMemCI} \\
		& \mc{4}{c}{\bench{tomcat}} \\
\end{tabular}
\medskip\\
\begin{tabular}{@{}l|Hl@{\;\;}l@{\;\;}l@{}}
		& w/ G & Unopt- & \FTO- & \REAbbrv- \\\hline 
\HB		& \rna							& \vci{\SLOWxalanHBMem}{\SLOWxalanHBMemCI}		& \vci{\FASTxalanFTOHBMem}{\FASTxalanFTOHBMemCI}		& \rna \\
\WCP	& \rna							& \vci{\SLOWxalanWCPMem}{\SLOWxalanWCPMemCI}		& \vci{\FASTxalanFTOWCPMem}{\FASTxalanFTOWCPMemCI}	& \vci{\FASTxalanREWCPMem}{\FASTxalanREWCPMemCI} \\
\DC		& \vci{\SLOWxalanDCExcMem}{\SLOWxalanDCExcMemCI}	& \vci{\SLOWxalanDCnoGExcMem}{\SLOWxalanDCnoGExcMemCI}	& \vci{\FASTxalanFTODCMem}{\FASTxalanFTODCMemCI}		& \vci{\FASTxalanREDCMem}{\FASTxalanREDCMemCI} \\
\CAPO	& \vci{\SLOWxalanCAPOExcMem}{\SLOWxalanCAPOExcMemCI}	& \vci{\SLOWxalanCAPOnoGExcMem}{\SLOWxalanCAPOnoGExcMemCI}& \vci{\FASTxalanFTOCAPOMem}{\FASTxalanFTOCAPOMemCI}	& \vci{\FASTxalanRECAPOMem}{\FASTxalanRECAPOMemCI} \\
\mc{1}{c}{} & \mc{4}{c}{\bench{xalan}} \\
\end{tabular}
\end{empty}
\caption{Memory usage, relative to uninstrumented execution, of various analyses for each evaluated program
with 95\% confidence intervals.}
\label{tab:memory:allDaCapo:CI}
\end{table*}

}{

\begin{table}[H]
\begin{empty}
\newcommand{\rzero}{0\xspace}
\newcommand{\roh}[1]{\ifthenelse{\equal{#1}{\rna}}{\rna}{#1$\;\!\times$}} 
\newcommand{\rna}{N/A}
\newcommand{\memna}{N/A} 
\newcommand{\st}[1]{(#1~s)} 
\newcommand{\dt}[1]{#1~s} 
\newcommand{\mem}[1]{\ifthenelse{\equal{#1}{\memna}}{\rna}{#1$\;\!\times$}}
\newcommand{\et}[1]{#1M} 
\newcommand{\enfp}[1]{(#1M)} 
\newcommand{\base}[1]{#1~s} 
\newcommand{\vci}[2]{\ifthenelse{\equal{#1}{\rna}}{\rna}{\roh{#1}{\;}\ci{#2}}}
\newcommand{\ci}[1]{\ensuremath{\pm}{\;}\roh{#1}}
\input{result-macros/PIP_slowTool_noCoresSet}
\input{result-macros/PIP_fastTool_extraOpt2Quiet}
\small
\centering
\begin{tabular}{@{}l|Hlll@{}}
        & w/ G & Unopt- & \FTO- & \REAbbrv- \\\hline 
\HB		& \rna							& \vci{\SLOWavroraHBMem}{\SLOWavroraHBMemCI}		& \vci{\FASTavroraFTOHBMem}{\FASTavroraFTOHBMemCI}		& \rna \\
\WCP	& \rna							& \vci{\SLOWavroraWCPMem}{\SLOWavroraWCPMemCI}		& \vci{\FASTavroraFTOWCPMem}{\FASTavroraFTOWCPMemCI}	& \vci{\FASTavroraREWCPMem}{\FASTavroraREWCPMemCI} \\
\DC		& \vci{\SLOWavroraDCExcMem}{\SLOWavroraDCExcMemCI}	& \vci{\SLOWavroraDCnoGExcMem}{\SLOWavroraDCnoGExcMemCI}	& \vci{\FASTavroraFTODCMem}{\FASTavroraFTODCMemCI}		& \vci{\FASTavroraREDCMem}{\FASTavroraREDCMemCI} \\
\CAPO	& \vci{\SLOWavroraCAPOExcMem}{\SLOWavroraCAPOExcMemCI}	& \vci{\SLOWavroraCAPOnoGExcMem}{\SLOWavroraCAPOnoGExcMemCI}& \vci{\FASTavroraFTOCAPOMem}{\FASTavroraFTOCAPOMemCI}	& \vci{\FASTavroraRECAPOMem}{\FASTavroraRECAPOMemCI} \\
\mc{1}{c}{} & \mc{4}{c}{\bench{avrora}} \\
\end{tabular}
\hfill
\begin{tabular}{@{}HHlll@{}}
Relation& w/ G & Unopt- & \FTO- & \REAbbrv- \\\hline 
\HB		& \rna							& \vci{\SLOWbatikHBMem}{\SLOWbatikHBMemCI}		& \vci{\FASTbatikFTOHBMem}{\FASTbatikFTOHBMemCI}		& \rna \\
\WCP	& \rna							& \vci{\SLOWbatikWCPMem}{\SLOWbatikWCPMemCI}		& \vci{\FASTbatikFTOWCPMem}{\FASTbatikFTOWCPMemCI}	& \vci{\FASTbatikREWCPMem}{\FASTbatikREWCPMemCI} \\
\DC		& \vci{\SLOWbatikDCExcMem}{\SLOWbatikDCExcMemCI}	& \vci{\SLOWbatikDCnoGExcMem}{\SLOWbatikDCnoGExcMemCI}	& \vci{\FASTbatikFTODCMem}{\FASTbatikFTODCMemCI}		& \vci{\FASTbatikREDCMem}{\FASTbatikREDCMemCI} \\
\CAPO	& \vci{\SLOWbatikCAPOExcMem}{\SLOWbatikCAPOExcMemCI}	& \vci{\SLOWbatikCAPOnoGExcMem}{\SLOWbatikCAPOnoGExcMemCI}& \vci{\FASTbatikFTOCAPOMem}{\FASTbatikFTOCAPOMemCI}	& \vci{\FASTbatikRECAPOMem}{\FASTbatikRECAPOMemCI} \\
		& \mc{4}{c}{\bench{batik}} \\
\end{tabular}
\medskip\\
\begin{tabular}{@{}l|Hlll@{}}
		& w/ G & Unopt- & \FTO- & \REAbbrv- \\\hline 
\HB		& \rna							& \vci{\SLOWhtwoHBMem}{\SLOWhtwoHBMemCI}		& \vci{\FASThtwoFTOHBMem}{\FASThtwoFTOHBMemCI}		& \rna \\
\WCP	& \rna							& \vci{\SLOWhtwoWCPMem}{\SLOWhtwoWCPMemCI}		& \vci{\FASThtwoFTOWCPMem}{\FASThtwoFTOWCPMemCI}	& \vci{\FASThtwoREWCPMem}{\FASThtwoREWCPMemCI} \\
\DC		& \vci{\SLOWhtwoDCExcMem}{\SLOWhtwoDCExcMemCI}	& \vci{\SLOWhtwoDCnoGExcMem}{\SLOWhtwoDCnoGExcMemCI}	& \vci{\FASThtwoFTODCMem}{\FASThtwoFTODCMemCI}		& \vci{\FASThtwoREDCMem}{\FASThtwoREDCMemCI} \\
\CAPO	& \vci{\SLOWhtwoCAPOExcMem}{\SLOWhtwoCAPOExcMemCI}	& \vci{\SLOWhtwoCAPOnoGExcMem}{\SLOWhtwoCAPOnoGExcMemCI}& \vci{\FASThtwoFTOCAPOMem}{\FASThtwoFTOCAPOMemCI}	& \vci{\FASThtwoRECAPOMem}{\FASThtwoRECAPOMemCI} \\
\mc{1}{c}{} & \mc{4}{c}{\bench{h2}} \\
\end{tabular}
\hfill
\begin{tabular}{@{}HHlll@{}}
		& w/ G & Unopt- & \FTO- & \REAbbrv- \\\hline 
\HB		& \rna							& \vci{\SLOWjythonHBMem}{\SLOWjythonHBMemCI}		& \vci{\FASTjythonFTOHBMem}{\FASTjythonFTOHBMemCI}		& \rna \\
\WCP	& \rna							& \vci{\SLOWjythonWCPMem}{\SLOWjythonWCPMemCI}		& \vci{\FASTjythonFTOWCPMem}{\FASTjythonFTOWCPMemCI}	& \vci{\FASTjythonREWCPMem}{\FASTjythonREWCPMemCI} \\
\DC		& \vci{\SLOWjythonDCExcMem}{\SLOWjythonDCExcMemCI}	& \vci{\SLOWjythonDCnoGExcMem}{\SLOWjythonDCnoGExcMemCI}	& \vci{\FASTjythonFTODCMem}{\FASTjythonFTODCMemCI}		& \vci{\FASTjythonREDCMem}{\FASTjythonREDCMemCI} \\
\CAPO	& \vci{\SLOWjythonCAPOExcMem}{\SLOWjythonCAPOExcMemCI}	& \vci{\SLOWjythonCAPOnoGExcMem}{\SLOWjythonCAPOnoGExcMemCI}& \vci{\FASTjythonFTOCAPOMem}{\FASTjythonFTOCAPOMemCI}	& \vci{\FASTjythonRECAPOMem}{\FASTjythonRECAPOMemCI} \\
		& \mc{4}{c}{\bench{jython}} \\
\end{tabular}
\medskip\\
\begin{tabular}{@{}l|Hlll@{}}
		& w/ G & Unopt- & \FTO- & \REAbbrv- \\\hline 
\HB		& \rna							& \vci{\SLOWluindexHBMem}{\SLOWluindexHBMemCI}		& \vci{\FASTluindexFTOHBMem}{\FASTluindexFTOHBMemCI}		& \rna \\
\WCP	& \rna							& \vci{\SLOWluindexWCPMem}{\SLOWluindexWCPMemCI}		& \vci{\FASTluindexFTOWCPMem}{\FASTluindexFTOWCPMemCI}	& \vci{\FASTluindexREWCPMem}{\FASTluindexREWCPMemCI} \\
\DC		& \vci{\SLOWluindexDCExcMem}{\SLOWluindexDCExcMemCI}	& \vci{\SLOWluindexDCnoGExcMem}{\SLOWluindexDCnoGExcMemCI}	& \vci{\FASTluindexFTODCMem}{\FASTluindexFTODCMemCI}		& \vci{\FASTluindexREDCMem}{\FASTluindexREDCMemCI} \\
\CAPO	& \vci{\SLOWluindexCAPOExcMem}{\SLOWluindexCAPOExcMemCI}	& \vci{\SLOWluindexCAPOnoGExcMem}{\SLOWluindexCAPOnoGExcMemCI}& \vci{\FASTluindexFTOCAPOMem}{\FASTluindexFTOCAPOMemCI}	& \vci{\FASTluindexRECAPOMem}{\FASTluindexRECAPOMemCI} \\
\mc{1}{c}{} & \mc{4}{c}{\bench{luindex}} \\
\end{tabular}
\hfill
\begin{tabular}{@{}HHlll@{}}
Relation& w/ G & Unopt- & \FTO- & \REAbbrv- \\\hline 
\HB		& \rna							& \vci{\SLOWlusearchHBMem}{\SLOWlusearchHBMemCI}		& \vci{\FASTlusearchFTOHBMem}{\FASTlusearchFTOHBMemCI}		& \rna \\
\WCP	& \rna							& \vci{\SLOWlusearchWCPMem}{\SLOWlusearchWCPMemCI}		& \vci{\FASTlusearchFTOWCPMem}{\FASTlusearchFTOWCPMemCI}	& \vci{\FASTlusearchREWCPMem}{\FASTlusearchREWCPMemCI} \\
\DC		& \vci{\SLOWlusearchDCExcMem}{\SLOWlusearchDCExcMemCI}	& \vci{\SLOWlusearchDCnoGExcMem}{\SLOWlusearchDCnoGExcMemCI}	& \vci{\FASTlusearchFTODCMem}{\FASTlusearchFTODCMemCI}		& \vci{\FASTlusearchREDCMem}{\FASTlusearchREDCMemCI} \\
\CAPO	& \vci{\SLOWlusearchCAPOExcMem}{\SLOWlusearchCAPOExcMemCI}	& \vci{\SLOWlusearchCAPOnoGExcMem}{\SLOWlusearchCAPOnoGExcMemCI}& \vci{\FASTlusearchFTOCAPOMem}{\FASTlusearchFTOCAPOMemCI}	& \vci{\FASTlusearchRECAPOMem}{\FASTlusearchRECAPOMemCI} \\
		& \mc{4}{c}{\bench{lusearch}} \\
\end{tabular}
\medskip\\
\begin{tabular}{@{}l|Hlll@{}}
		& w/ G & Unopt- & \FTO- & \REAbbrv- \\\hline 
\HB		& \rna							& \vci{\SLOWpmdHBMem}{\SLOWpmdHBMemCI}		& \vci{\FASTpmdFTOHBMem}{\FASTpmdFTOHBMemCI}		& \rna \\
\WCP	& \rna							& \vci{\SLOWpmdWCPMem}{\SLOWpmdWCPMemCI}		& \vci{\FASTpmdFTOWCPMem}{\FASTpmdFTOWCPMemCI}	& \vci{\FASTpmdREWCPMem}{\FASTpmdREWCPMemCI} \\
\DC		& \vci{\SLOWpmdDCExcMem}{\SLOWpmdDCExcMemCI}	& \vci{\SLOWpmdDCnoGExcMem}{\SLOWpmdDCnoGExcMemCI}	& \vci{\FASTpmdFTODCMem}{\FASTpmdFTODCMemCI}		& \vci{\FASTpmdREDCMem}{\FASTpmdREDCMemCI} \\
\CAPO	& \vci{\SLOWpmdCAPOExcMem}{\SLOWpmdCAPOExcMemCI}	& \vci{\SLOWpmdCAPOnoGExcMem}{\SLOWpmdCAPOnoGExcMemCI}& \vci{\FASTpmdFTOCAPOMem}{\FASTpmdFTOCAPOMemCI}	& \vci{\FASTpmdRECAPOMem}{\FASTpmdRECAPOMemCI} \\
\mc{1}{c}{} & \mc{4}{c}{\bench{pmd}} \\
\end{tabular}
\hfill
\begin{tabular}{@{}HHlll@{}}
Relation& w/ G & Unopt- & \FTO- & \REAbbrv- \\\hline 
\HB		& \rna							& \vci{\SLOWsunflowHBMem}{\SLOWsunflowHBMemCI}		& \vci{\FASTsunflowFTOHBMem}{\FASTsunflowFTOHBMemCI}		& \rna \\
\WCP	& \rna							& \vci{\SLOWsunflowWCPMem}{\SLOWsunflowWCPMemCI}		& \vci{\FASTsunflowFTOWCPMem}{\FASTsunflowFTOWCPMemCI}	& \vci{\FASTsunflowREWCPMem}{\FASTsunflowREWCPMemCI} \\
\DC		& \vci{\SLOWsunflowDCExcMem}{\SLOWsunflowDCExcMemCI}	& \vci{\SLOWsunflowDCnoGExcMem}{\SLOWsunflowDCnoGExcMemCI}	& \vci{\FASTsunflowFTODCMem}{\FASTsunflowFTODCMemCI}		& \vci{\FASTsunflowREDCMem}{\FASTsunflowREDCMemCI} \\
\CAPO	& \vci{\SLOWsunflowCAPOExcMem}{\SLOWsunflowCAPOExcMemCI}	& \vci{\SLOWsunflowCAPOnoGExcMem}{\SLOWsunflowCAPOnoGExcMemCI}& \vci{\FASTsunflowFTOCAPOMem}{\FASTsunflowFTOCAPOMemCI}	& \vci{\FASTsunflowRECAPOMem}{\FASTsunflowRECAPOMemCI} \\
		& \mc{4}{c}{\bench{sunflow}} \\
\end{tabular}
\medskip\\
\begin{tabular}{@{}l|Hlll@{}}
		& w/ G & Unopt- & \FTO- & \REAbbrv- \\\hline 
\HB		& \rna							& \vci{\SLOWtomcatHBMem}{\SLOWtomcatHBMemCI}		& \vci{\FASTtomcatFTOHBMem}{\FASTtomcatFTOHBMemCI}		& \rna \\
\WCP	& \rna							& \vci{\SLOWtomcatWCPMem}{\SLOWtomcatWCPMemCI}		& \vci{\FASTtomcatFTOWCPMem}{\FASTtomcatFTOWCPMemCI}	& \vci{\FASTtomcatREWCPMem}{\FASTtomcatREWCPMemCI} \\
\DC		& \vci{\SLOWtomcatDCExcMem}{\SLOWtomcatDCExcMemCI}	& \vci{\SLOWtomcatDCnoGExcMem}{\SLOWtomcatDCnoGExcMemCI}	& \vci{\FASTtomcatFTODCMem}{\FASTtomcatFTODCMemCI}		& \vci{\FASTtomcatREDCMem}{\FASTtomcatREDCMemCI} \\
\CAPO	& \vci{\SLOWtomcatCAPOExcMem}{\SLOWtomcatCAPOExcMemCI}	& \vci{\SLOWtomcatCAPOnoGExcMem}{\SLOWtomcatCAPOnoGExcMemCI}& \vci{\FASTtomcatFTOCAPOMem}{\FASTtomcatFTOCAPOMemCI}	& \vci{\FASTtomcatRECAPOMem}{\FASTtomcatRECAPOMemCI} \\
\mc{1}{c}{} & \mc{4}{c}{\bench{tomcat}} \\
\end{tabular}
\hfill
\begin{tabular}{@{}HHlll@{}}
		& w/ G & Unopt- & \FTO- & \REAbbrv- \\\hline 
\HB		& \rna							& \vci{\SLOWxalanHBMem}{\SLOWxalanHBMemCI}		& \vci{\FASTxalanFTOHBMem}{\FASTxalanFTOHBMemCI}		& \rna \\
\WCP	& \rna							& \vci{\SLOWxalanWCPMem}{\SLOWxalanWCPMemCI}		& \vci{\FASTxalanFTOWCPMem}{\FASTxalanFTOWCPMemCI}	& \vci{\FASTxalanREWCPMem}{\FASTxalanREWCPMemCI} \\
\DC		& \vci{\SLOWxalanDCExcMem}{\SLOWxalanDCExcMemCI}	& \vci{\SLOWxalanDCnoGExcMem}{\SLOWxalanDCnoGExcMemCI}	& \vci{\FASTxalanFTODCMem}{\FASTxalanFTODCMemCI}		& \vci{\FASTxalanREDCMem}{\FASTxalanREDCMemCI} \\
\CAPO	& \vci{\SLOWxalanCAPOExcMem}{\SLOWxalanCAPOExcMemCI}	& \vci{\SLOWxalanCAPOnoGExcMem}{\SLOWxalanCAPOnoGExcMemCI}& \vci{\FASTxalanFTOCAPOMem}{\FASTxalanFTOCAPOMemCI}	& \vci{\FASTxalanRECAPOMem}{\FASTxalanRECAPOMemCI} \\
		& \mc{4}{c}{\bench{xalan}} \\
\end{tabular}
\end{empty}
\caption{Memory usage, relative to uninstrumented execution, of various analyses for each evaluated program
with 95\% confidence intervals.}
\label{tab:memory:allDaCapo:CI}
\end{table}
}

\iftoggle{twoColumnText}{
\begin{table*}[t]
\newcommand{\rzero}{0\xspace}
\newcommand{\roh}[1]{\ifthenelse{\equal{#1}{\rna}}{\rna}{#1$\;\!\times$}} 
\newcommand{\rna}{N/A}
\newcommand{\memna}{N/A} 
\newcommand{\st}[1]{(#1~s)} 
\newcommand{\dt}[1]{#1~s} 
\newcommand{\mem}[1]{\ifthenelse{\equal{#1}{\memna}}{\memna}{#1$\;Mb$}} 
\newcommand{\et}[1]{#1M} 
\newcommand{\enfp}[1]{(#1M)} 
\newcommand{\base}[1]{#1~s} 
\newcommand{\sdr}[4]{\vci{#1}{#2}{\;}(\vci{#3}{#4})}
\newcommand{\vci}[2]{\ifthenelse{\equal{#1}{\rna}}{\rna}{#1{\;}\ci{#2}}}
\newcommand{\ci}[1]{\ensuremath{\pm}{\;}#1}
\input{result-macros/PIP_slowTool_noCoresSet}
\input{result-macros/PIP_fastTool_extraOpt2Quiet}
\smaller
\centering
\begin{tabular}{@{}l|Hl@{\;\;}l@{\;\;}l@{}}
        & w/ G & Unopt- & \FTO- & \REAbbrv- \\\hline
\HB		& \rna													& \sdr{\SLOWavroraHB}{\SLOWavroraHBCI}{\SLOWavroraHBDynamic}{\SLOWavroraHBDynamicCI}					& \sdr{\FASTavroraFTOHB}{\FASTavroraFTOHBCI}{\FASTavroraFTOHBDynamic}{\FASTavroraFTOHBDynamicCI}		& \rna \\
\WCP	& \rna													& \sdr{\SLOWavroraWCP}{\SLOWavroraWCPCI}{\SLOWavroraWCPDynamic}{\SLOWavroraWCPDynamicCI}				& \sdr{\FASTavroraFTOWCP}{\FASTavroraFTOWCPCI}{\FASTavroraFTOWCPDynamic}{\FASTavroraFTOWCPDynamicCI}		& \sdr{\FASTavroraREWCP}{\FASTavroraREWCPCI}{\FASTavroraREWCPDynamic}{\FASTavroraREWCPDynamicCI} \\
\DC		& \sdr{\SLOWavroraDCExc}{\SLOWavroraDCExcCI}{\SLOWavroraDCExcDynamic}{\SLOWavroraDCExcDynamicCI}		& \sdr{\SLOWavroraDCnoGExc}{\SLOWavroraDCnoGExcCI}{\SLOWavroraDCnoGExcDynamic}{\SLOWavroraDCnoGExcDynamicCI}	& \sdr{\FASTavroraFTODC}{\FASTavroraFTODCCI}{\FASTavroraFTODCDynamic}{\FASTavroraFTODCDynamicCI}		& \sdr{\FASTavroraREDC}{\FASTavroraREDCCI}{\FASTavroraREDCDynamic}{\FASTavroraREDCDynamicCI} \\
\CAPO	& \sdr{\SLOWavroraCAPOExc}{\SLOWavroraCAPOExcCI}{\SLOWavroraCAPOExcDynamic}{\SLOWavroraCAPOExcDynamicCI}	& \sdr{\SLOWavroraCAPOnoGExc}{\SLOWavroraCAPOnoGExcCI}{\SLOWavroraCAPOnoGExcDynamic}{\SLOWavroraCAPOnoGExcDynamicCI}	& \sdr{\FASTavroraFTOCAPO}{\FASTavroraFTOCAPOCI}{\FASTavroraFTOCAPODynamic}{\FASTavroraFTOCAPODynamicCI}	& \sdr{\FASTavroraRECAPO}{\FASTavroraRECAPOCI}{\FASTavroraRECAPODynamic}{\FASTavroraRECAPODynamicCI} \\
\mc{1}{c}{} & \mc{4}{c}{\bench{avrora}} \\
\end{tabular}
\medskip\\
\begin{tabular}{@{}l|Hl@{\;\;}l@{\;\;}l@{}}
		& w/ G & Unopt- & \FTO- & \REAbbrv- \\\hline
\HB		& \rna													& \sdr{\SLOWhtwoHB}{\SLOWhtwoHBCI}{\SLOWhtwoHBDynamic}{\SLOWhtwoHBDynamicCI}					& \sdr{\FASThtwoFTOHB}{\FASThtwoFTOHBCI}{\FASThtwoFTOHBDynamic}{\FASThtwoFTOHBDynamicCI}		& \rna \\
\WCP	& \rna													& \sdr{\SLOWhtwoWCP}{\SLOWhtwoWCPCI}{\SLOWhtwoWCPDynamic}{\SLOWhtwoWCPDynamicCI}				& \sdr{\FASThtwoFTOWCP}{\FASThtwoFTOWCPCI}{\FASThtwoFTOWCPDynamic}{\FASThtwoFTOWCPDynamicCI}		& \sdr{\FASThtwoREWCP}{\FASThtwoREWCPCI}{\FASThtwoREWCPDynamic}{\FASThtwoREWCPDynamicCI} \\
\DC		& \sdr{\SLOWhtwoDCExc}{\SLOWhtwoDCExcCI}{\SLOWhtwoDCExcDynamic}{\SLOWhtwoDCExcDynamicCI}		& \sdr{\SLOWhtwoDCnoGExc}{\SLOWhtwoDCnoGExcCI}{\SLOWhtwoDCnoGExcDynamic}{\SLOWhtwoDCnoGExcDynamicCI}	& \sdr{\FASThtwoFTODC}{\FASThtwoFTODCCI}{\FASThtwoFTODCDynamic}{\FASThtwoFTODCDynamicCI}		& \sdr{\FASThtwoREDC}{\FASThtwoREDCCI}{\FASThtwoREDCDynamic}{\FASThtwoREDCDynamicCI} \\
\CAPO	& \sdr{\SLOWhtwoCAPOExc}{\SLOWhtwoCAPOExcCI}{\SLOWhtwoCAPOExcDynamic}{\SLOWhtwoCAPOExcDynamicCI}	& \sdr{\SLOWhtwoCAPOnoGExc}{\SLOWhtwoCAPOnoGExcCI}{\SLOWhtwoCAPOnoGExcDynamic}{\SLOWhtwoCAPOnoGExcDynamicCI}	& \sdr{\FASThtwoFTOCAPO}{\FASThtwoFTOCAPOCI}{\FASThtwoFTOCAPODynamic}{\FASThtwoFTOCAPODynamicCI}	& \sdr{\FASThtwoRECAPO}{\FASThtwoRECAPOCI}{\FASThtwoRECAPODynamic}{\FASThtwoRECAPODynamicCI} \\
\mc{1}{c}{} & \mc{4}{c}{\bench{h2}} \\
\end{tabular}
\medskip\\
\begin{tabular}{@{}l|Hl@{\;\;}l@{\;\;}l@{}}
        & w/ G & Unopt- & \FTO- & \REAbbrv- \\\hline
\HB		& \rna													& \sdr{\SLOWjythonHB}{\SLOWjythonHBCI}{\SLOWjythonHBDynamic}{\SLOWjythonHBDynamicCI}					& \sdr{\FASTjythonFTOHB}{\FASTjythonFTOHBCI}{\FASTjythonFTOHBDynamic}{\FASTjythonFTOHBDynamicCI}		& \rna \\
\WCP	& \rna													& \sdr{\SLOWjythonWCP}{\SLOWjythonWCPCI}{\SLOWjythonWCPDynamic}{\SLOWjythonWCPDynamicCI}				& \sdr{\FASTjythonFTOWCP}{\FASTjythonFTOWCPCI}{\FASTjythonFTOWCPDynamic}{\FASTjythonFTOWCPDynamicCI}		& \sdr{\FASTjythonREWCP}{\FASTjythonREWCPCI}{\FASTjythonREWCPDynamic}{\FASTjythonREWCPDynamicCI} \\
\DC		& \sdr{\SLOWjythonDCExc}{\SLOWjythonDCExcCI}{\SLOWjythonDCExcDynamic}{\SLOWjythonDCExcDynamicCI}		& \sdr{\SLOWjythonDCnoGExc}{\SLOWjythonDCnoGExcCI}{\SLOWjythonDCnoGExcDynamic}{\SLOWjythonDCnoGExcDynamicCI}	& \sdr{\FASTjythonFTODC}{\FASTjythonFTODCCI}{\FASTjythonFTODCDynamic}{\FASTjythonFTODCDynamicCI}		& \sdr{\FASTjythonREDC}{\FASTjythonREDCCI}{\FASTjythonREDCDynamic}{\FASTjythonREDCDynamicCI} \\
\CAPO	& \sdr{\SLOWjythonCAPOExc}{\SLOWjythonCAPOExcCI}{\SLOWjythonCAPOExcDynamic}{\SLOWjythonCAPOExcDynamicCI}	& \sdr{\SLOWjythonCAPOnoGExc}{\SLOWjythonCAPOnoGExcCI}{\SLOWjythonCAPOnoGExcDynamic}{\SLOWjythonCAPOnoGExcDynamicCI}	& \sdr{\FASTjythonFTOCAPO}{\FASTjythonFTOCAPOCI}{\FASTjythonFTOCAPODynamic}{\FASTjythonFTOCAPODynamicCI}	& \sdr{\FASTjythonRECAPO}{\FASTjythonRECAPOCI}{\FASTjythonRECAPODynamic}{\FASTjythonRECAPODynamicCI} \\
\mc{1}{c}{} & \mc{4}{c}{\bench{jython}} \\
\end{tabular}
\hfill
\begin{tabular}{@{}HHl@{\;\;}l@{\;\;}l@{}}
        & w/ G & Unopt- & \FTO- & \REAbbrv- \\\hline
\HB		& \rna													& \sdr{\SLOWluindexHB}{\SLOWluindexHBCI}{\SLOWluindexHBDynamic}{\SLOWluindexHBDynamicCI}					& \sdr{\FASTluindexFTOHB}{\FASTluindexFTOHBCI}{\FASTluindexFTOHBDynamic}{\FASTluindexFTOHBDynamicCI}		& \rna \\
\WCP	& \rna													& \sdr{\SLOWluindexWCP}{\SLOWluindexWCPCI}{\SLOWluindexWCPDynamic}{\SLOWluindexWCPDynamicCI}				& \sdr{\FASTluindexFTOWCP}{\FASTluindexFTOWCPCI}{\FASTluindexFTOWCPDynamic}{\FASTluindexFTOWCPDynamicCI}		& \sdr{\FASTluindexREWCP}{\FASTluindexREWCPCI}{\FASTluindexREWCPDynamic}{\FASTluindexREWCPDynamicCI} \\
\DC		& \sdr{\SLOWluindexDCExc}{\SLOWluindexDCExcCI}{\SLOWluindexDCExcDynamic}{\SLOWluindexDCExcDynamicCI}		& \sdr{\SLOWluindexDCnoGExc}{\SLOWluindexDCnoGExcCI}{\SLOWluindexDCnoGExcDynamic}{\SLOWluindexDCnoGExcDynamicCI}	& \sdr{\FASTluindexFTODC}{\FASTluindexFTODCCI}{\FASTluindexFTODCDynamic}{\FASTluindexFTODCDynamicCI}		& \sdr{\FASTluindexREDC}{\FASTluindexREDCCI}{\FASTluindexREDCDynamic}{\FASTluindexREDCDynamicCI} \\
\CAPO	& \sdr{\SLOWluindexCAPOExc}{\SLOWluindexCAPOExcCI}{\SLOWluindexCAPOExcDynamic}{\SLOWluindexCAPOExcDynamicCI}	& \sdr{\SLOWluindexCAPOnoGExc}{\SLOWluindexCAPOnoGExcCI}{\SLOWluindexCAPOnoGExcDynamic}{\SLOWluindexCAPOnoGExcDynamicCI}	& \sdr{\FASTluindexFTOCAPO}{\FASTluindexFTOCAPOCI}{\FASTluindexFTOCAPODynamic}{\FASTluindexFTOCAPODynamicCI}	& \sdr{\FASTluindexRECAPO}{\FASTluindexRECAPOCI}{\FASTluindexRECAPODynamic}{\FASTluindexRECAPODynamicCI} \\
		& \mc{4}{c}{\bench{luindex}} \\
\end{tabular}
\medskip\\
\begin{tabular}{@{}l|Hl@{\;\;}l@{\;\;}l@{}}
        & w/ G & Unopt- & \FTO- & \REAbbrv- \\\hline
\HB		& \rna													& \sdr{\SLOWpmdHB}{\SLOWpmdHBCI}{\SLOWpmdHBDynamic}{\SLOWpmdHBDynamicCI}					& \sdr{\FASTpmdFTOHB}{\FASTpmdFTOHBCI}{\FASTpmdFTOHBDynamic}{\FASTpmdFTOHBDynamicCI}		& \rna \\
\WCP	& \rna													& \sdr{\SLOWpmdWCP}{\SLOWpmdWCPCI}{\SLOWpmdWCPDynamic}{\SLOWpmdWCPDynamicCI}				& \sdr{\FASTpmdFTOWCP}{\FASTpmdFTOWCPCI}{\FASTpmdFTOWCPDynamic}{\FASTpmdFTOWCPDynamicCI}		& \sdr{\FASTpmdREWCP}{\FASTpmdREWCPCI}{\FASTpmdREWCPDynamic}{\FASTpmdREWCPDynamicCI} \\
\DC		& \sdr{\SLOWpmdDCExc}{\SLOWpmdDCExcCI}{\SLOWpmdDCExcDynamic}{\SLOWpmdDCExcDynamicCI}		& \sdr{\SLOWpmdDCnoGExc}{\SLOWpmdDCnoGExcCI}{\SLOWpmdDCnoGExcDynamic}{\SLOWpmdDCnoGExcDynamicCI}	& \sdr{\FASTpmdFTODC}{\FASTpmdFTODCCI}{\FASTpmdFTODCDynamic}{\FASTpmdFTODCDynamicCI}		& \sdr{\FASTpmdREDC}{\FASTpmdREDCCI}{\FASTpmdREDCDynamic}{\FASTpmdREDCDynamicCI} \\
\CAPO	& \sdr{\SLOWpmdCAPOExc}{\SLOWpmdCAPOExcCI}{\SLOWpmdCAPOExcDynamic}{\SLOWpmdCAPOExcDynamicCI}	& \sdr{\SLOWpmdCAPOnoGExc}{\SLOWpmdCAPOnoGExcCI}{\SLOWpmdCAPOnoGExcDynamic}{\SLOWpmdCAPOnoGExcDynamicCI}	& \sdr{\FASTpmdFTOCAPO}{\FASTpmdFTOCAPOCI}{\FASTpmdFTOCAPODynamic}{\FASTpmdFTOCAPODynamicCI}	& \sdr{\FASTpmdRECAPO}{\FASTpmdRECAPOCI}{\FASTpmdRECAPODynamic}{\FASTpmdRECAPODynamicCI} \\
\mc{1}{c}{} & \mc{4}{c}{\bench{pmd}} \\
\end{tabular}
\hfill
\begin{tabular}{@{}HHl@{\;\;}l@{\;\;}l@{}}
        & w/ G & Unopt- & \FTO- & \REAbbrv- \\\hline
\HB		& \rna													& \sdr{\SLOWsunflowHB}{\SLOWsunflowHBCI}{\SLOWsunflowHBDynamic}{\SLOWsunflowHBDynamicCI}					& \sdr{\FASTsunflowFTOHB}{\FASTsunflowFTOHBCI}{\FASTsunflowFTOHBDynamic}{\FASTsunflowFTOHBDynamicCI}		& \rna \\
\WCP	& \rna													& \sdr{\SLOWsunflowWCP}{\SLOWsunflowWCPCI}{\SLOWsunflowWCPDynamic}{\SLOWsunflowWCPDynamicCI}				& \sdr{\FASTsunflowFTOWCP}{\FASTsunflowFTOWCPCI}{\FASTsunflowFTOWCPDynamic}{\FASTsunflowFTOWCPDynamicCI}		& \sdr{\FASTsunflowREWCP}{\FASTsunflowREWCPCI}{\FASTsunflowREWCPDynamic}{\FASTsunflowREWCPDynamicCI} \\
\DC		& \sdr{\SLOWsunflowDCExc}{\SLOWsunflowDCExcCI}{\SLOWsunflowDCExcDynamic}{\SLOWsunflowDCExcDynamicCI}		& \sdr{\SLOWsunflowDCnoGExc}{\SLOWsunflowDCnoGExcCI}{\SLOWsunflowDCnoGExcDynamic}{\SLOWsunflowDCnoGExcDynamicCI}	& \sdr{\FASTsunflowFTODC}{\FASTsunflowFTODCCI}{\FASTsunflowFTODCDynamic}{\FASTsunflowFTODCDynamicCI}		& \sdr{\FASTsunflowREDC}{\FASTsunflowREDCCI}{\FASTsunflowREDCDynamic}{\FASTsunflowREDCDynamicCI} \\
\CAPO	& \sdr{\SLOWsunflowCAPOExc}{\SLOWsunflowCAPOExcCI}{\SLOWsunflowCAPOExcDynamic}{\SLOWsunflowCAPOExcDynamicCI}	& \sdr{\SLOWsunflowCAPOnoGExc}{\SLOWsunflowCAPOnoGExcCI}{\SLOWsunflowCAPOnoGExcDynamic}{\SLOWsunflowCAPOnoGExcDynamicCI}	& \sdr{\FASTsunflowFTOCAPO}{\FASTsunflowFTOCAPOCI}{\FASTsunflowFTOCAPODynamic}{\FASTsunflowFTOCAPODynamicCI}	& \sdr{\FASTsunflowRECAPO}{\FASTsunflowRECAPOCI}{\FASTsunflowRECAPODynamic}{\FASTsunflowRECAPODynamicCI} \\
		& \mc{4}{c}{\bench{sunflow}} \\
\end{tabular}
\medskip\\
\begin{tabular}{@{}l|Hl@{\;\;}l@{\;\;}l@{}}
        & w/ G & Unopt- & \FTO- & \REAbbrv- \\\hline
\HB		& \rna													& \sdr{\SLOWtomcatHB}{\SLOWtomcatHBCI}{\SLOWtomcatHBDynamic}{\SLOWtomcatHBDynamicCI}					& \sdr{\FASTtomcatFTOHB}{\FASTtomcatFTOHBCI}{\FASTtomcatFTOHBDynamic}{\FASTtomcatFTOHBDynamicCI}		& \rna \\
\WCP	& \rna													& \sdr{\SLOWtomcatWCP}{\SLOWtomcatWCPCI}{\SLOWtomcatWCPDynamic}{\SLOWtomcatWCPDynamicCI}				& \sdr{\FASTtomcatFTOWCP}{\FASTtomcatFTOWCPCI}{\FASTtomcatFTOWCPDynamic}{\FASTtomcatFTOWCPDynamicCI}		& \sdr{\FASTtomcatREWCP}{\FASTtomcatREWCPCI}{\FASTtomcatREWCPDynamic}{\FASTtomcatREWCPDynamicCI} \\
\DC		& \sdr{\SLOWtomcatDCExc}{\SLOWtomcatDCExcCI}{\SLOWtomcatDCExcDynamic}{\SLOWtomcatDCExcDynamicCI}		& \sdr{\SLOWtomcatDCnoGExc}{\SLOWtomcatDCnoGExcCI}{\SLOWtomcatDCnoGExcDynamic}{\SLOWtomcatDCnoGExcDynamicCI}	& \sdr{\FASTtomcatFTODC}{\FASTtomcatFTODCCI}{\FASTtomcatFTODCDynamic}{\FASTtomcatFTODCDynamicCI}		& \sdr{\FASTtomcatREDC}{\FASTtomcatREDCCI}{\FASTtomcatREDCDynamic}{\FASTtomcatREDCDynamicCI} \\
\CAPO	& \sdr{\SLOWtomcatCAPOExc}{\SLOWtomcatCAPOExcCI}{\SLOWtomcatCAPOExcDynamic}{\SLOWtomcatCAPOExcDynamicCI}	& \sdr{\SLOWtomcatCAPOnoGExc}{\SLOWtomcatCAPOnoGExcCI}{\SLOWtomcatCAPOnoGExcDynamic}{\SLOWtomcatCAPOnoGExcDynamicCI}	& \sdr{\FASTtomcatFTOCAPO}{\FASTtomcatFTOCAPOCI}{\FASTtomcatFTOCAPODynamic}{\FASTtomcatFTOCAPODynamicCI}	& \sdr{\FASTtomcatRECAPO}{\FASTtomcatRECAPOCI}{\FASTtomcatRECAPODynamic}{\FASTtomcatRECAPODynamicCI} \\
\mc{1}{c}{} & \mc{4}{c}{\bench{tomcat}} \\
\end{tabular}
\medskip\\
\begin{tabular}{@{}l|Hl@{\;\;}l@{\;\;}l@{}}
        & w/ G & Unopt- & \FTO- & \REAbbrv- \\\hline
\HB		& \rna													& \sdr{\SLOWxalanHB}{\SLOWxalanHBCI}{\SLOWxalanHBDynamic}{\SLOWxalanHBDynamicCI}					& \sdr{\FASTxalanFTOHB}{\FASTxalanFTOHBCI}{\FASTxalanFTOHBDynamic}{\FASTxalanFTOHBDynamicCI}		& \rna \\
\WCP	& \rna													& \sdr{\SLOWxalanWCP}{\SLOWxalanWCPCI}{\SLOWxalanWCPDynamic}{\SLOWxalanWCPDynamicCI}				& \sdr{\FASTxalanFTOWCP}{\FASTxalanFTOWCPCI}{\FASTxalanFTOWCPDynamic}{\FASTxalanFTOWCPDynamicCI}		& \sdr{\FASTxalanREWCP}{\FASTxalanREWCPCI}{\FASTxalanREWCPDynamic}{\FASTxalanREWCPDynamicCI} \\
\DC		& \sdr{\SLOWxalanDCExc}{\SLOWxalanDCExcCI}{\SLOWxalanDCExcDynamic}{\SLOWxalanDCExcDynamicCI}		& \sdr{\SLOWxalanDCnoGExc}{\SLOWxalanDCnoGExcCI}{\SLOWxalanDCnoGExcDynamic}{\SLOWxalanDCnoGExcDynamicCI}	& \sdr{\FASTxalanFTODC}{\FASTxalanFTODCCI}{\FASTxalanFTODCDynamic}{\FASTxalanFTODCDynamicCI}		& \sdr{\FASTxalanREDC}{\FASTxalanREDCCI}{\FASTxalanREDCDynamic}{\FASTxalanREDCDynamicCI} \\
\CAPO	& \sdr{\SLOWxalanCAPOExc}{\SLOWxalanCAPOExcCI}{\SLOWxalanCAPOExcDynamic}{\SLOWxalanCAPOExcDynamicCI}	& \sdr{\SLOWxalanCAPOnoGExc}{\SLOWxalanCAPOnoGExcCI}{\SLOWxalanCAPOnoGExcDynamic}{\SLOWxalanCAPOnoGExcDynamicCI}	& \sdr{\FASTxalanFTOCAPO}{\FASTxalanFTOCAPOCI}{\FASTxalanFTOCAPODynamic}{\FASTxalanFTOCAPODynamicCI}	& \sdr{\FASTxalanRECAPO}{\FASTxalanRECAPOCI}{\FASTxalanRECAPODynamic}{\FASTxalanRECAPODynamicCI} \\
\mc{1}{c}{} & \mc{4}{c}{\bench{xalan}} \\
\end{tabular}

\caption{Average races reported by various analyses for each evaluated program
(excluding \bench{batik} and \bench{lusearch}, for which all analyses report no races).
In each cell, the first value is statically distinct races
(\ie, distinct program locations), and the second value, in parentheses, is total dynamic races,
both with 95\% confidence intervals.}
\label{tab:race:allDaCapo:CI}
\end{table*}

}{

\begin{table}[H]
\newcommand{\rzero}{0\xspace}
\newcommand{\roh}[1]{\ifthenelse{\equal{#1}{\rna}}{\rna}{#1$\;\!\times$}} 
\newcommand{\rna}{N/A}
\newcommand{\memna}{N/A} 
\newcommand{\st}[1]{(#1~s)} 
\newcommand{\dt}[1]{#1~s} 
\newcommand{\mem}[1]{\ifthenelse{\equal{#1}{\memna}}{\memna}{#1$\;Mb$}} 
\newcommand{\et}[1]{#1M} 
\newcommand{\enfp}[1]{(#1M)} 
\newcommand{\base}[1]{#1~s} 
\newcommand{\sdr}[4]{\vci{#1}{#2}{\;}(\vci{#3}{#4})}
\newcommand{\vci}[2]{\ifthenelse{\equal{#1}{\rna}}{\rna}{#1{\;}\ci{#2}}}
\newcommand{\ci}[1]{\ensuremath{\pm}{\;}#1}
\input{result-macros/PIP_slowTool_noCoresSet}
\input{result-macros/PIP_fastTool_extraOpt2Quiet}
\smaller
\centering
\begin{tabular}{@{}l|Hl@{\;\;}l@{\;\;}l@{}}
        & w/ G & Unopt- & \FTO- & \REAbbrv- \\\hline
\HB		& \rna													& \sdr{\SLOWavroraHB}{\SLOWavroraHBCI}{\SLOWavroraHBDynamic}{\SLOWavroraHBDynamicCI}					& \sdr{\FASTavroraFTOHB}{\FASTavroraFTOHBCI}{\FASTavroraFTOHBDynamic}{\FASTavroraFTOHBDynamicCI}		& \rna \\
\WCP	& \rna													& \sdr{\SLOWavroraWCP}{\SLOWavroraWCPCI}{\SLOWavroraWCPDynamic}{\SLOWavroraWCPDynamicCI}				& \sdr{\FASTavroraFTOWCP}{\FASTavroraFTOWCPCI}{\FASTavroraFTOWCPDynamic}{\FASTavroraFTOWCPDynamicCI}		& \sdr{\FASTavroraREWCP}{\FASTavroraREWCPCI}{\FASTavroraREWCPDynamic}{\FASTavroraREWCPDynamicCI} \\
\DC		& \sdr{\SLOWavroraDCExc}{\SLOWavroraDCExcCI}{\SLOWavroraDCExcDynamic}{\SLOWavroraDCExcDynamicCI}		& \sdr{\SLOWavroraDCnoGExc}{\SLOWavroraDCnoGExcCI}{\SLOWavroraDCnoGExcDynamic}{\SLOWavroraDCnoGExcDynamicCI}	& \sdr{\FASTavroraFTODC}{\FASTavroraFTODCCI}{\FASTavroraFTODCDynamic}{\FASTavroraFTODCDynamicCI}		& \sdr{\FASTavroraREDC}{\FASTavroraREDCCI}{\FASTavroraREDCDynamic}{\FASTavroraREDCDynamicCI} \\
\CAPO	& \sdr{\SLOWavroraCAPOExc}{\SLOWavroraCAPOExcCI}{\SLOWavroraCAPOExcDynamic}{\SLOWavroraCAPOExcDynamicCI}	& \sdr{\SLOWavroraCAPOnoGExc}{\SLOWavroraCAPOnoGExcCI}{\SLOWavroraCAPOnoGExcDynamic}{\SLOWavroraCAPOnoGExcDynamicCI}	& \sdr{\FASTavroraFTOCAPO}{\FASTavroraFTOCAPOCI}{\FASTavroraFTOCAPODynamic}{\FASTavroraFTOCAPODynamicCI}	& \sdr{\FASTavroraRECAPO}{\FASTavroraRECAPOCI}{\FASTavroraRECAPODynamic}{\FASTavroraRECAPODynamicCI} \\
\mc{1}{c}{} & \mc{4}{c}{\bench{avrora}} \\
\end{tabular}
\medskip\\
\begin{tabular}{@{}l|Hl@{\;\;}l@{\;\;}l@{}}
		& w/ G & Unopt- & \FTO- & \REAbbrv- \\\hline
\HB		& \rna													& \sdr{\SLOWhtwoHB}{\SLOWhtwoHBCI}{\SLOWhtwoHBDynamic}{\SLOWhtwoHBDynamicCI}					& \sdr{\FASThtwoFTOHB}{\FASThtwoFTOHBCI}{\FASThtwoFTOHBDynamic}{\FASThtwoFTOHBDynamicCI}		& \rna \\
\WCP	& \rna													& \sdr{\SLOWhtwoWCP}{\SLOWhtwoWCPCI}{\SLOWhtwoWCPDynamic}{\SLOWhtwoWCPDynamicCI}				& \sdr{\FASThtwoFTOWCP}{\FASThtwoFTOWCPCI}{\FASThtwoFTOWCPDynamic}{\FASThtwoFTOWCPDynamicCI}		& \sdr{\FASThtwoREWCP}{\FASThtwoREWCPCI}{\FASThtwoREWCPDynamic}{\FASThtwoREWCPDynamicCI} \\
\DC		& \sdr{\SLOWhtwoDCExc}{\SLOWhtwoDCExcCI}{\SLOWhtwoDCExcDynamic}{\SLOWhtwoDCExcDynamicCI}		& \sdr{\SLOWhtwoDCnoGExc}{\SLOWhtwoDCnoGExcCI}{\SLOWhtwoDCnoGExcDynamic}{\SLOWhtwoDCnoGExcDynamicCI}	& \sdr{\FASThtwoFTODC}{\FASThtwoFTODCCI}{\FASThtwoFTODCDynamic}{\FASThtwoFTODCDynamicCI}		& \sdr{\FASThtwoREDC}{\FASThtwoREDCCI}{\FASThtwoREDCDynamic}{\FASThtwoREDCDynamicCI} \\
\CAPO	& \sdr{\SLOWhtwoCAPOExc}{\SLOWhtwoCAPOExcCI}{\SLOWhtwoCAPOExcDynamic}{\SLOWhtwoCAPOExcDynamicCI}	& \sdr{\SLOWhtwoCAPOnoGExc}{\SLOWhtwoCAPOnoGExcCI}{\SLOWhtwoCAPOnoGExcDynamic}{\SLOWhtwoCAPOnoGExcDynamicCI}	& \sdr{\FASThtwoFTOCAPO}{\FASThtwoFTOCAPOCI}{\FASThtwoFTOCAPODynamic}{\FASThtwoFTOCAPODynamicCI}	& \sdr{\FASThtwoRECAPO}{\FASThtwoRECAPOCI}{\FASThtwoRECAPODynamic}{\FASThtwoRECAPODynamicCI} \\
\mc{1}{c}{} & \mc{4}{c}{\bench{h2}} \\
\end{tabular}
\medskip\\
\begin{tabular}{@{}l|Hl@{\;\;}l@{\;\;}l@{}}
        & w/ G & Unopt- & \FTO- & \REAbbrv- \\\hline
\HB		& \rna													& \sdr{\SLOWjythonHB}{\SLOWjythonHBCI}{\SLOWjythonHBDynamic}{\SLOWjythonHBDynamicCI}					& \sdr{\FASTjythonFTOHB}{\FASTjythonFTOHBCI}{\FASTjythonFTOHBDynamic}{\FASTjythonFTOHBDynamicCI}		& \rna \\
\WCP	& \rna													& \sdr{\SLOWjythonWCP}{\SLOWjythonWCPCI}{\SLOWjythonWCPDynamic}{\SLOWjythonWCPDynamicCI}				& \sdr{\FASTjythonFTOWCP}{\FASTjythonFTOWCPCI}{\FASTjythonFTOWCPDynamic}{\FASTjythonFTOWCPDynamicCI}		& \sdr{\FASTjythonREWCP}{\FASTjythonREWCPCI}{\FASTjythonREWCPDynamic}{\FASTjythonREWCPDynamicCI} \\
\DC		& \sdr{\SLOWjythonDCExc}{\SLOWjythonDCExcCI}{\SLOWjythonDCExcDynamic}{\SLOWjythonDCExcDynamicCI}		& \sdr{\SLOWjythonDCnoGExc}{\SLOWjythonDCnoGExcCI}{\SLOWjythonDCnoGExcDynamic}{\SLOWjythonDCnoGExcDynamicCI}	& \sdr{\FASTjythonFTODC}{\FASTjythonFTODCCI}{\FASTjythonFTODCDynamic}{\FASTjythonFTODCDynamicCI}		& \sdr{\FASTjythonREDC}{\FASTjythonREDCCI}{\FASTjythonREDCDynamic}{\FASTjythonREDCDynamicCI} \\
\CAPO	& \sdr{\SLOWjythonCAPOExc}{\SLOWjythonCAPOExcCI}{\SLOWjythonCAPOExcDynamic}{\SLOWjythonCAPOExcDynamicCI}	& \sdr{\SLOWjythonCAPOnoGExc}{\SLOWjythonCAPOnoGExcCI}{\SLOWjythonCAPOnoGExcDynamic}{\SLOWjythonCAPOnoGExcDynamicCI}	& \sdr{\FASTjythonFTOCAPO}{\FASTjythonFTOCAPOCI}{\FASTjythonFTOCAPODynamic}{\FASTjythonFTOCAPODynamicCI}	& \sdr{\FASTjythonRECAPO}{\FASTjythonRECAPOCI}{\FASTjythonRECAPODynamic}{\FASTjythonRECAPODynamicCI} \\
\mc{1}{c}{} & \mc{4}{c}{\bench{jython}} \\
\end{tabular}
\medskip\\
\begin{tabular}{@{}l|Hl@{\;\;}l@{\;\;}l@{}}
        & w/ G & Unopt- & \FTO- & \REAbbrv- \\\hline
\HB		& \rna													& \sdr{\SLOWluindexHB}{\SLOWluindexHBCI}{\SLOWluindexHBDynamic}{\SLOWluindexHBDynamicCI}					& \sdr{\FASTluindexFTOHB}{\FASTluindexFTOHBCI}{\FASTluindexFTOHBDynamic}{\FASTluindexFTOHBDynamicCI}		& \rna \\
\WCP	& \rna													& \sdr{\SLOWluindexWCP}{\SLOWluindexWCPCI}{\SLOWluindexWCPDynamic}{\SLOWluindexWCPDynamicCI}				& \sdr{\FASTluindexFTOWCP}{\FASTluindexFTOWCPCI}{\FASTluindexFTOWCPDynamic}{\FASTluindexFTOWCPDynamicCI}		& \sdr{\FASTluindexREWCP}{\FASTluindexREWCPCI}{\FASTluindexREWCPDynamic}{\FASTluindexREWCPDynamicCI} \\
\DC		& \sdr{\SLOWluindexDCExc}{\SLOWluindexDCExcCI}{\SLOWluindexDCExcDynamic}{\SLOWluindexDCExcDynamicCI}		& \sdr{\SLOWluindexDCnoGExc}{\SLOWluindexDCnoGExcCI}{\SLOWluindexDCnoGExcDynamic}{\SLOWluindexDCnoGExcDynamicCI}	& \sdr{\FASTluindexFTODC}{\FASTluindexFTODCCI}{\FASTluindexFTODCDynamic}{\FASTluindexFTODCDynamicCI}		& \sdr{\FASTluindexREDC}{\FASTluindexREDCCI}{\FASTluindexREDCDynamic}{\FASTluindexREDCDynamicCI} \\
\CAPO	& \sdr{\SLOWluindexCAPOExc}{\SLOWluindexCAPOExcCI}{\SLOWluindexCAPOExcDynamic}{\SLOWluindexCAPOExcDynamicCI}	& \sdr{\SLOWluindexCAPOnoGExc}{\SLOWluindexCAPOnoGExcCI}{\SLOWluindexCAPOnoGExcDynamic}{\SLOWluindexCAPOnoGExcDynamicCI}	& \sdr{\FASTluindexFTOCAPO}{\FASTluindexFTOCAPOCI}{\FASTluindexFTOCAPODynamic}{\FASTluindexFTOCAPODynamicCI}	& \sdr{\FASTluindexRECAPO}{\FASTluindexRECAPOCI}{\FASTluindexRECAPODynamic}{\FASTluindexRECAPODynamicCI} \\
\mc{1}{c}{} & \mc{4}{c}{\bench{luindex}} \\
\end{tabular}
\medskip\\
\begin{tabular}{@{}l|Hl@{\;\;}l@{\;\;}l@{}}
        & w/ G & Unopt- & \FTO- & \REAbbrv- \\\hline
\HB		& \rna													& \sdr{\SLOWpmdHB}{\SLOWpmdHBCI}{\SLOWpmdHBDynamic}{\SLOWpmdHBDynamicCI}					& \sdr{\FASTpmdFTOHB}{\FASTpmdFTOHBCI}{\FASTpmdFTOHBDynamic}{\FASTpmdFTOHBDynamicCI}		& \rna \\
\WCP	& \rna													& \sdr{\SLOWpmdWCP}{\SLOWpmdWCPCI}{\SLOWpmdWCPDynamic}{\SLOWpmdWCPDynamicCI}				& \sdr{\FASTpmdFTOWCP}{\FASTpmdFTOWCPCI}{\FASTpmdFTOWCPDynamic}{\FASTpmdFTOWCPDynamicCI}		& \sdr{\FASTpmdREWCP}{\FASTpmdREWCPCI}{\FASTpmdREWCPDynamic}{\FASTpmdREWCPDynamicCI} \\
\DC		& \sdr{\SLOWpmdDCExc}{\SLOWpmdDCExcCI}{\SLOWpmdDCExcDynamic}{\SLOWpmdDCExcDynamicCI}		& \sdr{\SLOWpmdDCnoGExc}{\SLOWpmdDCnoGExcCI}{\SLOWpmdDCnoGExcDynamic}{\SLOWpmdDCnoGExcDynamicCI}	& \sdr{\FASTpmdFTODC}{\FASTpmdFTODCCI}{\FASTpmdFTODCDynamic}{\FASTpmdFTODCDynamicCI}		& \sdr{\FASTpmdREDC}{\FASTpmdREDCCI}{\FASTpmdREDCDynamic}{\FASTpmdREDCDynamicCI} \\
\CAPO	& \sdr{\SLOWpmdCAPOExc}{\SLOWpmdCAPOExcCI}{\SLOWpmdCAPOExcDynamic}{\SLOWpmdCAPOExcDynamicCI}	& \sdr{\SLOWpmdCAPOnoGExc}{\SLOWpmdCAPOnoGExcCI}{\SLOWpmdCAPOnoGExcDynamic}{\SLOWpmdCAPOnoGExcDynamicCI}	& \sdr{\FASTpmdFTOCAPO}{\FASTpmdFTOCAPOCI}{\FASTpmdFTOCAPODynamic}{\FASTpmdFTOCAPODynamicCI}	& \sdr{\FASTpmdRECAPO}{\FASTpmdRECAPOCI}{\FASTpmdRECAPODynamic}{\FASTpmdRECAPODynamicCI} \\
\mc{1}{c}{} & \mc{4}{c}{\bench{pmd}} \\
\end{tabular}
\medskip\\
\begin{tabular}{@{}l|Hl@{\;\;}l@{\;\;}l@{}}
        & w/ G & Unopt- & \FTO- & \REAbbrv- \\\hline
\HB		& \rna													& \sdr{\SLOWsunflowHB}{\SLOWsunflowHBCI}{\SLOWsunflowHBDynamic}{\SLOWsunflowHBDynamicCI}					& \sdr{\FASTsunflowFTOHB}{\FASTsunflowFTOHBCI}{\FASTsunflowFTOHBDynamic}{\FASTsunflowFTOHBDynamicCI}		& \rna \\
\WCP	& \rna													& \sdr{\SLOWsunflowWCP}{\SLOWsunflowWCPCI}{\SLOWsunflowWCPDynamic}{\SLOWsunflowWCPDynamicCI}				& \sdr{\FASTsunflowFTOWCP}{\FASTsunflowFTOWCPCI}{\FASTsunflowFTOWCPDynamic}{\FASTsunflowFTOWCPDynamicCI}		& \sdr{\FASTsunflowREWCP}{\FASTsunflowREWCPCI}{\FASTsunflowREWCPDynamic}{\FASTsunflowREWCPDynamicCI} \\
\DC		& \sdr{\SLOWsunflowDCExc}{\SLOWsunflowDCExcCI}{\SLOWsunflowDCExcDynamic}{\SLOWsunflowDCExcDynamicCI}		& \sdr{\SLOWsunflowDCnoGExc}{\SLOWsunflowDCnoGExcCI}{\SLOWsunflowDCnoGExcDynamic}{\SLOWsunflowDCnoGExcDynamicCI}	& \sdr{\FASTsunflowFTODC}{\FASTsunflowFTODCCI}{\FASTsunflowFTODCDynamic}{\FASTsunflowFTODCDynamicCI}		& \sdr{\FASTsunflowREDC}{\FASTsunflowREDCCI}{\FASTsunflowREDCDynamic}{\FASTsunflowREDCDynamicCI} \\
\CAPO	& \sdr{\SLOWsunflowCAPOExc}{\SLOWsunflowCAPOExcCI}{\SLOWsunflowCAPOExcDynamic}{\SLOWsunflowCAPOExcDynamicCI}	& \sdr{\SLOWsunflowCAPOnoGExc}{\SLOWsunflowCAPOnoGExcCI}{\SLOWsunflowCAPOnoGExcDynamic}{\SLOWsunflowCAPOnoGExcDynamicCI}	& \sdr{\FASTsunflowFTOCAPO}{\FASTsunflowFTOCAPOCI}{\FASTsunflowFTOCAPODynamic}{\FASTsunflowFTOCAPODynamicCI}	& \sdr{\FASTsunflowRECAPO}{\FASTsunflowRECAPOCI}{\FASTsunflowRECAPODynamic}{\FASTsunflowRECAPODynamicCI} \\
\mc{1}{c}{} & \mc{4}{c}{\bench{sunflow}} \\
\end{tabular}
\medskip\\
\begin{tabular}{@{}l|Hl@{\;\;}l@{\;\;}l@{}}
        & w/ G & Unopt- & \FTO- & \REAbbrv- \\\hline
\HB		& \rna													& \sdr{\SLOWtomcatHB}{\SLOWtomcatHBCI}{\SLOWtomcatHBDynamic}{\SLOWtomcatHBDynamicCI}					& \sdr{\FASTtomcatFTOHB}{\FASTtomcatFTOHBCI}{\FASTtomcatFTOHBDynamic}{\FASTtomcatFTOHBDynamicCI}		& \rna \\
\WCP	& \rna													& \sdr{\SLOWtomcatWCP}{\SLOWtomcatWCPCI}{\SLOWtomcatWCPDynamic}{\SLOWtomcatWCPDynamicCI}				& \sdr{\FASTtomcatFTOWCP}{\FASTtomcatFTOWCPCI}{\FASTtomcatFTOWCPDynamic}{\FASTtomcatFTOWCPDynamicCI}		& \sdr{\FASTtomcatREWCP}{\FASTtomcatREWCPCI}{\FASTtomcatREWCPDynamic}{\FASTtomcatREWCPDynamicCI} \\
\DC		& \sdr{\SLOWtomcatDCExc}{\SLOWtomcatDCExcCI}{\SLOWtomcatDCExcDynamic}{\SLOWtomcatDCExcDynamicCI}		& \sdr{\SLOWtomcatDCnoGExc}{\SLOWtomcatDCnoGExcCI}{\SLOWtomcatDCnoGExcDynamic}{\SLOWtomcatDCnoGExcDynamicCI}	& \sdr{\FASTtomcatFTODC}{\FASTtomcatFTODCCI}{\FASTtomcatFTODCDynamic}{\FASTtomcatFTODCDynamicCI}		& \sdr{\FASTtomcatREDC}{\FASTtomcatREDCCI}{\FASTtomcatREDCDynamic}{\FASTtomcatREDCDynamicCI} \\
\CAPO	& \sdr{\SLOWtomcatCAPOExc}{\SLOWtomcatCAPOExcCI}{\SLOWtomcatCAPOExcDynamic}{\SLOWtomcatCAPOExcDynamicCI}	& \sdr{\SLOWtomcatCAPOnoGExc}{\SLOWtomcatCAPOnoGExcCI}{\SLOWtomcatCAPOnoGExcDynamic}{\SLOWtomcatCAPOnoGExcDynamicCI}	& \sdr{\FASTtomcatFTOCAPO}{\FASTtomcatFTOCAPOCI}{\FASTtomcatFTOCAPODynamic}{\FASTtomcatFTOCAPODynamicCI}	& \sdr{\FASTtomcatRECAPO}{\FASTtomcatRECAPOCI}{\FASTtomcatRECAPODynamic}{\FASTtomcatRECAPODynamicCI} \\
\mc{1}{c}{} & \mc{4}{c}{\bench{tomcat}} \\
\end{tabular}
\medskip\\
\begin{tabular}{@{}l|Hl@{\;\;}l@{\;\;}l@{}}
        & w/ G & Unopt- & \FTO- & \REAbbrv- \\\hline
\HB		& \rna													& \sdr{\SLOWxalanHB}{\SLOWxalanHBCI}{\SLOWxalanHBDynamic}{\SLOWxalanHBDynamicCI}					& \sdr{\FASTxalanFTOHB}{\FASTxalanFTOHBCI}{\FASTxalanFTOHBDynamic}{\FASTxalanFTOHBDynamicCI}		& \rna \\
\WCP	& \rna													& \sdr{\SLOWxalanWCP}{\SLOWxalanWCPCI}{\SLOWxalanWCPDynamic}{\SLOWxalanWCPDynamicCI}				& \sdr{\FASTxalanFTOWCP}{\FASTxalanFTOWCPCI}{\FASTxalanFTOWCPDynamic}{\FASTxalanFTOWCPDynamicCI}		& \sdr{\FASTxalanREWCP}{\FASTxalanREWCPCI}{\FASTxalanREWCPDynamic}{\FASTxalanREWCPDynamicCI} \\
\DC		& \sdr{\SLOWxalanDCExc}{\SLOWxalanDCExcCI}{\SLOWxalanDCExcDynamic}{\SLOWxalanDCExcDynamicCI}		& \sdr{\SLOWxalanDCnoGExc}{\SLOWxalanDCnoGExcCI}{\SLOWxalanDCnoGExcDynamic}{\SLOWxalanDCnoGExcDynamicCI}	& \sdr{\FASTxalanFTODC}{\FASTxalanFTODCCI}{\FASTxalanFTODCDynamic}{\FASTxalanFTODCDynamicCI}		& \sdr{\FASTxalanREDC}{\FASTxalanREDCCI}{\FASTxalanREDCDynamic}{\FASTxalanREDCDynamicCI} \\
\CAPO	& \sdr{\SLOWxalanCAPOExc}{\SLOWxalanCAPOExcCI}{\SLOWxalanCAPOExcDynamic}{\SLOWxalanCAPOExcDynamicCI}	& \sdr{\SLOWxalanCAPOnoGExc}{\SLOWxalanCAPOnoGExcCI}{\SLOWxalanCAPOnoGExcDynamic}{\SLOWxalanCAPOnoGExcDynamicCI}	& \sdr{\FASTxalanFTOCAPO}{\FASTxalanFTOCAPOCI}{\FASTxalanFTOCAPODynamic}{\FASTxalanFTOCAPODynamicCI}	& \sdr{\FASTxalanRECAPO}{\FASTxalanRECAPOCI}{\FASTxalanRECAPODynamic}{\FASTxalanRECAPODynamicCI} \\
\mc{1}{c}{} & \mc{4}{c}{\bench{xalan}} \\
\end{tabular}

\caption{Average races reported by various analyses for each evaluated program
(excluding \bench{batik} and \bench{lusearch}, for which all analyses report no races).
In each cell, the first value is statically distinct races
(\ie, distinct program locations), and the second value, in parentheses, is total dynamic races,
both with 95\% confidence intervals.}
\label{tab:race:allDaCapo:CI}
\end{table}
}

%% file: result-macros/PIP_slowTool_noWriteOrReadRaceEdge.tex
\newcommand{\SLOWavroraEvents}{1,400}
\newcommand{\SLOWavroraNoFPEvents}{160}
\newcommand{\SLOWavroraMaxLiveThreads}{7}
\newcommand{\SLOWavroraTotalThreads}{7}
\newcommand{\SLOWavroraBaseTime}{2.4}
\newcommand{\SLOWavroraBaseTimeCI}{18}
\newcommand{\SLOWavroraEmptyTime}{\rna}
\newcommand{\SLOWavroraEmptyTimeCI}{\rna}
\newcommand{\SLOWavroraEmptyTimeCIMIN}{\rna}
\newcommand{\SLOWavroraEmptyTimeCIMAX}{\rna}
\newcommand{\SLOWavroraFTTime}{\rna}
\newcommand{\SLOWavroraFTTimeCI}{\rna}
\newcommand{\SLOWavroraFTTimeCIMIN}{\rna}
\newcommand{\SLOWavroraFTTimeCIMAX}{\rna}
\newcommand{\SLOWavroraHBTime}{18}
\newcommand{\SLOWavroraHBTimeCI}{0.25}
\newcommand{\SLOWavroraWCPTime}{24}
\newcommand{\SLOWavroraWCPTimeCI}{0.92}
\newcommand{\SLOWavroraDCnoGExcTime}{22}
\newcommand{\SLOWavroraDCnoGExcTimeCI}{0.4}
\newcommand{\SLOWavroraDCnoGTime}{\rna}
\newcommand{\SLOWavroraDCnoGTimeCI}{\rna}
\newcommand{\SLOWavroraDCnoGTimeCIMIN}{\rna}
\newcommand{\SLOWavroraDCnoGTimeCIMAX}{\rna}
\newcommand{\SLOWavroraDCExcTime}{25}
\newcommand{\SLOWavroraDCExcTimeCI}{0.22}
\newcommand{\SLOWavroraDCTime}{\rna}
\newcommand{\SLOWavroraDCTimeCI}{\rna}
\newcommand{\SLOWavroraDCTimeCIMIN}{\rna}
\newcommand{\SLOWavroraDCTimeCIMAX}{\rna}
\newcommand{\SLOWavroraCAPOnoGExcTime}{21}
\newcommand{\SLOWavroraCAPOnoGExcTimeCI}{2.7}
\newcommand{\SLOWavroraCAPOnoGTime}{\rna}
\newcommand{\SLOWavroraCAPOnoGTimeCI}{\rna}
\newcommand{\SLOWavroraCAPOnoGTimeCIMIN}{\rna}
\newcommand{\SLOWavroraCAPOnoGTimeCIMAX}{\rna}
\newcommand{\SLOWavroraCAPOExcTime}{22}
\newcommand{\SLOWavroraCAPOExcTimeCI}{0.48}
\newcommand{\SLOWavroraCAPOTime}{\rna}
\newcommand{\SLOWavroraCAPOTimeCI}{\rna}
\newcommand{\SLOWavroraCAPOTimeCIMIN}{\rna}
\newcommand{\SLOWavroraCAPOTimeCIMAX}{\rna}
\newcommand{\SLOWavroraStaticTime}{\rzero}
\newcommand{\SLOWavroraDynamicTime}{\rzero}
\newcommand{\SLOWavroraBaseMem}{160}
\newcommand{\SLOWavroraBaseMemCI}{9.7}
\newcommand{\SLOWavroraHBMem}{32}
\newcommand{\SLOWavroraHBMemCI}{2.3}
\newcommand{\SLOWavroraFTMem}{\memna}
\newcommand{\SLOWavroraFTMemCI}{\memna}
\newcommand{\SLOWavroraFTMemCIMIN}{\memna}
\newcommand{\SLOWavroraFTMemCIMAX}{\memna}
\newcommand{\SLOWavroraWCPMem}{94}
\newcommand{\SLOWavroraWCPMemCI}{5.1}
\newcommand{\SLOWavroraDCnoGExcMem}{39}
\newcommand{\SLOWavroraDCnoGExcMemCI}{2.6}
\newcommand{\SLOWavroraDCnoGMem}{\memna}
\newcommand{\SLOWavroraDCnoGMemCI}{\memna}
\newcommand{\SLOWavroraDCnoGMemCIMIN}{\memna}
\newcommand{\SLOWavroraDCnoGMemCIMAX}{\memna}
\newcommand{\SLOWavroraDCExcMem}{69}
\newcommand{\SLOWavroraDCExcMemCI}{3.8}
\newcommand{\SLOWavroraDCMem}{\memna}
\newcommand{\SLOWavroraDCMemCI}{\memna}
\newcommand{\SLOWavroraDCMemCIMIN}{\memna}
\newcommand{\SLOWavroraDCMemCIMAX}{\memna}
\newcommand{\SLOWavroraCAPOnoGExcMem}{38}
\newcommand{\SLOWavroraCAPOnoGExcMemCI}{6.7}
\newcommand{\SLOWavroraCAPOnoGMem}{\memna}
\newcommand{\SLOWavroraCAPOnoGMemCI}{\memna}
\newcommand{\SLOWavroraCAPOnoGMemCIMIN}{\memna}
\newcommand{\SLOWavroraCAPOnoGMemCIMAX}{\memna}
\newcommand{\SLOWavroraCAPOExcMem}{69}
\newcommand{\SLOWavroraCAPOExcMemCI}{3.7}
\newcommand{\SLOWavroraCAPOMem}{\memna}
\newcommand{\SLOWavroraCAPOMemCI}{\memna}
\newcommand{\SLOWavroraCAPOMemCIMIN}{\memna}
\newcommand{\SLOWavroraCAPOMemCIMAX}{\memna}
\newcommand{\SLOWavroraEventsCI}{107,189}
\newcommand{\SLOWavroraEventsCIMIN}{1,438,491,676}
\newcommand{\SLOWavroraEventsCIMAX}{1,438,706,054}
\newcommand{\SLOWavroraNoFPEventsCI}{91,458}
\newcommand{\SLOWavroraNoFPEventsCIMIN}{156,345,614}
\newcommand{\SLOWavroraNoFPEventsCIMAX}{156,528,530}
\newcommand{\SLOWavroraHB}{6}
\newcommand{\SLOWavroraHBCI}{0}
\newcommand{\SLOWavroraHBCIMIN}{6}
\newcommand{\SLOWavroraHBCIMAX}{6}
\newcommand{\SLOWavroraHBDynamic}{522,985}
\newcommand{\SLOWavroraHBDynamicCI}{1,518}
\newcommand{\SLOWavroraHBDynamicCIMIN}{521,467}
\newcommand{\SLOWavroraHBDynamicCIMAX}{524,503}
\newcommand{\SLOWavroraFT}{\rna}
\newcommand{\SLOWavroraFTCI}{\rna}
\newcommand{\SLOWavroraFTCIMIN}{\rna}
\newcommand{\SLOWavroraFTCIMAX}{\rna}
\newcommand{\SLOWavroraFTDynamic}{\rna}
\newcommand{\SLOWavroraFTDynamicCI}{\rna}
\newcommand{\SLOWavroraFTDynamicCIMIN}{\rna}
\newcommand{\SLOWavroraFTDynamicCIMAX}{\rna}
\newcommand{\SLOWavroraWCP}{5}
\newcommand{\SLOWavroraWCPCI}{0}
\newcommand{\SLOWavroraWCPCIMIN}{5}
\newcommand{\SLOWavroraWCPCIMAX}{5}
\newcommand{\SLOWavroraWCPDynamic}{552,479}
\newcommand{\SLOWavroraWCPDynamicCI}{1,207}
\newcommand{\SLOWavroraWCPDynamicCIMIN}{551,272}
\newcommand{\SLOWavroraWCPDynamicCIMAX}{553,686}
\newcommand{\SLOWavroraDCnoGExc}{5}
\newcommand{\SLOWavroraDCnoGExcCI}{0}
\newcommand{\SLOWavroraDCnoGExcCIMIN}{5}
\newcommand{\SLOWavroraDCnoGExcCIMAX}{5}
\newcommand{\SLOWavroraDCnoGExcDynamic}{557,009}
\newcommand{\SLOWavroraDCnoGExcDynamicCI}{590}
\newcommand{\SLOWavroraDCnoGExcDynamicCIMIN}{556,419}
\newcommand{\SLOWavroraDCnoGExcDynamicCIMAX}{557,599}
\newcommand{\SLOWavroraDCnoG}{\rna}
\newcommand{\SLOWavroraDCnoGCI}{\rna}
\newcommand{\SLOWavroraDCnoGCIMIN}{\rna}
\newcommand{\SLOWavroraDCnoGCIMAX}{\rna}
\newcommand{\SLOWavroraDCnoGDynamic}{\rna}
\newcommand{\SLOWavroraDCnoGDynamicCI}{\rna}
\newcommand{\SLOWavroraDCnoGDynamicCIMIN}{\rna}
\newcommand{\SLOWavroraDCnoGDynamicCIMAX}{\rna}
\newcommand{\SLOWavroraDCExc}{6}
\newcommand{\SLOWavroraDCExcCI}{0}
\newcommand{\SLOWavroraDCExcCIMIN}{6}
\newcommand{\SLOWavroraDCExcCIMAX}{6}
\newcommand{\SLOWavroraDCExcDynamic}{154,673}
\newcommand{\SLOWavroraDCExcDynamicCI}{3,886}
\newcommand{\SLOWavroraDCExcDynamicCIMIN}{150,787}
\newcommand{\SLOWavroraDCExcDynamicCIMAX}{158,559}
\newcommand{\SLOWavroraDC}{\rna}
\newcommand{\SLOWavroraDCCI}{\rna}
\newcommand{\SLOWavroraDCCIMIN}{\rna}
\newcommand{\SLOWavroraDCCIMAX}{\rna}
\newcommand{\SLOWavroraDCDynamic}{\rna}
\newcommand{\SLOWavroraDCDynamicCI}{\rna}
\newcommand{\SLOWavroraDCDynamicCIMIN}{\rna}
\newcommand{\SLOWavroraDCDynamicCIMAX}{\rna}
\newcommand{\SLOWavroraCAPOnoGExc}{5}
\newcommand{\SLOWavroraCAPOnoGExcCI}{0}
\newcommand{\SLOWavroraCAPOnoGExcCIMIN}{5}
\newcommand{\SLOWavroraCAPOnoGExcCIMAX}{5}
\newcommand{\SLOWavroraCAPOnoGExcDynamic}{543,173}
\newcommand{\SLOWavroraCAPOnoGExcDynamicCI}{3,762}
\newcommand{\SLOWavroraCAPOnoGExcDynamicCIMIN}{539,411}
\newcommand{\SLOWavroraCAPOnoGExcDynamicCIMAX}{546,935}
\newcommand{\SLOWavroraCAPOnoG}{\rna}
\newcommand{\SLOWavroraCAPOnoGCI}{\rna}
\newcommand{\SLOWavroraCAPOnoGCIMIN}{\rna}
\newcommand{\SLOWavroraCAPOnoGCIMAX}{\rna}
\newcommand{\SLOWavroraCAPOnoGDynamic}{\rna}
\newcommand{\SLOWavroraCAPOnoGDynamicCI}{\rna}
\newcommand{\SLOWavroraCAPOnoGDynamicCIMIN}{\rna}
\newcommand{\SLOWavroraCAPOnoGDynamicCIMAX}{\rna}
\newcommand{\SLOWavroraCAPOExc}{6}
\newcommand{\SLOWavroraCAPOExcCI}{0}
\newcommand{\SLOWavroraCAPOExcCIMIN}{6}
\newcommand{\SLOWavroraCAPOExcCIMAX}{6}
\newcommand{\SLOWavroraCAPOExcDynamic}{217,518}
\newcommand{\SLOWavroraCAPOExcDynamicCI}{2,377}
\newcommand{\SLOWavroraCAPOExcDynamicCIMIN}{215,141}
\newcommand{\SLOWavroraCAPOExcDynamicCIMAX}{219,895}
\newcommand{\SLOWavroraCAPO}{\rna}
\newcommand{\SLOWavroraCAPOCI}{\rna}
\newcommand{\SLOWavroraCAPOCIMIN}{\rna}
\newcommand{\SLOWavroraCAPOCIMAX}{\rna}
\newcommand{\SLOWavroraCAPODynamic}{\rna}
\newcommand{\SLOWavroraCAPODynamicCI}{\rna}
\newcommand{\SLOWavroraCAPODynamicCIMIN}{\rna}
\newcommand{\SLOWavroraCAPODynamicCIMAX}{\rna}
\newcommand{\SLOWavroraPIP}{\rna}
\newcommand{\SLOWavroraPIPCI}{\rna}
\newcommand{\SLOWavroraPIPCIMIN}{\rna}
\newcommand{\SLOWavroraPIPCIMAX}{\rna}
\newcommand{\SLOWavroraPIPDynamic}{\rna}
\newcommand{\SLOWavroraPIPDynamicCI}{\rna}
\newcommand{\SLOWavroraPIPDynamicCIMIN}{\rna}
\newcommand{\SLOWavroraPIPDynamicCIMAX}{\rna}
\newcommand{\SLOWavroraHBUP}{3}
\newcommand{\SLOWavroraHBUPCI}{0}
\newcommand{\SLOWavroraHBUPCIMIN}{3}
\newcommand{\SLOWavroraHBUPCIMAX}{3}
\newcommand{\SLOWavroraHBDynamicUP}{522,985}
\newcommand{\SLOWavroraHBDynamicUPCI}{1,518}
\newcommand{\SLOWavroraHBDynamicUPCIMIN}{521,467}
\newcommand{\SLOWavroraHBDynamicUPCIMAX}{524,503}
\newcommand{\SLOWavroraWCPUP}{3}
\newcommand{\SLOWavroraWCPUPCI}{0}
\newcommand{\SLOWavroraWCPUPCIMIN}{3}
\newcommand{\SLOWavroraWCPUPCIMAX}{3}
\newcommand{\SLOWavroraWCPDynamicUP}{552,479}
\newcommand{\SLOWavroraWCPDynamicUPCI}{1,207}
\newcommand{\SLOWavroraWCPDynamicUPCIMIN}{551,272}
\newcommand{\SLOWavroraWCPDynamicUPCIMAX}{553,686}
\newcommand{\SLOWavroraWDCnoGUP}{\rna}
\newcommand{\SLOWavroraWDCnoGUPCI}{\rna}
\newcommand{\SLOWavroraWDCnoGUPCIMIN}{\rna}
\newcommand{\SLOWavroraWDCnoGUPCIMAX}{\rna}
\newcommand{\SLOWavroraWDCnoGDynamicUP}{\rna}
\newcommand{\SLOWavroraWDCnoGDynamicUPCI}{\rna}
\newcommand{\SLOWavroraWDCnoGDynamicUPCIMIN}{\rna}
\newcommand{\SLOWavroraWDCnoGDynamicUPCIMAX}{\rna}
\newcommand{\SLOWavroraWDCUP}{\rna}
\newcommand{\SLOWavroraWDCUPCI}{\rna}
\newcommand{\SLOWavroraWDCUPCIMIN}{\rna}
\newcommand{\SLOWavroraWDCUPCIMAX}{\rna}
\newcommand{\SLOWavroraWDCDynamicUP}{\rna}
\newcommand{\SLOWavroraWDCDynamicUPCI}{\rna}
\newcommand{\SLOWavroraWDCDynamicUPCIMIN}{\rna}
\newcommand{\SLOWavroraWDCDynamicUPCIMAX}{\rna}
\newcommand{\SLOWavroraCAPOnoGUP}{\rna}
\newcommand{\SLOWavroraCAPOnoGUPCI}{\rna}
\newcommand{\SLOWavroraCAPOnoGUPCIMIN}{\rna}
\newcommand{\SLOWavroraCAPOnoGUPCIMAX}{\rna}
\newcommand{\SLOWavroraCAPOnoGDynamicUP}{\rna}
\newcommand{\SLOWavroraCAPOnoGDynamicUPCI}{\rna}
\newcommand{\SLOWavroraCAPOnoGDynamicUPCIMIN}{\rna}
\newcommand{\SLOWavroraCAPOnoGDynamicUPCIMAX}{\rna}
\newcommand{\SLOWavroraCAPOUP}{\rna}
\newcommand{\SLOWavroraCAPOUPCI}{\rna}
\newcommand{\SLOWavroraCAPOUPCIMIN}{\rna}
\newcommand{\SLOWavroraCAPOUPCIMAX}{\rna}
\newcommand{\SLOWavroraCAPODynamicUP}{\rna}
\newcommand{\SLOWavroraCAPODynamicUPCI}{\rna}
\newcommand{\SLOWavroraCAPODynamicUPCIMIN}{\rna}
\newcommand{\SLOWavroraCAPODynamicUPCIMAX}{\rna}
\newcommand{\SLOWavroraPIPUP}{\rna}
\newcommand{\SLOWavroraPIPUPCI}{\rna}
\newcommand{\SLOWavroraPIPUPCIMIN}{\rna}
\newcommand{\SLOWavroraPIPUPCIMAX}{\rna}
\newcommand{\SLOWavroraPIPDynamicUP}{\rna}
\newcommand{\SLOWavroraPIPDynamicUPCI}{\rna}
\newcommand{\SLOWavroraPIPDynamicUPCIMIN}{\rna}
\newcommand{\SLOWavroraPIPDynamicUPCIMAX}{\rna}
\newcommand{\SLOWavroraPIPHB}{\rna}
\newcommand{\SLOWavroraPIPHBCI}{\rna}
\newcommand{\SLOWavroraPIPHBCIMIN}{\rna}
\newcommand{\SLOWavroraPIPHBCIMAX}{\rna}
\newcommand{\SLOWavroraPIPHBDynamic}{\rna}
\newcommand{\SLOWavroraPIPHBDynamicCI}{\rna}
\newcommand{\SLOWavroraPIPHBDynamicCIMIN}{\rna}
\newcommand{\SLOWavroraPIPHBDynamicCIMAX}{\rna}
\newcommand{\SLOWavroraPIPWCP}{\rna}
\newcommand{\SLOWavroraPIPWCPCI}{\rna}
\newcommand{\SLOWavroraPIPWCPCIMIN}{\rna}
\newcommand{\SLOWavroraPIPWCPCIMAX}{\rna}
\newcommand{\SLOWavroraPIPWCPDynamic}{\rna}
\newcommand{\SLOWavroraPIPWCPDynamicCI}{\rna}
\newcommand{\SLOWavroraPIPWCPDynamicCIMIN}{\rna}
\newcommand{\SLOWavroraPIPWCPDynamicCIMAX}{\rna}
\newcommand{\SLOWavroraPIPWDC}{\rna}
\newcommand{\SLOWavroraPIPWDCCI}{\rna}
\newcommand{\SLOWavroraPIPWDCCIMIN}{\rna}
\newcommand{\SLOWavroraPIPWDCCIMAX}{\rna}
\newcommand{\SLOWavroraPIPWDCDynamic}{\rna}
\newcommand{\SLOWavroraPIPWDCDynamicCI}{\rna}
\newcommand{\SLOWavroraPIPWDCDynamicCIMIN}{\rna}
\newcommand{\SLOWavroraPIPWDCDynamicCIMAX}{\rna}
\newcommand{\SLOWavroraPIPCAPO}{\rna}
\newcommand{\SLOWavroraPIPCAPOCI}{\rna}
\newcommand{\SLOWavroraPIPCAPOCIMIN}{\rna}
\newcommand{\SLOWavroraPIPCAPOCIMAX}{\rna}
\newcommand{\SLOWavroraPIPCAPODynamic}{\rna}
\newcommand{\SLOWavroraPIPCAPODynamicCI}{\rna}
\newcommand{\SLOWavroraPIPCAPODynamicCIMIN}{\rna}
\newcommand{\SLOWavroraPIPCAPODynamicCIMAX}{\rna}
\newcommand{\SLOWavroraPIPPIP}{\rna}
\newcommand{\SLOWavroraPIPPIPCI}{\rna}
\newcommand{\SLOWavroraPIPPIPCIMIN}{\rna}
\newcommand{\SLOWavroraPIPPIPCIMAX}{\rna}
\newcommand{\SLOWavroraPIPPIPDynamic}{\rna}
\newcommand{\SLOWavroraPIPPIPDynamicCI}{\rna}
\newcommand{\SLOWavroraPIPPIPDynamicCIMIN}{\rna}
\newcommand{\SLOWavroraPIPPIPDynamicCIMAX}{\rna}
\newcommand{\SLOWbatikEvents}{160}
\newcommand{\SLOWbatikNoFPEvents}{11}
\newcommand{\SLOWbatikMaxLiveThreads}{7}
\newcommand{\SLOWbatikTotalThreads}{7}
\newcommand{\SLOWbatikBaseTime}{2.6}
\newcommand{\SLOWbatikBaseTimeCI}{50}
\newcommand{\SLOWbatikEmptyTime}{\rna}
\newcommand{\SLOWbatikEmptyTimeCI}{\rna}
\newcommand{\SLOWbatikEmptyTimeCIMIN}{\rna}
\newcommand{\SLOWbatikEmptyTimeCIMAX}{\rna}
\newcommand{\SLOWbatikFTTime}{\rna}
\newcommand{\SLOWbatikFTTimeCI}{\rna}
\newcommand{\SLOWbatikFTTimeCIMIN}{\rna}
\newcommand{\SLOWbatikFTTimeCIMAX}{\rna}
\newcommand{\SLOWbatikHBTime}{7.9}
\newcommand{\SLOWbatikHBTimeCI}{0.15}
\newcommand{\SLOWbatikWCPTime}{12}
\newcommand{\SLOWbatikWCPTimeCI}{0.27}
\newcommand{\SLOWbatikDCnoGExcTime}{10}
\newcommand{\SLOWbatikDCnoGExcTimeCI}{0.28}
\newcommand{\SLOWbatikDCnoGTime}{\rna}
\newcommand{\SLOWbatikDCnoGTimeCI}{\rna}
\newcommand{\SLOWbatikDCnoGTimeCIMIN}{\rna}
\newcommand{\SLOWbatikDCnoGTimeCIMAX}{\rna}
\newcommand{\SLOWbatikDCExcTime}{12}
\newcommand{\SLOWbatikDCExcTimeCI}{0.12}
\newcommand{\SLOWbatikDCTime}{\rna}
\newcommand{\SLOWbatikDCTimeCI}{\rna}
\newcommand{\SLOWbatikDCTimeCIMIN}{\rna}
\newcommand{\SLOWbatikDCTimeCIMAX}{\rna}
\newcommand{\SLOWbatikCAPOnoGExcTime}{10}
\newcommand{\SLOWbatikCAPOnoGExcTimeCI}{0.27}
\newcommand{\SLOWbatikCAPOnoGTime}{\rna}
\newcommand{\SLOWbatikCAPOnoGTimeCI}{\rna}
\newcommand{\SLOWbatikCAPOnoGTimeCIMIN}{\rna}
\newcommand{\SLOWbatikCAPOnoGTimeCIMAX}{\rna}
\newcommand{\SLOWbatikCAPOExcTime}{11}
\newcommand{\SLOWbatikCAPOExcTimeCI}{0.22}
\newcommand{\SLOWbatikCAPOTime}{\rna}
\newcommand{\SLOWbatikCAPOTimeCI}{\rna}
\newcommand{\SLOWbatikCAPOTimeCIMIN}{\rna}
\newcommand{\SLOWbatikCAPOTimeCIMAX}{\rna}
\newcommand{\SLOWbatikStaticTime}{\rzero}
\newcommand{\SLOWbatikDynamicTime}{\rzero}
\newcommand{\SLOWbatikBaseMem}{220}
\newcommand{\SLOWbatikBaseMemCI}{6.3}
\newcommand{\SLOWbatikHBMem}{30}
\newcommand{\SLOWbatikHBMemCI}{0.82}
\newcommand{\SLOWbatikFTMem}{\memna}
\newcommand{\SLOWbatikFTMemCI}{\memna}
\newcommand{\SLOWbatikFTMemCIMIN}{\memna}
\newcommand{\SLOWbatikFTMemCIMAX}{\memna}
\newcommand{\SLOWbatikWCPMem}{50}
\newcommand{\SLOWbatikWCPMemCI}{1.8}
\newcommand{\SLOWbatikDCnoGExcMem}{43}
\newcommand{\SLOWbatikDCnoGExcMemCI}{3.3}
\newcommand{\SLOWbatikDCnoGMem}{\memna}
\newcommand{\SLOWbatikDCnoGMemCI}{\memna}
\newcommand{\SLOWbatikDCnoGMemCIMIN}{\memna}
\newcommand{\SLOWbatikDCnoGMemCIMAX}{\memna}
\newcommand{\SLOWbatikDCExcMem}{46}
\newcommand{\SLOWbatikDCExcMemCI}{1.6}
\newcommand{\SLOWbatikDCMem}{\memna}
\newcommand{\SLOWbatikDCMemCI}{\memna}
\newcommand{\SLOWbatikDCMemCIMIN}{\memna}
\newcommand{\SLOWbatikDCMemCIMAX}{\memna}
\newcommand{\SLOWbatikCAPOnoGExcMem}{41}
\newcommand{\SLOWbatikCAPOnoGExcMemCI}{3.0}
\newcommand{\SLOWbatikCAPOnoGMem}{\memna}
\newcommand{\SLOWbatikCAPOnoGMemCI}{\memna}
\newcommand{\SLOWbatikCAPOnoGMemCIMIN}{\memna}
\newcommand{\SLOWbatikCAPOnoGMemCIMAX}{\memna}
\newcommand{\SLOWbatikCAPOExcMem}{44}
\newcommand{\SLOWbatikCAPOExcMemCI}{1.6}
\newcommand{\SLOWbatikCAPOMem}{\memna}
\newcommand{\SLOWbatikCAPOMemCI}{\memna}
\newcommand{\SLOWbatikCAPOMemCIMIN}{\memna}
\newcommand{\SLOWbatikCAPOMemCIMAX}{\memna}
\newcommand{\SLOWbatikEventsCI}{0}
\newcommand{\SLOWbatikEventsCIMIN}{155,322,726}
\newcommand{\SLOWbatikEventsCIMAX}{155,322,726}
\newcommand{\SLOWbatikNoFPEventsCI}{0}
\newcommand{\SLOWbatikNoFPEventsCIMIN}{11,422,601}
\newcommand{\SLOWbatikNoFPEventsCIMAX}{11,422,601}
\newcommand{\SLOWbatikHB}{0}
\newcommand{\SLOWbatikHBCI}{0}
\newcommand{\SLOWbatikHBCIMIN}{0}
\newcommand{\SLOWbatikHBCIMAX}{0}
\newcommand{\SLOWbatikHBDynamic}{0}
\newcommand{\SLOWbatikHBDynamicCI}{0}
\newcommand{\SLOWbatikHBDynamicCIMIN}{0}
\newcommand{\SLOWbatikHBDynamicCIMAX}{0}
\newcommand{\SLOWbatikFT}{\rna}
\newcommand{\SLOWbatikFTCI}{\rna}
\newcommand{\SLOWbatikFTCIMIN}{\rna}
\newcommand{\SLOWbatikFTCIMAX}{\rna}
\newcommand{\SLOWbatikFTDynamic}{\rna}
\newcommand{\SLOWbatikFTDynamicCI}{\rna}
\newcommand{\SLOWbatikFTDynamicCIMIN}{\rna}
\newcommand{\SLOWbatikFTDynamicCIMAX}{\rna}
\newcommand{\SLOWbatikWCP}{0}
\newcommand{\SLOWbatikWCPCI}{0}
\newcommand{\SLOWbatikWCPCIMIN}{0}
\newcommand{\SLOWbatikWCPCIMAX}{0}
\newcommand{\SLOWbatikWCPDynamic}{0}
\newcommand{\SLOWbatikWCPDynamicCI}{0}
\newcommand{\SLOWbatikWCPDynamicCIMIN}{0}
\newcommand{\SLOWbatikWCPDynamicCIMAX}{0}
\newcommand{\SLOWbatikDCnoGExc}{0}
\newcommand{\SLOWbatikDCnoGExcCI}{0}
\newcommand{\SLOWbatikDCnoGExcCIMIN}{0}
\newcommand{\SLOWbatikDCnoGExcCIMAX}{0}
\newcommand{\SLOWbatikDCnoGExcDynamic}{0}
\newcommand{\SLOWbatikDCnoGExcDynamicCI}{0}
\newcommand{\SLOWbatikDCnoGExcDynamicCIMIN}{0}
\newcommand{\SLOWbatikDCnoGExcDynamicCIMAX}{0}
\newcommand{\SLOWbatikDCnoG}{\rna}
\newcommand{\SLOWbatikDCnoGCI}{\rna}
\newcommand{\SLOWbatikDCnoGCIMIN}{\rna}
\newcommand{\SLOWbatikDCnoGCIMAX}{\rna}
\newcommand{\SLOWbatikDCnoGDynamic}{\rna}
\newcommand{\SLOWbatikDCnoGDynamicCI}{\rna}
\newcommand{\SLOWbatikDCnoGDynamicCIMIN}{\rna}
\newcommand{\SLOWbatikDCnoGDynamicCIMAX}{\rna}
\newcommand{\SLOWbatikDCExc}{0}
\newcommand{\SLOWbatikDCExcCI}{0}
\newcommand{\SLOWbatikDCExcCIMIN}{0}
\newcommand{\SLOWbatikDCExcCIMAX}{0}
\newcommand{\SLOWbatikDCExcDynamic}{0}
\newcommand{\SLOWbatikDCExcDynamicCI}{0}
\newcommand{\SLOWbatikDCExcDynamicCIMIN}{0}
\newcommand{\SLOWbatikDCExcDynamicCIMAX}{0}
\newcommand{\SLOWbatikDC}{\rna}
\newcommand{\SLOWbatikDCCI}{\rna}
\newcommand{\SLOWbatikDCCIMIN}{\rna}
\newcommand{\SLOWbatikDCCIMAX}{\rna}
\newcommand{\SLOWbatikDCDynamic}{\rna}
\newcommand{\SLOWbatikDCDynamicCI}{\rna}
\newcommand{\SLOWbatikDCDynamicCIMIN}{\rna}
\newcommand{\SLOWbatikDCDynamicCIMAX}{\rna}
\newcommand{\SLOWbatikCAPOnoGExc}{0}
\newcommand{\SLOWbatikCAPOnoGExcCI}{0}
\newcommand{\SLOWbatikCAPOnoGExcCIMIN}{0}
\newcommand{\SLOWbatikCAPOnoGExcCIMAX}{0}
\newcommand{\SLOWbatikCAPOnoGExcDynamic}{0}
\newcommand{\SLOWbatikCAPOnoGExcDynamicCI}{0}
\newcommand{\SLOWbatikCAPOnoGExcDynamicCIMIN}{0}
\newcommand{\SLOWbatikCAPOnoGExcDynamicCIMAX}{0}
\newcommand{\SLOWbatikCAPOnoG}{\rna}
\newcommand{\SLOWbatikCAPOnoGCI}{\rna}
\newcommand{\SLOWbatikCAPOnoGCIMIN}{\rna}
\newcommand{\SLOWbatikCAPOnoGCIMAX}{\rna}
\newcommand{\SLOWbatikCAPOnoGDynamic}{\rna}
\newcommand{\SLOWbatikCAPOnoGDynamicCI}{\rna}
\newcommand{\SLOWbatikCAPOnoGDynamicCIMIN}{\rna}
\newcommand{\SLOWbatikCAPOnoGDynamicCIMAX}{\rna}
\newcommand{\SLOWbatikCAPOExc}{0}
\newcommand{\SLOWbatikCAPOExcCI}{0}
\newcommand{\SLOWbatikCAPOExcCIMIN}{0}
\newcommand{\SLOWbatikCAPOExcCIMAX}{0}
\newcommand{\SLOWbatikCAPOExcDynamic}{0}
\newcommand{\SLOWbatikCAPOExcDynamicCI}{0}
\newcommand{\SLOWbatikCAPOExcDynamicCIMIN}{0}
\newcommand{\SLOWbatikCAPOExcDynamicCIMAX}{0}
\newcommand{\SLOWbatikCAPO}{\rna}
\newcommand{\SLOWbatikCAPOCI}{\rna}
\newcommand{\SLOWbatikCAPOCIMIN}{\rna}
\newcommand{\SLOWbatikCAPOCIMAX}{\rna}
\newcommand{\SLOWbatikCAPODynamic}{\rna}
\newcommand{\SLOWbatikCAPODynamicCI}{\rna}
\newcommand{\SLOWbatikCAPODynamicCIMIN}{\rna}
\newcommand{\SLOWbatikCAPODynamicCIMAX}{\rna}
\newcommand{\SLOWbatikPIP}{\rna}
\newcommand{\SLOWbatikPIPCI}{\rna}
\newcommand{\SLOWbatikPIPCIMIN}{\rna}
\newcommand{\SLOWbatikPIPCIMAX}{\rna}
\newcommand{\SLOWbatikPIPDynamic}{\rna}
\newcommand{\SLOWbatikPIPDynamicCI}{\rna}
\newcommand{\SLOWbatikPIPDynamicCIMIN}{\rna}
\newcommand{\SLOWbatikPIPDynamicCIMAX}{\rna}
\newcommand{\SLOWbatikHBUP}{0}
\newcommand{\SLOWbatikHBUPCI}{0}
\newcommand{\SLOWbatikHBUPCIMIN}{0}
\newcommand{\SLOWbatikHBUPCIMAX}{0}
\newcommand{\SLOWbatikHBDynamicUP}{0}
\newcommand{\SLOWbatikHBDynamicUPCI}{0}
\newcommand{\SLOWbatikHBDynamicUPCIMIN}{0}
\newcommand{\SLOWbatikHBDynamicUPCIMAX}{0}
\newcommand{\SLOWbatikWCPUP}{0}
\newcommand{\SLOWbatikWCPUPCI}{0}
\newcommand{\SLOWbatikWCPUPCIMIN}{0}
\newcommand{\SLOWbatikWCPUPCIMAX}{0}
\newcommand{\SLOWbatikWCPDynamicUP}{0}
\newcommand{\SLOWbatikWCPDynamicUPCI}{0}
\newcommand{\SLOWbatikWCPDynamicUPCIMIN}{0}
\newcommand{\SLOWbatikWCPDynamicUPCIMAX}{0}
\newcommand{\SLOWbatikWDCnoGUP}{\rna}
\newcommand{\SLOWbatikWDCnoGUPCI}{\rna}
\newcommand{\SLOWbatikWDCnoGUPCIMIN}{\rna}
\newcommand{\SLOWbatikWDCnoGUPCIMAX}{\rna}
\newcommand{\SLOWbatikWDCnoGDynamicUP}{\rna}
\newcommand{\SLOWbatikWDCnoGDynamicUPCI}{\rna}
\newcommand{\SLOWbatikWDCnoGDynamicUPCIMIN}{\rna}
\newcommand{\SLOWbatikWDCnoGDynamicUPCIMAX}{\rna}
\newcommand{\SLOWbatikWDCUP}{\rna}
\newcommand{\SLOWbatikWDCUPCI}{\rna}
\newcommand{\SLOWbatikWDCUPCIMIN}{\rna}
\newcommand{\SLOWbatikWDCUPCIMAX}{\rna}
\newcommand{\SLOWbatikWDCDynamicUP}{\rna}
\newcommand{\SLOWbatikWDCDynamicUPCI}{\rna}
\newcommand{\SLOWbatikWDCDynamicUPCIMIN}{\rna}
\newcommand{\SLOWbatikWDCDynamicUPCIMAX}{\rna}
\newcommand{\SLOWbatikCAPOnoGUP}{\rna}
\newcommand{\SLOWbatikCAPOnoGUPCI}{\rna}
\newcommand{\SLOWbatikCAPOnoGUPCIMIN}{\rna}
\newcommand{\SLOWbatikCAPOnoGUPCIMAX}{\rna}
\newcommand{\SLOWbatikCAPOnoGDynamicUP}{\rna}
\newcommand{\SLOWbatikCAPOnoGDynamicUPCI}{\rna}
\newcommand{\SLOWbatikCAPOnoGDynamicUPCIMIN}{\rna}
\newcommand{\SLOWbatikCAPOnoGDynamicUPCIMAX}{\rna}
\newcommand{\SLOWbatikCAPOUP}{\rna}
\newcommand{\SLOWbatikCAPOUPCI}{\rna}
\newcommand{\SLOWbatikCAPOUPCIMIN}{\rna}
\newcommand{\SLOWbatikCAPOUPCIMAX}{\rna}
\newcommand{\SLOWbatikCAPODynamicUP}{\rna}
\newcommand{\SLOWbatikCAPODynamicUPCI}{\rna}
\newcommand{\SLOWbatikCAPODynamicUPCIMIN}{\rna}
\newcommand{\SLOWbatikCAPODynamicUPCIMAX}{\rna}
\newcommand{\SLOWbatikPIPUP}{\rna}
\newcommand{\SLOWbatikPIPUPCI}{\rna}
\newcommand{\SLOWbatikPIPUPCIMIN}{\rna}
\newcommand{\SLOWbatikPIPUPCIMAX}{\rna}
\newcommand{\SLOWbatikPIPDynamicUP}{\rna}
\newcommand{\SLOWbatikPIPDynamicUPCI}{\rna}
\newcommand{\SLOWbatikPIPDynamicUPCIMIN}{\rna}
\newcommand{\SLOWbatikPIPDynamicUPCIMAX}{\rna}
\newcommand{\SLOWbatikPIPHB}{\rna}
\newcommand{\SLOWbatikPIPHBCI}{\rna}
\newcommand{\SLOWbatikPIPHBCIMIN}{\rna}
\newcommand{\SLOWbatikPIPHBCIMAX}{\rna}
\newcommand{\SLOWbatikPIPHBDynamic}{\rna}
\newcommand{\SLOWbatikPIPHBDynamicCI}{\rna}
\newcommand{\SLOWbatikPIPHBDynamicCIMIN}{\rna}
\newcommand{\SLOWbatikPIPHBDynamicCIMAX}{\rna}
\newcommand{\SLOWbatikPIPWCP}{\rna}
\newcommand{\SLOWbatikPIPWCPCI}{\rna}
\newcommand{\SLOWbatikPIPWCPCIMIN}{\rna}
\newcommand{\SLOWbatikPIPWCPCIMAX}{\rna}
\newcommand{\SLOWbatikPIPWCPDynamic}{\rna}
\newcommand{\SLOWbatikPIPWCPDynamicCI}{\rna}
\newcommand{\SLOWbatikPIPWCPDynamicCIMIN}{\rna}
\newcommand{\SLOWbatikPIPWCPDynamicCIMAX}{\rna}
\newcommand{\SLOWbatikPIPWDC}{\rna}
\newcommand{\SLOWbatikPIPWDCCI}{\rna}
\newcommand{\SLOWbatikPIPWDCCIMIN}{\rna}
\newcommand{\SLOWbatikPIPWDCCIMAX}{\rna}
\newcommand{\SLOWbatikPIPWDCDynamic}{\rna}
\newcommand{\SLOWbatikPIPWDCDynamicCI}{\rna}
\newcommand{\SLOWbatikPIPWDCDynamicCIMIN}{\rna}
\newcommand{\SLOWbatikPIPWDCDynamicCIMAX}{\rna}
\newcommand{\SLOWbatikPIPCAPO}{\rna}
\newcommand{\SLOWbatikPIPCAPOCI}{\rna}
\newcommand{\SLOWbatikPIPCAPOCIMIN}{\rna}
\newcommand{\SLOWbatikPIPCAPOCIMAX}{\rna}
\newcommand{\SLOWbatikPIPCAPODynamic}{\rna}
\newcommand{\SLOWbatikPIPCAPODynamicCI}{\rna}
\newcommand{\SLOWbatikPIPCAPODynamicCIMIN}{\rna}
\newcommand{\SLOWbatikPIPCAPODynamicCIMAX}{\rna}
\newcommand{\SLOWbatikPIPPIP}{\rna}
\newcommand{\SLOWbatikPIPPIPCI}{\rna}
\newcommand{\SLOWbatikPIPPIPCIMIN}{\rna}
\newcommand{\SLOWbatikPIPPIPCIMAX}{\rna}
\newcommand{\SLOWbatikPIPPIPDynamic}{\rna}
\newcommand{\SLOWbatikPIPPIPDynamicCI}{\rna}
\newcommand{\SLOWbatikPIPPIPDynamicCIMIN}{\rna}
\newcommand{\SLOWbatikPIPPIPDynamicCIMAX}{\rna}
\newcommand{\SLOWhtwoEvents}{3,800}
\newcommand{\SLOWhtwoNoFPEvents}{460}
\newcommand{\SLOWhtwoMaxLiveThreads}{9}
\newcommand{\SLOWhtwoTotalThreads}{10}
\newcommand{\SLOWhtwoBaseTime}{4.9}
\newcommand{\SLOWhtwoBaseTimeCI}{94}
\newcommand{\SLOWhtwoEmptyTime}{\rna}
\newcommand{\SLOWhtwoEmptyTimeCI}{\rna}
\newcommand{\SLOWhtwoEmptyTimeCIMIN}{\rna}
\newcommand{\SLOWhtwoEmptyTimeCIMAX}{\rna}
\newcommand{\SLOWhtwoFTTime}{\rna}
\newcommand{\SLOWhtwoFTTimeCI}{\rna}
\newcommand{\SLOWhtwoFTTimeCIMIN}{\rna}
\newcommand{\SLOWhtwoFTTimeCIMAX}{\rna}
\newcommand{\SLOWhtwoHBTime}{29}
\newcommand{\SLOWhtwoHBTimeCI}{2.1}
\newcommand{\SLOWhtwoWCPTime}{82}
\newcommand{\SLOWhtwoWCPTimeCI}{7.8}
\newcommand{\SLOWhtwoDCnoGExcTime}{75}
\newcommand{\SLOWhtwoDCnoGExcTimeCI}{5.2}
\newcommand{\SLOWhtwoDCnoGTime}{\rna}
\newcommand{\SLOWhtwoDCnoGTimeCI}{\rna}
\newcommand{\SLOWhtwoDCnoGTimeCIMIN}{\rna}
\newcommand{\SLOWhtwoDCnoGTimeCIMAX}{\rna}
\newcommand{\SLOWhtwoDCExcTime}{77}
\newcommand{\SLOWhtwoDCExcTimeCI}{3.9}
\newcommand{\SLOWhtwoDCTime}{\rna}
\newcommand{\SLOWhtwoDCTimeCI}{\rna}
\newcommand{\SLOWhtwoDCTimeCIMIN}{\rna}
\newcommand{\SLOWhtwoDCTimeCIMAX}{\rna}
\newcommand{\SLOWhtwoCAPOnoGExcTime}{70}
\newcommand{\SLOWhtwoCAPOnoGExcTimeCI}{2.9}
\newcommand{\SLOWhtwoCAPOnoGTime}{\rna}
\newcommand{\SLOWhtwoCAPOnoGTimeCI}{\rna}
\newcommand{\SLOWhtwoCAPOnoGTimeCIMIN}{\rna}
\newcommand{\SLOWhtwoCAPOnoGTimeCIMAX}{\rna}
\newcommand{\SLOWhtwoCAPOExcTime}{77}
\newcommand{\SLOWhtwoCAPOExcTimeCI}{2.9}
\newcommand{\SLOWhtwoCAPOTime}{\rna}
\newcommand{\SLOWhtwoCAPOTimeCI}{\rna}
\newcommand{\SLOWhtwoCAPOTimeCIMIN}{\rna}
\newcommand{\SLOWhtwoCAPOTimeCIMAX}{\rna}
\newcommand{\SLOWhtwoStaticTime}{\rzero}
\newcommand{\SLOWhtwoDynamicTime}{\rzero}
\newcommand{\SLOWhtwoBaseMem}{1,800}
\newcommand{\SLOWhtwoBaseMemCI}{41.0}
\newcommand{\SLOWhtwoHBMem}{16}
\newcommand{\SLOWhtwoHBMemCI}{0.66}
\newcommand{\SLOWhtwoFTMem}{\memna}
\newcommand{\SLOWhtwoFTMemCI}{\memna}
\newcommand{\SLOWhtwoFTMemCIMIN}{\memna}
\newcommand{\SLOWhtwoFTMemCIMAX}{\memna}
\newcommand{\SLOWhtwoWCPMem}{62}
\newcommand{\SLOWhtwoWCPMemCI}{2.1}
\newcommand{\SLOWhtwoDCnoGExcMem}{56}
\newcommand{\SLOWhtwoDCnoGExcMemCI}{2.4}
\newcommand{\SLOWhtwoDCnoGMem}{\memna}
\newcommand{\SLOWhtwoDCnoGMemCI}{\memna}
\newcommand{\SLOWhtwoDCnoGMemCIMIN}{\memna}
\newcommand{\SLOWhtwoDCnoGMemCIMAX}{\memna}
\newcommand{\SLOWhtwoDCExcMem}{57}
\newcommand{\SLOWhtwoDCExcMemCI}{1.6}
\newcommand{\SLOWhtwoDCMem}{\memna}
\newcommand{\SLOWhtwoDCMemCI}{\memna}
\newcommand{\SLOWhtwoDCMemCIMIN}{\memna}
\newcommand{\SLOWhtwoDCMemCIMAX}{\memna}
\newcommand{\SLOWhtwoCAPOnoGExcMem}{56}
\newcommand{\SLOWhtwoCAPOnoGExcMemCI}{1.3}
\newcommand{\SLOWhtwoCAPOnoGMem}{\memna}
\newcommand{\SLOWhtwoCAPOnoGMemCI}{\memna}
\newcommand{\SLOWhtwoCAPOnoGMemCIMIN}{\memna}
\newcommand{\SLOWhtwoCAPOnoGMemCIMAX}{\memna}
\newcommand{\SLOWhtwoCAPOExcMem}{57}
\newcommand{\SLOWhtwoCAPOExcMemCI}{1.9}
\newcommand{\SLOWhtwoCAPOMem}{\memna}
\newcommand{\SLOWhtwoCAPOMemCI}{\memna}
\newcommand{\SLOWhtwoCAPOMemCIMIN}{\memna}
\newcommand{\SLOWhtwoCAPOMemCIMAX}{\memna}
\newcommand{\SLOWhtwoEventsCI}{1,258,655}
\newcommand{\SLOWhtwoEventsCIMIN}{3,801,406,538}
\newcommand{\SLOWhtwoEventsCIMAX}{3,803,923,848}
\newcommand{\SLOWhtwoNoFPEventsCI}{41,806}
\newcommand{\SLOWhtwoNoFPEventsCIMIN}{457,193,918}
\newcommand{\SLOWhtwoNoFPEventsCIMAX}{457,277,530}
\newcommand{\SLOWhtwoHB}{13}
\newcommand{\SLOWhtwoHBCI}{0}
\newcommand{\SLOWhtwoHBCIMIN}{13}
\newcommand{\SLOWhtwoHBCIMAX}{13}
\newcommand{\SLOWhtwoHBDynamic}{68,137}
\newcommand{\SLOWhtwoHBDynamicCI}{188}
\newcommand{\SLOWhtwoHBDynamicCIMIN}{67,949}
\newcommand{\SLOWhtwoHBDynamicCIMAX}{68,325}
\newcommand{\SLOWhtwoFT}{\rna}
\newcommand{\SLOWhtwoFTCI}{\rna}
\newcommand{\SLOWhtwoFTCIMIN}{\rna}
\newcommand{\SLOWhtwoFTCIMAX}{\rna}
\newcommand{\SLOWhtwoFTDynamic}{\rna}
\newcommand{\SLOWhtwoFTDynamicCI}{\rna}
\newcommand{\SLOWhtwoFTDynamicCIMIN}{\rna}
\newcommand{\SLOWhtwoFTDynamicCIMAX}{\rna}
\newcommand{\SLOWhtwoWCP}{13}
\newcommand{\SLOWhtwoWCPCI}{0}
\newcommand{\SLOWhtwoWCPCIMIN}{13}
\newcommand{\SLOWhtwoWCPCIMAX}{13}
\newcommand{\SLOWhtwoWCPDynamic}{67,946}
\newcommand{\SLOWhtwoWCPDynamicCI}{76}
\newcommand{\SLOWhtwoWCPDynamicCIMIN}{67,870}
\newcommand{\SLOWhtwoWCPDynamicCIMAX}{68,022}
\newcommand{\SLOWhtwoDCnoGExc}{13}
\newcommand{\SLOWhtwoDCnoGExcCI}{0}
\newcommand{\SLOWhtwoDCnoGExcCIMIN}{13}
\newcommand{\SLOWhtwoDCnoGExcCIMAX}{13}
\newcommand{\SLOWhtwoDCnoGExcDynamic}{68,694}
\newcommand{\SLOWhtwoDCnoGExcDynamicCI}{247}
\newcommand{\SLOWhtwoDCnoGExcDynamicCIMIN}{68,447}
\newcommand{\SLOWhtwoDCnoGExcDynamicCIMAX}{68,941}
\newcommand{\SLOWhtwoDCnoG}{\rna}
\newcommand{\SLOWhtwoDCnoGCI}{\rna}
\newcommand{\SLOWhtwoDCnoGCIMIN}{\rna}
\newcommand{\SLOWhtwoDCnoGCIMAX}{\rna}
\newcommand{\SLOWhtwoDCnoGDynamic}{\rna}
\newcommand{\SLOWhtwoDCnoGDynamicCI}{\rna}
\newcommand{\SLOWhtwoDCnoGDynamicCIMIN}{\rna}
\newcommand{\SLOWhtwoDCnoGDynamicCIMAX}{\rna}
\newcommand{\SLOWhtwoDCExc}{13}
\newcommand{\SLOWhtwoDCExcCI}{0}
\newcommand{\SLOWhtwoDCExcCIMIN}{13}
\newcommand{\SLOWhtwoDCExcCIMAX}{13}
\newcommand{\SLOWhtwoDCExcDynamic}{41,383}
\newcommand{\SLOWhtwoDCExcDynamicCI}{125}
\newcommand{\SLOWhtwoDCExcDynamicCIMIN}{41,258}
\newcommand{\SLOWhtwoDCExcDynamicCIMAX}{41,508}
\newcommand{\SLOWhtwoDC}{\rna}
\newcommand{\SLOWhtwoDCCI}{\rna}
\newcommand{\SLOWhtwoDCCIMIN}{\rna}
\newcommand{\SLOWhtwoDCCIMAX}{\rna}
\newcommand{\SLOWhtwoDCDynamic}{\rna}
\newcommand{\SLOWhtwoDCDynamicCI}{\rna}
\newcommand{\SLOWhtwoDCDynamicCIMIN}{\rna}
\newcommand{\SLOWhtwoDCDynamicCIMAX}{\rna}
\newcommand{\SLOWhtwoCAPOnoGExc}{13}
\newcommand{\SLOWhtwoCAPOnoGExcCI}{0}
\newcommand{\SLOWhtwoCAPOnoGExcCIMIN}{13}
\newcommand{\SLOWhtwoCAPOnoGExcCIMAX}{13}
\newcommand{\SLOWhtwoCAPOnoGExcDynamic}{68,774}
\newcommand{\SLOWhtwoCAPOnoGExcDynamicCI}{208}
\newcommand{\SLOWhtwoCAPOnoGExcDynamicCIMIN}{68,566}
\newcommand{\SLOWhtwoCAPOnoGExcDynamicCIMAX}{68,982}
\newcommand{\SLOWhtwoCAPOnoG}{\rna}
\newcommand{\SLOWhtwoCAPOnoGCI}{\rna}
\newcommand{\SLOWhtwoCAPOnoGCIMIN}{\rna}
\newcommand{\SLOWhtwoCAPOnoGCIMAX}{\rna}
\newcommand{\SLOWhtwoCAPOnoGDynamic}{\rna}
\newcommand{\SLOWhtwoCAPOnoGDynamicCI}{\rna}
\newcommand{\SLOWhtwoCAPOnoGDynamicCIMIN}{\rna}
\newcommand{\SLOWhtwoCAPOnoGDynamicCIMAX}{\rna}
\newcommand{\SLOWhtwoCAPOExc}{13}
\newcommand{\SLOWhtwoCAPOExcCI}{0}
\newcommand{\SLOWhtwoCAPOExcCIMIN}{13}
\newcommand{\SLOWhtwoCAPOExcCIMAX}{13}
\newcommand{\SLOWhtwoCAPOExcDynamic}{41,426}
\newcommand{\SLOWhtwoCAPOExcDynamicCI}{158}
\newcommand{\SLOWhtwoCAPOExcDynamicCIMIN}{41,268}
\newcommand{\SLOWhtwoCAPOExcDynamicCIMAX}{41,584}
\newcommand{\SLOWhtwoCAPO}{\rna}
\newcommand{\SLOWhtwoCAPOCI}{\rna}
\newcommand{\SLOWhtwoCAPOCIMIN}{\rna}
\newcommand{\SLOWhtwoCAPOCIMAX}{\rna}
\newcommand{\SLOWhtwoCAPODynamic}{\rna}
\newcommand{\SLOWhtwoCAPODynamicCI}{\rna}
\newcommand{\SLOWhtwoCAPODynamicCIMIN}{\rna}
\newcommand{\SLOWhtwoCAPODynamicCIMAX}{\rna}
\newcommand{\SLOWhtwoPIP}{\rna}
\newcommand{\SLOWhtwoPIPCI}{\rna}
\newcommand{\SLOWhtwoPIPCIMIN}{\rna}
\newcommand{\SLOWhtwoPIPCIMAX}{\rna}
\newcommand{\SLOWhtwoPIPDynamic}{\rna}
\newcommand{\SLOWhtwoPIPDynamicCI}{\rna}
\newcommand{\SLOWhtwoPIPDynamicCIMIN}{\rna}
\newcommand{\SLOWhtwoPIPDynamicCIMAX}{\rna}
\newcommand{\SLOWhtwoHBUP}{9}
\newcommand{\SLOWhtwoHBUPCI}{0}
\newcommand{\SLOWhtwoHBUPCIMIN}{9}
\newcommand{\SLOWhtwoHBUPCIMAX}{9}
\newcommand{\SLOWhtwoHBDynamicUP}{68,137}
\newcommand{\SLOWhtwoHBDynamicUPCI}{188}
\newcommand{\SLOWhtwoHBDynamicUPCIMIN}{67,949}
\newcommand{\SLOWhtwoHBDynamicUPCIMAX}{68,325}
\newcommand{\SLOWhtwoWCPUP}{9}
\newcommand{\SLOWhtwoWCPUPCI}{0}
\newcommand{\SLOWhtwoWCPUPCIMIN}{9}
\newcommand{\SLOWhtwoWCPUPCIMAX}{9}
\newcommand{\SLOWhtwoWCPDynamicUP}{67,946}
\newcommand{\SLOWhtwoWCPDynamicUPCI}{76}
\newcommand{\SLOWhtwoWCPDynamicUPCIMIN}{67,870}
\newcommand{\SLOWhtwoWCPDynamicUPCIMAX}{68,022}
\newcommand{\SLOWhtwoWDCnoGUP}{\rna}
\newcommand{\SLOWhtwoWDCnoGUPCI}{\rna}
\newcommand{\SLOWhtwoWDCnoGUPCIMIN}{\rna}
\newcommand{\SLOWhtwoWDCnoGUPCIMAX}{\rna}
\newcommand{\SLOWhtwoWDCnoGDynamicUP}{\rna}
\newcommand{\SLOWhtwoWDCnoGDynamicUPCI}{\rna}
\newcommand{\SLOWhtwoWDCnoGDynamicUPCIMIN}{\rna}
\newcommand{\SLOWhtwoWDCnoGDynamicUPCIMAX}{\rna}
\newcommand{\SLOWhtwoWDCUP}{\rna}
\newcommand{\SLOWhtwoWDCUPCI}{\rna}
\newcommand{\SLOWhtwoWDCUPCIMIN}{\rna}
\newcommand{\SLOWhtwoWDCUPCIMAX}{\rna}
\newcommand{\SLOWhtwoWDCDynamicUP}{\rna}
\newcommand{\SLOWhtwoWDCDynamicUPCI}{\rna}
\newcommand{\SLOWhtwoWDCDynamicUPCIMIN}{\rna}
\newcommand{\SLOWhtwoWDCDynamicUPCIMAX}{\rna}
\newcommand{\SLOWhtwoCAPOnoGUP}{\rna}
\newcommand{\SLOWhtwoCAPOnoGUPCI}{\rna}
\newcommand{\SLOWhtwoCAPOnoGUPCIMIN}{\rna}
\newcommand{\SLOWhtwoCAPOnoGUPCIMAX}{\rna}
\newcommand{\SLOWhtwoCAPOnoGDynamicUP}{\rna}
\newcommand{\SLOWhtwoCAPOnoGDynamicUPCI}{\rna}
\newcommand{\SLOWhtwoCAPOnoGDynamicUPCIMIN}{\rna}
\newcommand{\SLOWhtwoCAPOnoGDynamicUPCIMAX}{\rna}
\newcommand{\SLOWhtwoCAPOUP}{\rna}
\newcommand{\SLOWhtwoCAPOUPCI}{\rna}
\newcommand{\SLOWhtwoCAPOUPCIMIN}{\rna}
\newcommand{\SLOWhtwoCAPOUPCIMAX}{\rna}
\newcommand{\SLOWhtwoCAPODynamicUP}{\rna}
\newcommand{\SLOWhtwoCAPODynamicUPCI}{\rna}
\newcommand{\SLOWhtwoCAPODynamicUPCIMIN}{\rna}
\newcommand{\SLOWhtwoCAPODynamicUPCIMAX}{\rna}
\newcommand{\SLOWhtwoPIPUP}{\rna}
\newcommand{\SLOWhtwoPIPUPCI}{\rna}
\newcommand{\SLOWhtwoPIPUPCIMIN}{\rna}
\newcommand{\SLOWhtwoPIPUPCIMAX}{\rna}
\newcommand{\SLOWhtwoPIPDynamicUP}{\rna}
\newcommand{\SLOWhtwoPIPDynamicUPCI}{\rna}
\newcommand{\SLOWhtwoPIPDynamicUPCIMIN}{\rna}
\newcommand{\SLOWhtwoPIPDynamicUPCIMAX}{\rna}
\newcommand{\SLOWhtwoPIPHB}{\rna}
\newcommand{\SLOWhtwoPIPHBCI}{\rna}
\newcommand{\SLOWhtwoPIPHBCIMIN}{\rna}
\newcommand{\SLOWhtwoPIPHBCIMAX}{\rna}
\newcommand{\SLOWhtwoPIPHBDynamic}{\rna}
\newcommand{\SLOWhtwoPIPHBDynamicCI}{\rna}
\newcommand{\SLOWhtwoPIPHBDynamicCIMIN}{\rna}
\newcommand{\SLOWhtwoPIPHBDynamicCIMAX}{\rna}
\newcommand{\SLOWhtwoPIPWCP}{\rna}
\newcommand{\SLOWhtwoPIPWCPCI}{\rna}
\newcommand{\SLOWhtwoPIPWCPCIMIN}{\rna}
\newcommand{\SLOWhtwoPIPWCPCIMAX}{\rna}
\newcommand{\SLOWhtwoPIPWCPDynamic}{\rna}
\newcommand{\SLOWhtwoPIPWCPDynamicCI}{\rna}
\newcommand{\SLOWhtwoPIPWCPDynamicCIMIN}{\rna}
\newcommand{\SLOWhtwoPIPWCPDynamicCIMAX}{\rna}
\newcommand{\SLOWhtwoPIPWDC}{\rna}
\newcommand{\SLOWhtwoPIPWDCCI}{\rna}
\newcommand{\SLOWhtwoPIPWDCCIMIN}{\rna}
\newcommand{\SLOWhtwoPIPWDCCIMAX}{\rna}
\newcommand{\SLOWhtwoPIPWDCDynamic}{\rna}
\newcommand{\SLOWhtwoPIPWDCDynamicCI}{\rna}
\newcommand{\SLOWhtwoPIPWDCDynamicCIMIN}{\rna}
\newcommand{\SLOWhtwoPIPWDCDynamicCIMAX}{\rna}
\newcommand{\SLOWhtwoPIPCAPO}{\rna}
\newcommand{\SLOWhtwoPIPCAPOCI}{\rna}
\newcommand{\SLOWhtwoPIPCAPOCIMIN}{\rna}
\newcommand{\SLOWhtwoPIPCAPOCIMAX}{\rna}
\newcommand{\SLOWhtwoPIPCAPODynamic}{\rna}
\newcommand{\SLOWhtwoPIPCAPODynamicCI}{\rna}
\newcommand{\SLOWhtwoPIPCAPODynamicCIMIN}{\rna}
\newcommand{\SLOWhtwoPIPCAPODynamicCIMAX}{\rna}
\newcommand{\SLOWhtwoPIPPIP}{\rna}
\newcommand{\SLOWhtwoPIPPIPCI}{\rna}
\newcommand{\SLOWhtwoPIPPIPCIMIN}{\rna}
\newcommand{\SLOWhtwoPIPPIPCIMAX}{\rna}
\newcommand{\SLOWhtwoPIPPIPDynamic}{\rna}
\newcommand{\SLOWhtwoPIPPIPDynamicCI}{\rna}
\newcommand{\SLOWhtwoPIPPIPDynamicCIMIN}{\rna}
\newcommand{\SLOWhtwoPIPPIPDynamicCIMAX}{\rna}
\newcommand{\SLOWjythonEvents}{730}
\newcommand{\SLOWjythonNoFPEvents}{290}
\newcommand{\SLOWjythonMaxLiveThreads}{2}
\newcommand{\SLOWjythonTotalThreads}{2}
\newcommand{\SLOWjythonBaseTime}{4.0}
\newcommand{\SLOWjythonBaseTimeCI}{130}
\newcommand{\SLOWjythonEmptyTime}{\rna}
\newcommand{\SLOWjythonEmptyTimeCI}{\rna}
\newcommand{\SLOWjythonEmptyTimeCIMIN}{\rna}
\newcommand{\SLOWjythonEmptyTimeCIMAX}{\rna}
\newcommand{\SLOWjythonFTTime}{\rna}
\newcommand{\SLOWjythonFTTimeCI}{\rna}
\newcommand{\SLOWjythonFTTimeCIMIN}{\rna}
\newcommand{\SLOWjythonFTTimeCIMAX}{\rna}
\newcommand{\SLOWjythonHBTime}{23}
\newcommand{\SLOWjythonHBTimeCI}{1.2}
\newcommand{\SLOWjythonWCPTime}{32}
\newcommand{\SLOWjythonWCPTimeCI}{1.1}
\newcommand{\SLOWjythonDCnoGExcTime}{26}
\newcommand{\SLOWjythonDCnoGExcTimeCI}{0.94}
\newcommand{\SLOWjythonDCnoGTime}{\rna}
\newcommand{\SLOWjythonDCnoGTimeCI}{\rna}
\newcommand{\SLOWjythonDCnoGTimeCIMIN}{\rna}
\newcommand{\SLOWjythonDCnoGTimeCIMAX}{\rna}
\newcommand{\SLOWjythonDCExcTime}{31}
\newcommand{\SLOWjythonDCExcTimeCI}{1.1}
\newcommand{\SLOWjythonDCTime}{\rna}
\newcommand{\SLOWjythonDCTimeCI}{\rna}
\newcommand{\SLOWjythonDCTimeCIMIN}{\rna}
\newcommand{\SLOWjythonDCTimeCIMAX}{\rna}
\newcommand{\SLOWjythonCAPOnoGExcTime}{23}
\newcommand{\SLOWjythonCAPOnoGExcTimeCI}{0.69}
\newcommand{\SLOWjythonCAPOnoGTime}{\rna}
\newcommand{\SLOWjythonCAPOnoGTimeCI}{\rna}
\newcommand{\SLOWjythonCAPOnoGTimeCIMIN}{\rna}
\newcommand{\SLOWjythonCAPOnoGTimeCIMAX}{\rna}
\newcommand{\SLOWjythonCAPOExcTime}{28}
\newcommand{\SLOWjythonCAPOExcTimeCI}{0.82}
\newcommand{\SLOWjythonCAPOTime}{\rna}
\newcommand{\SLOWjythonCAPOTimeCI}{\rna}
\newcommand{\SLOWjythonCAPOTimeCIMIN}{\rna}
\newcommand{\SLOWjythonCAPOTimeCIMAX}{\rna}
\newcommand{\SLOWjythonStaticTime}{\rzero}
\newcommand{\SLOWjythonDynamicTime}{\rzero}
\newcommand{\SLOWjythonBaseMem}{730}
\newcommand{\SLOWjythonBaseMemCI}{4.2}
\newcommand{\SLOWjythonHBMem}{19}
\newcommand{\SLOWjythonHBMemCI}{0.74}
\newcommand{\SLOWjythonFTMem}{\memna}
\newcommand{\SLOWjythonFTMemCI}{\memna}
\newcommand{\SLOWjythonFTMemCIMIN}{\memna}
\newcommand{\SLOWjythonFTMemCIMAX}{\memna}
\newcommand{\SLOWjythonWCPMem}{19}
\newcommand{\SLOWjythonWCPMemCI}{0.58}
\newcommand{\SLOWjythonDCnoGExcMem}{18}
\newcommand{\SLOWjythonDCnoGExcMemCI}{1.1}
\newcommand{\SLOWjythonDCnoGMem}{\memna}
\newcommand{\SLOWjythonDCnoGMemCI}{\memna}
\newcommand{\SLOWjythonDCnoGMemCIMIN}{\memna}
\newcommand{\SLOWjythonDCnoGMemCIMAX}{\memna}
\newcommand{\SLOWjythonDCExcMem}{31}
\newcommand{\SLOWjythonDCExcMemCI}{0.87}
\newcommand{\SLOWjythonDCMem}{\memna}
\newcommand{\SLOWjythonDCMemCI}{\memna}
\newcommand{\SLOWjythonDCMemCIMIN}{\memna}
\newcommand{\SLOWjythonDCMemCIMAX}{\memna}
\newcommand{\SLOWjythonCAPOnoGExcMem}{16}
\newcommand{\SLOWjythonCAPOnoGExcMemCI}{0.75}
\newcommand{\SLOWjythonCAPOnoGMem}{\memna}
\newcommand{\SLOWjythonCAPOnoGMemCI}{\memna}
\newcommand{\SLOWjythonCAPOnoGMemCIMIN}{\memna}
\newcommand{\SLOWjythonCAPOnoGMemCIMAX}{\memna}
\newcommand{\SLOWjythonCAPOExcMem}{25}
\newcommand{\SLOWjythonCAPOExcMemCI}{0.47}
\newcommand{\SLOWjythonCAPOMem}{\memna}
\newcommand{\SLOWjythonCAPOMemCI}{\memna}
\newcommand{\SLOWjythonCAPOMemCIMIN}{\memna}
\newcommand{\SLOWjythonCAPOMemCIMAX}{\memna}
\newcommand{\SLOWjythonEventsCI}{211}
\newcommand{\SLOWjythonEventsCIMIN}{728,376,678}
\newcommand{\SLOWjythonEventsCIMAX}{728,377,100}
\newcommand{\SLOWjythonNoFPEventsCI}{18}
\newcommand{\SLOWjythonNoFPEventsCIMIN}{293,898,196}
\newcommand{\SLOWjythonNoFPEventsCIMAX}{293,898,232}
\newcommand{\SLOWjythonHB}{39}
\newcommand{\SLOWjythonHBCI}{2}
\newcommand{\SLOWjythonHBCIMIN}{37}
\newcommand{\SLOWjythonHBCIMAX}{41}
\newcommand{\SLOWjythonHBDynamic}{42}
\newcommand{\SLOWjythonHBDynamicCI}{4}
\newcommand{\SLOWjythonHBDynamicCIMIN}{38}
\newcommand{\SLOWjythonHBDynamicCIMAX}{46}
\newcommand{\SLOWjythonFT}{\rna}
\newcommand{\SLOWjythonFTCI}{\rna}
\newcommand{\SLOWjythonFTCIMIN}{\rna}
\newcommand{\SLOWjythonFTCIMAX}{\rna}
\newcommand{\SLOWjythonFTDynamic}{\rna}
\newcommand{\SLOWjythonFTDynamicCI}{\rna}
\newcommand{\SLOWjythonFTDynamicCIMIN}{\rna}
\newcommand{\SLOWjythonFTDynamicCIMAX}{\rna}
\newcommand{\SLOWjythonWCP}{39}
\newcommand{\SLOWjythonWCPCI}{0}
\newcommand{\SLOWjythonWCPCIMIN}{39}
\newcommand{\SLOWjythonWCPCIMAX}{39}
\newcommand{\SLOWjythonWCPDynamic}{44}
\newcommand{\SLOWjythonWCPDynamicCI}{0}
\newcommand{\SLOWjythonWCPDynamicCIMIN}{44}
\newcommand{\SLOWjythonWCPDynamicCIMAX}{44}
\newcommand{\SLOWjythonDCnoGExc}{50}
\newcommand{\SLOWjythonDCnoGExcCI}{0}
\newcommand{\SLOWjythonDCnoGExcCIMIN}{50}
\newcommand{\SLOWjythonDCnoGExcCIMAX}{50}
\newcommand{\SLOWjythonDCnoGExcDynamic}{61}
\newcommand{\SLOWjythonDCnoGExcDynamicCI}{0}
\newcommand{\SLOWjythonDCnoGExcDynamicCIMIN}{61}
\newcommand{\SLOWjythonDCnoGExcDynamicCIMAX}{61}
\newcommand{\SLOWjythonDCnoG}{\rna}
\newcommand{\SLOWjythonDCnoGCI}{\rna}
\newcommand{\SLOWjythonDCnoGCIMIN}{\rna}
\newcommand{\SLOWjythonDCnoGCIMAX}{\rna}
\newcommand{\SLOWjythonDCnoGDynamic}{\rna}
\newcommand{\SLOWjythonDCnoGDynamicCI}{\rna}
\newcommand{\SLOWjythonDCnoGDynamicCIMIN}{\rna}
\newcommand{\SLOWjythonDCnoGDynamicCIMAX}{\rna}
\newcommand{\SLOWjythonDCExc}{3}
\newcommand{\SLOWjythonDCExcCI}{0}
\newcommand{\SLOWjythonDCExcCIMIN}{3}
\newcommand{\SLOWjythonDCExcCIMAX}{3}
\newcommand{\SLOWjythonDCExcDynamic}{4}
\newcommand{\SLOWjythonDCExcDynamicCI}{1}
\newcommand{\SLOWjythonDCExcDynamicCIMIN}{3}
\newcommand{\SLOWjythonDCExcDynamicCIMAX}{5}
\newcommand{\SLOWjythonDC}{\rna}
\newcommand{\SLOWjythonDCCI}{\rna}
\newcommand{\SLOWjythonDCCIMIN}{\rna}
\newcommand{\SLOWjythonDCCIMAX}{\rna}
\newcommand{\SLOWjythonDCDynamic}{\rna}
\newcommand{\SLOWjythonDCDynamicCI}{\rna}
\newcommand{\SLOWjythonDCDynamicCIMIN}{\rna}
\newcommand{\SLOWjythonDCDynamicCIMAX}{\rna}
\newcommand{\SLOWjythonCAPOnoGExc}{50}
\newcommand{\SLOWjythonCAPOnoGExcCI}{1}
\newcommand{\SLOWjythonCAPOnoGExcCIMIN}{49}
\newcommand{\SLOWjythonCAPOnoGExcCIMAX}{51}
\newcommand{\SLOWjythonCAPOnoGExcDynamic}{61}
\newcommand{\SLOWjythonCAPOnoGExcDynamicCI}{1}
\newcommand{\SLOWjythonCAPOnoGExcDynamicCIMIN}{60}
\newcommand{\SLOWjythonCAPOnoGExcDynamicCIMAX}{62}
\newcommand{\SLOWjythonCAPOnoG}{\rna}
\newcommand{\SLOWjythonCAPOnoGCI}{\rna}
\newcommand{\SLOWjythonCAPOnoGCIMIN}{\rna}
\newcommand{\SLOWjythonCAPOnoGCIMAX}{\rna}
\newcommand{\SLOWjythonCAPOnoGDynamic}{\rna}
\newcommand{\SLOWjythonCAPOnoGDynamicCI}{\rna}
\newcommand{\SLOWjythonCAPOnoGDynamicCIMIN}{\rna}
\newcommand{\SLOWjythonCAPOnoGDynamicCIMAX}{\rna}
\newcommand{\SLOWjythonCAPOExc}{3}
\newcommand{\SLOWjythonCAPOExcCI}{0}
\newcommand{\SLOWjythonCAPOExcCIMIN}{3}
\newcommand{\SLOWjythonCAPOExcCIMAX}{3}
\newcommand{\SLOWjythonCAPOExcDynamic}{4}
\newcommand{\SLOWjythonCAPOExcDynamicCI}{1}
\newcommand{\SLOWjythonCAPOExcDynamicCIMIN}{3}
\newcommand{\SLOWjythonCAPOExcDynamicCIMAX}{5}
\newcommand{\SLOWjythonCAPO}{\rna}
\newcommand{\SLOWjythonCAPOCI}{\rna}
\newcommand{\SLOWjythonCAPOCIMIN}{\rna}
\newcommand{\SLOWjythonCAPOCIMAX}{\rna}
\newcommand{\SLOWjythonCAPODynamic}{\rna}
\newcommand{\SLOWjythonCAPODynamicCI}{\rna}
\newcommand{\SLOWjythonCAPODynamicCIMIN}{\rna}
\newcommand{\SLOWjythonCAPODynamicCIMAX}{\rna}
\newcommand{\SLOWjythonPIP}{\rna}
\newcommand{\SLOWjythonPIPCI}{\rna}
\newcommand{\SLOWjythonPIPCIMIN}{\rna}
\newcommand{\SLOWjythonPIPCIMAX}{\rna}
\newcommand{\SLOWjythonPIPDynamic}{\rna}
\newcommand{\SLOWjythonPIPDynamicCI}{\rna}
\newcommand{\SLOWjythonPIPDynamicCIMIN}{\rna}
\newcommand{\SLOWjythonPIPDynamicCIMAX}{\rna}
\newcommand{\SLOWjythonHBUP}{38}
\newcommand{\SLOWjythonHBUPCI}{2}
\newcommand{\SLOWjythonHBUPCIMIN}{36}
\newcommand{\SLOWjythonHBUPCIMAX}{40}
\newcommand{\SLOWjythonHBDynamicUP}{42}
\newcommand{\SLOWjythonHBDynamicUPCI}{4}
\newcommand{\SLOWjythonHBDynamicUPCIMIN}{38}
\newcommand{\SLOWjythonHBDynamicUPCIMAX}{46}
\newcommand{\SLOWjythonWCPUP}{38}
\newcommand{\SLOWjythonWCPUPCI}{0}
\newcommand{\SLOWjythonWCPUPCIMIN}{38}
\newcommand{\SLOWjythonWCPUPCIMAX}{38}
\newcommand{\SLOWjythonWCPDynamicUP}{44}
\newcommand{\SLOWjythonWCPDynamicUPCI}{0}
\newcommand{\SLOWjythonWCPDynamicUPCIMIN}{44}
\newcommand{\SLOWjythonWCPDynamicUPCIMAX}{44}
\newcommand{\SLOWjythonWDCnoGUP}{\rna}
\newcommand{\SLOWjythonWDCnoGUPCI}{\rna}
\newcommand{\SLOWjythonWDCnoGUPCIMIN}{\rna}
\newcommand{\SLOWjythonWDCnoGUPCIMAX}{\rna}
\newcommand{\SLOWjythonWDCnoGDynamicUP}{\rna}
\newcommand{\SLOWjythonWDCnoGDynamicUPCI}{\rna}
\newcommand{\SLOWjythonWDCnoGDynamicUPCIMIN}{\rna}
\newcommand{\SLOWjythonWDCnoGDynamicUPCIMAX}{\rna}
\newcommand{\SLOWjythonWDCUP}{\rna}
\newcommand{\SLOWjythonWDCUPCI}{\rna}
\newcommand{\SLOWjythonWDCUPCIMIN}{\rna}
\newcommand{\SLOWjythonWDCUPCIMAX}{\rna}
\newcommand{\SLOWjythonWDCDynamicUP}{\rna}
\newcommand{\SLOWjythonWDCDynamicUPCI}{\rna}
\newcommand{\SLOWjythonWDCDynamicUPCIMIN}{\rna}
\newcommand{\SLOWjythonWDCDynamicUPCIMAX}{\rna}
\newcommand{\SLOWjythonCAPOnoGUP}{\rna}
\newcommand{\SLOWjythonCAPOnoGUPCI}{\rna}
\newcommand{\SLOWjythonCAPOnoGUPCIMIN}{\rna}
\newcommand{\SLOWjythonCAPOnoGUPCIMAX}{\rna}
\newcommand{\SLOWjythonCAPOnoGDynamicUP}{\rna}
\newcommand{\SLOWjythonCAPOnoGDynamicUPCI}{\rna}
\newcommand{\SLOWjythonCAPOnoGDynamicUPCIMIN}{\rna}
\newcommand{\SLOWjythonCAPOnoGDynamicUPCIMAX}{\rna}
\newcommand{\SLOWjythonCAPOUP}{\rna}
\newcommand{\SLOWjythonCAPOUPCI}{\rna}
\newcommand{\SLOWjythonCAPOUPCIMIN}{\rna}
\newcommand{\SLOWjythonCAPOUPCIMAX}{\rna}
\newcommand{\SLOWjythonCAPODynamicUP}{\rna}
\newcommand{\SLOWjythonCAPODynamicUPCI}{\rna}
\newcommand{\SLOWjythonCAPODynamicUPCIMIN}{\rna}
\newcommand{\SLOWjythonCAPODynamicUPCIMAX}{\rna}
\newcommand{\SLOWjythonPIPUP}{\rna}
\newcommand{\SLOWjythonPIPUPCI}{\rna}
\newcommand{\SLOWjythonPIPUPCIMIN}{\rna}
\newcommand{\SLOWjythonPIPUPCIMAX}{\rna}
\newcommand{\SLOWjythonPIPDynamicUP}{\rna}
\newcommand{\SLOWjythonPIPDynamicUPCI}{\rna}
\newcommand{\SLOWjythonPIPDynamicUPCIMIN}{\rna}
\newcommand{\SLOWjythonPIPDynamicUPCIMAX}{\rna}
\newcommand{\SLOWjythonPIPHB}{\rna}
\newcommand{\SLOWjythonPIPHBCI}{\rna}
\newcommand{\SLOWjythonPIPHBCIMIN}{\rna}
\newcommand{\SLOWjythonPIPHBCIMAX}{\rna}
\newcommand{\SLOWjythonPIPHBDynamic}{\rna}
\newcommand{\SLOWjythonPIPHBDynamicCI}{\rna}
\newcommand{\SLOWjythonPIPHBDynamicCIMIN}{\rna}
\newcommand{\SLOWjythonPIPHBDynamicCIMAX}{\rna}
\newcommand{\SLOWjythonPIPWCP}{\rna}
\newcommand{\SLOWjythonPIPWCPCI}{\rna}
\newcommand{\SLOWjythonPIPWCPCIMIN}{\rna}
\newcommand{\SLOWjythonPIPWCPCIMAX}{\rna}
\newcommand{\SLOWjythonPIPWCPDynamic}{\rna}
\newcommand{\SLOWjythonPIPWCPDynamicCI}{\rna}
\newcommand{\SLOWjythonPIPWCPDynamicCIMIN}{\rna}
\newcommand{\SLOWjythonPIPWCPDynamicCIMAX}{\rna}
\newcommand{\SLOWjythonPIPWDC}{\rna}
\newcommand{\SLOWjythonPIPWDCCI}{\rna}
\newcommand{\SLOWjythonPIPWDCCIMIN}{\rna}
\newcommand{\SLOWjythonPIPWDCCIMAX}{\rna}
\newcommand{\SLOWjythonPIPWDCDynamic}{\rna}
\newcommand{\SLOWjythonPIPWDCDynamicCI}{\rna}
\newcommand{\SLOWjythonPIPWDCDynamicCIMIN}{\rna}
\newcommand{\SLOWjythonPIPWDCDynamicCIMAX}{\rna}
\newcommand{\SLOWjythonPIPCAPO}{\rna}
\newcommand{\SLOWjythonPIPCAPOCI}{\rna}
\newcommand{\SLOWjythonPIPCAPOCIMIN}{\rna}
\newcommand{\SLOWjythonPIPCAPOCIMAX}{\rna}
\newcommand{\SLOWjythonPIPCAPODynamic}{\rna}
\newcommand{\SLOWjythonPIPCAPODynamicCI}{\rna}
\newcommand{\SLOWjythonPIPCAPODynamicCIMIN}{\rna}
\newcommand{\SLOWjythonPIPCAPODynamicCIMAX}{\rna}
\newcommand{\SLOWjythonPIPPIP}{\rna}
\newcommand{\SLOWjythonPIPPIPCI}{\rna}
\newcommand{\SLOWjythonPIPPIPCIMIN}{\rna}
\newcommand{\SLOWjythonPIPPIPCIMAX}{\rna}
\newcommand{\SLOWjythonPIPPIPDynamic}{\rna}
\newcommand{\SLOWjythonPIPPIPDynamicCI}{\rna}
\newcommand{\SLOWjythonPIPPIPDynamicCIMIN}{\rna}
\newcommand{\SLOWjythonPIPPIPDynamicCIMAX}{\rna}
\newcommand{\SLOWluindexEvents}{400}
\newcommand{\SLOWluindexNoFPEvents}{45}
\newcommand{\SLOWluindexMaxLiveThreads}{3}
\newcommand{\SLOWluindexTotalThreads}{3}
\newcommand{\SLOWluindexBaseTime}{1.2}
\newcommand{\SLOWluindexBaseTimeCI}{38}
\newcommand{\SLOWluindexEmptyTime}{\rna}
\newcommand{\SLOWluindexEmptyTimeCI}{\rna}
\newcommand{\SLOWluindexEmptyTimeCIMIN}{\rna}
\newcommand{\SLOWluindexEmptyTimeCIMAX}{\rna}
\newcommand{\SLOWluindexFTTime}{\rna}
\newcommand{\SLOWluindexFTTimeCI}{\rna}
\newcommand{\SLOWluindexFTTimeCIMIN}{\rna}
\newcommand{\SLOWluindexFTTimeCIMAX}{\rna}
\newcommand{\SLOWluindexHBTime}{27}
\newcommand{\SLOWluindexHBTimeCI}{0.9}
\newcommand{\SLOWluindexWCPTime}{46}
\newcommand{\SLOWluindexWCPTimeCI}{1.9}
\newcommand{\SLOWluindexDCnoGExcTime}{39}
\newcommand{\SLOWluindexDCnoGExcTimeCI}{1.4}
\newcommand{\SLOWluindexDCnoGTime}{\rna}
\newcommand{\SLOWluindexDCnoGTimeCI}{\rna}
\newcommand{\SLOWluindexDCnoGTimeCIMIN}{\rna}
\newcommand{\SLOWluindexDCnoGTimeCIMAX}{\rna}
\newcommand{\SLOWluindexDCExcTime}{44}
\newcommand{\SLOWluindexDCExcTimeCI}{1.1}
\newcommand{\SLOWluindexDCTime}{\rna}
\newcommand{\SLOWluindexDCTimeCI}{\rna}
\newcommand{\SLOWluindexDCTimeCIMIN}{\rna}
\newcommand{\SLOWluindexDCTimeCIMAX}{\rna}
\newcommand{\SLOWluindexCAPOnoGExcTime}{39}
\newcommand{\SLOWluindexCAPOnoGExcTimeCI}{1.4}
\newcommand{\SLOWluindexCAPOnoGTime}{\rna}
\newcommand{\SLOWluindexCAPOnoGTimeCI}{\rna}
\newcommand{\SLOWluindexCAPOnoGTimeCIMIN}{\rna}
\newcommand{\SLOWluindexCAPOnoGTimeCIMAX}{\rna}
\newcommand{\SLOWluindexCAPOExcTime}{44}
\newcommand{\SLOWluindexCAPOExcTimeCI}{1.5}
\newcommand{\SLOWluindexCAPOTime}{\rna}
\newcommand{\SLOWluindexCAPOTimeCI}{\rna}
\newcommand{\SLOWluindexCAPOTimeCIMIN}{\rna}
\newcommand{\SLOWluindexCAPOTimeCIMAX}{\rna}
\newcommand{\SLOWluindexStaticTime}{\rzero}
\newcommand{\SLOWluindexDynamicTime}{\rzero}
\newcommand{\SLOWluindexBaseMem}{120}
\newcommand{\SLOWluindexBaseMemCI}{3.0}
\newcommand{\SLOWluindexHBMem}{35}
\newcommand{\SLOWluindexHBMemCI}{0.96}
\newcommand{\SLOWluindexFTMem}{\memna}
\newcommand{\SLOWluindexFTMemCI}{\memna}
\newcommand{\SLOWluindexFTMemCIMIN}{\memna}
\newcommand{\SLOWluindexFTMemCIMAX}{\memna}
\newcommand{\SLOWluindexWCPMem}{73}
\newcommand{\SLOWluindexWCPMemCI}{1.9}
\newcommand{\SLOWluindexDCnoGExcMem}{51}
\newcommand{\SLOWluindexDCnoGExcMemCI}{1.3}
\newcommand{\SLOWluindexDCnoGMem}{\memna}
\newcommand{\SLOWluindexDCnoGMemCI}{\memna}
\newcommand{\SLOWluindexDCnoGMemCIMIN}{\memna}
\newcommand{\SLOWluindexDCnoGMemCIMAX}{\memna}
\newcommand{\SLOWluindexDCExcMem}{66}
\newcommand{\SLOWluindexDCExcMemCI}{1.6}
\newcommand{\SLOWluindexDCMem}{\memna}
\newcommand{\SLOWluindexDCMemCI}{\memna}
\newcommand{\SLOWluindexDCMemCIMIN}{\memna}
\newcommand{\SLOWluindexDCMemCIMAX}{\memna}
\newcommand{\SLOWluindexCAPOnoGExcMem}{51}
\newcommand{\SLOWluindexCAPOnoGExcMemCI}{1.3}
\newcommand{\SLOWluindexCAPOnoGMem}{\memna}
\newcommand{\SLOWluindexCAPOnoGMemCI}{\memna}
\newcommand{\SLOWluindexCAPOnoGMemCIMIN}{\memna}
\newcommand{\SLOWluindexCAPOnoGMemCIMAX}{\memna}
\newcommand{\SLOWluindexCAPOExcMem}{66}
\newcommand{\SLOWluindexCAPOExcMemCI}{1.8}
\newcommand{\SLOWluindexCAPOMem}{\memna}
\newcommand{\SLOWluindexCAPOMemCI}{\memna}
\newcommand{\SLOWluindexCAPOMemCIMIN}{\memna}
\newcommand{\SLOWluindexCAPOMemCIMAX}{\memna}
\newcommand{\SLOWluindexEventsCI}{4}
\newcommand{\SLOWluindexEventsCIMIN}{396,268,397}
\newcommand{\SLOWluindexEventsCIMAX}{396,268,405}
\newcommand{\SLOWluindexNoFPEventsCI}{3}
\newcommand{\SLOWluindexNoFPEventsCIMIN}{45,411,630}
\newcommand{\SLOWluindexNoFPEventsCIMAX}{45,411,636}
\newcommand{\SLOWluindexHB}{1}
\newcommand{\SLOWluindexHBCI}{0}
\newcommand{\SLOWluindexHBCIMIN}{1}
\newcommand{\SLOWluindexHBCIMAX}{1}
\newcommand{\SLOWluindexHBDynamic}{1}
\newcommand{\SLOWluindexHBDynamicCI}{0}
\newcommand{\SLOWluindexHBDynamicCIMIN}{1}
\newcommand{\SLOWluindexHBDynamicCIMAX}{1}
\newcommand{\SLOWluindexFT}{\rna}
\newcommand{\SLOWluindexFTCI}{\rna}
\newcommand{\SLOWluindexFTCIMIN}{\rna}
\newcommand{\SLOWluindexFTCIMAX}{\rna}
\newcommand{\SLOWluindexFTDynamic}{\rna}
\newcommand{\SLOWluindexFTDynamicCI}{\rna}
\newcommand{\SLOWluindexFTDynamicCIMIN}{\rna}
\newcommand{\SLOWluindexFTDynamicCIMAX}{\rna}
\newcommand{\SLOWluindexWCP}{1}
\newcommand{\SLOWluindexWCPCI}{0}
\newcommand{\SLOWluindexWCPCIMIN}{1}
\newcommand{\SLOWluindexWCPCIMAX}{1}
\newcommand{\SLOWluindexWCPDynamic}{1}
\newcommand{\SLOWluindexWCPDynamicCI}{0}
\newcommand{\SLOWluindexWCPDynamicCIMIN}{1}
\newcommand{\SLOWluindexWCPDynamicCIMAX}{1}
\newcommand{\SLOWluindexDCnoGExc}{1}
\newcommand{\SLOWluindexDCnoGExcCI}{0}
\newcommand{\SLOWluindexDCnoGExcCIMIN}{1}
\newcommand{\SLOWluindexDCnoGExcCIMAX}{1}
\newcommand{\SLOWluindexDCnoGExcDynamic}{1}
\newcommand{\SLOWluindexDCnoGExcDynamicCI}{0}
\newcommand{\SLOWluindexDCnoGExcDynamicCIMIN}{1}
\newcommand{\SLOWluindexDCnoGExcDynamicCIMAX}{1}
\newcommand{\SLOWluindexDCnoG}{\rna}
\newcommand{\SLOWluindexDCnoGCI}{\rna}
\newcommand{\SLOWluindexDCnoGCIMIN}{\rna}
\newcommand{\SLOWluindexDCnoGCIMAX}{\rna}
\newcommand{\SLOWluindexDCnoGDynamic}{\rna}
\newcommand{\SLOWluindexDCnoGDynamicCI}{\rna}
\newcommand{\SLOWluindexDCnoGDynamicCIMIN}{\rna}
\newcommand{\SLOWluindexDCnoGDynamicCIMAX}{\rna}
\newcommand{\SLOWluindexDCExc}{1}
\newcommand{\SLOWluindexDCExcCI}{0}
\newcommand{\SLOWluindexDCExcCIMIN}{1}
\newcommand{\SLOWluindexDCExcCIMAX}{1}
\newcommand{\SLOWluindexDCExcDynamic}{1}
\newcommand{\SLOWluindexDCExcDynamicCI}{0}
\newcommand{\SLOWluindexDCExcDynamicCIMIN}{1}
\newcommand{\SLOWluindexDCExcDynamicCIMAX}{1}
\newcommand{\SLOWluindexDC}{\rna}
\newcommand{\SLOWluindexDCCI}{\rna}
\newcommand{\SLOWluindexDCCIMIN}{\rna}
\newcommand{\SLOWluindexDCCIMAX}{\rna}
\newcommand{\SLOWluindexDCDynamic}{\rna}
\newcommand{\SLOWluindexDCDynamicCI}{\rna}
\newcommand{\SLOWluindexDCDynamicCIMIN}{\rna}
\newcommand{\SLOWluindexDCDynamicCIMAX}{\rna}
\newcommand{\SLOWluindexCAPOnoGExc}{1}
\newcommand{\SLOWluindexCAPOnoGExcCI}{0}
\newcommand{\SLOWluindexCAPOnoGExcCIMIN}{1}
\newcommand{\SLOWluindexCAPOnoGExcCIMAX}{1}
\newcommand{\SLOWluindexCAPOnoGExcDynamic}{1}
\newcommand{\SLOWluindexCAPOnoGExcDynamicCI}{0}
\newcommand{\SLOWluindexCAPOnoGExcDynamicCIMIN}{1}
\newcommand{\SLOWluindexCAPOnoGExcDynamicCIMAX}{1}
\newcommand{\SLOWluindexCAPOnoG}{\rna}
\newcommand{\SLOWluindexCAPOnoGCI}{\rna}
\newcommand{\SLOWluindexCAPOnoGCIMIN}{\rna}
\newcommand{\SLOWluindexCAPOnoGCIMAX}{\rna}
\newcommand{\SLOWluindexCAPOnoGDynamic}{\rna}
\newcommand{\SLOWluindexCAPOnoGDynamicCI}{\rna}
\newcommand{\SLOWluindexCAPOnoGDynamicCIMIN}{\rna}
\newcommand{\SLOWluindexCAPOnoGDynamicCIMAX}{\rna}
\newcommand{\SLOWluindexCAPOExc}{1}
\newcommand{\SLOWluindexCAPOExcCI}{0}
\newcommand{\SLOWluindexCAPOExcCIMIN}{1}
\newcommand{\SLOWluindexCAPOExcCIMAX}{1}
\newcommand{\SLOWluindexCAPOExcDynamic}{1}
\newcommand{\SLOWluindexCAPOExcDynamicCI}{0}
\newcommand{\SLOWluindexCAPOExcDynamicCIMIN}{1}
\newcommand{\SLOWluindexCAPOExcDynamicCIMAX}{1}
\newcommand{\SLOWluindexCAPO}{\rna}
\newcommand{\SLOWluindexCAPOCI}{\rna}
\newcommand{\SLOWluindexCAPOCIMIN}{\rna}
\newcommand{\SLOWluindexCAPOCIMAX}{\rna}
\newcommand{\SLOWluindexCAPODynamic}{\rna}
\newcommand{\SLOWluindexCAPODynamicCI}{\rna}
\newcommand{\SLOWluindexCAPODynamicCIMIN}{\rna}
\newcommand{\SLOWluindexCAPODynamicCIMAX}{\rna}
\newcommand{\SLOWluindexPIP}{\rna}
\newcommand{\SLOWluindexPIPCI}{\rna}
\newcommand{\SLOWluindexPIPCIMIN}{\rna}
\newcommand{\SLOWluindexPIPCIMAX}{\rna}
\newcommand{\SLOWluindexPIPDynamic}{\rna}
\newcommand{\SLOWluindexPIPDynamicCI}{\rna}
\newcommand{\SLOWluindexPIPDynamicCIMIN}{\rna}
\newcommand{\SLOWluindexPIPDynamicCIMAX}{\rna}
\newcommand{\SLOWluindexHBUP}{1}
\newcommand{\SLOWluindexHBUPCI}{0}
\newcommand{\SLOWluindexHBUPCIMIN}{1}
\newcommand{\SLOWluindexHBUPCIMAX}{1}
\newcommand{\SLOWluindexHBDynamicUP}{1}
\newcommand{\SLOWluindexHBDynamicUPCI}{0}
\newcommand{\SLOWluindexHBDynamicUPCIMIN}{1}
\newcommand{\SLOWluindexHBDynamicUPCIMAX}{1}
\newcommand{\SLOWluindexWCPUP}{1}
\newcommand{\SLOWluindexWCPUPCI}{0}
\newcommand{\SLOWluindexWCPUPCIMIN}{1}
\newcommand{\SLOWluindexWCPUPCIMAX}{1}
\newcommand{\SLOWluindexWCPDynamicUP}{1}
\newcommand{\SLOWluindexWCPDynamicUPCI}{0}
\newcommand{\SLOWluindexWCPDynamicUPCIMIN}{1}
\newcommand{\SLOWluindexWCPDynamicUPCIMAX}{1}
\newcommand{\SLOWluindexWDCnoGUP}{\rna}
\newcommand{\SLOWluindexWDCnoGUPCI}{\rna}
\newcommand{\SLOWluindexWDCnoGUPCIMIN}{\rna}
\newcommand{\SLOWluindexWDCnoGUPCIMAX}{\rna}
\newcommand{\SLOWluindexWDCnoGDynamicUP}{\rna}
\newcommand{\SLOWluindexWDCnoGDynamicUPCI}{\rna}
\newcommand{\SLOWluindexWDCnoGDynamicUPCIMIN}{\rna}
\newcommand{\SLOWluindexWDCnoGDynamicUPCIMAX}{\rna}
\newcommand{\SLOWluindexWDCUP}{\rna}
\newcommand{\SLOWluindexWDCUPCI}{\rna}
\newcommand{\SLOWluindexWDCUPCIMIN}{\rna}
\newcommand{\SLOWluindexWDCUPCIMAX}{\rna}
\newcommand{\SLOWluindexWDCDynamicUP}{\rna}
\newcommand{\SLOWluindexWDCDynamicUPCI}{\rna}
\newcommand{\SLOWluindexWDCDynamicUPCIMIN}{\rna}
\newcommand{\SLOWluindexWDCDynamicUPCIMAX}{\rna}
\newcommand{\SLOWluindexCAPOnoGUP}{\rna}
\newcommand{\SLOWluindexCAPOnoGUPCI}{\rna}
\newcommand{\SLOWluindexCAPOnoGUPCIMIN}{\rna}
\newcommand{\SLOWluindexCAPOnoGUPCIMAX}{\rna}
\newcommand{\SLOWluindexCAPOnoGDynamicUP}{\rna}
\newcommand{\SLOWluindexCAPOnoGDynamicUPCI}{\rna}
\newcommand{\SLOWluindexCAPOnoGDynamicUPCIMIN}{\rna}
\newcommand{\SLOWluindexCAPOnoGDynamicUPCIMAX}{\rna}
\newcommand{\SLOWluindexCAPOUP}{\rna}
\newcommand{\SLOWluindexCAPOUPCI}{\rna}
\newcommand{\SLOWluindexCAPOUPCIMIN}{\rna}
\newcommand{\SLOWluindexCAPOUPCIMAX}{\rna}
\newcommand{\SLOWluindexCAPODynamicUP}{\rna}
\newcommand{\SLOWluindexCAPODynamicUPCI}{\rna}
\newcommand{\SLOWluindexCAPODynamicUPCIMIN}{\rna}
\newcommand{\SLOWluindexCAPODynamicUPCIMAX}{\rna}
\newcommand{\SLOWluindexPIPUP}{\rna}
\newcommand{\SLOWluindexPIPUPCI}{\rna}
\newcommand{\SLOWluindexPIPUPCIMIN}{\rna}
\newcommand{\SLOWluindexPIPUPCIMAX}{\rna}
\newcommand{\SLOWluindexPIPDynamicUP}{\rna}
\newcommand{\SLOWluindexPIPDynamicUPCI}{\rna}
\newcommand{\SLOWluindexPIPDynamicUPCIMIN}{\rna}
\newcommand{\SLOWluindexPIPDynamicUPCIMAX}{\rna}
\newcommand{\SLOWluindexPIPHB}{\rna}
\newcommand{\SLOWluindexPIPHBCI}{\rna}
\newcommand{\SLOWluindexPIPHBCIMIN}{\rna}
\newcommand{\SLOWluindexPIPHBCIMAX}{\rna}
\newcommand{\SLOWluindexPIPHBDynamic}{\rna}
\newcommand{\SLOWluindexPIPHBDynamicCI}{\rna}
\newcommand{\SLOWluindexPIPHBDynamicCIMIN}{\rna}
\newcommand{\SLOWluindexPIPHBDynamicCIMAX}{\rna}
\newcommand{\SLOWluindexPIPWCP}{\rna}
\newcommand{\SLOWluindexPIPWCPCI}{\rna}
\newcommand{\SLOWluindexPIPWCPCIMIN}{\rna}
\newcommand{\SLOWluindexPIPWCPCIMAX}{\rna}
\newcommand{\SLOWluindexPIPWCPDynamic}{\rna}
\newcommand{\SLOWluindexPIPWCPDynamicCI}{\rna}
\newcommand{\SLOWluindexPIPWCPDynamicCIMIN}{\rna}
\newcommand{\SLOWluindexPIPWCPDynamicCIMAX}{\rna}
\newcommand{\SLOWluindexPIPWDC}{\rna}
\newcommand{\SLOWluindexPIPWDCCI}{\rna}
\newcommand{\SLOWluindexPIPWDCCIMIN}{\rna}
\newcommand{\SLOWluindexPIPWDCCIMAX}{\rna}
\newcommand{\SLOWluindexPIPWDCDynamic}{\rna}
\newcommand{\SLOWluindexPIPWDCDynamicCI}{\rna}
\newcommand{\SLOWluindexPIPWDCDynamicCIMIN}{\rna}
\newcommand{\SLOWluindexPIPWDCDynamicCIMAX}{\rna}
\newcommand{\SLOWluindexPIPCAPO}{\rna}
\newcommand{\SLOWluindexPIPCAPOCI}{\rna}
\newcommand{\SLOWluindexPIPCAPOCIMIN}{\rna}
\newcommand{\SLOWluindexPIPCAPOCIMAX}{\rna}
\newcommand{\SLOWluindexPIPCAPODynamic}{\rna}
\newcommand{\SLOWluindexPIPCAPODynamicCI}{\rna}
\newcommand{\SLOWluindexPIPCAPODynamicCIMIN}{\rna}
\newcommand{\SLOWluindexPIPCAPODynamicCIMAX}{\rna}
\newcommand{\SLOWluindexPIPPIP}{\rna}
\newcommand{\SLOWluindexPIPPIPCI}{\rna}
\newcommand{\SLOWluindexPIPPIPCIMIN}{\rna}
\newcommand{\SLOWluindexPIPPIPCIMAX}{\rna}
\newcommand{\SLOWluindexPIPPIPDynamic}{\rna}
\newcommand{\SLOWluindexPIPPIPDynamicCI}{\rna}
\newcommand{\SLOWluindexPIPPIPDynamicCIMIN}{\rna}
\newcommand{\SLOWluindexPIPPIPDynamicCIMAX}{\rna}
\newcommand{\SLOWlusearchEvents}{1,400}
\newcommand{\SLOWlusearchNoFPEvents}{190}
\newcommand{\SLOWlusearchMaxLiveThreads}{10}
\newcommand{\SLOWlusearchTotalThreads}{10}
\newcommand{\SLOWlusearchBaseTime}{1.1}
\newcommand{\SLOWlusearchBaseTimeCI}{120}
\newcommand{\SLOWlusearchEmptyTime}{\rna}
\newcommand{\SLOWlusearchEmptyTimeCI}{\rna}
\newcommand{\SLOWlusearchEmptyTimeCIMIN}{\rna}
\newcommand{\SLOWlusearchEmptyTimeCIMAX}{\rna}
\newcommand{\SLOWlusearchFTTime}{\rna}
\newcommand{\SLOWlusearchFTTimeCI}{\rna}
\newcommand{\SLOWlusearchFTTimeCIMIN}{\rna}
\newcommand{\SLOWlusearchFTTimeCIMAX}{\rna}
\newcommand{\SLOWlusearchHBTime}{28}
\newcommand{\SLOWlusearchHBTimeCI}{3.8}
\newcommand{\SLOWlusearchWCPTime}{36}
\newcommand{\SLOWlusearchWCPTimeCI}{4.4}
\newcommand{\SLOWlusearchDCnoGExcTime}{27}
\newcommand{\SLOWlusearchDCnoGExcTimeCI}{2.9}
\newcommand{\SLOWlusearchDCnoGTime}{\rna}
\newcommand{\SLOWlusearchDCnoGTimeCI}{\rna}
\newcommand{\SLOWlusearchDCnoGTimeCIMIN}{\rna}
\newcommand{\SLOWlusearchDCnoGTimeCIMAX}{\rna}
\newcommand{\SLOWlusearchDCExcTime}{30}
\newcommand{\SLOWlusearchDCExcTimeCI}{3.1}
\newcommand{\SLOWlusearchDCTime}{\rna}
\newcommand{\SLOWlusearchDCTimeCI}{\rna}
\newcommand{\SLOWlusearchDCTimeCIMIN}{\rna}
\newcommand{\SLOWlusearchDCTimeCIMAX}{\rna}
\newcommand{\SLOWlusearchCAPOnoGExcTime}{27}
\newcommand{\SLOWlusearchCAPOnoGExcTimeCI}{3.0}
\newcommand{\SLOWlusearchCAPOnoGTime}{\rna}
\newcommand{\SLOWlusearchCAPOnoGTimeCI}{\rna}
\newcommand{\SLOWlusearchCAPOnoGTimeCIMIN}{\rna}
\newcommand{\SLOWlusearchCAPOnoGTimeCIMAX}{\rna}
\newcommand{\SLOWlusearchCAPOExcTime}{29}
\newcommand{\SLOWlusearchCAPOExcTimeCI}{2.8}
\newcommand{\SLOWlusearchCAPOTime}{\rna}
\newcommand{\SLOWlusearchCAPOTimeCI}{\rna}
\newcommand{\SLOWlusearchCAPOTimeCIMIN}{\rna}
\newcommand{\SLOWlusearchCAPOTimeCIMAX}{\rna}
\newcommand{\SLOWlusearchStaticTime}{\rzero}
\newcommand{\SLOWlusearchDynamicTime}{\rzero}
\newcommand{\SLOWlusearchBaseMem}{1,500}
\newcommand{\SLOWlusearchBaseMemCI}{180.0}
\newcommand{\SLOWlusearchHBMem}{12}
\newcommand{\SLOWlusearchHBMemCI}{2.6}
\newcommand{\SLOWlusearchFTMem}{\memna}
\newcommand{\SLOWlusearchFTMemCI}{\memna}
\newcommand{\SLOWlusearchFTMemCIMIN}{\memna}
\newcommand{\SLOWlusearchFTMemCIMAX}{\memna}
\newcommand{\SLOWlusearchWCPMem}{18}
\newcommand{\SLOWlusearchWCPMemCI}{4.5}
\newcommand{\SLOWlusearchDCnoGExcMem}{14}
\newcommand{\SLOWlusearchDCnoGExcMemCI}{2.7}
\newcommand{\SLOWlusearchDCnoGMem}{\memna}
\newcommand{\SLOWlusearchDCnoGMemCI}{\memna}
\newcommand{\SLOWlusearchDCnoGMemCIMIN}{\memna}
\newcommand{\SLOWlusearchDCnoGMemCIMAX}{\memna}
\newcommand{\SLOWlusearchDCExcMem}{14}
\newcommand{\SLOWlusearchDCExcMemCI}{3.0}
\newcommand{\SLOWlusearchDCMem}{\memna}
\newcommand{\SLOWlusearchDCMemCI}{\memna}
\newcommand{\SLOWlusearchDCMemCIMIN}{\memna}
\newcommand{\SLOWlusearchDCMemCIMAX}{\memna}
\newcommand{\SLOWlusearchCAPOnoGExcMem}{13}
\newcommand{\SLOWlusearchCAPOnoGExcMemCI}{2.9}
\newcommand{\SLOWlusearchCAPOnoGMem}{\memna}
\newcommand{\SLOWlusearchCAPOnoGMemCI}{\memna}
\newcommand{\SLOWlusearchCAPOnoGMemCIMIN}{\memna}
\newcommand{\SLOWlusearchCAPOnoGMemCIMAX}{\memna}
\newcommand{\SLOWlusearchCAPOExcMem}{14}
\newcommand{\SLOWlusearchCAPOExcMemCI}{2.6}
\newcommand{\SLOWlusearchCAPOMem}{\memna}
\newcommand{\SLOWlusearchCAPOMemCI}{\memna}
\newcommand{\SLOWlusearchCAPOMemCIMIN}{\memna}
\newcommand{\SLOWlusearchCAPOMemCIMAX}{\memna}
\newcommand{\SLOWlusearchEventsCI}{21}
\newcommand{\SLOWlusearchEventsCIMIN}{1,443,227,212}
\newcommand{\SLOWlusearchEventsCIMAX}{1,443,227,254}
\newcommand{\SLOWlusearchNoFPEventsCI}{22}
\newcommand{\SLOWlusearchNoFPEventsCIMIN}{194,058,044}
\newcommand{\SLOWlusearchNoFPEventsCIMAX}{194,058,088}
\newcommand{\SLOWlusearchHB}{0}
\newcommand{\SLOWlusearchHBCI}{0}
\newcommand{\SLOWlusearchHBCIMIN}{0}
\newcommand{\SLOWlusearchHBCIMAX}{0}
\newcommand{\SLOWlusearchHBDynamic}{0}
\newcommand{\SLOWlusearchHBDynamicCI}{0}
\newcommand{\SLOWlusearchHBDynamicCIMIN}{0}
\newcommand{\SLOWlusearchHBDynamicCIMAX}{0}
\newcommand{\SLOWlusearchFT}{\rna}
\newcommand{\SLOWlusearchFTCI}{\rna}
\newcommand{\SLOWlusearchFTCIMIN}{\rna}
\newcommand{\SLOWlusearchFTCIMAX}{\rna}
\newcommand{\SLOWlusearchFTDynamic}{\rna}
\newcommand{\SLOWlusearchFTDynamicCI}{\rna}
\newcommand{\SLOWlusearchFTDynamicCIMIN}{\rna}
\newcommand{\SLOWlusearchFTDynamicCIMAX}{\rna}
\newcommand{\SLOWlusearchWCP}{0}
\newcommand{\SLOWlusearchWCPCI}{0}
\newcommand{\SLOWlusearchWCPCIMIN}{0}
\newcommand{\SLOWlusearchWCPCIMAX}{0}
\newcommand{\SLOWlusearchWCPDynamic}{0}
\newcommand{\SLOWlusearchWCPDynamicCI}{0}
\newcommand{\SLOWlusearchWCPDynamicCIMIN}{0}
\newcommand{\SLOWlusearchWCPDynamicCIMAX}{0}
\newcommand{\SLOWlusearchDCnoGExc}{0}
\newcommand{\SLOWlusearchDCnoGExcCI}{0}
\newcommand{\SLOWlusearchDCnoGExcCIMIN}{0}
\newcommand{\SLOWlusearchDCnoGExcCIMAX}{0}
\newcommand{\SLOWlusearchDCnoGExcDynamic}{0}
\newcommand{\SLOWlusearchDCnoGExcDynamicCI}{0}
\newcommand{\SLOWlusearchDCnoGExcDynamicCIMIN}{0}
\newcommand{\SLOWlusearchDCnoGExcDynamicCIMAX}{0}
\newcommand{\SLOWlusearchDCnoG}{\rna}
\newcommand{\SLOWlusearchDCnoGCI}{\rna}
\newcommand{\SLOWlusearchDCnoGCIMIN}{\rna}
\newcommand{\SLOWlusearchDCnoGCIMAX}{\rna}
\newcommand{\SLOWlusearchDCnoGDynamic}{\rna}
\newcommand{\SLOWlusearchDCnoGDynamicCI}{\rna}
\newcommand{\SLOWlusearchDCnoGDynamicCIMIN}{\rna}
\newcommand{\SLOWlusearchDCnoGDynamicCIMAX}{\rna}
\newcommand{\SLOWlusearchDCExc}{0}
\newcommand{\SLOWlusearchDCExcCI}{0}
\newcommand{\SLOWlusearchDCExcCIMIN}{0}
\newcommand{\SLOWlusearchDCExcCIMAX}{0}
\newcommand{\SLOWlusearchDCExcDynamic}{0}
\newcommand{\SLOWlusearchDCExcDynamicCI}{0}
\newcommand{\SLOWlusearchDCExcDynamicCIMIN}{0}
\newcommand{\SLOWlusearchDCExcDynamicCIMAX}{0}
\newcommand{\SLOWlusearchDC}{\rna}
\newcommand{\SLOWlusearchDCCI}{\rna}
\newcommand{\SLOWlusearchDCCIMIN}{\rna}
\newcommand{\SLOWlusearchDCCIMAX}{\rna}
\newcommand{\SLOWlusearchDCDynamic}{\rna}
\newcommand{\SLOWlusearchDCDynamicCI}{\rna}
\newcommand{\SLOWlusearchDCDynamicCIMIN}{\rna}
\newcommand{\SLOWlusearchDCDynamicCIMAX}{\rna}
\newcommand{\SLOWlusearchCAPOnoGExc}{0}
\newcommand{\SLOWlusearchCAPOnoGExcCI}{0}
\newcommand{\SLOWlusearchCAPOnoGExcCIMIN}{0}
\newcommand{\SLOWlusearchCAPOnoGExcCIMAX}{0}
\newcommand{\SLOWlusearchCAPOnoGExcDynamic}{0}
\newcommand{\SLOWlusearchCAPOnoGExcDynamicCI}{0}
\newcommand{\SLOWlusearchCAPOnoGExcDynamicCIMIN}{0}
\newcommand{\SLOWlusearchCAPOnoGExcDynamicCIMAX}{0}
\newcommand{\SLOWlusearchCAPOnoG}{\rna}
\newcommand{\SLOWlusearchCAPOnoGCI}{\rna}
\newcommand{\SLOWlusearchCAPOnoGCIMIN}{\rna}
\newcommand{\SLOWlusearchCAPOnoGCIMAX}{\rna}
\newcommand{\SLOWlusearchCAPOnoGDynamic}{\rna}
\newcommand{\SLOWlusearchCAPOnoGDynamicCI}{\rna}
\newcommand{\SLOWlusearchCAPOnoGDynamicCIMIN}{\rna}
\newcommand{\SLOWlusearchCAPOnoGDynamicCIMAX}{\rna}
\newcommand{\SLOWlusearchCAPOExc}{0}
\newcommand{\SLOWlusearchCAPOExcCI}{0}
\newcommand{\SLOWlusearchCAPOExcCIMIN}{0}
\newcommand{\SLOWlusearchCAPOExcCIMAX}{0}
\newcommand{\SLOWlusearchCAPOExcDynamic}{0}
\newcommand{\SLOWlusearchCAPOExcDynamicCI}{0}
\newcommand{\SLOWlusearchCAPOExcDynamicCIMIN}{0}
\newcommand{\SLOWlusearchCAPOExcDynamicCIMAX}{0}
\newcommand{\SLOWlusearchCAPO}{\rna}
\newcommand{\SLOWlusearchCAPOCI}{\rna}
\newcommand{\SLOWlusearchCAPOCIMIN}{\rna}
\newcommand{\SLOWlusearchCAPOCIMAX}{\rna}
\newcommand{\SLOWlusearchCAPODynamic}{\rna}
\newcommand{\SLOWlusearchCAPODynamicCI}{\rna}
\newcommand{\SLOWlusearchCAPODynamicCIMIN}{\rna}
\newcommand{\SLOWlusearchCAPODynamicCIMAX}{\rna}
\newcommand{\SLOWlusearchPIP}{\rna}
\newcommand{\SLOWlusearchPIPCI}{\rna}
\newcommand{\SLOWlusearchPIPCIMIN}{\rna}
\newcommand{\SLOWlusearchPIPCIMAX}{\rna}
\newcommand{\SLOWlusearchPIPDynamic}{\rna}
\newcommand{\SLOWlusearchPIPDynamicCI}{\rna}
\newcommand{\SLOWlusearchPIPDynamicCIMIN}{\rna}
\newcommand{\SLOWlusearchPIPDynamicCIMAX}{\rna}
\newcommand{\SLOWlusearchHBUP}{0}
\newcommand{\SLOWlusearchHBUPCI}{0}
\newcommand{\SLOWlusearchHBUPCIMIN}{0}
\newcommand{\SLOWlusearchHBUPCIMAX}{0}
\newcommand{\SLOWlusearchHBDynamicUP}{0}
\newcommand{\SLOWlusearchHBDynamicUPCI}{0}
\newcommand{\SLOWlusearchHBDynamicUPCIMIN}{0}
\newcommand{\SLOWlusearchHBDynamicUPCIMAX}{0}
\newcommand{\SLOWlusearchWCPUP}{0}
\newcommand{\SLOWlusearchWCPUPCI}{0}
\newcommand{\SLOWlusearchWCPUPCIMIN}{0}
\newcommand{\SLOWlusearchWCPUPCIMAX}{0}
\newcommand{\SLOWlusearchWCPDynamicUP}{0}
\newcommand{\SLOWlusearchWCPDynamicUPCI}{0}
\newcommand{\SLOWlusearchWCPDynamicUPCIMIN}{0}
\newcommand{\SLOWlusearchWCPDynamicUPCIMAX}{0}
\newcommand{\SLOWlusearchWDCnoGUP}{\rna}
\newcommand{\SLOWlusearchWDCnoGUPCI}{\rna}
\newcommand{\SLOWlusearchWDCnoGUPCIMIN}{\rna}
\newcommand{\SLOWlusearchWDCnoGUPCIMAX}{\rna}
\newcommand{\SLOWlusearchWDCnoGDynamicUP}{\rna}
\newcommand{\SLOWlusearchWDCnoGDynamicUPCI}{\rna}
\newcommand{\SLOWlusearchWDCnoGDynamicUPCIMIN}{\rna}
\newcommand{\SLOWlusearchWDCnoGDynamicUPCIMAX}{\rna}
\newcommand{\SLOWlusearchWDCUP}{\rna}
\newcommand{\SLOWlusearchWDCUPCI}{\rna}
\newcommand{\SLOWlusearchWDCUPCIMIN}{\rna}
\newcommand{\SLOWlusearchWDCUPCIMAX}{\rna}
\newcommand{\SLOWlusearchWDCDynamicUP}{\rna}
\newcommand{\SLOWlusearchWDCDynamicUPCI}{\rna}
\newcommand{\SLOWlusearchWDCDynamicUPCIMIN}{\rna}
\newcommand{\SLOWlusearchWDCDynamicUPCIMAX}{\rna}
\newcommand{\SLOWlusearchCAPOnoGUP}{\rna}
\newcommand{\SLOWlusearchCAPOnoGUPCI}{\rna}
\newcommand{\SLOWlusearchCAPOnoGUPCIMIN}{\rna}
\newcommand{\SLOWlusearchCAPOnoGUPCIMAX}{\rna}
\newcommand{\SLOWlusearchCAPOnoGDynamicUP}{\rna}
\newcommand{\SLOWlusearchCAPOnoGDynamicUPCI}{\rna}
\newcommand{\SLOWlusearchCAPOnoGDynamicUPCIMIN}{\rna}
\newcommand{\SLOWlusearchCAPOnoGDynamicUPCIMAX}{\rna}
\newcommand{\SLOWlusearchCAPOUP}{\rna}
\newcommand{\SLOWlusearchCAPOUPCI}{\rna}
\newcommand{\SLOWlusearchCAPOUPCIMIN}{\rna}
\newcommand{\SLOWlusearchCAPOUPCIMAX}{\rna}
\newcommand{\SLOWlusearchCAPODynamicUP}{\rna}
\newcommand{\SLOWlusearchCAPODynamicUPCI}{\rna}
\newcommand{\SLOWlusearchCAPODynamicUPCIMIN}{\rna}
\newcommand{\SLOWlusearchCAPODynamicUPCIMAX}{\rna}
\newcommand{\SLOWlusearchPIPUP}{\rna}
\newcommand{\SLOWlusearchPIPUPCI}{\rna}
\newcommand{\SLOWlusearchPIPUPCIMIN}{\rna}
\newcommand{\SLOWlusearchPIPUPCIMAX}{\rna}
\newcommand{\SLOWlusearchPIPDynamicUP}{\rna}
\newcommand{\SLOWlusearchPIPDynamicUPCI}{\rna}
\newcommand{\SLOWlusearchPIPDynamicUPCIMIN}{\rna}
\newcommand{\SLOWlusearchPIPDynamicUPCIMAX}{\rna}
\newcommand{\SLOWlusearchPIPHB}{\rna}
\newcommand{\SLOWlusearchPIPHBCI}{\rna}
\newcommand{\SLOWlusearchPIPHBCIMIN}{\rna}
\newcommand{\SLOWlusearchPIPHBCIMAX}{\rna}
\newcommand{\SLOWlusearchPIPHBDynamic}{\rna}
\newcommand{\SLOWlusearchPIPHBDynamicCI}{\rna}
\newcommand{\SLOWlusearchPIPHBDynamicCIMIN}{\rna}
\newcommand{\SLOWlusearchPIPHBDynamicCIMAX}{\rna}
\newcommand{\SLOWlusearchPIPWCP}{\rna}
\newcommand{\SLOWlusearchPIPWCPCI}{\rna}
\newcommand{\SLOWlusearchPIPWCPCIMIN}{\rna}
\newcommand{\SLOWlusearchPIPWCPCIMAX}{\rna}
\newcommand{\SLOWlusearchPIPWCPDynamic}{\rna}
\newcommand{\SLOWlusearchPIPWCPDynamicCI}{\rna}
\newcommand{\SLOWlusearchPIPWCPDynamicCIMIN}{\rna}
\newcommand{\SLOWlusearchPIPWCPDynamicCIMAX}{\rna}
\newcommand{\SLOWlusearchPIPWDC}{\rna}
\newcommand{\SLOWlusearchPIPWDCCI}{\rna}
\newcommand{\SLOWlusearchPIPWDCCIMIN}{\rna}
\newcommand{\SLOWlusearchPIPWDCCIMAX}{\rna}
\newcommand{\SLOWlusearchPIPWDCDynamic}{\rna}
\newcommand{\SLOWlusearchPIPWDCDynamicCI}{\rna}
\newcommand{\SLOWlusearchPIPWDCDynamicCIMIN}{\rna}
\newcommand{\SLOWlusearchPIPWDCDynamicCIMAX}{\rna}
\newcommand{\SLOWlusearchPIPCAPO}{\rna}
\newcommand{\SLOWlusearchPIPCAPOCI}{\rna}
\newcommand{\SLOWlusearchPIPCAPOCIMIN}{\rna}
\newcommand{\SLOWlusearchPIPCAPOCIMAX}{\rna}
\newcommand{\SLOWlusearchPIPCAPODynamic}{\rna}
\newcommand{\SLOWlusearchPIPCAPODynamicCI}{\rna}
\newcommand{\SLOWlusearchPIPCAPODynamicCIMIN}{\rna}
\newcommand{\SLOWlusearchPIPCAPODynamicCIMAX}{\rna}
\newcommand{\SLOWlusearchPIPPIP}{\rna}
\newcommand{\SLOWlusearchPIPPIPCI}{\rna}
\newcommand{\SLOWlusearchPIPPIPCIMIN}{\rna}
\newcommand{\SLOWlusearchPIPPIPCIMAX}{\rna}
\newcommand{\SLOWlusearchPIPPIPDynamic}{\rna}
\newcommand{\SLOWlusearchPIPPIPDynamicCI}{\rna}
\newcommand{\SLOWlusearchPIPPIPDynamicCIMIN}{\rna}
\newcommand{\SLOWlusearchPIPPIPDynamicCIMAX}{\rna}
\newcommand{\SLOWpmdEvents}{200}
\newcommand{\SLOWpmdNoFPEvents}{25}
\newcommand{\SLOWpmdMaxLiveThreads}{9}
\newcommand{\SLOWpmdTotalThreads}{9}
\newcommand{\SLOWpmdBaseTime}{1.4}
\newcommand{\SLOWpmdBaseTimeCI}{28}
\newcommand{\SLOWpmdEmptyTime}{\rna}
\newcommand{\SLOWpmdEmptyTimeCI}{\rna}
\newcommand{\SLOWpmdEmptyTimeCIMIN}{\rna}
\newcommand{\SLOWpmdEmptyTimeCIMAX}{\rna}
\newcommand{\SLOWpmdFTTime}{\rna}
\newcommand{\SLOWpmdFTTimeCI}{\rna}
\newcommand{\SLOWpmdFTTimeCIMIN}{\rna}
\newcommand{\SLOWpmdFTTimeCIMAX}{\rna}
\newcommand{\SLOWpmdHBTime}{13}
\newcommand{\SLOWpmdHBTimeCI}{0.38}
\newcommand{\SLOWpmdWCPTime}{16}
\newcommand{\SLOWpmdWCPTimeCI}{0.46}
\newcommand{\SLOWpmdDCnoGExcTime}{14}
\newcommand{\SLOWpmdDCnoGExcTimeCI}{0.56}
\newcommand{\SLOWpmdDCnoGTime}{\rna}
\newcommand{\SLOWpmdDCnoGTimeCI}{\rna}
\newcommand{\SLOWpmdDCnoGTimeCIMIN}{\rna}
\newcommand{\SLOWpmdDCnoGTimeCIMAX}{\rna}
\newcommand{\SLOWpmdDCExcTime}{15}
\newcommand{\SLOWpmdDCExcTimeCI}{0.4}
\newcommand{\SLOWpmdDCTime}{\rna}
\newcommand{\SLOWpmdDCTimeCI}{\rna}
\newcommand{\SLOWpmdDCTimeCIMIN}{\rna}
\newcommand{\SLOWpmdDCTimeCIMAX}{\rna}
\newcommand{\SLOWpmdCAPOnoGExcTime}{14}
\newcommand{\SLOWpmdCAPOnoGExcTimeCI}{0.42}
\newcommand{\SLOWpmdCAPOnoGTime}{\rna}
\newcommand{\SLOWpmdCAPOnoGTimeCI}{\rna}
\newcommand{\SLOWpmdCAPOnoGTimeCIMIN}{\rna}
\newcommand{\SLOWpmdCAPOnoGTimeCIMAX}{\rna}
\newcommand{\SLOWpmdCAPOExcTime}{15}
\newcommand{\SLOWpmdCAPOExcTimeCI}{0.45}
\newcommand{\SLOWpmdCAPOTime}{\rna}
\newcommand{\SLOWpmdCAPOTimeCI}{\rna}
\newcommand{\SLOWpmdCAPOTimeCIMIN}{\rna}
\newcommand{\SLOWpmdCAPOTimeCIMAX}{\rna}
\newcommand{\SLOWpmdStaticTime}{\rzero}
\newcommand{\SLOWpmdDynamicTime}{\rzero}
\newcommand{\SLOWpmdBaseMem}{590}
\newcommand{\SLOWpmdBaseMemCI}{9.9}
\newcommand{\SLOWpmdHBMem}{11}
\newcommand{\SLOWpmdHBMemCI}{0.32}
\newcommand{\SLOWpmdFTMem}{\memna}
\newcommand{\SLOWpmdFTMemCI}{\memna}
\newcommand{\SLOWpmdFTMemCIMIN}{\memna}
\newcommand{\SLOWpmdFTMemCIMAX}{\memna}
\newcommand{\SLOWpmdWCPMem}{17}
\newcommand{\SLOWpmdWCPMemCI}{0.4}
\newcommand{\SLOWpmdDCnoGExcMem}{12}
\newcommand{\SLOWpmdDCnoGExcMemCI}{0.95}
\newcommand{\SLOWpmdDCnoGMem}{\memna}
\newcommand{\SLOWpmdDCnoGMemCI}{\memna}
\newcommand{\SLOWpmdDCnoGMemCIMIN}{\memna}
\newcommand{\SLOWpmdDCnoGMemCIMAX}{\memna}
\newcommand{\SLOWpmdDCExcMem}{13}
\newcommand{\SLOWpmdDCExcMemCI}{0.26}
\newcommand{\SLOWpmdDCMem}{\memna}
\newcommand{\SLOWpmdDCMemCI}{\memna}
\newcommand{\SLOWpmdDCMemCIMIN}{\memna}
\newcommand{\SLOWpmdDCMemCIMAX}{\memna}
\newcommand{\SLOWpmdCAPOnoGExcMem}{12}
\newcommand{\SLOWpmdCAPOnoGExcMemCI}{0.83}
\newcommand{\SLOWpmdCAPOnoGMem}{\memna}
\newcommand{\SLOWpmdCAPOnoGMemCI}{\memna}
\newcommand{\SLOWpmdCAPOnoGMemCIMIN}{\memna}
\newcommand{\SLOWpmdCAPOnoGMemCIMAX}{\memna}
\newcommand{\SLOWpmdCAPOExcMem}{13}
\newcommand{\SLOWpmdCAPOExcMemCI}{0.22}
\newcommand{\SLOWpmdCAPOMem}{\memna}
\newcommand{\SLOWpmdCAPOMemCI}{\memna}
\newcommand{\SLOWpmdCAPOMemCIMIN}{\memna}
\newcommand{\SLOWpmdCAPOMemCIMAX}{\memna}
\newcommand{\SLOWpmdEventsCI}{455}
\newcommand{\SLOWpmdEventsCIMIN}{200,051,915}
\newcommand{\SLOWpmdEventsCIMAX}{200,052,825}
\newcommand{\SLOWpmdNoFPEventsCI}{55}
\newcommand{\SLOWpmdNoFPEventsCIMIN}{25,405,134}
\newcommand{\SLOWpmdNoFPEventsCIMAX}{25,405,244}
\newcommand{\SLOWpmdHB}{8}
\newcommand{\SLOWpmdHBCI}{0}
\newcommand{\SLOWpmdHBCIMIN}{8}
\newcommand{\SLOWpmdHBCIMAX}{8}
\newcommand{\SLOWpmdHBDynamic}{2,773}
\newcommand{\SLOWpmdHBDynamicCI}{565}
\newcommand{\SLOWpmdHBDynamicCIMIN}{2,208}
\newcommand{\SLOWpmdHBDynamicCIMAX}{3,338}
\newcommand{\SLOWpmdFT}{\rna}
\newcommand{\SLOWpmdFTCI}{\rna}
\newcommand{\SLOWpmdFTCIMIN}{\rna}
\newcommand{\SLOWpmdFTCIMAX}{\rna}
\newcommand{\SLOWpmdFTDynamic}{\rna}
\newcommand{\SLOWpmdFTDynamicCI}{\rna}
\newcommand{\SLOWpmdFTDynamicCIMIN}{\rna}
\newcommand{\SLOWpmdFTDynamicCIMAX}{\rna}
\newcommand{\SLOWpmdWCP}{8}
\newcommand{\SLOWpmdWCPCI}{0}
\newcommand{\SLOWpmdWCPCIMIN}{8}
\newcommand{\SLOWpmdWCPCIMAX}{8}
\newcommand{\SLOWpmdWCPDynamic}{1,436}
\newcommand{\SLOWpmdWCPDynamicCI}{638}
\newcommand{\SLOWpmdWCPDynamicCIMIN}{798}
\newcommand{\SLOWpmdWCPDynamicCIMAX}{2,074}
\newcommand{\SLOWpmdDCnoGExc}{12}
\newcommand{\SLOWpmdDCnoGExcCI}{0}
\newcommand{\SLOWpmdDCnoGExcCIMIN}{12}
\newcommand{\SLOWpmdDCnoGExcCIMAX}{12}
\newcommand{\SLOWpmdDCnoGExcDynamic}{9,618}
\newcommand{\SLOWpmdDCnoGExcDynamicCI}{3,490}
\newcommand{\SLOWpmdDCnoGExcDynamicCIMIN}{6,128}
\newcommand{\SLOWpmdDCnoGExcDynamicCIMAX}{13,108}
\newcommand{\SLOWpmdDCnoG}{\rna}
\newcommand{\SLOWpmdDCnoGCI}{\rna}
\newcommand{\SLOWpmdDCnoGCIMIN}{\rna}
\newcommand{\SLOWpmdDCnoGCIMAX}{\rna}
\newcommand{\SLOWpmdDCnoGDynamic}{\rna}
\newcommand{\SLOWpmdDCnoGDynamicCI}{\rna}
\newcommand{\SLOWpmdDCnoGDynamicCIMIN}{\rna}
\newcommand{\SLOWpmdDCnoGDynamicCIMAX}{\rna}
\newcommand{\SLOWpmdDCExc}{5}
\newcommand{\SLOWpmdDCExcCI}{0}
\newcommand{\SLOWpmdDCExcCIMIN}{5}
\newcommand{\SLOWpmdDCExcCIMAX}{5}
\newcommand{\SLOWpmdDCExcDynamic}{28}
\newcommand{\SLOWpmdDCExcDynamicCI}{3}
\newcommand{\SLOWpmdDCExcDynamicCIMIN}{25}
\newcommand{\SLOWpmdDCExcDynamicCIMAX}{31}
\newcommand{\SLOWpmdDC}{\rna}
\newcommand{\SLOWpmdDCCI}{\rna}
\newcommand{\SLOWpmdDCCIMIN}{\rna}
\newcommand{\SLOWpmdDCCIMAX}{\rna}
\newcommand{\SLOWpmdDCDynamic}{\rna}
\newcommand{\SLOWpmdDCDynamicCI}{\rna}
\newcommand{\SLOWpmdDCDynamicCIMIN}{\rna}
\newcommand{\SLOWpmdDCDynamicCIMAX}{\rna}
\newcommand{\SLOWpmdCAPOnoGExc}{12}
\newcommand{\SLOWpmdCAPOnoGExcCI}{0}
\newcommand{\SLOWpmdCAPOnoGExcCIMIN}{12}
\newcommand{\SLOWpmdCAPOnoGExcCIMAX}{12}
\newcommand{\SLOWpmdCAPOnoGExcDynamic}{10,631}
\newcommand{\SLOWpmdCAPOnoGExcDynamicCI}{3,995}
\newcommand{\SLOWpmdCAPOnoGExcDynamicCIMIN}{6,636}
\newcommand{\SLOWpmdCAPOnoGExcDynamicCIMAX}{14,626}
\newcommand{\SLOWpmdCAPOnoG}{\rna}
\newcommand{\SLOWpmdCAPOnoGCI}{\rna}
\newcommand{\SLOWpmdCAPOnoGCIMIN}{\rna}
\newcommand{\SLOWpmdCAPOnoGCIMAX}{\rna}
\newcommand{\SLOWpmdCAPOnoGDynamic}{\rna}
\newcommand{\SLOWpmdCAPOnoGDynamicCI}{\rna}
\newcommand{\SLOWpmdCAPOnoGDynamicCIMIN}{\rna}
\newcommand{\SLOWpmdCAPOnoGDynamicCIMAX}{\rna}
\newcommand{\SLOWpmdCAPOExc}{6}
\newcommand{\SLOWpmdCAPOExcCI}{0}
\newcommand{\SLOWpmdCAPOExcCIMIN}{6}
\newcommand{\SLOWpmdCAPOExcCIMAX}{6}
\newcommand{\SLOWpmdCAPOExcDynamic}{32}
\newcommand{\SLOWpmdCAPOExcDynamicCI}{3}
\newcommand{\SLOWpmdCAPOExcDynamicCIMIN}{29}
\newcommand{\SLOWpmdCAPOExcDynamicCIMAX}{35}
\newcommand{\SLOWpmdCAPO}{\rna}
\newcommand{\SLOWpmdCAPOCI}{\rna}
\newcommand{\SLOWpmdCAPOCIMIN}{\rna}
\newcommand{\SLOWpmdCAPOCIMAX}{\rna}
\newcommand{\SLOWpmdCAPODynamic}{\rna}
\newcommand{\SLOWpmdCAPODynamicCI}{\rna}
\newcommand{\SLOWpmdCAPODynamicCIMIN}{\rna}
\newcommand{\SLOWpmdCAPODynamicCIMAX}{\rna}
\newcommand{\SLOWpmdPIP}{\rna}
\newcommand{\SLOWpmdPIPCI}{\rna}
\newcommand{\SLOWpmdPIPCIMIN}{\rna}
\newcommand{\SLOWpmdPIPCIMAX}{\rna}
\newcommand{\SLOWpmdPIPDynamic}{\rna}
\newcommand{\SLOWpmdPIPDynamicCI}{\rna}
\newcommand{\SLOWpmdPIPDynamicCIMIN}{\rna}
\newcommand{\SLOWpmdPIPDynamicCIMAX}{\rna}
\newcommand{\SLOWpmdHBUP}{8}
\newcommand{\SLOWpmdHBUPCI}{0}
\newcommand{\SLOWpmdHBUPCIMIN}{8}
\newcommand{\SLOWpmdHBUPCIMAX}{8}
\newcommand{\SLOWpmdHBDynamicUP}{2,772}
\newcommand{\SLOWpmdHBDynamicUPCI}{564}
\newcommand{\SLOWpmdHBDynamicUPCIMIN}{2,208}
\newcommand{\SLOWpmdHBDynamicUPCIMAX}{3,336}
\newcommand{\SLOWpmdWCPUP}{8}
\newcommand{\SLOWpmdWCPUPCI}{0}
\newcommand{\SLOWpmdWCPUPCIMIN}{8}
\newcommand{\SLOWpmdWCPUPCIMAX}{8}
\newcommand{\SLOWpmdWCPDynamicUP}{1,436}
\newcommand{\SLOWpmdWCPDynamicUPCI}{638}
\newcommand{\SLOWpmdWCPDynamicUPCIMIN}{798}
\newcommand{\SLOWpmdWCPDynamicUPCIMAX}{2,074}
\newcommand{\SLOWpmdWDCnoGUP}{\rna}
\newcommand{\SLOWpmdWDCnoGUPCI}{\rna}
\newcommand{\SLOWpmdWDCnoGUPCIMIN}{\rna}
\newcommand{\SLOWpmdWDCnoGUPCIMAX}{\rna}
\newcommand{\SLOWpmdWDCnoGDynamicUP}{\rna}
\newcommand{\SLOWpmdWDCnoGDynamicUPCI}{\rna}
\newcommand{\SLOWpmdWDCnoGDynamicUPCIMIN}{\rna}
\newcommand{\SLOWpmdWDCnoGDynamicUPCIMAX}{\rna}
\newcommand{\SLOWpmdWDCUP}{\rna}
\newcommand{\SLOWpmdWDCUPCI}{\rna}
\newcommand{\SLOWpmdWDCUPCIMIN}{\rna}
\newcommand{\SLOWpmdWDCUPCIMAX}{\rna}
\newcommand{\SLOWpmdWDCDynamicUP}{\rna}
\newcommand{\SLOWpmdWDCDynamicUPCI}{\rna}
\newcommand{\SLOWpmdWDCDynamicUPCIMIN}{\rna}
\newcommand{\SLOWpmdWDCDynamicUPCIMAX}{\rna}
\newcommand{\SLOWpmdCAPOnoGUP}{\rna}
\newcommand{\SLOWpmdCAPOnoGUPCI}{\rna}
\newcommand{\SLOWpmdCAPOnoGUPCIMIN}{\rna}
\newcommand{\SLOWpmdCAPOnoGUPCIMAX}{\rna}
\newcommand{\SLOWpmdCAPOnoGDynamicUP}{\rna}
\newcommand{\SLOWpmdCAPOnoGDynamicUPCI}{\rna}
\newcommand{\SLOWpmdCAPOnoGDynamicUPCIMIN}{\rna}
\newcommand{\SLOWpmdCAPOnoGDynamicUPCIMAX}{\rna}
\newcommand{\SLOWpmdCAPOUP}{\rna}
\newcommand{\SLOWpmdCAPOUPCI}{\rna}
\newcommand{\SLOWpmdCAPOUPCIMIN}{\rna}
\newcommand{\SLOWpmdCAPOUPCIMAX}{\rna}
\newcommand{\SLOWpmdCAPODynamicUP}{\rna}
\newcommand{\SLOWpmdCAPODynamicUPCI}{\rna}
\newcommand{\SLOWpmdCAPODynamicUPCIMIN}{\rna}
\newcommand{\SLOWpmdCAPODynamicUPCIMAX}{\rna}
\newcommand{\SLOWpmdPIPUP}{\rna}
\newcommand{\SLOWpmdPIPUPCI}{\rna}
\newcommand{\SLOWpmdPIPUPCIMIN}{\rna}
\newcommand{\SLOWpmdPIPUPCIMAX}{\rna}
\newcommand{\SLOWpmdPIPDynamicUP}{\rna}
\newcommand{\SLOWpmdPIPDynamicUPCI}{\rna}
\newcommand{\SLOWpmdPIPDynamicUPCIMIN}{\rna}
\newcommand{\SLOWpmdPIPDynamicUPCIMAX}{\rna}
\newcommand{\SLOWpmdPIPHB}{\rna}
\newcommand{\SLOWpmdPIPHBCI}{\rna}
\newcommand{\SLOWpmdPIPHBCIMIN}{\rna}
\newcommand{\SLOWpmdPIPHBCIMAX}{\rna}
\newcommand{\SLOWpmdPIPHBDynamic}{\rna}
\newcommand{\SLOWpmdPIPHBDynamicCI}{\rna}
\newcommand{\SLOWpmdPIPHBDynamicCIMIN}{\rna}
\newcommand{\SLOWpmdPIPHBDynamicCIMAX}{\rna}
\newcommand{\SLOWpmdPIPWCP}{\rna}
\newcommand{\SLOWpmdPIPWCPCI}{\rna}
\newcommand{\SLOWpmdPIPWCPCIMIN}{\rna}
\newcommand{\SLOWpmdPIPWCPCIMAX}{\rna}
\newcommand{\SLOWpmdPIPWCPDynamic}{\rna}
\newcommand{\SLOWpmdPIPWCPDynamicCI}{\rna}
\newcommand{\SLOWpmdPIPWCPDynamicCIMIN}{\rna}
\newcommand{\SLOWpmdPIPWCPDynamicCIMAX}{\rna}
\newcommand{\SLOWpmdPIPWDC}{\rna}
\newcommand{\SLOWpmdPIPWDCCI}{\rna}
\newcommand{\SLOWpmdPIPWDCCIMIN}{\rna}
\newcommand{\SLOWpmdPIPWDCCIMAX}{\rna}
\newcommand{\SLOWpmdPIPWDCDynamic}{\rna}
\newcommand{\SLOWpmdPIPWDCDynamicCI}{\rna}
\newcommand{\SLOWpmdPIPWDCDynamicCIMIN}{\rna}
\newcommand{\SLOWpmdPIPWDCDynamicCIMAX}{\rna}
\newcommand{\SLOWpmdPIPCAPO}{\rna}
\newcommand{\SLOWpmdPIPCAPOCI}{\rna}
\newcommand{\SLOWpmdPIPCAPOCIMIN}{\rna}
\newcommand{\SLOWpmdPIPCAPOCIMAX}{\rna}
\newcommand{\SLOWpmdPIPCAPODynamic}{\rna}
\newcommand{\SLOWpmdPIPCAPODynamicCI}{\rna}
\newcommand{\SLOWpmdPIPCAPODynamicCIMIN}{\rna}
\newcommand{\SLOWpmdPIPCAPODynamicCIMAX}{\rna}
\newcommand{\SLOWpmdPIPPIP}{\rna}
\newcommand{\SLOWpmdPIPPIPCI}{\rna}
\newcommand{\SLOWpmdPIPPIPCIMIN}{\rna}
\newcommand{\SLOWpmdPIPPIPCIMAX}{\rna}
\newcommand{\SLOWpmdPIPPIPDynamic}{\rna}
\newcommand{\SLOWpmdPIPPIPDynamicCI}{\rna}
\newcommand{\SLOWpmdPIPPIPDynamicCIMIN}{\rna}
\newcommand{\SLOWpmdPIPPIPDynamicCIMAX}{\rna}
\newcommand{\SLOWsunflowEvents}{9,700}
\newcommand{\SLOWsunflowNoFPEvents}{760}
\newcommand{\SLOWsunflowMaxLiveThreads}{17}
\newcommand{\SLOWsunflowTotalThreads}{17}
\newcommand{\SLOWsunflowBaseTime}{1.4}
\newcommand{\SLOWsunflowBaseTimeCI}{43}
\newcommand{\SLOWsunflowEmptyTime}{\rna}
\newcommand{\SLOWsunflowEmptyTimeCI}{\rna}
\newcommand{\SLOWsunflowEmptyTimeCIMIN}{\rna}
\newcommand{\SLOWsunflowEmptyTimeCIMAX}{\rna}
\newcommand{\SLOWsunflowFTTime}{\rna}
\newcommand{\SLOWsunflowFTTimeCI}{\rna}
\newcommand{\SLOWsunflowFTTimeCIMIN}{\rna}
\newcommand{\SLOWsunflowFTTimeCIMAX}{\rna}
\newcommand{\SLOWsunflowHBTime}{85}
\newcommand{\SLOWsunflowHBTimeCI}{3.2}
\newcommand{\SLOWsunflowWCPTime}{110}
\newcommand{\SLOWsunflowWCPTimeCI}{3.2}
\newcommand{\SLOWsunflowDCnoGExcTime}{95}
\newcommand{\SLOWsunflowDCnoGExcTimeCI}{12}
\newcommand{\SLOWsunflowDCnoGTime}{\rna}
\newcommand{\SLOWsunflowDCnoGTimeCI}{\rna}
\newcommand{\SLOWsunflowDCnoGTimeCIMIN}{\rna}
\newcommand{\SLOWsunflowDCnoGTimeCIMAX}{\rna}
\newcommand{\SLOWsunflowDCExcTime}{100}
\newcommand{\SLOWsunflowDCExcTimeCI}{11}
\newcommand{\SLOWsunflowDCTime}{\rna}
\newcommand{\SLOWsunflowDCTimeCI}{\rna}
\newcommand{\SLOWsunflowDCTimeCIMIN}{\rna}
\newcommand{\SLOWsunflowDCTimeCIMAX}{\rna}
\newcommand{\SLOWsunflowCAPOnoGExcTime}{85}
\newcommand{\SLOWsunflowCAPOnoGExcTimeCI}{3.2}
\newcommand{\SLOWsunflowCAPOnoGTime}{\rna}
\newcommand{\SLOWsunflowCAPOnoGTimeCI}{\rna}
\newcommand{\SLOWsunflowCAPOnoGTimeCIMIN}{\rna}
\newcommand{\SLOWsunflowCAPOnoGTimeCIMAX}{\rna}
\newcommand{\SLOWsunflowCAPOExcTime}{100}
\newcommand{\SLOWsunflowCAPOExcTimeCI}{12}
\newcommand{\SLOWsunflowCAPOTime}{\rna}
\newcommand{\SLOWsunflowCAPOTimeCI}{\rna}
\newcommand{\SLOWsunflowCAPOTimeCIMIN}{\rna}
\newcommand{\SLOWsunflowCAPOTimeCIMAX}{\rna}
\newcommand{\SLOWsunflowStaticTime}{\rzero}
\newcommand{\SLOWsunflowDynamicTime}{\rzero}
\newcommand{\SLOWsunflowBaseMem}{610}
\newcommand{\SLOWsunflowBaseMemCI}{0.78}
\newcommand{\SLOWsunflowHBMem}{23}
\newcommand{\SLOWsunflowHBMemCI}{1.1}
\newcommand{\SLOWsunflowFTMem}{\memna}
\newcommand{\SLOWsunflowFTMemCI}{\memna}
\newcommand{\SLOWsunflowFTMemCIMIN}{\memna}
\newcommand{\SLOWsunflowFTMemCIMAX}{\memna}
\newcommand{\SLOWsunflowWCPMem}{54}
\newcommand{\SLOWsunflowWCPMemCI}{0.7}
\newcommand{\SLOWsunflowDCnoGExcMem}{25}
\newcommand{\SLOWsunflowDCnoGExcMemCI}{0.78}
\newcommand{\SLOWsunflowDCnoGMem}{\memna}
\newcommand{\SLOWsunflowDCnoGMemCI}{\memna}
\newcommand{\SLOWsunflowDCnoGMemCIMIN}{\memna}
\newcommand{\SLOWsunflowDCnoGMemCIMAX}{\memna}
\newcommand{\SLOWsunflowDCExcMem}{26}
\newcommand{\SLOWsunflowDCExcMemCI}{0.72}
\newcommand{\SLOWsunflowDCMem}{\memna}
\newcommand{\SLOWsunflowDCMemCI}{\memna}
\newcommand{\SLOWsunflowDCMemCIMIN}{\memna}
\newcommand{\SLOWsunflowDCMemCIMAX}{\memna}
\newcommand{\SLOWsunflowCAPOnoGExcMem}{25}
\newcommand{\SLOWsunflowCAPOnoGExcMemCI}{0.52}
\newcommand{\SLOWsunflowCAPOnoGMem}{\memna}
\newcommand{\SLOWsunflowCAPOnoGMemCI}{\memna}
\newcommand{\SLOWsunflowCAPOnoGMemCIMIN}{\memna}
\newcommand{\SLOWsunflowCAPOnoGMemCIMAX}{\memna}
\newcommand{\SLOWsunflowCAPOExcMem}{26}
\newcommand{\SLOWsunflowCAPOExcMemCI}{0.9}
\newcommand{\SLOWsunflowCAPOMem}{\memna}
\newcommand{\SLOWsunflowCAPOMemCI}{\memna}
\newcommand{\SLOWsunflowCAPOMemCIMIN}{\memna}
\newcommand{\SLOWsunflowCAPOMemCIMAX}{\memna}
\newcommand{\SLOWsunflowEventsCI}{1}
\newcommand{\SLOWsunflowEventsCIMIN}{9,681,321,388}
\newcommand{\SLOWsunflowEventsCIMAX}{9,681,321,390}
\newcommand{\SLOWsunflowNoFPEventsCI}{5}
\newcommand{\SLOWsunflowNoFPEventsCIMIN}{764,388,873}
\newcommand{\SLOWsunflowNoFPEventsCIMAX}{764,388,883}
\newcommand{\SLOWsunflowHB}{5}
\newcommand{\SLOWsunflowHBCI}{0}
\newcommand{\SLOWsunflowHBCIMIN}{5}
\newcommand{\SLOWsunflowHBCIMAX}{5}
\newcommand{\SLOWsunflowHBDynamic}{42}
\newcommand{\SLOWsunflowHBDynamicCI}{5}
\newcommand{\SLOWsunflowHBDynamicCIMIN}{37}
\newcommand{\SLOWsunflowHBDynamicCIMAX}{47}
\newcommand{\SLOWsunflowFT}{\rna}
\newcommand{\SLOWsunflowFTCI}{\rna}
\newcommand{\SLOWsunflowFTCIMIN}{\rna}
\newcommand{\SLOWsunflowFTCIMAX}{\rna}
\newcommand{\SLOWsunflowFTDynamic}{\rna}
\newcommand{\SLOWsunflowFTDynamicCI}{\rna}
\newcommand{\SLOWsunflowFTDynamicCIMIN}{\rna}
\newcommand{\SLOWsunflowFTDynamicCIMAX}{\rna}
\newcommand{\SLOWsunflowWCP}{26}
\newcommand{\SLOWsunflowWCPCI}{0}
\newcommand{\SLOWsunflowWCPCIMIN}{26}
\newcommand{\SLOWsunflowWCPCIMAX}{26}
\newcommand{\SLOWsunflowWCPDynamic}{742}
\newcommand{\SLOWsunflowWCPDynamicCI}{12}
\newcommand{\SLOWsunflowWCPDynamicCIMIN}{730}
\newcommand{\SLOWsunflowWCPDynamicCIMAX}{754}
\newcommand{\SLOWsunflowDCnoGExc}{28}
\newcommand{\SLOWsunflowDCnoGExcCI}{0}
\newcommand{\SLOWsunflowDCnoGExcCIMIN}{28}
\newcommand{\SLOWsunflowDCnoGExcCIMAX}{28}
\newcommand{\SLOWsunflowDCnoGExcDynamic}{1,462}
\newcommand{\SLOWsunflowDCnoGExcDynamicCI}{317}
\newcommand{\SLOWsunflowDCnoGExcDynamicCIMIN}{1,145}
\newcommand{\SLOWsunflowDCnoGExcDynamicCIMAX}{1,779}
\newcommand{\SLOWsunflowDCnoG}{\rna}
\newcommand{\SLOWsunflowDCnoGCI}{\rna}
\newcommand{\SLOWsunflowDCnoGCIMIN}{\rna}
\newcommand{\SLOWsunflowDCnoGCIMAX}{\rna}
\newcommand{\SLOWsunflowDCnoGDynamic}{\rna}
\newcommand{\SLOWsunflowDCnoGDynamicCI}{\rna}
\newcommand{\SLOWsunflowDCnoGDynamicCIMIN}{\rna}
\newcommand{\SLOWsunflowDCnoGDynamicCIMAX}{\rna}
\newcommand{\SLOWsunflowDCExc}{3}
\newcommand{\SLOWsunflowDCExcCI}{0}
\newcommand{\SLOWsunflowDCExcCIMIN}{3}
\newcommand{\SLOWsunflowDCExcCIMAX}{3}
\newcommand{\SLOWsunflowDCExcDynamic}{14}
\newcommand{\SLOWsunflowDCExcDynamicCI}{0}
\newcommand{\SLOWsunflowDCExcDynamicCIMIN}{14}
\newcommand{\SLOWsunflowDCExcDynamicCIMAX}{14}
\newcommand{\SLOWsunflowDC}{\rna}
\newcommand{\SLOWsunflowDCCI}{\rna}
\newcommand{\SLOWsunflowDCCIMIN}{\rna}
\newcommand{\SLOWsunflowDCCIMAX}{\rna}
\newcommand{\SLOWsunflowDCDynamic}{\rna}
\newcommand{\SLOWsunflowDCDynamicCI}{\rna}
\newcommand{\SLOWsunflowDCDynamicCIMIN}{\rna}
\newcommand{\SLOWsunflowDCDynamicCIMAX}{\rna}
\newcommand{\SLOWsunflowCAPOnoGExc}{28}
\newcommand{\SLOWsunflowCAPOnoGExcCI}{0}
\newcommand{\SLOWsunflowCAPOnoGExcCIMIN}{28}
\newcommand{\SLOWsunflowCAPOnoGExcCIMAX}{28}
\newcommand{\SLOWsunflowCAPOnoGExcDynamic}{1,323}
\newcommand{\SLOWsunflowCAPOnoGExcDynamicCI}{236}
\newcommand{\SLOWsunflowCAPOnoGExcDynamicCIMIN}{1,087}
\newcommand{\SLOWsunflowCAPOnoGExcDynamicCIMAX}{1,559}
\newcommand{\SLOWsunflowCAPOnoG}{\rna}
\newcommand{\SLOWsunflowCAPOnoGCI}{\rna}
\newcommand{\SLOWsunflowCAPOnoGCIMIN}{\rna}
\newcommand{\SLOWsunflowCAPOnoGCIMAX}{\rna}
\newcommand{\SLOWsunflowCAPOnoGDynamic}{\rna}
\newcommand{\SLOWsunflowCAPOnoGDynamicCI}{\rna}
\newcommand{\SLOWsunflowCAPOnoGDynamicCIMIN}{\rna}
\newcommand{\SLOWsunflowCAPOnoGDynamicCIMAX}{\rna}
\newcommand{\SLOWsunflowCAPOExc}{3}
\newcommand{\SLOWsunflowCAPOExcCI}{0}
\newcommand{\SLOWsunflowCAPOExcCIMIN}{3}
\newcommand{\SLOWsunflowCAPOExcCIMAX}{3}
\newcommand{\SLOWsunflowCAPOExcDynamic}{14}
\newcommand{\SLOWsunflowCAPOExcDynamicCI}{0}
\newcommand{\SLOWsunflowCAPOExcDynamicCIMIN}{14}
\newcommand{\SLOWsunflowCAPOExcDynamicCIMAX}{14}
\newcommand{\SLOWsunflowCAPO}{\rna}
\newcommand{\SLOWsunflowCAPOCI}{\rna}
\newcommand{\SLOWsunflowCAPOCIMIN}{\rna}
\newcommand{\SLOWsunflowCAPOCIMAX}{\rna}
\newcommand{\SLOWsunflowCAPODynamic}{\rna}
\newcommand{\SLOWsunflowCAPODynamicCI}{\rna}
\newcommand{\SLOWsunflowCAPODynamicCIMIN}{\rna}
\newcommand{\SLOWsunflowCAPODynamicCIMAX}{\rna}
\newcommand{\SLOWsunflowPIP}{\rna}
\newcommand{\SLOWsunflowPIPCI}{\rna}
\newcommand{\SLOWsunflowPIPCIMIN}{\rna}
\newcommand{\SLOWsunflowPIPCIMAX}{\rna}
\newcommand{\SLOWsunflowPIPDynamic}{\rna}
\newcommand{\SLOWsunflowPIPDynamicCI}{\rna}
\newcommand{\SLOWsunflowPIPDynamicCIMIN}{\rna}
\newcommand{\SLOWsunflowPIPDynamicCIMAX}{\rna}
\newcommand{\SLOWsunflowHBUP}{5}
\newcommand{\SLOWsunflowHBUPCI}{0}
\newcommand{\SLOWsunflowHBUPCIMIN}{5}
\newcommand{\SLOWsunflowHBUPCIMAX}{5}
\newcommand{\SLOWsunflowHBDynamicUP}{42}
\newcommand{\SLOWsunflowHBDynamicUPCI}{5}
\newcommand{\SLOWsunflowHBDynamicUPCIMIN}{37}
\newcommand{\SLOWsunflowHBDynamicUPCIMAX}{47}
\newcommand{\SLOWsunflowWCPUP}{26}
\newcommand{\SLOWsunflowWCPUPCI}{0}
\newcommand{\SLOWsunflowWCPUPCIMIN}{26}
\newcommand{\SLOWsunflowWCPUPCIMAX}{26}
\newcommand{\SLOWsunflowWCPDynamicUP}{743}
\newcommand{\SLOWsunflowWCPDynamicUPCI}{12}
\newcommand{\SLOWsunflowWCPDynamicUPCIMIN}{731}
\newcommand{\SLOWsunflowWCPDynamicUPCIMAX}{755}
\newcommand{\SLOWsunflowWDCnoGUP}{\rna}
\newcommand{\SLOWsunflowWDCnoGUPCI}{\rna}
\newcommand{\SLOWsunflowWDCnoGUPCIMIN}{\rna}
\newcommand{\SLOWsunflowWDCnoGUPCIMAX}{\rna}
\newcommand{\SLOWsunflowWDCnoGDynamicUP}{\rna}
\newcommand{\SLOWsunflowWDCnoGDynamicUPCI}{\rna}
\newcommand{\SLOWsunflowWDCnoGDynamicUPCIMIN}{\rna}
\newcommand{\SLOWsunflowWDCnoGDynamicUPCIMAX}{\rna}
\newcommand{\SLOWsunflowWDCUP}{\rna}
\newcommand{\SLOWsunflowWDCUPCI}{\rna}
\newcommand{\SLOWsunflowWDCUPCIMIN}{\rna}
\newcommand{\SLOWsunflowWDCUPCIMAX}{\rna}
\newcommand{\SLOWsunflowWDCDynamicUP}{\rna}
\newcommand{\SLOWsunflowWDCDynamicUPCI}{\rna}
\newcommand{\SLOWsunflowWDCDynamicUPCIMIN}{\rna}
\newcommand{\SLOWsunflowWDCDynamicUPCIMAX}{\rna}
\newcommand{\SLOWsunflowCAPOnoGUP}{\rna}
\newcommand{\SLOWsunflowCAPOnoGUPCI}{\rna}
\newcommand{\SLOWsunflowCAPOnoGUPCIMIN}{\rna}
\newcommand{\SLOWsunflowCAPOnoGUPCIMAX}{\rna}
\newcommand{\SLOWsunflowCAPOnoGDynamicUP}{\rna}
\newcommand{\SLOWsunflowCAPOnoGDynamicUPCI}{\rna}
\newcommand{\SLOWsunflowCAPOnoGDynamicUPCIMIN}{\rna}
\newcommand{\SLOWsunflowCAPOnoGDynamicUPCIMAX}{\rna}
\newcommand{\SLOWsunflowCAPOUP}{\rna}
\newcommand{\SLOWsunflowCAPOUPCI}{\rna}
\newcommand{\SLOWsunflowCAPOUPCIMIN}{\rna}
\newcommand{\SLOWsunflowCAPOUPCIMAX}{\rna}
\newcommand{\SLOWsunflowCAPODynamicUP}{\rna}
\newcommand{\SLOWsunflowCAPODynamicUPCI}{\rna}
\newcommand{\SLOWsunflowCAPODynamicUPCIMIN}{\rna}
\newcommand{\SLOWsunflowCAPODynamicUPCIMAX}{\rna}
\newcommand{\SLOWsunflowPIPUP}{\rna}
\newcommand{\SLOWsunflowPIPUPCI}{\rna}
\newcommand{\SLOWsunflowPIPUPCIMIN}{\rna}
\newcommand{\SLOWsunflowPIPUPCIMAX}{\rna}
\newcommand{\SLOWsunflowPIPDynamicUP}{\rna}
\newcommand{\SLOWsunflowPIPDynamicUPCI}{\rna}
\newcommand{\SLOWsunflowPIPDynamicUPCIMIN}{\rna}
\newcommand{\SLOWsunflowPIPDynamicUPCIMAX}{\rna}
\newcommand{\SLOWsunflowPIPHB}{\rna}
\newcommand{\SLOWsunflowPIPHBCI}{\rna}
\newcommand{\SLOWsunflowPIPHBCIMIN}{\rna}
\newcommand{\SLOWsunflowPIPHBCIMAX}{\rna}
\newcommand{\SLOWsunflowPIPHBDynamic}{\rna}
\newcommand{\SLOWsunflowPIPHBDynamicCI}{\rna}
\newcommand{\SLOWsunflowPIPHBDynamicCIMIN}{\rna}
\newcommand{\SLOWsunflowPIPHBDynamicCIMAX}{\rna}
\newcommand{\SLOWsunflowPIPWCP}{\rna}
\newcommand{\SLOWsunflowPIPWCPCI}{\rna}
\newcommand{\SLOWsunflowPIPWCPCIMIN}{\rna}
\newcommand{\SLOWsunflowPIPWCPCIMAX}{\rna}
\newcommand{\SLOWsunflowPIPWCPDynamic}{\rna}
\newcommand{\SLOWsunflowPIPWCPDynamicCI}{\rna}
\newcommand{\SLOWsunflowPIPWCPDynamicCIMIN}{\rna}
\newcommand{\SLOWsunflowPIPWCPDynamicCIMAX}{\rna}
\newcommand{\SLOWsunflowPIPWDC}{\rna}
\newcommand{\SLOWsunflowPIPWDCCI}{\rna}
\newcommand{\SLOWsunflowPIPWDCCIMIN}{\rna}
\newcommand{\SLOWsunflowPIPWDCCIMAX}{\rna}
\newcommand{\SLOWsunflowPIPWDCDynamic}{\rna}
\newcommand{\SLOWsunflowPIPWDCDynamicCI}{\rna}
\newcommand{\SLOWsunflowPIPWDCDynamicCIMIN}{\rna}
\newcommand{\SLOWsunflowPIPWDCDynamicCIMAX}{\rna}
\newcommand{\SLOWsunflowPIPCAPO}{\rna}
\newcommand{\SLOWsunflowPIPCAPOCI}{\rna}
\newcommand{\SLOWsunflowPIPCAPOCIMIN}{\rna}
\newcommand{\SLOWsunflowPIPCAPOCIMAX}{\rna}
\newcommand{\SLOWsunflowPIPCAPODynamic}{\rna}
\newcommand{\SLOWsunflowPIPCAPODynamicCI}{\rna}
\newcommand{\SLOWsunflowPIPCAPODynamicCIMIN}{\rna}
\newcommand{\SLOWsunflowPIPCAPODynamicCIMAX}{\rna}
\newcommand{\SLOWsunflowPIPPIP}{\rna}
\newcommand{\SLOWsunflowPIPPIPCI}{\rna}
\newcommand{\SLOWsunflowPIPPIPCIMIN}{\rna}
\newcommand{\SLOWsunflowPIPPIPCIMAX}{\rna}
\newcommand{\SLOWsunflowPIPPIPDynamic}{\rna}
\newcommand{\SLOWsunflowPIPPIPDynamicCI}{\rna}
\newcommand{\SLOWsunflowPIPPIPDynamicCIMIN}{\rna}
\newcommand{\SLOWsunflowPIPPIPDynamicCIMAX}{\rna}
\newcommand{\SLOWtomcatEvents}{47}
\newcommand{\SLOWtomcatNoFPEvents}{18}
\newcommand{\SLOWtomcatMaxLiveThreads}{33}
\newcommand{\SLOWtomcatTotalThreads}{35}
\newcommand{\SLOWtomcatBaseTime}{1.0}
\newcommand{\SLOWtomcatBaseTimeCI}{190}
\newcommand{\SLOWtomcatEmptyTime}{\rna}
\newcommand{\SLOWtomcatEmptyTimeCI}{\rna}
\newcommand{\SLOWtomcatEmptyTimeCIMIN}{\rna}
\newcommand{\SLOWtomcatEmptyTimeCIMAX}{\rna}
\newcommand{\SLOWtomcatFTTime}{\rna}
\newcommand{\SLOWtomcatFTTimeCI}{\rna}
\newcommand{\SLOWtomcatFTTimeCIMIN}{\rna}
\newcommand{\SLOWtomcatFTTimeCIMAX}{\rna}
\newcommand{\SLOWtomcatHBTime}{20}
\newcommand{\SLOWtomcatHBTimeCI}{2.1}
\newcommand{\SLOWtomcatWCPTime}{25}
\newcommand{\SLOWtomcatWCPTimeCI}{3.9}
\newcommand{\SLOWtomcatDCnoGExcTime}{19}
\newcommand{\SLOWtomcatDCnoGExcTimeCI}{2.8}
\newcommand{\SLOWtomcatDCnoGTime}{\rna}
\newcommand{\SLOWtomcatDCnoGTimeCI}{\rna}
\newcommand{\SLOWtomcatDCnoGTimeCIMIN}{\rna}
\newcommand{\SLOWtomcatDCnoGTimeCIMAX}{\rna}
\newcommand{\SLOWtomcatDCExcTime}{13}
\newcommand{\SLOWtomcatDCExcTimeCI}{1.7}
\newcommand{\SLOWtomcatDCTime}{\rna}
\newcommand{\SLOWtomcatDCTimeCI}{\rna}
\newcommand{\SLOWtomcatDCTimeCIMIN}{\rna}
\newcommand{\SLOWtomcatDCTimeCIMAX}{\rna}
\newcommand{\SLOWtomcatCAPOnoGExcTime}{19}
\newcommand{\SLOWtomcatCAPOnoGExcTimeCI}{2.4}
\newcommand{\SLOWtomcatCAPOnoGTime}{\rna}
\newcommand{\SLOWtomcatCAPOnoGTimeCI}{\rna}
\newcommand{\SLOWtomcatCAPOnoGTimeCIMIN}{\rna}
\newcommand{\SLOWtomcatCAPOnoGTimeCIMAX}{\rna}
\newcommand{\SLOWtomcatCAPOExcTime}{8.6}
\newcommand{\SLOWtomcatCAPOExcTimeCI}{1.0}
\newcommand{\SLOWtomcatCAPOTime}{\rna}
\newcommand{\SLOWtomcatCAPOTimeCI}{\rna}
\newcommand{\SLOWtomcatCAPOTimeCIMIN}{\rna}
\newcommand{\SLOWtomcatCAPOTimeCIMAX}{\rna}
\newcommand{\SLOWtomcatStaticTime}{\rzero}
\newcommand{\SLOWtomcatDynamicTime}{\rzero}
\newcommand{\SLOWtomcatBaseMem}{510}
\newcommand{\SLOWtomcatBaseMemCI}{3.9}
\newcommand{\SLOWtomcatHBMem}{68}
\newcommand{\SLOWtomcatHBMemCI}{5.4}
\newcommand{\SLOWtomcatFTMem}{\memna}
\newcommand{\SLOWtomcatFTMemCI}{\memna}
\newcommand{\SLOWtomcatFTMemCIMIN}{\memna}
\newcommand{\SLOWtomcatFTMemCIMAX}{\memna}
\newcommand{\SLOWtomcatWCPMem}{88}
\newcommand{\SLOWtomcatWCPMemCI}{11.0}
\newcommand{\SLOWtomcatDCnoGExcMem}{58}
\newcommand{\SLOWtomcatDCnoGExcMemCI}{5.2}
\newcommand{\SLOWtomcatDCnoGMem}{\memna}
\newcommand{\SLOWtomcatDCnoGMemCI}{\memna}
\newcommand{\SLOWtomcatDCnoGMemCIMIN}{\memna}
\newcommand{\SLOWtomcatDCnoGMemCIMAX}{\memna}
\newcommand{\SLOWtomcatDCExcMem}{23}
\newcommand{\SLOWtomcatDCExcMemCI}{0.77}
\newcommand{\SLOWtomcatDCMem}{\memna}
\newcommand{\SLOWtomcatDCMemCI}{\memna}
\newcommand{\SLOWtomcatDCMemCIMIN}{\memna}
\newcommand{\SLOWtomcatDCMemCIMAX}{\memna}
\newcommand{\SLOWtomcatCAPOnoGExcMem}{70}
\newcommand{\SLOWtomcatCAPOnoGExcMemCI}{8.5}
\newcommand{\SLOWtomcatCAPOnoGMem}{\memna}
\newcommand{\SLOWtomcatCAPOnoGMemCI}{\memna}
\newcommand{\SLOWtomcatCAPOnoGMemCIMIN}{\memna}
\newcommand{\SLOWtomcatCAPOnoGMemCIMAX}{\memna}
\newcommand{\SLOWtomcatCAPOExcMem}{21}
\newcommand{\SLOWtomcatCAPOExcMemCI}{0.97}
\newcommand{\SLOWtomcatCAPOMem}{\memna}
\newcommand{\SLOWtomcatCAPOMemCI}{\memna}
\newcommand{\SLOWtomcatCAPOMemCIMIN}{\memna}
\newcommand{\SLOWtomcatCAPOMemCIMAX}{\memna}
\newcommand{\SLOWtomcatEventsCI}{131,702}
\newcommand{\SLOWtomcatEventsCIMIN}{46,958,604}
\newcommand{\SLOWtomcatEventsCIMAX}{47,222,008}
\newcommand{\SLOWtomcatNoFPEventsCI}{21,346}
\newcommand{\SLOWtomcatNoFPEventsCIMIN}{18,266,398}
\newcommand{\SLOWtomcatNoFPEventsCIMAX}{18,309,090}
\newcommand{\SLOWtomcatHB}{1,310}
\newcommand{\SLOWtomcatHBCI}{43}
\newcommand{\SLOWtomcatHBCIMIN}{1,267}
\newcommand{\SLOWtomcatHBCIMAX}{1,353}
\newcommand{\SLOWtomcatHBDynamic}{10,294,951}
\newcommand{\SLOWtomcatHBDynamicCI}{648,018}
\newcommand{\SLOWtomcatHBDynamicCIMIN}{9,646,933}
\newcommand{\SLOWtomcatHBDynamicCIMAX}{10,942,969}
\newcommand{\SLOWtomcatFT}{\rna}
\newcommand{\SLOWtomcatFTCI}{\rna}
\newcommand{\SLOWtomcatFTCIMIN}{\rna}
\newcommand{\SLOWtomcatFTCIMAX}{\rna}
\newcommand{\SLOWtomcatFTDynamic}{\rna}
\newcommand{\SLOWtomcatFTDynamicCI}{\rna}
\newcommand{\SLOWtomcatFTDynamicCIMIN}{\rna}
\newcommand{\SLOWtomcatFTDynamicCIMAX}{\rna}
\newcommand{\SLOWtomcatWCP}{1,287}
\newcommand{\SLOWtomcatWCPCI}{53}
\newcommand{\SLOWtomcatWCPCIMIN}{1,234}
\newcommand{\SLOWtomcatWCPCIMAX}{1,340}
\newcommand{\SLOWtomcatWCPDynamic}{8,032,131}
\newcommand{\SLOWtomcatWCPDynamicCI}{1,107,283}
\newcommand{\SLOWtomcatWCPDynamicCIMIN}{6,924,848}
\newcommand{\SLOWtomcatWCPDynamicCIMAX}{9,139,414}
\newcommand{\SLOWtomcatDCnoGExc}{1,252}
\newcommand{\SLOWtomcatDCnoGExcCI}{40}
\newcommand{\SLOWtomcatDCnoGExcCIMIN}{1,212}
\newcommand{\SLOWtomcatDCnoGExcCIMAX}{1,292}
\newcommand{\SLOWtomcatDCnoGExcDynamic}{6,553,393}
\newcommand{\SLOWtomcatDCnoGExcDynamicCI}{1,222,637}
\newcommand{\SLOWtomcatDCnoGExcDynamicCIMIN}{5,330,756}
\newcommand{\SLOWtomcatDCnoGExcDynamicCIMAX}{7,776,030}
\newcommand{\SLOWtomcatDCnoG}{\rna}
\newcommand{\SLOWtomcatDCnoGCI}{\rna}
\newcommand{\SLOWtomcatDCnoGCIMIN}{\rna}
\newcommand{\SLOWtomcatDCnoGCIMAX}{\rna}
\newcommand{\SLOWtomcatDCnoGDynamic}{\rna}
\newcommand{\SLOWtomcatDCnoGDynamicCI}{\rna}
\newcommand{\SLOWtomcatDCnoGDynamicCIMIN}{\rna}
\newcommand{\SLOWtomcatDCnoGDynamicCIMAX}{\rna}
\newcommand{\SLOWtomcatDCExc}{65}
\newcommand{\SLOWtomcatDCExcCI}{1}
\newcommand{\SLOWtomcatDCExcCIMIN}{64}
\newcommand{\SLOWtomcatDCExcCIMAX}{66}
\newcommand{\SLOWtomcatDCExcDynamic}{11,991}
\newcommand{\SLOWtomcatDCExcDynamicCI}{577}
\newcommand{\SLOWtomcatDCExcDynamicCIMIN}{11,414}
\newcommand{\SLOWtomcatDCExcDynamicCIMAX}{12,568}
\newcommand{\SLOWtomcatDC}{\rna}
\newcommand{\SLOWtomcatDCCI}{\rna}
\newcommand{\SLOWtomcatDCCIMIN}{\rna}
\newcommand{\SLOWtomcatDCCIMAX}{\rna}
\newcommand{\SLOWtomcatDCDynamic}{\rna}
\newcommand{\SLOWtomcatDCDynamicCI}{\rna}
\newcommand{\SLOWtomcatDCDynamicCIMIN}{\rna}
\newcommand{\SLOWtomcatDCDynamicCIMAX}{\rna}
\newcommand{\SLOWtomcatCAPOnoGExc}{1,325}
\newcommand{\SLOWtomcatCAPOnoGExcCI}{54}
\newcommand{\SLOWtomcatCAPOnoGExcCIMIN}{1,271}
\newcommand{\SLOWtomcatCAPOnoGExcCIMAX}{1,379}
\newcommand{\SLOWtomcatCAPOnoGExcDynamic}{8,954,449}
\newcommand{\SLOWtomcatCAPOnoGExcDynamicCI}{1,439,076}
\newcommand{\SLOWtomcatCAPOnoGExcDynamicCIMIN}{7,515,373}
\newcommand{\SLOWtomcatCAPOnoGExcDynamicCIMAX}{10,393,525}
\newcommand{\SLOWtomcatCAPOnoG}{\rna}
\newcommand{\SLOWtomcatCAPOnoGCI}{\rna}
\newcommand{\SLOWtomcatCAPOnoGCIMIN}{\rna}
\newcommand{\SLOWtomcatCAPOnoGCIMAX}{\rna}
\newcommand{\SLOWtomcatCAPOnoGDynamic}{\rna}
\newcommand{\SLOWtomcatCAPOnoGDynamicCI}{\rna}
\newcommand{\SLOWtomcatCAPOnoGDynamicCIMIN}{\rna}
\newcommand{\SLOWtomcatCAPOnoGDynamicCIMAX}{\rna}
\newcommand{\SLOWtomcatCAPOExc}{65}
\newcommand{\SLOWtomcatCAPOExcCI}{3}
\newcommand{\SLOWtomcatCAPOExcCIMIN}{62}
\newcommand{\SLOWtomcatCAPOExcCIMAX}{68}
\newcommand{\SLOWtomcatCAPOExcDynamic}{13,225}
\newcommand{\SLOWtomcatCAPOExcDynamicCI}{228}
\newcommand{\SLOWtomcatCAPOExcDynamicCIMIN}{12,997}
\newcommand{\SLOWtomcatCAPOExcDynamicCIMAX}{13,453}
\newcommand{\SLOWtomcatCAPO}{\rna}
\newcommand{\SLOWtomcatCAPOCI}{\rna}
\newcommand{\SLOWtomcatCAPOCIMIN}{\rna}
\newcommand{\SLOWtomcatCAPOCIMAX}{\rna}
\newcommand{\SLOWtomcatCAPODynamic}{\rna}
\newcommand{\SLOWtomcatCAPODynamicCI}{\rna}
\newcommand{\SLOWtomcatCAPODynamicCIMIN}{\rna}
\newcommand{\SLOWtomcatCAPODynamicCIMAX}{\rna}
\newcommand{\SLOWtomcatPIP}{\rna}
\newcommand{\SLOWtomcatPIPCI}{\rna}
\newcommand{\SLOWtomcatPIPCIMIN}{\rna}
\newcommand{\SLOWtomcatPIPCIMAX}{\rna}
\newcommand{\SLOWtomcatPIPDynamic}{\rna}
\newcommand{\SLOWtomcatPIPDynamicCI}{\rna}
\newcommand{\SLOWtomcatPIPDynamicCIMIN}{\rna}
\newcommand{\SLOWtomcatPIPDynamicCIMAX}{\rna}
\newcommand{\SLOWtomcatHBUP}{1,728}
\newcommand{\SLOWtomcatHBUPCI}{69}
\newcommand{\SLOWtomcatHBUPCIMIN}{1,659}
\newcommand{\SLOWtomcatHBUPCIMAX}{1,797}
\newcommand{\SLOWtomcatHBDynamicUP}{10,287,316}
\newcommand{\SLOWtomcatHBDynamicUPCI}{652,725}
\newcommand{\SLOWtomcatHBDynamicUPCIMIN}{9,634,591}
\newcommand{\SLOWtomcatHBDynamicUPCIMAX}{10,940,041}
\newcommand{\SLOWtomcatWCPUP}{1,728}
\newcommand{\SLOWtomcatWCPUPCI}{96}
\newcommand{\SLOWtomcatWCPUPCIMIN}{1,632}
\newcommand{\SLOWtomcatWCPUPCIMAX}{1,824}
\newcommand{\SLOWtomcatWCPDynamicUP}{8,032,393}
\newcommand{\SLOWtomcatWCPDynamicUPCI}{1,107,245}
\newcommand{\SLOWtomcatWCPDynamicUPCIMIN}{6,925,148}
\newcommand{\SLOWtomcatWCPDynamicUPCIMAX}{9,139,638}
\newcommand{\SLOWtomcatWDCnoGUP}{\rna}
\newcommand{\SLOWtomcatWDCnoGUPCI}{\rna}
\newcommand{\SLOWtomcatWDCnoGUPCIMIN}{\rna}
\newcommand{\SLOWtomcatWDCnoGUPCIMAX}{\rna}
\newcommand{\SLOWtomcatWDCnoGDynamicUP}{\rna}
\newcommand{\SLOWtomcatWDCnoGDynamicUPCI}{\rna}
\newcommand{\SLOWtomcatWDCnoGDynamicUPCIMIN}{\rna}
\newcommand{\SLOWtomcatWDCnoGDynamicUPCIMAX}{\rna}
\newcommand{\SLOWtomcatWDCUP}{\rna}
\newcommand{\SLOWtomcatWDCUPCI}{\rna}
\newcommand{\SLOWtomcatWDCUPCIMIN}{\rna}
\newcommand{\SLOWtomcatWDCUPCIMAX}{\rna}
\newcommand{\SLOWtomcatWDCDynamicUP}{\rna}
\newcommand{\SLOWtomcatWDCDynamicUPCI}{\rna}
\newcommand{\SLOWtomcatWDCDynamicUPCIMIN}{\rna}
\newcommand{\SLOWtomcatWDCDynamicUPCIMAX}{\rna}
\newcommand{\SLOWtomcatCAPOnoGUP}{\rna}
\newcommand{\SLOWtomcatCAPOnoGUPCI}{\rna}
\newcommand{\SLOWtomcatCAPOnoGUPCIMIN}{\rna}
\newcommand{\SLOWtomcatCAPOnoGUPCIMAX}{\rna}
\newcommand{\SLOWtomcatCAPOnoGDynamicUP}{\rna}
\newcommand{\SLOWtomcatCAPOnoGDynamicUPCI}{\rna}
\newcommand{\SLOWtomcatCAPOnoGDynamicUPCIMIN}{\rna}
\newcommand{\SLOWtomcatCAPOnoGDynamicUPCIMAX}{\rna}
\newcommand{\SLOWtomcatCAPOUP}{\rna}
\newcommand{\SLOWtomcatCAPOUPCI}{\rna}
\newcommand{\SLOWtomcatCAPOUPCIMIN}{\rna}
\newcommand{\SLOWtomcatCAPOUPCIMAX}{\rna}
\newcommand{\SLOWtomcatCAPODynamicUP}{\rna}
\newcommand{\SLOWtomcatCAPODynamicUPCI}{\rna}
\newcommand{\SLOWtomcatCAPODynamicUPCIMIN}{\rna}
\newcommand{\SLOWtomcatCAPODynamicUPCIMAX}{\rna}
\newcommand{\SLOWtomcatPIPUP}{\rna}
\newcommand{\SLOWtomcatPIPUPCI}{\rna}
\newcommand{\SLOWtomcatPIPUPCIMIN}{\rna}
\newcommand{\SLOWtomcatPIPUPCIMAX}{\rna}
\newcommand{\SLOWtomcatPIPDynamicUP}{\rna}
\newcommand{\SLOWtomcatPIPDynamicUPCI}{\rna}
\newcommand{\SLOWtomcatPIPDynamicUPCIMIN}{\rna}
\newcommand{\SLOWtomcatPIPDynamicUPCIMAX}{\rna}
\newcommand{\SLOWtomcatPIPHB}{\rna}
\newcommand{\SLOWtomcatPIPHBCI}{\rna}
\newcommand{\SLOWtomcatPIPHBCIMIN}{\rna}
\newcommand{\SLOWtomcatPIPHBCIMAX}{\rna}
\newcommand{\SLOWtomcatPIPHBDynamic}{\rna}
\newcommand{\SLOWtomcatPIPHBDynamicCI}{\rna}
\newcommand{\SLOWtomcatPIPHBDynamicCIMIN}{\rna}
\newcommand{\SLOWtomcatPIPHBDynamicCIMAX}{\rna}
\newcommand{\SLOWtomcatPIPWCP}{\rna}
\newcommand{\SLOWtomcatPIPWCPCI}{\rna}
\newcommand{\SLOWtomcatPIPWCPCIMIN}{\rna}
\newcommand{\SLOWtomcatPIPWCPCIMAX}{\rna}
\newcommand{\SLOWtomcatPIPWCPDynamic}{\rna}
\newcommand{\SLOWtomcatPIPWCPDynamicCI}{\rna}
\newcommand{\SLOWtomcatPIPWCPDynamicCIMIN}{\rna}
\newcommand{\SLOWtomcatPIPWCPDynamicCIMAX}{\rna}
\newcommand{\SLOWtomcatPIPWDC}{\rna}
\newcommand{\SLOWtomcatPIPWDCCI}{\rna}
\newcommand{\SLOWtomcatPIPWDCCIMIN}{\rna}
\newcommand{\SLOWtomcatPIPWDCCIMAX}{\rna}
\newcommand{\SLOWtomcatPIPWDCDynamic}{\rna}
\newcommand{\SLOWtomcatPIPWDCDynamicCI}{\rna}
\newcommand{\SLOWtomcatPIPWDCDynamicCIMIN}{\rna}
\newcommand{\SLOWtomcatPIPWDCDynamicCIMAX}{\rna}
\newcommand{\SLOWtomcatPIPCAPO}{\rna}
\newcommand{\SLOWtomcatPIPCAPOCI}{\rna}
\newcommand{\SLOWtomcatPIPCAPOCIMIN}{\rna}
\newcommand{\SLOWtomcatPIPCAPOCIMAX}{\rna}
\newcommand{\SLOWtomcatPIPCAPODynamic}{\rna}
\newcommand{\SLOWtomcatPIPCAPODynamicCI}{\rna}
\newcommand{\SLOWtomcatPIPCAPODynamicCIMIN}{\rna}
\newcommand{\SLOWtomcatPIPCAPODynamicCIMAX}{\rna}
\newcommand{\SLOWtomcatPIPPIP}{\rna}
\newcommand{\SLOWtomcatPIPPIPCI}{\rna}
\newcommand{\SLOWtomcatPIPPIPCIMIN}{\rna}
\newcommand{\SLOWtomcatPIPPIPCIMAX}{\rna}
\newcommand{\SLOWtomcatPIPPIPDynamic}{\rna}
\newcommand{\SLOWtomcatPIPPIPDynamicCI}{\rna}
\newcommand{\SLOWtomcatPIPPIPDynamicCIMIN}{\rna}
\newcommand{\SLOWtomcatPIPPIPDynamicCIMAX}{\rna}
\newcommand{\SLOWxalanEvents}{630}
\newcommand{\SLOWxalanNoFPEvents}{270}
\newcommand{\SLOWxalanMaxLiveThreads}{9}
\newcommand{\SLOWxalanTotalThreads}{9}
\newcommand{\SLOWxalanBaseTime}{2.0}
\newcommand{\SLOWxalanBaseTimeCI}{130}
\newcommand{\SLOWxalanEmptyTime}{\rna}
\newcommand{\SLOWxalanEmptyTimeCI}{\rna}
\newcommand{\SLOWxalanEmptyTimeCIMIN}{\rna}
\newcommand{\SLOWxalanEmptyTimeCIMAX}{\rna}
\newcommand{\SLOWxalanFTTime}{\rna}
\newcommand{\SLOWxalanFTTimeCI}{\rna}
\newcommand{\SLOWxalanFTTimeCIMIN}{\rna}
\newcommand{\SLOWxalanFTTimeCIMAX}{\rna}
\newcommand{\SLOWxalanHBTime}{19}
\newcommand{\SLOWxalanHBTimeCI}{5.3}
\newcommand{\SLOWxalanWCPTime}{48}
\newcommand{\SLOWxalanWCPTimeCI}{3.7}
\newcommand{\SLOWxalanDCnoGExcTime}{47}
\newcommand{\SLOWxalanDCnoGExcTimeCI}{3.6}
\newcommand{\SLOWxalanDCnoGTime}{\rna}
\newcommand{\SLOWxalanDCnoGTimeCI}{\rna}
\newcommand{\SLOWxalanDCnoGTimeCIMIN}{\rna}
\newcommand{\SLOWxalanDCnoGTimeCIMAX}{\rna}
\newcommand{\SLOWxalanDCExcTime}{57}
\newcommand{\SLOWxalanDCExcTimeCI}{4.9}
\newcommand{\SLOWxalanDCTime}{\rna}
\newcommand{\SLOWxalanDCTimeCI}{\rna}
\newcommand{\SLOWxalanDCTimeCIMIN}{\rna}
\newcommand{\SLOWxalanDCTimeCIMAX}{\rna}
\newcommand{\SLOWxalanCAPOnoGExcTime}{44}
\newcommand{\SLOWxalanCAPOnoGExcTimeCI}{2.8}
\newcommand{\SLOWxalanCAPOnoGTime}{\rna}
\newcommand{\SLOWxalanCAPOnoGTimeCI}{\rna}
\newcommand{\SLOWxalanCAPOnoGTimeCIMIN}{\rna}
\newcommand{\SLOWxalanCAPOnoGTimeCIMAX}{\rna}
\newcommand{\SLOWxalanCAPOExcTime}{51}
\newcommand{\SLOWxalanCAPOExcTimeCI}{3.8}
\newcommand{\SLOWxalanCAPOTime}{\rna}
\newcommand{\SLOWxalanCAPOTimeCI}{\rna}
\newcommand{\SLOWxalanCAPOTimeCIMIN}{\rna}
\newcommand{\SLOWxalanCAPOTimeCIMAX}{\rna}
\newcommand{\SLOWxalanStaticTime}{\rzero}
\newcommand{\SLOWxalanDynamicTime}{\rzero}
\newcommand{\SLOWxalanBaseMem}{670}
\newcommand{\SLOWxalanBaseMemCI}{8.6}
\newcommand{\SLOWxalanHBMem}{21}
\newcommand{\SLOWxalanHBMemCI}{0.91}
\newcommand{\SLOWxalanFTMem}{\memna}
\newcommand{\SLOWxalanFTMemCI}{\memna}
\newcommand{\SLOWxalanFTMemCIMIN}{\memna}
\newcommand{\SLOWxalanFTMemCIMAX}{\memna}
\newcommand{\SLOWxalanWCPMem}{64}
\newcommand{\SLOWxalanWCPMemCI}{2.3}
\newcommand{\SLOWxalanDCnoGExcMem}{55}
\newcommand{\SLOWxalanDCnoGExcMemCI}{1.6}
\newcommand{\SLOWxalanDCnoGMem}{\memna}
\newcommand{\SLOWxalanDCnoGMemCI}{\memna}
\newcommand{\SLOWxalanDCnoGMemCIMIN}{\memna}
\newcommand{\SLOWxalanDCnoGMemCIMAX}{\memna}
\newcommand{\SLOWxalanDCExcMem}{61}
\newcommand{\SLOWxalanDCExcMemCI}{1.0}
\newcommand{\SLOWxalanDCMem}{\memna}
\newcommand{\SLOWxalanDCMemCI}{\memna}
\newcommand{\SLOWxalanDCMemCIMIN}{\memna}
\newcommand{\SLOWxalanDCMemCIMAX}{\memna}
\newcommand{\SLOWxalanCAPOnoGExcMem}{54}
\newcommand{\SLOWxalanCAPOnoGExcMemCI}{1.7}
\newcommand{\SLOWxalanCAPOnoGMem}{\memna}
\newcommand{\SLOWxalanCAPOnoGMemCI}{\memna}
\newcommand{\SLOWxalanCAPOnoGMemCIMIN}{\memna}
\newcommand{\SLOWxalanCAPOnoGMemCIMAX}{\memna}
\newcommand{\SLOWxalanCAPOExcMem}{57}
\newcommand{\SLOWxalanCAPOExcMemCI}{1.5}
\newcommand{\SLOWxalanCAPOMem}{\memna}
\newcommand{\SLOWxalanCAPOMemCI}{\memna}
\newcommand{\SLOWxalanCAPOMemCIMIN}{\memna}
\newcommand{\SLOWxalanCAPOMemCIMAX}{\memna}
\newcommand{\SLOWxalanEventsCI}{18}
\newcommand{\SLOWxalanEventsCIMIN}{626,353,750}
\newcommand{\SLOWxalanEventsCIMAX}{626,353,786}
\newcommand{\SLOWxalanNoFPEventsCI}{15}
\newcommand{\SLOWxalanNoFPEventsCIMIN}{273,861,865}
\newcommand{\SLOWxalanNoFPEventsCIMAX}{273,861,895}
\newcommand{\SLOWxalanHB}{9}
\newcommand{\SLOWxalanHBCI}{0}
\newcommand{\SLOWxalanHBCIMIN}{9}
\newcommand{\SLOWxalanHBCIMAX}{9}
\newcommand{\SLOWxalanHBDynamic}{1,221}
\newcommand{\SLOWxalanHBDynamicCI}{31}
\newcommand{\SLOWxalanHBDynamicCIMIN}{1,190}
\newcommand{\SLOWxalanHBDynamicCIMAX}{1,252}
\newcommand{\SLOWxalanFT}{\rna}
\newcommand{\SLOWxalanFTCI}{\rna}
\newcommand{\SLOWxalanFTCIMIN}{\rna}
\newcommand{\SLOWxalanFTCIMAX}{\rna}
\newcommand{\SLOWxalanFTDynamic}{\rna}
\newcommand{\SLOWxalanFTDynamicCI}{\rna}
\newcommand{\SLOWxalanFTDynamicCIMIN}{\rna}
\newcommand{\SLOWxalanFTDynamicCIMAX}{\rna}
\newcommand{\SLOWxalanWCP}{150}
\newcommand{\SLOWxalanWCPCI}{0}
\newcommand{\SLOWxalanWCPCIMIN}{150}
\newcommand{\SLOWxalanWCPCIMAX}{150}
\newcommand{\SLOWxalanWCPDynamic}{9,994,327}
\newcommand{\SLOWxalanWCPDynamicCI}{26,675}
\newcommand{\SLOWxalanWCPDynamicCIMIN}{9,967,652}
\newcommand{\SLOWxalanWCPDynamicCIMAX}{10,021,002}
\newcommand{\SLOWxalanDCnoGExc}{162}
\newcommand{\SLOWxalanDCnoGExcCI}{0}
\newcommand{\SLOWxalanDCnoGExcCIMIN}{162}
\newcommand{\SLOWxalanDCnoGExcCIMAX}{162}
\newcommand{\SLOWxalanDCnoGExcDynamic}{12,316,030}
\newcommand{\SLOWxalanDCnoGExcDynamicCI}{14,382}
\newcommand{\SLOWxalanDCnoGExcDynamicCIMIN}{12,301,648}
\newcommand{\SLOWxalanDCnoGExcDynamicCIMAX}{12,330,412}
\newcommand{\SLOWxalanDCnoG}{\rna}
\newcommand{\SLOWxalanDCnoGCI}{\rna}
\newcommand{\SLOWxalanDCnoGCIMIN}{\rna}
\newcommand{\SLOWxalanDCnoGCIMAX}{\rna}
\newcommand{\SLOWxalanDCnoGDynamic}{\rna}
\newcommand{\SLOWxalanDCnoGDynamicCI}{\rna}
\newcommand{\SLOWxalanDCnoGDynamicCIMIN}{\rna}
\newcommand{\SLOWxalanDCnoGDynamicCIMAX}{\rna}
\newcommand{\SLOWxalanDCExc}{8}
\newcommand{\SLOWxalanDCExcCI}{0}
\newcommand{\SLOWxalanDCExcCIMIN}{8}
\newcommand{\SLOWxalanDCExcCIMAX}{8}
\newcommand{\SLOWxalanDCExcDynamic}{262,419}
\newcommand{\SLOWxalanDCExcDynamicCI}{841}
\newcommand{\SLOWxalanDCExcDynamicCIMIN}{261,578}
\newcommand{\SLOWxalanDCExcDynamicCIMAX}{263,260}
\newcommand{\SLOWxalanDC}{\rna}
\newcommand{\SLOWxalanDCCI}{\rna}
\newcommand{\SLOWxalanDCCIMIN}{\rna}
\newcommand{\SLOWxalanDCCIMAX}{\rna}
\newcommand{\SLOWxalanDCDynamic}{\rna}
\newcommand{\SLOWxalanDCDynamicCI}{\rna}
\newcommand{\SLOWxalanDCDynamicCIMIN}{\rna}
\newcommand{\SLOWxalanDCDynamicCIMAX}{\rna}
\newcommand{\SLOWxalanCAPOnoGExc}{162}
\newcommand{\SLOWxalanCAPOnoGExcCI}{0}
\newcommand{\SLOWxalanCAPOnoGExcCIMIN}{162}
\newcommand{\SLOWxalanCAPOnoGExcCIMAX}{162}
\newcommand{\SLOWxalanCAPOnoGExcDynamic}{12,302,548}
\newcommand{\SLOWxalanCAPOnoGExcDynamicCI}{11,428}
\newcommand{\SLOWxalanCAPOnoGExcDynamicCIMIN}{12,291,120}
\newcommand{\SLOWxalanCAPOnoGExcDynamicCIMAX}{12,313,976}
\newcommand{\SLOWxalanCAPOnoG}{\rna}
\newcommand{\SLOWxalanCAPOnoGCI}{\rna}
\newcommand{\SLOWxalanCAPOnoGCIMIN}{\rna}
\newcommand{\SLOWxalanCAPOnoGCIMAX}{\rna}
\newcommand{\SLOWxalanCAPOnoGDynamic}{\rna}
\newcommand{\SLOWxalanCAPOnoGDynamicCI}{\rna}
\newcommand{\SLOWxalanCAPOnoGDynamicCIMIN}{\rna}
\newcommand{\SLOWxalanCAPOnoGDynamicCIMAX}{\rna}
\newcommand{\SLOWxalanCAPOExc}{7}
\newcommand{\SLOWxalanCAPOExcCI}{0}
\newcommand{\SLOWxalanCAPOExcCIMIN}{7}
\newcommand{\SLOWxalanCAPOExcCIMAX}{7}
\newcommand{\SLOWxalanCAPOExcDynamic}{259,955}
\newcommand{\SLOWxalanCAPOExcDynamicCI}{736}
\newcommand{\SLOWxalanCAPOExcDynamicCIMIN}{259,219}
\newcommand{\SLOWxalanCAPOExcDynamicCIMAX}{260,691}
\newcommand{\SLOWxalanCAPO}{\rna}
\newcommand{\SLOWxalanCAPOCI}{\rna}
\newcommand{\SLOWxalanCAPOCIMIN}{\rna}
\newcommand{\SLOWxalanCAPOCIMAX}{\rna}
\newcommand{\SLOWxalanCAPODynamic}{\rna}
\newcommand{\SLOWxalanCAPODynamicCI}{\rna}
\newcommand{\SLOWxalanCAPODynamicCIMIN}{\rna}
\newcommand{\SLOWxalanCAPODynamicCIMAX}{\rna}
\newcommand{\SLOWxalanPIP}{\rna}
\newcommand{\SLOWxalanPIPCI}{\rna}
\newcommand{\SLOWxalanPIPCIMIN}{\rna}
\newcommand{\SLOWxalanPIPCIMAX}{\rna}
\newcommand{\SLOWxalanPIPDynamic}{\rna}
\newcommand{\SLOWxalanPIPDynamicCI}{\rna}
\newcommand{\SLOWxalanPIPDynamicCIMIN}{\rna}
\newcommand{\SLOWxalanPIPDynamicCIMAX}{\rna}
\newcommand{\SLOWxalanHBUP}{34}
\newcommand{\SLOWxalanHBUPCI}{0}
\newcommand{\SLOWxalanHBUPCIMIN}{34}
\newcommand{\SLOWxalanHBUPCIMAX}{34}
\newcommand{\SLOWxalanHBDynamicUP}{1,195}
\newcommand{\SLOWxalanHBDynamicUPCI}{28}
\newcommand{\SLOWxalanHBDynamicUPCIMIN}{1,167}
\newcommand{\SLOWxalanHBDynamicUPCIMAX}{1,223}
\newcommand{\SLOWxalanWCPUP}{238}
\newcommand{\SLOWxalanWCPUPCI}{0}
\newcommand{\SLOWxalanWCPUPCIMIN}{238}
\newcommand{\SLOWxalanWCPUPCIMAX}{238}
\newcommand{\SLOWxalanWCPDynamicUP}{9,994,062}
\newcommand{\SLOWxalanWCPDynamicUPCI}{26,485}
\newcommand{\SLOWxalanWCPDynamicUPCIMIN}{9,967,577}
\newcommand{\SLOWxalanWCPDynamicUPCIMAX}{10,020,547}
\newcommand{\SLOWxalanWDCnoGUP}{\rna}
\newcommand{\SLOWxalanWDCnoGUPCI}{\rna}
\newcommand{\SLOWxalanWDCnoGUPCIMIN}{\rna}
\newcommand{\SLOWxalanWDCnoGUPCIMAX}{\rna}
\newcommand{\SLOWxalanWDCnoGDynamicUP}{\rna}
\newcommand{\SLOWxalanWDCnoGDynamicUPCI}{\rna}
\newcommand{\SLOWxalanWDCnoGDynamicUPCIMIN}{\rna}
\newcommand{\SLOWxalanWDCnoGDynamicUPCIMAX}{\rna}
\newcommand{\SLOWxalanWDCUP}{\rna}
\newcommand{\SLOWxalanWDCUPCI}{\rna}
\newcommand{\SLOWxalanWDCUPCIMIN}{\rna}
\newcommand{\SLOWxalanWDCUPCIMAX}{\rna}
\newcommand{\SLOWxalanWDCDynamicUP}{\rna}
\newcommand{\SLOWxalanWDCDynamicUPCI}{\rna}
\newcommand{\SLOWxalanWDCDynamicUPCIMIN}{\rna}
\newcommand{\SLOWxalanWDCDynamicUPCIMAX}{\rna}
\newcommand{\SLOWxalanCAPOnoGUP}{\rna}
\newcommand{\SLOWxalanCAPOnoGUPCI}{\rna}
\newcommand{\SLOWxalanCAPOnoGUPCIMIN}{\rna}
\newcommand{\SLOWxalanCAPOnoGUPCIMAX}{\rna}
\newcommand{\SLOWxalanCAPOnoGDynamicUP}{\rna}
\newcommand{\SLOWxalanCAPOnoGDynamicUPCI}{\rna}
\newcommand{\SLOWxalanCAPOnoGDynamicUPCIMIN}{\rna}
\newcommand{\SLOWxalanCAPOnoGDynamicUPCIMAX}{\rna}
\newcommand{\SLOWxalanCAPOUP}{\rna}
\newcommand{\SLOWxalanCAPOUPCI}{\rna}
\newcommand{\SLOWxalanCAPOUPCIMIN}{\rna}
\newcommand{\SLOWxalanCAPOUPCIMAX}{\rna}
\newcommand{\SLOWxalanCAPODynamicUP}{\rna}
\newcommand{\SLOWxalanCAPODynamicUPCI}{\rna}
\newcommand{\SLOWxalanCAPODynamicUPCIMIN}{\rna}
\newcommand{\SLOWxalanCAPODynamicUPCIMAX}{\rna}
\newcommand{\SLOWxalanPIPUP}{\rna}
\newcommand{\SLOWxalanPIPUPCI}{\rna}
\newcommand{\SLOWxalanPIPUPCIMIN}{\rna}
\newcommand{\SLOWxalanPIPUPCIMAX}{\rna}
\newcommand{\SLOWxalanPIPDynamicUP}{\rna}
\newcommand{\SLOWxalanPIPDynamicUPCI}{\rna}
\newcommand{\SLOWxalanPIPDynamicUPCIMIN}{\rna}
\newcommand{\SLOWxalanPIPDynamicUPCIMAX}{\rna}
\newcommand{\SLOWxalanPIPHB}{\rna}
\newcommand{\SLOWxalanPIPHBCI}{\rna}
\newcommand{\SLOWxalanPIPHBCIMIN}{\rna}
\newcommand{\SLOWxalanPIPHBCIMAX}{\rna}
\newcommand{\SLOWxalanPIPHBDynamic}{\rna}
\newcommand{\SLOWxalanPIPHBDynamicCI}{\rna}
\newcommand{\SLOWxalanPIPHBDynamicCIMIN}{\rna}
\newcommand{\SLOWxalanPIPHBDynamicCIMAX}{\rna}
\newcommand{\SLOWxalanPIPWCP}{\rna}
\newcommand{\SLOWxalanPIPWCPCI}{\rna}
\newcommand{\SLOWxalanPIPWCPCIMIN}{\rna}
\newcommand{\SLOWxalanPIPWCPCIMAX}{\rna}
\newcommand{\SLOWxalanPIPWCPDynamic}{\rna}
\newcommand{\SLOWxalanPIPWCPDynamicCI}{\rna}
\newcommand{\SLOWxalanPIPWCPDynamicCIMIN}{\rna}
\newcommand{\SLOWxalanPIPWCPDynamicCIMAX}{\rna}
\newcommand{\SLOWxalanPIPWDC}{\rna}
\newcommand{\SLOWxalanPIPWDCCI}{\rna}
\newcommand{\SLOWxalanPIPWDCCIMIN}{\rna}
\newcommand{\SLOWxalanPIPWDCCIMAX}{\rna}
\newcommand{\SLOWxalanPIPWDCDynamic}{\rna}
\newcommand{\SLOWxalanPIPWDCDynamicCI}{\rna}
\newcommand{\SLOWxalanPIPWDCDynamicCIMIN}{\rna}
\newcommand{\SLOWxalanPIPWDCDynamicCIMAX}{\rna}
\newcommand{\SLOWxalanPIPCAPO}{\rna}
\newcommand{\SLOWxalanPIPCAPOCI}{\rna}
\newcommand{\SLOWxalanPIPCAPOCIMIN}{\rna}
\newcommand{\SLOWxalanPIPCAPOCIMAX}{\rna}
\newcommand{\SLOWxalanPIPCAPODynamic}{\rna}
\newcommand{\SLOWxalanPIPCAPODynamicCI}{\rna}
\newcommand{\SLOWxalanPIPCAPODynamicCIMIN}{\rna}
\newcommand{\SLOWxalanPIPCAPODynamicCIMAX}{\rna}
\newcommand{\SLOWxalanPIPPIP}{\rna}
\newcommand{\SLOWxalanPIPPIPCI}{\rna}
\newcommand{\SLOWxalanPIPPIPCIMIN}{\rna}
\newcommand{\SLOWxalanPIPPIPCIMAX}{\rna}
\newcommand{\SLOWxalanPIPPIPDynamic}{\rna}
\newcommand{\SLOWxalanPIPPIPDynamicCI}{\rna}
\newcommand{\SLOWxalanPIPPIPDynamicCIMIN}{\rna}
\newcommand{\SLOWxalanPIPPIPDynamicCIMAX}{\rna}
\newcommand{\SLOWBaseTimeGeoMean}{1900}
\newcommand{\SLOWEmptyTimeGeoMean}{\rna}
\newcommand{\SLOWFTTimeGeoMean}{\rna}
\newcommand{\SLOWHBTimeGeoMean}{22}
\newcommand{\SLOWWCPTimeGeoMean}{35}
\newcommand{\SLOWDCnoGExcTimeGeoMean}{30}
\newcommand{\SLOWDCnoGTimeGeoMean}{\rna}
\newcommand{\SLOWDCExcTimeGeoMean}{32}
\newcommand{\SLOWDCTimeGeoMean}{\rna}
\newcommand{\SLOWCAPOnoGExcTimeGeoMean}{29}
\newcommand{\SLOWCAPOnoGTimeGeoMean}{\rna}
\newcommand{\SLOWCAPOExcTimeGeoMean}{29}
\newcommand{\SLOWCAPOTimeGeoMean}{\rna}
\newcommand{\SLOWBaseMemGeoMean}{500}
\newcommand{\SLOWEmptyMemGeoMean}{0.0}
\newcommand{\SLOWFTMemGeoMean}{\memna}
\newcommand{\SLOWHBMemGeoMean}{23}
\newcommand{\SLOWWCPMemGeoMean}{46}
\newcommand{\SLOWDCnoGExcMemGeoMean}{32}
\newcommand{\SLOWDCnoGMemGeoMean}{\memna}
\newcommand{\SLOWDCExcMemGeoMean}{35}
\newcommand{\SLOWDCMemGeoMean}{\memna}
\newcommand{\SLOWCAPOnoGExcMemGeoMean}{32}
\newcommand{\SLOWCAPOnoGMemGeoMean}{\memna}
\newcommand{\SLOWCAPOExcMemGeoMean}{33}
\newcommand{\SLOWCAPOMemGeoMean}{\memna}
\newcommand{\SLOWWCPDynamicTotal}{18,649,106}
\newcommand{\SLOWHBDynamicUPTotal}{10,882,490}
\newcommand{\SLOWPIPWDCDynamicTotal}{0}
\newcommand{\SLOWCAPOTotal}{0}
\newcommand{\SLOWPIPWCPTotal}{0}
\newcommand{\SLOWWCPTotal}{1529}
\newcommand{\SLOWWDCnoGDynamicUPTotal}{0}
\newcommand{\SLOWCAPOnoGExcDynamicTotal}{21,880,960}
\newcommand{\SLOWCAPOExcTotal}{104}
\newcommand{\SLOWDCExcTotal}{104}
\newcommand{\SLOWDCnoGExcTotal}{1523}
\newcommand{\SLOWPIPDynamicUPTotal}{0}
\newcommand{\SLOWPIPPIPTotal}{0}
\newcommand{\SLOWFTTotal}{0}
\newcommand{\SLOWHBDynamicTotal}{10,890,152}
\newcommand{\SLOWCAPOnoGDynamicUPTotal}{0}
\newcommand{\SLOWDCnoGExcDynamicTotal}{19,506,268}
\newcommand{\SLOWCAPOnoGTotal}{0}
\newcommand{\SLOWPIPWDCTotal}{0}
\newcommand{\SLOWCAPOnoGExcTotal}{1596}
\newcommand{\SLOWPIPHBDynamicTotal}{0}
\newcommand{\SLOWWDCUPTotal}{0}
\newcommand{\SLOWDCnoGTotal}{0}
\newcommand{\SLOWWCPDynamicUPTotal}{18,649,104}
\newcommand{\SLOWPIPUPTotal}{0}
\newcommand{\SLOWCAPODynamicTotal}{0}
\newcommand{\SLOWCAPOnoGUPTotal}{0}
\newcommand{\SLOWPIPCAPOTotal}{0}
\newcommand{\SLOWCAPOnoGDynamicTotal}{0}
\newcommand{\SLOWFTDynamicTotal}{0}
\newcommand{\SLOWPIPCAPODynamicTotal}{0}
\newcommand{\SLOWDCnoGDynamicTotal}{0}
\newcommand{\SLOWWDCnoGUPTotal}{0}
\newcommand{\SLOWDCExcDynamicTotal}{470,513}
\newcommand{\SLOWWDCDynamicUPTotal}{0}
\newcommand{\SLOWHBUPTotal}{1826}
\newcommand{\SLOWCAPOExcDynamicTotal}{532,175}
\newcommand{\SLOWPIPTotal}{0}
\newcommand{\SLOWWCPUPTotal}{2051}
\newcommand{\SLOWCAPOUPTotal}{0}
\newcommand{\SLOWPIPHBTotal}{0}
\newcommand{\SLOWHBTotal}{1391}
\newcommand{\SLOWDCDynamicTotal}{0}
\newcommand{\SLOWPIPDynamicTotal}{0}
\newcommand{\SLOWCAPODynamicUPTotal}{0}
\newcommand{\SLOWPIPPIPDynamicTotal}{0}
\newcommand{\SLOWPIPWCPDynamicTotal}{0}
\newcommand{\SLOWDCTotal}{0}

%% file: result-macros/PIP_fastTool_noWriteOrReadRaceEdge_accessMetaUpdate.tex
\newcommand{\FASTavroraEvents}{0}
\newcommand{\FASTavroraNoFPEvents}{0}
\newcommand{\FASTavroraMaxLiveThreads}{7}
\newcommand{\FASTavroraTotalThreads}{7}
\newcommand{\FASTavroraBaseTime}{2.4}
\newcommand{\FASTavroraBaseTimeCI}{9.1}
\newcommand{\FASTavroraEmptyTime}{\rna}
\newcommand{\FASTavroraEmptyTimeCI}{\rna}
\newcommand{\FASTavroraEmptyTimeCIMIN}{\rna}
\newcommand{\FASTavroraEmptyTimeCIMAX}{\rna}
\newcommand{\FASTavroraFTTime}{7.1}
\newcommand{\FASTavroraFTTimeCI}{0.096}
\newcommand{\FASTavroraHBTime}{4.2}
\newcommand{\FASTavroraHBTimeCI}{0.033}
\newcommand{\FASTavroraFTOHBTime}{4.2}
\newcommand{\FASTavroraFTOHBTimeCI}{0.041}
\newcommand{\FASTavroraWCPTime}{\rna}
\newcommand{\FASTavroraWCPTimeCI}{\rna}
\newcommand{\FASTavroraWCPTimeCIMIN}{\rna}
\newcommand{\FASTavroraWCPTimeCIMAX}{\rna}
\newcommand{\FASTavroraFTOWCPTime}{8.5}
\newcommand{\FASTavroraFTOWCPTimeCI}{0.11}
\newcommand{\FASTavroraREWCPTime}{6.5}
\newcommand{\FASTavroraREWCPTimeCI}{0.066}
\newcommand{\FASTavroraDCTime}{\rna}
\newcommand{\FASTavroraDCTimeCI}{\rna}
\newcommand{\FASTavroraDCTimeCIMIN}{\rna}
\newcommand{\FASTavroraDCTimeCIMAX}{\rna}
\newcommand{\FASTavroraFTODCTime}{9.0}
\newcommand{\FASTavroraFTODCTimeCI}{0.11}
\newcommand{\FASTavroraREDCTime}{7.0}
\newcommand{\FASTavroraREDCTimeCI}{0.11}
\newcommand{\FASTavroraCAPOTime}{\rna}
\newcommand{\FASTavroraCAPOTimeCI}{\rna}
\newcommand{\FASTavroraCAPOTimeCIMIN}{\rna}
\newcommand{\FASTavroraCAPOTimeCIMAX}{\rna}
\newcommand{\FASTavroraFTOCAPOTime}{6.7}
\newcommand{\FASTavroraFTOCAPOTimeCI}{0.11}
\newcommand{\FASTavroraRECAPOTime}{5.0}
\newcommand{\FASTavroraRECAPOTimeCI}{0.088}
\newcommand{\FASTavroraAGGCAPOTime}{\rna}
\newcommand{\FASTavroraAGGCAPOTimeCI}{\rna}
\newcommand{\FASTavroraAGGCAPOTimeCIMIN}{\rna}
\newcommand{\FASTavroraAGGCAPOTimeCIMAX}{\rna}
\newcommand{\FASTavroraStaticTime}{\rzero}
\newcommand{\FASTavroraDynamicTime}{\rzero}
\newcommand{\FASTavroraBaseMem}{150}
\newcommand{\FASTavroraBaseMemCI}{7.9}
\newcommand{\FASTavroraFTMem}{22}
\newcommand{\FASTavroraFTMemCI}{4.7}
\newcommand{\FASTavroraHBMem}{4.5}
\newcommand{\FASTavroraHBMemCI}{0.25}
\newcommand{\FASTavroraFTOHBMem}{4.6}
\newcommand{\FASTavroraFTOHBMemCI}{0.24}
\newcommand{\FASTavroraWCPMem}{\memna}
\newcommand{\FASTavroraWCPMemCI}{\memna}
\newcommand{\FASTavroraWCPMemCIMIN}{\memna}
\newcommand{\FASTavroraWCPMemCIMAX}{\memna}
\newcommand{\FASTavroraFTOWCPMem}{13}
\newcommand{\FASTavroraFTOWCPMemCI}{0.65}
\newcommand{\FASTavroraREWCPMem}{9.1}
\newcommand{\FASTavroraREWCPMemCI}{0.46}
\newcommand{\FASTavroraDCMem}{\memna}
\newcommand{\FASTavroraDCMemCI}{\memna}
\newcommand{\FASTavroraDCMemCIMIN}{\memna}
\newcommand{\FASTavroraDCMemCIMAX}{\memna}
\newcommand{\FASTavroraFTODCMem}{13}
\newcommand{\FASTavroraFTODCMemCI}{0.64}
\newcommand{\FASTavroraREDCMem}{8.4}
\newcommand{\FASTavroraREDCMemCI}{0.44}
\newcommand{\FASTavroraCAPOMem}{\memna}
\newcommand{\FASTavroraCAPOMemCI}{\memna}
\newcommand{\FASTavroraCAPOMemCIMIN}{\memna}
\newcommand{\FASTavroraCAPOMemCIMAX}{\memna}
\newcommand{\FASTavroraFTOCAPOMem}{9.7}
\newcommand{\FASTavroraFTOCAPOMemCI}{0.47}
\newcommand{\FASTavroraRECAPOMem}{5.9}
\newcommand{\FASTavroraRECAPOMemCI}{0.28}
\newcommand{\FASTavroraAGGCAPOMem}{\memna}
\newcommand{\FASTavroraAGGCAPOMemCI}{\memna}
\newcommand{\FASTavroraAGGCAPOMemCIMIN}{\memna}
\newcommand{\FASTavroraAGGCAPOMemCIMAX}{\memna}
\newcommand{\FASTavroraEventsCI}{0}
\newcommand{\FASTavroraEventsCIMIN}{0}
\newcommand{\FASTavroraEventsCIMAX}{0}
\newcommand{\FASTavroraNoFPEventsCI}{0}
\newcommand{\FASTavroraNoFPEventsCIMIN}{0}
\newcommand{\FASTavroraNoFPEventsCIMAX}{0}
\newcommand{\FASTavroraFT}{3}
\newcommand{\FASTavroraFTCI}{0}
\newcommand{\FASTavroraFTCIMIN}{3}
\newcommand{\FASTavroraFTCIMAX}{3}
\newcommand{\FASTavroraFTDynamic}{755,846}
\newcommand{\FASTavroraFTDynamicCI}{2,296}
\newcommand{\FASTavroraFTDynamicCIMIN}{753,550}
\newcommand{\FASTavroraFTDynamicCIMAX}{758,142}
\newcommand{\FASTavroraHB}{6}
\newcommand{\FASTavroraHBCI}{0}
\newcommand{\FASTavroraHBCIMIN}{6}
\newcommand{\FASTavroraHBCIMAX}{6}
\newcommand{\FASTavroraHBDynamic}{426,801}
\newcommand{\FASTavroraHBDynamicCI}{350}
\newcommand{\FASTavroraHBDynamicCIMIN}{426,451}
\newcommand{\FASTavroraHBDynamicCIMAX}{427,151}
\newcommand{\FASTavroraFTOHB}{6}
\newcommand{\FASTavroraFTOHBCI}{0}
\newcommand{\FASTavroraFTOHBCIMIN}{6}
\newcommand{\FASTavroraFTOHBCIMAX}{6}
\newcommand{\FASTavroraFTOHBDynamic}{407,783}
\newcommand{\FASTavroraFTOHBDynamicCI}{428}
\newcommand{\FASTavroraFTOHBDynamicCIMIN}{407,355}
\newcommand{\FASTavroraFTOHBDynamicCIMAX}{408,211}
\newcommand{\FASTavroraWCP}{\rna}
\newcommand{\FASTavroraWCPCI}{\rna}
\newcommand{\FASTavroraWCPCIMIN}{\rna}
\newcommand{\FASTavroraWCPCIMAX}{\rna}
\newcommand{\FASTavroraWCPDynamic}{\rna}
\newcommand{\FASTavroraWCPDynamicCI}{\rna}
\newcommand{\FASTavroraWCPDynamicCIMIN}{\rna}
\newcommand{\FASTavroraWCPDynamicCIMAX}{\rna}
\newcommand{\FASTavroraFTOWCP}{6}
\newcommand{\FASTavroraFTOWCPCI}{0}
\newcommand{\FASTavroraFTOWCPCIMIN}{6}
\newcommand{\FASTavroraFTOWCPCIMAX}{6}
\newcommand{\FASTavroraFTOWCPDynamic}{404,826}
\newcommand{\FASTavroraFTOWCPDynamicCI}{148}
\newcommand{\FASTavroraFTOWCPDynamicCIMIN}{404,678}
\newcommand{\FASTavroraFTOWCPDynamicCIMAX}{404,974}
\newcommand{\FASTavroraREWCP}{6}
\newcommand{\FASTavroraREWCPCI}{0}
\newcommand{\FASTavroraREWCPCIMIN}{6}
\newcommand{\FASTavroraREWCPCIMAX}{6}
\newcommand{\FASTavroraREWCPDynamic}{406,667}
\newcommand{\FASTavroraREWCPDynamicCI}{233}
\newcommand{\FASTavroraREWCPDynamicCIMIN}{406,434}
\newcommand{\FASTavroraREWCPDynamicCIMAX}{406,900}
\newcommand{\FASTavroraDC}{\rna}
\newcommand{\FASTavroraDCCI}{\rna}
\newcommand{\FASTavroraDCCIMIN}{\rna}
\newcommand{\FASTavroraDCCIMAX}{\rna}
\newcommand{\FASTavroraDCDynamic}{\rna}
\newcommand{\FASTavroraDCDynamicCI}{\rna}
\newcommand{\FASTavroraDCDynamicCIMIN}{\rna}
\newcommand{\FASTavroraDCDynamicCIMAX}{\rna}
\newcommand{\FASTavroraFTODC}{6}
\newcommand{\FASTavroraFTODCCI}{0}
\newcommand{\FASTavroraFTODCCIMIN}{6}
\newcommand{\FASTavroraFTODCCIMAX}{6}
\newcommand{\FASTavroraFTODCDynamic}{404,260}
\newcommand{\FASTavroraFTODCDynamicCI}{325}
\newcommand{\FASTavroraFTODCDynamicCIMIN}{403,935}
\newcommand{\FASTavroraFTODCDynamicCIMAX}{404,585}
\newcommand{\FASTavroraREDC}{6}
\newcommand{\FASTavroraREDCCI}{0}
\newcommand{\FASTavroraREDCCIMIN}{6}
\newcommand{\FASTavroraREDCCIMAX}{6}
\newcommand{\FASTavroraREDCDynamic}{407,104}
\newcommand{\FASTavroraREDCDynamicCI}{102}
\newcommand{\FASTavroraREDCDynamicCIMIN}{407,002}
\newcommand{\FASTavroraREDCDynamicCIMAX}{407,206}
\newcommand{\FASTavroraCAPO}{\rna}
\newcommand{\FASTavroraCAPOCI}{\rna}
\newcommand{\FASTavroraCAPOCIMIN}{\rna}
\newcommand{\FASTavroraCAPOCIMAX}{\rna}
\newcommand{\FASTavroraCAPODynamic}{\rna}
\newcommand{\FASTavroraCAPODynamicCI}{\rna}
\newcommand{\FASTavroraCAPODynamicCIMIN}{\rna}
\newcommand{\FASTavroraCAPODynamicCIMAX}{\rna}
\newcommand{\FASTavroraFTOCAPO}{6}
\newcommand{\FASTavroraFTOCAPOCI}{0.2}
\newcommand{\FASTavroraFTOCAPOCIMIN}{6}
\newcommand{\FASTavroraFTOCAPOCIMAX}{6}
\newcommand{\FASTavroraFTOCAPODynamic}{406,348}
\newcommand{\FASTavroraFTOCAPODynamicCI}{125}
\newcommand{\FASTavroraFTOCAPODynamicCIMIN}{406,223}
\newcommand{\FASTavroraFTOCAPODynamicCIMAX}{406,473}
\newcommand{\FASTavroraRECAPO}{6}
\newcommand{\FASTavroraRECAPOCI}{0}
\newcommand{\FASTavroraRECAPOCIMIN}{6}
\newcommand{\FASTavroraRECAPOCIMAX}{6}
\newcommand{\FASTavroraRECAPODynamic}{408,677}
\newcommand{\FASTavroraRECAPODynamicCI}{218}
\newcommand{\FASTavroraRECAPODynamicCIMIN}{408,459}
\newcommand{\FASTavroraRECAPODynamicCIMAX}{408,895}
\newcommand{\FASTavroraAGGCAPO}{\rna}
\newcommand{\FASTavroraAGGCAPOCI}{\rna}
\newcommand{\FASTavroraAGGCAPOCIMIN}{\rna}
\newcommand{\FASTavroraAGGCAPOCIMAX}{\rna}
\newcommand{\FASTavroraAGGCAPODynamic}{\rna}
\newcommand{\FASTavroraAGGCAPODynamicCI}{\rna}
\newcommand{\FASTavroraAGGCAPODynamicCIMIN}{\rna}
\newcommand{\FASTavroraAGGCAPODynamicCIMAX}{\rna}
\newcommand{\FASTbatikEvents}{0}
\newcommand{\FASTbatikNoFPEvents}{0}
\newcommand{\FASTbatikMaxLiveThreads}{2}
\newcommand{\FASTbatikTotalThreads}{7}
\newcommand{\FASTbatikBaseTime}{2.6}
\newcommand{\FASTbatikBaseTimeCI}{41}
\newcommand{\FASTbatikEmptyTime}{\rna}
\newcommand{\FASTbatikEmptyTimeCI}{\rna}
\newcommand{\FASTbatikEmptyTimeCIMIN}{\rna}
\newcommand{\FASTbatikEmptyTimeCIMAX}{\rna}
\newcommand{\FASTbatikFTTime}{4.2}
\newcommand{\FASTbatikFTTimeCI}{0.10}
\newcommand{\FASTbatikHBTime}{4.1}
\newcommand{\FASTbatikHBTimeCI}{0.075}
\newcommand{\FASTbatikFTOHBTime}{4.1}
\newcommand{\FASTbatikFTOHBTimeCI}{0.082}
\newcommand{\FASTbatikWCPTime}{\rna}
\newcommand{\FASTbatikWCPTimeCI}{\rna}
\newcommand{\FASTbatikWCPTimeCIMIN}{\rna}
\newcommand{\FASTbatikWCPTimeCIMAX}{\rna}
\newcommand{\FASTbatikFTOWCPTime}{7.0}
\newcommand{\FASTbatikFTOWCPTimeCI}{0.15}
\newcommand{\FASTbatikREWCPTime}{4.7}
\newcommand{\FASTbatikREWCPTimeCI}{0.092}
\newcommand{\FASTbatikDCTime}{\rna}
\newcommand{\FASTbatikDCTimeCI}{\rna}
\newcommand{\FASTbatikDCTimeCIMIN}{\rna}
\newcommand{\FASTbatikDCTimeCIMAX}{\rna}
\newcommand{\FASTbatikFTODCTime}{7.1}
\newcommand{\FASTbatikFTODCTimeCI}{0.12}
\newcommand{\FASTbatikREDCTime}{4.7}
\newcommand{\FASTbatikREDCTimeCI}{0.088}
\newcommand{\FASTbatikCAPOTime}{\rna}
\newcommand{\FASTbatikCAPOTimeCI}{\rna}
\newcommand{\FASTbatikCAPOTimeCIMIN}{\rna}
\newcommand{\FASTbatikCAPOTimeCIMAX}{\rna}
\newcommand{\FASTbatikFTOCAPOTime}{7.1}
\newcommand{\FASTbatikFTOCAPOTimeCI}{0.081}
\newcommand{\FASTbatikRECAPOTime}{4.7}
\newcommand{\FASTbatikRECAPOTimeCI}{0.044}
\newcommand{\FASTbatikAGGCAPOTime}{\rna}
\newcommand{\FASTbatikAGGCAPOTimeCI}{\rna}
\newcommand{\FASTbatikAGGCAPOTimeCIMIN}{\rna}
\newcommand{\FASTbatikAGGCAPOTimeCIMAX}{\rna}
\newcommand{\FASTbatikStaticTime}{\rzero}
\newcommand{\FASTbatikDynamicTime}{\rzero}
\newcommand{\FASTbatikBaseMem}{220}
\newcommand{\FASTbatikBaseMemCI}{6.5}
\newcommand{\FASTbatikFTMem}{5.5}
\newcommand{\FASTbatikFTMemCI}{0.17}
\newcommand{\FASTbatikHBMem}{5.5}
\newcommand{\FASTbatikHBMemCI}{0.18}
\newcommand{\FASTbatikFTOHBMem}{5.5}
\newcommand{\FASTbatikFTOHBMemCI}{0.14}
\newcommand{\FASTbatikWCPMem}{\memna}
\newcommand{\FASTbatikWCPMemCI}{\memna}
\newcommand{\FASTbatikWCPMemCIMIN}{\memna}
\newcommand{\FASTbatikWCPMemCIMAX}{\memna}
\newcommand{\FASTbatikFTOWCPMem}{14}
\newcommand{\FASTbatikFTOWCPMemCI}{0.38}
\newcommand{\FASTbatikREWCPMem}{11}
\newcommand{\FASTbatikREWCPMemCI}{0.32}
\newcommand{\FASTbatikDCMem}{\memna}
\newcommand{\FASTbatikDCMemCI}{\memna}
\newcommand{\FASTbatikDCMemCIMIN}{\memna}
\newcommand{\FASTbatikDCMemCIMAX}{\memna}
\newcommand{\FASTbatikFTODCMem}{14}
\newcommand{\FASTbatikFTODCMemCI}{0.46}
\newcommand{\FASTbatikREDCMem}{11}
\newcommand{\FASTbatikREDCMemCI}{0.29}
\newcommand{\FASTbatikCAPOMem}{\memna}
\newcommand{\FASTbatikCAPOMemCI}{\memna}
\newcommand{\FASTbatikCAPOMemCIMIN}{\memna}
\newcommand{\FASTbatikCAPOMemCIMAX}{\memna}
\newcommand{\FASTbatikFTOCAPOMem}{14}
\newcommand{\FASTbatikFTOCAPOMemCI}{0.57}
\newcommand{\FASTbatikRECAPOMem}{11}
\newcommand{\FASTbatikRECAPOMemCI}{0.30}
\newcommand{\FASTbatikAGGCAPOMem}{\memna}
\newcommand{\FASTbatikAGGCAPOMemCI}{\memna}
\newcommand{\FASTbatikAGGCAPOMemCIMIN}{\memna}
\newcommand{\FASTbatikAGGCAPOMemCIMAX}{\memna}
\newcommand{\FASTbatikEventsCI}{0}
\newcommand{\FASTbatikEventsCIMIN}{0}
\newcommand{\FASTbatikEventsCIMAX}{0}
\newcommand{\FASTbatikNoFPEventsCI}{0}
\newcommand{\FASTbatikNoFPEventsCIMIN}{0}
\newcommand{\FASTbatikNoFPEventsCIMAX}{0}
\newcommand{\FASTbatikFT}{0}
\newcommand{\FASTbatikFTCI}{0}
\newcommand{\FASTbatikFTCIMIN}{0}
\newcommand{\FASTbatikFTCIMAX}{0}
\newcommand{\FASTbatikFTDynamic}{0}
\newcommand{\FASTbatikFTDynamicCI}{0}
\newcommand{\FASTbatikFTDynamicCIMIN}{0}
\newcommand{\FASTbatikFTDynamicCIMAX}{0}
\newcommand{\FASTbatikHB}{0}
\newcommand{\FASTbatikHBCI}{0}
\newcommand{\FASTbatikHBCIMIN}{0}
\newcommand{\FASTbatikHBCIMAX}{0}
\newcommand{\FASTbatikHBDynamic}{0}
\newcommand{\FASTbatikHBDynamicCI}{0}
\newcommand{\FASTbatikHBDynamicCIMIN}{0}
\newcommand{\FASTbatikHBDynamicCIMAX}{0}
\newcommand{\FASTbatikFTOHB}{0}
\newcommand{\FASTbatikFTOHBCI}{0}
\newcommand{\FASTbatikFTOHBCIMIN}{0}
\newcommand{\FASTbatikFTOHBCIMAX}{0}
\newcommand{\FASTbatikFTOHBDynamic}{0}
\newcommand{\FASTbatikFTOHBDynamicCI}{0}
\newcommand{\FASTbatikFTOHBDynamicCIMIN}{0}
\newcommand{\FASTbatikFTOHBDynamicCIMAX}{0}
\newcommand{\FASTbatikWCP}{\rna}
\newcommand{\FASTbatikWCPCI}{\rna}
\newcommand{\FASTbatikWCPCIMIN}{\rna}
\newcommand{\FASTbatikWCPCIMAX}{\rna}
\newcommand{\FASTbatikWCPDynamic}{\rna}
\newcommand{\FASTbatikWCPDynamicCI}{\rna}
\newcommand{\FASTbatikWCPDynamicCIMIN}{\rna}
\newcommand{\FASTbatikWCPDynamicCIMAX}{\rna}
\newcommand{\FASTbatikFTOWCP}{0}
\newcommand{\FASTbatikFTOWCPCI}{0}
\newcommand{\FASTbatikFTOWCPCIMIN}{0}
\newcommand{\FASTbatikFTOWCPCIMAX}{0}
\newcommand{\FASTbatikFTOWCPDynamic}{0}
\newcommand{\FASTbatikFTOWCPDynamicCI}{0}
\newcommand{\FASTbatikFTOWCPDynamicCIMIN}{0}
\newcommand{\FASTbatikFTOWCPDynamicCIMAX}{0}
\newcommand{\FASTbatikREWCP}{0}
\newcommand{\FASTbatikREWCPCI}{0}
\newcommand{\FASTbatikREWCPCIMIN}{0}
\newcommand{\FASTbatikREWCPCIMAX}{0}
\newcommand{\FASTbatikREWCPDynamic}{0}
\newcommand{\FASTbatikREWCPDynamicCI}{0}
\newcommand{\FASTbatikREWCPDynamicCIMIN}{0}
\newcommand{\FASTbatikREWCPDynamicCIMAX}{0}
\newcommand{\FASTbatikDC}{\rna}
\newcommand{\FASTbatikDCCI}{\rna}
\newcommand{\FASTbatikDCCIMIN}{\rna}
\newcommand{\FASTbatikDCCIMAX}{\rna}
\newcommand{\FASTbatikDCDynamic}{\rna}
\newcommand{\FASTbatikDCDynamicCI}{\rna}
\newcommand{\FASTbatikDCDynamicCIMIN}{\rna}
\newcommand{\FASTbatikDCDynamicCIMAX}{\rna}
\newcommand{\FASTbatikFTODC}{0}
\newcommand{\FASTbatikFTODCCI}{0}
\newcommand{\FASTbatikFTODCCIMIN}{0}
\newcommand{\FASTbatikFTODCCIMAX}{0}
\newcommand{\FASTbatikFTODCDynamic}{0}
\newcommand{\FASTbatikFTODCDynamicCI}{0}
\newcommand{\FASTbatikFTODCDynamicCIMIN}{0}
\newcommand{\FASTbatikFTODCDynamicCIMAX}{0}
\newcommand{\FASTbatikREDC}{0}
\newcommand{\FASTbatikREDCCI}{0}
\newcommand{\FASTbatikREDCCIMIN}{0}
\newcommand{\FASTbatikREDCCIMAX}{0}
\newcommand{\FASTbatikREDCDynamic}{0}
\newcommand{\FASTbatikREDCDynamicCI}{0}
\newcommand{\FASTbatikREDCDynamicCIMIN}{0}
\newcommand{\FASTbatikREDCDynamicCIMAX}{0}
\newcommand{\FASTbatikCAPO}{\rna}
\newcommand{\FASTbatikCAPOCI}{\rna}
\newcommand{\FASTbatikCAPOCIMIN}{\rna}
\newcommand{\FASTbatikCAPOCIMAX}{\rna}
\newcommand{\FASTbatikCAPODynamic}{\rna}
\newcommand{\FASTbatikCAPODynamicCI}{\rna}
\newcommand{\FASTbatikCAPODynamicCIMIN}{\rna}
\newcommand{\FASTbatikCAPODynamicCIMAX}{\rna}
\newcommand{\FASTbatikFTOCAPO}{0}
\newcommand{\FASTbatikFTOCAPOCI}{0}
\newcommand{\FASTbatikFTOCAPOCIMIN}{0}
\newcommand{\FASTbatikFTOCAPOCIMAX}{0}
\newcommand{\FASTbatikFTOCAPODynamic}{0}
\newcommand{\FASTbatikFTOCAPODynamicCI}{0}
\newcommand{\FASTbatikFTOCAPODynamicCIMIN}{0}
\newcommand{\FASTbatikFTOCAPODynamicCIMAX}{0}
\newcommand{\FASTbatikRECAPO}{0}
\newcommand{\FASTbatikRECAPOCI}{0}
\newcommand{\FASTbatikRECAPOCIMIN}{0}
\newcommand{\FASTbatikRECAPOCIMAX}{0}
\newcommand{\FASTbatikRECAPODynamic}{0}
\newcommand{\FASTbatikRECAPODynamicCI}{0}
\newcommand{\FASTbatikRECAPODynamicCIMIN}{0}
\newcommand{\FASTbatikRECAPODynamicCIMAX}{0}
\newcommand{\FASTbatikAGGCAPO}{\rna}
\newcommand{\FASTbatikAGGCAPOCI}{\rna}
\newcommand{\FASTbatikAGGCAPOCIMIN}{\rna}
\newcommand{\FASTbatikAGGCAPOCIMAX}{\rna}
\newcommand{\FASTbatikAGGCAPODynamic}{\rna}
\newcommand{\FASTbatikAGGCAPODynamicCI}{\rna}
\newcommand{\FASTbatikAGGCAPODynamicCIMIN}{\rna}
\newcommand{\FASTbatikAGGCAPODynamicCIMAX}{\rna}
\newcommand{\FASThtwoEvents}{0}
\newcommand{\FASThtwoNoFPEvents}{0}
\newcommand{\FASThtwoMaxLiveThreads}{9}
\newcommand{\FASThtwoTotalThreads}{10}
\newcommand{\FASThtwoBaseTime}{4.8}
\newcommand{\FASThtwoBaseTimeCI}{51}
\newcommand{\FASThtwoEmptyTime}{\rna}
\newcommand{\FASThtwoEmptyTimeCI}{\rna}
\newcommand{\FASThtwoEmptyTimeCIMIN}{\rna}
\newcommand{\FASThtwoEmptyTimeCIMAX}{\rna}
\newcommand{\FASThtwoFTTime}{9.6}
\newcommand{\FASThtwoFTTimeCI}{0.42}
\newcommand{\FASThtwoHBTime}{8.8}
\newcommand{\FASThtwoHBTimeCI}{0.17}
\newcommand{\FASThtwoFTOHBTime}{8.8}
\newcommand{\FASThtwoFTOHBTimeCI}{0.15}
\newcommand{\FASThtwoWCPTime}{\rna}
\newcommand{\FASThtwoWCPTimeCI}{\rna}
\newcommand{\FASThtwoWCPTimeCIMIN}{\rna}
\newcommand{\FASThtwoWCPTimeCIMAX}{\rna}
\newcommand{\FASThtwoFTOWCPTime}{60}
\newcommand{\FASThtwoFTOWCPTimeCI}{4.8}
\newcommand{\FASThtwoREWCPTime}{12}
\newcommand{\FASThtwoREWCPTimeCI}{0.34}
\newcommand{\FASThtwoDCTime}{\rna}
\newcommand{\FASThtwoDCTimeCI}{\rna}
\newcommand{\FASThtwoDCTimeCIMIN}{\rna}
\newcommand{\FASThtwoDCTimeCIMAX}{\rna}
\newcommand{\FASThtwoFTODCTime}{60}
\newcommand{\FASThtwoFTODCTimeCI}{5.2}
\newcommand{\FASThtwoREDCTime}{11}
\newcommand{\FASThtwoREDCTimeCI}{0.36}
\newcommand{\FASThtwoCAPOTime}{\rna}
\newcommand{\FASThtwoCAPOTimeCI}{\rna}
\newcommand{\FASThtwoCAPOTimeCIMIN}{\rna}
\newcommand{\FASThtwoCAPOTimeCIMAX}{\rna}
\newcommand{\FASThtwoFTOCAPOTime}{57}
\newcommand{\FASThtwoFTOCAPOTimeCI}{5.3}
\newcommand{\FASThtwoRECAPOTime}{11}
\newcommand{\FASThtwoRECAPOTimeCI}{0.14}
\newcommand{\FASThtwoAGGCAPOTime}{\rna}
\newcommand{\FASThtwoAGGCAPOTimeCI}{\rna}
\newcommand{\FASThtwoAGGCAPOTimeCIMIN}{\rna}
\newcommand{\FASThtwoAGGCAPOTimeCIMAX}{\rna}
\newcommand{\FASThtwoStaticTime}{\rzero}
\newcommand{\FASThtwoDynamicTime}{\rzero}
\newcommand{\FASThtwoBaseMem}{1,800}
\newcommand{\FASThtwoBaseMemCI}{51.0}
\newcommand{\FASThtwoFTMem}{3.1}
\newcommand{\FASThtwoFTMemCI}{0.076}
\newcommand{\FASThtwoHBMem}{3.1}
\newcommand{\FASThtwoHBMemCI}{0.093}
\newcommand{\FASThtwoFTOHBMem}{3.1}
\newcommand{\FASThtwoFTOHBMemCI}{0.12}
\newcommand{\FASThtwoWCPMem}{\memna}
\newcommand{\FASThtwoWCPMemCI}{\memna}
\newcommand{\FASThtwoWCPMemCIMIN}{\memna}
\newcommand{\FASThtwoWCPMemCIMAX}{\memna}
\newcommand{\FASThtwoFTOWCPMem}{42}
\newcommand{\FASThtwoFTOWCPMemCI}{2.2}
\newcommand{\FASThtwoREWCPMem}{7.9}
\newcommand{\FASThtwoREWCPMemCI}{0.22}
\newcommand{\FASThtwoDCMem}{\memna}
\newcommand{\FASThtwoDCMemCI}{\memna}
\newcommand{\FASThtwoDCMemCIMIN}{\memna}
\newcommand{\FASThtwoDCMemCIMAX}{\memna}
\newcommand{\FASThtwoFTODCMem}{41}
\newcommand{\FASThtwoFTODCMemCI}{2.1}
\newcommand{\FASThtwoREDCMem}{7.9}
\newcommand{\FASThtwoREDCMemCI}{0.28}
\newcommand{\FASThtwoCAPOMem}{\memna}
\newcommand{\FASThtwoCAPOMemCI}{\memna}
\newcommand{\FASThtwoCAPOMemCIMIN}{\memna}
\newcommand{\FASThtwoCAPOMemCIMAX}{\memna}
\newcommand{\FASThtwoFTOCAPOMem}{40}
\newcommand{\FASThtwoFTOCAPOMemCI}{2.7}
\newcommand{\FASThtwoRECAPOMem}{7.3}
\newcommand{\FASThtwoRECAPOMemCI}{0.26}
\newcommand{\FASThtwoAGGCAPOMem}{\memna}
\newcommand{\FASThtwoAGGCAPOMemCI}{\memna}
\newcommand{\FASThtwoAGGCAPOMemCIMIN}{\memna}
\newcommand{\FASThtwoAGGCAPOMemCIMAX}{\memna}
\newcommand{\FASThtwoEventsCI}{0}
\newcommand{\FASThtwoEventsCIMIN}{0}
\newcommand{\FASThtwoEventsCIMAX}{0}
\newcommand{\FASThtwoNoFPEventsCI}{0}
\newcommand{\FASThtwoNoFPEventsCIMIN}{0}
\newcommand{\FASThtwoNoFPEventsCIMAX}{0}
\newcommand{\FASThtwoFT}{11}
\newcommand{\FASThtwoFTCI}{0}
\newcommand{\FASThtwoFTCIMIN}{11}
\newcommand{\FASThtwoFTCIMAX}{11}
\newcommand{\FASThtwoFTDynamic}{92,976}
\newcommand{\FASThtwoFTDynamicCI}{483}
\newcommand{\FASThtwoFTDynamicCIMIN}{92,493}
\newcommand{\FASThtwoFTDynamicCIMAX}{93,459}
\newcommand{\FASThtwoHB}{13}
\newcommand{\FASThtwoHBCI}{0.43}
\newcommand{\FASThtwoHBCIMIN}{13}
\newcommand{\FASThtwoHBCIMAX}{13}
\newcommand{\FASThtwoHBDynamic}{68,528}
\newcommand{\FASThtwoHBDynamicCI}{1,085}
\newcommand{\FASThtwoHBDynamicCIMIN}{67,443}
\newcommand{\FASThtwoHBDynamicCIMAX}{69,613}
\newcommand{\FASThtwoFTOHB}{13}
\newcommand{\FASThtwoFTOHBCI}{0.2}
\newcommand{\FASThtwoFTOHBCIMIN}{13}
\newcommand{\FASThtwoFTOHBCIMAX}{13}
\newcommand{\FASThtwoFTOHBDynamic}{51,472}
\newcommand{\FASThtwoFTOHBDynamicCI}{2,499}
\newcommand{\FASThtwoFTOHBDynamicCIMIN}{48,973}
\newcommand{\FASThtwoFTOHBDynamicCIMAX}{53,971}
\newcommand{\FASThtwoWCP}{\rna}
\newcommand{\FASThtwoWCPCI}{\rna}
\newcommand{\FASThtwoWCPCIMIN}{\rna}
\newcommand{\FASThtwoWCPCIMAX}{\rna}
\newcommand{\FASThtwoWCPDynamic}{\rna}
\newcommand{\FASThtwoWCPDynamicCI}{\rna}
\newcommand{\FASThtwoWCPDynamicCIMIN}{\rna}
\newcommand{\FASThtwoWCPDynamicCIMAX}{\rna}
\newcommand{\FASThtwoFTOWCP}{13}
\newcommand{\FASThtwoFTOWCPCI}{0.3}
\newcommand{\FASThtwoFTOWCPCIMIN}{13}
\newcommand{\FASThtwoFTOWCPCIMAX}{13}
\newcommand{\FASThtwoFTOWCPDynamic}{79,543}
\newcommand{\FASThtwoFTOWCPDynamicCI}{236}
\newcommand{\FASThtwoFTOWCPDynamicCIMIN}{79,307}
\newcommand{\FASThtwoFTOWCPDynamicCIMAX}{79,779}
\newcommand{\FASThtwoREWCP}{13}
\newcommand{\FASThtwoREWCPCI}{0.2}
\newcommand{\FASThtwoREWCPCIMIN}{13}
\newcommand{\FASThtwoREWCPCIMAX}{13}
\newcommand{\FASThtwoREWCPDynamic}{68,714}
\newcommand{\FASThtwoREWCPDynamicCI}{4,274}
\newcommand{\FASThtwoREWCPDynamicCIMIN}{64,440}
\newcommand{\FASThtwoREWCPDynamicCIMAX}{72,988}
\newcommand{\FASThtwoDC}{\rna}
\newcommand{\FASThtwoDCCI}{\rna}
\newcommand{\FASThtwoDCCIMIN}{\rna}
\newcommand{\FASThtwoDCCIMAX}{\rna}
\newcommand{\FASThtwoDCDynamic}{\rna}
\newcommand{\FASThtwoDCDynamicCI}{\rna}
\newcommand{\FASThtwoDCDynamicCIMIN}{\rna}
\newcommand{\FASThtwoDCDynamicCIMAX}{\rna}
\newcommand{\FASThtwoFTODC}{13}
\newcommand{\FASThtwoFTODCCI}{0}
\newcommand{\FASThtwoFTODCCIMIN}{13}
\newcommand{\FASThtwoFTODCCIMAX}{13}
\newcommand{\FASThtwoFTODCDynamic}{79,966}
\newcommand{\FASThtwoFTODCDynamicCI}{140}
\newcommand{\FASThtwoFTODCDynamicCIMIN}{79,826}
\newcommand{\FASThtwoFTODCDynamicCIMAX}{80,106}
\newcommand{\FASThtwoREDC}{13}
\newcommand{\FASThtwoREDCCI}{0}
\newcommand{\FASThtwoREDCCIMIN}{13}
\newcommand{\FASThtwoREDCCIMAX}{13}
\newcommand{\FASThtwoREDCDynamic}{71,201}
\newcommand{\FASThtwoREDCDynamicCI}{4,168}
\newcommand{\FASThtwoREDCDynamicCIMIN}{67,033}
\newcommand{\FASThtwoREDCDynamicCIMAX}{75,369}
\newcommand{\FASThtwoCAPO}{\rna}
\newcommand{\FASThtwoCAPOCI}{\rna}
\newcommand{\FASThtwoCAPOCIMIN}{\rna}
\newcommand{\FASThtwoCAPOCIMAX}{\rna}
\newcommand{\FASThtwoCAPODynamic}{\rna}
\newcommand{\FASThtwoCAPODynamicCI}{\rna}
\newcommand{\FASThtwoCAPODynamicCIMIN}{\rna}
\newcommand{\FASThtwoCAPODynamicCIMAX}{\rna}
\newcommand{\FASThtwoFTOCAPO}{13}
\newcommand{\FASThtwoFTOCAPOCI}{0}
\newcommand{\FASThtwoFTOCAPOCIMIN}{13}
\newcommand{\FASThtwoFTOCAPOCIMAX}{13}
\newcommand{\FASThtwoFTOCAPODynamic}{79,934}
\newcommand{\FASThtwoFTOCAPODynamicCI}{119}
\newcommand{\FASThtwoFTOCAPODynamicCIMIN}{79,815}
\newcommand{\FASThtwoFTOCAPODynamicCIMAX}{80,053}
\newcommand{\FASThtwoRECAPO}{13}
\newcommand{\FASThtwoRECAPOCI}{0}
\newcommand{\FASThtwoRECAPOCIMIN}{13}
\newcommand{\FASThtwoRECAPOCIMAX}{13}
\newcommand{\FASThtwoRECAPODynamic}{56,735}
\newcommand{\FASThtwoRECAPODynamicCI}{3,210}
\newcommand{\FASThtwoRECAPODynamicCIMIN}{53,525}
\newcommand{\FASThtwoRECAPODynamicCIMAX}{59,945}
\newcommand{\FASThtwoAGGCAPO}{\rna}
\newcommand{\FASThtwoAGGCAPOCI}{\rna}
\newcommand{\FASThtwoAGGCAPOCIMIN}{\rna}
\newcommand{\FASThtwoAGGCAPOCIMAX}{\rna}
\newcommand{\FASThtwoAGGCAPODynamic}{\rna}
\newcommand{\FASThtwoAGGCAPODynamicCI}{\rna}
\newcommand{\FASThtwoAGGCAPODynamicCIMIN}{\rna}
\newcommand{\FASThtwoAGGCAPODynamicCIMAX}{\rna}
\newcommand{\FASTjythonEvents}{0}
\newcommand{\FASTjythonNoFPEvents}{0}
\newcommand{\FASTjythonMaxLiveThreads}{2}
\newcommand{\FASTjythonTotalThreads}{2}
\newcommand{\FASTjythonBaseTime}{3.9}
\newcommand{\FASTjythonBaseTimeCI}{110}
\newcommand{\FASTjythonEmptyTime}{\rna}
\newcommand{\FASTjythonEmptyTimeCI}{\rna}
\newcommand{\FASTjythonEmptyTimeCIMIN}{\rna}
\newcommand{\FASTjythonEmptyTimeCIMAX}{\rna}
\newcommand{\FASTjythonFTTime}{8.4}
\newcommand{\FASTjythonFTTimeCI}{0.25}
\newcommand{\FASTjythonHBTime}{8.7}
\newcommand{\FASTjythonHBTimeCI}{0.30}
\newcommand{\FASTjythonFTOHBTime}{8.5}
\newcommand{\FASTjythonFTOHBTimeCI}{0.21}
\newcommand{\FASTjythonWCPTime}{\rna}
\newcommand{\FASTjythonWCPTimeCI}{\rna}
\newcommand{\FASTjythonWCPTimeCIMIN}{\rna}
\newcommand{\FASTjythonWCPTimeCIMAX}{\rna}
\newcommand{\FASTjythonFTOWCPTime}{11}
\newcommand{\FASTjythonFTOWCPTimeCI}{0.24}
\newcommand{\FASTjythonREWCPTime}{12}
\newcommand{\FASTjythonREWCPTimeCI}{0.37}
\newcommand{\FASTjythonDCTime}{\rna}
\newcommand{\FASTjythonDCTimeCI}{\rna}
\newcommand{\FASTjythonDCTimeCIMIN}{\rna}
\newcommand{\FASTjythonDCTimeCIMAX}{\rna}
\newcommand{\FASTjythonFTODCTime}{11}
\newcommand{\FASTjythonFTODCTimeCI}{0.24}
\newcommand{\FASTjythonREDCTime}{12}
\newcommand{\FASTjythonREDCTimeCI}{0.34}
\newcommand{\FASTjythonCAPOTime}{\rna}
\newcommand{\FASTjythonCAPOTimeCI}{\rna}
\newcommand{\FASTjythonCAPOTimeCIMIN}{\rna}
\newcommand{\FASTjythonCAPOTimeCIMAX}{\rna}
\newcommand{\FASTjythonFTOCAPOTime}{8.7}
\newcommand{\FASTjythonFTOCAPOTimeCI}{0.19}
\newcommand{\FASTjythonRECAPOTime}{9.4}
\newcommand{\FASTjythonRECAPOTimeCI}{0.39}
\newcommand{\FASTjythonAGGCAPOTime}{\rna}
\newcommand{\FASTjythonAGGCAPOTimeCI}{\rna}
\newcommand{\FASTjythonAGGCAPOTimeCIMIN}{\rna}
\newcommand{\FASTjythonAGGCAPOTimeCIMAX}{\rna}
\newcommand{\FASTjythonStaticTime}{\rzero}
\newcommand{\FASTjythonDynamicTime}{\rzero}
\newcommand{\FASTjythonBaseMem}{730}
\newcommand{\FASTjythonBaseMemCI}{5.1}
\newcommand{\FASTjythonFTMem}{5.9}
\newcommand{\FASTjythonFTMemCI}{0.18}
\newcommand{\FASTjythonHBMem}{6.8}
\newcommand{\FASTjythonHBMemCI}{0.31}
\newcommand{\FASTjythonFTOHBMem}{6.3}
\newcommand{\FASTjythonFTOHBMemCI}{0.12}
\newcommand{\FASTjythonWCPMem}{\memna}
\newcommand{\FASTjythonWCPMemCI}{\memna}
\newcommand{\FASTjythonWCPMemCIMIN}{\memna}
\newcommand{\FASTjythonWCPMemCIMAX}{\memna}
\newcommand{\FASTjythonFTOWCPMem}{12}
\newcommand{\FASTjythonFTOWCPMemCI}{0.39}
\newcommand{\FASTjythonREWCPMem}{15}
\newcommand{\FASTjythonREWCPMemCI}{0.68}
\newcommand{\FASTjythonDCMem}{\memna}
\newcommand{\FASTjythonDCMemCI}{\memna}
\newcommand{\FASTjythonDCMemCIMIN}{\memna}
\newcommand{\FASTjythonDCMemCIMAX}{\memna}
\newcommand{\FASTjythonFTODCMem}{13}
\newcommand{\FASTjythonFTODCMemCI}{0.22}
\newcommand{\FASTjythonREDCMem}{15}
\newcommand{\FASTjythonREDCMemCI}{0.93}
\newcommand{\FASTjythonCAPOMem}{\memna}
\newcommand{\FASTjythonCAPOMemCI}{\memna}
\newcommand{\FASTjythonCAPOMemCIMIN}{\memna}
\newcommand{\FASTjythonCAPOMemCIMAX}{\memna}
\newcommand{\FASTjythonFTOCAPOMem}{9.0}
\newcommand{\FASTjythonFTOCAPOMemCI}{0.20}
\newcommand{\FASTjythonRECAPOMem}{12}
\newcommand{\FASTjythonRECAPOMemCI}{0.37}
\newcommand{\FASTjythonAGGCAPOMem}{\memna}
\newcommand{\FASTjythonAGGCAPOMemCI}{\memna}
\newcommand{\FASTjythonAGGCAPOMemCIMIN}{\memna}
\newcommand{\FASTjythonAGGCAPOMemCIMAX}{\memna}
\newcommand{\FASTjythonEventsCI}{0}
\newcommand{\FASTjythonEventsCIMIN}{0}
\newcommand{\FASTjythonEventsCIMAX}{0}
\newcommand{\FASTjythonNoFPEventsCI}{0}
\newcommand{\FASTjythonNoFPEventsCIMIN}{0}
\newcommand{\FASTjythonNoFPEventsCIMAX}{0}
\newcommand{\FASTjythonFT}{24}
\newcommand{\FASTjythonFTCI}{1.2}
\newcommand{\FASTjythonFTCIMIN}{23}
\newcommand{\FASTjythonFTCIMAX}{25}
\newcommand{\FASTjythonFTDynamic}{48}
\newcommand{\FASTjythonFTDynamicCI}{1.2}
\newcommand{\FASTjythonFTDynamicCIMIN}{47}
\newcommand{\FASTjythonFTDynamicCIMAX}{49}
\newcommand{\FASTjythonHB}{24}
\newcommand{\FASTjythonHBCI}{1}
\newcommand{\FASTjythonHBCIMIN}{23}
\newcommand{\FASTjythonHBCIMAX}{25}
\newcommand{\FASTjythonHBDynamic}{27}
\newcommand{\FASTjythonHBDynamicCI}{1.5}
\newcommand{\FASTjythonHBDynamicCIMIN}{25}
\newcommand{\FASTjythonHBDynamicCIMAX}{29}
\newcommand{\FASTjythonFTOHB}{24}
\newcommand{\FASTjythonFTOHBCI}{1}
\newcommand{\FASTjythonFTOHBCIMIN}{23}
\newcommand{\FASTjythonFTOHBCIMAX}{25}
\newcommand{\FASTjythonFTOHBDynamic}{26}
\newcommand{\FASTjythonFTOHBDynamicCI}{1}
\newcommand{\FASTjythonFTOHBDynamicCIMIN}{25}
\newcommand{\FASTjythonFTOHBDynamicCIMAX}{27}
\newcommand{\FASTjythonWCP}{\rna}
\newcommand{\FASTjythonWCPCI}{\rna}
\newcommand{\FASTjythonWCPCIMIN}{\rna}
\newcommand{\FASTjythonWCPCIMAX}{\rna}
\newcommand{\FASTjythonWCPDynamic}{\rna}
\newcommand{\FASTjythonWCPDynamicCI}{\rna}
\newcommand{\FASTjythonWCPDynamicCIMIN}{\rna}
\newcommand{\FASTjythonWCPDynamicCIMAX}{\rna}
\newcommand{\FASTjythonFTOWCP}{19}
\newcommand{\FASTjythonFTOWCPCI}{0.3}
\newcommand{\FASTjythonFTOWCPCIMIN}{19}
\newcommand{\FASTjythonFTOWCPCIMAX}{19}
\newcommand{\FASTjythonFTOWCPDynamic}{19}
\newcommand{\FASTjythonFTOWCPDynamicCI}{0.5}
\newcommand{\FASTjythonFTOWCPDynamicCIMIN}{19}
\newcommand{\FASTjythonFTOWCPDynamicCIMAX}{19}
\newcommand{\FASTjythonREWCP}{23}
\newcommand{\FASTjythonREWCPCI}{2}
\newcommand{\FASTjythonREWCPCIMIN}{21}
\newcommand{\FASTjythonREWCPCIMAX}{25}
\newcommand{\FASTjythonREWCPDynamic}{24}
\newcommand{\FASTjythonREWCPDynamicCI}{2}
\newcommand{\FASTjythonREWCPDynamicCIMIN}{22}
\newcommand{\FASTjythonREWCPDynamicCIMAX}{26}
\newcommand{\FASTjythonDC}{\rna}
\newcommand{\FASTjythonDCCI}{\rna}
\newcommand{\FASTjythonDCCIMIN}{\rna}
\newcommand{\FASTjythonDCCIMAX}{\rna}
\newcommand{\FASTjythonDCDynamic}{\rna}
\newcommand{\FASTjythonDCDynamicCI}{\rna}
\newcommand{\FASTjythonDCDynamicCIMIN}{\rna}
\newcommand{\FASTjythonDCDynamicCIMAX}{\rna}
\newcommand{\FASTjythonFTODC}{27}
\newcommand{\FASTjythonFTODCCI}{0}
\newcommand{\FASTjythonFTODCCIMIN}{27}
\newcommand{\FASTjythonFTODCCIMAX}{27}
\newcommand{\FASTjythonFTODCDynamic}{29}
\newcommand{\FASTjythonFTODCDynamicCI}{0}
\newcommand{\FASTjythonFTODCDynamicCIMIN}{29}
\newcommand{\FASTjythonFTODCDynamicCIMAX}{29}
\newcommand{\FASTjythonREDC}{29}
\newcommand{\FASTjythonREDCCI}{1}
\newcommand{\FASTjythonREDCCIMIN}{28}
\newcommand{\FASTjythonREDCCIMAX}{30}
\newcommand{\FASTjythonREDCDynamic}{32}
\newcommand{\FASTjythonREDCDynamicCI}{1}
\newcommand{\FASTjythonREDCDynamicCIMIN}{31}
\newcommand{\FASTjythonREDCDynamicCIMAX}{33}
\newcommand{\FASTjythonCAPO}{\rna}
\newcommand{\FASTjythonCAPOCI}{\rna}
\newcommand{\FASTjythonCAPOCIMIN}{\rna}
\newcommand{\FASTjythonCAPOCIMAX}{\rna}
\newcommand{\FASTjythonCAPODynamic}{\rna}
\newcommand{\FASTjythonCAPODynamicCI}{\rna}
\newcommand{\FASTjythonCAPODynamicCIMIN}{\rna}
\newcommand{\FASTjythonCAPODynamicCIMAX}{\rna}
\newcommand{\FASTjythonFTOCAPO}{28}
\newcommand{\FASTjythonFTOCAPOCI}{1}
\newcommand{\FASTjythonFTOCAPOCIMIN}{27}
\newcommand{\FASTjythonFTOCAPOCIMAX}{29}
\newcommand{\FASTjythonFTOCAPODynamic}{30}
\newcommand{\FASTjythonFTOCAPODynamicCI}{1}
\newcommand{\FASTjythonFTOCAPODynamicCIMIN}{29}
\newcommand{\FASTjythonFTOCAPODynamicCIMAX}{31}
\newcommand{\FASTjythonRECAPO}{29}
\newcommand{\FASTjythonRECAPOCI}{1}
\newcommand{\FASTjythonRECAPOCIMIN}{28}
\newcommand{\FASTjythonRECAPOCIMAX}{30}
\newcommand{\FASTjythonRECAPODynamic}{32}
\newcommand{\FASTjythonRECAPODynamicCI}{1}
\newcommand{\FASTjythonRECAPODynamicCIMIN}{31}
\newcommand{\FASTjythonRECAPODynamicCIMAX}{33}
\newcommand{\FASTjythonAGGCAPO}{\rna}
\newcommand{\FASTjythonAGGCAPOCI}{\rna}
\newcommand{\FASTjythonAGGCAPOCIMIN}{\rna}
\newcommand{\FASTjythonAGGCAPOCIMAX}{\rna}
\newcommand{\FASTjythonAGGCAPODynamic}{\rna}
\newcommand{\FASTjythonAGGCAPODynamicCI}{\rna}
\newcommand{\FASTjythonAGGCAPODynamicCIMIN}{\rna}
\newcommand{\FASTjythonAGGCAPODynamicCIMAX}{\rna}
\newcommand{\FASTluindexEvents}{0}
\newcommand{\FASTluindexNoFPEvents}{0}
\newcommand{\FASTluindexMaxLiveThreads}{3}
\newcommand{\FASTluindexTotalThreads}{3}
\newcommand{\FASTluindexBaseTime}{1.2}
\newcommand{\FASTluindexBaseTimeCI}{95}
\newcommand{\FASTluindexEmptyTime}{\rna}
\newcommand{\FASTluindexEmptyTimeCI}{\rna}
\newcommand{\FASTluindexEmptyTimeCIMIN}{\rna}
\newcommand{\FASTluindexEmptyTimeCIMAX}{\rna}
\newcommand{\FASTluindexFTTime}{7.9}
\newcommand{\FASTluindexFTTimeCI}{0.50}
\newcommand{\FASTluindexHBTime}{7.3}
\newcommand{\FASTluindexHBTimeCI}{0.45}
\newcommand{\FASTluindexFTOHBTime}{7.4}
\newcommand{\FASTluindexFTOHBTimeCI}{0.37}
\newcommand{\FASTluindexWCPTime}{\rna}
\newcommand{\FASTluindexWCPTimeCI}{\rna}
\newcommand{\FASTluindexWCPTimeCIMIN}{\rna}
\newcommand{\FASTluindexWCPTimeCIMAX}{\rna}
\newcommand{\FASTluindexFTOWCPTime}{24}
\newcommand{\FASTluindexFTOWCPTimeCI}{1.3}
\newcommand{\FASTluindexREWCPTime}{8.8}
\newcommand{\FASTluindexREWCPTimeCI}{0.52}
\newcommand{\FASTluindexDCTime}{\rna}
\newcommand{\FASTluindexDCTimeCI}{\rna}
\newcommand{\FASTluindexDCTimeCIMIN}{\rna}
\newcommand{\FASTluindexDCTimeCIMAX}{\rna}
\newcommand{\FASTluindexFTODCTime}{23}
\newcommand{\FASTluindexFTODCTimeCI}{1.2}
\newcommand{\FASTluindexREDCTime}{8.6}
\newcommand{\FASTluindexREDCTimeCI}{0.49}
\newcommand{\FASTluindexCAPOTime}{\rna}
\newcommand{\FASTluindexCAPOTimeCI}{\rna}
\newcommand{\FASTluindexCAPOTimeCIMIN}{\rna}
\newcommand{\FASTluindexCAPOTimeCIMAX}{\rna}
\newcommand{\FASTluindexFTOCAPOTime}{23}
\newcommand{\FASTluindexFTOCAPOTimeCI}{1.2}
\newcommand{\FASTluindexRECAPOTime}{8.4}
\newcommand{\FASTluindexRECAPOTimeCI}{0.52}
\newcommand{\FASTluindexAGGCAPOTime}{\rna}
\newcommand{\FASTluindexAGGCAPOTimeCI}{\rna}
\newcommand{\FASTluindexAGGCAPOTimeCIMIN}{\rna}
\newcommand{\FASTluindexAGGCAPOTimeCIMAX}{\rna}
\newcommand{\FASTluindexStaticTime}{\rzero}
\newcommand{\FASTluindexDynamicTime}{\rzero}
\newcommand{\FASTluindexBaseMem}{120}
\newcommand{\FASTluindexBaseMemCI}{8.9}
\newcommand{\FASTluindexFTMem}{4.9}
\newcommand{\FASTluindexFTMemCI}{0.35}
\newcommand{\FASTluindexHBMem}{4.9}
\newcommand{\FASTluindexHBMemCI}{0.30}
\newcommand{\FASTluindexFTOHBMem}{4.9}
\newcommand{\FASTluindexFTOHBMemCI}{0.31}
\newcommand{\FASTluindexWCPMem}{\memna}
\newcommand{\FASTluindexWCPMemCI}{\memna}
\newcommand{\FASTluindexWCPMemCIMIN}{\memna}
\newcommand{\FASTluindexWCPMemCIMAX}{\memna}
\newcommand{\FASTluindexFTOWCPMem}{24}
\newcommand{\FASTluindexFTOWCPMemCI}{1.5}
\newcommand{\FASTluindexREWCPMem}{9.0}
\newcommand{\FASTluindexREWCPMemCI}{0.98}
\newcommand{\FASTluindexDCMem}{\memna}
\newcommand{\FASTluindexDCMemCI}{\memna}
\newcommand{\FASTluindexDCMemCIMIN}{\memna}
\newcommand{\FASTluindexDCMemCIMAX}{\memna}
\newcommand{\FASTluindexFTODCMem}{24}
\newcommand{\FASTluindexFTODCMemCI}{1.5}
\newcommand{\FASTluindexREDCMem}{9.8}
\newcommand{\FASTluindexREDCMemCI}{1.2}
\newcommand{\FASTluindexCAPOMem}{\memna}
\newcommand{\FASTluindexCAPOMemCI}{\memna}
\newcommand{\FASTluindexCAPOMemCIMIN}{\memna}
\newcommand{\FASTluindexCAPOMemCIMAX}{\memna}
\newcommand{\FASTluindexFTOCAPOMem}{24}
\newcommand{\FASTluindexFTOCAPOMemCI}{1.6}
\newcommand{\FASTluindexRECAPOMem}{8.1}
\newcommand{\FASTluindexRECAPOMemCI}{0.52}
\newcommand{\FASTluindexAGGCAPOMem}{\memna}
\newcommand{\FASTluindexAGGCAPOMemCI}{\memna}
\newcommand{\FASTluindexAGGCAPOMemCIMIN}{\memna}
\newcommand{\FASTluindexAGGCAPOMemCIMAX}{\memna}
\newcommand{\FASTluindexEventsCI}{0}
\newcommand{\FASTluindexEventsCIMIN}{0}
\newcommand{\FASTluindexEventsCIMAX}{0}
\newcommand{\FASTluindexNoFPEventsCI}{0}
\newcommand{\FASTluindexNoFPEventsCIMIN}{0}
\newcommand{\FASTluindexNoFPEventsCIMAX}{0}
\newcommand{\FASTluindexFT}{1}
\newcommand{\FASTluindexFTCI}{0}
\newcommand{\FASTluindexFTCIMIN}{1}
\newcommand{\FASTluindexFTCIMAX}{1}
\newcommand{\FASTluindexFTDynamic}{1}
\newcommand{\FASTluindexFTDynamicCI}{0}
\newcommand{\FASTluindexFTDynamicCIMIN}{1}
\newcommand{\FASTluindexFTDynamicCIMAX}{1}
\newcommand{\FASTluindexHB}{1}
\newcommand{\FASTluindexHBCI}{0}
\newcommand{\FASTluindexHBCIMIN}{1}
\newcommand{\FASTluindexHBCIMAX}{1}
\newcommand{\FASTluindexHBDynamic}{1}
\newcommand{\FASTluindexHBDynamicCI}{0}
\newcommand{\FASTluindexHBDynamicCIMIN}{1}
\newcommand{\FASTluindexHBDynamicCIMAX}{1}
\newcommand{\FASTluindexFTOHB}{1}
\newcommand{\FASTluindexFTOHBCI}{0}
\newcommand{\FASTluindexFTOHBCIMIN}{1}
\newcommand{\FASTluindexFTOHBCIMAX}{1}
\newcommand{\FASTluindexFTOHBDynamic}{1}
\newcommand{\FASTluindexFTOHBDynamicCI}{0}
\newcommand{\FASTluindexFTOHBDynamicCIMIN}{1}
\newcommand{\FASTluindexFTOHBDynamicCIMAX}{1}
\newcommand{\FASTluindexWCP}{\rna}
\newcommand{\FASTluindexWCPCI}{\rna}
\newcommand{\FASTluindexWCPCIMIN}{\rna}
\newcommand{\FASTluindexWCPCIMAX}{\rna}
\newcommand{\FASTluindexWCPDynamic}{\rna}
\newcommand{\FASTluindexWCPDynamicCI}{\rna}
\newcommand{\FASTluindexWCPDynamicCIMIN}{\rna}
\newcommand{\FASTluindexWCPDynamicCIMAX}{\rna}
\newcommand{\FASTluindexFTOWCP}{1}
\newcommand{\FASTluindexFTOWCPCI}{0}
\newcommand{\FASTluindexFTOWCPCIMIN}{1}
\newcommand{\FASTluindexFTOWCPCIMAX}{1}
\newcommand{\FASTluindexFTOWCPDynamic}{1}
\newcommand{\FASTluindexFTOWCPDynamicCI}{0}
\newcommand{\FASTluindexFTOWCPDynamicCIMIN}{1}
\newcommand{\FASTluindexFTOWCPDynamicCIMAX}{1}
\newcommand{\FASTluindexREWCP}{1}
\newcommand{\FASTluindexREWCPCI}{0}
\newcommand{\FASTluindexREWCPCIMIN}{1}
\newcommand{\FASTluindexREWCPCIMAX}{1}
\newcommand{\FASTluindexREWCPDynamic}{1}
\newcommand{\FASTluindexREWCPDynamicCI}{0}
\newcommand{\FASTluindexREWCPDynamicCIMIN}{1}
\newcommand{\FASTluindexREWCPDynamicCIMAX}{1}
\newcommand{\FASTluindexDC}{\rna}
\newcommand{\FASTluindexDCCI}{\rna}
\newcommand{\FASTluindexDCCIMIN}{\rna}
\newcommand{\FASTluindexDCCIMAX}{\rna}
\newcommand{\FASTluindexDCDynamic}{\rna}
\newcommand{\FASTluindexDCDynamicCI}{\rna}
\newcommand{\FASTluindexDCDynamicCIMIN}{\rna}
\newcommand{\FASTluindexDCDynamicCIMAX}{\rna}
\newcommand{\FASTluindexFTODC}{1}
\newcommand{\FASTluindexFTODCCI}{0}
\newcommand{\FASTluindexFTODCCIMIN}{1}
\newcommand{\FASTluindexFTODCCIMAX}{1}
\newcommand{\FASTluindexFTODCDynamic}{1}
\newcommand{\FASTluindexFTODCDynamicCI}{0}
\newcommand{\FASTluindexFTODCDynamicCIMIN}{1}
\newcommand{\FASTluindexFTODCDynamicCIMAX}{1}
\newcommand{\FASTluindexREDC}{1}
\newcommand{\FASTluindexREDCCI}{0}
\newcommand{\FASTluindexREDCCIMIN}{1}
\newcommand{\FASTluindexREDCCIMAX}{1}
\newcommand{\FASTluindexREDCDynamic}{1}
\newcommand{\FASTluindexREDCDynamicCI}{0}
\newcommand{\FASTluindexREDCDynamicCIMIN}{1}
\newcommand{\FASTluindexREDCDynamicCIMAX}{1}
\newcommand{\FASTluindexCAPO}{\rna}
\newcommand{\FASTluindexCAPOCI}{\rna}
\newcommand{\FASTluindexCAPOCIMIN}{\rna}
\newcommand{\FASTluindexCAPOCIMAX}{\rna}
\newcommand{\FASTluindexCAPODynamic}{\rna}
\newcommand{\FASTluindexCAPODynamicCI}{\rna}
\newcommand{\FASTluindexCAPODynamicCIMIN}{\rna}
\newcommand{\FASTluindexCAPODynamicCIMAX}{\rna}
\newcommand{\FASTluindexFTOCAPO}{1}
\newcommand{\FASTluindexFTOCAPOCI}{0}
\newcommand{\FASTluindexFTOCAPOCIMIN}{1}
\newcommand{\FASTluindexFTOCAPOCIMAX}{1}
\newcommand{\FASTluindexFTOCAPODynamic}{1}
\newcommand{\FASTluindexFTOCAPODynamicCI}{0}
\newcommand{\FASTluindexFTOCAPODynamicCIMIN}{1}
\newcommand{\FASTluindexFTOCAPODynamicCIMAX}{1}
\newcommand{\FASTluindexRECAPO}{1}
\newcommand{\FASTluindexRECAPOCI}{0}
\newcommand{\FASTluindexRECAPOCIMIN}{1}
\newcommand{\FASTluindexRECAPOCIMAX}{1}
\newcommand{\FASTluindexRECAPODynamic}{1}
\newcommand{\FASTluindexRECAPODynamicCI}{0}
\newcommand{\FASTluindexRECAPODynamicCIMIN}{1}
\newcommand{\FASTluindexRECAPODynamicCIMAX}{1}
\newcommand{\FASTluindexAGGCAPO}{\rna}
\newcommand{\FASTluindexAGGCAPOCI}{\rna}
\newcommand{\FASTluindexAGGCAPOCIMIN}{\rna}
\newcommand{\FASTluindexAGGCAPOCIMAX}{\rna}
\newcommand{\FASTluindexAGGCAPODynamic}{\rna}
\newcommand{\FASTluindexAGGCAPODynamicCI}{\rna}
\newcommand{\FASTluindexAGGCAPODynamicCIMIN}{\rna}
\newcommand{\FASTluindexAGGCAPODynamicCIMAX}{\rna}
\newcommand{\FASTlusearchEvents}{0}
\newcommand{\FASTlusearchNoFPEvents}{0}
\newcommand{\FASTlusearchMaxLiveThreads}{10}
\newcommand{\FASTlusearchTotalThreads}{10}
\newcommand{\FASTlusearchBaseTime}{0.93}
\newcommand{\FASTlusearchBaseTimeCI}{33}
\newcommand{\FASTlusearchEmptyTime}{\rna}
\newcommand{\FASTlusearchEmptyTimeCI}{\rna}
\newcommand{\FASTlusearchEmptyTimeCIMIN}{\rna}
\newcommand{\FASTlusearchEmptyTimeCIMAX}{\rna}
\newcommand{\FASTlusearchFTTime}{13}
\newcommand{\FASTlusearchFTTimeCI}{0.50}
\newcommand{\FASTlusearchHBTime}{12}
\newcommand{\FASTlusearchHBTimeCI}{0.44}
\newcommand{\FASTlusearchFTOHBTime}{12}
\newcommand{\FASTlusearchFTOHBTimeCI}{0.46}
\newcommand{\FASTlusearchWCPTime}{\rna}
\newcommand{\FASTlusearchWCPTimeCI}{\rna}
\newcommand{\FASTlusearchWCPTimeCIMIN}{\rna}
\newcommand{\FASTlusearchWCPTimeCIMAX}{\rna}
\newcommand{\FASTlusearchFTOWCPTime}{15}
\newcommand{\FASTlusearchFTOWCPTimeCI}{0.53}
\newcommand{\FASTlusearchREWCPTime}{14}
\newcommand{\FASTlusearchREWCPTimeCI}{0.44}
\newcommand{\FASTlusearchDCTime}{\rna}
\newcommand{\FASTlusearchDCTimeCI}{\rna}
\newcommand{\FASTlusearchDCTimeCIMIN}{\rna}
\newcommand{\FASTlusearchDCTimeCIMAX}{\rna}
\newcommand{\FASTlusearchFTODCTime}{16}
\newcommand{\FASTlusearchFTODCTimeCI}{0.59}
\newcommand{\FASTlusearchREDCTime}{15}
\newcommand{\FASTlusearchREDCTimeCI}{0.66}
\newcommand{\FASTlusearchCAPOTime}{\rna}
\newcommand{\FASTlusearchCAPOTimeCI}{\rna}
\newcommand{\FASTlusearchCAPOTimeCIMIN}{\rna}
\newcommand{\FASTlusearchCAPOTimeCIMAX}{\rna}
\newcommand{\FASTlusearchFTOCAPOTime}{14}
\newcommand{\FASTlusearchFTOCAPOTimeCI}{0.44}
\newcommand{\FASTlusearchRECAPOTime}{13}
\newcommand{\FASTlusearchRECAPOTimeCI}{0.42}
\newcommand{\FASTlusearchAGGCAPOTime}{\rna}
\newcommand{\FASTlusearchAGGCAPOTimeCI}{\rna}
\newcommand{\FASTlusearchAGGCAPOTimeCIMIN}{\rna}
\newcommand{\FASTlusearchAGGCAPOTimeCIMAX}{\rna}
\newcommand{\FASTlusearchStaticTime}{\rzero}
\newcommand{\FASTlusearchDynamicTime}{\rzero}
\newcommand{\FASTlusearchBaseMem}{1,600}
\newcommand{\FASTlusearchBaseMemCI}{11.0}
\newcommand{\FASTlusearchFTMem}{8.8}
\newcommand{\FASTlusearchFTMemCI}{0.37}
\newcommand{\FASTlusearchHBMem}{8.8}
\newcommand{\FASTlusearchHBMemCI}{0.39}
\newcommand{\FASTlusearchFTOHBMem}{8.1}
\newcommand{\FASTlusearchFTOHBMemCI}{0.36}
\newcommand{\FASTlusearchWCPMem}{\memna}
\newcommand{\FASTlusearchWCPMemCI}{\memna}
\newcommand{\FASTlusearchWCPMemCIMIN}{\memna}
\newcommand{\FASTlusearchWCPMemCIMAX}{\memna}
\newcommand{\FASTlusearchFTOWCPMem}{9.4}
\newcommand{\FASTlusearchFTOWCPMemCI}{0.36}
\newcommand{\FASTlusearchREWCPMem}{9.4}
\newcommand{\FASTlusearchREWCPMemCI}{0.32}
\newcommand{\FASTlusearchDCMem}{\memna}
\newcommand{\FASTlusearchDCMemCI}{\memna}
\newcommand{\FASTlusearchDCMemCIMIN}{\memna}
\newcommand{\FASTlusearchDCMemCIMAX}{\memna}
\newcommand{\FASTlusearchFTODCMem}{9.4}
\newcommand{\FASTlusearchFTODCMemCI}{0.29}
\newcommand{\FASTlusearchREDCMem}{10}
\newcommand{\FASTlusearchREDCMemCI}{0.35}
\newcommand{\FASTlusearchCAPOMem}{\memna}
\newcommand{\FASTlusearchCAPOMemCI}{\memna}
\newcommand{\FASTlusearchCAPOMemCIMIN}{\memna}
\newcommand{\FASTlusearchCAPOMemCIMAX}{\memna}
\newcommand{\FASTlusearchFTOCAPOMem}{9.4}
\newcommand{\FASTlusearchFTOCAPOMemCI}{0.30}
\newcommand{\FASTlusearchRECAPOMem}{9.4}
\newcommand{\FASTlusearchRECAPOMemCI}{0.20}
\newcommand{\FASTlusearchAGGCAPOMem}{\memna}
\newcommand{\FASTlusearchAGGCAPOMemCI}{\memna}
\newcommand{\FASTlusearchAGGCAPOMemCIMIN}{\memna}
\newcommand{\FASTlusearchAGGCAPOMemCIMAX}{\memna}
\newcommand{\FASTlusearchEventsCI}{0}
\newcommand{\FASTlusearchEventsCIMIN}{0}
\newcommand{\FASTlusearchEventsCIMAX}{0}
\newcommand{\FASTlusearchNoFPEventsCI}{0}
\newcommand{\FASTlusearchNoFPEventsCIMIN}{0}
\newcommand{\FASTlusearchNoFPEventsCIMAX}{0}
\newcommand{\FASTlusearchFT}{0}
\newcommand{\FASTlusearchFTCI}{0}
\newcommand{\FASTlusearchFTCIMIN}{0}
\newcommand{\FASTlusearchFTCIMAX}{0}
\newcommand{\FASTlusearchFTDynamic}{0}
\newcommand{\FASTlusearchFTDynamicCI}{0}
\newcommand{\FASTlusearchFTDynamicCIMIN}{0}
\newcommand{\FASTlusearchFTDynamicCIMAX}{0}
\newcommand{\FASTlusearchHB}{0}
\newcommand{\FASTlusearchHBCI}{0}
\newcommand{\FASTlusearchHBCIMIN}{0}
\newcommand{\FASTlusearchHBCIMAX}{0}
\newcommand{\FASTlusearchHBDynamic}{0}
\newcommand{\FASTlusearchHBDynamicCI}{0}
\newcommand{\FASTlusearchHBDynamicCIMIN}{0}
\newcommand{\FASTlusearchHBDynamicCIMAX}{0}
\newcommand{\FASTlusearchFTOHB}{0}
\newcommand{\FASTlusearchFTOHBCI}{0}
\newcommand{\FASTlusearchFTOHBCIMIN}{0}
\newcommand{\FASTlusearchFTOHBCIMAX}{0}
\newcommand{\FASTlusearchFTOHBDynamic}{0}
\newcommand{\FASTlusearchFTOHBDynamicCI}{0}
\newcommand{\FASTlusearchFTOHBDynamicCIMIN}{0}
\newcommand{\FASTlusearchFTOHBDynamicCIMAX}{0}
\newcommand{\FASTlusearchWCP}{\rna}
\newcommand{\FASTlusearchWCPCI}{\rna}
\newcommand{\FASTlusearchWCPCIMIN}{\rna}
\newcommand{\FASTlusearchWCPCIMAX}{\rna}
\newcommand{\FASTlusearchWCPDynamic}{\rna}
\newcommand{\FASTlusearchWCPDynamicCI}{\rna}
\newcommand{\FASTlusearchWCPDynamicCIMIN}{\rna}
\newcommand{\FASTlusearchWCPDynamicCIMAX}{\rna}
\newcommand{\FASTlusearchFTOWCP}{0}
\newcommand{\FASTlusearchFTOWCPCI}{0}
\newcommand{\FASTlusearchFTOWCPCIMIN}{0}
\newcommand{\FASTlusearchFTOWCPCIMAX}{0}
\newcommand{\FASTlusearchFTOWCPDynamic}{0}
\newcommand{\FASTlusearchFTOWCPDynamicCI}{0}
\newcommand{\FASTlusearchFTOWCPDynamicCIMIN}{0}
\newcommand{\FASTlusearchFTOWCPDynamicCIMAX}{0}
\newcommand{\FASTlusearchREWCP}{0}
\newcommand{\FASTlusearchREWCPCI}{0}
\newcommand{\FASTlusearchREWCPCIMIN}{0}
\newcommand{\FASTlusearchREWCPCIMAX}{0}
\newcommand{\FASTlusearchREWCPDynamic}{0}
\newcommand{\FASTlusearchREWCPDynamicCI}{0}
\newcommand{\FASTlusearchREWCPDynamicCIMIN}{0}
\newcommand{\FASTlusearchREWCPDynamicCIMAX}{0}
\newcommand{\FASTlusearchDC}{\rna}
\newcommand{\FASTlusearchDCCI}{\rna}
\newcommand{\FASTlusearchDCCIMIN}{\rna}
\newcommand{\FASTlusearchDCCIMAX}{\rna}
\newcommand{\FASTlusearchDCDynamic}{\rna}
\newcommand{\FASTlusearchDCDynamicCI}{\rna}
\newcommand{\FASTlusearchDCDynamicCIMIN}{\rna}
\newcommand{\FASTlusearchDCDynamicCIMAX}{\rna}
\newcommand{\FASTlusearchFTODC}{0}
\newcommand{\FASTlusearchFTODCCI}{0}
\newcommand{\FASTlusearchFTODCCIMIN}{0}
\newcommand{\FASTlusearchFTODCCIMAX}{0}
\newcommand{\FASTlusearchFTODCDynamic}{0}
\newcommand{\FASTlusearchFTODCDynamicCI}{0}
\newcommand{\FASTlusearchFTODCDynamicCIMIN}{0}
\newcommand{\FASTlusearchFTODCDynamicCIMAX}{0}
\newcommand{\FASTlusearchREDC}{0}
\newcommand{\FASTlusearchREDCCI}{0}
\newcommand{\FASTlusearchREDCCIMIN}{0}
\newcommand{\FASTlusearchREDCCIMAX}{0}
\newcommand{\FASTlusearchREDCDynamic}{0}
\newcommand{\FASTlusearchREDCDynamicCI}{0}
\newcommand{\FASTlusearchREDCDynamicCIMIN}{0}
\newcommand{\FASTlusearchREDCDynamicCIMAX}{0}
\newcommand{\FASTlusearchCAPO}{\rna}
\newcommand{\FASTlusearchCAPOCI}{\rna}
\newcommand{\FASTlusearchCAPOCIMIN}{\rna}
\newcommand{\FASTlusearchCAPOCIMAX}{\rna}
\newcommand{\FASTlusearchCAPODynamic}{\rna}
\newcommand{\FASTlusearchCAPODynamicCI}{\rna}
\newcommand{\FASTlusearchCAPODynamicCIMIN}{\rna}
\newcommand{\FASTlusearchCAPODynamicCIMAX}{\rna}
\newcommand{\FASTlusearchFTOCAPO}{0}
\newcommand{\FASTlusearchFTOCAPOCI}{0}
\newcommand{\FASTlusearchFTOCAPOCIMIN}{0}
\newcommand{\FASTlusearchFTOCAPOCIMAX}{0}
\newcommand{\FASTlusearchFTOCAPODynamic}{0}
\newcommand{\FASTlusearchFTOCAPODynamicCI}{0}
\newcommand{\FASTlusearchFTOCAPODynamicCIMIN}{0}
\newcommand{\FASTlusearchFTOCAPODynamicCIMAX}{0}
\newcommand{\FASTlusearchRECAPO}{0}
\newcommand{\FASTlusearchRECAPOCI}{0}
\newcommand{\FASTlusearchRECAPOCIMIN}{0}
\newcommand{\FASTlusearchRECAPOCIMAX}{0}
\newcommand{\FASTlusearchRECAPODynamic}{0}
\newcommand{\FASTlusearchRECAPODynamicCI}{0}
\newcommand{\FASTlusearchRECAPODynamicCIMIN}{0}
\newcommand{\FASTlusearchRECAPODynamicCIMAX}{0}
\newcommand{\FASTlusearchAGGCAPO}{\rna}
\newcommand{\FASTlusearchAGGCAPOCI}{\rna}
\newcommand{\FASTlusearchAGGCAPOCIMIN}{\rna}
\newcommand{\FASTlusearchAGGCAPOCIMAX}{\rna}
\newcommand{\FASTlusearchAGGCAPODynamic}{\rna}
\newcommand{\FASTlusearchAGGCAPODynamicCI}{\rna}
\newcommand{\FASTlusearchAGGCAPODynamicCIMIN}{\rna}
\newcommand{\FASTlusearchAGGCAPODynamicCIMAX}{\rna}
\newcommand{\FASTpmdEvents}{0}
\newcommand{\FASTpmdNoFPEvents}{0}
\newcommand{\FASTpmdMaxLiveThreads}{9}
\newcommand{\FASTpmdTotalThreads}{9}
\newcommand{\FASTpmdBaseTime}{1.4}
\newcommand{\FASTpmdBaseTimeCI}{25}
\newcommand{\FASTpmdEmptyTime}{\rna}
\newcommand{\FASTpmdEmptyTimeCI}{\rna}
\newcommand{\FASTpmdEmptyTimeCIMIN}{\rna}
\newcommand{\FASTpmdEmptyTimeCIMAX}{\rna}
\newcommand{\FASTpmdFTTime}{6.1}
\newcommand{\FASTpmdFTTimeCI}{0.26}
\newcommand{\FASTpmdHBTime}{5.7}
\newcommand{\FASTpmdHBTimeCI}{0.16}
\newcommand{\FASTpmdFTOHBTime}{5.8}
\newcommand{\FASTpmdFTOHBTimeCI}{0.16}
\newcommand{\FASTpmdWCPTime}{\rna}
\newcommand{\FASTpmdWCPTimeCI}{\rna}
\newcommand{\FASTpmdWCPTimeCIMIN}{\rna}
\newcommand{\FASTpmdWCPTimeCIMAX}{\rna}
\newcommand{\FASTpmdFTOWCPTime}{6.0}
\newcommand{\FASTpmdFTOWCPTimeCI}{0.15}
\newcommand{\FASTpmdREWCPTime}{6.9}
\newcommand{\FASTpmdREWCPTimeCI}{0.12}
\newcommand{\FASTpmdDCTime}{\rna}
\newcommand{\FASTpmdDCTimeCI}{\rna}
\newcommand{\FASTpmdDCTimeCIMIN}{\rna}
\newcommand{\FASTpmdDCTimeCIMAX}{\rna}
\newcommand{\FASTpmdFTODCTime}{6.0}
\newcommand{\FASTpmdFTODCTimeCI}{0.12}
\newcommand{\FASTpmdREDCTime}{7.0}
\newcommand{\FASTpmdREDCTimeCI}{0.16}
\newcommand{\FASTpmdCAPOTime}{\rna}
\newcommand{\FASTpmdCAPOTimeCI}{\rna}
\newcommand{\FASTpmdCAPOTimeCIMIN}{\rna}
\newcommand{\FASTpmdCAPOTimeCIMAX}{\rna}
\newcommand{\FASTpmdFTOCAPOTime}{5.9}
\newcommand{\FASTpmdFTOCAPOTimeCI}{0.16}
\newcommand{\FASTpmdRECAPOTime}{6.8}
\newcommand{\FASTpmdRECAPOTimeCI}{0.15}
\newcommand{\FASTpmdAGGCAPOTime}{\rna}
\newcommand{\FASTpmdAGGCAPOTimeCI}{\rna}
\newcommand{\FASTpmdAGGCAPOTimeCIMIN}{\rna}
\newcommand{\FASTpmdAGGCAPOTimeCIMAX}{\rna}
\newcommand{\FASTpmdStaticTime}{\rzero}
\newcommand{\FASTpmdDynamicTime}{\rzero}
\newcommand{\FASTpmdBaseMem}{590}
\newcommand{\FASTpmdBaseMemCI}{12.0}
\newcommand{\FASTpmdFTMem}{2.5}
\newcommand{\FASTpmdFTMemCI}{0.18}
\newcommand{\FASTpmdHBMem}{2.5}
\newcommand{\FASTpmdHBMemCI}{0.14}
\newcommand{\FASTpmdFTOHBMem}{2.7}
\newcommand{\FASTpmdFTOHBMemCI}{0.17}
\newcommand{\FASTpmdWCPMem}{\memna}
\newcommand{\FASTpmdWCPMemCI}{\memna}
\newcommand{\FASTpmdWCPMemCIMIN}{\memna}
\newcommand{\FASTpmdWCPMemCIMAX}{\memna}
\newcommand{\FASTpmdFTOWCPMem}{2.9}
\newcommand{\FASTpmdFTOWCPMemCI}{0.19}
\newcommand{\FASTpmdREWCPMem}{4.9}
\newcommand{\FASTpmdREWCPMemCI}{0.13}
\newcommand{\FASTpmdDCMem}{\memna}
\newcommand{\FASTpmdDCMemCI}{\memna}
\newcommand{\FASTpmdDCMemCIMIN}{\memna}
\newcommand{\FASTpmdDCMemCIMAX}{\memna}
\newcommand{\FASTpmdFTODCMem}{2.9}
\newcommand{\FASTpmdFTODCMemCI}{0.17}
\newcommand{\FASTpmdREDCMem}{5.1}
\newcommand{\FASTpmdREDCMemCI}{0.17}
\newcommand{\FASTpmdCAPOMem}{\memna}
\newcommand{\FASTpmdCAPOMemCI}{\memna}
\newcommand{\FASTpmdCAPOMemCIMIN}{\memna}
\newcommand{\FASTpmdCAPOMemCIMAX}{\memna}
\newcommand{\FASTpmdFTOCAPOMem}{2.7}
\newcommand{\FASTpmdFTOCAPOMemCI}{0.17}
\newcommand{\FASTpmdRECAPOMem}{4.9}
\newcommand{\FASTpmdRECAPOMemCI}{0.19}
\newcommand{\FASTpmdAGGCAPOMem}{\memna}
\newcommand{\FASTpmdAGGCAPOMemCI}{\memna}
\newcommand{\FASTpmdAGGCAPOMemCIMIN}{\memna}
\newcommand{\FASTpmdAGGCAPOMemCIMAX}{\memna}
\newcommand{\FASTpmdEventsCI}{0}
\newcommand{\FASTpmdEventsCIMIN}{0}
\newcommand{\FASTpmdEventsCIMAX}{0}
\newcommand{\FASTpmdNoFPEventsCI}{0}
\newcommand{\FASTpmdNoFPEventsCIMIN}{0}
\newcommand{\FASTpmdNoFPEventsCIMAX}{0}
\newcommand{\FASTpmdFT}{18}
\newcommand{\FASTpmdFTCI}{0}
\newcommand{\FASTpmdFTCIMIN}{18}
\newcommand{\FASTpmdFTCIMAX}{18}
\newcommand{\FASTpmdFTDynamic}{4,283}
\newcommand{\FASTpmdFTDynamicCI}{714}
\newcommand{\FASTpmdFTDynamicCIMIN}{3,569}
\newcommand{\FASTpmdFTDynamicCIMAX}{4,997}
\newcommand{\FASTpmdHB}{18}
\newcommand{\FASTpmdHBCI}{0}
\newcommand{\FASTpmdHBCIMIN}{18}
\newcommand{\FASTpmdHBCIMAX}{18}
\newcommand{\FASTpmdHBDynamic}{1,803}
\newcommand{\FASTpmdHBDynamicCI}{183}
\newcommand{\FASTpmdHBDynamicCIMIN}{1,620}
\newcommand{\FASTpmdHBDynamicCIMAX}{1,986}
\newcommand{\FASTpmdFTOHB}{18}
\newcommand{\FASTpmdFTOHBCI}{0}
\newcommand{\FASTpmdFTOHBCIMIN}{18}
\newcommand{\FASTpmdFTOHBCIMAX}{18}
\newcommand{\FASTpmdFTOHBDynamic}{1,612}
\newcommand{\FASTpmdFTOHBDynamicCI}{90}
\newcommand{\FASTpmdFTOHBDynamicCIMIN}{1,522}
\newcommand{\FASTpmdFTOHBDynamicCIMAX}{1,702}
\newcommand{\FASTpmdWCP}{\rna}
\newcommand{\FASTpmdWCPCI}{\rna}
\newcommand{\FASTpmdWCPCIMIN}{\rna}
\newcommand{\FASTpmdWCPCIMAX}{\rna}
\newcommand{\FASTpmdWCPDynamic}{\rna}
\newcommand{\FASTpmdWCPDynamicCI}{\rna}
\newcommand{\FASTpmdWCPDynamicCIMIN}{\rna}
\newcommand{\FASTpmdWCPDynamicCIMAX}{\rna}
\newcommand{\FASTpmdFTOWCP}{18}
\newcommand{\FASTpmdFTOWCPCI}{0}
\newcommand{\FASTpmdFTOWCPCIMIN}{18}
\newcommand{\FASTpmdFTOWCPCIMAX}{18}
\newcommand{\FASTpmdFTOWCPDynamic}{1,552}
\newcommand{\FASTpmdFTOWCPDynamicCI}{51}
\newcommand{\FASTpmdFTOWCPDynamicCIMIN}{1,501}
\newcommand{\FASTpmdFTOWCPDynamicCIMAX}{1,603}
\newcommand{\FASTpmdREWCP}{18}
\newcommand{\FASTpmdREWCPCI}{0}
\newcommand{\FASTpmdREWCPCIMIN}{18}
\newcommand{\FASTpmdREWCPCIMAX}{18}
\newcommand{\FASTpmdREWCPDynamic}{1,679}
\newcommand{\FASTpmdREWCPDynamicCI}{97}
\newcommand{\FASTpmdREWCPDynamicCIMIN}{1,582}
\newcommand{\FASTpmdREWCPDynamicCIMAX}{1,776}
\newcommand{\FASTpmdDC}{\rna}
\newcommand{\FASTpmdDCCI}{\rna}
\newcommand{\FASTpmdDCCIMIN}{\rna}
\newcommand{\FASTpmdDCCIMAX}{\rna}
\newcommand{\FASTpmdDCDynamic}{\rna}
\newcommand{\FASTpmdDCDynamicCI}{\rna}
\newcommand{\FASTpmdDCDynamicCIMIN}{\rna}
\newcommand{\FASTpmdDCDynamicCIMAX}{\rna}
\newcommand{\FASTpmdFTODC}{18}
\newcommand{\FASTpmdFTODCCI}{0}
\newcommand{\FASTpmdFTODCCIMIN}{18}
\newcommand{\FASTpmdFTODCCIMAX}{18}
\newcommand{\FASTpmdFTODCDynamic}{3,095}
\newcommand{\FASTpmdFTODCDynamicCI}{71}
\newcommand{\FASTpmdFTODCDynamicCIMIN}{3,024}
\newcommand{\FASTpmdFTODCDynamicCIMAX}{3,166}
\newcommand{\FASTpmdREDC}{18}
\newcommand{\FASTpmdREDCCI}{0}
\newcommand{\FASTpmdREDCCIMIN}{18}
\newcommand{\FASTpmdREDCCIMAX}{18}
\newcommand{\FASTpmdREDCDynamic}{3,173}
\newcommand{\FASTpmdREDCDynamicCI}{99}
\newcommand{\FASTpmdREDCDynamicCIMIN}{3,074}
\newcommand{\FASTpmdREDCDynamicCIMAX}{3,272}
\newcommand{\FASTpmdCAPO}{\rna}
\newcommand{\FASTpmdCAPOCI}{\rna}
\newcommand{\FASTpmdCAPOCIMIN}{\rna}
\newcommand{\FASTpmdCAPOCIMAX}{\rna}
\newcommand{\FASTpmdCAPODynamic}{\rna}
\newcommand{\FASTpmdCAPODynamicCI}{\rna}
\newcommand{\FASTpmdCAPODynamicCIMIN}{\rna}
\newcommand{\FASTpmdCAPODynamicCIMAX}{\rna}
\newcommand{\FASTpmdFTOCAPO}{18}
\newcommand{\FASTpmdFTOCAPOCI}{0}
\newcommand{\FASTpmdFTOCAPOCIMIN}{18}
\newcommand{\FASTpmdFTOCAPOCIMAX}{18}
\newcommand{\FASTpmdFTOCAPODynamic}{2,982}
\newcommand{\FASTpmdFTOCAPODynamicCI}{83}
\newcommand{\FASTpmdFTOCAPODynamicCIMIN}{2,899}
\newcommand{\FASTpmdFTOCAPODynamicCIMAX}{3,065}
\newcommand{\FASTpmdRECAPO}{18}
\newcommand{\FASTpmdRECAPOCI}{0}
\newcommand{\FASTpmdRECAPOCIMIN}{18}
\newcommand{\FASTpmdRECAPOCIMAX}{18}
\newcommand{\FASTpmdRECAPODynamic}{3,085}
\newcommand{\FASTpmdRECAPODynamicCI}{92}
\newcommand{\FASTpmdRECAPODynamicCIMIN}{2,993}
\newcommand{\FASTpmdRECAPODynamicCIMAX}{3,177}
\newcommand{\FASTpmdAGGCAPO}{\rna}
\newcommand{\FASTpmdAGGCAPOCI}{\rna}
\newcommand{\FASTpmdAGGCAPOCIMIN}{\rna}
\newcommand{\FASTpmdAGGCAPOCIMAX}{\rna}
\newcommand{\FASTpmdAGGCAPODynamic}{\rna}
\newcommand{\FASTpmdAGGCAPODynamicCI}{\rna}
\newcommand{\FASTpmdAGGCAPODynamicCIMIN}{\rna}
\newcommand{\FASTpmdAGGCAPODynamicCIMAX}{\rna}
\newcommand{\FASTsunflowEvents}{0}
\newcommand{\FASTsunflowNoFPEvents}{0}
\newcommand{\FASTsunflowMaxLiveThreads}{17}
\newcommand{\FASTsunflowTotalThreads}{17}
\newcommand{\FASTsunflowBaseTime}{1.4}
\newcommand{\FASTsunflowBaseTimeCI}{69}
\newcommand{\FASTsunflowEmptyTime}{\rna}
\newcommand{\FASTsunflowEmptyTimeCI}{\rna}
\newcommand{\FASTsunflowEmptyTimeCIMIN}{\rna}
\newcommand{\FASTsunflowEmptyTimeCIMAX}{\rna}
\newcommand{\FASTsunflowFTTime}{22}
\newcommand{\FASTsunflowFTTimeCI}{1.1}
\newcommand{\FASTsunflowHBTime}{22}
\newcommand{\FASTsunflowHBTimeCI}{1.1}
\newcommand{\FASTsunflowFTOHBTime}{21}
\newcommand{\FASTsunflowFTOHBTimeCI}{1.2}
\newcommand{\FASTsunflowWCPTime}{\rna}
\newcommand{\FASTsunflowWCPTimeCI}{\rna}
\newcommand{\FASTsunflowWCPTimeCIMIN}{\rna}
\newcommand{\FASTsunflowWCPTimeCIMAX}{\rna}
\newcommand{\FASTsunflowFTOWCPTime}{23}
\newcommand{\FASTsunflowFTOWCPTimeCI}{1.1}
\newcommand{\FASTsunflowREWCPTime}{24}
\newcommand{\FASTsunflowREWCPTimeCI}{1.5}
\newcommand{\FASTsunflowDCTime}{\rna}
\newcommand{\FASTsunflowDCTimeCI}{\rna}
\newcommand{\FASTsunflowDCTimeCIMIN}{\rna}
\newcommand{\FASTsunflowDCTimeCIMAX}{\rna}
\newcommand{\FASTsunflowFTODCTime}{22}
\newcommand{\FASTsunflowFTODCTimeCI}{1.1}
\newcommand{\FASTsunflowREDCTime}{24}
\newcommand{\FASTsunflowREDCTimeCI}{1.2}
\newcommand{\FASTsunflowCAPOTime}{\rna}
\newcommand{\FASTsunflowCAPOTimeCI}{\rna}
\newcommand{\FASTsunflowCAPOTimeCIMIN}{\rna}
\newcommand{\FASTsunflowCAPOTimeCIMAX}{\rna}
\newcommand{\FASTsunflowFTOCAPOTime}{23}
\newcommand{\FASTsunflowFTOCAPOTimeCI}{0.96}
\newcommand{\FASTsunflowRECAPOTime}{25}
\newcommand{\FASTsunflowRECAPOTimeCI}{1.1}
\newcommand{\FASTsunflowAGGCAPOTime}{\rna}
\newcommand{\FASTsunflowAGGCAPOTimeCI}{\rna}
\newcommand{\FASTsunflowAGGCAPOTimeCIMIN}{\rna}
\newcommand{\FASTsunflowAGGCAPOTimeCIMAX}{\rna}
\newcommand{\FASTsunflowStaticTime}{\rzero}
\newcommand{\FASTsunflowDynamicTime}{\rzero}
\newcommand{\FASTsunflowBaseMem}{610}
\newcommand{\FASTsunflowBaseMemCI}{4.6}
\newcommand{\FASTsunflowFTMem}{8.2}
\newcommand{\FASTsunflowFTMemCI}{0.062}
\newcommand{\FASTsunflowHBMem}{8.2}
\newcommand{\FASTsunflowHBMemCI}{0.090}
\newcommand{\FASTsunflowFTOHBMem}{8.2}
\newcommand{\FASTsunflowFTOHBMemCI}{0.064}
\newcommand{\FASTsunflowWCPMem}{\memna}
\newcommand{\FASTsunflowWCPMemCI}{\memna}
\newcommand{\FASTsunflowWCPMemCIMIN}{\memna}
\newcommand{\FASTsunflowWCPMemCIMAX}{\memna}
\newcommand{\FASTsunflowFTOWCPMem}{9.0}
\newcommand{\FASTsunflowFTOWCPMemCI}{0.083}
\newcommand{\FASTsunflowREWCPMem}{38}
\newcommand{\FASTsunflowREWCPMemCI}{0.28}
\newcommand{\FASTsunflowDCMem}{\memna}
\newcommand{\FASTsunflowDCMemCI}{\memna}
\newcommand{\FASTsunflowDCMemCIMIN}{\memna}
\newcommand{\FASTsunflowDCMemCIMAX}{\memna}
\newcommand{\FASTsunflowFTODCMem}{9.0}
\newcommand{\FASTsunflowFTODCMemCI}{0.065}
\newcommand{\FASTsunflowREDCMem}{38}
\newcommand{\FASTsunflowREDCMemCI}{0.27}
\newcommand{\FASTsunflowCAPOMem}{\memna}
\newcommand{\FASTsunflowCAPOMemCI}{\memna}
\newcommand{\FASTsunflowCAPOMemCIMIN}{\memna}
\newcommand{\FASTsunflowCAPOMemCIMAX}{\memna}
\newcommand{\FASTsunflowFTOCAPOMem}{9.0}
\newcommand{\FASTsunflowFTOCAPOMemCI}{0.063}
\newcommand{\FASTsunflowRECAPOMem}{38}
\newcommand{\FASTsunflowRECAPOMemCI}{0.24}
\newcommand{\FASTsunflowAGGCAPOMem}{\memna}
\newcommand{\FASTsunflowAGGCAPOMemCI}{\memna}
\newcommand{\FASTsunflowAGGCAPOMemCIMIN}{\memna}
\newcommand{\FASTsunflowAGGCAPOMemCIMAX}{\memna}
\newcommand{\FASTsunflowEventsCI}{0}
\newcommand{\FASTsunflowEventsCIMIN}{0}
\newcommand{\FASTsunflowEventsCIMAX}{0}
\newcommand{\FASTsunflowNoFPEventsCI}{0}
\newcommand{\FASTsunflowNoFPEventsCIMIN}{0}
\newcommand{\FASTsunflowNoFPEventsCIMAX}{0}
\newcommand{\FASTsunflowFT}{5}
\newcommand{\FASTsunflowFTCI}{0}
\newcommand{\FASTsunflowFTCIMIN}{5}
\newcommand{\FASTsunflowFTCIMAX}{5}
\newcommand{\FASTsunflowFTDynamic}{61}
\newcommand{\FASTsunflowFTDynamicCI}{4.4}
\newcommand{\FASTsunflowFTDynamicCIMIN}{57}
\newcommand{\FASTsunflowFTDynamicCIMAX}{65}
\newcommand{\FASTsunflowHB}{6}
\newcommand{\FASTsunflowHBCI}{0}
\newcommand{\FASTsunflowHBCIMIN}{6}
\newcommand{\FASTsunflowHBCIMAX}{6}
\newcommand{\FASTsunflowHBDynamic}{32}
\newcommand{\FASTsunflowHBDynamicCI}{2.3}
\newcommand{\FASTsunflowHBDynamicCIMIN}{30}
\newcommand{\FASTsunflowHBDynamicCIMAX}{34}
\newcommand{\FASTsunflowFTOHB}{6}
\newcommand{\FASTsunflowFTOHBCI}{0}
\newcommand{\FASTsunflowFTOHBCIMIN}{6}
\newcommand{\FASTsunflowFTOHBCIMAX}{6}
\newcommand{\FASTsunflowFTOHBDynamic}{28}
\newcommand{\FASTsunflowFTOHBDynamicCI}{1}
\newcommand{\FASTsunflowFTOHBDynamicCIMIN}{27}
\newcommand{\FASTsunflowFTOHBDynamicCIMAX}{29}
\newcommand{\FASTsunflowWCP}{\rna}
\newcommand{\FASTsunflowWCPCI}{\rna}
\newcommand{\FASTsunflowWCPCIMIN}{\rna}
\newcommand{\FASTsunflowWCPCIMAX}{\rna}
\newcommand{\FASTsunflowWCPDynamic}{\rna}
\newcommand{\FASTsunflowWCPDynamicCI}{\rna}
\newcommand{\FASTsunflowWCPDynamicCIMIN}{\rna}
\newcommand{\FASTsunflowWCPDynamicCIMAX}{\rna}
\newcommand{\FASTsunflowFTOWCP}{18}
\newcommand{\FASTsunflowFTOWCPCI}{0}
\newcommand{\FASTsunflowFTOWCPCIMIN}{18}
\newcommand{\FASTsunflowFTOWCPCIMAX}{18}
\newcommand{\FASTsunflowFTOWCPDynamic}{112}
\newcommand{\FASTsunflowFTOWCPDynamicCI}{4}
\newcommand{\FASTsunflowFTOWCPDynamicCIMIN}{108}
\newcommand{\FASTsunflowFTOWCPDynamicCIMAX}{116}
\newcommand{\FASTsunflowREWCP}{19}
\newcommand{\FASTsunflowREWCPCI}{0}
\newcommand{\FASTsunflowREWCPCIMIN}{19}
\newcommand{\FASTsunflowREWCPCIMAX}{19}
\newcommand{\FASTsunflowREWCPDynamic}{136}
\newcommand{\FASTsunflowREWCPDynamicCI}{1}
\newcommand{\FASTsunflowREWCPDynamicCIMIN}{135}
\newcommand{\FASTsunflowREWCPDynamicCIMAX}{137}
\newcommand{\FASTsunflowDC}{\rna}
\newcommand{\FASTsunflowDCCI}{\rna}
\newcommand{\FASTsunflowDCCIMIN}{\rna}
\newcommand{\FASTsunflowDCCIMAX}{\rna}
\newcommand{\FASTsunflowDCDynamic}{\rna}
\newcommand{\FASTsunflowDCDynamicCI}{\rna}
\newcommand{\FASTsunflowDCDynamicCIMIN}{\rna}
\newcommand{\FASTsunflowDCDynamicCIMAX}{\rna}
\newcommand{\FASTsunflowFTODC}{19}
\newcommand{\FASTsunflowFTODCCI}{0}
\newcommand{\FASTsunflowFTODCCIMIN}{19}
\newcommand{\FASTsunflowFTODCCIMAX}{19}
\newcommand{\FASTsunflowFTODCDynamic}{225}
\newcommand{\FASTsunflowFTODCDynamicCI}{14}
\newcommand{\FASTsunflowFTODCDynamicCIMIN}{211}
\newcommand{\FASTsunflowFTODCDynamicCIMAX}{239}
\newcommand{\FASTsunflowREDC}{19}
\newcommand{\FASTsunflowREDCCI}{0}
\newcommand{\FASTsunflowREDCCIMIN}{19}
\newcommand{\FASTsunflowREDCCIMAX}{19}
\newcommand{\FASTsunflowREDCDynamic}{236}
\newcommand{\FASTsunflowREDCDynamicCI}{4}
\newcommand{\FASTsunflowREDCDynamicCIMIN}{232}
\newcommand{\FASTsunflowREDCDynamicCIMAX}{240}
\newcommand{\FASTsunflowCAPO}{\rna}
\newcommand{\FASTsunflowCAPOCI}{\rna}
\newcommand{\FASTsunflowCAPOCIMIN}{\rna}
\newcommand{\FASTsunflowCAPOCIMAX}{\rna}
\newcommand{\FASTsunflowCAPODynamic}{\rna}
\newcommand{\FASTsunflowCAPODynamicCI}{\rna}
\newcommand{\FASTsunflowCAPODynamicCIMIN}{\rna}
\newcommand{\FASTsunflowCAPODynamicCIMAX}{\rna}
\newcommand{\FASTsunflowFTOCAPO}{19}
\newcommand{\FASTsunflowFTOCAPOCI}{0}
\newcommand{\FASTsunflowFTOCAPOCIMIN}{19}
\newcommand{\FASTsunflowFTOCAPOCIMAX}{19}
\newcommand{\FASTsunflowFTOCAPODynamic}{231}
\newcommand{\FASTsunflowFTOCAPODynamicCI}{6}
\newcommand{\FASTsunflowFTOCAPODynamicCIMIN}{225}
\newcommand{\FASTsunflowFTOCAPODynamicCIMAX}{237}
\newcommand{\FASTsunflowRECAPO}{19}
\newcommand{\FASTsunflowRECAPOCI}{0}
\newcommand{\FASTsunflowRECAPOCIMIN}{19}
\newcommand{\FASTsunflowRECAPOCIMAX}{19}
\newcommand{\FASTsunflowRECAPODynamic}{234}
\newcommand{\FASTsunflowRECAPODynamicCI}{2}
\newcommand{\FASTsunflowRECAPODynamicCIMIN}{232}
\newcommand{\FASTsunflowRECAPODynamicCIMAX}{236}
\newcommand{\FASTsunflowAGGCAPO}{\rna}
\newcommand{\FASTsunflowAGGCAPOCI}{\rna}
\newcommand{\FASTsunflowAGGCAPOCIMIN}{\rna}
\newcommand{\FASTsunflowAGGCAPOCIMAX}{\rna}
\newcommand{\FASTsunflowAGGCAPODynamic}{\rna}
\newcommand{\FASTsunflowAGGCAPODynamicCI}{\rna}
\newcommand{\FASTsunflowAGGCAPODynamicCIMIN}{\rna}
\newcommand{\FASTsunflowAGGCAPODynamicCIMAX}{\rna}
\newcommand{\FASTtomcatEvents}{0}
\newcommand{\FASTtomcatNoFPEvents}{0}
\newcommand{\FASTtomcatMaxLiveThreads}{35}
\newcommand{\FASTtomcatTotalThreads}{35}
\newcommand{\FASTtomcatBaseTime}{0.93}
\newcommand{\FASTtomcatBaseTimeCI}{7.7}
\newcommand{\FASTtomcatEmptyTime}{\rna}
\newcommand{\FASTtomcatEmptyTimeCI}{\rna}
\newcommand{\FASTtomcatEmptyTimeCIMIN}{\rna}
\newcommand{\FASTtomcatEmptyTimeCIMAX}{\rna}
\newcommand{\FASTtomcatFTTime}{40}
\newcommand{\FASTtomcatFTTimeCI}{2.5}
\newcommand{\FASTtomcatHBTime}{4.4}
\newcommand{\FASTtomcatHBTimeCI}{0.20}
\newcommand{\FASTtomcatFTOHBTime}{4.3}
\newcommand{\FASTtomcatFTOHBTimeCI}{0.24}
\newcommand{\FASTtomcatWCPTime}{\rna}
\newcommand{\FASTtomcatWCPTimeCI}{\rna}
\newcommand{\FASTtomcatWCPTimeCIMIN}{\rna}
\newcommand{\FASTtomcatWCPTimeCIMAX}{\rna}
\newcommand{\FASTtomcatFTOWCPTime}{6.4}
\newcommand{\FASTtomcatFTOWCPTimeCI}{0.31}
\newcommand{\FASTtomcatREWCPTime}{7.4}
\newcommand{\FASTtomcatREWCPTimeCI}{0.62}
\newcommand{\FASTtomcatDCTime}{\rna}
\newcommand{\FASTtomcatDCTimeCI}{\rna}
\newcommand{\FASTtomcatDCTimeCIMIN}{\rna}
\newcommand{\FASTtomcatDCTimeCIMAX}{\rna}
\newcommand{\FASTtomcatFTODCTime}{7.8}
\newcommand{\FASTtomcatFTODCTimeCI}{0.27}
\newcommand{\FASTtomcatREDCTime}{8.1}
\newcommand{\FASTtomcatREDCTimeCI}{0.37}
\newcommand{\FASTtomcatCAPOTime}{\rna}
\newcommand{\FASTtomcatCAPOTimeCI}{\rna}
\newcommand{\FASTtomcatCAPOTimeCIMIN}{\rna}
\newcommand{\FASTtomcatCAPOTimeCIMAX}{\rna}
\newcommand{\FASTtomcatFTOCAPOTime}{4.7}
\newcommand{\FASTtomcatFTOCAPOTimeCI}{0.29}
\newcommand{\FASTtomcatRECAPOTime}{5.9}
\newcommand{\FASTtomcatRECAPOTimeCI}{0.38}
\newcommand{\FASTtomcatAGGCAPOTime}{\rna}
\newcommand{\FASTtomcatAGGCAPOTimeCI}{\rna}
\newcommand{\FASTtomcatAGGCAPOTimeCIMIN}{\rna}
\newcommand{\FASTtomcatAGGCAPOTimeCIMAX}{\rna}
\newcommand{\FASTtomcatStaticTime}{\rzero}
\newcommand{\FASTtomcatDynamicTime}{\rzero}
\newcommand{\FASTtomcatBaseMem}{510}
\newcommand{\FASTtomcatBaseMemCI}{2.5}
\newcommand{\FASTtomcatFTMem}{76}
\newcommand{\FASTtomcatFTMemCI}{18}
\newcommand{\FASTtomcatHBMem}{2.7}
\newcommand{\FASTtomcatHBMemCI}{0.030}
\newcommand{\FASTtomcatFTOHBMem}{2.7}
\newcommand{\FASTtomcatFTOHBMemCI}{0.041}
\newcommand{\FASTtomcatWCPMem}{\memna}
\newcommand{\FASTtomcatWCPMemCI}{\memna}
\newcommand{\FASTtomcatWCPMemCIMIN}{\memna}
\newcommand{\FASTtomcatWCPMemCIMAX}{\memna}
\newcommand{\FASTtomcatFTOWCPMem}{6.1}
\newcommand{\FASTtomcatFTOWCPMemCI}{0.52}
\newcommand{\FASTtomcatREWCPMem}{7.8}
\newcommand{\FASTtomcatREWCPMemCI}{0.71}
\newcommand{\FASTtomcatDCMem}{\memna}
\newcommand{\FASTtomcatDCMemCI}{\memna}
\newcommand{\FASTtomcatDCMemCIMIN}{\memna}
\newcommand{\FASTtomcatDCMemCIMAX}{\memna}
\newcommand{\FASTtomcatFTODCMem}{8.2}
\newcommand{\FASTtomcatFTODCMemCI}{0.36}
\newcommand{\FASTtomcatREDCMem}{9.2}
\newcommand{\FASTtomcatREDCMemCI}{0.13}
\newcommand{\FASTtomcatCAPOMem}{\memna}
\newcommand{\FASTtomcatCAPOMemCI}{\memna}
\newcommand{\FASTtomcatCAPOMemCIMIN}{\memna}
\newcommand{\FASTtomcatCAPOMemCIMAX}{\memna}
\newcommand{\FASTtomcatFTOCAPOMem}{3.9}
\newcommand{\FASTtomcatFTOCAPOMemCI}{0.23}
\newcommand{\FASTtomcatRECAPOMem}{5.5}
\newcommand{\FASTtomcatRECAPOMemCI}{0.17}
\newcommand{\FASTtomcatAGGCAPOMem}{\memna}
\newcommand{\FASTtomcatAGGCAPOMemCI}{\memna}
\newcommand{\FASTtomcatAGGCAPOMemCIMIN}{\memna}
\newcommand{\FASTtomcatAGGCAPOMemCIMAX}{\memna}
\newcommand{\FASTtomcatEventsCI}{0}
\newcommand{\FASTtomcatEventsCIMIN}{0}
\newcommand{\FASTtomcatEventsCIMAX}{0}
\newcommand{\FASTtomcatNoFPEventsCI}{0}
\newcommand{\FASTtomcatNoFPEventsCIMIN}{0}
\newcommand{\FASTtomcatNoFPEventsCIMAX}{0}
\newcommand{\FASTtomcatFT}{415}
\newcommand{\FASTtomcatFTCI}{4.8}
\newcommand{\FASTtomcatFTCIMIN}{410}
\newcommand{\FASTtomcatFTCIMAX}{420}
\newcommand{\FASTtomcatFTDynamic}{4,732,109}
\newcommand{\FASTtomcatFTDynamicCI}{361,536}
\newcommand{\FASTtomcatFTDynamicCIMIN}{4,370,573}
\newcommand{\FASTtomcatFTDynamicCIMAX}{5,093,645}
\newcommand{\FASTtomcatHB}{632}
\newcommand{\FASTtomcatHBCI}{2.7}
\newcommand{\FASTtomcatHBCIMIN}{629}
\newcommand{\FASTtomcatHBCIMAX}{635}
\newcommand{\FASTtomcatHBDynamic}{1,377,438}
\newcommand{\FASTtomcatHBDynamicCI}{337,252}
\newcommand{\FASTtomcatHBDynamicCIMIN}{1,040,186}
\newcommand{\FASTtomcatHBDynamicCIMAX}{1,714,690}
\newcommand{\FASTtomcatFTOHB}{597}
\newcommand{\FASTtomcatFTOHBCI}{4}
\newcommand{\FASTtomcatFTOHBCIMIN}{593}
\newcommand{\FASTtomcatFTOHBCIMAX}{601}
\newcommand{\FASTtomcatFTOHBDynamic}{1,479,945}
\newcommand{\FASTtomcatFTOHBDynamicCI}{351,609}
\newcommand{\FASTtomcatFTOHBDynamicCIMIN}{1,128,336}
\newcommand{\FASTtomcatFTOHBDynamicCIMAX}{1,831,554}
\newcommand{\FASTtomcatWCP}{\rna}
\newcommand{\FASTtomcatWCPCI}{\rna}
\newcommand{\FASTtomcatWCPCIMIN}{\rna}
\newcommand{\FASTtomcatWCPCIMAX}{\rna}
\newcommand{\FASTtomcatWCPDynamic}{\rna}
\newcommand{\FASTtomcatWCPDynamicCI}{\rna}
\newcommand{\FASTtomcatWCPDynamicCIMIN}{\rna}
\newcommand{\FASTtomcatWCPDynamicCIMAX}{\rna}
\newcommand{\FASTtomcatFTOWCP}{602}
\newcommand{\FASTtomcatFTOWCPCI}{16}
\newcommand{\FASTtomcatFTOWCPCIMIN}{586}
\newcommand{\FASTtomcatFTOWCPCIMAX}{618}
\newcommand{\FASTtomcatFTOWCPDynamic}{1,138,992}
\newcommand{\FASTtomcatFTOWCPDynamicCI}{21,753}
\newcommand{\FASTtomcatFTOWCPDynamicCIMIN}{1,117,239}
\newcommand{\FASTtomcatFTOWCPDynamicCIMAX}{1,160,745}
\newcommand{\FASTtomcatREWCP}{600}
\newcommand{\FASTtomcatREWCPCI}{14}
\newcommand{\FASTtomcatREWCPCIMIN}{586}
\newcommand{\FASTtomcatREWCPCIMAX}{614}
\newcommand{\FASTtomcatREWCPDynamic}{1,261,229}
\newcommand{\FASTtomcatREWCPDynamicCI}{168,878}
\newcommand{\FASTtomcatREWCPDynamicCIMIN}{1,092,351}
\newcommand{\FASTtomcatREWCPDynamicCIMAX}{1,430,107}
\newcommand{\FASTtomcatDC}{\rna}
\newcommand{\FASTtomcatDCCI}{\rna}
\newcommand{\FASTtomcatDCCIMIN}{\rna}
\newcommand{\FASTtomcatDCCIMAX}{\rna}
\newcommand{\FASTtomcatDCDynamic}{\rna}
\newcommand{\FASTtomcatDCDynamicCI}{\rna}
\newcommand{\FASTtomcatDCDynamicCIMIN}{\rna}
\newcommand{\FASTtomcatDCDynamicCIMAX}{\rna}
\newcommand{\FASTtomcatFTODC}{588}
\newcommand{\FASTtomcatFTODCCI}{19}
\newcommand{\FASTtomcatFTODCCIMIN}{569}
\newcommand{\FASTtomcatFTODCCIMAX}{607}
\newcommand{\FASTtomcatFTODCDynamic}{1,062,329}
\newcommand{\FASTtomcatFTODCDynamicCI}{17,645}
\newcommand{\FASTtomcatFTODCDynamicCIMIN}{1,044,684}
\newcommand{\FASTtomcatFTODCDynamicCIMAX}{1,079,974}
\newcommand{\FASTtomcatREDC}{579}
\newcommand{\FASTtomcatREDCCI}{19}
\newcommand{\FASTtomcatREDCCIMIN}{560}
\newcommand{\FASTtomcatREDCCIMAX}{598}
\newcommand{\FASTtomcatREDCDynamic}{1,084,849}
\newcommand{\FASTtomcatREDCDynamicCI}{19,396}
\newcommand{\FASTtomcatREDCDynamicCIMIN}{1,065,453}
\newcommand{\FASTtomcatREDCDynamicCIMAX}{1,104,245}
\newcommand{\FASTtomcatCAPO}{\rna}
\newcommand{\FASTtomcatCAPOCI}{\rna}
\newcommand{\FASTtomcatCAPOCIMIN}{\rna}
\newcommand{\FASTtomcatCAPOCIMAX}{\rna}
\newcommand{\FASTtomcatCAPODynamic}{\rna}
\newcommand{\FASTtomcatCAPODynamicCI}{\rna}
\newcommand{\FASTtomcatCAPODynamicCIMIN}{\rna}
\newcommand{\FASTtomcatCAPODynamicCIMAX}{\rna}
\newcommand{\FASTtomcatFTOCAPO}{610}
\newcommand{\FASTtomcatFTOCAPOCI}{10}
\newcommand{\FASTtomcatFTOCAPOCIMIN}{600}
\newcommand{\FASTtomcatFTOCAPOCIMAX}{620}
\newcommand{\FASTtomcatFTOCAPODynamic}{1,041,181}
\newcommand{\FASTtomcatFTOCAPODynamicCI}{81,077}
\newcommand{\FASTtomcatFTOCAPODynamicCIMIN}{960,104}
\newcommand{\FASTtomcatFTOCAPODynamicCIMAX}{1,122,258}
\newcommand{\FASTtomcatRECAPO}{601}
\newcommand{\FASTtomcatRECAPOCI}{3}
\newcommand{\FASTtomcatRECAPOCIMIN}{598}
\newcommand{\FASTtomcatRECAPOCIMAX}{604}
\newcommand{\FASTtomcatRECAPODynamic}{1,269,137}
\newcommand{\FASTtomcatRECAPODynamicCI}{292,262}
\newcommand{\FASTtomcatRECAPODynamicCIMIN}{976,875}
\newcommand{\FASTtomcatRECAPODynamicCIMAX}{1,561,399}
\newcommand{\FASTtomcatAGGCAPO}{\rna}
\newcommand{\FASTtomcatAGGCAPOCI}{\rna}
\newcommand{\FASTtomcatAGGCAPOCIMIN}{\rna}
\newcommand{\FASTtomcatAGGCAPOCIMAX}{\rna}
\newcommand{\FASTtomcatAGGCAPODynamic}{\rna}
\newcommand{\FASTtomcatAGGCAPODynamicCI}{\rna}
\newcommand{\FASTtomcatAGGCAPODynamicCIMIN}{\rna}
\newcommand{\FASTtomcatAGGCAPODynamicCIMAX}{\rna}
\newcommand{\FASTxalanEvents}{0}
\newcommand{\FASTxalanNoFPEvents}{0}
\newcommand{\FASTxalanMaxLiveThreads}{9}
\newcommand{\FASTxalanTotalThreads}{9}
\newcommand{\FASTxalanBaseTime}{2.1}
\newcommand{\FASTxalanBaseTimeCI}{65}
\newcommand{\FASTxalanEmptyTime}{\rna}
\newcommand{\FASTxalanEmptyTimeCI}{\rna}
\newcommand{\FASTxalanEmptyTimeCIMIN}{\rna}
\newcommand{\FASTxalanEmptyTimeCIMAX}{\rna}
\newcommand{\FASTxalanFTTime}{4.7}
\newcommand{\FASTxalanFTTimeCI}{0.12}
\newcommand{\FASTxalanHBTime}{4.7}
\newcommand{\FASTxalanHBTimeCI}{0.20}
\newcommand{\FASTxalanFTOHBTime}{4.6}
\newcommand{\FASTxalanFTOHBTimeCI}{0.22}
\newcommand{\FASTxalanWCPTime}{\rna}
\newcommand{\FASTxalanWCPTimeCI}{\rna}
\newcommand{\FASTxalanWCPTimeCIMIN}{\rna}
\newcommand{\FASTxalanWCPTimeCIMAX}{\rna}
\newcommand{\FASTxalanFTOWCPTime}{29}
\newcommand{\FASTxalanFTOWCPTimeCI}{1.5}
\newcommand{\FASTxalanREWCPTime}{8.2}
\newcommand{\FASTxalanREWCPTimeCI}{0.34}
\newcommand{\FASTxalanDCTime}{\rna}
\newcommand{\FASTxalanDCTimeCI}{\rna}
\newcommand{\FASTxalanDCTimeCIMIN}{\rna}
\newcommand{\FASTxalanDCTimeCIMAX}{\rna}
\newcommand{\FASTxalanFTODCTime}{29}
\newcommand{\FASTxalanFTODCTimeCI}{1.7}
\newcommand{\FASTxalanREDCTime}{8.7}
\newcommand{\FASTxalanREDCTimeCI}{0.32}
\newcommand{\FASTxalanCAPOTime}{\rna}
\newcommand{\FASTxalanCAPOTimeCI}{\rna}
\newcommand{\FASTxalanCAPOTimeCIMIN}{\rna}
\newcommand{\FASTxalanCAPOTimeCIMAX}{\rna}
\newcommand{\FASTxalanFTOCAPOTime}{27}
\newcommand{\FASTxalanFTOCAPOTimeCI}{0.98}
\newcommand{\FASTxalanRECAPOTime}{5.9}
\newcommand{\FASTxalanRECAPOTimeCI}{0.29}
\newcommand{\FASTxalanAGGCAPOTime}{\rna}
\newcommand{\FASTxalanAGGCAPOTimeCI}{\rna}
\newcommand{\FASTxalanAGGCAPOTimeCIMIN}{\rna}
\newcommand{\FASTxalanAGGCAPOTimeCIMAX}{\rna}
\newcommand{\FASTxalanStaticTime}{\rzero}
\newcommand{\FASTxalanDynamicTime}{\rzero}
\newcommand{\FASTxalanBaseMem}{670}
\newcommand{\FASTxalanBaseMemCI}{7.6}
\newcommand{\FASTxalanFTMem}{6.6}
\newcommand{\FASTxalanFTMemCI}{0.081}
\newcommand{\FASTxalanHBMem}{6.7}
\newcommand{\FASTxalanHBMemCI}{0.083}
\newcommand{\FASTxalanFTOHBMem}{6.7}
\newcommand{\FASTxalanFTOHBMemCI}{0.080}
\newcommand{\FASTxalanWCPMem}{\memna}
\newcommand{\FASTxalanWCPMemCI}{\memna}
\newcommand{\FASTxalanWCPMemCIMIN}{\memna}
\newcommand{\FASTxalanWCPMemCIMAX}{\memna}
\newcommand{\FASTxalanFTOWCPMem}{31}
\newcommand{\FASTxalanFTOWCPMemCI}{1.2}
\newcommand{\FASTxalanREWCPMem}{18}
\newcommand{\FASTxalanREWCPMemCI}{0.87}
\newcommand{\FASTxalanDCMem}{\memna}
\newcommand{\FASTxalanDCMemCI}{\memna}
\newcommand{\FASTxalanDCMemCIMIN}{\memna}
\newcommand{\FASTxalanDCMemCIMAX}{\memna}
\newcommand{\FASTxalanFTODCMem}{30}
\newcommand{\FASTxalanFTODCMemCI}{0.89}
\newcommand{\FASTxalanREDCMem}{18}
\newcommand{\FASTxalanREDCMemCI}{0.49}
\newcommand{\FASTxalanCAPOMem}{\memna}
\newcommand{\FASTxalanCAPOMemCI}{\memna}
\newcommand{\FASTxalanCAPOMemCIMIN}{\memna}
\newcommand{\FASTxalanCAPOMemCIMAX}{\memna}
\newcommand{\FASTxalanFTOCAPOMem}{28}
\newcommand{\FASTxalanFTOCAPOMemCI}{0.86}
\newcommand{\FASTxalanRECAPOMem}{13}
\newcommand{\FASTxalanRECAPOMemCI}{0.52}
\newcommand{\FASTxalanAGGCAPOMem}{\memna}
\newcommand{\FASTxalanAGGCAPOMemCI}{\memna}
\newcommand{\FASTxalanAGGCAPOMemCIMIN}{\memna}
\newcommand{\FASTxalanAGGCAPOMemCIMAX}{\memna}
\newcommand{\FASTxalanEventsCI}{0}
\newcommand{\FASTxalanEventsCIMIN}{0}
\newcommand{\FASTxalanEventsCIMAX}{0}
\newcommand{\FASTxalanNoFPEventsCI}{0}
\newcommand{\FASTxalanNoFPEventsCIMIN}{0}
\newcommand{\FASTxalanNoFPEventsCIMAX}{0}
\newcommand{\FASTxalanFT}{8}
\newcommand{\FASTxalanFTCI}{0}
\newcommand{\FASTxalanFTCIMIN}{8}
\newcommand{\FASTxalanFTCIMAX}{8}
\newcommand{\FASTxalanFTDynamic}{1,484}
\newcommand{\FASTxalanFTDynamicCI}{0}
\newcommand{\FASTxalanFTDynamicCIMIN}{1,484}
\newcommand{\FASTxalanFTDynamicCIMAX}{1,484}
\newcommand{\FASTxalanHB}{8}
\newcommand{\FASTxalanHBCI}{0}
\newcommand{\FASTxalanHBCIMIN}{8}
\newcommand{\FASTxalanHBCIMAX}{8}
\newcommand{\FASTxalanHBDynamic}{1,402}
\newcommand{\FASTxalanHBDynamicCI}{0}
\newcommand{\FASTxalanHBDynamicCIMIN}{1,402}
\newcommand{\FASTxalanHBDynamicCIMAX}{1,402}
\newcommand{\FASTxalanFTOHB}{8}
\newcommand{\FASTxalanFTOHBCI}{0}
\newcommand{\FASTxalanFTOHBCIMIN}{8}
\newcommand{\FASTxalanFTOHBCIMAX}{8}
\newcommand{\FASTxalanFTOHBDynamic}{1,402}
\newcommand{\FASTxalanFTOHBDynamicCI}{0}
\newcommand{\FASTxalanFTOHBDynamicCIMIN}{1,402}
\newcommand{\FASTxalanFTOHBDynamicCIMAX}{1,402}
\newcommand{\FASTxalanWCP}{\rna}
\newcommand{\FASTxalanWCPCI}{\rna}
\newcommand{\FASTxalanWCPCIMIN}{\rna}
\newcommand{\FASTxalanWCPCIMAX}{\rna}
\newcommand{\FASTxalanWCPDynamic}{\rna}
\newcommand{\FASTxalanWCPDynamicCI}{\rna}
\newcommand{\FASTxalanWCPDynamicCIMIN}{\rna}
\newcommand{\FASTxalanWCPDynamicCIMAX}{\rna}
\newcommand{\FASTxalanFTOWCP}{42}
\newcommand{\FASTxalanFTOWCPCI}{0.9}
\newcommand{\FASTxalanFTOWCPCIMIN}{41}
\newcommand{\FASTxalanFTOWCPCIMAX}{43}
\newcommand{\FASTxalanFTOWCPDynamic}{3,119,436}
\newcommand{\FASTxalanFTOWCPDynamicCI}{3,338}
\newcommand{\FASTxalanFTOWCPDynamicCIMIN}{3,116,098}
\newcommand{\FASTxalanFTOWCPDynamicCIMAX}{3,122,774}
\newcommand{\FASTxalanREWCP}{49}
\newcommand{\FASTxalanREWCPCI}{0}
\newcommand{\FASTxalanREWCPCIMIN}{49}
\newcommand{\FASTxalanREWCPCIMAX}{49}
\newcommand{\FASTxalanREWCPDynamic}{3,337,291}
\newcommand{\FASTxalanREWCPDynamicCI}{2,587}
\newcommand{\FASTxalanREWCPDynamicCIMIN}{3,334,704}
\newcommand{\FASTxalanREWCPDynamicCIMAX}{3,339,878}
\newcommand{\FASTxalanDC}{\rna}
\newcommand{\FASTxalanDCCI}{\rna}
\newcommand{\FASTxalanDCCIMIN}{\rna}
\newcommand{\FASTxalanDCCIMAX}{\rna}
\newcommand{\FASTxalanDCDynamic}{\rna}
\newcommand{\FASTxalanDCDynamicCI}{\rna}
\newcommand{\FASTxalanDCDynamicCIMIN}{\rna}
\newcommand{\FASTxalanDCDynamicCIMAX}{\rna}
\newcommand{\FASTxalanFTODC}{52}
\newcommand{\FASTxalanFTODCCI}{0.4}
\newcommand{\FASTxalanFTODCCIMIN}{52}
\newcommand{\FASTxalanFTODCCIMAX}{52}
\newcommand{\FASTxalanFTODCDynamic}{3,442,567}
\newcommand{\FASTxalanFTODCDynamicCI}{6,707}
\newcommand{\FASTxalanFTODCDynamicCIMIN}{3,435,860}
\newcommand{\FASTxalanFTODCDynamicCIMAX}{3,449,274}
\newcommand{\FASTxalanREDC}{50}
\newcommand{\FASTxalanREDCCI}{0.4}
\newcommand{\FASTxalanREDCCIMIN}{50}
\newcommand{\FASTxalanREDCCIMAX}{50}
\newcommand{\FASTxalanREDCDynamic}{3,628,421}
\newcommand{\FASTxalanREDCDynamicCI}{2,345}
\newcommand{\FASTxalanREDCDynamicCIMIN}{3,626,076}
\newcommand{\FASTxalanREDCDynamicCIMAX}{3,630,766}
\newcommand{\FASTxalanCAPO}{\rna}
\newcommand{\FASTxalanCAPOCI}{\rna}
\newcommand{\FASTxalanCAPOCIMIN}{\rna}
\newcommand{\FASTxalanCAPOCIMAX}{\rna}
\newcommand{\FASTxalanCAPODynamic}{\rna}
\newcommand{\FASTxalanCAPODynamicCI}{\rna}
\newcommand{\FASTxalanCAPODynamicCIMIN}{\rna}
\newcommand{\FASTxalanCAPODynamicCIMAX}{\rna}
\newcommand{\FASTxalanFTOCAPO}{51}
\newcommand{\FASTxalanFTOCAPOCI}{0.8}
\newcommand{\FASTxalanFTOCAPOCIMIN}{50}
\newcommand{\FASTxalanFTOCAPOCIMAX}{52}
\newcommand{\FASTxalanFTOCAPODynamic}{3,409,871}
\newcommand{\FASTxalanFTOCAPODynamicCI}{8,701}
\newcommand{\FASTxalanFTOCAPODynamicCIMIN}{3,401,170}
\newcommand{\FASTxalanFTOCAPODynamicCIMAX}{3,418,572}
\newcommand{\FASTxalanRECAPO}{51}
\newcommand{\FASTxalanRECAPOCI}{1}
\newcommand{\FASTxalanRECAPOCIMIN}{50}
\newcommand{\FASTxalanRECAPOCIMAX}{52}
\newcommand{\FASTxalanRECAPODynamic}{3,620,883}
\newcommand{\FASTxalanRECAPODynamicCI}{1,884}
\newcommand{\FASTxalanRECAPODynamicCIMIN}{3,618,999}
\newcommand{\FASTxalanRECAPODynamicCIMAX}{3,622,767}
\newcommand{\FASTxalanAGGCAPO}{\rna}
\newcommand{\FASTxalanAGGCAPOCI}{\rna}
\newcommand{\FASTxalanAGGCAPOCIMIN}{\rna}
\newcommand{\FASTxalanAGGCAPOCIMAX}{\rna}
\newcommand{\FASTxalanAGGCAPODynamic}{\rna}
\newcommand{\FASTxalanAGGCAPODynamicCI}{\rna}
\newcommand{\FASTxalanAGGCAPODynamicCIMIN}{\rna}
\newcommand{\FASTxalanAGGCAPODynamicCIMAX}{\rna}
\newcommand{\FASTBaseTimeGeoMean}{1900}
\newcommand{\FASTEmptyTimeGeoMean}{\rna}
\newcommand{\FASTFTTimeGeoMean}{9.5}
\newcommand{\FASTHBTimeGeoMean}{7.1}
\newcommand{\FASTFTOHBTimeGeoMean}{7.0}
\newcommand{\FASTWCPTimeGeoMean}{\rna}
\newcommand{\FASTFTOWCPTimeGeoMean}{14}
\newcommand{\FASTREWCPTimeGeoMean}{9.4}
\newcommand{\FASTDCTimeGeoMean}{\rna}
\newcommand{\FASTFTODCTimeGeoMean}{15}
\newcommand{\FASTREDCTimeGeoMean}{9.6}
\newcommand{\FASTCAPOTimeGeoMean}{\rna}
\newcommand{\FASTFTOCAPOTimeGeoMean}{13}
\newcommand{\FASTRECAPOTimeGeoMean}{8.3}
\newcommand{\FASTAGGCAPOTimeGeoMean}{\rna}
\newcommand{\FASTBaseMemGeoMean}{500}
\newcommand{\FASTEmptyMemGeoMean}{0}
\newcommand{\FASTFTMemGeoMean}{7.9}
\newcommand{\FASTHBMemGeoMean}{4.9}
\newcommand{\FASTFTOHBMemGeoMean}{4.9}
\newcommand{\FASTWCPMemGeoMean}{\memna}
\newcommand{\FASTFTOWCPMemGeoMean}{13}
\newcommand{\FASTREWCPMemGeoMean}{11}
\newcommand{\FASTDCMemGeoMean}{\memna}
\newcommand{\FASTFTODCMemGeoMean}{13}
\newcommand{\FASTREDCMemGeoMean}{11}
\newcommand{\FASTCAPOMemGeoMean}{\memna}
\newcommand{\FASTFTOCAPOMemGeoMean}{11}
\newcommand{\FASTRECAPOMemGeoMean}{9.5}
\newcommand{\FASTAGGCAPOMemGeoMean}{\memna}
\newcommand{\FASTWCPDynamicTotal}{0}
\newcommand{\FASTFTODCTotal}{724}
\newcommand{\FASTCAPOTotal}{0}
\newcommand{\FASTWCPTotal}{0}
\newcommand{\FASTAGGCAPODynamicTotal}{0}
\newcommand{\FASTRECAPOTotal}{738}
\newcommand{\FASTFTTotal}{485}
\newcommand{\FASTHBDynamicTotal}{1,876,032}
\newcommand{\FASTREDCTotal}{715}
\newcommand{\FASTREWCPTotal}{729}
\newcommand{\FASTFTOWCPDynamicTotal}{4,744,481}
\newcommand{\FASTRECAPODynamicTotal}{5,358,784}
\newcommand{\FASTFTOWCPTotal}{719}
\newcommand{\FASTAGGCAPOTotal}{0}
\newcommand{\FASTCAPODynamicTotal}{0}
\newcommand{\FASTFTOCAPOTotal}{746}
\newcommand{\FASTFTODCDynamicTotal}{4,992,472}
\newcommand{\FASTFTDynamicTotal}{5,586,808}
\newcommand{\FASTREWCPDynamicTotal}{5,075,741}
\newcommand{\FASTREDCDynamicTotal}{5,195,017}
\newcommand{\FASTFTOCAPODynamicTotal}{4,940,578}
\newcommand{\FASTFTOHBDynamicTotal}{1,942,269}
\newcommand{\FASTHBTotal}{708}
\newcommand{\FASTFTOHBTotal}{673}
\newcommand{\FASTDCDynamicTotal}{0}
\newcommand{\FASTDCTotal}{0}

%% file: re-stats-table.tex
\section{\RE Run-Time Characteristics}
\label{appendix:smarttrack-stats}

Table~\ref{tab:characteristic:event-type} reports frequencies of each \FTO case for \RECAPO analysis,
averaged over the 10 trials. The
\col{Total} column counts the \emph{non-same-epoch} reads and writes, \ie,
all read and write events that do \emph{not} take \ReadSameEpoch, \SharedSameEpoch, or
\WriteSameEpoch cases.
Each value in the remaining columns represents, for a specific read or
write case, the percentage of the total non-same-epoch reads or writes,
respectively.

\iftoggle{twoColumnText}{
\begin{table*}[t]
\newcommand{\rzero}{0\xspace}
\newcommand{\roh}[1]{\ifthenelse{\equal{#1}{\rna}}{\rna}{#1$\;\!\times$}} 
\newcommand{\rna}{N/A}
\newcommand{\memna}{N/A} 
\newcommand{\st}[1]{(#1~s)} 
\newcommand{\dt}[1]{#1~s} 
\newcommand{\mem}[1]{\ifthenelse{\equal{#1}{\memna}}{\memna}{#1$\;\!Mb$}} 
\newcommand{\et}[1]{#1{\;}M} 
\newcommand{\enfp}[1]{(#1M)} 
\newcommand{\base}[1]{#1~s} 
\newcommand{\per}[1]{\ifthenelse{\equal{#1}{\cna}}{\cna}{#1\%}}
\newcommand{\cna}{---} 
\input{result-macros/PIP_fastTool_extraOpt2Stat}
\small
\centering
\renewcommand\bench[1]{\multirow{2}{*}{\code{#1}}}
\begin{tabular}{@{}l|l|r|HHr|r|r|r|r@{}ZZ}
&Event	&       & \mc{2}{Z}{Same} & \mc{2}{c|}{Owned} & \mc{3}{c@{}}{Unowned} & \mc{2}{Z}{Locks} \\
&type	& Total & Epoch & Shr Epoch & Excl & Shared & Excl & Share & Shared & Held & Not Held \\\hline
\bench{avrora}	&Read	& \et{\avroraRECAPONoFPReadTotal}	& \per{\avroraRECAPOReadSameEp}		& \per{\avroraRECAPOReadSharedSameEp}	& \per{\avroraRECAPOReadOwned} 		& \per{\avroraRECAPOReadSharedOwned} 	& \per{\avroraRECAPOReadExclusive} 		& \per{\avroraRECAPOReadShare} 	& \per{\avroraRECAPOReadShared} 	& \per{\avroraRECAPOReadInCS} 	& \per{\avroraRECAPOReadOutCS} \\
				&Write	& \et{\avroraRECAPONoFPWriteTotal}	& \per{\avroraRECAPOWriteSameEp}	& \rna									& \per{\avroraRECAPOWriteOwned}		& \rna									& \per{\avroraRECAPOWriteExclusive}		& \rna							& \per{\avroraRECAPOWriteShared}	& \per{\avroraRECAPOWriteInCS} 	& \per{\avroraRECAPOWriteOutCS} \\\hline
\bench{batik}	&Read	& \et{\batikRECAPONoFPReadTotal}	& \per{\batikRECAPOReadSameEp}		& \per{\batikRECAPOReadSharedSameEp}	& \per{\batikRECAPOReadOwned} 		& \per{\batikRECAPOReadSharedOwned} 	& \per{\batikRECAPOReadExclusive} 		& \per{\batikRECAPOReadShare} 	& \per{\batikRECAPOReadShared} 		& \per{\batikRECAPOReadInCS} 	& \per{\batikRECAPOReadOutCS} \\
				&Write	& \et{\batikRECAPONoFPWriteTotal}	& \per{\batikRECAPOWriteSameEp}		& \rna									& \per{\batikRECAPOWriteOwned}		& \rna									& \per{\batikRECAPOWriteExclusive}		& \rna							& \per{\batikRECAPOWriteShared}		& \per{\batikRECAPOWriteInCS} 	& \per{\batikRECAPOWriteOutCS} \\\hline
\bench{h2}		&Read	& \et{\htwoRECAPONoFPReadTotal}		& \per{\htwoRECAPOReadSameEp}		& \per{\htwoRECAPOReadSharedSameEp}		& \per{\htwoRECAPOReadOwned} 		& \per{\htwoRECAPOReadSharedOwned} 		& \per{\htwoRECAPOReadExclusive} 		& \per{\htwoRECAPOReadShare} 	& \per{\htwoRECAPOReadShared} 		& \per{\htwoRECAPOReadInCS} 	& \per{\htwoRECAPOReadOutCS} \\
				&Write	& \et{\htwoRECAPONoFPWriteTotal}	& \per{\htwoRECAPOWriteSameEp}		& \rna									& \per{\htwoRECAPOWriteOwned}		& \rna									& \per{\htwoRECAPOWriteExclusive}		& \rna							& \per{\htwoRECAPOWriteShared}		& \per{\htwoRECAPOWriteInCS} 	& \per{\htwoRECAPOWriteOutCS} \\\hline
\bench{jython}	&Read	& \et{\jythonRECAPONoFPReadTotal}	& \per{\jythonRECAPOReadSameEp}		& \per{\jythonRECAPOReadSharedSameEp}	& \per{\jythonRECAPOReadOwned} 		& \per{\jythonRECAPOReadSharedOwned} 	& \per{\jythonRECAPOReadExclusive} 		& \per{\jythonRECAPOReadShare} 	& \per{\jythonRECAPOReadShared} 	& \per{\jythonRECAPOReadInCS} 	& \per{\jythonRECAPOReadOutCS} \\
				&Write	& \et{\jythonRECAPONoFPWriteTotal}	& \per{\jythonRECAPOWriteSameEp}	& \rna									& \per{\jythonRECAPOWriteOwned}		& \rna									& \per{\jythonRECAPOWriteExclusive}		& \rna							& \per{\jythonRECAPOWriteShared}	& \per{\jythonRECAPOWriteInCS} 	& \per{\jythonRECAPOWriteOutCS} \\\hline
\bench{luindex}	&Read	& \et{\luindexRECAPONoFPReadTotal}	& \per{\luindexRECAPOReadSameEp}	& \per{\luindexRECAPOReadSharedSameEp}	& \per{\luindexRECAPOReadOwned} 	& \per{\luindexRECAPOReadSharedOwned} 	& \per{\luindexRECAPOReadExclusive} 	& \per{\luindexRECAPOReadShare} & \per{\luindexRECAPOReadShared} 	& \per{\luindexRECAPOReadInCS} 	& \per{\luindexRECAPOReadOutCS} \\
				&Write	& \et{\luindexRECAPONoFPWriteTotal}	& \per{\luindexRECAPOWriteSameEp}	& \rna									& \per{\luindexRECAPOWriteOwned}	& \rna									& \per{\luindexRECAPOWriteExclusive}	& \rna							& \per{\luindexRECAPOWriteShared}	& \per{\luindexRECAPOWriteInCS} & \per{\luindexRECAPOWriteOutCS} \\\hline
\bench{lusearch}&Read	& \et{\lusearchRECAPONoFPReadTotal}	& \per{\lusearchRECAPOReadSameEp}	& \per{\lusearchRECAPOReadSharedSameEp}	& \per{\lusearchRECAPOReadOwned} 	& \per{\lusearchRECAPOReadSharedOwned} 	& \per{\lusearchRECAPOReadExclusive} 	& \per{\lusearchRECAPOReadShare}& \per{\lusearchRECAPOReadShared} 	& \per{\lusearchRECAPOReadInCS} & \per{\lusearchRECAPOReadOutCS} \\
				&Write	& \et{\lusearchRECAPONoFPWriteTotal}& \per{\lusearchRECAPOWriteSameEp}	& \rna									& \per{\lusearchRECAPOWriteOwned}	& \rna									& \per{\lusearchRECAPOWriteExclusive}	& \rna							& \per{\lusearchRECAPOWriteShared}	& \per{\lusearchRECAPOWriteInCS}& \per{\lusearchRECAPOWriteOutCS} \\\hline
\bench{pmd}		&Read	& \et{\pmdRECAPONoFPReadTotal}		& \per{\pmdRECAPOReadSameEp}		& \per{\pmdRECAPOReadSharedSameEp}		& \per{\pmdRECAPOReadOwned} 		& \per{\pmdRECAPOReadSharedOwned} 		& \per{\pmdRECAPOReadExclusive} 		& \per{\pmdRECAPOReadShare} 	& \per{\pmdRECAPOReadShared} 		& \per{\pmdRECAPOReadInCS} 		& \per{\pmdRECAPOReadOutCS} \\
				&Write	& \et{\pmdRECAPONoFPWriteTotal}		& \per{\pmdRECAPOWriteSameEp}		& \rna									& \per{\pmdRECAPOWriteOwned}		& \rna									& \per{\pmdRECAPOWriteExclusive}		& \rna							& \per{\pmdRECAPOWriteShared}		& \per{\pmdRECAPOWriteInCS} 	& \per{\pmdRECAPOWriteOutCS} \\\hline
\bench{sunflow}	&Read	& \et{\sunflowRECAPONoFPReadTotal}	& \per{\sunflowRECAPOReadSameEp}	& \per{\sunflowRECAPOReadSharedSameEp}	& \per{\sunflowRECAPOReadOwned} 	& \per{\sunflowRECAPOReadSharedOwned} 	& \per{\sunflowRECAPOReadExclusive} 	& \per{\sunflowRECAPOReadShare} & \per{\sunflowRECAPOReadShared} 	& \per{\sunflowRECAPOReadInCS} 	& \per{\sunflowRECAPOReadOutCS} \\
				&Write	& \et{\sunflowRECAPONoFPWriteTotal}	& \per{\sunflowRECAPOWriteSameEp}	& \rna									& \per{\sunflowRECAPOWriteOwned}	& \rna									& \per{\sunflowRECAPOWriteExclusive}	& \rna							& \per{\sunflowRECAPOWriteShared}	& \per{\sunflowRECAPOWriteInCS} & \per{\sunflowRECAPOWriteOutCS} \\\hline
\bench{tomcat}	&Read	& \et{\tomcatRECAPONoFPReadTotal}	& \per{\tomcatRECAPOReadSameEp}		& \per{\tomcatRECAPOReadSharedSameEp}	& \per{\tomcatRECAPOReadOwned} 		& \per{\tomcatRECAPOReadSharedOwned} 	& \per{\tomcatRECAPOReadExclusive} 		& \per{\tomcatRECAPOReadShare} 	& \per{\tomcatRECAPOReadShared} 	& \per{\tomcatRECAPOReadInCS} 	& \per{\tomcatRECAPOReadOutCS} \\
				&Write	& \et{\tomcatRECAPONoFPWriteTotal}	& \per{\tomcatRECAPOWriteSameEp}	& \rna									& \per{\tomcatRECAPOWriteOwned}		& \rna									& \per{\tomcatRECAPOWriteExclusive}		& \rna							& \per{\tomcatRECAPOWriteShared}	& \per{\tomcatRECAPOWriteInCS} 	& \per{\tomcatRECAPOWriteOutCS} \\\hline
\bench{xalan}	&Read	& \et{\xalanRECAPONoFPReadTotal}	& \per{\xalanRECAPOReadSameEp}		& \per{\xalanRECAPOReadSharedSameEp}	& \per{\xalanRECAPOReadOwned} 		& \per{\xalanRECAPOReadSharedOwned} 	& \per{\xalanRECAPOReadExclusive} 		& \per{\xalanRECAPOReadShare} 	& \per{\xalanRECAPOReadShared} 		& \per{\xalanRECAPOReadInCS} 	& \per{\xalanRECAPOReadOutCS} \\
				&Write	& \et{\xalanRECAPONoFPWriteTotal}	& \per{\xalanRECAPOWriteSameEp}		& \rna									& \per{\xalanRECAPOWriteOwned}		& \rna									& \per{\xalanRECAPOWriteExclusive}		& \rna							& \per{\xalanRECAPOWriteShared}		& \per{\xalanRECAPOWriteInCS} 	& \per{\xalanRECAPOWriteOutCS} \\
\end{tabular}
\caption{Frequencies of \emph{non-same-epoch} reads and writes for \RE-\CAPO, for each evaluated program.}
\label{tab:characteristic:event-type}
\end{table*}

}{

\begin{table}[H]
\newcommand{\rzero}{0\xspace}
\newcommand{\roh}[1]{\ifthenelse{\equal{#1}{\rna}}{\rna}{#1$\;\!\times$}} 
\newcommand{\rna}{N/A}
\newcommand{\memna}{N/A} 
\newcommand{\st}[1]{(#1~s)} 
\newcommand{\dt}[1]{#1~s} 
\newcommand{\mem}[1]{\ifthenelse{\equal{#1}{\memna}}{\memna}{#1$\;\!Mb$}} 
\newcommand{\et}[1]{#1{\;}M} 
\newcommand{\enfp}[1]{(#1M)} 
\newcommand{\base}[1]{#1~s} 
\newcommand{\per}[1]{\ifthenelse{\equal{#1}{\cna}}{\cna}{#1\%}}
\newcommand{\cna}{---} 
\input{result-macros/PIP_fastTool_extraOpt2Stat}
\small
\centering
\renewcommand\bench[1]{\multirow{2}{*}{\code{#1}}}
\begin{tabular}{@{}l|l|r|HHr|r|r|r|r@{}ZZ}
&Event	&       & \mc{2}{Z}{Same} & \mc{2}{c|}{Owned} & \mc{3}{c@{}}{Unowned} & \mc{2}{Z}{Locks} \\
&type	& Total & Epoch & Shr Epoch & Excl & Shared & Excl & Share & Shared & Held & Not Held \\\hline
\bench{avrora}	&Read	& \et{\avroraRECAPONoFPReadTotal}	& \per{\avroraRECAPOReadSameEp}		& \per{\avroraRECAPOReadSharedSameEp}	& \per{\avroraRECAPOReadOwned} 		& \per{\avroraRECAPOReadSharedOwned} 	& \per{\avroraRECAPOReadExclusive} 		& \per{\avroraRECAPOReadShare} 	& \per{\avroraRECAPOReadShared} 	& \per{\avroraRECAPOReadInCS} 	& \per{\avroraRECAPOReadOutCS} \\
				&Write	& \et{\avroraRECAPONoFPWriteTotal}	& \per{\avroraRECAPOWriteSameEp}	& \rna									& \per{\avroraRECAPOWriteOwned}		& \rna									& \per{\avroraRECAPOWriteExclusive}		& \rna							& \per{\avroraRECAPOWriteShared}	& \per{\avroraRECAPOWriteInCS} 	& \per{\avroraRECAPOWriteOutCS} \\\hline
\bench{batik}	&Read	& \et{\batikRECAPONoFPReadTotal}	& \per{\batikRECAPOReadSameEp}		& \per{\batikRECAPOReadSharedSameEp}	& \per{\batikRECAPOReadOwned} 		& \per{\batikRECAPOReadSharedOwned} 	& \per{\batikRECAPOReadExclusive} 		& \per{\batikRECAPOReadShare} 	& \per{\batikRECAPOReadShared} 		& \per{\batikRECAPOReadInCS} 	& \per{\batikRECAPOReadOutCS} \\
				&Write	& \et{\batikRECAPONoFPWriteTotal}	& \per{\batikRECAPOWriteSameEp}		& \rna									& \per{\batikRECAPOWriteOwned}		& \rna									& \per{\batikRECAPOWriteExclusive}		& \rna							& \per{\batikRECAPOWriteShared}		& \per{\batikRECAPOWriteInCS} 	& \per{\batikRECAPOWriteOutCS} \\\hline
\bench{h2}		&Read	& \et{\htwoRECAPONoFPReadTotal}		& \per{\htwoRECAPOReadSameEp}		& \per{\htwoRECAPOReadSharedSameEp}		& \per{\htwoRECAPOReadOwned} 		& \per{\htwoRECAPOReadSharedOwned} 		& \per{\htwoRECAPOReadExclusive} 		& \per{\htwoRECAPOReadShare} 	& \per{\htwoRECAPOReadShared} 		& \per{\htwoRECAPOReadInCS} 	& \per{\htwoRECAPOReadOutCS} \\
				&Write	& \et{\htwoRECAPONoFPWriteTotal}	& \per{\htwoRECAPOWriteSameEp}		& \rna									& \per{\htwoRECAPOWriteOwned}		& \rna									& \per{\htwoRECAPOWriteExclusive}		& \rna							& \per{\htwoRECAPOWriteShared}		& \per{\htwoRECAPOWriteInCS} 	& \per{\htwoRECAPOWriteOutCS} \\\hline
\bench{jython}	&Read	& \et{\jythonRECAPONoFPReadTotal}	& \per{\jythonRECAPOReadSameEp}		& \per{\jythonRECAPOReadSharedSameEp}	& \per{\jythonRECAPOReadOwned} 		& \per{\jythonRECAPOReadSharedOwned} 	& \per{\jythonRECAPOReadExclusive} 		& \per{\jythonRECAPOReadShare} 	& \per{\jythonRECAPOReadShared} 	& \per{\jythonRECAPOReadInCS} 	& \per{\jythonRECAPOReadOutCS} \\
				&Write	& \et{\jythonRECAPONoFPWriteTotal}	& \per{\jythonRECAPOWriteSameEp}	& \rna									& \per{\jythonRECAPOWriteOwned}		& \rna									& \per{\jythonRECAPOWriteExclusive}		& \rna							& \per{\jythonRECAPOWriteShared}	& \per{\jythonRECAPOWriteInCS} 	& \per{\jythonRECAPOWriteOutCS} \\\hline
\bench{luindex}	&Read	& \et{\luindexRECAPONoFPReadTotal}	& \per{\luindexRECAPOReadSameEp}	& \per{\luindexRECAPOReadSharedSameEp}	& \per{\luindexRECAPOReadOwned} 	& \per{\luindexRECAPOReadSharedOwned} 	& \per{\luindexRECAPOReadExclusive} 	& \per{\luindexRECAPOReadShare} & \per{\luindexRECAPOReadShared} 	& \per{\luindexRECAPOReadInCS} 	& \per{\luindexRECAPOReadOutCS} \\
				&Write	& \et{\luindexRECAPONoFPWriteTotal}	& \per{\luindexRECAPOWriteSameEp}	& \rna									& \per{\luindexRECAPOWriteOwned}	& \rna									& \per{\luindexRECAPOWriteExclusive}	& \rna							& \per{\luindexRECAPOWriteShared}	& \per{\luindexRECAPOWriteInCS} & \per{\luindexRECAPOWriteOutCS} \\\hline
\bench{lusearch}&Read	& \et{\lusearchRECAPONoFPReadTotal}	& \per{\lusearchRECAPOReadSameEp}	& \per{\lusearchRECAPOReadSharedSameEp}	& \per{\lusearchRECAPOReadOwned} 	& \per{\lusearchRECAPOReadSharedOwned} 	& \per{\lusearchRECAPOReadExclusive} 	& \per{\lusearchRECAPOReadShare}& \per{\lusearchRECAPOReadShared} 	& \per{\lusearchRECAPOReadInCS} & \per{\lusearchRECAPOReadOutCS} \\
				&Write	& \et{\lusearchRECAPONoFPWriteTotal}& \per{\lusearchRECAPOWriteSameEp}	& \rna									& \per{\lusearchRECAPOWriteOwned}	& \rna									& \per{\lusearchRECAPOWriteExclusive}	& \rna							& \per{\lusearchRECAPOWriteShared}	& \per{\lusearchRECAPOWriteInCS}& \per{\lusearchRECAPOWriteOutCS} \\\hline
\bench{pmd}		&Read	& \et{\pmdRECAPONoFPReadTotal}		& \per{\pmdRECAPOReadSameEp}		& \per{\pmdRECAPOReadSharedSameEp}		& \per{\pmdRECAPOReadOwned} 		& \per{\pmdRECAPOReadSharedOwned} 		& \per{\pmdRECAPOReadExclusive} 		& \per{\pmdRECAPOReadShare} 	& \per{\pmdRECAPOReadShared} 		& \per{\pmdRECAPOReadInCS} 		& \per{\pmdRECAPOReadOutCS} \\
				&Write	& \et{\pmdRECAPONoFPWriteTotal}		& \per{\pmdRECAPOWriteSameEp}		& \rna									& \per{\pmdRECAPOWriteOwned}		& \rna									& \per{\pmdRECAPOWriteExclusive}		& \rna							& \per{\pmdRECAPOWriteShared}		& \per{\pmdRECAPOWriteInCS} 	& \per{\pmdRECAPOWriteOutCS} \\\hline
\bench{sunflow}	&Read	& \et{\sunflowRECAPONoFPReadTotal}	& \per{\sunflowRECAPOReadSameEp}	& \per{\sunflowRECAPOReadSharedSameEp}	& \per{\sunflowRECAPOReadOwned} 	& \per{\sunflowRECAPOReadSharedOwned} 	& \per{\sunflowRECAPOReadExclusive} 	& \per{\sunflowRECAPOReadShare} & \per{\sunflowRECAPOReadShared} 	& \per{\sunflowRECAPOReadInCS} 	& \per{\sunflowRECAPOReadOutCS} \\
				&Write	& \et{\sunflowRECAPONoFPWriteTotal}	& \per{\sunflowRECAPOWriteSameEp}	& \rna									& \per{\sunflowRECAPOWriteOwned}	& \rna									& \per{\sunflowRECAPOWriteExclusive}	& \rna							& \per{\sunflowRECAPOWriteShared}	& \per{\sunflowRECAPOWriteInCS} & \per{\sunflowRECAPOWriteOutCS} \\\hline
\bench{tomcat}	&Read	& \et{\tomcatRECAPONoFPReadTotal}	& \per{\tomcatRECAPOReadSameEp}		& \per{\tomcatRECAPOReadSharedSameEp}	& \per{\tomcatRECAPOReadOwned} 		& \per{\tomcatRECAPOReadSharedOwned} 	& \per{\tomcatRECAPOReadExclusive} 		& \per{\tomcatRECAPOReadShare} 	& \per{\tomcatRECAPOReadShared} 	& \per{\tomcatRECAPOReadInCS} 	& \per{\tomcatRECAPOReadOutCS} \\
				&Write	& \et{\tomcatRECAPONoFPWriteTotal}	& \per{\tomcatRECAPOWriteSameEp}	& \rna									& \per{\tomcatRECAPOWriteOwned}		& \rna									& \per{\tomcatRECAPOWriteExclusive}		& \rna							& \per{\tomcatRECAPOWriteShared}	& \per{\tomcatRECAPOWriteInCS} 	& \per{\tomcatRECAPOWriteOutCS} \\\hline
\bench{xalan}	&Read	& \et{\xalanRECAPONoFPReadTotal}	& \per{\xalanRECAPOReadSameEp}		& \per{\xalanRECAPOReadSharedSameEp}	& \per{\xalanRECAPOReadOwned} 		& \per{\xalanRECAPOReadSharedOwned} 	& \per{\xalanRECAPOReadExclusive} 		& \per{\xalanRECAPOReadShare} 	& \per{\xalanRECAPOReadShared} 		& \per{\xalanRECAPOReadInCS} 	& \per{\xalanRECAPOReadOutCS} \\
				&Write	& \et{\xalanRECAPONoFPWriteTotal}	& \per{\xalanRECAPOWriteSameEp}		& \rna									& \per{\xalanRECAPOWriteOwned}		& \rna									& \per{\xalanRECAPOWriteExclusive}		& \rna							& \per{\xalanRECAPOWriteShared}		& \per{\xalanRECAPOWriteInCS} 	& \per{\xalanRECAPOWriteOutCS} \\
\end{tabular}
\caption{Frequencies of \FTO cases for \emph{non-same-epoch} reads and writes, for \RE-\CAPO on each evaluated program.}
\label{tab:characteristic:event-type}
\end{table}
}